\documentclass[12pt]{amsbook}
\usepackage{amssymb}
\usepackage[all]{xy}

\usepackage[colorlinks=true,linkcolor=magenta,citecolor=magenta]{hyperref}
\makeindex

\newtheorem{theorem}{Theorem}[chapter]
\newtheorem{advertisement}[theorem]{Advertisement}
\newtheorem{answer}[theorem]{Answer}
\newtheorem{cat}[theorem]{Cat}
\newtheorem{cats}[theorem]{Cats}
\newtheorem{claim}[theorem]{Claim}
\newtheorem{comment}[theorem]{Comment}
\newtheorem{conclusion}[theorem]{Conclusion}

\newtheorem{definition}[theorem]{Definition}
\newtheorem{dillema}[theorem]{Dillema}
\newtheorem{example}[theorem]{Example}
\newtheorem{exercise}[theorem]{Exercise}
\newtheorem{fact}[theorem]{Fact}
\newtheorem{kitty}[theorem]{Kitty}
\newtheorem{lemma}[theorem]{Lemma}
\newtheorem{objective}[theorem]{Objective}
\newtheorem{plan}[theorem]{Plan}
\newtheorem{principle}[theorem]{Principle}
\newtheorem{problem}[theorem]{Problem}
\newtheorem{proposition}[theorem]{Proposition}
\newtheorem{question}[theorem]{Question}

\newtheorem{speculation}[theorem]{Speculation}
\newtheorem{strategy}[theorem]{Strategy}

\begin{document}

\title{Measure and integration}

\author{Teo Banica}
\address{Department of Mathematics, University of Cergy-Pontoise, F-95000 Cergy-Pontoise, France. {\tt teo.banica@gmail.com}}


\begin{abstract}
This is an introduction to measure theory, integration and function spaces, with all the needed preliminaries included, and with some applications included as well. We first discuss some basic motivations, coming from discrete probability, that we develop in detail, as a preliminary to general measure theory. Then we discuss measure theory, integration and function spaces, all developed in a standard way, and with emphasis on the explicit computation of various integrals. Finally, we come back to probability, discrete and continuous, with a more advanced discussion, of quantum flavor.
\end{abstract}

\maketitle

\chapter*{Preface}

You certainly know some calculus, in one and several variables, along with some basic algebra, geometry and probability. For most questions in science and engineering, that knowledge is enough, although learning new methods and tricks, from time to time, say from older colleagues that you will work with, will always improve your science.

\bigskip

This being said, why not learning some new methods and tricks right away, now that you are still studying. You would perhaps say, why bothering with this, I already had enough troubles with math, and learned enough of it, as to get started in life. But let me ask you a few questions, about your current knowledge, and judge yourself:

\bigskip

(1) The probability for a randomly picked $x\in\mathbb R$ to be $x=1$ is, and you surely know that, $P=0$. But what does this exactly mean, mathematically speaking?

\bigskip

(2) Along the same lines, but more concerning, everything that happens in the real life, happens with probability $P=0$. And this is something that we must clarify.

\bigskip

(3) Even worse, in certain physics disciplines like quantum mechanics, or statistical mechanics, $P$ is the only tool available, and $P=0$, your daily nightmare. 

\bigskip

(4) Another interesting thing that you surely know is that the reals are uncountable, $|\mathbb R|>|\mathbb N|=\infty$. So, how many possible $\infty$ beasts are there, in mathematics?

\bigskip

(5) What about arbitrarily small infinitesimals, how many $1/\infty$ beasts are there? And with the comment that these can be, technically speaking, very useful.

\bigskip

(6) Less philosophically now, you are surely a heavy user of $dxdy=dydx$, called Fubini theorem, for integrating 2-variable functions. But, does Fubini always work?

\bigskip

(7) Also, when using polar coordinates, the formula is $dxdy=Jdrdt$, with $J=r$. But no one ever fully proves this in class, so is this formula really correct, and why?

\bigskip

(8) You surely know about the Dirac $\delta_x$ function from electrostatics, or other physics. But, what is the precise mathematics behind this delta function?

\bigskip

(9) In fact, even mathematicians can pull out tricks based on $\delta_x$. For instance they have a theory where the basic step function is differentiable, with derivative $\delta_0$.

\bigskip

(10) Still talking classical analysis, you might know from Fourier transform theory that functions can be of type $L^1,L^2,L^\infty$ and so on, and clarifying all this matters.

\bigskip

(11) As even worse classical analysis, arbitrary functions can be non-integrable at all, at least in theory. And with these being non-negligible in quantum mechanics.

\bigskip

(12) In fact, talking worse things that can happen, these are wrong formulae, obtained via correct proofs. So, make sure that you know as much theory as possible.

\bigskip 

In the hope that I convinced you that learning some more analysis would be a great investment, and the present book will be here for that. The above questions all fall under the banner ``measure theory'', and we will discuss here measure theory, with an introduction to the subject, and answers, from partial to full, to most of the above questions.

\bigskip

This book was concieved as a complement to my calculus book \cite{ba1}, or to any other modern calculus book, which is typically weak on measure theory. Of course, if you are brave enough for learning full calculus directly, the old way, with measure theory included, go with some classical Cold War books, such as those of Rudin \cite{ru1}, \cite{ru2}.

\bigskip

Many thanks go to my students, who had to endure several beta versions of the material presented here, once for instance, long ago, with a measure theory class featuring a full proof for the Zorn Lemma. Thanks as well go to my cats, who seem to deal with everything, successfully, just by using fast binary logic. But we are not all that fast.

\bigskip

\

{\em Cergy, October 2026}

\smallskip

{\em Teo Banica}

\baselineskip=15.95pt
\tableofcontents
\baselineskip=14pt

\part{Discrete laws}

\ \vskip50mm

\begin{center}
{\em Low lie 

The fields of Athenry

Where once we watched

The small free birds fly}
\end{center}

\chapter{Real numbers}

\section*{1a. Real numbers}

Welcome to measure theory. What we will be doing in this book, namely measure theory and function spaces, is standard advanced undergraduate mathematics, sometimes erring on the graduate side, and can be best described as being a mixture of ``advanced calculus'' and ``rigorous calculus''. That is, you are supposed to know reasonably well basic calculus, in one and several variables, learned from one of the many calculus books available, and so many possible good choices here, and looking for a continuation of that learning, which is more advanced of course, but more rigorous too.

\bigskip

I can hear you thinking ``why bothering with more rigorous calculus, what I want to learn is more advanced calculus, that is what looks the most useful''. Good point, and if you are really into this, I can only recommend you, as a continuation of your calculus experience, to start reading some physics. You have here the famous course of Feynman \cite{fe1}, \cite{fe2}, \cite{fe3}, or the lovely books of Griffiths \cite{gr1}, \cite{gr2}, \cite{gr3}, or those of Weinberg \cite{we1}, \cite{we2}, \cite{we3}, if you already know some physics. So go with them, for advanced calculus, physicists know best, that is their job, to come up via advanced calculus to all sorts of amazing formulae, and you will certainly learn many interesting things from them. 

\bigskip

This being said, even if you are genuinely interested in physics, a bit of rigor, that you will not necessarily learn from Feynman, Griffiths or Weinberg, can help. To make my point, let me formulate the following question, which is something very natural:

\begin{question}
The probability for a randomly picked $x\in\mathbb R$ to be $x=1$ is obviously $P=0$. But what does this mean, mathematically speaking?
\end{question} 

And I hope that you agree with me, this is a very good question, for anyone having a bit of calculus background, and looking to learn more, regardless of the precise goals. For instance, don't expect to understand any sort of truly advanced physics, which often relies on probabilistic techniques, without having such things perfectly understood.

\bigskip

We will discuss this question several times in this book, first at the end of the present chapter, with a quick but rather advanced explanation, and then on several occasions afterwards, once our knowledge in measure theory and probability improves.

\bigskip

Getting started now, let us first talk about numbers. The commonly accepted fact, by most humans, is that the set $\mathbb N=\{0,1,2,3,\ldots\}$ was invented by God. According however to certain mathematicians, God only invented the empty set $\emptyset$, and then $\mathbb N$ naturally came afterwards, somehow by inventing itself, according to the following scheme:
$$|\emptyset|=0$$
$$|\{\emptyset\}|=1$$
$$|\{\emptyset,\{\emptyset\}\}|=2$$
$$|\{\emptyset,\{\emptyset\},\{\emptyset,\{\emptyset\}\}\}|=3$$
$$\vdots$$

Along the same lines, also alternatively, according to certain physicists, God only invented the Big Bang, and everything including $\mathbb N$ naturally came afterwards. But shall we trust all this modern science, better stick with good old traditional religion.

\bigskip

Once you have $\mathbb N$, solving $a+b=c$ naturally leads you to the set of all integers $\mathbb Z$. Then, once you have $\mathbb Z$, solving $ab=c$ naturally leads you to the set of all rationals $\mathbb Q$. So, this will be our starting point, the set of rationals $\mathbb Q$, defined as follows:

\index{rational number}

\begin{definition}
The rational numbers are the quotients $r=a/b$, with $a,b\in\mathbb Z$ and $b\neq0$, identified according to the usual rule for quotients, namely:
$$\frac{a}{b}=\frac{c}{d}\iff ad=bc$$
These quotients add, multiply and invert according to the following formulae:
$$\frac{a}{b}+\frac{c}{d}=\frac{ad+bc}{bd}\quad,\quad 
\frac{a}{b}\cdot\frac{c}{d}=\frac{ac}{bd}\quad,\quad
\left(\frac{a}{b}\right)^{-1}=\frac{b}{a}$$
We denote the set of rational numbers by $\mathbb Q$, standing for ``quotients''.
\end{definition}

As a comment here, in the hope that you love fractions. This is what makes the difference between science and humanities, in science we all love fractions, and are particularly excited whenever we must do any computations with them, useful or unuseful.

\bigskip

In more advanced mathematical terms, the above operations, namely sum, product and inversion, tell us that $\mathbb Q$ is a field, in the following sense: 

\index{field}

\begin{definition}
A field is a set $F$ with a sum operation $+$ and a product operation $\times$, subject to the following conditions:
\begin{enumerate}
\item $a+b=b+a$, $a+(b+c)=(a+b)+c$, there exists $0\in F$ such that $a+0=0$, and any $a\in F$ has an inverse $-a\in F$, satisfying $a+(-a)=0$.

\item $ab=ba$, $a(bc)=(ab)c$, there exists $1\in F$ such that $a1=a$, and any $a\neq0$ has a multiplicative inverse $a^{-1}\in F$, satisfying $aa^{-1}=1$.

\item The sum and product are compatible via $a(b+c)=ab+ac$.
\end{enumerate}
\end{definition}

Apparently, what we did so far, with our philosophical discussion regarding creation, $\diamondsuit\to\mathbb N\to\mathbb Z\to\mathbb Q$, was to construct the simplest possible field, $\mathbb Q$. However, this is not exactly true, because by a strange twist of fate, the numbers $0,1$, whose presence in a field is mandatory, $0,1\in F$, can form themselves a field, with addition as follows:
$$1+1=0$$

To be more precise, according to our field axioms, we certainly must have:
$$0+0=0\times0=0\times1=1\times0=0$$
$$0+1=1+0=1\times1=1$$

Thus, everything regarding the addition and multiplication of $0,1$ is uniquely determined, except for the value of $1+1$. And here, you would say that we should normally set $1+1=2$, with $2\neq0$ being a new field element, but the point is that $1+1=0$ is something natural too, this being the addition modulo 2. And, what we get is a field:
$$\mathbb F_2=\{0,1\}$$

Let us summarize this finding, along with a bit more, obtained by suitably replacing our 2, used for addition, with an arbitrary prime number $p$, as follows:

\index{finite field}

\begin{theorem}
The following happen:
\begin{enumerate}
\item $\mathbb Q$ is the simplest field having the property $1+\ldots+1\neq0$, in the sense that any field $F$ having this property must contain it, $\mathbb Q\subset F$.

\item The property $1+\ldots+1\neq0$ can hold or not, and if not, the smallest number of terms needed for having $1+\ldots+1=0$ is a certain prime number $p$.

\item $\mathbb F_p=\{0,1,\ldots,p-1\}$, with $p$ prime, is the simplest field having the property $1+\ldots+1=0$, with $p$ terms, in the sense that this implies $\mathbb F_p\subset F$.
\end{enumerate}
\end{theorem}

\begin{proof}
All this is basic number theory, the idea being as follows:

\medskip

(1) This is clear, because $1+\ldots+1\neq0$ tells us that we have an embedding $\mathbb N\subset F$, and then by taking inverses with respect to $+$ and $\times$ we obtain $\mathbb Q\subset F$. 

\medskip

(2) Again, this is clear, because assuming $1+\ldots+1=0$, with $p=ab$ terms, chosen minimal, we would have a formula as follows, which is a contradiction:
$$(\underbrace{1+\ldots+1}_{a\ terms})(\underbrace{1+\ldots+1}_{b\ terms})=0$$

(3) This follows a bit as in (1), with the copy $\mathbb F_p\subset F$ consisting by definition of the various sums of type $1+\ldots+1$, which must cycle modulo $p$, as shown by (2).
\end{proof}

Getting back now to our philosophical discussion regarding creation, what we have in Theorem 1.4 is not exactly good news, suggesting that, on purely mathematical grounds, there is a certain rivalry between $\mathbb Q$ and $\mathbb F_p$, as being the simplest field. So, which of them shall we study, as being created first? Not an easy question, and as answer, we have:

\begin{answer}
Ignoring what pure mathematics might say, and trusting instead physics and chemistry, we will choose to trust in $\mathbb Q$, as being the simplest field.
\end{answer}

In short, welcome to analysis, and with this being something quite natural, analysis being the main topic of the present book. Moving ahead now, many things can be done with $\mathbb Q$, but getting straight to the point, one thing that fails is solving $x^2=2$:

\index{square root}

\begin{theorem}
The field $\mathbb Q$ does not contain a square root of $2$:
$$\sqrt{2}\notin\mathbb Q$$
In fact, among integers, only the squares, $n=m^2$ with $m\in\mathbb N$, have square roots in $\mathbb Q$.
\end{theorem}

\begin{proof}
This is something very standard, the idea being as follows:

\medskip

(1) In what regards $\sqrt{2}$, assuming that $r=a/b$ with $a,b\in\mathbb N$ prime to each other satisfies $r^2=2$, we have $a^2=2b^2$, and so $a\in2\mathbb N$. But then by using again $a^2=2b^2$ we obtain $b\in2\mathbb N$ as well, which contradicts our assumption $(a,b)=1$.

\medskip

(2) Along the same lines, any prime number $p\in\mathbb N$ has the property $\sqrt{p}\notin\mathbb Q$, with the proof here being as the above one for $p=2$, by congruence and contradiction.

\medskip

(3) More generally, our claim is that any $n\in\mathbb N$ which is not a square has the property $\sqrt{n}\notin\mathbb Q$. Indeed, we can argue here that our number decomposes as $n=p_1^{a_1}\ldots p_k^{a_k}$, with $p_1,\ldots,p_k$ distinct primes, and our assumption that $n$ is not a square tells us that one of the exponents $a_1,\ldots,a_k\in\mathbb N$ must be odd. Moreover, by extracting all the obvious squares from $n$, we can in fact assume $a_1=\ldots=a_k=1$. But with this done, we can set $p=p_1$, and the congruence argument from (2) applies, and gives $\sqrt{n}\notin\mathbb Q$, as desired. 
\end{proof}

In short, in order to advance with our mathematics, we are in need to introduce the field of real numbers $\mathbb R$. You would probably say that this is very easy, via decimal writing, like everyone does, but before doing that, let me ask you a few questions:

\bigskip

(1) Honestly, do you really like the addition of real numbers, using the decimal form? Let us take, as example, the following computation:
$$\ \ \ \,12.456\,783\,872$$
$$+\ 27.536\,678\,377$$

This computation can surely be done, but, annoyingly, it must be done from right to left, instead of left to right, as we would prefer. I mean, personally I would be most interested in knowing first what happens at left, if the integer part is 39 or 40, but go do all the computation, starting from the right, in order to figure out that. In short, my feeling is that this addition algorithm, while certainly good, is a bit deceiving.

\bigskip

(2) What about multiplication. Here things become even more complicated, imagine for instance that Mars attacks, with $\delta$-rays, which are something unknown to us, and $100,000$ stronger than $\gamma$-rays, and which have paralyzed all our electronics, and that in order to protect Planet Earth, you must do the following multiplication by hand:
$$\ \ \ \,12.456\,783\,872$$
$$\times\ 27.536\,678\,377$$

This does not look very inviting, doesn't it. In short, as before with the addition, there is a bit of a bug with all this, the algorithm being too complicated.

\bigskip

(3) Getting now to the problem that we were interested in, namely extracting the square root of 2, here the algorithm is as follows, not very inviting either:
$$1.4^2<2<1.5^2\implies\sqrt{2}=1.4\ldots$$
$$1.41^2<2<1.42^2\implies\sqrt{2}=1.41\ldots$$
$$1.414^2<2<1.415^2\implies\sqrt{2}=1.414\ldots$$
$$1.4142^2<2<1.4143^2\implies\sqrt{2}=1.4142\ldots$$
$$\vdots$$

In short, quite concerning all this, and don't count on such things, mathematics of the decimal form, if Mars attacks. Let us record these findings as follows:

\begin{fact}
The real numbers $x\in\mathbb R$ can be certainly introduced via their decimal form, but with this, the field structure of $\mathbb R$ remains something a bit unclear.
\end{fact}

Well, it looks like we are a bit stuck. Fortunately, there is a clever solution to this, due to Dedekind. His definition for the real numbers is as follows:

\index{real number}
\index{Dedekind cut}

\begin{definition}
The real numbers $x\in\mathbb R$ are formal cuts in the set of rationals,
$$\mathbb Q=A_x\sqcup B_x$$
with such a cut being by definition subject to the following conditions:
$$p\in A_x\ ,\ q\in B_x\implies p<q\qquad,\qquad\inf B_x\notin B_x$$
These numbers add and multiply by adding and multiplying the corresponding cuts.
\end{definition}

This might look quite original, but believe me, there is some genius behind this definition. As a first observation, we have an inclusion $\mathbb Q\subset\mathbb R$, obtained by identifying each rational number $r\in\mathbb Q$ with the obvious cut that it produces, namely:
$$A_r=\left\{p\in\mathbb Q\Big|p\leq r\right\}
\quad,\quad
B_r=\left\{q\in\mathbb Q\Big|q>r\right\}$$

As a second observation, the addition and multiplication of real numbers, obtained by adding and multiplying the corresponding cuts, in the obvious way, is something very simple. To be more precise, in what regards the addition, the formula is as follows:
$$A_{x+y}=A_x+A_y$$ 

As for the multiplication, the formula here is similar, namely $A_{xy}=A_xA_y$, up to some issues with positives and negatives, which are quite easy to untangle, and with this being a good exercise. We can also talk about order between real numbers, as follows:
$$x\leq y\iff A_x\subset A_y$$

But let us perhaps leave more abstractions for later, and go back to more concrete things. As a first success of our theory, we can formulate the following theorem:

\begin{theorem}
The equation $x^2=2$ has two solutions over the real numbers, namely the positive solution, denoted $\sqrt{2}$, and its negative counterpart, which is $-\sqrt{2}$.
\end{theorem}

\begin{proof}
By using $x\to-x$, it is enough to prove that $x^2=2$ has exactly one positive solution $\sqrt{2}$. But this is clear, because $\sqrt{2}$ can only come from the following cut:
$$A_{\sqrt{2}}=\mathbb Q_-\bigsqcup\left\{p\in\mathbb Q_+\Big|p^2<2\right\}\quad,\quad B_{\sqrt{2}}=\left\{q\in\mathbb Q_+\Big|q^2>2\right\}$$

Thus, we are led to the conclusion in the statement.
\end{proof}

More generally, the same method works in order to extract the square root $\sqrt{r}$ of any number $r\in\mathbb Q_+$, or even of any number $r\in\mathbb R_+$, and we have the following result:

\begin{theorem}
The solutions of $ax^2+bx+c=0$ with $a,b,c\in\mathbb R$ are
$$x_{1,2}=\frac{-b\pm\sqrt{b^2-4ac}}{2a}$$
provided that $b^2-4ac\geq0$. In the case $b^2-4ac<0$, there are no solutions.
\end{theorem}

\begin{proof}
We can write our equation in the following way:
\begin{eqnarray*}
ax^2+bx+c=0
&\iff&x^2+\frac{b}{a}x+\frac{c}{a}=0\\
&\iff&\left(x+\frac{b}{2a}\right)^2=\frac{b^2-4ac}{4a^2}
\end{eqnarray*}

Thus, we are led to the conclusion in the statement.
\end{proof}

Summarizing, we have a nice abstract definition for the real numbers, that we can certainly do some mathematics with. As a first general result now, which is something very useful, and puts us back into real life, and science and engineering, we have:

\begin{theorem}
The real numbers $x\in\mathbb R$ can be written in decimal form,
$$x=\pm a_1\ldots a_n.b_1b_2b_3\ldots\ldots$$
with $a_i,b_i\in\{0,1,\ldots,9\}$, with the convention $\ldots b999\ldots=\ldots(b+1)000\ldots$
\end{theorem}

\begin{proof}
This is something non-trivial, even for the rationals $x\in\mathbb Q$ themselves, which require some work in order to be put in decimal form, the idea being as follows:

\medskip

(1) First of all, our precise claim is that any $x\in\mathbb R$ can be written in the form in the statement, with the integer $\pm a_1\ldots a_n$ and then each of the digits $b_1,b_2,b_3,\ldots$ providing the best approximation of $x$, at that stage of the approximation. 

\medskip

(2) Moreover, we have a second claim as well, namely that any expression of type $x=\pm a_1\ldots a_n.b_1b_2b_3\ldots\ldots$ corresponds to a real number $x\in\mathbb R$, and that with the convention $\ldots b999\ldots=\ldots(b+1)000\ldots\,$, the correspondence is bijective.

\medskip

(3) In order to prove now these two assertions, our first claim is that we can restrict the attention to the case $x\in[0,1)$, and with this meaning of course $0\leq x<1$, with respect to the order relation for the reals discussed in the above.

\medskip

(4) Getting started now, let $x\in\mathbb R$, coming from a cut $\mathbb Q=A_x\sqcup B_x$. Since the set $A_x\cap\mathbb Z$ consists of integers, and is bounded from above by any element $q\in B_x$ of your choice, this set has a maximal element, that we can denote $[x]$:
$$[x]=\max\left(A_x\cap\mathbb Z\right)$$

It follows from definitions that $[x]$ has the usual properties of the integer part, namely:
$$[x]\leq x<[x]+1$$

Thus we have $x=[x]+y$ with $[x]\in\mathbb Z$ and $y\in[0,1)$, and getting back now to what we want to prove, namely (1,2) above, it is clear that it is enough to prove these assertions for the remainder $y\in[0,1)$. Thus, we have proved (3), and we can assume $x\in[0,1)$.

\medskip

(5) So, assume $x\in[0,1)$. We are first looking for a best approximation from below of type $0.b_1$, with $b_1\in\{0,\ldots,9\}$, and it is clear that such an approximation exists, simply by comparing $x$ with the numbers $0.0,0.1,\ldots,0.9$. Thus, we have our first digit $b_1$, and then we can construct the second digit $b_2$ as well, by comparing $x$ with the numbers $0.b_10,0.b_11,\ldots,0.b_19$. And so on, which finishes the proof of our claim (1).

\medskip

(6) In order to prove now the remaining claim (2), let us restrict again the attention, as explained in (4), to the  case $x\in[0,1)$. First, it is clear that any expression of type $x=0.b_1b_2b_3\ldots$ defines a real number $x\in[0,1]$, simply by declaring that the corresponding cut $\mathbb Q=A_x\sqcup B_x$ comes from the following set, and its complement:
$$A_x=\bigcup_{n\geq1}\left\{p\in\mathbb Q\Big|p\leq 0.b_1\ldots b_n\right\}$$

(7) Thus, we have our correspondence between real numbers as cuts, and real numbers as decimal expressions, and we are left with the question of investigating the bijectivity of this correspondence. But here, the only bug that happens is that numbers of type $x=\ldots b999\ldots$, which produce reals $x\in\mathbb R$ via (6), do not come from reals $x\in\mathbb R$ via (5). So, in order to finish our proof, we must investigate such numbers.

\medskip

(8) So, consider an expression of type $\ldots b999\ldots$ Going back to the construction in (6), we are led to the conclusion that we have the following equality:
$$A_{b999\ldots}=B_{(b+1)000\ldots}$$

Thus, at the level of the real numbers defined as cuts, we have:
$$\ldots b999\ldots=\ldots(b+1)000\ldots$$

But this solves our problem, because by identifying $\ldots b999\ldots=\ldots(b+1)000\ldots$ the bijectivity issue of our correspondence is fixed, and we are done.
\end{proof}

We have as well the following result, coming as a useful complement to the above:

\index{probability 0}

\begin{theorem}
The following happen, in relation with $\mathbb Q\to\mathbb R$:
\begin{enumerate}
\item $\mathbb Q$ is countable, while $\mathbb R$ is not countable.

\item $r\in\mathbb R$ is rational precisely when its decimal writing is periodic.

\item The probability for a randomly picked $x\in\mathbb R$ to be rational is $0$.
\end{enumerate}
\end{theorem}

\begin{proof}
We have several things to be proved, the idea being as follows:

\medskip

(1) We can count the positive rationals, with some redundancies, as follows:
$$\xymatrix@R=10pt@C=10pt{
1/1\ar[r]&1/2\ar[dl]&1/3\ar[r]&1/4\ar[dl]&\ldots\\
2/1\ar[d]&2/2\ar[ur]&2/3\ar[dl]&2/4&\ldots\\
3/1\ar[ur]&3/2\ar[dl]&3/3&3/4&\ldots\\
4/1&4/2&4/3&4/4&\ldots}$$

Now after eliminating the redundancies, and then adding the negatives, which must be countable too, say via an alternating $+/-$ scheme, countability of $\mathbb Q$ proved. Regarding now the reals, assume by contradiction that $[0,1]$ is countable, with its elements listed as follows, with the convention that the writings of type $\ldots999\ldots$ are avoided:
$$x_1=0.a_1a_2a_3\ldots$$
$$x_2=0.b_1b_2b_3\ldots$$
$$x_3=0.c_1c_2c_3\ldots$$
$$\ldots$$

Now pick digits $\sigma_1\neq a_1$, $\sigma_2\neq b_2$, $\sigma_3\neq c_3$ and so on, again with a technical convention here, that these are different from 9, and define $x\in\mathbb R$ as follows:
$$x=0.\sigma_1\sigma_2\sigma_3\ldots$$

We have then $x\in[0,1]$, and since $x$ is obviously not on the above list, this is a contradiction. Thus $[0,1]$ is not countable, and it follows that $\mathbb R$ is not countable either.

\medskip

(2) Assuming that $r\in\mathbb R$ has periodic decimal writing, the following computation, based on $(10^p-1)(\sum_k10^{-kp})=1$, obtained by multiplying, gives $r\in\mathbb Q$:
\begin{eqnarray*}
r
&=&\pm a_1\ldots a_m.b_1\ldots b_nc_1\ldots c_pc_1\ldots c_p\ldots\\
&=&\pm\frac{1}{10^n}\left(a_1\ldots a_mb_1\ldots b_n+c_1\ldots c_p\left(\frac{1}{10^p}+\frac{1}{10^{2p}}+\ldots\right)\right)\\
&=&\pm\frac{1}{10^n}\left(a_1\ldots a_mb_1\ldots b_n+\frac{c_1\ldots c_p}{10^p-1}\right)
\end{eqnarray*}

Conversely, given a rational number $r=k/l$, we can find its decimal writing by performing the usual division algorithm, $k$ divided by $l$. But this algorithm will be surely periodic, after some time, so the decimal writing of $r$ is indeed periodic.

\medskip

(3) Finally, regarding the probability assertion, in order to avoid some troubles, we will prove that the probability for a real number $x\in[0,1]$ to be rational is 0. So, let us write the rational numbers $r\in[0,1]$ in the form of a sequence $r_1,r_2,r_3\ldots\,$ as follows:
$$\mathbb Q\cap[0,1]=\big\{r_1,r_2,r_3,\ldots\big\}$$

Let us also pick a number $c>0$. Since the probability of having $x=r_1$ is certainly smaller than $c/2$, then the probability of having $x=r_2$ is certainly smaller than $c/4$, and so on, the probability for $x$ to be rational satisfies the following inequality:
$$P\leq\frac{c}{2}+\frac{c}{4}+\frac{c}{8}+\ldots=c\left(\frac{1}{2}+\frac{1}{4}+\frac{1}{8}+\ldots\right)=c$$

Now since the number $c>0$ was arbitrary, we obtain $P=0$, as desired.
\end{proof}

Moving on, we have the following alternative definition for the real numbers:

\begin{theorem}
The field of real numbers $\mathbb R$ can be defined as well as the completion of $\mathbb Q$ with respect to the usual distance on the rationals, namely
$$d\left(\frac{a}{b}\,,\,\frac{c}{d}\right)=\left|\frac{a}{b}-\frac{c}{d}\right|$$
and with the operations on $\mathbb R$ coming from those on $\mathbb Q$, via Cauchy sequences.
\end{theorem} 

\begin{proof}
There are several things going on here, the idea being as follows:

\medskip

(1) Getting back to Definition 1.2, we know from there what the rational numbers are. As a continuation of the material there, we can talk about the distance between such rational numbers, given by the formula in the statement.

\medskip

(2) Very good, so let us get now into Cauchy sequences. We say that a sequence of rational numbers $\{r_n\}\subset\mathbb Q$ is Cauchy when the following condition is satisfied:
$$\forall\varepsilon>0,\exists N\in\mathbb N, m,n\geq N\implies d(r_m,r_n)<\varepsilon$$

Here of course $\varepsilon\in\mathbb Q$, because we do not know yet what the real numbers are.

\medskip

(3) With this notion in hand, the idea will be to define the reals $x\in\mathbb R$ as being the limits of the Cauchy sequences $\{r_n\}\subset\mathbb Q$. But since these limits are not known yet to exist to us, precisely because they are real, we must employ a trick. So, let us define instead the reals $x\in\mathbb R$ as being the Cauchy sequences $\{r_n\}\subset\mathbb Q$ themselves.

\medskip

(4) The question is now, will this work. As a first observation, we have an inclusion $\mathbb Q\subset\mathbb R$, obtained by identifying each rational $r\in\mathbb Q$ with the constant sequence $r_n=r$. Also, we can sum and multiply our real numbers in the obvious way, namely:
$$(r_n)+(p_n)=(r_n+p_n)\quad,\quad (r_n)(p_n)=(r_np_n)$$

We can also talk about the order between such reals, as follows:
$$(r_n)<(p_n)\iff\exists N,n\geq N\implies r_n<p_n$$

(5) However, there is still a bug with our theory, because there are obviously more Cauchy sequences of rationals, than real numbers. In order to fix this, let us go back to the end of step (3) above, and make the following convention:
$$(r_n)=(p_n)\iff d(r_n,p_n)\to0$$

(6) But, with this convention made, we have our theory. Indeed, the considerations in (4) apply again, with this change, and we obtain an ordered field $\mathbb R$, containing $\mathbb Q$, and the equivalence with the Dedekind cuts is something which is easy to establish.
\end{proof}

Finally, we have as well a third possible definition for the real numbers, namely:

\begin{theorem}
The real numbers $\mathbb R$ can be defined as well via the decimal form
$$x=\pm a_1\ldots a_n.a_{n+1}a_{n+2}a_{n+3}\ldots\ldots$$
with $a_i\in\{0,1,\ldots,9\}$, with the usual convention for such numbers, namely
$$\ldots a999\ldots=\ldots(a+1)000\ldots$$
and with the sum and multiplication coming by writing such numbers as
$$x=\pm\sum_{k\in\mathbb Z}a_k10^{-k}$$
and then summing and multiplying, in the obvious way.
\end{theorem} 

\begin{proof}
Indeed, forgetting about the precise decimal writing in the statement, we can define the real numbers $x\in\mathbb R$ as being formal sums as follows, with the sum being over integers $k\in\mathbb Z$ assumed to be greater than a certain integer, $k\geq k_0$:
$$x=\pm\sum_{k\in\mathbb Z}a_k10^{-k}$$

Now by truncating, we can see that what we have here are certain Cauchy sequences of rationals, and with a bit more work, we conclude that the $\mathbb R$ that we constructed here coincides with the $\mathbb R$ that we constructed in Theorem 1.13. Thus, we get the result.
\end{proof}

\section*{1b. Equations, roots}

Is the field $\mathbb R$ that we constructed good enough, for doing analysis? Certainly yes for most questions, as we both know, but not so quick with conclusions, because we have:

\begin{fact}
It would be desirable to have complex numbers, $z=a+ib$ with $a,b\in\mathbb R$ and with $i^2=-1$, as for the following dreams to become true:
\begin{enumerate}
\item The polynomial $X^2+1$, and other polynomials $P\in\mathbb R[X]$ too, to have roots.

\item In particular, characteristic polynomials $P(x)=\det(A-x)$ to have roots.

\item And so, to have more matrices $A\in M_N(\mathbb R)$ which are diagonalizable.

\item With these including beasts like $A=(df_i/dx_j)_{ij}$, with $f:\mathbb R^N\to\mathbb R^N$.

\item Or other analytic beasts, such as $A=(d^2f/dx_idx_j)_{ij}$, with $f:\mathbb R^N\to\mathbb R$.
\end{enumerate}
\end{fact}

In short, and you get the point I hope, multivariable calculus begs for complex numbers. So, let us introduce the complex numbers $\mathbb C$. Many interesting things can be said here, and you certainly know all of them, with their summary being as follows:

\index{complex number}
\index{root of unity}

\begin{theorem}
The complex numbers, $z=a+ib$ with $a,b\in\mathbb R$ and with $i$ being a formal number satisying $i^2=-1$, form a field $\mathbb C$. Moreover:
\begin{enumerate}
\item We have a field embedding $\mathbb R\subset\mathbb C$, given by $a\to a+0\cdot i$.

\item Additively, we have $\mathbb C\simeq\mathbb R^2$, with $z=a+ib$ corresponding to $(a,b)$.

\item The length of vectors $r=|z|$, with $z=a+ib$, is given by $r=\sqrt{a^2+b^2}$.

\item With $z=r(\cos t+i\sin t)$, the products $z=z'z''$ are given by $r=r'r''$, $t=t'+t''$.

\item We have $e^{it}=\cos t+i\sin t$, so we can write $z=re^{it}$.

\item There are $N$ solutions to the equation $z^N=1$, called $N$-th roots of unity.

\item Any degree $2$ equation with complex coefficients has both roots in $\mathbb C$.
\end{enumerate}
\end{theorem}

\begin{proof}
We have a field, with $z^{-1}=(a-ib)/(a^2+b^2)$, and regarding the rest:

\medskip

(1) This is clear.

\medskip

(2) Again, this is clear.

\medskip

(3) Again, this is clear. Observe also that we have $r^2=z\bar{z}$, with $\bar{z}=a-ib$.

\medskip

(4) We have indeed the following computation, which gives the result:
\begin{eqnarray*}
&&(\cos s+i\sin s)(\cos t+i\sin t)\\
&=&(\cos s\cos t-\sin s\sin t)+i(\sin s\cos t+\cos s\sin t)\\
&=&\cos(s+t)+i\sin(s+t)
\end{eqnarray*}

(5) In order to prove $e^{it}=\cos t+i\sin t$, consider the following function $f:\mathbb R\to\mathbb C$:
$$f(t)=\frac{\cos t+i\sin t}{e^{it}}$$

By using $\sin'=\cos$ and $\cos'=-\sin$, we have the following computation:
\begin{eqnarray*}
f'(t)
&=&(e^{-it}(\cos t+i\sin t))'\\
&=&-ie^{-it}(\cos t+i\sin t)+e^{-it}(-\sin t+i\cos t)\\
&=&e^{-it}(-i\cos t+\sin t)+e^{-it}(-\sin t+i\cos t)\\
&=&0
\end{eqnarray*}

We conclude that $f:\mathbb R\to\mathbb C$ is constant, equal to $f(0)=1$, as desired.

\medskip

(6) This is clear from (5), with $z=w^k$, with $w=e^{2\pi i/N}$ and $k=0,1,\ldots,N-1$.

\medskip

(7) This follows in the usual way, with $\sqrt{re^{it}}=\pm\sqrt{r}e^{it/2}$ at the end, using (5). 
\end{proof}

Quite remarkably, we have in fact the following result, generalizing what we know in degree 2, and telling us that with $\mathbb C$ we are safe, no need to look for more:

\begin{theorem}
Any polynomial $P\in\mathbb C[X]$ decomposes as
$$P=c(X-a_1)\ldots (X-a_N)$$
with $c\in\mathbb C$ and with $a_1,\ldots,a_N\in\mathbb C$.
\end{theorem}

\begin{proof}
The problem is that of proving that our polynomial has at least one root, because afterwards we can proceed by recurrence. We prove this by contradiction. So, assume that $P$ has no roots, and pick a number $z\in\mathbb C$ where $|P|$ attains its minimum:
$$|P(z)|=\min_{x\in\mathbb C}|P(x)|>0$$ 

Since $Q(t)=P(z+t)-P(z)$ is a polynomial which vanishes at $t=0$, this polynomial must be of the form $ct^k$ + higher terms, with $c\neq0$, and with $k\geq1$ being an integer. We obtain from this that, with $t\in\mathbb C$ small, we have the following estimate:
$$P(z+t)\simeq P(z)+ct^k$$

Now let us write $t=rw$, with $r>0$ small, and with $|w|=1$. Our estimate becomes:
$$P(z+rw)\simeq P(z)+cr^kw^k$$

Now recall that we have assumed $P(z)\neq0$. We can therefore choose $w\in\mathbb T$ such that $cw^k$ points in the opposite direction to that of $P(z)$, and we obtain in this way:
$$|P(z+rw)|
\simeq|P(z)+cr^kw^k|
=|P(z)|(1-|c|r^k)$$

Now by choosing $r>0$ small enough, as for the error in the first estimate to be small, and overcame by the negative quantity $-|c|r^k$, we obtain from this:
$$|P(z+rw)|<|P(z)|$$

But this contradicts our definition of $z\in\mathbb C$, as a point where $|P|$ attains its minimum. Thus $P$ has a root, and by recurrence it has $N$ roots, as stated.
\end{proof}

As a concrete question, let us try to solve a degree 3 equation. By using linear transformations we can assume that our equation is of the form $x^3+3px+2q=0$, and regarding such equations, we have the following famous result, due to Cardano:

\index{degree 3 equation}
\index{Cardano formula}
\index{cubic root}
\index{discriminant}

\begin{theorem}
For a normalized degree $3$ equation, namely 
$$x^3+3px+2q=0$$
the discriminant is $\Delta=-108(p^3+q^2)$. Assuming $p,q\in\mathbb R$ and $\Delta<0$, the numbers
$$z=w\sqrt[3]{-q+\sqrt{p^3+q^2}}+w^2\sqrt[3]{-q-\sqrt{p^3+q^2}}$$
with $w=1,e^{2\pi i/3},e^{4\pi i/3}$ are the solutions of our equation.
\end{theorem}

\begin{proof}
The formula of $\Delta$ comes the theory of the discriminant, and if you know that, good, and if you don't, no worries, because we will only need $p^3+q^2>0$, in what follows. With $z$ as in the statement, by using $(a+b)^3=a^3+b^3+3ab(a+b)$, we have:
\begin{eqnarray*}
z^3
&=&\left(w\sqrt[3]{-q+\sqrt{p^3+q^2}}+w^2\sqrt[3]{-q-\sqrt{p^3+q^2}}\right)^3\\
&=&-2q+3\sqrt[3]{-q+\sqrt{p^3+q^2}}\cdot\sqrt[3]{-q-\sqrt{p^3+q^2}}\cdot z\\
&=&-2q+3\sqrt[3]{q^2-p^3-q^2}\cdot z\\
&=&-2q-3pz
\end{eqnarray*}

Thus, we are led to the conclusion in the statement.
\end{proof}

In degree 4 now, we have the following result, which is famous as well:

\index{degree 4 equation}
\index{Cardano formula}

\begin{theorem}
The roots of a normalized degree $4$ equation, written as
$$x^4+6px^2+4qx+3r=0$$ 
are as follows, with $y$ satisfying the equation $(y^2-3r)(y-3p)=2q^2$,
$$x_1=\frac{1}{\sqrt{2}}\left(-\sqrt{y-3p}+\sqrt{-y-3p+\frac{4q}{\sqrt{2y-6p}}}\right)$$
$$x_2=\frac{1}{\sqrt{2}}\left(-\sqrt{y-3p}-\sqrt{-y-3p+\frac{4q}{\sqrt{2y-6p}}}\right)$$
$$x_3=\frac{1}{\sqrt{2}}\left(\sqrt{y-3p}+\sqrt{-y-3p-\frac{4q}{\sqrt{2y-6p}}}\right)$$
$$x_4=\frac{1}{\sqrt{2}}\left(\sqrt{y-3p}-\sqrt{-y-3p-\frac{4q}{\sqrt{2y-6p}}}\right)$$
and with $y$ being computable via the Cardano formula.
\end{theorem}

\begin{proof}
This is something quite tricky, the idea being as follows:

\medskip

(1) To start with, let us write our equation in the following form:
$$x^4=-6px^2-4qx-3r$$

Now assume that we have a number $y$ satisfying the following equation:
$$(y^2-3r)(y-3p)=2q^2$$

With this magic number $y$ in hand, our equation takes the following form:
\begin{eqnarray*}
(x^2+y)^2
&=&x^4+2x^2y+y^2\\
&=&-6px^2-4qx-3r+2x^2y+y^2\\
&=&(2y-6p)x^2-4qx+y^2-3r\\
&=&(2y-6p)x^2-4qx+\frac{2q^2}{y-3p}\\
&=&\left(\sqrt{2y-6p}\cdot x-\frac{2q}{\sqrt{2y-6p}}\right)^2
\end{eqnarray*}

(2) Which looks very good, leading us to the following degree 2 equations:
$$x^2+y+\sqrt{2y-6p}\cdot x-\frac{2q}{\sqrt{2y-6p}}=0$$
$$x^2+y-\sqrt{2y-6p}\cdot x+\frac{2q}{\sqrt{2y-6p}}=0$$

(3) Regarding the first equation, the solutions there are as follows:
$$x_1=\frac{1}{2}\left(-\sqrt{2y-6p}+\sqrt{-2y-6p+\frac{8q}{\sqrt{2y-6p}}}\right)$$
$$x_2=\frac{1}{2}\left(-\sqrt{2y-6p}-\sqrt{-2y-6p+\frac{8q}{\sqrt{2y-6p}}}\right)$$

As for the second equation, the solutions there are as follows:
$$x_3=\frac{1}{2}\left(\sqrt{2y-6p}+\sqrt{-2y-6p-\frac{8q}{\sqrt{2y-6p}}}\right)$$
$$x_4=\frac{1}{2}\left(\sqrt{2y-6p}-\sqrt{-2y-6p-\frac{8q}{\sqrt{2y-6p}}}\right)$$

(4) Now by cutting a $\sqrt{2}$ factor from everything, this gives the formulae in the statement. As for the last claim, regarding the nature of $y$, this comes from Cardano.
\end{proof}

We still have to compute the number $y$ appearing in the above via Cardano, and the result here, adding to what we already have in Theorem 1.19, is as follows:

\begin{theorem}[continuation]
The value of $y$ in the previous theorem is
$$y=t+p+\frac{a}{t}$$
where the number $t$ is given by the formula
$$t=\sqrt[3]{b+\sqrt{b^2-a^3}}$$
with $a=p^2+r$ and $b=2p^2-3pr+q^2$.
\end{theorem}

\begin{proof}
The legend has it that this is what comes from Cardano, but depressing and normalizing and solving $(y^2-3r)(y-3p)=2q^2$ makes it for too many operations, so the most pragmatic way is to simply check this equation. With $y$ as above, we have:
\begin{eqnarray*}
y^2-3r
&=&t^2+2pt+(p^2+2a)+\frac{2pa}{t}+\frac{a^2}{t^2}-3r\\
&=&t^2+2pt+(3p^2-r)+\frac{2pa}{t}+\frac{a^2}{t^2}
\end{eqnarray*}

With this in hand, we have the following computation:
\begin{eqnarray*}
(y^2-3r)(y-3p)
&=&\left(t^2+2pt+(3p^2-r)+\frac{2pa}{t}+\frac{a^2}{t^2}\right)\left(t-2p+\frac{a}{t}\right)\\
&=&t^3+(a-4p^2+3p^2-r)t+(2pa-6p^3+2pr+2pa)\\
&&+(3p^2a-ra-4p^2a+a^2)\frac{1}{t}+\frac{a^3}{t^3}\\
&=&t^3+(a-p^2-r)t+2p(2a-3p^2+r)+a(a-p^2-r)\frac{1}{t}+\frac{a^3}{t^3}\\
&=&t^3+2p(-p^2+3r)+\frac{a^3}{t^3}
\end{eqnarray*}

Now by using the formula of $t$ in the statement, this gives:
\begin{eqnarray*}
(y^2-3r)(y-3p)
&=&b+\sqrt{b^2-a^3}-4p^2+6pr+\frac{a^3}{b+\sqrt{b^2-a^3}}\\
&=&b+\sqrt{b^2-a^3}-4p^2+6pr+b-\sqrt{b^2-a^3}\\
&=&2b-4p^2+6pr\\
&=&2(2p^2-3pr+q^2)-4p^2+6pr\\
&=&2q^2
\end{eqnarray*}

Thus, we are led to the conclusion in the statement.
\end{proof}

In degree 5 and more, things become fairly complicated, and we have:

\index{degree 5 polynomial}
\index{roots}

\begin{theorem}
There is no general formula for the roots of polynomials of degree $N=5$ and higher, with the reason for this, coming from Galois theory, being that the group $S_5$ is not solvable. The simplest numeric example is $P=X^5-X-1$.
\end{theorem}

\begin{proof}
This is something quite tricky, the idea being as follows:

\medskip

(1) Given a field $F$, assume that the roots of $P\in F[X]$ can be computed by using iterated roots, a bit as for the degree 2 equation, or the degree 3 and 4 equations. Then, algebrically speaking, this gives rise to a tower of fields as folows, with $F_0=F$, and each $F_{i+1}$ being obtained from $F_i$ by adding a root, $F_{i+1}=F_i(x_i)$, with $x_i^{n_i}\in F_i$:
$$F_0\subset F_1\subset\ldots\subset F_k$$

(2) In order for Galois theory to apply to this situation, we must make all the extensions normal, which amounts in replacing each $F_{i+1}=F_i(x_i)$ by its extension $K_i(x_i)$, with $K_i$ extending $F_i$ by adding a $n_i$-th root of unity. Thus, with this replacement, we can assume that the tower in (1) in normal, meaning that all Galois groups are cyclic.

\medskip

(3) Now by Galois theory, at the level of the corresponding Galois groups we obtain a tower of groups as follows as follows, which is a resolution of the last group $G_k$, the Galois group of $P$, in the sense of group theory, in the sense that all quotients are cyclic:
$$G_1\subset G_2\subset\ldots\subset G_k$$

As a conclusion, Galois theory tells us that if the roots of a polynomial $P\in F[X]$ can be computed by using iterated roots, then its Galois group $G=G_k$ must be solvable.

\medskip

(4) In the generic case, the conclusion is that Galois theory tells us that, in order for all polynomials of degree 5 to be solvable, via square roots, the group $S_5$, which appears there as Galois group, must be solvable, in the sense of group theory. But this is wrong, because the alternating subgroup $A_5\subset S_5$ is simple, and therefore not solvable.

\medskip

(5) Finally, regarding the polynomial $P=X^5-X-1$, some elementary computations here, based on arithmetic over $\mathbb F_2,\mathbb F_3$, and involving various cycles of length $2,3,5$, show that its Galois group is $S_5$. Thus, we have our counterexample.
\end{proof}

There is a lot of further interesting theory that can be developed here, following Galois and others. For more on all this, we recommend any number theory book.

\section*{1c. Random numbers}

All that we have been talking about so far, while certainly beautiful and potentially useful, was a bit abstract. So, what about true applications, say for the scientist or engineer using a computer for everything that is heavy mathematics?

\bigskip

So, let us talk now about integrating functions, which is a favorite pastime of our friends, the computers. As definition for the integral, which is something simple and straightforward, that our computer friend will certainly appreciate, we have:

\index{Riemann integration}

\begin{theorem}
We have the Riemann integration formula,
$$\int_a^bf(x)dx=(b-a)\times\lim_{N\to\infty}\frac{1}{N}\sum_{k=1}^Nf\left(a+\frac{b-a}{N}\cdot k\right)$$
which can serve as a definition for the integral, for the continuous functions.
\end{theorem}

\begin{proof}
This is standard, by drawing rectangles. We have indeed the following formula, which can stand as a definition for the signed area below the graph of $f$:
$$\int_a^bf(x)dx=\lim_{N\to\infty}\sum_{k=1}^N\frac{b-a}{N}\cdot f\left(a+\frac{b-a}{N}\cdot k\right)$$

Thus, we are led to the formula in the statement.
\end{proof}

Observe that the above formula suggests that $\int_a^bf(x)dx$ is the length of the interval $[a,b]$, namely $b-a$, times the average of $f$ on the interval $[a,b]$. Thinking a bit, this is indeed something true, with no need for Riemann sums, coming directly from definitions, because area means side times average height. Thus, we can also formulate:

\index{average of function}

\begin{theorem}
The integral of a function $f:[a,b]\to\mathbb R$ is given by
$$\int_a^bf(x)dx=(b-a)\times A(f)$$
where $A(f)$ is the average of $f$ over the interval $[a,b]$.
\end{theorem}

\begin{proof}
As explained above, this is clear from definitions, via some geometric thinking. Alternatively, this is something which certainly comes from Theorem 1.22.
\end{proof}

Going ahead with more interpretations of the integral, we have:

\index{Monte Carlo integration}
\index{random number}

\begin{theorem}
We have the Monte Carlo integration formula,
$$\int_a^bf(x)dx=(b-a)\times\lim_{N\to\infty}\frac{1}{N}\sum_{k=1}^Nf(x_k)$$
with $x_1,\ldots,x_N\in[a,b]$ being random.
\end{theorem}

\begin{proof}
We recall from Theorem 1.22 that the idea is to use a formula as follows, with the points $x_1,\ldots,x_N\in[a,b]$ being uniformly distributed: 
$$\int_a^bf(x)dx=(b-a)\times\lim_{N\to\infty}\frac{1}{N}\sum_{k=1}^Nf(x_k)$$

But this works as well when the points $x_1,\ldots,x_N\in[a,b]$ are randomly distributed, for somewhat obvious reasons, and this gives the result.
\end{proof}

As a key observation now, the Monte Carlo integration works better than Riemann integration, for instance when trying to improve the estimate, via:
$$N\to N+1$$

Indeed, in the context of computing integrals via the original Riemann method, from Theorem 1.22, assume that we managed to find an estimate as follows, which in practice requires computing $N$ values of our function $f$, and making their average:
$$\int_a^bf(x)dx\simeq\frac{b-a}{N}\sum_{k=1}^Nf\left(a+\frac{b-a}{N}\cdot k\right)$$

In order to improve this estimate, any extra computed value of our function $f(y)$ will be unuseful. For improving our formula, what we need are $N$ extra values of our function, $f(y_1),\ldots,f(y_N)$, with the points $y_1,\ldots,y_N$ being precisely the midpoints of the previous division of $[a,b]$, so that we can write an improvement of our formula, as follows:
$$\int_a^bf(x)dx\simeq\frac{b-a}{2N}\sum_{k=1}^{2N}f\left(a+\frac{b-a}{2N}\cdot k\right)$$

With Monte Carlo, things are far more flexible. Assume indeed that we managed to find an estimate as follows, which again requires computing $N$ values of our function:
$$\int_a^bf(x)dx\simeq\frac{b-a}{N}\sum_{k=1}^Nf(x_k)$$

Now if we want to improve this, any extra computed value of our function $f(y)$ will be helpful, because we can set $x_{N+1}=y$, and improve our estimate as follows:
$$\int_a^bf(x)dx\simeq\frac{b-a}{N+1}\sum_{k=1}^{N+1}f(x_k)$$

And isn't this potentially useful, and powerful, when thinking at practically computing integrals, either by hand, or by using a computer. Let us record this finding as follows:

\begin{conclusion}
Monte Carlo integration works better than Riemann integration, when it comes to computing as usual, by estimating, and refining the estimate.
\end{conclusion}

As another interesting feature of Monte Carlo integration, this works better than Riemann integration for functions having various symmetries, because Riemann integration can get ``fooled'' by these symmetries, while Monte Carlo remains strong.

\bigskip

As an example for this phenomenon, chosen to be quite drastic, let us attempt to integrate, via both Riemann and Monte Carlo, the following function $f:[0,\pi]\to\mathbb R$:
$$f(x)=\Big|\sin(120x)\Big|$$

The first few Riemann sums for this function are then as follows:
$$I_2(f)=\frac{\pi}{2}(|\sin 0|+|\sin 60\pi|)=0$$
$$I_3(f)=\frac{\pi}{3}(|\sin 0|+|\sin 40\pi|+|\sin 80\pi|)=0$$
$$I_4(f)=\frac{\pi}{4}(|\sin 0|+|\sin 30\pi|+|\sin 60\pi|+|\sin 90\pi|)=0$$
$$I_5(f)=\frac{\pi}{5}(|\sin 0|+|\sin 24\pi|+|\sin 48\pi|+|\sin 72\pi|+|\sin 96\pi|)=0$$
$$\vdots$$

Based on this evidence, we will conclude, obviously, that $\int_0^\pi f(x)dx=0$. With Monte Carlo, however, such things cannot happen. Indeed, since there are only finitely many points $x\in[0,\pi]$ having the property $\sin(120 x)=0$, a random point $x\in[0,\pi]$ will have the property $|\sin(120 x)|>0$, so the Monte Carlo method will give, at any $N\in\mathbb N$:
$$\int_0^\pi f(x)dx\simeq\frac{b-a}{N}\sum_{k=1}^Nf(x_k)>0$$

Again, this is something interesting, when practically computing integrals, either by hand, or by using a computer. So, let us record, as a complement to Conclusion 1.25:

\begin{conclusion}
Monte Carlo integration is smarter than Riemann integration, because the symmetries of the function can fool Riemann, but not Monte Carlo.
\end{conclusion}

All this is good to know, and in practice now, we are led into the following question, coming from these Monte Carlo considerations, and from a myriad other reasons:

\begin{question}
How to produce random numbers?
\end{question}

And good question this is. The first thought goes into producing first random digits $x\in\{0,1,\ldots,9\}$, that we can combine afterwards into random real numbers, by using the decimal form. But, how to produce these random digits $x\in\{0,1,\ldots,9\}$?

\bigskip

You would say, easy, just pick the digits of some reasonably complicated real number $r\in\mathbb R$. In practice, however, this is not easy at all, the reasons being as follows:

\bigskip

(1) To start with, any rational number $r\in\mathbb Q$, be that a very complicated one, will not do, because the decimal expansion of rationals is periodic.

\bigskip

(2) Then, we can try beasts as $\sqrt{2}$, or more generally, roots of polynomials $P\in\mathbb Q[X]$. But here, we potentially run again into troubles, because, say in the context of a Monte Carlo application, if our function $f$ to be integrated happens to be related to $P$, using random numbers coming from solutions of $P=0$ might flaw our computation.

\bigskip

(3) So, we are seemingly led into transcendental numbers, those which are not roots of polynomials $P\in\mathbb Q[X]$. But here, again, lots of problems waiting for us. For instance being transcendental is certainly not enough, because there are such numbers having only $0,1$ as digits. Also, the simplest such numbers are $e,\pi$, and we don't want to use these for integrating our functions $f$, which usually have something with do with $e,\pi$.

\bigskip

So, what to do? Not clear. Question 1.27 is actually one of the points where modern mathematics fails, and for telling you everything, what we scientists do is to use for our simulations whatever random numbers are provided by the software. And, in case the output starts going completely wrong, pick random numbers from another batch.

\section*{1d. Probability zero}

We have now a quite good knowledge of numbers, and randomness. So, let us go back to the question that we raised in the beginning of this chapter, namely:

\index{probability zero}
\index{probability paradox}

\begin{question}
The probability for a randomly picked $x\in\mathbb R$ to be $x=1$ is obviously $P=0$. But what does this exactly mean, mathematically speaking?
\end{question}

In answer, we already talked about such things in Theorem 1.12, with a stronger result, namely that the probability for a randomly picked $x\in\mathbb R$ to be rational is 0. However, our proof there was not $100\%$ honest, based on the fact, which is somehow obvious, but can be subject to debate, that we can restrict the attention from $x\in\mathbb R$ to $x\in[0,1]$.

\bigskip

So, can we answer Question 1.28 directly, by reasoning over $\mathbb R$? And here, surprise, things are quite tricky, because the uniform measure on $\mathbb R$, that you can first think of, when it comes to randomly pick numbers $x\in\mathbb R$, has mass $\infty$, so is not a probability measure. So our formula $P(x=1)=0$ looks more like a ``social science fact'', coming from our human methods of picking random numbers $x\in\mathbb R$, than something mathematical.

\bigskip

So, in order to answer our question, here we are into some deep thinking, both in social science, and in abstract mathematics. And the situation here is as follows:

\bigskip

(1) Humanities first. If you ask people on the street for random numbers $x\in\mathbb R$, it is most likely that most of them will choose small numbers, say $x\in[-1000,1000]$, and basically no one will think at choices of type $x=872,563,267,127,167,176$.

\bigskip

(2) Of course, there is a similar phenomenon involving decimals too, but let us ignore this. Our finding is that $P$ should come from some sort of bell-shaped curve around the origin 0, and that can only be, by the Central Limit Theorem, a normal law.

\bigskip

(3) Regarding now the mathematics, this is quite clear. Once we have our normal law, the probability $P(x=1)$ that we want to compute is obtained by integrating the density of this law between 1 and 1. And we obtain of course 0, no matter what $t>0$ is.

\bigskip

As a conclusion to this, mystery solved, our answer to Question 1.28 being:

\index{normal law}

\begin{answer}
The probability for a randomly picked $x\in\mathbb R$ to be $x=1$ is
$$P=\frac{1}{\sqrt{2\pi t}}\int_1^1e^{-x^2/2t}dx$$
which means, by computing the integral, $P=0$.
\end{answer}

Well done, but that wasn't trivial, wasn't it. Which leads us now into a bit of a philosophy scare, so if dealing with such a simple question like Question 1.28 was non-trivial, what about more complicated questions of the same type, based on the obvious fact that ``everything that happens in the real life, happens with probability $P=0$''.

\bigskip

Quite concerning all this, hope you agree with me. In short, we have to review now all our basic probability knowledge, by putting that on a fully rigorous basis, in order to never ever get annoyed by such $P=0$ things, and have these fully understood, in a fully rigorous way. And good news, this is what we will be doing in this book, by systematically developing measure theory. So, let us record this finding, as follows:

\begin{fact}
Everything that happens in the real life, happens with probability $P=0$, and in order to clarify this, we must develop measure theory.
\end{fact}

Before developing measure theory, however, let us further explore Question 1.28, with the aim of bringing even more mess to what we have, namely Answer 1.29. Indeed, there are in fact some bugs with that Answer 1.29, the rationale being as follows:

\bigskip

(1) If you ask people on the street for random numbers $x\in\mathbb R$, frankly, you will certainly have some answering $x=1$. Thus, we have in fact $P(x=1)>0$.

\bigskip

(2) Looking back at our reasoning leading to Answer 1.29, the modeling error there was by ignoring what happens to the decimals. People don't like them, for sure.

\bigskip

(3) Thus, by simplifying, we must rather understand what happens when picking random integers $x\in\mathbb Z$. Or even better, by picking numbers $x\in\mathbb N$.

\bigskip

In view of this, let us formulate a more realistic version of Question 1.28, taken of course from the perspective of social science, as follows:

\begin{question}
The probability for a randomly picked $x\in\mathbb N$ to be $x=1$ is obviously $P>0$. But what does this mean, mathematically speaking, and what exactly is $P$?
\end{question} 

So, here we go again with modeling questions, belonging to both social science, and to mathematics. To be more precise, from a pure mathematics viewpoint, the first measure that comes to mind, namely the uniform measure on $\mathbb N$, has mass $\infty$, and so is not a probability measure. Thus, wrong way, and we must throw in a bit of social science.

\bigskip

But social science tells us that, exactly as before in the continuous setting, people prefer small numbers, say $x\in[0,1000]$, to beasts such as $x=872,563,267,127,167,176$. Thus, we can only expect to have a density decreasing from 0 to $\infty$, and a bit of thinking, based on the Poisson Limit Theorem, tells us that what we need is a Poisson law. And with this agreed upon, we are led to the following answer to Question 1.31: 

\index{Poisson law}

\begin{answer}
The probability for a randomly picked $x\in\mathbb N$ to be $x=1$ is
$$P=\frac{1}{e^t}\int_1^1\sum_{k=0}^\infty\frac{t^k\delta_k}{k!}$$
which means, by computing the integral, $P=t/e^t$.
\end{answer}

Of course, this was something a bit mathematical, with the $P>0$ phenomenon being definitely explained, but with the actual figure $P=t/e^t$ still depending on a parameter $t>0$, whose realistic value remains to be computed, based on social science considerations. But we will not get here into this latter question, which looks quite difficult.

\bigskip

As a conclusion to this, let us complement Fact 1.30 with:

\begin{fact}
Some things happen in the real life with probability $P>0$, but even these need, mathematically speaking, measure theory, for dealing with the Dirac masses.
\end{fact}

Long story short, no matter what we want to do advanced probability, or even basic probability, to be fully honest, we must develop measure theory first.

\section*{1e. Exercises}

This was a quite crowded preliminary chapter, and as exercises, we have:

\begin{exercise}
Learn more about fields, and their characteristic.
\end{exercise}

\begin{exercise}
Write down a self-contained approach to $\mathbb R$, using the decimal writing.
\end{exercise}

\begin{exercise}
Learn about what can be done with numeration bases $\neq10$.
\end{exercise}

\begin{exercise}
Check all the details for the Cardano formulae in degree $3$ and $4$.
\end{exercise}

\begin{exercise}
Learn more about fields, and their Galois theory.
\end{exercise}

\begin{exercise}
Prove that $e$ is irrational, then look up the fact that $e$ is transcendental.
\end{exercise}

\begin{exercise}
Learn about random numbers, and how to produce them.
\end{exercise}

\begin{exercise}
Experiment a bit with Monte Carlo, on a computer.
\end{exercise}

As bonus exercise, work more on the material from the last section.

\chapter{Finite laws}

\section*{2a. Coins and dice}

We have seen in chapter 1 that many basic analysis questions, and even simple philosophical questions regarding the real numbers $x\in\mathbb R$ themselves, suggest systematically developing measure theory. So, time to do this, and with the entirety of this book being dedicated to this, measure theory and applications, we will take a very slow approach to all this, going from finite to discrete to general, our plan being as follows:

\begin{plan}
We will develop measure theory gradually, as follows:
\begin{enumerate}
\item In this chapter we will talk about measure theory on finite spaces $X$.

\item In chapter 3 we will talk about measure theory on countable spaces $X$.

\item In chapter 4 we will explore how continuity can appear, out of the blue.

\item And in chapter 5 and afterwards we will develop measure theory, in general.
\end{enumerate}
\end{plan}

Getting started now, you might probably wonder, is there enough mathematics to be done in relation with measure theory, over the finite spaces $X$? Well, in answer, here is a question that I have for you, obviously of measure theory flavor, over finite spaces, and if you know how to solve it on the spot, congratulations, and feel free of course to skip the present chapter, or even this whole Part I, and go directly to Part II:

\begin{question}
What is the distribution of the sum, when rolling a die $n$ times?
\end{question}

So, this is the kind of situation that produces life and mathematics over the finite spaces $X$, and that we would like to understand, before doing measure theory in the continuous setting. Let us start our study with a basic definition, as follows:

\begin{definition}
A discrete probability space is a set $X$, usually finite or countable, whose elements $x\in X$ are called events, together with a function
$$P:X\to[0,\infty)$$
called probability function, which is subject to the condition
$$\sum_{x\in X}P(x)=1$$
telling us that the overall probability for something to happen is $1$.
\end{definition}

Which sounds quite neat, this is definitely a rigorous mathematical definition, that we can build our theory upon, and we will do so, first in the present chapter, in the case where $X$ is finite, and then in the next chapter, in the case where $X$ is countable. Here are now a few technical comments, in relation with the above definition:

\bigskip

(1) As a first comment, our condition $\sum_{x\in X}P(x)=1$ perfectly makes sense, and this even if $X$ is uncountable, because the sum of positive numbers is always defined, as a number in $[0,\infty]$, and this no matter how many positive numbers we have.

\bigskip

(2) As a second comment, we have chosen in the above not to assume that $X$ is finite or countable, and this for instance because we want to be able to regard any probability function on $\mathbb N$ as a probability function on $\mathbb R$, by setting $P(x)=0$ for $x\notin\mathbb N$. 

\bigskip

(3) As a third comment, standing as a complement to Definition 2.3, once we have a probability function $P:X\to[0,\infty)$, given a  set of events $Y\subset X$, we can say that the probability for the events of type $y\in Y$ to happen is $P(Y)=\sum_{y\in Y}P(y)$.

\bigskip

As our first result now, making the link with our intuition from real life, we have:

\begin{proposition}
In the simplest case, that where $X$ is finite and $P$ is constant,
$$P(Y)=\frac{{\rm number\ of\ times}\ Y\ {\rm happens}}{{\rm total\ number\ of\ possibilities}}$$
for any set of events $Y\subset X$.
\end{proposition}

\begin{proof}
Assume indeed that our probability function $P:X\to[0,\infty)$ is constant. But then, $\sum_{x\in X}P(x)=1$ implies that $X$ must be finite, and that we must have:
$$P(x)=\frac{1}{|X|}\quad,\quad\forall x\in X$$

Now by summing this over the elements $y\in Y$ of a subset $Y\subset X$, we obtain:
$$P(Y)=\frac{|Y|}{|X|}\quad,\quad\forall Y\subset X$$

We are therefore led to the conclusion in the statement.
\end{proof}

Time now for some examples? The basic examples will come of course from basic games, and at the beginning of everything, we certainly have flipping coins:

\begin{example}
Flipping coins.
\end{example}

Here things are simple and clear, because when you flip a coin the corresponding discrete probability space, together with its probability measure, is as follows:
$$X=\big\{{\rm heads},\,{\rm tails}\big\}\quad,\quad P({\rm heads})=P({\rm tails})=\frac{1}{2}$$

In the case where the coin is biased, as to land on heads with probability $2/3$, and on tails with probability $1/3$, the corresponding probability space is as follows:
$$X=\big\{{\rm heads},\,{\rm tails}\big\}\quad,\quad P({\rm heads})=\frac{2}{3}\quad,\quad P({\rm tails})=\frac{1}{3}$$

More generally, given any number $p\in[0,1]$, we have an abstract probability space as follows, where we have replaced heads and tails by win and lose:
$$X=\big\{{\rm win},\,{\rm lose}\big\}\quad,\quad P({\rm win})=p\quad,\quad P({\rm lose})=1-p$$

Things become even more interesting when flipping a coin, biased or not, several times in a row. In the case of a biased coin, landing on heads with probability $p_1$ and on tails with probability $p_2=1-p_2$, thrown $n$ times, the probability space is:
$$X=\big\{1,2\}^n\quad,\quad P(i_1\ldots i_n)=p_{i_1}\ldots p_{i_n}\quad,\quad p_i\geq0\ ,\ p_1+p_2=1$$

Finally, in relation with this latter example, let us check that the sum 1 condition in Definition 2.3 is indeed satisfied. But this is indeed the case, as shown by:
$$\sum_{i\in X}P(i)
=\sum_{i_1,\ldots,i_n}p_{i_1}\ldots p_{i_n}
=\sum_{i_1}p_{i_1}\ldots\sum_{i_n}p_{i_n}
=1$$

Summarizing, coins reasonably understood, or at least the mathematical formalism of the coin game, reasonably understood. However, in what regards the mathematics itself of the coin game, many interesting things can be said. More on this later.

\bigskip

Moving on, at a more advanced level we can talk about rolling dice, as follows:

\begin{example}
Rolling dice.
\end{example}

Again, things here are simple and clear, because when you throw a die the corresponding probability space, together with its probability measure, is as follows:
$$X=\big\{1,\ldots,6\big\}\quad,\quad P(i)=\frac{1}{6}\ ,\ \forall i$$

As before with the coins, we can further complicate this by assuming that the die is biased, say landing on face $i$ with probability $p_i\in[0,1]$. In this case the corresponding probability space, together with its probability measure, is as follows:
$$X=\big\{1,\ldots,6\big\}\quad,\quad P(i)=p_i\quad,\quad p_i\geq0\ ,\ \sum_ip_i=1$$

Also as before with coins, things become more interesting when throwing a die several times in a row, or equivalently, when throwing several identical dice at the same time. In this latter case, with $n$ identically biased dice, the probability space is as follows:
$$X=\big\{1,\ldots,6\big\}^n\quad,\quad P(i_1\ldots i_n)=p_{i_1}\ldots p_{i_n}\quad,\quad p_i\geq0\ ,\ \sum_ip_i=1$$

Observe that the sum 1 condition in Definition 2.3 is indeed satisfied, and with this proving that our dice modeling is bug-free, due to the following computation:
$$\sum_{i\in X}P(i)
=\sum_{i_1,\ldots,i_n}p_{i_1}\ldots p_{i_n}
=\sum_{i_1}p_{i_1}\ldots\sum_{i_n}p_{i_n}
=1$$

In fact, this computation is identical to the one that we did before, for coins.

\bigskip

Getting back now to theory, in the general context of Definition 2.3, we can see that what we have there is very close to the biased die, from Example 2.6. Indeed, in the general context of Definition 2.3, we can say that what happens is that we have a die with $|X|$ faces, which is biased such that it lands on face $i$ with probability $P(i)$. 

\bigskip

Which is something quite interesting, theoretically speaking, in relation with the discrete probability theory that we want to develop in this chapter, because we have now some intuition on all this. So, as a conclusion, let us record this finding as follows:

\begin{principle}
Discrete probability can be understood as being about throwing a general die, having an arbitrary number of faces, and which is arbitrarily biased too:
$$\xymatrix@R=7pt@C=7pt{
\ar@{-}[rr]\ar@{-}[dd]&&\ar@{-}[dd]&&\ar@{-}[rr]\ar@{-}[dd]&&\ar@{-}[dd]&&\ar@{-}[rr]\ar@{-}[dd]&&\ar@{-}[dd]&&\ar@{-}[rr]\ar@{-}[dd]&&\ar@{-}[dd]\\
&p_x&&&&p_y&&&&p_z&&&&p_t&&&&\ldots\\
\ar@{-}[rr]&&&&\ar@{-}[rr]&&&&\ar@{-}[rr]&&&&\ar@{-}[rr]&&}$$
To be more precise, when writing our probability space as $X=\{x,y,z,t,\ldots\}$, with the corresponding probabilities being $p_x,p_y,p_z,p_t,\ldots\,$, the picture is the one above.
\end{principle}

Moving on, in order to do some mathematics, in the context of Definition 2.3, we need a random variable $f:X\to\mathbb R$, and the mathematics will consist in computing the expectation of this variable, $E(f)\in\mathbb R$. Let us axiomatize this situation as follows:

\index{random variable}
\index{expectation}

\begin{definition}
A random variable on a probability space $X$ is a function
$$f:X\to\mathbb R$$
and the expectation of such a random variable is the quantity
$$E(f)=\sum_{x\in X}f(x)P(x)$$
which is best thought as being the average gain, when the game is played.
\end{definition}

Here the word ``game'' refers to the probability space interpretation from Principle 2.7. Indeed, in that context, with our discrete set of events $X$ being thought of as corresponding to a generalized die, and by thinking of $f:X\to\mathbb R$ as representing some sort of money, the above quantity $E(f)$ is what we win, on average, when playing the game.

\bigskip

Getting back now to the examples, as our first result, we have:

\begin{theorem}
When flipping a usual coin $n$ times in a row:
\begin{enumerate}
\item $E({\rm heads})=n/2$.

\item $E({\rm tails})=n/2$.

\item $E(a\times{\rm heads}+b\times{\rm tails})=n(a+b)/2$.
\end{enumerate} 
\end{theorem}

\begin{proof}
Let us first fix some mathematical notation, in the spirit of what we have in Definition 2.8. Consider the variables in the statement, namely:
$$f:\big\{{\rm heads},\,{\rm tails}\big\}^n\to\mathbb R\quad,\quad f={\rm number\ of\ heads}$$ 
$$g:\big\{{\rm heads},\,{\rm tails}\big\}^n\to\mathbb R\quad,\quad g={\rm number\ of\ tails}$$ 
$$h:\big\{{\rm heads},\,{\rm tails}\big\}^n\to\mathbb R\quad,\quad h=a\times{\rm heads}+b\times{\rm tails}$$ 

With a bit of probabilistic know-how, we can compute the expectations of these variables, by using several methods, which are all instructive, as follows:

\medskip

(1) \underline{Symmetry proof}. Observe that, by using the obvious symmetry between heads and tails, at the level of the corresponding expectations, we have:
$$E(f)=E(g)$$

On the other hand, we also have $f+g=n$, which gives the following formula:
$$E(f)+E(g)=n$$

Thus $E(f)=E(g)=n/2$, and then $E(h)=n(a+b)/2$ follows by linearity.

\medskip

(2) \underline{Additive proof}. When flipping our coin several times in a row, the expectations will obviously sum up. Thus, if we denote by $f_n,g_n,h_n$ our variables, we have:
$$E(f_n)=nE(f_1)\quad,\quad E(g_n)=nE(g_1)\quad,\quad E(h_n)=nE(h_1)$$

But together with $E(f_1)=E(g_1)=1/2$ and $E(h_1)=(a+b)/2$, we are done.

\medskip

(3) \underline{Algebraic proof}. Here is as well a third proof, by brainless computation:
\begin{eqnarray*}
E(f)
&=&\sum_{x\in\{{\rm heads,\, tails}\}^n}\#({\rm heads}\in x)\times\frac{1}{2^n}\\
&=&\sum_{s=1}^n\binom{n}{s}\times s\times\frac{1}{2^n}\\
&=&\frac{n}{2^n}\sum_{t=0}^{n-1}\binom{n-1}{t}\\
&=&\frac{n}{2}
\end{eqnarray*}

Similarly, we have $E(g)=n/2$, and then $E(h)=n(a+b)/2$ comes by linearity.
\end{proof}

Getting now to the more general case of a biased coin, we have here:

\begin{theorem}
When flipping a biased coin $n$ times in a row:
\begin{enumerate}
\item $E({\rm heads})=np$.

\item $E({\rm tails})=n(1-p)$.

\item $E(a\times{\rm heads}+b\times{\rm tails})=n(ap+b(1-p))$.
\end{enumerate} 
\end{theorem}

\begin{proof}
As before with the usual coins, we can denote by $f,g,h$ our random variables, and we have 3 possible approaches here, the situation with them being as follows:

\medskip

(1) \underline{Symmetry proof}. Normally, the idea here would be to exploit the symmetry between heads and tails, with the goal of reaching to the following formula:
$$(1-p)E(f)=pE(g)$$

Indeed, together with $E(f)+E(g)=n$, coming from $f+g=n$, this would finish the proof. However, the symmetry study is not obvious at $p\neq1/2$. So, wrong way. 

\medskip

(2) \underline{Additive proof}. As before, when flipping our coin several times in a row, the expectations will sum up. Thus, if we denote by $f_n,g_n,h_n$ our variables, we have:
$$E(f_n)=nE(f_1)\quad,\quad E(g_n)=nE(g_1)\quad,\quad E(h_n)=nE(h_1)$$

But together with $E(f_1)=p$, $E(g_1)=1-p$, $E(h_1)=ap+b(1-p)$, we are done.

\medskip

(3) \underline{Algebraic proof}. Here is as well the algebraic proof, your old man's favorite:
\begin{eqnarray*}
E(f)
&=&\sum_{x\in\{{\rm heads,\, tails}\}^n}\#({\rm heads}\in x)\cdot p^{\#({\rm heads}\in x)}(1-p)^{\#({\rm tails}\in x)}\\
&=&\sum_{s=1}^n\binom{n}{s}sp^s(1-p)^{n-s}\\
&=&np\sum_{t=0}^{n-1}\binom{n-1}{t}p^t(1-p)^{n-t-1}\\
&=&np
\end{eqnarray*}

Similarly, $E(g)=n(1-p)$, and then $E(h)=n(ap+b(1-p))$ comes by linearity.
\end{proof}

Next, come the dice. Here there are several possible natural choices for our random variable, and regarding the essentials, in the unbiased die case, we have:

\begin{theorem}
When rolling a usual die $n$ times in a row:
\begin{enumerate}
\item For $f_i=\#i$, we have $E(f_i)=n/6$.

\item For $g_i=i\#i$, we have $E(g_i)=ni/6$.

\item For $h={\rm sum}$, we have $E(h)=3.5n$.
\end{enumerate} 
\end{theorem}

\begin{proof}
With respect to our study for the usual coins, the situation is as follows:

\medskip

(1) \underline{Symmetry proof}. This obviously works, exactly as for unbiased coins.

\medskip

(2) \underline{Additive proof}. Again this works, obviously, as for the unbiased coins.

\medskip

(3) \underline{Algebraic proof}. Things are a bit more tricky here, depending on the exact variables that we have in mind, the situation being as follows:

\medskip

-- Normally, we can compute indeed $E(f_i)$ algebrically, a bit as before for the coins, by using binomial coefficients and some summing, and I will leave this as an exercise for you, and then pass to $E(g_i)$, and then to $E(h)$, by linearity.

\medskip

-- However, if we want to compute for instance $E(h)$ directly, algebrically, we are a bit in trouble, because, as you can convince yourself by doing some computations, finding the formula of the distribution of the sum $h$ is a non-trivial question.
\end{proof}

Finally, in the case of a biased die, the result is as follows:

\begin{theorem}
When rolling a biased die $n$ times in a row:
\begin{enumerate}
\item For $f_i=\#i$, we have $E(f_i)=np_i$.

\item For $g_i=i\#i$, we have $E(g_i)=nip_i$.

\item For $h={\rm sum}$, we have $E(h)=n\sum_iip_i$.
\end{enumerate} 
\end{theorem}

\begin{proof}
This is something hybrid between Theorems 2.10 and 2.11:

\medskip

(1) \underline{Symmetry proof}. This won't quite work for a biased die, for the technical reasons explained in the proof of Theorem 2.10, dealing with a biased coin.

\medskip

(2) \underline{Additive proof}. This works, as usual, because if we denote by $f_i^n,g_i^n,h^n$ our variables, it is clear that we have the following formulae:
$$E(f_i^n)=nE(f_i^1)\quad,\quad E(g_i^n)=nE(g_i^1)\quad,\quad E(h^n)=nE(h^1)$$

Thus, together with $E(f_i^1)=p_i$, $E(g_i^1)=ip_i$, $E(h^1)=\sum_iip_i$, we are done.

\medskip

(3) \underline{Algebraic proof}. This won't work well for a biased die, for the technical reasons explained in the proof of Theorem 2.11, dealing with a usual die.
\end{proof}

\section*{2b. General theory}

We have already seen some good illustrations for Definition 2.8, so time now to get into more delicate aspects. Imagine that you want to set up some sort of business, with your variable $f:X\to\mathbb R$. You are of course mostly interested in the expectation $E(f)\in\mathbb R$, but passed that, the way this expectation comes in matters too. For instance:

\medskip

-- When your variable is constant, $f=c$, you certainly have $E(f)=c$, and your business will run smoothly, with not so many surprises on the way.

\medskip

-- On the opposite, for a complicated variable satisfying $E(f)=c$, your business will be more bumpy, with wins and loses on the way, depending on your skills.

\medskip

In short, and extrapolating now from business to mathematics, physics, chemistry and everything else, we must complement Definition 2.8 with something finer, regarding the ``quality'' of the expectation $E(f)\in\mathbb R$ appearing there. And the first thought here, which is the correct one, goes to the following definition, of the variance of our variable:

\index{variance}

\begin{definition}
The variance of a variable $f:X\to\mathbb R$ is the quantity
\begin{eqnarray*}
V(f)
&=&E\left([f-E(f)]^2\right)\\
&=&E(f^2)-E(f)^2
\end{eqnarray*}
intuitively measuring how far is $f$ from a constant variable.
\end{definition}

As a first observation, which is in tune with what we were saying in the above, the variance is 0 precisely when the variable is constant, equal to its expectation:
\begin{eqnarray*}
V(f)=0
&\iff&f={\rm constant}\\
&\iff&f=E(f)
\end{eqnarray*}

As a complement now to Definition 2.13, or rather as an alternative to it, which can bring some intuition too, with this being a matter of taste, let us formulate as well:

\index{standard deviation}

\begin{definition}
The standard deviation of a variable $f:X\to\mathbb R$ is the quantity
$$\sigma(f)=\sqrt{V(f)}$$
intuitively measuring as well how far is $f$ from a constant variable.
\end{definition}

Moving on, but still regarding expectations and variances of variables, as constructed above, at the abstract level, following Markov, we have the following result:

\index{Markov inequality}

\begin{proposition}
We have the following Markov inequality,
$$P\big(|f|\geq a\big)\leq\frac{E(f)}{a}$$
valid for any random variable $f:X\to\mathbb R$, and any $a>0$.
\end{proposition}

\begin{proof}
This is something trivial, coming from definitions. Indeed, we have:
$$E(f)
\geq\sum_{|f(x)|\geq a}aP(x)
=aP\big(|f|\geq a\big)$$

Thus, we are led to the Markov inequality in the statement.
\end{proof}

Next, following Chebycheff, we have the following key estimate:

\index{Chebycheff inequality}

\begin{theorem}
We have the following Chebycheff inequality,
$$P\big(|g-E|\geq b\big)=\frac{V}{b^2}$$
valid for any random variable $g:X\to\mathbb R$, having mean $E$ and variance $V$.
\end{theorem}

\begin{proof}
Given a random variable $g:X\to\mathbb R$, having mean $E$ and variance $V$, we can use the Markov inequality above with $f=(g-E)^2$, and we obtain in this way:
$$P\big(|g-E|^2\geq b^2\big)
\leq\frac{E((g-E)^2)}{b^2}
=\frac{V}{b^2}$$

Thus, we are led to the Chebycheff inequality in the statement.
\end{proof}

Getting back now to the real-life considerations from the beginning of this section, good that we have the notion of variance, for understanding how to run our business, but let us not stop here. For a total control of your business, be that of financial, mathematical, physical or chemical type, you will certainly want to know more about your variable $f:X\to\mathbb R$. Which leads us into general moments, constructed as follows:

\index{moments}

\begin{definition}
The moments of a variable $f:X\to\mathbb R$ are the numbers
$$M_k=E(f^k)$$
which satisfy $M_0=1$, then $M_1=E(f)$, and then $V(f)=M_2-M_1^2$.
\end{definition}

And with this, good news, we have now all the needed tools in our bag for doing some good business, in what follows. To put things in a very compacted way:

\medskip

-- $M_0$ is about foundations.

\medskip

-- $M_1$ is about running some business.

\medskip

-- $M_2$ is about running that business well.

\medskip

-- $M_3$ and higher are advanced level, about ruining all the competing businesses. 

\medskip

As a next piece of discussion, still regarding the moments, we can formulate the following version of Definition 2.17, making a more clear link with the variance:

\index{central moments}

\begin{definition}
The central moments of a variable $f:X\to\mathbb R$ are the numbers
$$M_k'=E((f-E)^k)$$
with $E=E(f)$, which satisfy $M_0'=1$, $M_1'=0$, $M_2'=V(f)$.
\end{definition}

And I will leave it to you, to think a bit at this, and decide whether you prefer this over Definition 2.17. Along the same lines, we can in fact do better, as follows:

\index{normalized central moments}

\begin{definition}
The normalized central moments of $f:X\to\mathbb R$ are the numbers
$$M''_k=E\left(\left(\frac{f-E}{\sigma}\right)^k\right)$$
with $\sigma=\sigma(f)$, which satisfy $M''_0=1$, $M''_1=0$, $M''_2=1$.
\end{definition}

Here we assume of course $\sigma>0$, which is the same as saying that $f:X\to\mathbb R$ is not constant. Of particular interest are the cases $k=3,4$, where we can formulate:

\index{skewness}
\index{kurtosis}

\begin{definition}
The third and fourth normalized central moments
$$\gamma=M_3''\quad,\quad 
\kappa=M_4''$$
are called skewness and kurtosis of the variable $f:X\to\mathbb R$.
\end{definition}

As a conclusion to all this, a random variable $f:X\to\mathbb R$ is quite reasonably described by the data $(E,V,\gamma,\kappa)$, whose knowledge is the same as that of $M_1,M_2,M_3,M_4$. And for more delicate questions, we still have the higher moments, $M_k$ with $k\geq5$.

\bigskip

Done with the theory? Not really, because we still have to formulate one more key result, which is actually the most important one, in basic probability, as follows:

\index{law of variable}

\begin{theorem}
If we define the law of $f:X\to\mathbb R$ as being the real measure
$$\mu=\sum_{x\in X}P(x)\delta_{f(x)}$$
then $\mu(\mathbb R)=1$, and we have the following formula, for any function $\phi:\mathbb R\to\mathbb R$,
$$E(\phi(f))=\int_\mathbb R\phi(y)d\mu(y)$$
with the usual convention that each Dirac mass integrates up to $1$.
\end{theorem} 

\begin{proof}
To start with, we can certainly talk about the law $\mu$, as being the linear combination of Dirac masses in the statement. Observe that, alternatively, we have:
$$\mu=\sum_{y\in\mathbb R}P(f=y)\delta_y$$

As yet another alternative definition for $\mu$, we have the following formula, with $\nu$ being the probability measure on $X$, and with $*$ being the push-forward operation:
$$\mu=f_*\nu$$

Getting now to the expectation formula in the statement, this is clear, as follows:
$$E(\phi(f))
=\sum_{x\in X}P(x)\phi(f(x))
=\int_\mathbb R\phi(y)d\mu(y)$$

We are therefore led to the conclusions in the statement.
\end{proof}

Next, quite remarkably, the sequence of moments uniquely determines the law:

\begin{theorem}
The sequence of moments of a variable $f:X\to\mathbb R$,
$$M_k=E(f^k)$$
uniquely determines the law of the variable.
\end{theorem}

\begin{proof}
By Theorem 2.21, we have the following formula, for the moments:
$$\mu=\sum_i\lambda_i\delta_{x_i}\implies M_k=\sum_i\lambda_ix_i^k$$

But it is then a standard question to recover the numbers $\lambda_i,x_i\in\mathbb R$, and so the measure $\mu$, out of the sequence of numbers $M_k$. Indeed, assuming that the numbers $x_i$ are $0<x_1<\ldots<x_n$ for simplifying, with $k\to\infty$ we have the following estimate:
$$M_k\sim\lambda_nx_n^k$$

Thus, we got the parameters $\lambda_n,x_n\in\mathbb R$ of our measure $\mu$, and by substracting them and doing an obvious recurrence, we get the other parameters $\lambda_i,x_i\in\mathbb R$ as well.
\end{proof}

Switching topics, inspired now by what happens for coins and dice, let us formulate: 

\index{independence}
\index{independent variables}

\begin{definition}
We say that two variables $f,g:X\to\mathbb R$ are independent when
$$P(f=x,g=y)=P(f=x)P(g=y)$$
happens, for any $x,y\in\mathbb R$.
\end{definition}

As mentioned, this is something quite intuitive, inspired by what happens for coins and dice. More on this, and on other examples available, later. For the moment, let us develop some theory. As our first result regarding independence, we have:

\index{mixed moments}

\begin{theorem}
Assuming that $f,g:X\to\mathbb R$ are independent, we have
$$E(f^kg^l)=E(f^k)E(g^l)$$
and the converse holds, in the sense that this implies the independence of $f,g$.
\end{theorem}

\begin{proof}
This is something very standard, the idea being as follows:

\medskip

(1) In one sense, we have the following computation, for the mixed moments:
\begin{eqnarray*}
E(f^kg^l)
&=&\sum_{xy}x^ky^lP(f=x,g=y)\\
&=&\sum_{xy}x^ky^lP(f=x)P(g=y)\\
&=&\sum_xx^kP(f=x)\sum_yy^lP(g=y)\\
&=&E(f^k)E(g^l)
\end{eqnarray*}

(2) Conversely, the moment condition $E(f^kg^l)=E(f^k)E(g^l)$ reformulates as:
$$\sum_{xy}c(x,y)x^ky^l=0\quad,\quad 
c(x,y)=P(f=x,g=y)-P(f=x)P(g=y)$$

So, let us examine this. Our variables $f,g:X\to\mathbb R$ being discrete, let $x_1<\ldots<x_n$ and $y_1<\ldots<y_m$ be their images. With $c_{ij}=c(x_i,x_j)$, our formula reads:
$$\sum_{ij}c_{ij}x_i^ky_j^l=0\quad,\quad\forall k,l\in\mathbb N$$

Assuming now $x_i>0$ for simplifying, we have, with $k\to\infty$, and with $l$ fixed:
$$\sum_{ij}c_{ij}x_i^ky_j^l\sim x_n^k\sum_jc_{nj}y_j^l$$

Thus $\sum_jc_{nj}y_j^l=0$, so $c_{nj}=0$. But with this, again with $k\to\infty$, and $l$ fixed:
$$\sum_{ij}c_{ij}x_i^ky_j^l\sim x_{n-1}^k\sum_jc_{n-1,j}y_j^l$$

And so on, the idea being that we obtain by recurrence $c_{ij}=0$, as desired.
\end{proof}

Getting now to sums of independent variables, we have here:

\begin{theorem}
Assuming that $f,g:X\to\mathbb R$ are independent, we have
$$\mu_{f+g}=\mu_f*\mu_g$$
where $*$ is the convolution operation, defined by $\delta_x*\delta_y=\delta_{x+y}$ and linearity.
\end{theorem}

\begin{proof}
We have indeed the following straightforward computation:
\begin{eqnarray*}
\mu_{f+g}
&=&\sum_{x\in\mathbb R}P(f+g=x)\delta_x\\
&=&\sum_{y,z\in\mathbb R}P(f=y,g=z)\delta_{y+z}\\
&=&\sum_{y,z\in\mathbb R}P(f=y)P(g=z)\delta_y*\delta_z\\
&=&\left(\sum_{y\in\mathbb R}P(f=y)\delta_y\right)*\left(\sum_{z\in\mathbb R}P(g=z)\delta_z\right)\\
&=&\mu_f*\mu_g
\end{eqnarray*}

Thus, we are led to the conclusion in the statement.
\end{proof}

As a first application of all this, we can now solve Question 2.2, as follows:

\begin{theorem}
When rolling a die $n$ times, the distribution of the sum is
$$P(s)=\frac{1}{6^n}\sum_{k=0}^{[\frac{s-n}{6}]}(-1)^k\binom{n}{k}\binom{s-6k-1}{n-1}$$
with this coming from independence, and the binomial formula for negative exponents.
\end{theorem}

\begin{proof}
This is something quite tricky, the idea being as follows:

\medskip

(1) According to our independence theory above, the law of the sum is:
$$\mu_n=\frac{1}{6^n}(\delta_1+\delta_2+\delta_3+\delta_4+\delta_5+\delta_6)^{*n}$$

(2) Equivalently, when thinking a bit, we have the following formula for the law of the sum, and I will leave it to you, to clarify the details here:
$$\sum_{s=n}^{6n}P(s)x^s=\frac{1}{6^n}(x+x^2+x^3+x^4+x^5+x^6)^n$$

(3) So, let us compute the function on the right. This function is given by:
$$f(x)
=\frac{x^n}{6^n}\cdot\frac{(1-x^6)^n}{(1-x)^n}$$

(4) Next, let us recall that the generalized binomial formula at negative integer exponents takes the following form, and with this being in fact something which is elementary, easy to establish by recurrence on $n$, by using the Pascal formula for binomials:
$$\frac{1}{(1-x)^n}=\sum_{l=0}^\infty\binom{n+l-1}{n-1}x^l$$

(5) Now by using the usual binomial formula for the numerator $(1-x^6)^n$, and the generalized  binomial formula for the fraction $1/(1-x)^n$, we obtain:
\begin{eqnarray*}
f(x)
&=&\frac{x^n}{6^n}\sum_{k=0}^n(-1)^k\binom{n}{k}x^{6k}\sum_{l=0}^\infty\binom{n+l-1}{n-1}x^l\\
&=&\frac{1}{6^n}\sum_{k=0}^n\sum_{l=0}^\infty(-1)^k\binom{n}{k}\binom{n+l-1}{n-1}x^{n+6k+l}\\
&=&\frac{1}{6^n}\sum_{s=0}^\infty\sum_{k=0}^n(-1)^k\binom{n}{k}\binom{s-6k-1}{n-1}x^s
\end{eqnarray*}

(6) Finally, remember that $f$ was a polynomial of degree $6n$, divisible by $x^n$. Thus, the true range of the first summing index is $s=n,\ldots,6n$. Also, from $s=n+6k+l$ in the above computation, we get $s\geq n+6k$, and so $k\leq[(s-n)/6]$, integer part. Thus, our formula in (5), written by ignoring zero coefficients appearing from cancellation, is:
$$f(x)=\frac{1}{6^n}\sum_{s=n}^{6n}\sum_{k=0}^{[\frac{s-n}{6}]}(-1)^k\binom{n}{k}\binom{s-6k-1}{n-1}x^s$$

But with this, we are led via (2) to the formula in the statement.
\end{proof}

\section*{2c. Binomial laws}

With Question 2.2 solved, what is next? Simpler things, I guess, and getting now to variables and laws that we can study, at the beginning of everything, we have:

\index{Bernoulli law}
\index{binomial law}

\begin{theorem}
When playing with a biased coin, your winning law is
$$b_p=(1-p)\delta_0+p\delta_1$$
called Bernoulli law of parameter $p\in[0,1]$. When playing $n$ times, your winning law is
$$b_{np}=\sum_{s=0}^n\binom{n}{s}p^s(1-p)^{n-s}\delta_s$$
called binomial law of parameter $p\in[0,1]$. We have the formula $b_{np}=b_p^{*n}$.
\end{theorem}

\begin{proof}
This is something coming from our independence technology, as follows:

\medskip

(1) The first assertion, regarding the Bernoulli law $b_p$, is more of a definition, for this law. Next, when playing $n$ times, in order to have $s$ wins, we must select the $s$ occurrences of heads, among the $n$ throws, and there are $\binom{n}{s}$ choices here. And then, we must sum over these $\binom{n}{s}$ choices the common probability $p^s(1-p)^{n-s}$, for $s$ heads and $n-s$ tails. Thus, we are led to the formula for the binomial law $b_{np}$ in the statement.

\medskip

(2) The formula $b_{np}=b_p^{*n}$ comes from the fact that the Bernoulli laws $b_p$ produce the binomial laws $b_{np}$, by iterating the game $n$ times, via the independence of the throws. Equivalently, this latter formula comes from the following formal computation:
\begin{eqnarray*}
b_p^{*n}
&=&\big((1-p)\delta_0+p\delta_1\big)^{*n}\\
&=&\sum_{s=0}^np^s(1-p)^{n-s}\binom{n}{s}\delta_0^{*(n-s)}*\delta_1^{*s}\\
&=&\sum_{s=0}^np^s(1-p)^{n-s}\binom{n}{s}\delta_s\\
&=&b_{np}
\end{eqnarray*}

Thus, we are led to the conclusions in the statement.
\end{proof}

Getting now to the study of the binomial laws, we first have the following result:

\index{mean of binomial law}
\index{variance of binomial law}
\index{moments of binomial law}

\begin{theorem}
The binomial law $b_{np}$ has the following properties:
\begin{enumerate}
\item The mean is $E=np$.

\item The variance is $V=np(1-p)$.
\end{enumerate}
\end{theorem}

\begin{proof}
Regarding the mean of a binomial variable $f:X\to\mathbb R$, we have:
\begin{eqnarray*}
E(f)
&=&\sum_{s=1}^ns\binom{n}{s}p^s(1-p)^{n-s}\\
&=&np\sum_{s=1}^n\frac{(n-1)!}{(s-1)!(n-s)!}\,p^{s-1}(1-p)^{n-s}\\
&=&np\sum_{t=0}^{n-1}\binom{n-1}{t}p^t(1-p)^{n-t-1}\\
&=&np(p+1-p)^{n-1}\\
&=&np
\end{eqnarray*}

Coming next, with the same trick, we can compute the following quantity:
\begin{eqnarray*}
E(f^2)-E(f)
&=&\sum_{s=2}^n(s^2-s)\binom{n}{s}p^s(1-p)^{n-s}\\
&=&n(n-1)p^2\sum_{s=2}^n\frac{(n-2)!}{(s-2)!(n-s)!}p^{s-2}(1-p)^{n-s}\\
&=&n(n-1)p^2\sum_{t=0}^{n-2}\binom{n-2}{t}p^t(1-p)^{n-t-2}\\
&=&n(n-1)p^2(p+1-p)^{n-2}\\
&=&n(n-1)p^2
\end{eqnarray*}

Now by using the above formulae, we conclude that the variance is given by:
\begin{eqnarray*}
V(f)
&=&[E(f^2)-E(f)]+[E(f)-E(f)^2]\\
&=&n(n-1)p^2+np-(np)^2\\
&=&np(1-p)
\end{eqnarray*}

We are therefore led to the conclusions in the statement.
\end{proof}

Regarding now the higher moments, we have the following result, about them:

\index{number of blocks}
\index{Bell numbers}
\index{Stirling numbers}

\begin{theorem}
The moments of the binomial law $b_{np}$ are given by
$$M_k=\sum_{\pi\in P(k)}\frac{n!}{(n-|\pi|)!}\,p^{|\pi|}$$
where $P(k)$ is the set of partitions of $\{1,\ldots,k\}$, and $|.|$ is the number of blocks.
\end{theorem}

\begin{proof}
This is something very standard, the idea being as follows:

\medskip

(1) Some numerics first. At $k=1$ we only have one partition, namely $|$\,, and the formula in the statement holds indeed, as shown by the following computation:
$$M_1=\sum_{\pi\in P(1)}\frac{n!}{(n-|\pi|)!}\,p^{|\pi|}
=\frac{n!}{(n-1)!}\,p^1
=np$$

At $k=2$ now, we have two partitions, namely $|\,|$ and $\sqcap$, and the moment is:
$$M_2=\frac{n!}{(n-2)!}\,p^2+\frac{n!}{(n-1)!}\,p^1
=n(n-1)p^2+np$$

At $k=3$ the partitions are $|\ |\ |$\,, $\sqcap\ |$\,, $\sqcap\hskip-3.2mm{\ }_|$\,\,, $|\ \sqcap$\,, $\sqcap\hskip-0.7mm\sqcap$, and the moment is:
$$M_3=n(n-1)(n-2)p^3+3n(n-1)p^2+np$$

At $k=4$ we have 15 partitions, with the list of relevant partitions, and their multiplicities up to permutations, which is self-explanatory, being as follows:
$$|\ |\ |\ |\times1\quad,\quad \sqcap\ |\ |\times 6\quad,\quad \sqcap\sqcap\times 3
\quad,\quad \sqcap\hskip-1.6mm\sqcap|\times4\quad,\quad \sqcap\hskip-1.6mm\sqcap\hskip-1.6mm\sqcap\times1$$

Now by doing the computation, as before, the formula of the moment is:
$$M_4=n(n-1)(n-2)(n-3)p^4+6n(n-1)(n-2)p^3+7n(n-1)p^2+np$$

(2) In general now, in order to prove the result, we can use the trick from the proof of Theorem 2.28. Indeed, assume that we managed to find a formula as follows:
$$s^k=\sum_{r=1}^kc_r\cdot s(s-1)\ldots(s-r+1)$$

We have then the following computation, based on this formula:
\begin{eqnarray*}
M_k
&=&\sum_{r=1}^kc_r\sum_{s=r}^ns(s-1)\ldots(s-r+1)\binom{n}{s}p^s(1-p)^{n-s}\\
&=&\sum_{r=1}^kc_r\cdot n(n-1)\ldots(n-r+1)\sum_{s=r}^n\binom{n-r}{s-r}p^s(1-p)^{n-s}\\
&=&\sum_{r=1}^kc_r\cdot n(n-1)\ldots(n-r+1)\cdot p^r
\end{eqnarray*}

(3) The problem is now, how to find that magic formula from (2)? And the answer here comes from partitions. Indeed, with the standard convention that for a multi-index $i=(i_1,\ldots,i_k)$, its kernel is the partition $\ker i\in P(k)$ joining equal indices, we have:
\begin{eqnarray*}
s^k
&=&\sum_{i_1=1}^s\ldots\sum_{i_k=1}^s1\\
&=&\sum_{\pi\in P(k)}\#\left\{(i_1,\ldots,i_k)\in\{1,\ldots,s\}^k\Big|\ker i=\pi\right\}\\
&=&\sum_{\pi\in P(k)}\frac{s!}{(s-|\pi|)!}
\end{eqnarray*}

Thus, good news, we have the magic formula that we need as data in (2), and according now to that computation in (2), we are led to the following formula, for the moment:
$$M_k=\sum_{\pi\in P(k)}\frac{n!}{(n-|\pi|)!}\,p^{|\pi|}$$

Thus, we have indeed the formula in the statement.
\end{proof}

Let us end this discussion with the following result, capturing the essentials:

\index{mean}
\index{variance}
\index{skewness}
\index{kurtosis}

\begin{theorem}
The mean, variance, skewness and kurtosis of $b_{np}$ are
$$E=np\quad,\quad V=npq\quad,\quad\gamma=\frac{q-p}{\sqrt{npq}}\quad,\quad\kappa=3+\frac{1-6pq}{npq}$$
with the convention $q=1-p$.
\end{theorem}

\begin{proof}
We know the first two formulae, from Theorem 2.28. Next, according to Definition 2.18, the central moments of a variable $f:X\to\mathbb R$ are given by:
\begin{eqnarray*}
M_k'
&=&E((f-E)^k)\\
&=&\sum_{r=0}^k(-1)^r\binom{k}{r}E^rM_{k-r}\\
&=&\sum_{r=0}^{k-2}(-1)^r\binom{k}{r}E^rM_{k-r}+(-1)^{k-1}(k-1)E^k
\end{eqnarray*}

At $k=3$, by using the formulae from the proof of Theorem 2.29, we obtain:
\begin{eqnarray*}
M_3'
&=&M_3-3EM_2+2E^3\\
&=&[n(n-1)(n-2)p^3+3n(n-1)p^2+np]-3np(n(n-1)p^2+np)+2(np)^3\\
&=&np(1-p)(1-2p)
\end{eqnarray*}

At $k=4$, again by using the formulae from the proof of Theorem 2.29, we have:
\begin{eqnarray*}
&&M_4'\\
&=&M_4-4EM_3+6E^2M_2-3E^4\\
&=&n(n-1)(n-2)(n-3)p^4+6n(n-1)(n-2)p^3+7n(n-1)p^2+np\\
&&\!\!-4np[n(n-1)(n-2)p^3+3n(n-1)p^2+np]+6(np)^2(n(n-1)p^2+np)-3(np)^4\\
&=&np(1-p)(1+(3n-6)p(1-p))
\end{eqnarray*}

Observe now that with the convention $q=1-p$, our formulae for $M_3',M_4'$ read:
$$M_3'=npq(q-p)\quad,\quad M_4'=npq(1+(3n-6)pq)$$

Now by dividing respectively by $\sigma^3=V\sqrt{V}$ and $\sigma^4=V^2$, we obtain the formulae in the statement for the skewness $\gamma$ and the kurtosis $\kappa$, as desired.
\end{proof}

\section*{2d. Further laws}

At a more advanced level, still talking measure theory over finite spaces, we can talk about the hypergeometric laws, which are something interesting too, as follows:

\index{hypergeometric law}
\index{Vandermonde formula}

\begin{theorem}
Given a population of size $N\in\mathbb N$, with $m\in\{0,\ldots,N\}$ of these objects having a certain feature, the probability of having $s$ successes, when performing $p\in\{0,\ldots,N\}$ draws without replacement, and looking for that feature, is
$$P(s)=\frac{\binom{m}{s}\binom{N-m}{p-s}}{\binom{N}{p}}$$
with this being called hypergeometric law of parameters $(N,m,p)$. When doing the same thing, but with replacement, we obtain a binomial law of parameter $m/N$.
\end{theorem}

\begin{proof}
Many things can be said here, the idea being as follows:

\medskip

(1) To start with, the main assertion is clear, because we have a total of $\binom{N}{p}$ possibilities for our $p$ draws without replacement, and among these draws, those amounting to $s$ successes come by multiplying $\binom{m}{s}$, standing for the $s$ successes, and $\binom{N-m}{p-s}$, standing for the $p-s$ fails. Thus, we are led to the formula of $P(s)$ in the statement.

\medskip

(2) Regarding now the fact that we have indeed a probability measure, this is something which is plainly clear, because when performing our $p$ draws, something must happen, I mean we have a well-defined number of successes $s\in\mathbb N$, and $\sum_sP(s)=1$.

\medskip

(3) However, let us comment more on this. To start with, we must have:
$$0\leq s\leq m\quad,\quad 0\leq p-s\leq N-m$$

In other words, the number of successes $s\in\mathbb N$, where $P(s)>0$, is subject to:
$$s\in\{\max(0,p+m-N),\ldots,\min(m,p)\}$$

(4) Getting now to the mass 1 formula $\sum_sP(s)=1$, this looks in practice as follows, with the bounds for the summing parameter $s$ being those found above:
$$\sum_s\binom{m}{s}\binom{N-m}{p-s}=\binom{N}{p}$$

But this comes from the Vandermonde formula, which is as follows, coming from $(1+x)^m(1+x)^n=(1+x)^{m+n}$, by looking at the coefficient of $x^p$, on both sides:
$$\sum_{s=0}^p\binom{m}{s}\binom{n}{p-s}=\binom{m+n}{p}$$

(5) Finally, in what regards the last assertion of the theorem, assume that we are doing our $p$ draws, but this time with replacement. In this case, for having $s$ successes, we have $\binom{p}{s}$ choices for the occurences of these successes, and then $(m/N)^s$ probability for the successes, and $(1-m/N)^{n-s}$ probability for the fails, so we obtain:
$$P(s)=\binom{p}{s}\left(\frac{m}{N}\right)^s\left(1-\frac{m}{N}\right)^{n-s}$$

But this is a binomial law of parameter $m/N$, as stated.
\end{proof}

Still at the theoretical level, as an interesting observation, we have:

\begin{proposition}
The hypergeometric law of parameters $(N,m,p)$, given by
$$P(s)=\frac{m!p!(N-m)!(N-p)!}{s!(m-s)!(p-s)!N!(N-m-p+s)!}$$
is symmetric in $m,p$, so is equal to the hypergeometric law of parameters $(N,p,m)$.
\end{proposition}

\begin{proof}
This is indeed something self-explanatory, coming from definitions.
\end{proof}

Getting now to our usual business, computation of moments, regarding the mean and variance of the hypergeometric laws, these are as follows:

\begin{theorem}
The mean and variance of the hypergeometric laws are
$$E=\frac{mp}{N}\quad,\quad V=\frac{mp(N-m)(N-p)}{N^2(N-1)}$$
with $(N,m,p)$ standing as usual for the parameters.
\end{theorem}

\begin{proof}
The computations are similar to those for the binomial laws, as follows:

\medskip

(1) Regarding the mean, we can compute it is as follows, by making use of the Vandermonde formula discussed above, that is, by making use of $\sum_sP(s)=1$, for a certain related hypergeometric law, with slightly different parameters:
\begin{eqnarray*}
E
&=&\frac{1}{\binom{N}{p}}\sum_ss\binom{m}{s}\binom{N-m}{p-s}\\
&=&\frac{1}{\binom{N}{p}}\sum_sm\binom{m-1}{s-1}\binom{N-m}{p-s}\\
&=&\frac{m}{\binom{N}{p}}\binom{N-1}{p-1}\\
&=&\frac{mp}{N}
\end{eqnarray*}

(2) Next, with the same trick we can compute $M_2-M_1$, as follows:
\begin{eqnarray*}
M_2-M_1
&=&\frac{1}{\binom{N}{p}}\sum_ss(s-1)\binom{m}{s}\binom{N-m}{p-s}\\
&=&\frac{1}{\binom{N}{p}}\sum_sm(m-1)\binom{m-2}{s-2}\binom{N-m}{p-s}\\
&=&\frac{m(m-1)}{\binom{N}{p}}\binom{N-2}{p-2}\\
&=&\frac{mp(m-1)(p-1)}{N(N-1)}
\end{eqnarray*}

(3) Thus, the second moment is given by the following formula:
\begin{eqnarray*}
M_2
&=&\frac{mp(m-1)(p-1)}{N(N-1)}+\frac{mp}{N}\\
&=&\frac{mp}{N}\left(\frac{(m-1)(p-1)}{N-1}+1\right)\\
&=&\frac{mp(mp-m-p+N)}{N(N-1)}
\end{eqnarray*}

(4) We conclude that the variance is given by the following formula:
\begin{eqnarray*}
V
&=&\frac{mp(mp-m-p+N)}{N(N-1)}-\frac{m^2p^2}{N^2}\\
&=&\frac{mp}{N}\left(\frac{mp-m-p+N}{N-1}-\frac{mp}{N}\right)\\
&=&\frac{mp(N-m)(N-p)}{N^2(N-1)}
\end{eqnarray*}

Thus, we are led to the conclusions in the statement.
\end{proof}

In order to solve the moment problem, in general, we can use some algebra. Let us regard the symmetric group $S_N$ as being the permutation group of the $N$ coordinate axes of $\mathbb R^N$. We obtain an embedding $S_N\subset O_N$, given by the following formula:
$$\sigma(e_i)=e_{\sigma(i)}$$

Now let us look at the corresponding $N^2$ coordinate functions $S_N\subset O_N\subset M_N(\mathbb R)$. According to our formula above, these are the functions $u_{ij}:S_N\to\{0,1\}$ given by:
$$u_{ij}(\sigma)=\delta_{\sigma(j)i}$$

These $N^2$ coordinate functions form a $N\times N$ square matrix $u=(u_{ij})$, and given two integers $m,p\leq N$, we can write this matrix as follows:
$$u=\begin{pmatrix}
u_{11}&\ldots&u_{1p}&\ldots&\ldots&u_{1N}\\
\vdots&\bigstar&\vdots&&&\vdots\\
u_{m1}&\ldots&u_{mp}&\ldots&\ldots&u_{mN}\\
\vdots&&\vdots&&&\vdots\\
\vdots&&\vdots&&&\vdots\\
u_{N1}&\ldots&u_{Np}&\ldots&\ldots&u_{NN}
\end{pmatrix}$$

The point now is that the sum of the entries of the rectangular matrix $\bigstar$ appearing in the upper left corner is hypergeometric, as shown by the following result:

\index{hypergeometric variable}

\begin{theorem}
The following variable over the symmetric group $S_N\subset O_N$,
$$S_{mp}=\sum_{i=1}^m\sum_{j=1}^pu_{ij}$$
is hypergeometric, having parameters $(N,m,p)$.
\end{theorem}

\begin{proof}
We know that the coordinates of the symmetric group, viewed as group of permutation matrices, $S_N\subset O_N$, are given by the following formula:
$$u_{ij}=\chi\left(\sigma\in S_N\Big|\sigma(j)=i\right)$$

Thus, the variable in the statement is given by the following formula:
$$S_{mp}=\sum_{i=1}^m\sum_{j=1}^p\chi\left(\sigma\in S_N\Big|\sigma(j)=i\right)$$

Let us compute now the law of this variable. We have the following formula:
\begin{eqnarray*}
P(S_{mp}=s)
&=&\frac{1}{N!}\,\#\left(\sigma\in S_N\Big|S_{mp}(\sigma)=s\right)\\
&=&\frac{1}{N!}\,\#\left(\sigma\in S_N\Big|\#\left\{i\leq m,j\leq p\big|\sigma(j)=i\right\}=s\right)\\
&=&\frac{1}{N!}\,\#\left(\sigma\in S_N\Big|\#\left\{j\leq p\big|\sigma(j)\leq m\right\}=s\right)
\end{eqnarray*}

Now let us try to count the permutations on the right. Such a permutation $\sigma\in S_N$ comes from 5 operations involved, as follows:

\medskip

(1) We first have to pick a subset $X\subset\{1,\ldots,p\}$, with $|X|=s$.

\medskip

(2) We also have to pick a subset $Y\subset\{1,\ldots,m\}$, with $|Y|=s$.

\medskip

(3) Then, we have to bijectively map $X\to Y$.

\medskip

(4) Next, we have to map injectively $\{1,\ldots,p\}-X\to\{m+1,\ldots,N\}$.

\medskip

(5) And finally, we have to bijectively map $\{p+1,\ldots,N\}$ to what is left.

\medskip

Regarding now the precise numbers of choices for these 5 operations involved, these are easy to compute, as follows:

\medskip

(1) Here we have $\binom{p}{s}$ choices.

\medskip

(2) Here we have $\binom{m}{s}$ choices.

\medskip

(3) Here we have $s!$ choices.

\medskip

(4) Here we have $\frac{(N-m)!}{(N-m-p+s)!}$ choices.

\medskip

(5) Here we have $(N-p)!$ choices.

\medskip

We can now finish our probability computation started above, as follows:
\begin{eqnarray*}
P(S_{mp}=s)
&=&\frac{1}{N!}\binom{p}{s}\binom{m}{s}s!\,\frac{(N-m)!}{(N-m-p+s)!}(N-p)!\\
&=&\frac{1}{N!}\cdot\frac{p!}{s!(p-s)!}\cdot\frac{m!}{s!(m-s)!}\cdot s!\cdot\frac{(N-m)!}{(N-m-p+s)!}(N-p)!\\
&=&\frac{p!m!(N-m)!(N-p)!}{N!s!(p-s)!(m-s)!(N-m-p+s)!}\\
&=&\frac{m!}{s!(m-s)!}\cdot\frac{(N-m)!}{(p-s)!(N-m-p+s)!}\cdot\frac{p!(N-p)!}{N!}\\
&=&\frac{\binom{m}{s}\binom{N-m}{p-s}}{\binom{N}{p}}
\end{eqnarray*}

We are therefore led to the conclusion in the statement.
\end{proof}

As a first observation, the above trivializes the symmetry observation from Proposition 2.32. In order to discuss now the moment problematics, we will need:

\begin{proposition}
The integrals over the symmetric group $S_N$ are given by
$$\int_{S_N}u_{i_1j_1}\ldots u_{i_kj_k}=\begin{cases}
\frac{(N-|\ker i|)!}{N!}&{\rm if}\ \ker i=\ker j\\
0&{\rm otherwise}
\end{cases}$$
where $\ker i$ denotes as usual the partition of $\{1,\ldots,k\}$ whose blocks collect the equal indices of $i$, and where $|.|$ denotes the number of blocks.
\end{proposition}

\begin{proof}
Observe first that the formula in the statement computes indeed all the integrals over $S_N$, and this due to Stone-Weierstrass, because the standard coordinates $u_{ij}$ separate the points of $S_N$. Regarding now the proof of the formula, according to the definition of the coordinates $u_{ij}$, the integrals in the statement are given by:
$$\int_{S_N}u_{i_1j_1}\ldots u_{i_kj_k}=\frac{1}{N!}\#\left\{\sigma\in S_N\Big|\sigma(j_1)=i_1,\ldots,\sigma(j_k)=i_k\right\}$$

Now observe that the existence of $\sigma\in S_N$ as above requires:
$$i_m=i_n\iff j_m=j_n$$

Thus, the above integral vanishes when the following condition is satisfied:
$$\ker i\neq\ker j$$

Regarding now the case $\ker i=\ker j$, if we denote by $b\in\{1,\ldots,k\}$ the number of blocks of this partition $\ker i=\ker j$, we have $N-b$ points to be sent bijectively to $N-b$ points, and so $(N-b)!$ solutions, and the integral is $\frac{(N-b)!}{N!}$, as claimed.
\end{proof}

Getting back now to the hypergeometric variables, we have:

\index{moments of hypergeometric law}

\begin{theorem}
For the hypergeometric law of parameters $(N,m,p)$ we have
$$M_k=\sum_{\pi\in P(k)}\frac{m!}{(m-|\pi|)!}\cdot\frac{p!}{(p-|\pi|)!}\cdot\frac{(N-|\pi|)!}{N!}$$
where $|.|$ denotes as usual the number of blocks of partitions, and
$$E=\frac{mp}{N}\quad,\quad V=\frac{mp(N-m)(N-p)}{N^2(N-1)}\quad,\quad 
\gamma=\frac{(N-2m)(N-2p)\sqrt{N-1}}{(N-2)\sqrt{mp(N-m)(N-p)}}$$
$$\kappa=3+\frac{\begin{pmatrix}N^2(N-1)[N(N+1)-6m(N-m)-6p(N-p)]\\
+6mp(N-m)(N-p)(5N-6)\end{pmatrix}}{mp(N-m)(N-p)(N-2)(N-3)}$$
and the corresponding mean, variance, skewness and kurtosis.
\end{theorem}

\begin{proof}
We have the following computation, using the model in Theorem 2.34:
\begin{eqnarray*}
M_k
&=&\int_{S_N}\left(\sum_{i=1}^m\sum_{j=1}^pu_{ij}\right)^k\\
&=&\int_{S_N}\sum_{i_1=1}^m\ldots\sum_{i_k=1}^m\sum_{j_1=1}^p\ldots\sum_{j_k=1}^pu_{i_1j_1}\ldots u_{i_kj_k}\\
&=&\sum_{i_1=1}^m\ldots\sum_{i_k=1}^m\sum_{j_1=1}^p\ldots\sum_{j_k=1}^p\int_{S_N}u_{i_1j_1}\ldots u_{i_kj_k}\\
&=&\sum_{i_1=1}^m\ldots\sum_{i_k=1}^m\sum_{j_1=1}^p\ldots\sum_{j_k=1}^p\delta_{\ker i,\ker j}\frac{(N-|\ker i|)!}{N!}\\
&=&\sum_{\pi\in P(k)}\sum_{i_1=1}^m\ldots\sum_{i_k=1}^m\sum_{j_1=1}^p\ldots\sum_{j_k=1}^p\delta_{\ker i,\ker j,\pi}\frac{(N-|\pi|)!}{N!}\\
&=&\sum_{\pi\in P(k)}\left(\sum_{i_1=1}^m\ldots\sum_{i_k=1}^m\delta_{\ker i,\pi}\right)
\left(\sum_{j_1=1}^p\ldots\sum_{j_k=1}^p\delta_{\ker j,\pi}\right)\frac{(N-|\pi|)!}{N!}\\
&=&\sum_{\pi\in P(k)}\frac{m!}{(m-|\pi|)!}\cdot\frac{p!}{(p-|\pi|)!}\cdot\frac{(N-|\pi|)!}{N!}
\end{eqnarray*}

As for the numerics, I will leave them to you as an exercise, enjoy.
\end{proof}

\section*{2e. Exercises}

We had a lot on interesting theory in this chapter, and as exercises, we have:

\begin{exercise}
Do some computations for card games, that we forgot to talk about.
\end{exercise}

\begin{exercise}
Further meditate at the symmetry question, for biased coins and dice.
\end{exercise}

\begin{exercise}
Learn more about the Dirac masses, from the physicists.
\end{exercise}

\begin{exercise}
Think a bit at measures, and at the push-forward formula $\mu=f_*\nu$.
\end{exercise}

\begin{exercise}
Learn the binomial formula with arbitrary real exponents.
\end{exercise}

\begin{exercise}
Meditate on the variance $V$, and its probabilistic meaning.
\end{exercise}

\begin{exercise}
Learn more about the hypergeometric laws, and their occurrences.
\end{exercise}

\begin{exercise}
Compute the skewness and kurtosis of the hypergeometric laws.
\end{exercise}

As bonus exercise, set up a small business, with the theory that you learned here.

\chapter{Discrete laws}

\section*{3a. Geometric laws}

We have seen that there is life in finite probability, $|X|<\infty$, with the coin tosses and other Bernoulli trials producing several interesting laws. Also, we have seen that the moment mathematics of these laws can be something non-trivial, leading us into subtle combinatorial objects, like partitions $\pi\in P(k)$ and permutations $\sigma\in S_N$.

\bigskip

In this chapter we explore what happens when allowing the discrete space $X$ to be infinite and countable, as for instance are the spaces $X=\mathbb N$ and $X=\mathbb Z$. We will see that several interesting new laws appear, with again their moment combinatorics leading us into partitions $\pi\in P(k)$ and permutations $\sigma\in S_N$. All this knowledge will be of course useful later, when fighting with measure theory and probability, in the general case.

\bigskip

Let us start with an interesting ``negative'' analogue of the binomial law, again obtained by flipping coins, but having this time the whole $\mathbb N$ as support:

\begin{theorem}
When flipping a $p$-biased coin until reaching to $m$ heads,
$$P(s\ {\rm tails})=\binom{s+m-1}{s}(1-p)^sp^m$$
with this being called negative binomial law $c_{mp}$ of parameters $m\in\mathbb N$ and $p\in[0,1]$.
\end{theorem}

\begin{proof}
When computing $P(s\ {\rm tails})$, we certainly have the $(1-p)^sp^m$ factor, coming from the $m$ heads and $s$ tails. As for the multiplicity, this is the binomial coefficient in the statement, coming from choosing the positions of the $s$ tails, among the total of $s+m-1$ attempts, up to the last one, which does not count, because it must be heads. 
\end{proof}

The terminology in the above result comes from the following fact:

\index{Pascal law}
\index{negative binomial law}
\index{generalized binomial coefficient}

\begin{proposition}
The negative binomial law $c_{mp}$ is best viewed as
$$P(s)=\binom{-m}{s}(p-1)^sp^m$$
with $\binom{-m}{s}$ being a generalized binomial coefficient.
\end{proposition}

\begin{proof}
To start with, let us recall that the usual binomial formula generalizes in the following way, to the case of arbitrary exponents $n\in\mathbb R$, and variables $|x|<1$:
$$(1+x)^n=\sum_{s\geq0}\binom{n}{s}x^s\qquad,\qquad\binom{n}{s}=\frac{n(n-1)\ldots(n-s+1)}{s!}$$

To be more precise, what we have here is the Taylor series of $(1+x)^n$, and the fact that this Taylor approximation is exact, at any $|x|<1$, comes for instance from the fact that the derivative of the function $f(x)=(1+x)^{-n}S(x)$, with $S(x)$ being the above series, vanishes. Now in relation with what our statement says, this comes from:
\begin{eqnarray*}
\binom{-m}{s}
&=&\frac{-m(-m-1)\ldots(-m-s+1)}{s!}\\
&=&(-1)^s\,\frac{m(m+1)\ldots(m+s-1)}{s!}\\
&=&(-1)^s\binom{s+m-1}{s}
\end{eqnarray*}

Observe now that in this picture, the mass 1 property is obvious, coming from:
\begin{eqnarray*}
\sum_{s\geq0}P(s)
&=&\sum_{s\geq0}\binom{-m}{s}(p-1)^sp^m\\
&=&p^m\sum_{s\geq0}\binom{-m}{s}(p-1)^s\\
&=&p^m[1+(p-1)]^{-m}\\
&=&p^mp^{-m}\\
&=&1
\end{eqnarray*}

Thus, we are led to the conclusions in the statement.
\end{proof}

In analogy now with the formula $b_{np}=b_p^{*n}$ for the usual binomial laws, we have:

\begin{theorem}
The negative binomial laws $c_p=c_{1p}$ of exponent $m=1$ are
$$P(s)=(1-p)^sp$$
with these being called geometric laws, and we have $c_{mp}=c_p^{*m}$, for any $m\in\mathbb N$.
\end{theorem}

\begin{proof}
The first assertion is clear from definitions, and with the name geometric laws coming from the geometric series which produces them, namely:
$$\sum_{s\geq0}P(s)=\sum_{s\geq0}(1-p)^sp=\frac{p}{1-(1-p)}=1$$

As for the second assertion, this comes from the following computation:
\begin{eqnarray*}
c_p^{*m}
&=&\left(\sum_{a_1\geq0}(1-p)^{a_1}p\delta_{a_1}\right)*\ldots*\left(\sum_{a_m\geq0}(1-p)^{a_m}p\delta_{a_m}\right)\\
&=&\sum_{a_1,\ldots,a_m\geq0}(1-p)^{a_1+\ldots+a_m}p^m\delta_{a_1+\ldots+a_m}\\
&=&\sum_{s\geq0}\#\left\{a_1,\ldots,a_m\geq0\Big|a_1+\ldots+a_m=s\right\}(1-p)^sp^m\delta_s\\
&=&\sum_{s\geq0}\#\left\{b_1,\ldots,b_m\geq1\Big|b_1+\ldots+b_m=s+m\right\}(1-p)^sp^m\delta_s\\
&=&\sum_{s\geq0}\binom{s+m-1}{m-1}(1-p)^sp^m\delta_s\\
&=&\sum_{s\geq0}\binom{s+m-1}{s}(1-p)^sp^m\delta_s\\
&=&c_{mp}
\end{eqnarray*}

To be more precise here, the count at the end comes from the fact that, in order to partition $s+m$, best thought as a sequence of $s+m$ boxes, as $s+m=b_1+\ldots+b_m$, we have to insert $m-1$ markers between these boxes, at different locations, at the $s+m-1$ positions available. Thus, we are led to the conclusion in the statement.
\end{proof}

Regarding now the moments, again inspired from what we did before for the usual binomial laws, we have the following result, making appear partitions:

\begin{theorem}
The moments of $c_{mp}$ are given by the following formula,
$$M_k=\sum_{\pi\in P(k)}\frac{(m+|\pi|-1)!}{(m-1)!}\left(\frac{1-p}{p}\right)^{|\pi|}$$
with $|.|$ standing as usual for the number of blocks of partitions, and 
$$E=\frac{m(1-p)}{p}\ \ ,\  \ V=\frac{m(1-p)}{p^2}\ \ ,\ \ \gamma=\frac{2-p}{\sqrt{m(1-p)}}\ \ ,\ \ \kappa=3+\frac{6}{m}+\frac{p^2}{m(1-p)}$$
are the corresponding mean, variance, skewness and kurtosis.
\end{theorem}

\begin{proof}
Regarding the mean, this can be computed as follows:
\begin{eqnarray*}
E
&=&\sum_{s\geq1}s\binom{-m}{s}(p-1)^sp^m\\
&=&\sum_{s\geq1}-m\binom{-m-1}{s-1}(p-1)^sp^m\\
&=&-mp^m(p-1)\sum_{s\geq1}\binom{-m-1}{s-1}(p-1)^{s-1}\\
&=&mp^m(1-p)[1+(p-1)]^{-m-1}\\
&=&mp^m(1-p)p^{-m-1}\\
&=&\frac{m(1-p)}{p}
\end{eqnarray*}

With a similar trick, we can compute the difference $M_2-M_1$, as follows:
\begin{eqnarray*}
M_2-M_1
&=&\sum_{s\geq2}s(s-1)\binom{-m}{s}(p-1)^sp^m\\
&=&\sum_{s\geq2}m(m+1)\binom{-m-2}{s-2}(p-1)^sp^m\\
&=&m(m+1)p^m(p-1)^2\sum_{s\geq2}\binom{-m-2}{s-2}(p-1)^{s-2}\\
&=&m(m+1)p^m(1-p)^2[1+(p-1)]^{-m-2}\\
&=&m(m+1)p^m(1-p)^2p^{-m-2}\\
&=&\frac{m(m+1)(1-p)^2}{p^2}
\end{eqnarray*}

We conclude that the second moment is given by the following formula:
$$M_2=\frac{m(1-p)(1+m-mp)}{p^2}$$

Thus, we are led to the variance formula in the statement, namely:
$$V=\frac{m(1-p)(1+m-mp)}{p^2}-\frac{m^2(1-p)^2}{p^2}=\frac{m(1-p)}{p^2}$$

In order to compute now the higher moments, with the same trick, we can use the following formula, that we know well since chapter 2, from our study there of $b_{np}$:
$$s^k=\sum_{\pi\in P(k)}\frac{s!}{(s-|\pi|)!}$$

Indeed, with this as input for our moment computations, we are led to:
$$M_k=\sum_{\pi\in P(k)}\frac{(m+|\pi|-1)!}{(m-1)!}\left(\frac{1-p}{p}\right)^{|\pi|}$$

Getting now to $M_3,M_4$, these do not look very good, but their central versions do:
$$M_3'=\frac{m(1-p)(2-p)}{p^3}$$
$$M_4'=\frac{m(1-p)((3m+6)(1-p)+p^2)}{p^4}$$

And with this, we are led to the formulae of $\gamma,\kappa$ in the statement, as desired.
\end{proof}

Next, we can talk as well about negative hypergeometric laws, as follows:

\index{negative hypergeometric law}
\index{negative binomial law}

\begin{theorem}
Given a population of size $N\in\mathbb N$, with $m\in\{0,\ldots,N\}$ of these objects having a certain feature, the probability of having $s$ successes, when performing draws without replacement, and stopping after $r\in\{0,\ldots,N-m\}$ failures, is
$$P(s)=\frac{\binom{s+r-1}{s}\binom{N-r-s}{m-s}}{\binom{N}{m}}$$
with this being called negative hypergeometric law of parameters $(N,m,r)$. When doing this with replacement, we obtain a negative binomial law of parameter $m/N$.
\end{theorem}

\begin{proof}
Many things can be said here, the idea being as follows:

\medskip

(1) Observe first the obvious similarity with the definition of the usual hypergeometric laws, from chapter 2. To be more precise, the common part is that, in both these cases, we have a population of size $N\in\mathbb N$, with $m\in\{0,\ldots,N\}$ of these objects having a certain feature, and we are looking for the probability of having $s$ successes, when performing draws with replacement. The difference comes from the stopping convention, for the hypergeometric laws this being after $p\in\{0,\ldots,N\}$ draws, and for the negative hypergeometric laws this being after $r\in\{0,\ldots,N-m\}$ failures.

\medskip

(2) In practice, based on this, we can develop the theory of negative hypergeometric laws by using the previously developed theory of the usual hypergeometric laws. Indeed, to start with, the probability $P(s)$ that we are interested in comes by multiplying a probability coming from a usual hypergeometric law, corresponding to the $m$ succeses in the first $m+r-1$ draws, and a basic fraction, corresponding to the failure required at the $(m+r)$-th draw. In practice, this gives the following formula, as desired:
\begin{eqnarray*}
P(s)
&=&\frac{\binom{m}{s}\binom{N-m}{s+r-1-s}}{\binom{N}{s+r-1}}\cdot\frac{N-m-(r-1)}{N-(s+r-1)}\\
&=&\frac{\binom{m}{s}\binom{N-m}{r-1}}{\binom{N}{s+r-1}}\cdot\frac{N-m-r+1}{N-s-r+1}\\
&=&\frac{m!(N-m)!(s+r-1)!(N-s-r+1)!(N-m-r+1)}{s!(m-s)!(r-1)!N!(N-m-r+1)!(N-s-r+1)}\\
&=&\frac{m!(N-m)!(s+r-1)!(N-s-r)!}{s!(m-s)!(r-1)!N!(N-m-r)!}\\
&=&\frac{(s+r-1)!(N-r-s)!(N-m)!m!}{s!(r-1)!(m-s)!(N-r-m)!N!}\\
&=&\frac{\binom{s+r-1}{s}\binom{N-r-s}{m-s}}{\binom{N}{m}}
\end{eqnarray*}

(3) Regarding now the fact that we have indeed a probability measure, $\sum_sP(s)=1$, this is something which is plainly clear from definitions, but which can be deduced as well from the Vandermonde formula, along with the identity $\binom{n}{k}=(-1)^k\binom{k-n-1}{k}$:
\begin{eqnarray*}
\sum_s\binom{s+r-1}{s}\binom{N-r-s}{m-s}
&=&\sum_s(-1)^s\binom{-r}{s}(-1)^{m-s}\binom{m+r-N-1}{m-s}\\
&=&(-1)^m\sum_s\binom{-r}{s}\binom{m+r-N-1}{m-s}\\
&=&(-1)^m\binom{m-N-1}{m}\\
&=&\binom{N}{m}
\end{eqnarray*}

(4) Finally, in what regards the second assertion of the theorem, assume that we are doing our draws, but this time with replacement. In this case, for having $s$ successes, we have $\binom{s+r-1}{s}$ choices for the occurences of these successes, and then $(m/N)^s$ probability for the successes, and $(1-m/N)^r$ probability for the fails, so we obtain:
$$P(s)=\binom{s+r-1}{s}\left(\frac{m}{N}\right)^s\left(1-\frac{m}{N}\right)^r$$

But this is a negative binomial law of parameter $m/N$, as stated.
\end{proof}

As a main result now regarding the negative hypergeometric laws, we have:

\begin{theorem}
For the negative hypergeometric law of parameters $(N,m,r)$,
$$M_k=\sum_{\pi\in P(k)}\frac{m!}{(m-|\pi|)!}\cdot\frac{(r+|\pi|-1)!}{(r-1)!}\cdot\frac{(N-m)!}{(N-m+|\pi|)!}$$
where $|.|$ denotes as usual the number of blocks of partitions. Also, we have
$$E=\frac{mr}{N-m+1}\quad,\quad V=\frac{mr(N+1)(N-m-r+1)}{(N-m+1)^2(N-m+2)}$$
$$\gamma=\frac{(N+m+1)(N-m-2r+1)}{N-m+3}\sqrt{\frac{N-m+2}{mr(N+1)(N-m-r+1)}}$$
and with the notations $c=N-m+1$ and $d=r(N-m-r+1)$, 
$$\kappa=\frac{(c+1)(c^2[c^2+(6m-1)c+6m^2]+3d[(m-2)c^2+m(m-6)c-6m^2])}{md(c+2)(c+3)(c+m)}$$
is the corresponding kurtosis.
\end{theorem}

\begin{proof}
This is something quite standard, the idea being that the formula in the statement follows by recurrence, by using our standard trick, namely: 
$$s^k=\sum_{\pi\in P(k)}\frac{s!}{(s-|\pi|)!}$$

We will leave the computations here, as well as the remaining discussion, regarding the mean, variance, skewness and kurtosis, as an instructive exercise. Enjoy.
\end{proof}

\section*{3b. Poisson limits}

We would like to discuss now the central objects in discrete probability, which are the Poisson laws $p_t$, appearing via the Poisson Limit Theorem (PLT). Let us start with the following definition, which is something quite natural, in view of $e^t=\sum_kt^k/k!$:

\index{Poisson law}
\index{Poisson variable}

\begin{definition}
The Poisson law of parameter $t>0$ is the following measure,
$$p_t=e^{-t}\sum_{k\geq0}\frac{t^k}{k!}\,\delta_k$$
with the letter ``p'' standing for Poisson.
\end{definition}

We will see in the moment why the above laws appear a bit everywhere, in discrete contexts, the reasons for this coming from the Poisson Limit Theorem. Let us first develop some straightforward general theory. We first have the following result:

\index{mean of Poisson law}
\index{variance of Poisson law}

\begin{proposition}
The mean and variance of $p_t$ are given by:
$$E=t\quad,\quad V=t$$
In particular for the Poisson law $p_1$ we have $E=1,V=1$.
\end{proposition}

\begin{proof}
Regarding the mean, this can be computed as follows:
$$E
=e^{-t}\sum_{k\geq0}\frac{t^k}{k!}\cdot k
=e^{-t}\sum_{k\geq1}\frac{t^k}{(k-1)!}
=te^{-t}\sum_{l\geq0}\frac{t^l}{l!}
=t$$

For the variance, we can first compute the second moment, as follows:
\begin{eqnarray*}
M_2
&=&e^{-t}\sum_{k\geq1}\frac{t^kk}{(k-1)!}\\
&=&te^{-t}\sum_{l\geq0}\frac{t^ll}{l!}+te^{-t}\sum_{l\geq0}\frac{t^l}{l!}\\
&=&te^{-t}\sum_{l\geq1}\frac{t^l}{(l-1)!}+t\\
&=&t^2+t
\end{eqnarray*}

Thus the variance is given by $V=(t^2+t)-t^2=t$, as claimed.
\end{proof}

Next, we have the following quite conceptual result, in relation with convolution:

\index{convolution semigroup}

\begin{theorem}
We have the following formula, for any $s,t>0$,
$$p_s*p_t=p_{s+t}$$
so the Poisson laws form a convolution semigroup.
\end{theorem}

\begin{proof}
By using $\delta_k*\delta_l=\delta_{k+l}$ and the binomial formula, we obtain:
\begin{eqnarray*}
p_s*p_t
&=&e^{-s}\sum_{k\geq0}\frac{s^k}{k!}\,\delta_k*e^{-t}\sum_{l\geq0}\frac{t^l}{l!}\,\delta_l\\
&=&e^{-s-t}\sum_{n\geq0}\delta_n\sum_{k+l=n}\frac{s^kt^l}{k!l!}\\
&=&e^{-s-t}\sum_{n\geq0}\frac{\delta_n}{n!}\sum_{k+l=n}\frac{n!}{k!l!}s^kt^l\\\
&=&e^{-s-t}\sum_{n\geq0}\frac{(s+t)^n}{n!}\,\delta_n\\
&=&p_{s+t}
\end{eqnarray*}

Thus, we are led to the conclusion in the statement.
\end{proof}

Next in line, we have the following result, which is fundamental as well:

\index{convolution exponential}
\index{formal exponential}

\begin{theorem}
The Poisson laws appear as formal exponentials
$$p_t=\sum_{k\geq0}\frac{t^k(\delta_1-\delta_0)^{*k}}{k!}$$
with respect to the convolution of measures $*$.
\end{theorem}

\begin{proof}
By using the binomial formula, the measure on the right is:
\begin{eqnarray*}
\mu
&=&\sum_{k\geq0}t^k\sum_{r+s=k}(-1)^s\frac{\delta_r}{r!s!}\\
&=&\sum_{r\geq0}\frac{t^r\delta_r}{r!}\sum_{s\geq0}\frac{(-1)^s}{s!}\\
&=&\frac{1}{e}\sum_{r\geq0}\frac{t^r\delta_r}{r!}\\
&=&p_t
\end{eqnarray*}

Thus, we are led to the conclusion in the statement.
\end{proof}

Getting now to what we wanted to do, Poisson limits, it is convenient to use, as our main tool for dealing with independence, the Fourier transform, coming via:

\index{independence}
\index{Fourier transform}

\begin{theorem}
Assuming that $f,g:X\to\mathbb R$ are independent, we have
$$F_{f+g}=F_fF_g$$
where $F_f(x)=E(e^{ixf})$ is the Fourier transform.
\end{theorem}

\begin{proof}
We have indeed the following computation, using $\mu_{f+g}=\mu_f*\mu_g$:
\begin{eqnarray*}
F_{f+g}(x)
&=&\int_Xe^{ixz}d\mu_{f+g}(z)\\
&=&\int_Xe^{ixz}d(\mu_f*\mu_g)(z)\\
&=&\int_{X\times X}e^{ix(z+t)}d\mu_f(z)d\mu_g(t)\\
&=&\int_Xe^{ixz}d\mu_f(z)\int_Xe^{ixt}d\mu_g(t)\\
&=&F_f(x)F_g(x)
\end{eqnarray*}

Thus, we are led to the conclusion in the statement.
\end{proof}

Regarding the Fourier transform computation for $p_t$, this is as follows:

\index{Fourier transform}

\begin{theorem}
The Fourier transform of $p_t$ is given by
$$F_{p_t}(x)=\exp\left((e^{ix}-1)t\right)$$
for any $t>0$, with its logarithm being linear in $t$, as it should.
\end{theorem}

\begin{proof}
We have indeed the following computation, which gives the result:
\begin{eqnarray*}
F_{p_t}(x)
&=&e^{-t}\sum_{k\geq0}\frac{t^k}{k!}F_{\delta_k}(x)\\
&=&e^{-t}\sum_{k\geq0}\frac{t^k}{k!}\,e^{ikx}\\
&=&e^{-t}\sum_{k\geq0}\frac{(e^{ix}t)^k}{k!}\\
&=&\exp(-t)\exp(e^{ix}t)\\
&=&\exp\left((e^{ix}-1)t\right)
\end{eqnarray*}

As for the last assertion, this is just a remark, to be related to Theorem 3.11.
\end{proof}

Good news, we can now establish the Poisson Limit Theorem, as follows:

\index{PLT}
\index{Poisson Limit Theorem}
\index{Bernoulli laws}
\index{Poisson limit}

\begin{theorem}[PLT]
We have the following convergence, in moments,
$$\left(\left(1-\frac{t}{n}\right)\delta_0+\frac{t}{n}\,\delta_1\right)^{*n}\to p_t$$
for any $t>0$. Equivalently, $b_{np}\to p_t$, when $p=t/n$ with $t>0$ fixed.
\end{theorem}

\begin{proof}
Let us denote by $\mu_n$ the Bernoulli law under the convolution sign:
$$\mu_n=\left(1-\frac{t}{n}\right)\delta_0+\frac{t}{n}\,\delta_1$$

We have the following computation, for the Fourier transform of the limit: 
\begin{eqnarray*}
F_{\delta_r}(x)=e^{irx}
&\implies&F_{\mu_n}(x)=\left(1-\frac{t}{n}\right)+\frac{t}{n}\,e^{ix}\\
&\implies&F_{\mu_n^{*n}}(x)=\left(\left(1-\frac{t}{n}\right)+\frac{t}{n}\,e^{ix}\right)^n\\
&\implies&F_{\mu_n^{*n}}(x)=\left(1+\frac{(e^{ix}-1)t}{n}\right)^n\\
&\implies&F(x)=\exp\left((e^{ix}-1)t\right)
\end{eqnarray*}

Thus, we obtain indeed the Fourier transform of $p_t$, as desired.
\end{proof}

We have as well a negative Poisson Limit Theorem, as follows:

\index{PLT}
\index{Poisson Limit Theorem}
\index{Bernoulli laws}
\index{Poisson limit}
\index{negative PLT}
\index{NPLT}
\index{geometric law}

\begin{theorem}[negative PLT]
We have the following convergence, in moments,
$$\left(\sum_{k\geq0}\left(\frac{t}{m+t}\right)^k\frac{m}{m+t}\,\delta_k\right)^{*m}\to p_t$$
for any $t>0$. Equivalently, $c_{mp}\to p_t$, when $p=m/(m+t)$ with $t>0$ fixed.
\end{theorem}

\begin{proof}
Let us denote by $\nu_m$ the geometric law under the convolution sign:
$$\nu_m=\sum_{k\geq0}\left(\frac{t}{m+t}\right)^k\frac{m}{m+t}\,\delta_k$$

We have the following computation, for the Fourier transform of the limit: 
\begin{eqnarray*}
F_{\delta_k}(x)=e^{ikx}
&\implies&F_{\nu_m}(x)=\sum_{k\geq0}\left(\frac{t}{m+t}\right)^k\frac{m}{m+t}\,e^{ikx}\\
&\implies&F_{\nu_m}(x)=\frac{m}{m+t-te^{ix}}\\
&\implies&F_{\nu_m^{*m}}(x)=\left(\frac{m}{m+t-te^{ix}}\right)^m\\
&\implies&F_{\nu_m^{*m}}(x)=\left(1+\frac{t(e^{ix}-1)}{m+t-te^{ix}}\right)^m\\
&\implies&F(x)=\exp\left(t(e^{ix}-1)\right)
\end{eqnarray*}

Thus, we obtain indeed the Fourier transform of $p_t$, as desired.
\end{proof}

In practice, Theorems 3.13 and 3.14 lead to many occurrences of the Poisson laws, in relation with continuous phenomena from the real life. For instance, we have:

\begin{theorem}
The number of chewing gum pieces on a single tile of a sidewalk
$$\xymatrix@R=5pt@C=7pt{
\ar@{-}[rrrrrrrrrr]\ar@{-}[dddddd]&&\ar@{-}[dddddd]&&\ar@{-}[dddddd]&&\ar@{-}[dddddd]&&\ar@{-}[dddddd]&&\ar@{-}[dddddd]\\
&\bullet&&&&\bullet&&\ \,\bullet\bullet\,\ &&&&\ldots\\
\ar@{-}[rrrrrrrrrr]&&&&&&&&&&\\
&&&\,\bullet\bullet\bullet\,&&&&\bullet&&\ \,\bullet\bullet\,\ &&\ldots\\
\ar@{-}[rrrrrrrrrr]&&&&&&&&&&\\
&\ \ \, \bullet\ \ \,&&&&\ \,\bullet\bullet\,\ &&\bullet&&&&\ldots\\
\ar@{-}[rrrrrrrrrr]&&&&&&&&&&
}$$
follows a Poisson law.
\end{theorem}

\begin{proof}
It is pretty much clear, by thinking a bit, that the law that we are looking for, that of the number of chewing gum pieces on a single tile of a sidewalk, must be a discrete law, supported by $\mathbb N$, and appearing as a suitable limit of binomial laws:
$$b_{np}\to\sum_{k\geq 0}c_k\delta_k$$

But, how does this limit exactly appear? In answer, since the total number of pieces of chewing gum must not blow up with $n\to\infty$, and more specifically, must remain approximately constant, we must impose the condition $p=t/n$, for some fixed $t>0$. But with this, we are led into the Poisson limiting formula from Theorem 3.13, namely:
$$\left(\left(1-\frac{t}{n}\right)\delta_0+\frac{t}{n}\,\delta_1\right)^{*n}\to p_t$$

Thus, we are led to the conclusion in the statement. And to the above picture too, the data there coming, up to an epsilon, from the Poisson law of parameter $t=1$.
\end{proof}

Regarding now the moments of the Poisson laws, the result here is as follows:

\index{number of blocks}
\index{Stirling numbers}

\begin{theorem}
The moments of $p_t$ with $t>0$ are given by:
$$M_k(p_t)=\sum_{\pi\in P(k)}t^{|\pi|}$$
In particular at $t=1$ we have $M_k(p_1)=|P(k)|$, the Bell numbers.
\end{theorem}

\begin{proof}
We have the following recurrence formula for the moments of $p_t$:
\begin{eqnarray*}
M_{k+1}
&=&e^{-t}\sum_{n\geq1}\frac{t^nn^{k+1}}{n!}\\
&=&e^{-t}\sum_{m\geq0}\frac{t^{m+1}(m+1)^k}{m!}\\
&=&e^{-t}\sum_{m\geq0}\frac{t^{m+1}m^k}{m!}\left(1+\frac{1}{m}\right)^k\\
&=&e^{-t}\sum_{m\geq0}\frac{t^{m+1}m^k}{m!}\sum_{s=0}^k\binom{k}{s}m^{-s}\\
&=&\sum_{s=0}^k\binom{k}{s}\cdot e^{-t}\sum_{m\geq0}\frac{t^{m+1}m^{k-s}}{m!}\\
&=&t\sum_{s=0}^k\binom{k}{s}M_{k-s}
\end{eqnarray*}

On the other hand, consider the numbers $S_k=\sum_{\pi\in P(k)}t^{|\pi|}$ in the statement. Since a partition of $\{1,\ldots,k+1\}$ appears by choosing $s$ partners for $1$, among the $k$ numbers available, and then partitioning the $k-s$ elements left, we have the following formula:
$$S_{k+1}=t\sum_{s=0}^k\binom{k}{s}S_{k-s}$$

Thus the numbers $M_k,S_k$ satisfy the same recurrence, and with the initial values matching, namely $M_1=S_1=t$ and $M_2=S_2=t+t^2$, we obtain $M_k=S_k$, as stated.
\end{proof}

As a question now that you might have, is it possible to simplify the moment combinatorics of the Poisson laws? In answer, we can use the following definition:

\begin{definition}
The factorial moments of $f:X\to\mathbb R$ are the numbers
$$\widetilde{M}_k=E[f(f-1)\ldots(f-k+1)]$$
with the convention $\widetilde{M}_0=1$.
\end{definition}

The point indeed is that the above manipulation ``kills the underlying partitions'', with the result for the Poisson laws being something remarkable, as follows:

\begin{theorem}
The factorial moments of the Poisson law $p_t$ are
$$\widetilde{M_k}=t^k$$
and with this being something infinitely simpler than the formula of $M_k$. 
\end{theorem}

\begin{proof}
We are in need of a conversion formula, connecting the usual and factorial moments. So, recall the following magic formula, valid for any number $s\in\mathbb N$:
$$s^k=\sum_{\pi\in P(k)}\frac{s!}{(s-|\pi|)!}$$

Now since this is an equality of polynomials in $s$, we can replace our $s\in\mathbb N$ by anything, and in particular, by an arbitrary random variable $f:X\to\mathbb R$. We obtain:
$$f^k=\sum_{\pi\in P(k)}f(f-1)\ldots(f-|\pi|+1)$$

By taking now the expectation, on both sides, we obtain the following conversion formula, connecting the usual and the factorial moments of a variable $f:X\to\mathbb R$:
$$M_k=\sum_{\pi\in P(k)}\widetilde{M}_{|\pi|}$$

But with this in hand, we can convert our formula in Theorem 3.16 for the moments of $p_t$ into a formula for the factorial moments, and we obtain in this way:
$$\widetilde{M_k}(p_t)=t^k$$

Thus, we are led to the conclusion in the statement.
\end{proof}

Let us end this discussion with the following result, capturing the essentials:

\begin{theorem}
For the Poisson law $p_t$ of parameter $t>0$,
$$E=t\quad,\quad  V=t\quad ,\quad \gamma=\frac{1}{\sqrt{t}}\quad ,\quad \kappa=3+\frac{1}{t}$$
are the mean, variance, skewness and kurtosis.
\end{theorem}

\begin{proof}
Regarding the formulae of $E,V$, we know these from Proposition 3.8. As for the formulae of $\gamma,\kappa$, these follow from Theorem 3.16. Indeed, that formula gives:
$$M_3=t+3t^2+t^3\quad,\quad 
M_4=t+7t^2+6t^3+t^4$$

By passing now to central moments, we obtain the following formulae:
$$M_3'=t\quad,\quad 
M_4'=t+3t^2$$

Thus, by normalizing, we are led to the formulae of $\gamma$ and $\kappa$ in the statement.
\end{proof}

\section*{3c. Derangements}

What is next? Some more advanced combinatorics, I guess, in relation with the Poisson laws. Let us start with the following result, which is something very useful:

\begin{proposition}
We have the following formula,
$$\left|\left(\bigcup_iA_i\right)^c\right|=|A|-\sum_i|A_i|+\sum_{i<j}|A_i\cap A_j|-\sum_{i<j<k}|A_i\cap A_j\cap A_k|+\ldots$$
for any subsets $A_i$ of a certain set $A$, called inclusion-exclusion principle.
\end{proposition}

\begin{proof}
This is indeed quite clear, by thinking a bit, as follows:

\medskip

(1) In order to count $(\cup_iA_i)^c$, we certainly have to start with $|A|$.

\medskip

(2) Then, we obviously have to remove each $|A_i|$, and so remove $\sum_i|A_i|$.

\medskip

(3) But then, we have to put back each $|A_i\cap A_j|$, and so put back $\sum_{i<j}|A_i\cap A_j|$.

\medskip

$\vdots$

\medskip

(4) And so on, which leads to the formula in the statement.
\end{proof}

Getting back now to probability, we first have the following key result:

\index{permutation}
\index{derangement}
\index{fixed points}
\index{random permutation}

\begin{theorem}
The probability for a random permutation $\sigma\in S_N$ to be a derangement, that is, to have no fixed points, is given by the following formula:
$$P=1-\frac{1}{1!}+\frac{1}{2!}-\frac{1}{3!}+\ldots+(-1)^N\frac{1}{N!}$$
Thus we have the estimate $P\simeq1/e$, in the $N\to\infty$ limit.
\end{theorem}

\begin{proof}
Consider the sets $S_N^i=\{\sigma\in S_N|\sigma(i)=i\}$. According to the above inclusion-exclusion principle, the probability that we are interested in is given by:
\begin{eqnarray*}
P
&=&\frac{1}{N!}\left(|S_N|-\sum_i|S_N^i|+\sum_{i<j}|S_N^i\cap S_N^j|-\sum_{i<j<k}|S_N^i\cap S_N^j\cap S_N^k|+\ldots\right)\\
&=&\frac{1}{N!}\sum_{k=0}^N(-1)^k\sum_{i_1<\ldots<i_k}|S_N^{i_1}\cap\ldots\cap S_N^{i_k}|\\
&=&\frac{1}{N!}\sum_{k=0}^N(-1)^k\binom{N}{k}(N-k)!\\
&=&\sum_{k=0}^N\frac{(-1)^k}{k!}
\end{eqnarray*}

Since at the end we have the expansion of $1/e$, we obtain the result.
\end{proof}

Let us discuss now, more generally, what happens when counting permutations having exactly $k$ fixed points. The result here, extending Theorem 3.21, is as follows:

\begin{theorem}
The probability for a random permutation $\sigma\in S_N$ to have exactly $k$ fixed points is given by the following formula:
$$P=\frac{1}{k!}\left(1-\frac{1}{1!}+\frac{1}{2!}-\ldots+(-1)^{N-1}\frac{1}{(N-1)!}+(-1)^N\frac{1}{N!}\right)$$
Thus we have the estimate $P\simeq1/(ek!)$, in the $N\to\infty$ limit.
\end{theorem}

\begin{proof}
We have to count the permutations $\sigma\in S_N$ having exactly $k$ points. Since having such a permutation amounts in choosing $k$ points among $1,\ldots,N$, and then permuting the $N-k$ points left, without fixed points allowed, we have:
\begin{eqnarray*}
\#\left\{\sigma\in S_N\Big|\chi(\sigma)=k\right\}
&=&\binom{N}{k}\#\left\{\sigma\in S_{N-k}\Big|\chi(\sigma)=0\right\}\\
&=&\frac{N!}{k!(N-k)!}\#\left\{\sigma\in S_{N-k}\Big|\chi(\sigma)=0\right\}\\
&=&N!\times\frac{1}{k!}\times\frac{\#\left\{\sigma\in S_{N-k}\Big|\chi(\sigma)=0\right\}}{(N-k)!}
\end{eqnarray*}

Now by dividing everything by $N!$, we obtain from this the following formula:
$$\frac{\#\left\{\sigma\in S_N\Big|\chi(\sigma)=k\right\}}{N!}=\frac{1}{k!}\times\frac{\#\left\{\sigma\in S_{N-k}\Big|\chi(\sigma)=0\right\}}{(N-k)!}$$

By using now the computation at $k=0$, from Theorem 3.21, it follows that with $N\to\infty$ we have the following estimate, $\chi$ being the number of fixed points:
$$P(\chi=k)
\simeq\frac{1}{k!}\cdot P(\chi=0)
\simeq\frac{1}{k!}\cdot\frac{1}{e}$$

Thus, we are led to the conclusion in the statement.
\end{proof}

Now observe that Theorem 3.22 reformulates into something very simple, namely:

\begin{theorem}
The number of fixed points of permutations, $\chi:S_N\to\mathbb N$ given by
$$\chi(\sigma)=\#\left\{i\in\{1,\ldots,N\}\Big|\sigma(i)=i\right\}$$ 
follows the Poisson law of parameter $1$, in the $N\to\infty$ limit.
\end{theorem}

\begin{proof}
We know from Theorem 3.22 that with $N\to\infty$ we have, for any $k\in\mathbb N$:
$$P(\chi=k)\simeq\frac{1}{ek!}$$

But this tells us precisely that $\chi:S_N\to\mathbb N$ is asymptotically Poisson (1), as stated.
\end{proof}

More generally now, and quite remarkably, we have in fact the following result:

\index{symmetric group}
\index{Poisson law}

\begin{theorem}
The number of partial fixed points, $\chi_t:S_N\to\mathbb N$ given by
$$\chi_t(\sigma)=\#\left\{i\in\{1,\ldots,[tN]\}\Big|\sigma(i)=i\right\}$$ 
follows the Poisson law of parameter $t\in(0,1]$, in the $N\to\infty$ limit.
\end{theorem}

\begin{proof}
We already know from Theorem 3.23 that the result holds at $t=1$. In general, the proof is similar, the idea being as follows:

\medskip

(1) In order to compute $P(\chi_t=0)$, consider the sets $S_N^i=\{\sigma\in S_N|\sigma(i)=i\}$. As before in the proof of Theorem 3.21, we obtain by inclusion-exclusion that:
\begin{eqnarray*}
P(\chi_t=0)
&=&\frac{1}{N!}\sum_{k=0}^{[tN]}(-1)^k\sum_{i_1<\ldots<i_k<[tN]}|S_N^{i_1\ldots i_k}|\\
&=&\frac{1}{N!}\sum_{k=0}^{[tN]}(-1)^k\sum_{i_1<\ldots<i_k<[tN]}(N-k)!\\
&=&\frac{1}{N!}\sum_{k=0}^{[tN]}(-1)^k\binom{[tN]}{k}(N-k)!\\
&=&\sum_{k=0}^{[tN]}\frac{(-1)^k}{k!}\cdot\frac{[tN]!(N-k)!}{N!([tN]-k)!}
\end{eqnarray*}

But with $N\to\infty$, we obtain from this the following estimate, as desired:
$$P(\chi_t=0)
\simeq\sum_{k=0}^{[tN]}\frac{(-1)^k}{k!}\cdot t^k
=\sum_{k=0}^{[tN]}\frac{(-t)^k}{k!}
\simeq e^{-t}$$

(2) More generally now, by counting the permutations $\sigma\in S_N$ having exactly $k$ fixed points among $1,\ldots,[tN]$, as in the proof of Theorem 3.22, our claim is that we get:
$$P(\chi_t=k)\simeq\frac{t^k}{k!e^t}$$

Indeed, we already know from (1) that this formula holds at $k=0$. In general now, we have to count the permutations $\sigma\in S_N$ having exactly $k$ fixed points among $1,\ldots,[tN]$. Since having such a permutation amounts in choosing $k$ points among $1,\ldots,[tN]$, and then permuting the $N-k$ points left, without fixed points among $1,\ldots,[tN]$ allowed, we obtain the following formula, where $s\in(0,1]$ is such that $[s(N-k)]=[tN]-k$:
\begin{eqnarray*}
\#\left\{\sigma\in S_N\Big|\chi_t(\sigma)=k\right\}
&=&\binom{[tN]}{k}\#\left\{\sigma\in S_{N-k}\Big|\chi_s(\sigma)=0\right\}\\
&=&\frac{[tN]!}{k!([tN]-k)!}\#\left\{\sigma\in S_{N-k}\Big|\chi_s(\sigma)=0\right\}\\
&=&\frac{1}{k!}\times\frac{[tN]!(N-k)!}{([tN]-k)!}\times\frac{\#\left\{\sigma\in S_{N-k}\Big|\chi_s(\sigma)=0\right\}}{(N-k)!}
\end{eqnarray*}

Now by dividing everything by $N!$, we obtain from this the following formula:
$$\frac{\#\left\{\sigma\in S_N\Big|\chi_t(\sigma)=k\right\}}{N!}=\frac{1}{k!}\times\frac{[tN]!(N-k)!}{N!([tN]-k)!}\times\frac{\#\left\{\sigma\in S_{N-k}\Big|\chi_s(\sigma)=0\right\}}{(N-k)!}$$

By using now the computation at $k=0$, that we already have, from (1) above, it follows that with $N\to\infty$ we have the following estimate:
$$P(\chi_t=k)
\simeq\frac{1}{k!}\times\frac{[tN]!(N-k)!}{N!([tN]-k)!}\cdot P(\chi_s=0)
\simeq\frac{t^k}{k!}\cdot\frac{1}{e^s}$$

Now recall that the parameter $s\in(0,1]$ was chosen such that $[s(N-k)]=[tN]-k$. Thus in the $N\to\infty$ limit we have $s=t$, and so we obtain, as claimed:
$$P(\chi_t=k)\simeq\frac{t^k}{k!}\cdot\frac{1}{e^t}$$

It follows that we obtain in the limit a Poisson law of parameter $t$, as stated.
\end{proof}

And with this, end of the story? Not yet, because with a bit of group theory know-how, we can formulate the following definition, which is something quite conceptual:

\begin{definition}
Given a group of matrices $G\subset_u O_N$, the function
$$\chi:G\to\mathbb C\quad,\quad 
\chi=\sum_{i=1}^Nu_{ii}$$
is called main character of $G$. More generally, the function
$$\chi_t:G\to\mathbb C\quad,\quad 
\chi_t=\sum_{i=1}^{[tN]}u_{ii}$$
is called main truncated character of $G$, of parameter $t\in(0,1]$.
\end{definition}

These notions are quite interesting in relation with group theory, and more on them later in this book. In relation with our questions the point is that, in terms of these, our main result, Theorem 3.24, reformulates in a quite conceptual way, as follows:

\begin{theorem}
For the symmetric group $S_N\subset O_N$ we have
$$\chi_t\sim p_t$$
in the $N\to\infty$ limit, for any $t\in(0,1]$.
\end{theorem} 

\begin{proof}
As we already know from chapter 2, the standard coordinates on $S_N\subset O_N$, which are the coefficients of the permutation matrices, are given by:
$$u_{ij}(\sigma)=\delta_{\sigma(j)i}$$

Thus the sums of diagonal coordinates $u_{ii}$ count the fixed points, and with this observation in hand, Theorem 3.24 reformulates as in the statement.
\end{proof}

Good to have this done, and more on this, group characters, later in this book.

\section*{3d. Poisson sums}

As a continuation of the above material, Definition 3.25 and Theorem 3.26 suggest doing some computations for other interesting groups $G\subset O_N$ that we know. And here, I don't know about you, but my favorite example is the hyperoctahedral group:

\begin{theorem}
Consider the hyperoctahedral group $H_N$, which appears as the symmetry group of the $N$-cube, or the symmetry group of the $N$ coordinate axes of $\mathbb R^N$:
$$S_N\subset H_N\subset O_N$$
In matrix terms, $H_N$ consists of the permutation-type matrices having $\pm1$ as nonzero entries, and we have a wreath product decomposition as follows:
$$H_N=\mathbb Z_2\wr S_N$$
In this picture, the main character counts the signed number of fixed points, among the coordinate axes, and its truncations count the truncations of such numbers.
\end{theorem}

\begin{proof}
Many things going on here, the idea being as follows:

\medskip

(1) To start with, we can certainly talk about the hyperoctahedral group $H_N$, as being the symmetry group of the $N$-cube, viewed as a graph, or as a metric space:
$$\xymatrix@R=20pt@C=22pt{
&\bullet\ar@{-}[rr]&&\bullet\\
\bullet\ar@{-}[rr]\ar@{-}[ur]&&\bullet\ar@{-}[ur]\\
&\bullet\ar@{-}[rr]\ar@{-}[uu]&&\bullet\ar@{-}[uu]\\
\bullet\ar@{-}[uu]\ar@{-}[ur]\ar@{-}[rr]&&\bullet\ar@{-}[uu]\ar@{-}[ur]
}$$

(2) As a first observation, in small dimensions we have $H_1=\mathbb Z_2$, which is obvious, then $H_2=D_4=\mathbb Z_4\rtimes\mathbb Z_2$, the symmetry group of the square, and then $H_3=S_4\times\mathbb Z_2$, with the copy of $S_4$ best viewed as permuting the main diagonals of the cube.

\medskip

(3) Next, by centering the $N$-cube at the origin, its symmetry group $H_N$ appears as the symmetry group of the $N$ coordinate axes of $\mathbb R^N$. Which is a fruitful viewpoint, proving right away $H_N=M_N(-1,0,1)\cap O_N$, and giving as well the following formula:
$$|H_N|
=|S_N|\cdot|\mathbb Z_2^N|
=N!\cdot2^N$$ 

(4) In algebraic terms, this shows that we are in the framework of the crossed products, and we conclude that we have a crossed product decomposition, as follows:
$$H_N=\mathbb Z_2^N\rtimes S_N$$

Which means that we have a decomposition $H_N=\mathbb Z_2\wr S_N$, as stated.

\medskip

(5) Finally, in what regards the assertions at the end, concerning the main character and its truncations, these are clear from definitions, exactly as previously for $S_N$.
\end{proof}

Regarding now the character laws, we can compute them by using the same method as for the symmetric group $S_N$, namely inclusion-exclusion, and we have:

\index{main character}

\begin{theorem}
For the hyperoctahedral group $H_N\subset O_N$, the law of the variable
$$\chi_t=\sum_{i=1}^{[tN]}u_{ii}$$
with $u_{ij}$ being the standard coordinates, becomes in the $N\to\infty$ limit the measure
$$p^2_t=e^{-t}\sum_{k=-\infty}^\infty\delta_k\sum_{p=0}^\infty \frac{(t/2)^{|k|+2p}}{(|k|+p)!p!}$$ 
where $\delta_k$ is the Dirac mass at $k\in\mathbb Z$.
\end{theorem}

\begin{proof}
We can regard the group $H_N$ as being the symmetry group of the graph $I_N=\{I^1,\ldots ,I^N\}$ formed by $N$ segments. The diagonal coefficients are given by:
$$u_{ii}(g)=\begin{cases}
\ 0\ \mbox{ if $g$ moves $I^i$}\\
\ 1\ \mbox{ if $g$ fixes $I^i$}\\
-1\mbox{ if $g$ returns $I^i$}
\end{cases}$$

We denote by $\uparrow g,\downarrow g$ the number of segments among $\{I^1,\ldots ,I^s\}$ which are fixed, respectively returned by an element $g\in H_N$. With this notation, we have:
$$u_{11}+\ldots+u_{ss}=\uparrow g-\downarrow g$$

Let us denote by $P_N$ probabilities computed over the group $H_N$. The density of the law of $u_{11}+\ldots+u_{ss}$ at a point $k\geq 0$ is then given by the following formula:
\begin{eqnarray*}
D(k)
&=&P_N(\uparrow g-\downarrow g=k)\\
&=&\sum_{p=0}^\infty P_N(\uparrow g=k+p, \downarrow g=p)
\end{eqnarray*}

Assume first that we have $t=1$. By using the fact, that we know from before, that the probability of $\sigma\in S_N$ to have $m$ fixed points is $P_m\simeq 1/(em!)$, we obtain:
\begin{eqnarray*}
\lim_{N\to\infty}D(k)
&=&\lim_{N\to\infty}\sum_{p=0}^\infty(1/2)^{k+2p}\begin{pmatrix}k+2p\\ k+p\end{pmatrix} P_N(\uparrow g+\downarrow g=k+2p)\\ 
&=&\sum_{p=0}^\infty(1/2)^{k+2p}\begin{pmatrix}k+2p\\
k+p\end{pmatrix}\frac{1}{e(k+2p)!}\\
&=&\frac{1}{e}\sum_{p=0}^\infty \frac{(1/2)^{k+2p}}{(k+p)!p!}
\end{eqnarray*}

As for the general case $0<t\leq 1$, here the result follows by performing some modifications in the above computation. The asymptotic density is computed as follows:
\begin{eqnarray*}
\lim_{N\to\infty}D(k)
&=&\lim_{N\to\infty}\sum_{p=0}^\infty(1/2)^{k+2p}\begin{pmatrix}k+2p\\ k+p\end{pmatrix} P_N(\uparrow g+\downarrow g=k+2p)\\
&=&\sum_{p=0}^\infty(1/2)^{k+2p}\begin{pmatrix}k+2p\\
k+p\end{pmatrix}\frac{t^{k+2p}}{e^t(k+2p)!}\\
&=&e^{-t}\sum_{p=0}^\infty \frac{(t/2)^{k+2p}}{(k+p)!p!}
\end{eqnarray*}

Together with $D(-k)=D(k)$, this gives the formula in the statement.
\end{proof}

The above result is quite interesting, and based on it, we can formulate:

\index{Bessel function}
\index{Bessel law}

\begin{definition}
The Bessel law of parameter $t>0$ is the measure
$$p^2_t=e^{-t}\sum_{k=-\infty}^\infty\delta_k\,f_k(t/2)\quad,\quad 
f_k(t)=\sum_{p=0}^\infty \frac{t^{|k|+2p}}{(|k|+p)!p!}$$
with $f_k$ being the Bessel function of the first kind, from classical analysis.
\end{definition}

Let us study now these Bessel laws that we found. In analogy with what we know about the Poisson laws $p_t$, from before, we first have the following result:

\index{convolution}
\index{convolution semigroup}

\begin{theorem}
The Bessel laws $p^2_t$ have the property
$$p^2_s*p^2_t=p^2_{s+t}$$
so they form a truncated one-parameter semigroup with respect to convolution.
\end{theorem}

\begin{proof}
We use the formula in Definition 3.29. The Fourier transform is:
$$F(x)=e^{-t}\sum_{k=-\infty}^\infty e^{ikx}\,f_k(t/2)$$

We can compute the derivative of $F$ with respect to $t$, as follows:
\begin{eqnarray*}
F(x)'
&=&-e^{-t}\sum_{k=-\infty}^\infty e^{ikx}\,f_k(t/2)+\frac{e^{-t}}{2}\sum_{k=-\infty}^\infty e^{ikx}\,f_k'(t/2)\\
&=&-F(x)+\frac{e^{-t}}{2}\sum_{k=-\infty}^\infty e^{ikx}\,f_k'(t/2)
\end{eqnarray*}

On the other hand, the derivative of $f_k$ with $k\geq 1$ is given by:
\begin{eqnarray*}
f_k'(t)
&=&\sum_{p=0}^\infty \frac{(k+2p)t^{k+2p-1}}{(k+p)!p!}\\
&=&\sum_{p=0}^\infty \frac{(k+p)t^{k+2p-1}}{(k+p)!p!}+\sum_{p=0}^\infty\frac{p\,t^{k+2p-1}}{(k+p)!p!}\\
&=&\sum_{p=0}^\infty \frac{t^{k+2p-1}}{(k+p-1)!p!}+\sum_{p=1}^\infty\frac{t^{k+2p-1}}{(k+p)!(p-1)!}\\
&=&\sum_{p=0}^\infty \frac{t^{(k-1)+2p}}{((k-1)+p)!p!}+\sum_{p=1}^\infty\frac{t^{(k+1)+2(p-1)}}{((k+1)+(p-1))!(p-1)!}\\
&=&f_{k-1}(t)+f_{k+1}(t)
\end{eqnarray*}

This computation works in fact for any $k\in\mathbb Z$, so we get:
\begin{eqnarray*}
F(x)'
&=&-F(x)+\frac{e^{-t}}{2}
\sum_{k=-\infty}^\infty e^{ikx} (f_{k-1}(t/2)+f_{k+1}(t/2))\\
&=&-F(x)+\frac{e^{-t}}{2} \sum_{k=-\infty}^\infty
e^{i(k+1)x}f_{k}(t/2)+e^{i(k-1)x}f_{k}(t/2)\\
&=&-F(x)+\frac{e^{ix}+e^{-ix}}{2}\,F(x)\\
&=&(\cos x-1)F(x)
\end{eqnarray*}

Thus $F(x)=\exp((\cos x-1)t)$, whose log is linear in $t$, and we get the result.
\end{proof}

In order to unify now the theory of the Poisson and Bessel laws, we have the following notion, extending the Poisson limit theory developed earlier in this chapter:

\index{compound Poisson law}

\begin{definition}
Associated to any discrete positive measure $\nu$ on $\mathbb R$, not necessarily of mass 1, is the probability measure
$$p_\nu=\lim_{n\to\infty}\left(\left(1-\frac{c}{n}\right)\delta_0+\frac{1}{n}\,\nu\right)^{*n}$$
where $c=mass(\nu)$, called compound Poisson law.
\end{definition}

In other words, what we are doing here is to generalize the construction in the Poisson Limit Theorem, by allowing the only parameter there, which was the positive real number $t>0$, to be replaced by a certain probability measure $\nu$, of arbitrary mass $c>0$.

\bigskip

We first have the following result, allowing us to detect such laws:

\index{Fourier transform}

\begin{proposition}
For a discrete measure, $\nu=\sum_{i=1}^sc_i\delta_{z_i}$ with $c_i>0$ and $z_i\in\mathbb R$,
$$F_{p_\nu}(x)=\exp\left(\sum_{i=1}^sc_i(e^{ixz_i}-1)\right)$$
where $F$ denotes as usual the Fourier transform.
\end{proposition}

\begin{proof}
With $\mu_n$ being the measure in Definition 3.31, under the $*$ sign, we have:
\begin{eqnarray*}
F_{\mu_n}(x)=\left(1-\frac{c}{n}\right)+\frac{1}{n}\sum_{i=1}^sc_ie^{ixz_i}
&\implies&F_{\mu_n^{*n}}(x)=\left(\left(1-\frac{c}{n}\right)+\frac{1}{n}\sum_{i=1}^sc_ie^{ixz_i}\right)^n\\
&\implies&F_{p_\nu}(x)=\exp\left(\sum_{i=1}^sc_i(e^{ixz_i}-1)\right)
\end{eqnarray*}

Thus, we have obtained the formula in the statement.
\end{proof}

Next, we have the following key result, providing an alternative to Definition 3.31, and which will be our formulation here of the Compound Poisson Limit Theorem:

\index{compound Poisson Limit theorem}
\index{CPLT}
\index{compound PLT}
\index{compound Poisson limit}

\begin{theorem}[CPLT]
For a discrete measure, $\nu=\sum_{i=1}^sc_i\delta_{z_i}$ with $c_i>0$ and $z_i\in\mathbb C$, we have the following formula,
$$p_\nu={\rm law}\left(\sum_{i=1}^sz_i\alpha_i\right)$$
with the variables $\alpha_i$ being Poisson $(c_i)$, and independent.
\end{theorem}

\begin{proof}
With $\alpha$ being the sum of Poisson variables in the statement, we have:
\begin{eqnarray*}
F_{\alpha_i}(x)=\exp(c_i(e^{ix}-1))
&\implies&F_{z_i\alpha_i}(x)=\exp(c_i(e^{ixz_i}-1))\\
&\implies&F_\alpha(x)=\exp\left(\sum_{i=1}^sc_i(e^{ixz_i}-1)\right)
\end{eqnarray*}

Thus we have the same formula as in Proposition 3.32, as desired.
\end{proof}

Getting back now to the Bessel laws, we have the following result:

\index{Bessel law}

\begin{theorem}
The Bessel laws $p^2_t$ are compound Poisson laws, given by
$$p^2_t=p_{t\varepsilon}$$
where $\varepsilon=\frac{1}{2}(\delta_{-1}+\delta_1)$ is the uniform measure on $\mathbb Z_2$.
\end{theorem}

\begin{proof}
We recall from the proof of Theorem 3.30 that the Fourier transform of the Bessel law $p^2_t$, with respect to a variable $x$, is given by the following formula:
$$F(x)=\exp\left(\left(\cos x-1\right)t\right)$$

But, according to Proposition 3.32, this is exactly the Fourier transform of the compound Poisson law $p_{t\varepsilon}$ in the statement. Thus, we have $p^2_t=p_{t\varepsilon}$, as claimed.
\end{proof}

Finally, regarding the moments of the Bessel laws, the result is as follows:

\begin{theorem}
The moments of the Bessel law $p^2_t$ are the numbers
$$M_k=\sum_{\pi\in P_{even}(k)}t^{|\pi|}$$
where $P_{even}(k)$ is the set of partitions of $\{1,\ldots,k\}$ having even blocks, and
$$E=0\quad,\quad V=t\quad,\quad\gamma=0\quad,\quad\kappa=3+1/t$$
are the expectation, variance, skewness and kurtotis.
\end{theorem}

\begin{proof}
This is something more tricky, the idea being as follows:

\medskip

(1) In what regards the moment formula in the statement, this can be certainly deduced from the formula of the Fourier transform, $F(x)=\exp((\cos x-1)t)$, via some computations, which are not exactly trivial, and that we will leave as an exercise. 

\medskip

(2) Alternatively, we can use the density formula $p^2_t=e^{-t}\sum_{k=-\infty}^\infty\delta_k\,f_k(t/2)$, with $f_k$ being the Bessel function of the first kind, with again the computations here, which are not exactly trivial, say left as an exercise for you, leading to the result.

\medskip

(3) As yet another proof, which is more advanced, with no computations, we can use the representation theory of $H_N$, and more specifically the fact that this group is easy, coming from the category $P_{even}$. And more on this, easiness, later in this book.

\medskip

(4) Finally, in what regards the values of $E,V,\gamma,\kappa$, the formulae of $E,V,\gamma$ are clear, and in what regards the kurtosis, which comes from $P_{even}(4)=\{\cap\,\cap,\cap\!\!\cap,\Cap,\sqcap\hskip-1.6mm\sqcap\hskip-1.6mm\sqcap\}$, which gives $M_4=3t^2+t$, which after centering and normalization gives $\kappa=3+1/t$.
\end{proof}

Quite nice all this, and as a continuation, we can talk, more generally, about complex reflection groups $G\subset U_N$, which according to a theorem of Shephard and Todd consist of a two-parameter series $H_N^{sd}$, and a number of exceptional examples. And, in what regards the series $H_N^{sd}$, the character computations can be done, a bit as before for $S_N,H_N$, with the conclusion that the parameter $d$ is irrelevant, and that we obtain as limiting laws the compound Poisson laws $p_{t\varepsilon}$, with $\varepsilon$ being the uniform measure on the $s$-th roots of unity. For more on all this, story and details, you can check my book \cite{ba2}.

\section*{3e. Exercises}

Here are some exercises on the above, all combinatorics, of various flavors:

\begin{exercise}
Review if needed the basics of the generalized binomial formula.
\end{exercise}

\begin{exercise}
Do the full computations, for the negative hypergeometric laws.
\end{exercise}

\begin{exercise}
Clarify what we said, about real-life occurences of the Poisson laws.
\end{exercise}

\begin{exercise}
Learn more about the Bell and Stirling numbers, and their properties.
\end{exercise}

\begin{exercise}
Can you exactly compute the law of $\chi_t$, for the symmetric group $S_N$?
\end{exercise}

\begin{exercise}
Learn a bit about group representations, and their characters.
\end{exercise}

\begin{exercise}
Learn more about the compound Poisson laws.
\end{exercise}

\begin{exercise}
Do the moment computations for the Bessel laws $p^2_t$.
\end{exercise}

As bonus exercise, learn as well about Catalan numbers. We will meet them later.

\chapter{Central limits}

\section*{4a. Central limits}

We have seen that some interesting theory can be developed for the discrete measures, notably with a lot of exciting results regarding the Poisson laws, and their versions. Normally this would be the end of what we have to say, in the discrete case, and our job now would be that of getting into the continuous case, first with measure theory systematically developed, and then with some applications of this, to continuous probability.

\bigskip

However, we cannot leave basic probability without talking, in one form or another, about central limits. You have certainly heard about bell-shaped curves, and perhaps even observed them in physics or chemistry class, because any routine measurement leads to such curves. Mathematically, here is the question that we would like to solve:

\index{CLT}
\index{Central Limit theorem}
\index{normal variable}
\index{Gaussian variable}

\begin{question}
Given random variables $f_1,f_2,f_3,\ldots$, say taken discrete, which are i.i.d., centered, and with common variance $t>0$, do we have
$$\frac{1}{\sqrt{n}}\sum_{i=1}^nf_i\sim g_t$$
in the $n\to\infty$ limit, for some bell-shaped density $g_t$? And, what is the formula of $g_t$? 
\end{question}

Observe that this question perfectly makes sense, with the probability theory that we know, by assuming that our random variables $f_1,f_2,f_3,\ldots$ are discrete, as said above. As for the $1/\sqrt{n}$ factor, there is certainly need for a normalization factor there, as for things to have a chance to converge, and the good factor is $1/\sqrt{n}$, as shown by:

\begin{proposition}
In order for a sum of the following type to have a chance to converge, with $f_1,f_2,f_3,\ldots$ being i.i.d., centered, and with common variance $t>0$,
$$S=\sum_{i=1}^nf_i$$
we must normalize this sum by a $1/\sqrt{n}$ factor, as in Question 4.1.
\end{proposition}

\begin{proof}
The idea here is to look at the moments of $S$. Since all variables $f_i$ are centered, $E(f_i)=0$, so is their sum, $E(S)=0$, and no contradiction here. However, when looking at the variance of $S$, which equals the second moment, due to $E(S)=0$, things become interesting, due to the following computation:
$$V(S)
=\sum_iE(f_i^2)+\sum_{i\neq j}E(f_i)E(f_j)
=\sum_iE(f_i^2)
=nt$$

Thus, we are in need a normalization factor $\alpha$, in order for our sum to have a chance to converge. But, repeating the computation with $S$ replaced by $\alpha S$ gives:
$$V(\alpha S)=\alpha^2nt$$

Thus, the good normalization factor is $\alpha=1/\sqrt{n}$, as claimed.
\end{proof}

So far, so good, we have a nice problem above, and time now to make a plan, in order to solve it. With the tools that we have, from chapters 2-3, here is such a plan:

\begin{plan}
In order to solve our central limiting question, we have to:
\begin{enumerate}
\item Apply Fourier and let $n\to\infty$, as to compute the Fourier transform of $g_t$.

\item Do some combinatorics and calculus, as to compute the moments of $g_t$.

\item Recover $g_t$ out of its moments, again via combinatorics and calculus.
\end{enumerate}
\end{plan}

Which sounds very good, but since we will be navigating here in troubled waters, with $g_t$ being continuous, as opposed to the discrete densities that we are used to, perhaps time to ask the cat, before embarking on our trip. And good news, cat is already onboard, hired to catch mice on our ship, so let's ask him what he has to say:

\begin{cat}
Capitain, that $g_t$ we are looking for is something two-dimensional, and all this normally comes after multivariable calculus, well done. But we can still go ahead, as long as there's enough booze for you guys, and mice for me, things fine.
\end{cat}

Nice advice, especially for the last part, but in what regards the first part, I must admit that I am completely puzzled. What are that two dimensions cat is talking about? Hope he has not confused mice with booze. So thanks cat, but please don't touch human food during our trip, mice only for you, and let's leave now, time to go.

\bigskip

So, back to our Plan 4.3, let us start with (1) there. Things are quickly done here, by using the linearization results for convolution from chapter 2, which lead to:

\begin{theorem}
Given discrete variables $f_1,f_2,f_3,\ldots$, which are i.i.d., centered, and with common variance $t>0$, we have
$$\frac{1}{\sqrt{n}}\sum_{i=1}^nf_i\sim g_t$$
with $n\to\infty$, with $g_t$ being the law having $F(x)=e^{-tx^2/2}$ as Fourier transform.
\end{theorem}

\begin{proof}
Observe first that in terms of moments, the Fourier transform of an arbitrary random variable $f:X\to\mathbb R$ is given by the following formula:
\begin{eqnarray*}
F_f(x)
&=&E(e^{ixf})\\
&=&E\left(\sum_{k=0}^\infty\frac{(ixf)^k}{k!}\right)\\
&=&\sum_{k=0}^\infty\frac{(ix)^kE(f^k)}{k!}\\
&=&\sum_{k=0}^\infty\frac{i^kM_k(f)}{k!}\,x^k
\end{eqnarray*}

In particular, in the case of a centered variable, $E(f)=0$, as those that we are interested in, the Fourier transform formula that we get is as follows:
$$F_f(x)=1-\frac{M_2(f)}{2}\cdot x^2-i\,\frac{M_3(f)}{6}\cdot x^3+\ldots$$

Moreover, by further assuming that the Fourier variable is small, $x\simeq0$, the Fourier transform formula that we get, that we will use in what follows, becomes:
$$F_f(x)=1-\frac{M_2(f)}{2}\cdot x^2+O(x^2)$$

In addition to this, we will also need to know what happens to the Fourier transform when rescaling. But the formula here is very easy to find, as follows:
$$F_{\alpha f}(x)
=E(e^{ix\alpha f})
=E(e^{i\alpha xf})
=F_f(\alpha x)$$

Good news, we can now do our computation. By using the above formulae, the Fourier transform of the variable in the statement is given by:
\begin{eqnarray*}
F(x)
&=&\left[F_f\left(\frac{x}{\sqrt{n}}\right)\right]^n\\
&=&\left[1-\frac{tx^2}{2n}+O(n^{-2})\right]^n\\
&\simeq&\left[1-\frac{tx^2}{2n}\right]^n\\
&\simeq&e^{-tx^2/2}
\end{eqnarray*}

We are therefore led to the conclusion in the statement, modulo the fact that we do not know yet that a density $g_t$ having as Fourier transform $F(x)=e^{-tx^2/2}$ really exists, plus perhaps some other abstract issues, related to the continuous measures, to be discussed too. But too late to go back, both cat and sailors are happy, we will go ahead. So, theorem proved, modulo finding that law $g_t$, which still remains to be done.
\end{proof}

Getting now to step (2) of our Plan 4.3, that is easy to work out too, via some elementary one-variable calculus, with the result here being as follows:

\begin{theorem}
The ``normal'' law $g_t$, having as Fourier transform 
$$F(x)=e^{-tx^2/2}$$
must have all odd moments zero, and its even moments must be the numbers
$$M_k(g_t)=t^{k/2}\times k!!$$
where $k!!=(k-1)(k-3)(k-5)\ldots$, for $k\in2\mathbb N$. 
\end{theorem}

\begin{proof}
Again, several things going on here, the idea being as follows:

\medskip

(1) To start with, at the level of formalism and notations, in view of Question 4.1 and of Theorem 4.5, we have adopted the term ``normal'' for the mysterious law $g_t$ that we are looking for, the one having $F(x)=e^{-tx^2/2}$ as Fourier transform. 

\medskip

(2) Getting towards the computation of the moments, as a first useful observation, according to Theorem 4.5 this normal law $g_t$ must be centered, as shown by:
\begin{eqnarray*}
f_i={\rm centered}
&\implies&\sum_{i=1}^nf_i={\rm centered}\\
&\implies&\frac{1}{\sqrt{n}}\sum_{i=1}^nf_i={\rm centered}\\
&\implies&g_t={\rm centered}
\end{eqnarray*}

Moreover, the same argument works by replacing ``centered'' with ``having an even function as density'', and this shows, via some standard calculus, that we will leave here as an exercise, that the odd moments of our normal law must vanish:
$$M_{2l+1}(g_t)=0$$

Thus, first assertion proved, and we only have to care about the even moments. 

\medskip

(3) As a comment here, as we will see in a moment, our study below of the moments computes in fact the odd moments too, as being all equal to 0, this time without making reference to Theorem 4.5. Thus, definitely no worries with the odd moments.

\medskip

(4) Getting to work now, we must reformulate the equation $F(x)=e^{-tx^2/2}$, in terms of moments. We know from the proof of Theorem 4.5 that we have:
$$F(x)=\sum_{k=0}^\infty\frac{i^kM_k(g_t)}{k!}\,x^k$$

On the other hand, we have the following formula, for the exponential:
$$e^{-tx^2/2}
=\sum_{r=0}^\infty
(-1)^r\frac{t^rx^{2r}}{2^rr!}$$

Thus, our equation $F(x)=e^{-tx^2/2}$ takes the following form:
$$\sum_{k=0}^\infty\frac{i^kM_k(g_t)}{k!}\,x^k=\sum_{r=0}^\infty
(-1)^r\frac{t^rx^{2r}}{2^rr!}$$

(5) As a first observation, the odd moments must vanish, as said in (2) above. As for the even moments, these can be computed as follows:
\begin{eqnarray*}
M_k(g_t)
&=&k!\times\frac{t^{k/2}}{2^{k/2}(k/2)!}\\
&=&t^{k/2}\times\frac{k!}{2^{k/2}(k/2)!}\\
&=&t^{k/2}\times\frac{2\cdot3\cdot4\ldots (k-1)\cdot k}{2\cdot 4\cdot 6\ldots (k-2)\cdot k}\\
&=&t^{k/2}\times 3\cdot 5\ldots(k-3)(k-1)\\
&=&t^{k/2}\times k!!
\end{eqnarray*}

Thus, we are led to the formula in the statement.
\end{proof}

The moment formula that we found is quite interesting, and before going ahead with step (3) of our Plan 4.3, let us look a bit at this, and see what we can further say. To be more precise, in analogy with what we know about the Poisson laws, and about the Bessel laws too, making reference to interesting combinatorics and partitions, when it comes to moments, we have the following result, regarding the normal laws:

\begin{theorem}
The moments of the normal law $g_t$ are given by
$$M_k(g_t)=t^{k/2}|P_2(k)|$$
for any $k\in\mathbb N$, with $P_2(k)$ standing for the pairings of $\{1,\ldots,k\}$.
\end{theorem}

\begin{proof}
This is a reformulation of Theorem 4.6, the idea being as follows:

\medskip

(1) We know from Theorem 4.6 that the moments of the normal law $M_k=M_k(g_t)$ that we are interested in are given by the following formula, with the convention $k!!=0$ for $k$ odd, and $k!!=(k-1)(k-3)(k-5)\ldots$ for $k$ even, for the double factorials:
$$M_k(g_t)=t^{k/2}\times k!!$$

Now observe that, according to our above convention for the double factorials, these are subject to the following recurrence relation, with initial data $1!!=0,2!!=1$:
$$k!!=(k-1)(k-2)!!$$

We conclude that the moments of the normal law $M_k=M_k(g_t)$ are subject to the following recurrence relation, with initial data $M_1=0,M_2=t$:
$$M_k=t(k-1)M_{k-2}$$

(2) On the other hand, let us first count the pairings of the set $\{1,\ldots,k\}$. In order to have such a pairing, we must pair $1$ with one of the numbers $2,\ldots,k$, and then use a pairing of the remaining $k-2$ numbers. Thus, we have the following recurrence formula for the number $P_k$ of such pairings, with the initial data $P_1=0,P_2=1$:
$$P_k=(k-1)P_{k-2}$$

Now by multiplying by $t^{k/2}$, the resulting numbers $N_k=t^{k/2}P_k$ will be subject to the following recurrence relation, with initial data $N_1=0,N_2=t$:
$$N_k=t(k-1)N_{k-2}$$

(3) Thus, the moments $M_k=M_k(g_t)$ and the numbers $N_k=t^{k/2}P_k$ are subject to the same recurrence relation, with the same initial data, so they are equal, as claimed.
\end{proof}

Still in analogy with what we know about the Poisson laws, and about the Bessel laws too, we can further process what we found in Theorem 4.7, and we are led to:

\begin{theorem}
The moments of the normal law $g_t$ are given by
$$M_k(g_t)=\sum_{\pi\in P_2(k)}t^{|\pi|}$$
where $P_2(k)$ is the set of pairings of $\{1,\ldots,k\}$, and $|.|$ is the number of blocks.
\end{theorem}

\begin{proof}
This is a quick reformulation of Theorem 4.7, with the number of blocks of a pairing of $\{1,\ldots,k\}$ being trivially $k/2$, independently of the pairing.
\end{proof}

As a philosophical conclusion now to all this, let us formulate:

\begin{conclusion}
The normal laws $g_t$ have properties which are quite similar to those of the Poisson laws $p_t$, and combinatorially, the passage
$$p_t\to g_t$$
appears by replacing the partitions with the pairings.
\end{conclusion}

Which sounds quite conceptual, hope you agree with me. According to Question 4.1, compared to the Poisson Limit Theorem, the normal law $g_t$ should be rather the ``continuous version'' of the Poisson law $p_t$, and here we are into some sort of quite advanced mathematics, telling us that the passage from discrete to continuous appears, combinatorially speaking, by replacing the partitions with the pairings.

\bigskip

We will explore such things, which are indeed quite advanced, later in this book. In the meantime, however, we still need to know what the density of $g_t$ is.

\section*{4b. Normal laws}

So, let us get now to step (3) of our Plan 4.3. This does not look obvious at all, but some partial integration know-how leads us to the following statement:

\index{double factorial}

\begin{theorem}
The normal laws are given by
$$g_t=\frac{1}{\sqrt{2t}\cdot I}\,e^{-x^2/2t}dx$$
with the constant on the bottom being $I=\int_\mathbb Re^{-x^2}dx$.
\end{theorem}

\begin{proof}
This comes from partial integration, as follows:

\medskip

(1) Let us first do a naive computation. Consider the following quantities:
$$M_k=\int_\mathbb Rx^ke^{-x^2}dx$$

It is quite obvious that by partial integration we will get a recurrence formula for these numbers, similar to the one that we have for the moments of the normal laws. So, let us do this. By partial integration we obtain the following formula, for any $k\in\mathbb N$:
\begin{eqnarray*}
M_k
&=&-\frac{1}{2}\int_\mathbb Rx^{k-1}\left(e^{-x^2}\right)'dx\\
&=&\frac{1}{2}\int_\mathbb R(k-1)x^{k-2}e^{-x^2}dx\\
&=&\frac{k-1}{2}\cdot M_{k-2}
\end{eqnarray*}

(2) Thus, we are on the good way, with the recurrence formula that we got being the same as that for the moments of $g_{1/2}$. Now let us fine-tune this, as to reach to the same recurrence as for the moments of $g_t$. Consider the following quantities:
$$N_k=\int_\mathbb Rx^ke^{-x^2/2t}dx$$

By partial integration as before, we obtain the following formula:
\begin{eqnarray*}
N_k
&=&\int_\mathbb R(tx^{k-1})\left(-e^{-x^2/2t}\right)'dx\\
&=&\int_\mathbb Rt(k-1)x^{k-2}e^{-x^2/2t}dx\\
&=&t(k-1)\int_\mathbb Rx^{k-2}e^{-x^2/2t}dx\\
&=&t(k-1)N_{k-2}
\end{eqnarray*}

(3) Thus, almost done, and it remains to discuss normalization. We know from the above that we must have a formula as follows, with $I_t$ being a certain constant:
$$g_t=\frac{1}{I_t}\cdot e^{-x^2/2t}\,dx$$

But the constant $I_t$ must be the one making $g_t$ of mass 1, and so:
\begin{eqnarray*}
I_t
&=&\int_\mathbb Re^{-x^2/2t}\,dx\\
&=&\int_\mathbb Re^{-2ty^2/2t}\,\sqrt{2t}dy\\
&=&\sqrt{2t}\int_\mathbb Re^{-y^2}dy
\end{eqnarray*}

Thus, we are led to the formula in the statement.
\end{proof}

What we did in the above is good work, and it remains to compute the constant $I$ appearing in Theorem 4.10, given by the following formula, and called Gauss integral:
$$I=\int_\mathbb Re^{-x^2}dx$$

Which does not look obvious at all, so time to ask again the cat. Unfortunately cat is nowhere to be found, probably gone with the sailors, for some partying, now that we are again on land, for a short break. This being said, walking through this debauchery, I can spot a small kitty, behind a pile of harbor garbage. She's really small, and can barely meow, but never knows, let's ask her about mathematics. And kitty answers:

\begin{kitty}
The normal laws are naturally two-dimensional, and it is
$$I^2=\int_\mathbb R\int_\mathbb Re^{-x^2-y^2}dxdy$$
the quantity that you should compute. I bet you can do that.
\end{kitty}

Thanks kitty, this sounds wise, so again that mysterious two-dimensionality, that cat was talking about. So, I will try this, and by the way, come with me. Cat is surely very good at both catching mice and helping with trigonometry and other math, but these days he's learning more and more bad things from the sailors, I am a bit worried about this, and in any case, we need a backup. You'll like it, plenty of mice on our ship.

\bigskip

So, following now kitty's advice, we are led to the following result:

\begin{theorem}
We have the following formula,
$$\int_\mathbb Re^{-x^2}dx=\sqrt{\pi}$$
called Gauss integral formula.
\end{theorem}

\begin{proof}
As already mentioned, this is something which is nearly impossible to prove, with bare hands. However, this can be proved by using two dimensions, as follows:
\begin{eqnarray*}
\int_\mathbb R\int_\mathbb Re^{-x^2-y^2}dxdy
&=&4\int_0^\infty\int_0^\infty e^{-x^2-y^2}dxdy\\
&=&4\int_0^\infty\int_0^\infty e^{-t^2y^2-y^2}ydtdy\\
&=&4\int_0^\infty\int_0^\infty ye^{-y^2(1+t^2)}dydt\\
&=&2\int_0^\infty\int_0^\infty\left(-\frac{e^{-y^2(1+t^2)}}{1+t^2}\right)'dydt\\
&=&2\int_0^\infty\frac{dt}{1+t^2}\\
&=&2\int_0^\infty(\arctan t)'dt\\
&=&\pi
\end{eqnarray*}

Thus, we are led to the conclusion in the statement.
\end{proof}

Very nice, so as a final conclusion to our study, started long ago, in the beginning of this chapter, we can now formulate the Central Limit Theorem (CLT), as follows:

\index{CLT}
\index{Central Limit theorem}
\index{normal variable}
\index{Gaussian variable}

\begin{theorem}[CLT]
Given discrete random variables $f_1,f_2,f_3,\ldots$, which are i.i.d., centered, and with common variance $t>0$, we have
$$\frac{1}{\sqrt{n}}\sum_{i=1}^nf_i\sim g_t$$
in the $n\to\infty$ limit, in moments, with the limiting mesure being
$$g_t=\frac{1}{\sqrt{2\pi t}}\,e^{-x^2/2t}dx$$
called normal, or Gaussian law of parameter $t>0$.
\end{theorem}

\begin{proof}
This follows indeed from our various results above, and more specifically from Theorem 4.5, Theorem 4.6 for the terminology, Theorem 4.10 and Theorem 4.12.
\end{proof}

Let us study now more in detail the laws that we found. Normally we already have everything that is needed, but it is instructive at this point to do some computations, based on the explicit formula of $g_t$ found above, and on Theorem 4.12. We first have: 

\index{variance}

\begin{proposition}
We have the variance formula
$$V(g_t)=t$$
valid for any $t>0$.
\end{proposition}

\begin{proof}
We already know this, but we can establish this as well directly, starting from our formula of $g_t$ from Theorem 4.13. Indeed, the first moment is 0, because our normal law $g_t$ is centered. As for the second moment, this can be computed as follows:
\begin{eqnarray*}
M_2
&=&\frac{1}{\sqrt{2\pi t}}\int_\mathbb Rx^2e^{-x^2/2t}dx\\
&=&\frac{1}{\sqrt{2\pi t}}\int_\mathbb R(tx)\left(-e^{-x^2/2t}\right)'dx\\
&=&\frac{1}{\sqrt{2\pi t}}\int_\mathbb Rte^{-x^2/2t}dx\\
&=&t
\end{eqnarray*}

We conclude from this that the variance is $V=M_2=t$, as claimed.
\end{proof}

More generally, we can recover in this way the computation of all moments:

\begin{theorem}
The even moments of the normal law are the numbers
$$M_k(g_t)=t^{k/2}\times k!!$$
where $k!!=(k-1)(k-3)(k-5)\ldots\,$, and the odd moments vanish. 
\end{theorem}

\begin{proof}
Again, we already know this, but we can establish this as well directly, starting from our formula above of $g_t$. Indeed, we have the following computation:
\begin{eqnarray*}
M_k
&=&\frac{1}{\sqrt{2\pi t}}\int_\mathbb Ry^ke^{-y^2/2t}dy\\
&=&\frac{1}{\sqrt{2\pi t}}\int_\mathbb R(ty^{k-1})\left(-e^{-y^2/2t}\right)'dy\\
&=&\frac{1}{\sqrt{2\pi t}}\int_\mathbb Rt(k-1)y^{k-2}e^{-y^2/2t}dy\\
&=&t(k-1)\times\frac{1}{\sqrt{2\pi t}}\int_\mathbb Ry^{k-2}e^{-y^2/2t}dy\\
&=&t(k-1)M_{k-2}
\end{eqnarray*}

Thus by recurrence, we are led to the formula in the statement.
\end{proof}

Here is another result, which is the key one for the study of the normal laws:

\index{convolution semigroup}

\begin{theorem}
We have the following formula, valid for any $t>0$:
$$F_{g_t}(x)=e^{-tx^2/2}$$
In particular, the normal laws satisfy $g_s*g_t=g_{s+t}$, for any $s,t>0$.
\end{theorem}

\begin{proof}
As before, we already know this, but we can establish now the Fourier transform formula as well directly, by using the explicit formula of $g_t$, as follows:
\begin{eqnarray*}
F_{g_t}(x)
&=&\frac{1}{\sqrt{2\pi t}}\int_\mathbb Re^{-y^2/2t+ixy}dy\\
&=&\frac{1}{\sqrt{2\pi t}}\int_\mathbb Re^{-(y/\sqrt{2t}-\sqrt{t/2}ix)^2-tx^2/2}dy\\
&=&\frac{1}{\sqrt{2\pi t}}\int_\mathbb Re^{-z^2-tx^2/2}\sqrt{2t}dz\\
&=&\frac{1}{\sqrt{\pi}}e^{-tx^2/2}\int_\mathbb Re^{-z^2}dz\\
&=&e^{-tx^2/2}
\end{eqnarray*}

As for the last assertion, this follows from the fact that $\log F_{g_t}$ is linear in $t$.
\end{proof}

Observe that, thinking retrospectively, the above computation formally solves the question raised by Theorem 4.5, and so could have been used there, afterwards. However, and here comes the point, all this is based on Theorem 4.12, and also, crucially, on our work from Theorem 4.10, which in turn was based on moments and so on.

\bigskip

Thus, no really way of doing things in a simpler way. There is however one simplification, because if you know well multivariable calculus, the Gauss formula comes right away via polar coordinates, with the computation being as follows:
$$\int_\mathbb R\int_\mathbb Re^{-x^2-y^2}dxdy
=\int_0^{2\pi}\int_0^\infty e^{-r^2}rdrdt
=2\pi\times\frac{1}{2}
=\pi$$

And, based on this, it is quite natural to introduce the normal laws $g_t$ directly, then compute Fourier, and prove the CLT, as probabilists do. More on this later, once we will know well multivariable calculus, and in particular the formula $dxdy=rdrdt$, which was crucially used in the above computation, we will certainly come back to this.

\bigskip

Moving on, completing our basic study of the normal laws, we have:

\begin{theorem}
The even normalized central moments of the normal law are
$$M_k''(g_t)=k!!$$
with $k!!=(k-1)(k-3)\ldots\,$ as usual, and the odd such moments vanish. Also,
$$E=0\quad,\quad V=t\quad,\quad \gamma=0\quad,\quad \kappa=3$$
are the expectation, variance, skewness and kurtosis.
\end{theorem}

\begin{proof}
The normal laws being centered, their central moments equal the plain ones, $M_k'=M_k$. Since the variance is $V=t$, the normalized central moments are given by:
$$M_k''=\frac{M'_k}{t^{k/2}}=\frac{M_k}{t^{k/2}}=\frac{\delta_{2|k}t^{k/2}\times k!!}{t^{k/2}}=\delta_{2|k}k!!$$

Finally, at $k=3$ this gives $\gamma=0$, and at $k=4$ this gives $\kappa=4!!=3$.
\end{proof}

In practice now, the Gaussian bells quite often appear shifted by a parameter $a\in\mathbb R$, I mean with symmetry axis there at $a$. So, we must study these versions too:

\index{shifted normal law}

\begin{theorem}
The normal law $g_t$, shifted to the right by $a\in\mathbb R$, has density
$$g_t^a=\frac{1}{\sqrt{2\pi t}}\,e^{-(x-a)^2/2t}\,dx$$
and the basic properties of this shifted law $g_t^a$ are as follows:
\begin{enumerate}
\item We have $f\sim g_t\implies f+a\sim g_t^a$.

\item The mean is $E=a$, the variance is $V=t$.

\item The moments are $M_k=\sum_{r=0}^{[k/2]}\binom{k}{2r}(2r)!!t^ra^{k-2r}$.

\item The moments satisfy $M_k=aM_{k-1}+(k-1)tM_{k-2}$.

\item The central moments are $M_k'=\delta_{2|k}t^{k/2}k!!$.

\item The normalized central moments are $M_k''=\delta_{2|k}k!!$.

\item The skewness is $\gamma=0$, the kurtosis is $\kappa=3$.

\item The Fourier transform is $F(x)=\exp(iax-tx^2/2)$.

\item We have the convolution formula $g_s^a*g_t^b=g_{s+t}^{a+b}$.
\end{enumerate}
\end{theorem}

\begin{proof}
These are things that we know well at $a=0$, and the general case follows from this. To start with, let us draw a picture, that will help us in our work:
$$\xymatrix@R=13pt@C=10pt{
&&&&&\\
&&&&&\frac{1}{\sqrt{2\pi t}}\ar@.[u]&&\ar@/^/@{-}[drr]\ar@/_/@{-}[dll]&\\
&&&&&\ar@/^/@{-}[dllll]&&&&\ar@/_/@{-}[drrrr]&\\
\ar@{.}[rrrrr]&&&&&0\ar@{.}[uu]\ar@{.}[rr]&&a\ar@.[rrrrrrr]&&&&&&&}$$

(1) This is actually the definition of $g_t^a$, and the density formula in the statement comes from the following computation, assuming $f\sim g_t$, valid for any $\varphi:\mathbb R\to\mathbb R$:
\begin{eqnarray*}
E(\varphi(f+a))
&=&\frac{1}{\sqrt{2\pi t}}\int_\mathbb R\varphi(x+a)e^{-x^2/2t}\,dx\\
&=&\frac{1}{\sqrt{2\pi t}}\int_\mathbb R\varphi(x)e^{-(x-a)^2/2t}\,dx
\end{eqnarray*}

Indeed, the density of $f+a\sim g_t^a$ follows to be $1/\sqrt{2\pi t}\,e^{-(x-a)^2/2t}\,dx$, as claimed.

\medskip

(2) Regarding the mean, that is clearly $E=a$ on the picture, because the symmetry axis lies there, at $a\in\mathbb R$. As for the variance, by partial integration, we have:
\begin{eqnarray*}
M_2-aM_1
&=&\frac{1}{\sqrt{2\pi t}}\int_\mathbb R(x^2-ax)e^{-(x-a)^2/2t}dx\\
&=&\frac{1}{\sqrt{2\pi t}}\int_\mathbb R(tx)\left(-e^{-(x-a)^2/2t}\right)'dx\\
&=&\frac{1}{\sqrt{2\pi t}}\int_\mathbb Rte^{-(x-a)^2/2t}dx\\
&=&t
\end{eqnarray*}

Thus $M_2=a^2+t$, and the variance is $V=(a^2+t)-a^2=t$, as stated.

\medskip

(3) We have indeed the following computation, using a variable $f\sim g_t$:
\begin{eqnarray*}
M_k
&=&E((f+a)^k)\\
&=&\sum_{r=0}^{[k/2]}\binom{k}{2r}E(f^{2r})a^{k-2r}\\
&=&\sum_{r=0}^{[k/2]}\binom{k}{2r}(2r)!!t^ra^{k-2r}
\end{eqnarray*}

(4) By using the same trick as in (2), we obtain the following recurrence:
\begin{eqnarray*}
M_k-aM_{k-1}
&=&\frac{1}{\sqrt{2\pi t}}\int_\mathbb R(x^k-ax^{k-1})e^{-(x-a)^2/2t}dx\\
&=&\frac{1}{\sqrt{2\pi t}}\int_\mathbb R(tx^{k-1})\left(-e^{-(x-a)^2/2t}\right)'dx\\
&=&\frac{1}{\sqrt{2\pi t}}\int_\mathbb R(k-1)tx^{k-2}\cdot e^{-(x-a)^2/2t}dx\\
&=&(k-1)tM_{k-2}
\end{eqnarray*}

In practice, here is the list of the first few moments, obtained in this way:
$$M_1=a$$
$$M_2=a^2+t$$
$$M_3=a^3+3at$$
$$M_4=a^4+6a^2t+3t^2$$
$$M_5=a^5+10a^3t+15at^2$$
$$M_6=a^6+15a^4t+45a^2t^2+15t^3$$
$$M_7=a^7+21a^5t+105a^3t^2+105at^3$$
$$M_8=a^8+28a^6t+210a^4t^2+420a^2t^3+105t^4$$

(5) Let us look now at the central moments. And here, fortunately, we obtain the same moments as for $g_t$, as shown by the following computation:
\begin{eqnarray*}
M_k'(g_t^a)
&=&\frac{1}{\sqrt{2\pi t}}\int_\mathbb R(x-a)^ke^{-(x-a)^2/2t}dx\\
&=&\frac{1}{\sqrt{2\pi t}}\int_\mathbb Rx^ke^{-x^2/2t}dx\\
&=&M_k(g_t)
\end{eqnarray*}

Thus, by using now the formula for the moments of $g_t$, we have:
$$M_k'(g_t^a)=\delta_{2|k}t^{k/2}k!!$$

(6) Next, since the central moments are the same for $g_t^a$ and $g_t$, and since the variances are also the same, the normalized central moments must all coincide:
$$M_k''(g^t_a)=M_k''(g_t)$$

(7) In particular, we deduce from this formula that the skewness is $\gamma=0$, and that the kurtosis is $\kappa=3$, exactly as for the laws $g_t$ before, as claimed.

\medskip

(8) Regarding now the Fourier transform computation, we have:
\begin{eqnarray*}
F_{g_t^a}(x)
&=&\frac{1}{\sqrt{2\pi t}}\int_\mathbb Re^{-(y-a)^2/2t+ixy}dy\\
&=&\frac{1}{\sqrt{2\pi t}}\int_\mathbb Re^{-y^2/2t+ix(y+a)}dy\\
&=&\frac{e^{iax}}{\sqrt{2\pi t}}\int_\mathbb Re^{-y^2/2t+ixy}dy\\
&=&e^{iax}F_{g_t}(x)
\end{eqnarray*}

Now by using the formula of the Fourier transform of $g_t$, we get:
$$F_{g_t^a}(x)=e^{iax}e^{-tx^2/2}=e^{iax-tx^2/2}$$
 
(9) Observe now that this Fourier transform that we just computed satisfies:
\begin{eqnarray*}
F_{g_s^a}(x)F_{g_t^b}(x)
&=&e^{iax-sx^2/2}e^{ibx-tx^2/2}\\
&=&e^{i(a+b)x-(s+t)x^2/2}\\
&=&F_{g_{s+t}^{a+b}}(x)
\end{eqnarray*}

Thus, at the level of the corresponding laws, we have the following formula:
$$g_s^a*g_t^b=g_{s+t}^{a+b}$$

And with this, done, we are led to the conclusions in the statement.
\end{proof}

Finally, a few words about the relation with groups, which are still missing from our picture. In analogy with what we know about the Poisson and Bessel laws, we have:

\begin{fact}
For the orthogonal group $O_N$, the main character
$$\chi(U)=\sum_{i=1}^NU_{ii}$$
follows with $N\to\infty$ the normal law $g_1$. More generally, the truncated character
$$\chi(U)=\sum_{i=1}^{[tN]}U_{ii}$$
follows with $N\to\infty$ the normal law $g_t$, for any $t\in(0,1]$. 
\end{fact}

Which looks very good, and in tune with what we know about the Poisson and Bessel laws, with the overall conclusion being that the passage $p_t\to g_t$, from discrete to continuous, amounts in performing a passage $S_N\to O_N$ at the group theory level. 

\bigskip

However, and here comes the point, proving such things is quite difficult, because unlike in the finite group case, where we were able to get away with elementary counts on $S_N,H_N$ and other reflection groups, in order to deal with continuous groups like $O_N$, we have to know more about the uniform integration on such groups, and this is something non-trivial. So, all this will have to wait a bit, we will be back to it in chapter 14.

\section*{4c. Complex variables}

Let us discuss now the complex analogues of all the above, with a notion of complex normal, or Gaussian law. To start with, we have the following definition:

\begin{definition}
A complex random variable is a variable $f:X\to\mathbb C$. In the discrete case, the law of such a variable is the complex probability measure
$$\mu=\sum_i\alpha_i\delta_{z_i}\quad,\quad\alpha_i\geq0\quad,\quad\sum_i\alpha_i=1\quad,\quad z_i\in\mathbb C$$
given by the following formula, with $P$ being the probability over $X$,
$$\mu=\sum_{z\in\mathbb C}P(f=z)\delta_z$$
with the sum being finite or countable, as per our discretness assumption.
\end{definition}

Observe the similarity with the analogous notions introduced in chapter 2, for the real variables $f:X\to\mathbb R$. In fact, what we are doing here is to extend the formalism from chapter 2, from real to complex, in a straightforward way. As a basic example for this, any real variable $f:X\to\mathbb R$ can be regarded as a complex variable $f:X\to\mathbb C$.

\bigskip

In order to understand the precise relation with the real theory, from chapter 2, we can decompose any complex variable $f:X\to\mathbb C$ as a sum, as follows:
$$f=g+ih\quad,\quad g=Re(f),\ h=Im(f)$$

With this done, we have the following computation, for the corresponding law:
\begin{eqnarray*}
\mu
&=&\sum_{z\in\mathbb C}P(f=z)\delta_z\\
&=&\sum_{x,y\in\mathbb R}P(f=x+iy)\delta_{x+iy}\\
&=&\sum_{x,y\in\mathbb R}P(g+ih=x+iy)\delta_{x+iy}\\
&=&\sum_{x,y\in\mathbb R}P(g=x,h=y)\delta_{x+iy}
\end{eqnarray*}

In the case where the real and imaginary parts $g,h:X\to\mathbb R$ are independent, we can say more about this, with the above computation having the following continuation:
\begin{eqnarray*}
\mu
&=&\sum_{x,y\in\mathbb R}P(g=x,h=y)\delta_{x+iy}\\
&=&\sum_{x,y\in\mathbb R}P(g=x)P(h=y)\delta_{x+iy}\\
&=&\sum_{x,y\in\mathbb R}P(g=x)P(h=y)\delta_x*\delta_{iy}\\
&=&\left(\sum_{x\in\mathbb R}P(g=x)\delta_x\right)*\left(\sum_{y\in\mathbb R}P(h=y)\delta_{iy}\right)\\
&=&\mu_g*i\mu_h
\end{eqnarray*}

To be more precise, we have used here in the beginning the independence of the variables $h,g:X\to\mathbb R$, and at the end we have denoted the measure on the right, which is obtained from $\mu_h$ by putting this measure on the imaginary axis, by $i\mu_h$.

\bigskip

All this is quite interesting, going beyond what we know so far about basic probability, in the real case, so let us record this finding, along with a bit more, as follows:

\begin{theorem}
For a discrete complex random variable $f:X\to\mathbb C$, decomposed into real and imaginary parts as $f=g+ih$, and with $g,h$ assumed independent, we have
$$\mu_f=\mu_g*i\mu_h$$
with $*$ being the usual convolution operation, $\delta_z*\delta_t=\delta_{z+t}$, and with $\mu\to i\mu$ denoting the rotated version, $\mathbb R\to i\mathbb R$. If $g,h$ are not independent, this formula does not hold.
\end{theorem}

\begin{proof}
We already know that the first assertion holds, as explained in the above. As for the second assertion, this follows by carefully examining the above computation. Indeed, we have used only at one point the independence of $g,h$, so for the formula $\mu_f=\mu_g*i\mu_h$ to hold, the equality used at that point, which is as follows, must hold:
$$\sum_{x,y\in\mathbb R}P(g=x,h=y)\delta_{x+iy}
=\sum_{x,y\in\mathbb R}P(g=x)P(h=y)\delta_{x+iy}$$

But this is the same as saying that the following must hold, for any $x,y$:
$$P(g=x,h=y)=P(g=x)P(h=y)$$

We conclude that, in order for the decomposition formula $\mu_f=\mu_g*i\mu_h$ to hold, the real and imaginary parts $g,h:X\to\mathbb R$ must be independent, as stated.
\end{proof}

Many other things can be said, along these lines, inspired by the basic theory of the complex numbers. Going now straight to the point, probabilistic limiting theorems, let us discuss the complex analogue of the CLT. We first have the following statement:

\begin{theorem}
Given discrete complex variables $f_1,f_2,f_3,\ldots$ whose real and imaginary parts are i.i.d., centered, and with common variance $t>0$, we have
$$\frac{1}{\sqrt{n}}\sum_{i=1}^nf_i\sim C_t$$
with $n\to\infty$, in moments, where $C_t$ is the law of a complex variable whose real and imaginary parts are independent, and each following the law $g_t$.
\end{theorem}

\begin{proof}
This follows indeed from the real CLT, established in Theorem 4.13, simply by taking the real and imaginary parts of all the variables involved.
\end{proof}

It is tempting at this point to call Theorem 4.22 the complex CLT, or CCLT, but before doing that, let us study a bit more all this. We would like to have a better understanding of the limiting law $C_t$ at the end, and for this purpose, let us look at a sum as follows, with $a,b$ being real independent variables, both following the normal law $g_t$:
$$c=a+ib$$

To start with, this variable is centered, in a complex sense, because we have:
\begin{eqnarray*}
E(c)
&=&E(a+ib)\\
&=&E(a)+iE(b)\\
&=&0+i\cdot 0\\
&=&0
\end{eqnarray*}

Regarding now the variance, things are more complicated, because the usual variance formula from the real case, which is $V(c)=E(c^2)$ in the centered case, will not provide us with a positive number, in the case where our variable is not real. So, in order to have a variance which is real, and positive too, we must rather use a formula of type $V(c)=E(|c|^2)$, in the centered case. And, with this convention for the variance, we have then the following computation, for the variance of the above variable $c$:
\begin{eqnarray*}
V(c)
&=&E(|c|^2)\\
&=&E(a^2+b^2)\\
&=&E(a^2)+E(b^2)\\
&=&V(a^2)+V(b^2)\\
&=&t+t\\
&=&2t
\end{eqnarray*}

But this suggests to divide everything by $\sqrt{2}$, as to have in the end a variable having complex variance $t$, in our sense, and we are led in this way into:

\index{complex normal law}
\index{complex Gaussian law}

\begin{definition}
The complex normal, or Gaussian law of parameter $t>0$ is
$$G_t=law\left(\frac{1}{\sqrt{2}}(a+ib)\right)$$
where $a,b$ are real and independent, each following the law $g_t$.
\end{definition}

In short, the complex normal laws appear as natural complexifications of the real normal laws. As in the real case, these measures form convolution semigroups:

\index{convolution semigroup}

\begin{proposition}
The complex Gaussian laws have the property
$$G_s*G_t=G_{s+t}$$
for any $s,t>0$, and so they form a convolution semigroup.
\end{proposition}

\begin{proof}
This follows indeed from the real result, namely $g_s*g_t=g_{s+t}$, established in Theorem 4.16, simply by taking real and imaginary parts.
\end{proof}

We have as well the following complex analogue of the CLT:

\index{CCLT}
\index{Complex CLT}
\index{Complex Central Limit Theorem}
\index{complex variables}
\index{complex normal law}

\begin{theorem}[CCLT]
Given discrete complex variables $f_1,f_2,f_3,\ldots$ whose real and imaginary parts are i.i.d. and centered, and having variance $t>0$, we have
$$\frac{1}{\sqrt{n}}\sum_{i=1}^nf_i\sim G_t$$
with $n\to\infty$, in moments.
\end{theorem}

\begin{proof}
This follows indeed from our previous CCLT result, from Theorem 4.22, by dividing everything by $\sqrt{2}$, as explained in the above.
\end{proof}

Finally, in relation with groups, we have the following analogue of Fact 4.19:

\begin{fact}
For the unitary group $U_N$, the main character
$$\chi(U)=\sum_{i=1}^NU_{ii}$$
follows with $N\to\infty$ the normal law $G_1$. More generally, the truncated character
$$\chi(U)=\sum_{i=1}^{[tN]}U_{ii}$$
follows with $N\to\infty$ the normal law $G_t$, for any $t\in(0,1]$. 
\end{fact}

Which looks of course good and conceptual, but as before with Fact 4.19, when it comes to prove such things, this is no easy business. More on this later in this book.

\section*{4d. Wick formula}

Regarding now the moments, the situation here is more complicated than in the real case, because in order to have good results, we have to deal with both the complex variables, and their conjugates. Let us formulate the following definition:

\index{colored integers}
\index{colored moments}

\begin{definition}
The moments a complex variable $f\in L^\infty(X)$ are the numbers
$$M_k=E(f^k)$$
depending on colored integers $k=\circ\bullet\bullet\circ\ldots\,$, with the conventions
$$f^\emptyset=1\quad,\quad f^\circ=f\quad,\quad f^\bullet=\bar{f}$$
and multiplicativity, in order to define the colored powers $f^k$.
\end{definition}

As an illustration for this notion, which is something very intuitive, here are the formulae of the four possible order 2 moments of a complex variable $f$:
$$M_{\circ\circ}=E(f^2)\quad,\quad M_{\circ\bullet}=E(f\bar{f})$$
$$M_{\bullet\circ}=E(\bar{f}f)\quad,\quad M_{\bullet\bullet}=E(\bar{f}^2)$$

Observe that, since $f,\bar{f}$ commute, we have the following identity, which shows that there is a bit of redundancy in our above definition, as formulated:
$$M_{\circ\bullet}=M_{\bullet\circ}$$

In fact, again since $f,\bar{f}$ commute, we can permute terms, in the general context of Definition 4.27, and restrict the attention to exponents of the following type:
$$k=\ldots\circ\circ\circ\bullet\bullet\bullet\bullet\ldots$$

However, our results about the complex Gaussian laws, and other complex laws, later on, not to talk about laws of matrices, random matrices and other noncommuting variables, that will appear later too, will look better without doing this. So, we will use Definition 4.27 as stated. Getting to work now, we first have the following result:

\begin{theorem}
The moments of the complex normal law are given by
$$M_k(G_t)=\begin{cases}
t^pp!&(k\ {\rm uniform, of\ length}\ 2p)\\
0&(k\ {\rm not\ uniform})
\end{cases}$$
where $k=\circ\bullet\bullet\circ\ldots$ is called uniform when it contains the same number of $\circ$ and $\bullet$.
\end{theorem}

\begin{proof}
We must compute the moments, with respect to colored integer exponents $k=\circ\bullet\bullet\circ\ldots$ as above, of the variable from Definition 4.23, namely:
$$f=\frac{1}{\sqrt{2}}(a+ib)$$

We can assume that we are in the case $t=1$, and the proof here goes as follows:

\medskip

(1) As a first observation, in the case where our exponent $k=\circ\bullet\bullet\circ\ldots$ is not uniform, a standard rotation argument shows that the corresponding moment of $f$ vanishes. To be more precise, the variable $f'=wf$ is complex Gaussian too, for any complex number $w\in\mathbb T$, and from $M_k(f)=M_k(f')$ we obtain $M_k(f)=0$, in this case.

\medskip

(2) In the uniform case now, where the exponent $k=\circ\bullet\bullet\circ\ldots$ consists of $p$ copies of $\circ$ and $p$ copies of $\bullet$\,, the corresponding moment can be computed as follows:
\begin{eqnarray*}
M_k
&=&\frac{1}{2^p}E\left[(a^2+b^2)^p\right]\\
&=&\frac{1}{2^p}\sum_r\binom{p}{r}E(a^{2r})E(b^{2p-2r})\\
&=&\frac{1}{2^p}\sum_r\binom{p}{r}(2r)!!(2p-2r)!!\\
&=&\frac{1}{2^p}\sum_r\frac{p!}{r!(p-r)!}\cdot\frac{(2r)!}{2^rr!}\cdot\frac{(2p-2r)!}{2^{p-r}(p-r)!}\\
&=&\frac{p!}{4^p}\sum_r\binom{2r}{r}\binom{2p-2r}{p-r}
\end{eqnarray*}

(3) In order to finish now the computation, let us recall that we have the following formula, coming from the generalized binomial formula, at exponent $-1/2$:
$$\frac{1}{\sqrt{1+t}}=\sum_{q=0}^\infty\binom{2q}{q}\left(\frac{-t}{4}\right)^q$$

By taking the square of this series, we obtain the following formula:
$$\frac{1}{1+t}
=\sum_p\left(\frac{-t}{4}\right)^p\sum_r\binom{2r}{r}\binom{2p-2r}{p-r}$$

Now by looking at the coefficient of $t^p$ on both sides, we conclude that the sum on the right equals $4^p$. Thus, we can finish the moment computation in (2), as follows:
$$M_k=\frac{p!}{4^p}\times 4^p=p!$$

We are therefore led to the conclusion in the statement.
\end{proof}

Before going further, let us record the following consequence, of the above:

\begin{theorem}
The moments of the Rayleigh law, given by
$$R_t=law(|G_t|)$$
are given by the following formula, at the parameter value $t=1$,
$$M_p=p!$$ 
and are given by the formula $M_p=t^pp!$, in general.
\end{theorem} 

\begin{proof}
This follows indeed from Theorem 4.28, or simply from the computation (2) in the above proof, at $t=1$, and then by rescaling by $t$, in general.
\end{proof}

Many other things can be said about the Rayleigh laws, which are quite interesting mathematical objects, notably with some further formulae, regarding the Fourier transform, and the density too, which can be obtained as a consequence of Theorem 4.29. We will be back to these remarkable laws, on several occasions, in what follows.

\bigskip

Let us record, however, the following statement, called Rayleigh Central Limiting Theorem, which is something of theoretical importance:

\begin{theorem}[RCLT]
Given discrete complex variables $f_1,f_2,f_3,\ldots$ whose real and imaginary parts are i.i.d. and centered, and having variance $t>0$, we have
$$\left|\frac{1}{\sqrt{n}}\sum_{i=1}^nf_i\right|\sim R_t$$
with $n\to\infty$, in moments.
\end{theorem}

\begin{proof}
This follows indeed from our previous Central Limiting result, namely the CCLT from Theorem 4.25, by taking absolute values on both sides.
\end{proof}

As a comment here, observe that the Rayleigh laws, while being certainly real, capture the essentials of the two-dimensional nature of the normal laws. So, probably time now for some further thinking, in relation with what cat and kitty were saying. But with this being something a bit complicated, and with our advisors being not really available, one with a hangover, and the other chasing mice, we will not further explore this subject.

\bigskip

Getting back now to the complex normal laws, what we have in Theorem 4.28 about them is usually what is needed in practice, when dealing with moments. But, as before with the real Gaussian laws, or even further before with the Poisson and Bessel laws, a better-looking statement regarding the moments is in terms of partitions.

\bigskip

Indeed, given a colored integer $k=\circ\bullet\bullet\circ\ldots\,$, let us say that $\pi\in P_2(k)$ is matching when it pairs $\circ-\bullet$ symbols. With this convention, we have the following result:

\index{matching pairings}

\begin{theorem}
The moments of the complex normal law are the numbers
$$M_k(G_t)=\sum_{\pi\in\mathcal P_2(k)}t^{|\pi|}$$
where $\mathcal P_2(k)$ are the matching pairings of $\{1,\ldots,k\}$, and $|.|$ is the number of blocks.
\end{theorem}

\begin{proof}
This is a reformulation of Theorem 4.28. Indeed, we can assume that we are in the case $t=1$, and here we know from Theorem 4.28 that the moments are:
$$M_k=\begin{cases}
(|k|/2)!&(k\ {\rm uniform})\\
0&(k\ {\rm not\ uniform})
\end{cases}$$

On the other hand, the numbers $|\mathcal P_2(k)|$ are given by exactly the same formula. Indeed, in order to have a matching pairing of $k$, our exponent $k=\circ\bullet\bullet\circ\ldots$ must be uniform, consisting of $p$ copies of $\circ$ and $p$ copies of $\bullet$, with $p=|k|/2$. But then the matching pairings of $k$ correspond to the permutations of the $\bullet$ symbols, as to be matched with $\circ$ symbols, and so we have $p!$ such pairings. Thus, we have the same formula as for the moments of $f$, and we are led to the conclusion in the statement.
\end{proof}

In practice, we also need to know how to compute joint moments. We have here:

\index{joint moments}
\index{Wick formula}

\begin{theorem}[Wick formula]
Given independent variables $f_i$, each following the complex normal law $G_t$, with $t>0$ being a fixed parameter, we have the formula
$$E\left(f_{i_1}^{k_1}\ldots f_{i_s}^{k_s}\right)=t^{s/2}\#\left\{\pi\in\mathcal P_2(k)\Big|\pi\leq\ker i\right\}$$
where $k=k_1\ldots k_s$ and $i=i_1\ldots i_s$, for the joint moments of these variables, where $\pi\leq\ker i$ means that the indices of $i$ must fit into the blocks of $\pi$, in the obvious way.
\end{theorem}

\begin{proof}
This is something well-known, which can be proved as follows:

\medskip

(1) Let us first discuss the case where we have a single variable $f$, which amounts in taking $f_i=f$ for any $i$ in the formula in the statement. What we have to compute here are the moments of $f$, with respect to colored integer exponents $k=\circ\bullet\bullet\circ\ldots\,$, and the formula in the statement tells us that these moments must be:
$$E(f^k)=t^{|k|/2}|\mathcal P_2(k)|$$

But this is the formula in Theorem 4.31, so we are done with this case.

\medskip

(2) In general now, when expanding the product $f_{i_1}^{k_1}\ldots f_{i_s}^{k_s}$ and rearranging the terms, we are left with doing a number of computations as in (1), and then making the product of the expectations that we found. But this amounts in counting the partitions in the statement, with the condition $\pi\leq\ker i$ there standing for the fact that we are doing the various type (1) computations independently, and then making the product.
\end{proof}

The above statement is one of the possible formulations of the Wick formula, and there are many more formulations, which are all useful. For instance, we have:

\index{Wick formula}

\begin{theorem}[Wick formula 2]
Given independent variables $f_i$, each following the complex normal law $G_t$, with $t>0$ being a fixed parameter, we have the formula
$$E\left(f_{i_1}\ldots f_{i_k}f_{j_1}^*\ldots f_{j_k}^*\right)=t^k\#\left\{\pi\in S_k\Big|i_{\pi(r)}=j_r,\forall r\right\}$$
for the non-vanishing joint moments of these variables.
\end{theorem}

\begin{proof}
This follows from the usual Wick formula, from Theorem 4.32. With some changes in the indices and notations, the formula there reads:
$$E\left(f_{I_1}^{K_1}\ldots f_{I_s}^{K_s}\right)=t^{s/2}\#\left\{\sigma\in\mathcal P_2(K)\Big|\sigma\leq\ker I\right\}$$

Now observe that we have $\mathcal P_2(K)=\emptyset$, unless the colored integer $K=K_1\ldots K_s$ is uniform, in the sense that it contains the same number of $\circ$ and $\bullet$ symbols. Up to permutations, the non-trivial case, where the moment is non-vanishing, is the case where the colored integer $K=K_1\ldots K_s$ is of the following special form:
$$K=\underbrace{\circ\circ\ldots\circ}_k\ \underbrace{\bullet\bullet\ldots\bullet}_k$$

So, let us focus on this case, which is the non-trivial one. Here we have $s=2k$, and we can write the multi-index $I=I_1\ldots I_s$ in the following way:
$$I=i_1\ldots i_k\ j_1\ldots j_k$$

With these changes made, the above usual Wick formula reads:
$$E\left(f_{i_1}\ldots f_{i_k}f_{j_1}^*\ldots f_{j_k}^*\right)=t^k\#\left\{\sigma\in\mathcal P_2(K)\Big|\sigma\leq\ker(ij)\right\}$$

The point now is that the matching pairings $\sigma\in\mathcal P_2(K)$, with $K=\circ\ldots\circ\bullet\ldots\bullet\,$, of length $2k$, as above, correspond to the permutations $\pi\in S_k$, in the obvious way. With this identification made, the above modified usual Wick formula becomes:
$$E\left(f_{i_1}\ldots f_{i_k}f_{j_1}^*\ldots f_{j_k}^*\right)=t^k\#\left\{\pi\in S_k\Big|i_{\pi(r)}=j_r,\forall r\right\}$$

Thus, we have reached to the formula in the statement, and we are done.
\end{proof}

Finally, here is one more formulation of the Wick formula, useful as well:

\index{Wick formula}

\begin{theorem}[Wick formula 3]
Given independent variables $f_i$, each following the complex normal law $G_t$, with $t>0$ being a fixed parameter, we have the formula
$$E\left(f_{i_1}f_{j_1}^*\ldots f_{i_k}f_{j_k}^*\right)=t^k\#\left\{\pi\in S_k\Big|i_{\pi(r)}=j_r,\forall r\right\}$$
for the non-vanishing joint moments of these variables.
\end{theorem}

\begin{proof}
This follows from our second Wick formula, from Theorem 4.33, simply by permuting the terms, as to have an alternating sequence of plain and conjugate variables. Alternatively, we can start with Theorem 4.32, and then perform the same manipulations as in the proof of Theorem 4.33, but with the exponent being this time as follows: 
$$K=\underbrace{\circ\bullet\circ\bullet\ldots\ldots\circ\bullet}_{2k}$$

Thus, we are led to the conclusion in the statement.
\end{proof}

Many other things can be said about the normal laws, and also about Rayleigh laws. We will be back to this, notably in chapter 15 below, when doing random matrices.

\section*{4e. Exercises}

Strange and exciting chapter that we had here, with our explorations involving ships, drunken sailors and mysterious cats, and as exercises on all this, we have:

\begin{exercise}
Do some abstract simulations, of your choice, for the CLT.
\end{exercise}

\begin{exercise}
Do as well some computer simulations for the CLT.
\end{exercise}

\begin{exercise}
Prove that the density is even when the odd moments vanish.
\end{exercise}

\begin{exercise}
Find a genius method for dealing with the Gauss integral.
\end{exercise}

\begin{exercise}
How to best integrate $g_t$, over a bounded interval $[a,b]$?
\end{exercise}

\begin{exercise}
Learn more about the Rayleigh laws, and their properties.
\end{exercise}

\begin{exercise}
Work out our RCLT program, outlined in the above.
\end{exercise}

\begin{exercise}
Do some probabilistic computations for $O_N,U_N$.
\end{exercise}

As bonus exercise, learn some systematic probability, say from probabilists, or from physicists. Be aware tough that the books there are also based on some cat knowledge.

\part{Measure theory}

\ \vskip50mm

\begin{center}
{\em And it's no nay never

No nay never no more

Will I play the wild rover

No never no more}
\end{center}

\chapter{Measure theory}

\section*{5a. Metric spaces}

Welcome to measure theory, take two. We have seen in Part I a lot of interesting mathematics, mostly of probabilistic nature, coming as a complement to the basic calculus that you surely know, and often pushing us into developing measure theory.

\bigskip

So, this is what we will do here, in this present Part II, developing mesure theory, in a standard, rigorous and mathematical way. Getting started now, before even talking measures, we need spaces. And here, we have the following question:

\begin{question}
What is a space?
\end{question}

And good question this is. There are of course a myriad answers to it, but remembering that we are analysts, let us consider instead a more modest question, as follows:

\begin{question}
What are the spaces where we can do analysis?
\end{question}

But, this looks again like an excessively general question. Indeed, examples of such spaces include for instance $\mathbb R,\mathbb C$, then the powers $\mathbb R^N,\mathbb C^N$, and then all sorts of curves, surfaces and so on $X\subset\mathbb R^N,X\subset\mathbb C^N$. And with the story being not over here, because we should normally be able to allow $N=\infty$ too in all this. Plus, there might be other spaces $X$ where we can do analysis, which do not necessarily embed into $\mathbb R^N,\mathbb C^N$. 

\bigskip

In order to see more clearly what is going on, let us forget about spaces, and think instead at what doing analysis means. And here, for doing analysis, we obviously need convergence, with the usual definition of convergence being as follows:
$$x_n\to x\iff d(x,x_n)\to0$$

Long story short, for doing analysis on a space $X$, we need a distance on $X$. Thus, as answer to Question 5.2, which is something very intuitive, we have:

\begin{answer}
In order to be able to do analysis on $X$, we need a distance on $X$:
$$d:X\times X\to\mathbb R_+$$
Indeed, once we have $d$, we can talk about convergence, then continuity and so on.
\end{answer}

Very good all this, and the question is now, what are the exact axioms on $(X,d)$. And here, a bit of thinking leads to the following definition, fine-tuning Answer 5.3:

\index{distance}
\index{metric space}

\begin{definition}
A metric space is a set $X$ with a distance function $d:X\times X\to\mathbb R_+$, having the following properties:
\begin{enumerate}
\item $d(x,y)>0$ if $x\neq y$, and $d(x,x)=0$.

\item $d(x,y)=d(y,x)$.

\item $d(x,y)\leq d(x,z)+d(y,z)$.
\end{enumerate}
\end{definition}

As a basic example, we have $\mathbb R^N$, as well as any of its subsets $X\subset\mathbb R^N$. Indeed, the first two axioms are clear, and for the third axiom, we must prove that:
$$\sqrt{\sum_i(a_i+b_i)^2}\leq\sqrt{\sum_i a_i^2}+\sqrt{\sum_i b_i^2}$$

Now by raising to the square, this is the same as proving that:
$$\left(\sum_ia_ib_i\right)^2\leq\left(\sum_ia_i^2\right)\left(\sum_ib_i^2\right)$$

But this latter inequality is one of the many equivalent formulations of the Cauchy-Schwarz inequality, that we know well from calculus, and which follows by using the fact that $f(t)=\sum_i(a_i+tb_i)^2$ being positive, its discriminant must be negative.

\bigskip

As another example, we have $\mathbb C^N$, as well as any of its subsets $X\subset\mathbb C^N$. Indeed, this follows either from $\mathbb C^N\simeq\mathbb R^{2N}$, or directly, along the lines of the above proof for $\mathbb R^N$. To be more precise, after some algebra, we are left with proving the following inequality:
$$\left|\sum_ia_i\bar{b}_i\right|^2\leq\left(\sum_i|a_i|^2\right)\left(\sum_i|b_i|^2\right)$$

But this is the complex version of the Cauchy-Schwarz inequality, that we know well from calculus, and which follows by using the fact that the function $f(t)=\sum_i|a_i+twb_i|^2$ with $t\in\mathbb R$ and $|w|=1$ being positive, its discriminant must be negative.

\bigskip

Here is now another example, which at first looks new and interesting, but is in fact not new, because it appears as a subspace of some suitable $\mathbb R^N$:

\begin{proposition}
Given a finite set $X$, the following function is a metric on it, called discrete metric:
$$d(x,y)=\begin{cases}
1&{\rm if}\ x\neq y\\
0&{\rm if}\ x=y
\end{cases}$$
This metric space is in fact the $N$-simplex, which can be realized as a subspace of $\mathbb R^{N-1}$, or, more conveniently, as a subspace of $\mathbb R^N$. 
\end{proposition}

\begin{proof}
There are several things going on here, the idea being as follows:

\medskip

(1) First of all, the axioms from Definition 5.4 are trivially satisfied, and with the main axiom, namely the triangle inequality, basically coming from:
$$1\leq1+1$$

(2) At the level of examples, at $|X|=1$ we obtain a point, at $|X|=2$ we obtain a segment, at $|X|=3$ we obtain an equilateral triangle, at $|X|=4$ we obtain a regular tetrahedron, and so on. Thus, what we have in general, at $|X|=N$, is the arbitrary dimensional generalization of this series of geometric objects, called $N$-simplex.

\medskip

(3) In what regards now the geometric generalization of the $N$-simplex, our above examples suggest to look for an embedding $X\subset\mathbb R^{N-1}$. Which is something which is certainly possible, but the computations here are quite complicated.

\medskip

(4) However, mathematics, or perhaps physics come to the rescue, via the idea ``add a dimension, for getting smarter''. Indeed, when looking for an embedding $X\subset\mathbb R^N$ things drastically simplify, because we can simply take $X$ to be the standard basis of $\mathbb R^N$:
$$X=\{e_1,\ldots,e_N\}$$

Indeed, we have by definition $d(e_i,e_j)=1$ for any $i\neq j$. So, we solved our embedding problem, just like that, without doing any computations and trigonometry.
\end{proof}

Moving ahead now with some theory, and allowing us a bit of slopiness, we have:

\begin{proposition}
We can talk about limits inside metric spaces $X$, by saying that
$$x_n\to x\iff d(x_n,x)\to0$$
and we can talk as well about continuous functions $f:X\to Y$, by requiring that
$$x_n\to x\implies f(x_n)\to f(x)$$
and with these notions in hand, all the basic results from the cases $X=\mathbb R,\mathbb C$ extend.
\end{proposition}

\begin{proof}
All this is very standard, and we will leave this as an exercise, namely carefully checking what you learned in basic calculus, in relation with limits and continuity, in the cases $X=\mathbb R,\mathbb C$, and working out the metric space extensions of this.
\end{proof}

More interestingly now, we can talk about open and closed sets inside metric spaces $X$, again in analogy with what we did for $X=\mathbb R,\mathbb C$, but with a whole lot of interesting new phenomena appearing. So, we will do this in detail. Let us start with:

\begin{definition}
Let $X$ be a metric space.
\begin{enumerate}
\item The open balls are the sets $B_x(r)=\{y\in X|d(x,y)<r\}$.

\item The closed balls are the sets $\bar{B}_x(r)=\{y\in X|d(x,y)\leq r\}$.

\item $U\subset X$ is called open if for any $x\in U$ we have a ball $B_x(r)\subset U$.

\item $F\subset X$ is called closed if its complement $F^c\subset X$ is open.
\end{enumerate}
\end{definition}

At the level of examples, you can quickly convince yourself, by working out a few of them, that our notions above coincide with the usual ones, that we know well, in the case $X=\mathbb R,\mathbb C$. We will be back to this later, with some general results in this sense, confirming all this. But for the moment, let us work out the basics. We first have the following result, clarifying some terminology issues from Definition 5.7:

\begin{proposition}
The open balls are open, and the closed balls are closed.
\end{proposition}

\begin{proof}
This might sound like a joke, but it is not one, because this is the kind of thing that we have to check. Fortunately, all this is elementary, as follows:

\medskip

(1) Given an open ball $B_x(r)$ and a point $y\in B_x(r)$, by using the triangle inequality we have $B_y(r')\subset B_x(r)$, with $r'=r-d(x,y)$. Thus, $B_x(r)$ is indeed open.

\medskip

(2) Given a closed ball $\bar{B}_x(r)$ and a point $y\in B_x(r)^c$, by using the triangle inequality we have $B_y(r')\subset B_x(r)^c$, with $r'=d(x,y)-r$. Thus, $\bar{B}_x(r)$ is indeed closed.
\end{proof}

Here is now something more interesting, making the link with our intuitive understanding of the notion of closedness, coming from our experience so far with analysis:

\begin{theorem}
For a subset $F\subset X$, the following are equivalent:
\begin{enumerate}
\item $F$ is closed in our sense, meaning that $F^c$ is open.

\item We have $x_n\to x,x_n\in F\implies x\in F$.
\end{enumerate}
\end{theorem}

\begin{proof}
We can prove this by double implication, as follows:

\medskip

$(1)\implies(2)$ Assume by contradiction $x_n\to x,x_n\in F$ with $x\notin F$. Since we have $x\in F^c$, which is open, we can pick a ball $B_x(r)\subset F^c$. But this contradicts our convergence assumption $x_n\to x$, so we are done with this implication. 

\medskip

$(2)\implies(1)$ Assume by contradiction that $F$ is not closed in our sense, meaning that $F^c$ is not open. Thus, we can find $x\in F^c$ such that there is no ball $B_x(r)\subset F^c$. But with $r=1/n$ this provides us with a point $x_n\in B_x(1/n)\cap F$, and since we have $x_n\to x$, this contradicts our assumption (2). Thus, we are done with this implication too.
\end{proof}

Here is another basic theorem about open and closed sets:

\begin{theorem}
Let $X$ be a metric space.
\begin{enumerate}
\item If $U_i$ are open, then $\cup_iU_i$ is open.

\item If $F_i$ are closed, then $\cap_iF_i$ is closed.

\item If $U_1,\ldots,U_n$ are open, then $\cap_iU_i$ is open.

\item If $F_1,\ldots,F_n$ are closed, then $\cup_iF_i$ is closed.
\end{enumerate}
Moreover, both $(3)$ and $(4)$ can fail for infinite intersections and unions.
\end{theorem}

\begin{proof}
We have several things to be proved, the idea being as follows:

\medskip

(1) This is clear from definitions, because any point $x\in\cup_iU_i$ must satisfy $x\in U_i$ for some $i$, and so has a ball around it belonging to $U_i$, and so to $\cup_iU_i$.

\medskip

(2) This follows from (1), by using the following well-known set theory formula:
$$\left(\bigcup_iU_i\right)^c=\bigcap_iU_i^c$$

(3) Given an arbitrary point $x\in\cap_iU_i$, we have $x\in U_i$ for any $i$, and so we have a ball $B_x(r_i)\subset U_i$ for any $i$. Now with this in hand, let us set:
$$B=B_x(r_1)\cap\ldots\cap B_x(r_n)$$

As a first observation, this is a ball around $x$, $B=B_x(r)$, of radius given by:
$$r=\min(r_1,\ldots,r_n)$$

But this ball belongs to all the $U_i$, and so belongs to their intersection $\cap_iU_i$. We conclude that the intersection $\cap_iU_i$ is open, as desired.

\medskip

(4) This follows from (3), by using the following well-known set theory formula:
$$\left(\bigcap_iU_i\right)^c=\bigcup_iU_i^c$$

(5) Finally, in what regards the counterexamples at the end, these can be both found on $\mathbb R$, and we will leave this as an instructive exercise.
\end{proof}

Still in relation with the open and closed sets, let us formulate as well:

\index{interior of set}
\index{boundary of set}
\index{closure of set}

\begin{definition}
Let $X$ be a metric space, and $E\subset X$ be a subset.
\begin{enumerate}
\item The interior $E^\circ\subset E$ is the set of points $x\in E$ which admit around them open balls $B_x(r)\subset E$.

\item The closure $E\subset\bar{E}$ is the set of points $x\in X$ which appear as limits of sequences $x_n\to x$, with $x\in E$.
\end{enumerate}
\end{definition}

These notions are quite interesting, because they make sense for any set $E$. That is, when $E$ is open, that is open and end of the story, and when $E$ is closed, that is closed and end of the story. In general, however, a set $E\subset X$ is not open or closed, and what we can best do to it, in order to study it with our tools, is to ``squeeze'' it, as follows:
$$E^\circ\subset E\subset\bar{E}$$

In practice now, in order to use the above notions, we need to know a number of things, including that fact that $E$ open implies $E^\circ=E$, the fact that $E$ closed implies $\bar{E}=E$, and many more such results. But all this can be done, and the useful statement here, summarizing all that we need to know about interiors and closures, is as follows:

\begin{theorem}
Let $X$ be a metric space, and $E\subset X$ be a subset.
\begin{enumerate}
\item The interior $E^\circ\subset E$ is the biggest open set contained in $E$.

\item The closure $E\subset\bar{E}$ is the smallest closed set containing $E$.
\end{enumerate}
\end{theorem}

\begin{proof}
We have several things to be proved, the idea being as follows:

\medskip

(1) Let us first prove that the interior $E^\circ$ is open. For this purpose, pick $x\in E^\circ$. We know that we have a ball $B_x(r)\subset E$, and since this ball is open, it follows that we have  $B_x(r)\subset E^\circ$. Thus, the interior $E^\circ$ is open, as claimed.

\medskip

(2) Let us prove now that the closure $\bar{E}$ is closed. For this purpose, we will prove that the complement $\bar{E}^c$ is open. So, pick $x\in\bar{E}^c$. Then $x$ cannot appear as a limit of a sequence $x_n\to x$ with $x_n\in E$, so we have a ball $B_x(r)\subset \bar{E}^c$, as desired.

\medskip

(3) Finally, the maximality and minimality assertions regarding $E^\circ$ and $\bar{E}$ are both routine too, coming from definitions, and we will leave them as exercises.
\end{proof}

As a continuation of this, we can talk as well about density, as follows:

\begin{definition}
We say that a subset $E\subset X$ is dense when:
$$\bar{E}=X$$
That is, any point of $X$ must appear as a limit of points of $E$.
\end{definition}

Obviously, this is something which is in tune with what we know so far from this book, and with the intuitive notion of density. As a basic example, we have $\bar{\mathbb Q}=\mathbb R$.

\section*{5b. Compactness}

Again in analogy with what we know about $X=\mathbb R,\mathbb C$, we can talk about compact sets. However, things here are quite tricky, in the general metric space framework, substantially deviating from what we know, and we will do this in detail. Let us start with:

\begin{definition}
A set $K\subset X$ is called compact if any cover with open sets
$$K\subset\bigcup_iU_i$$
has a finite subcover, $K\subset(U_{i_1}\cup\ldots\cup U_{i_n})$.
\end{definition}

This definition, which is probably new to you, might seem overly abstract, but our claim is that this is the correct definition, and that there is no way of doing otherwise. Let us start with some examples, with $X=\mathbb R$. The situation here is as follows:

\bigskip

(1) A point is obviously compact, and we can choose that finite subcover with $n=1$. Similarly, 2 points are compact, and we can choose the subcover with $n=2$. More generally, $N$ points are compact, and we can choose the subcover with $n=N$.

\bigskip

(2) In contrast, the set $\mathbb N\subset\mathbb R$ is not compact, because we can cover it with of a suitable union of small open intervals around each point, and this open cover has no finite subcover. For the same reasons, the set $\{1/n|n\in\mathbb N\}$ is not compact either.

\bigskip

(3) However, and here comes an interesting point, the following set is compact:
$$K=\left\{\frac{1}{n}\Big|n\in\mathbb N\right\}\cup\{0\}$$

Indeed, any open cover of it $\cup_iU_i$ has to cover 0, and by selecting an open set $U_i$ covering 0, this set $U_i$ will cover the whole $K$, except for finitely many points, due to $1/n\to0$. But these finitely points left, say $N$ of them, can be covered by suitable sets $U_{i_1},\ldots,U_{i_N}$, and by adding this family to the set $U_i$, we have our finite subcover.

\bigskip

As a conclusion to this, Definition 5.14 seems to be in tune with what we know about the compact subsets $K\subset\mathbb R$, namely that these are the sets which are closed and bounded. However, and here comes our point, such things are wrong in general, due to:

\begin{proposition}
Given an infinite set $X$ with the discrete distance on it, namely $d(p,q)=1-\delta_{pq}$, which can be modeled as the basis of a suitable Hilbert space,
$$X=\{e_x\}_{x\in X}\subset l^2(X)$$
this set is closed and bounded, but not compact.
\end{proposition}

\begin{proof}
Here the first part, regarding the modeling of $X$, that we will actually not need here, is something that we already know. Regarding now the second part:

\medskip

(1) $X$ being the total space, it is closed. In fact, since the points of $X$ are open, any subset $E\subset X$ is open, and by taking complements, any set $E\subset X$ is closed as well.

\medskip

(2) $X$ is also bounded, because all distances are smaller than $1$.

\medskip

(3) However, our set $X$ is not compact, because its points being open, as noted above, $X=\cup_{x\in X}\{x\}$ is an open cover, having no finite subcover.
\end{proof}

Let us develop now the theory of compact sets, and see what we get. We first have:

\begin{proposition}
The following hold:
\begin{enumerate}
\item Compact implies closed.

\item Closed inside compact is compact.

\item Compact intersected with closed is compact.
\end{enumerate}
\end{proposition}

\begin{proof}
These assertions are all clear from definitions, as follows:

\medskip

(1) Assume that $K\subset X$ is compact, and let us prove that $K$ is closed. For this purpose, we will prove that $K^c$ is open. So, pick $p\in K^c$. For any $q\in K$ we set $r=d(p,q)/3$, and we consider the following balls, separating $p$ and $q$:
$$U_q=B_p(r)\quad,\quad V_q=B_q(r)$$

We have then $K\subset\cup_{q\in K}V_q$, so we can pick a finite subcover, as follows:
$$K\subset\left(V_{q_1}\cup\ldots\cup V_{q_n}\right)$$

With this done, consider the following intersection:
$$U=U_{q_1}\cap\ldots\cap U_{q_n}$$

This intersection is then a ball around $p$, and since this ball avoids $V_{q_1},\ldots,V_{q_n}$, it avoids the whole $K$. Thus, we have proved that $K^c$ is open at $p$, as desired.

\medskip

(2) Assume that $F\subset K$ is closed, with $K\subset X$ being compact. For proving our result, we can assume, by replacing $X$ with $K$, that we have $X=K$. In order to prove now that $F$ is compact, consider an open cover of it, as follows:
$$F\subset\bigcup_iU_i$$

By adding the set $F^c$, which is open, to this cover, we obtain a cover of $K$. Now since $K$ is compact, we can extract from this a finite subcover $\Omega$, and there are two cases:

\medskip

-- If $F^c\in\Omega$, by removing $F^c$ from $\Omega$ we obtain a finite cover of $F$, as desired.

\medskip

-- If $F^c\notin\Omega$, we are done too, because in this case $\Omega$ is a finite cover of $F$.

\medskip

(3) This follows from (1) and (2), because if $K\subset X$ is compact, and $F\subset X$ is closed, then $K\cap F\subset K$ is closed inside a compact, so it is compact.
\end{proof}

As a second batch of results, which are useful as well, we have:

\begin{proposition}
The following hold:
\begin{enumerate}
\item If $K_i\subset X$ are compact, satisfying $K_{i_1}\cap\ldots\cap K_{i_n}\neq\emptyset$, then $\cap_iK_i\neq\emptyset$.

\item If $K_1\supset K_2\supset K_3\supset\ldots$ are non-empty compacts, then $\cap_iK_i\neq\emptyset$.

\item If $K$ is compact, and $E\subset K$ is infinite, then $E$ has a limit point in $K$.

\item If $K$ is compact, any sequence $\{x_n\}\subset K$ has a limit point in $K$.

\item If $K$ is compact, any $\{x_n\}\subset K$ has a subsequence which converges in $K$.
\end{enumerate}
\end{proposition}

\begin{proof}
Again, these are elementary results, which can be proved as follows:

\medskip

(1) Assume by contradiction $\cap_iK_i=\emptyset$, and let us pick $K_1\in\{K_i\}$. Since any $x\in K_1$ is not in $\cap_iK_i$, there is an index $i$ such that $x\in K_i^c$, and we conclude that we have:
$$K_1\subset\bigcup_{i\neq 1}K_i^c$$

But this can be regarded as being an open cover of $K_1$, that we know to be compact, so we can extract from it a finite subcover, as follows:
$$K_1\subset\left(K_{i_1}^c\cup\ldots\cup K_{i_n}^c\right)$$

But this contradicts our non-empty intersection assumption, and we are done.

\medskip

(2) This is a particular case of (1), proved above.

\medskip

(3) We prove this by contradiction. So, assume that $E$ has no limit point in $K$. This means that any $p\in K$ can be isolated from the rest of $E$ by a certain open ball $V_p=B_p(r)$, and in both the cases that can appear, $p\in E$ or $p\notin E$, we have:
$$|V_p\cap E|=0,1$$

Now observe that these sets $V_p$ form an open cover of $K$, and so of $E$. But due to $|V_p\cap E|=0,1$ and to $|E|=\infty$, this open cover of $E$ has no finite subcover. Thus the same cover, regarded now as cover of $K$, has no finite subcover either, contradiction.

\medskip

(4) This follows from (3) that we just proved, with $E=\{x_n\}$.

\medskip

(5) This is a reformulation of (4), that we just proved.
\end{proof}

Getting now to more exciting theory, here is a key result about compactness:

\index{compact set}
\index{closed and bounded}

\begin{theorem}
For a subset $K\subset\mathbb R^N$, the following are equivalent:
\begin{enumerate}
\item $K$ is closed and bounded.

\item $K$ is compact.

\item Any infinite subset $E\subset K$ has a limiting point in $K$.
\end{enumerate}
\end{theorem}

\begin{proof}
This is something quite tricky, the idea being as follows:

\medskip

$(1)\implies(2)$ As a first task, let us prove that any product of closed intervals is indeed compact. We can assume by linearity that we are dealing with the unit cube:
$$C_1=\prod_{i=1}^N[0,1]\subset\mathbb R^N$$

In order to prove that $C_1$ is compact, we proceed by contradiction. So, assume that we have an open cover as follows, having no finite subcover:
$$C_1\subset\bigcup_iU_i$$

Now let us cut $C_1$ into $2^N$ small cubes, in the obvious way, over the $N$ coordinate axes. Then at least one of these small cubes, which are all covered by $\cup_iU_i$ too, has no finite subcover. So, let us call $C_2\subset C_1$ one of these small cubes, having no finite subcover:
$$C_2\subset\bigcup_iU_i$$

We can then cut $C_2$ into $2^N$ small cubes, and by the same reasoning, we obtain a smaller cube $C_3\subset C_2$ having no finite subcover. And so on by recurrence, and we end up with a decreasing sequence of cubes, as follows, having no finite subcover:
$$C_1\supset C_2\supset C_3\supset\ldots$$

Now since these decreasing cubes have edge size $1,1/2,1/4,\ldots\,$, their intersection must be a point. So, let us call $p$ this point, defined by the following formula:
$$\{p\}=\bigcap_kC_k$$

But this point $p$ must be covered by $\cup_iU_i$, so we can find an index $i$ such that $p\in U_i$. Now observe that $U_i$ must contain a whole ball around $p$, and so starting from a certain $K\in\mathbb N$, all the cubes $C_k$ will be contained in this ball, and so in $U_i$:
$$C_k\subset E_i\quad,\quad\forall k\geq K$$

But this is a contradiction, because $C_K$, and in fact the smaller cubes $C_k$ with $k>K$ as well, were assumed to have no finite subcover. Thus, we proved our claim. But with this claim in hand, the result is now clear. Indeed, assuming that $K\subset\mathbb R^N$ is closed and bounded, we can view it as a subset as a suitable big cube, of the following form:
$$K\subset \prod_{i=1}^N[-M,M]\subset\mathbb R^N$$

But, what we have here is a closed subset inside a compact set, which by Proposition 5.16 follows to be compact, as desired.

\medskip

$(2)\implies(3)$ This is something that we already know, not needing $K\subset\mathbb R^N$.

\medskip

$(3)\implies(1)$ We have to prove that $K$ as in the statement is both closed and bounded, and we can do both these things by contradiction, as follows:

\medskip

-- Assume first that $K$ is not closed. But this means that we can find a point $x\notin K$ which is a limiting point of $K$. Now let us pick $x_n\in K$, with $x_n\to x$, and consider the set $E=\{x_n\}$. According to our assumption, $E$ must have a limiting point in $K$. But this limiting point can only be $x$, which is not in $K$, contradiction.

\medskip

-- Assume now that $K$ is not bounded. But this means that we can find points $x_n\in K$ satisfying $||x_n||\to\infty$, and if we consider the set $E=\{x_n\}$, then again this set must have a limiting point in $K$, which is impossible, so we have our contradiction, as desired.
\end{proof}

So long for compactness. As a last piece of general topology, in our metric space framework, we can talk as well about connectedness, as follows:

\index{connected set}

\begin{definition}
We can talk about connected sets $E\subset X$, as follows:
\begin{enumerate}
\item We say that $E$ is connected if it cannot be separated as $E=E_1\cup E_2$, with the components $E_1,E_2$ satisfying $E_1\cap\bar{E}_2=\bar{E}_1\cap E_2=\emptyset$.

\item We say that $E$ is path connected if any two points $p,q\in E$ can be joined by a path, meaning a continuous $f:[0,1]\to X$, with $f(0)=p$, $f(1)=q$.
\end{enumerate}
\end{definition}

All this looks a bit technical, and indeed it is. To start with, (1) is something quite natural, but the separation condition there $E_1\cap\bar{E}_2=\bar{E}_1\cap E_2=\emptyset$ can be weakened into $E_1\cap E_2=\emptyset$, or strengthened into $\bar{E}_1\cap\bar{E}_2=\emptyset$, depeding on purposes, and with our (1) as formulated being the good compromise, for most purposes. As for (2), this condition is obviously something stronger, and we have in fact the following implications:
$${\rm convex}\implies{\rm path\ connected}\implies {\rm connected}$$

Anyway, leaving aside the discussion here, once all these questions clarified, the idea is that any set $E$ can be written as a disjoint union of connected components:
$$E=\bigsqcup_i E_i$$

Getting back now to more concrete things, remember that we are here in this book for studying functions, and doing calculus. And, regarding functions, we have:

\index{continuous function}

\begin{theorem}
Assuming that $f:X\to Y$ is continuous, the following happen:
\begin{enumerate}
\item If $O$ is open, then $f^{-1}(O)$ is open. 

\item If $C$ is closed, then $f^{-1}(C)$ is closed. 

\item If $K$ is compact, then $f(K)$ is compact. 

\item If $E$ is connected, then $f(E)$ is connected. 
\end{enumerate}
\end{theorem}

\begin{proof}
This is something fundamental, which can be proved as follows:

\medskip

(1) This is clear from the definition of continuity, written with $\varepsilon,\delta$. In fact, the converse holds too, in the sense that if $f^{-1}({\rm open})={\rm open}$, then $f$ must be continuous.

\medskip

(2) This follows from (1), by taking complements. And again, the converse holds too, in the sense that if $f^{-1}({\rm closed})={\rm closed}$, then $f$ must be continuous.

\medskip

(3) Indeed, given an open cover $f(K)\subset\cup_iU_i$, we have an open cover $K\subset\cup_if^{-1}(U_i)$, and so by compactness of $K$, a finite subcover $K\subset f^{-1}(U_{i_1})\cup\ldots\cup f^{-1}(U_{i_n})$, and so finally a finite subcover $f(K)\subset U_{i_1}\cup\ldots\cup U_{i_n}$, as desired.

\medskip

(4) This can be proved via the same trick as for (3). Indeed, any separation of $f(E)$ into two parts can be returned via $f^{-1}$ into a separation of $E$ into two parts, contradiction.
\end{proof}

Observe the power of (3,4) in the above result, which among others prove the mean value theorem, which is something non-trivial. Good mathematics that we have here.

\section*{5c. Measured spaces}

Now that we have spaces, which are of quite general type, we are ready to attack measure theory. And here, as a first bad surprise on the way, we have:

\begin{proposition}
In the continuous setting, some sets cannot be measurable.
\end{proposition}

\begin{proof}
Obviously, this cannot be something rigorous, because we have not talked yet about the precise definition of the measured spaces. But listen to what I have to say, all this will be very useful, precisely for this goal, axiomatizing the measured spaces. Consider the unit circle $\mathbb T$, measured as usual, the total mass being $\mu(\mathbb T)=2\pi$, and then consider the group $G$ consisting of the rational rotations of this unit circle $\mathbb T$:
$$G=\left\{z\to e^{2r\pi i}z\Big|r\in\mathbb Q\cap[0,1)\right\}$$

The action $G\curvearrowright\mathbb T$ partitions then $\mathbb T$ into a certain number of orbits, which is uncountable, and this because $G$ is obviously countable, while $\mathbb T$ itself is uncountable. Now if we choose $A\subset\mathbb T$ containing exactly one point from each orbit, we have:
$$\mathbb T=\bigsqcup_{r\in\mathbb Q\cap[0,1)}e^{2r\pi i}A$$  

But this shows that $A$ is not measurable. Indeed, with the convention that we list our group as $\mathbb Q\cap[0,1)=\{r_1,r_2,r_3,\ldots\}$, we have two possible cases, as follows:

\medskip

(1) Assuming $\mu(A)=c>0$, we would get from this $\mu(\mathbb T)=\infty$, which is contradiction, as shown by the following computation, which is something quite self-explanatory:
$$\mu(\mathbb T)=\sum_{n=1}^\infty\mu(e^{2r_n\pi i}A)=\sum_{n=1}^\infty c=c\cdot\infty=\infty$$

(2) Assuming $\mu(A)=0$, we would get from this $\mu(\mathbb T)=0$, which is a contradiction again, as shown by the following estimate, self-explanatory too, valid for any $\varepsilon>0$:
$$\mu(\mathbb T)=\sum_{n=1}^\infty\mu(e^{2r_n\pi i}A)<\sum_{n=1}^\infty\frac{\varepsilon}{2^n}=\varepsilon$$

Summarizing, save for some discussion of the axiomatics, and more on this in a moment, we are led, just via some common sense, to the conclusion in the statement.
\end{proof}

In view of Proposition 5.21, as a first task in our axiomatization work, we must start with a set $X$, and come up with some axioms for the measurable subsets $Y\subset X$. And here, a bit of thinking and common sense lead to the following definition:

\index{measurable set}

\begin{definition}
A measurable space is a set $X$, given with a set of subsets $M\subset P(X)$, called measurable sets, which form an algebra, in the sense that:
\begin{enumerate}
\item $\emptyset,X\in M$.

\item $E\in M\implies E^c\in M$.

\item $M$ is stable under countable unions and intersections.
\end{enumerate}
\end{definition}

Observe that some of the axioms above are redundant, because assuming (2) we have $\emptyset\in M\iff X\in M$, which can help in verifying (1). The same goes for (3), with only one of the conditions there being in need to be verified, due to $\left(\bigcup_iE_i\right)^c=\bigcap_iE_i^c$.

\bigskip

At the level of examples, these usually come via the following simple fact:

\begin{proposition}
Given a set of subsets $S\subset P(X)$, there is a smallest algebra
$$M=\bar{S}$$
containing it, called algebra generated by $X$.
\end{proposition}

\begin{proof}
This can be viewed in two possible ways, as follows:

\medskip

(1) According to the axioms in Definition 5.22, what we have to do in order to construct $M=\bar{S}$ is to start with $S$, then add $\emptyset,X$ to it, along with the complements $E^c$ of all the sets $E\in S$, and then take countable unions and intersections of such sets. And, some elementary verifications show that what we get in this way is indeed an algebra.

\medskip

(2) Alternatively, we can define $M=\bar{S}$ as being the intersection of all algebras containing $S$, and with the remark that we have at least one such algebra, namely $P(X)$ itself. It is then clear from definitions that $M$ is an algebra, as desired.
\end{proof}

Getting now to the concrete examples of measurable spaces, we have:

\index{Borel set}

\begin{definition}
Any metric space $X$ is automatically a measurable space, with the algebra of measurable sets being 
$$B=\bar{O}$$
that is, the smallest algebra containing the open sets, called Borel algebra of $X$.
\end{definition}

Observe that the Borel sets include all open sets, all closed sets, as well as all countable unions of closed sets, and all countable intersections of open sets. As an example here, in the case $X=\mathbb R$, with its usual topology, all kinds of intervals are Borel sets:
$$(a,b),\ [a,b],\ (a,b],\ [a,b)\in B$$

Indeed, the first interval is open, and the second one is closed, so these are certainly Borel sets. As for the third and fourth intervals, these appear as countable unions of closed intervals, or as countable intersections of open intervals, so they are Borel too.

\bigskip

Getting back now to the general case, as a complement to Definition 5.22, we have:

\begin{definition}
Given a measurable space $(X,M)$, a measure on it is a function $\mu:M\to[0,\infty]$ which is countably additive, in the sense that
$$\mu\left(\bigcup_{i=1}^\infty E_i\right)=\sum_{i=1}^\infty\mu(E_i)$$
for any countable family of disjoint measurable sets $E_i\in M$. In this case, we say that $(X,M,\mu)$ is a measured space.
\end{definition}

Observe that we did not assume that our measures are finite, and this for including spaces like $X=\mathbb R$. In the case where $\mu$ happens to be bounded, up to a rescaling we can assume $\mu(X)=1$, and we say in this case that we have a probability measure on $X$.

\bigskip

Looking at what we have so far, Definition 5.24 and Definition 5.25, many natural questions appear, and leaving aside anything too specialized, we are led to:

\begin{question}
Given a metric space $X$, such as $X=\mathbb R$, or $X=\mathbb R^N$, how to construct a measure on it? Also, once we have such a measure, how to integrate the functions $f:X\to\mathbb R$, or $f:X\to\mathbb C$, with respect to this measure?
\end{question}

As you can see, we have two questions here, and none is trivial:

\bigskip

(1) Indeed, regarding the first question, this is something that we do not know yet how to solve, even this even for very simple spaces like $X=\mathbb R$. Indeed, while we certainly know how to measure the real intervals, simply by setting $\mu(a,b)=b-a$, and then unions of such intervals too, by decomposing then into disjoint unions of intervals, and making sums, nothing guarantees that we can measure any Borel set $E\in B$, in this way. 

\bigskip

(2) As for the second question, experience with the usual Riemann integral, that you know well from calculus, shows that such things can be quite tricky too. Observe also that this second question is more general than the first one, because we have $\mu(E)=\int\chi_E$ for any $E\in M$, so if we know how to integrate, we know how to measure. 

\bigskip

In short, we are facing non-trivial questions here, and we need a plan. And, perhaps a bit surprisingly, the best plan in order to deal with Question 5.26 is as follows:

\begin{plan}
We will jointly develop measure and integration theory, as follows:
\begin{enumerate}
\item We will first keep staying abstract, and understand how the functions $f:X\to\mathbb R$, or $f:X\to\mathbb C$, can be integrated, with respect to an abstract measure.

\item With this understood, we will conclude that the integration over $X=\mathbb R$ can only be a straightforward extension of the usual Riemann integration.

\item But, in the end, this will enable us to both measure all the Borel sets $E\subset\mathbb R$, and to integrate the measurable functions $f:\mathbb R\to\mathbb C$.

\item Finally, we will discuss how to deal with $X=\mathbb R^N$ too, and with other product spaces $X=Y\times Z$, both measure theory and integration.
\end{enumerate}
\end{plan}

As you can see, all this is quite tricky, the main idea behind this plan being functional analysis, that is, using functions and their integrals, instead of just fighting with abstract measure theory first, and looking at functions and their integrals afterwards.

\bigskip

Getting started now, let us first talk about measurable functions, in the general context of Definition 5.22, with no measure involved. We have here the following notion:

\begin{definition}
Given a measurable space $X$, and a metric space $Y$, a function
$$f:X\to Y$$
is called measurable when it satisfies the following condition:
$$U\in O\implies f^{-1}(U)\in M$$
When $X$ comes with a measure, we also call such functions integrable.
\end{definition}

Obviously, this is something simplified, because for doing abstract measurability theory, our spaces should be measurable spaces in the sense of Definition 5.22, and so the functions that we should normally care about should be functions $f:X\to Y$, with both $X,Y$ being abstract measured spaces. However, in view of Definition 5.24, this is more or less that what we are doing here, by restricting the attention to the target spaces $Y$ which are metric, with this being the case that really matters.

\bigskip

Many things can be said about the measurable functions. We first have:

\begin{proposition}
The measurable functions have the following properties:
\begin{enumerate}
\item If $f:X\to Y$ is measurable and $g:Y\to Z$ is continuous, $g\circ f$ is measurable.

\item If $f,g:X\to\mathbb R$ are both measurable and $h:\mathbb R^2\to Y$ is continuous, then the function $x\to h(f(x),g(x))$ is measurable. 

\item $f:X\to\mathbb C$ is measurable precisely when $Re(f),Im(f):X\to\mathbb R$ are measurable. In this case, the function $|f|:X\to\mathbb R$ is measurable too.

\item If $f,g:X\to\mathbb C$ are measurable, then so are $f+g,fg:X\to\mathbb C$. 
\end{enumerate}
\end{proposition}

\begin{proof}
This is something very standard, the idea being as follows:

\medskip

(1) This is clear from definitions, because we have:
$$U\in O\implies g^{-1}(U)\in O\implies f^{-1}(g^{-1}(U))\in M$$

(2) By using (1), it is enough to check that the function $k(x)=(f(x),g(x))$ is measurable. But this follows by writing any open set $U\subset\mathbb R^2$ as a union of rectangles. Indeed, the preminage of each rectangle $R=I\times J$ is measurable, as shown by:
$$k^{-1}(R)
=(f,g)^{-1}(I\times J)
=f^{-1}(I)\cap g^{-1}(J)
\in M$$

But then, with this in hand, since any open set $U\subset\mathbb R^2$ can be written as a  union of rectangles, it follows that $k^{-1}(U)$ is measurable, as desired.

\medskip

(3) This follows indeed by using (2), with the following functions:
$$h(z)=z,Re(z),Im(z),|z|$$

(4) In the real case, $f,g:X\to\mathbb R$, the result follows by using (2), with:
$$h(x,y)=x+y,xy$$

Then, the result can be extended to the complex case, $f,g:X\to\mathbb C$, by using (3).
\end{proof}

Next, we have the following useful characterization of the real measurable functions:

\begin{proposition}
A function $f:X\to[-\infty,\infty]$ is mesurable precisely when 
$$f^{-1}(I)\in M$$
for any interval of type $I=(a,\infty]$, with $a\in\mathbb R$. 
\end{proposition}

\begin{proof}
Consider the following set, which is easily seen to be an algebra:
$$\Omega=\left\{E\subset[-\infty,\infty]\Big|f^{-1}(E)\in M\right\}$$

We want to prove that $\Omega$ contains all the open sets, and this can be done as follows:

\medskip

(1) Pick $a\in\mathbb R$, and then pick an increasing sequence $a_n\to a$. We have then the following formula, which shows that we have $[-\infty,a)\in\Omega$:
$$[-\infty,a)=\bigcup_n[-\infty,a_n)=\bigcup_n(a_n,\infty]^c\in\Omega$$

(2) But with this in hand, we obtain, for any $a<b$, that we have:
$$(a,b)=[-\infty,b)\cap(a,\infty]\in\Omega$$

(3) Now since any open set $U\subset[-\infty,\infty]$ can be written as a union of open intervals, we conclude that we have $U\in\Omega$, as desired.
\end{proof}

Getting now to limits of measurable functions, we have the following result:

\begin{proposition}
Given an abstract measured space $X$, the measurable functions $f:X\to[-\infty,\infty]$ have the following properties:
\begin{enumerate}
\item If $f_n$ are measurable, so are $g=\sup_nf_n$, and $h=\limsup_nf_n$.

\item If $f_n$ are measurable, so are $k=\inf_nf_n$, and $l=\liminf_nf_n$.

\item If $f_n\to f$ and $f_n$ are measurable, then $f$ is measurable.

\item If $f,g$ are measurable, so are $h=\min(f,g)$ and $k=\max(f,g)$.

\item If $f$ is measurable, so are $f^+=\max(f,0)$ and $f^-=-\min(f,0)$.
\end{enumerate}
\end{proposition}

\begin{proof}
This is again something very standard, the idea being as follows:

\medskip

(1) For the function $g=\sup_nf_n$ we can use the measurability criterion from Proposition 5.30, along with the following fact, valid for any $a\in\mathbb R$:
$$g=\sup_nf_n\implies g^{-1}(a,\infty]=\bigcup_nf_n^{-1}(a,\infty]$$

By symmetry we obtain that the function $k=\inf_nf_n$ is measurable as well. But with these results in hand, the last assertion follows too, by using the following formula:
$$\limsup_nf_n=\inf_k\left(\sup_{n\geq k}f_n\right)$$

(2) Here the fact that $k=\inf_nf_n$ is measurable was already proved in the above, and for $l=\liminf_nf_n$ we can use the same argument, symmetry, or the following formula:
$$\liminf_nf_n=\sup_k\left(\inf_{n\geq k}f_n\right)$$

(3) This follows indeed from (1), or from (2), and from the following fact:
$$f_n\to f\implies f=\inf_nf_n=\sup_nf_n$$

(4) This is a trivial application of (1) and (2).

\medskip

(5) This follows from (4), and from the fact that if $f$ is measurable, so is $-f$.
\end{proof}

As a main result now regarding the measurable functions, which will be of key importance in what follows, we have the following statement, with the convention that a step function means a function which takes finitely many values:

\begin{theorem}
Given a measurable function $f:X\to[0,\infty]$, we can write
$$f(x)=\lim_{n\to\infty}\varphi_n(x)$$
with $0\leq\varphi_1\leq\varphi_2\leq\ldots\leq f$, and with each $\varphi_i$ being a measurable step function.
\end{theorem}

\begin{proof}
As a first observation, the converse holds too, thanks to the results in Proposition 5.31. Regarding now the proof, this goes as follows:

\medskip

(1) First, it is clear by drawing a picture that we can approximate the identity of $[0,\infty]$ with step functions as in the statement. That is, we can obviously find an increasing sequence of measurable step functions $0\leq\psi_1\leq\psi_2\leq\ldots\leq id$, satisfying:
$$\lim_{n\to\infty}\psi_n(x)=x$$ 

(2) Now let us set $\varphi_n=\psi_n\circ f$. According to the limiting formula above, we have:
$$\lim_{n\to\infty}\varphi_n(x)=f(x)$$

On the other hand, by using the measurability criterion from Proposition 5.30, it follows that our truncation functions $\varphi_n=\psi_n\circ f$ are measurable, as desired.
\end{proof}

\section*{5d. Integration theory} 

Good news, with the above general theory understood, we can now integrate functions, by following the good old method of Riemann, that you know well from one-variable calculus. To be more precise, we have the following result, which is of course stated a bit informally, with some of the details being left to you, as an instructive exercise:

\begin{theorem}
We can integrate the measurable functions $f:X\to\mathbb R_+$ by setting
$$\int_Xf(x)d\mu(x)=\sup_{0\leq\varphi\leq f}\int_X\varphi(x)d\mu(x)$$
with sup over measurable step functions, then extend this by linearity.
\end{theorem}

\begin{proof}
This is something very standard, and we will leave the clarification of all this, both precise statement, and proof, as an instructive exercise. To be more precise:

\medskip

(1) We can certainly integrate the step functions $\varphi:X\to\mathbb R_+$, by writing each such function as a linear combination of characteristic functions, as follows:
$$\varphi=\sum_i\lambda_i\chi_{E_i}$$

Indeed, with this formula in hand, we can integrate our function $\varphi$, as follows:
$$\int_X\varphi(x)\,d\mu(x)=\sum_i\lambda_i\mu(E_i)$$

The integral of step functions constructed in this way has then all the linearity and positivity properties that you might expect, and behaves well with respect to limits.

\medskip

(2) Next, consider an arbitrary measurable function $f:X\to[0,\infty]$. We know from Theorem 5.32 that we can write this function as an increasing limit, as follows, with $0\leq\varphi_1\leq\varphi_2\leq\ldots\leq f$, and with each $\varphi_i$ being a measurable step function:
$$f(x)=\lim_{n\to\infty}\varphi_n(x)$$

But this suggests to define the integral of $f$ by the formula in the statement, namely:
$$\int_Xf(x)d\mu(x)=\sup_{0\leq\varphi\leq f}\int_X\varphi(x)d\mu(x)$$

Indeed, we can see that the integral constructed in this way has all the linearity and positivity properties that you might expect, and behaves well with respect to limits.
\end{proof}

More in detail now, here are some basic properties of the integrals, which are very similar to those of the Riemann integral, that you know from calculus:

\begin{proposition}
The integrals of measurable functions $f:X\to\mathbb R_+$ have the following properties:
\begin{enumerate}
\item $f\leq g$ implies $\int f\leq\int g$.

\item $E\subset F$ implies $\int_Ef\leq\int_Ff$.

\item $\int f+g=\int f+\int g$.

\item $\int\lambda f=\lambda\int f$.
\end{enumerate}
\end{proposition}

\begin{proof}
All this is indeed very standard, all routine verifications.
\end{proof}

Summarizing, we know how to integrate the real positive functions, in our abstract measure theory setting. In general, we can use the following formula:
$$\int_X(f-g)(x)\,d\mu(x)=\int_Xf(x)d\mu(x)-\int_Xg(x)d\mu(x)$$

We can integrate as well the complex functions, by setting:
$$\int_X(f+ig)(x)\,d\mu(x)=\int_Xf(x)d\mu(x)+i\int_Xg(x)d\mu(x)$$

All this is, indeed, very standard, exactly as for the Riemann integral. Let us record these findings as an upgrade of Theorem 5.33, as follows, once again with the statement being a bit informal, and with some of the details being left as instructive exercises:

\begin{theorem}
We can integrate the measurable functions $f:X\to\mathbb C$ by setting
$$\int_Xf(x)d\mu(x)=\sup_{0\leq\varphi\leq f}\int_X\varphi(x)d\mu(x)$$
with sup over measurable step functions, for $f:X\to\mathbb R_+$ then extend this by linearity.
\end{theorem}

\begin{proof}
This follows indeed from what we already know from Theorem 5.33, and from the above discussion, exactly as for the Riemann integral.
\end{proof}

Many other things can be said here, following Lebesgue, Fatou and others. We first have the following result, regarding the monotone convergence, due to Lebesgue:

\begin{theorem}[Lebesgue]
Given an increasing sequence of measurable functions 
$$0\leq f_1\leq f_2\leq\ldots\leq\infty$$
which converges pointwise, $f_n\to f$, their limit is measurable, and we have
$$\int_Xf_n(x)d\mu(x)\to\int_Xf(x)d\mu(x)$$
for any positive measure on $X$.
\end{theorem}

\begin{proof}
This is indeed something very standard, the idea being as follows:

\medskip

(1) We first have the following obvious implication, showing that the sequence of integrals on the right converges, to a certain number in $[0,\infty]$: 
$$f_1\leq f_2\leq\ldots\implies\int_Xf_1(x)d\mu(x)\leq\int_Xf_2(x)d\mu(x)\leq\ldots$$

Moreover, since we have $f_n\to f$, it follows that we have the following inequality:
$$\lim_{n\to\infty}\int f_n(x)d\mu(x)\leq\int_Xf(x)d\mu(x)$$

(2) In order to prove now the reverse inequality, pick a measurable simple function $0\leq\varphi\leq f$, pick also a number $c\in(0,1)$, and consider the following sets:
$$E_n=\left\{x\in X\Big|f_n(x)\geq c\varphi(x)\right\}$$

These sets are then measurable, we have $E_1\subset E_2\subset\ldots$, and our claim is that:
$$X=\bigcup_nE_n$$

Indeed, given $x\in X$, if $f(x)=0$ then $x\in E_1$ and things fine. Otherwise, we have $f(x)>0$, and so $f(x)>c\varphi(x)$, since $c<1$, and so $x\in E_n$ for some $n$, as desired.

\medskip

(3) Now observe that, with $0\leq\varphi\leq f$ and $c\in(0,1)$ as above, we have:
$$\int_Xf_n(x)d\mu(x)\geq\int_{E_n}f_n(x)d\mu(x)\geq c\int_{E_n}\varphi(x)d\mu(x)$$

By taking the limit of this estimate, with $n\to\infty$, we obtain:
$$\lim_{n\to\infty}\int f_n(x)d\mu(x)\geq c\int_X\varphi(x)d\mu(x)$$

Now with $c\to1$, we obtain from this the following estimate:
$$\lim_{n\to\infty}\int f_n(x)d\mu(x)\geq\int_X\varphi(x)d\mu(x)$$

But this being true for any simple function $0\leq\varphi\leq f$, we conclude that we have:
$$\lim_{n\to\infty}\int f_n(x)d\mu(x)\geq\int_X f(x)d\mu(x)$$

Thus we have the reverse of the estimate found in (1), which finishes the proof.
\end{proof}

Regarding now the series of functions, again following Lebesgue, we have:

\begin{theorem}
Given measurable functions $f_n:X\to[0,\infty]$, their sum 
$$f(x)=\sum_{n=1}^\infty f_n(x)$$
is measurable, and we have the formula
$$\int_Xf(x)d\mu(x)=\sum_{n=1}^\infty\int_Xf_n(x)d\mu(x)$$
for any positive measure on $X$.
\end{theorem}

\begin{proof}
This follows indeed from Theorem 5.36, applied to the partial sums.
\end{proof}

Following now Fatou, we have as well the following key result:

\begin{theorem}[Fatou]
Given measurable functions $f_n:X\to[0,\infty]$, their limit 
$$f(x)=\liminf_nf_n(x)$$
is measurable, and we have the formula
$$\int_Xf(x)d\mu(x)\leq\liminf_n\int_Xf_n(x)d\mu(x)$$
for any positive measure on $X$.
\end{theorem}

\begin{proof}
The first assertion is something that we know, from Proposition 5.31. As for the second assertion, this can be proved by using the same trick as there, namely: 
$$\liminf_nf_n=\sup_k\left(\inf_{n\geq k}f_n\right)$$

Indeed, we can apply Theorem 5.36 to the following sequence of functions:
$$g_k(x)=\inf_{n\geq k}f_n(x)$$

Thus, we are led to the conclusion in the statement.
\end{proof}

Regarding the above result of Fatou, let us mention that there are examples where the inequality can be strict, with the standard example here being as follows:
$$f_n=\begin{cases}
\chi_E&(n\ {\rm odd})\\
1-\chi_E&(n\ {\rm even})
\end{cases}$$

Finally, as a last basic result, due to Lebesgue again, we have:

\begin{theorem}[Lebesgue]
Assuming $f_n\to f$ pointwise, and assuming too 
$$|f_n|\leq g$$
with $\int g<\infty$, the following happen:
\begin{enumerate}
\item $\int|f|<\infty$.

\item $\int|f_n-f|\to0$.

\item $\int f_n\to\int f$.
\end{enumerate}
\end{theorem}

\begin{proof}
This is something very standard, using Fatou, the idea being as follows:

\medskip

(1) This follows from $|f_n|\leq g$, which in the limit gives $|f|<g$, so $\int|f|<\infty$.

\medskip

(2) Since $|f_n-f|\leq 2g$, we can apply Theorem 5.38 to the following functions:
$$h_n=2g-|f_n-f|$$

We obtain in this way the following estimate:
\begin{eqnarray*}
\int_X2g
&\leq&\liminf_n\int_X2g-|f_n-f|\\
&=&\int_X2g+\liminf_n\left(-\int_X|f_n-f|\right)\\
&=&\int_X2g-\limsup_n\int_X|f_n-f|
\end{eqnarray*}

Now by substracting $\int_X2g$, this estimate gives the following formula:
$$\limsup_n\int_X|f_n-f|\leq0$$

We conclude that this limit must be 0, as claimed in (2).

\medskip

(3) This follows indeed from (2).
\end{proof}

Many other things can be said, along these lines, which are more specialized, and such knowledge can be very useful when doing applied probability, and statistics. For our purposes here, which will be mostly pure mathematical, with a touch of theoretical physics, the above general theory, and results of Lebesgue and Fatou, will do.

\section*{5e. Exercises}

This was a very standard chapter, all fundamentals of advanced analysis, and as exercises on all this, numerous, and all mandatory, we have:

\begin{exercise}
Master Cauchy-Schwarz, and all its generalizations.
\end{exercise}

\begin{exercise}
Find the coordinates of the regular $N$-simplex, inside $\mathbb R^{N-1}$.
\end{exercise}

\begin{exercise}
Learn more about connectedness, and connected components.
\end{exercise}

\begin{exercise}
Learn about topological spaces, generalizing the metric spaces.
\end{exercise}

\begin{exercise}
Learn also about measure theory, on such topological spaces.
\end{exercise}

\begin{exercise}
Have some fun with Borel and non-Borel sets, on $\mathbb R$.
\end{exercise}

\begin{exercise}
Check all the details for our Riemann integration method.
\end{exercise}

\begin{exercise}
Look up and solve exercises, using Lebesgue and Fatou.
\end{exercise}

As bonus exercise, as already suggested, have a look into what probabilists and statisticians say too, about Lebesgue, Fatou and related topics. All this is good learning.

\chapter{Main theorems}

\section*{6a. Preliminaries}

Welcome to measure theory, again. What we did in the previous chapter, while certainly useful, was not very connected to life and reality, because even for simple measured spaces like $X=\mathbb R$, we still don't know what the measure $\mu$ exactly is.

\bigskip

So, time to discuss such questions, following Riesz, Lebesgue and others. All this will be quite tricky and technical, and as a guiding principle, we will have:

\begin{principle}
The following are the same thing:
\begin{enumerate}
\item Measure theory.

\item Integration theory.

\item Probability theory.
\end{enumerate}
\end{principle}

Now back to our question, constructing measures $\mu$ on spaces like $X=\mathbb R$, or more generally $X=\mathbb R^N$, and perhaps on some other metric spaces $X$ too, this does not look obvious to do, with bare hands. However, and here comes the point, in view of Principle 6.1, it is normally enough to know how to integrate, or equivalently, to know how to do probability, on our spaces $X=\mathbb R$, or more generally $X=\mathbb R^N$, and so on.

\bigskip

And here, we certainly know from calculus how to integrate over $X=\mathbb R$, and that is excellent knowledge, that we should normally be able to covert into systematic measure theory on $\mathbb R$, in particular with $\mu$ fully constructed. As for $X=\mathbb R^N$, and more complicated spaces, such as various curves, surfaces and other manifolds $X\subset\mathbb R^N$, here we certainly have some integration knowledge too, from multivariable calculus, so with a bit of luck, we should be able to improve that knowledge, and convert it into measure theory too.

\bigskip

But probably too much talking, let us make a concrete plan, based on this:

\begin{plan}
In order to fully develop measure theory:
\begin{enumerate}
\item Following Riesz, we will first need to fully understand, abstractly, the precise relation between measure and integration.

\item Then, following Lebesgue, we will develop measure theory on $X=\mathbb R$, and more generally on $X=\mathbb R^N$, based on our integration knowledge there.
\end{enumerate}
And afterwards, we can have a look at more complicated spaces $X\subset\mathbb R^N$ too.
\end{plan}

Getting to work now, what can be the ``Riesz theorem'', answering (1) above? Ask Riesz, of course, or intercept a cat, and ask the cat, or simply meditate on all this, so many techniques are available to us. In practice, however, simplest is to have a look at a good measure theory book, such as Rudin \cite{ru2}, which reveals the following answer:

\begin{fact}[Riesz theorem]
Any positive functional $I:C_c(X)\to\mathbb R$ comes by integrating with respect to a certain measure on $X$,
$$I(f)=\int_Xf(x)d\mu(x)$$
which is unique, modulo null sets. In addition, the following happen:
\begin{enumerate}
\item $\mu(K)<\infty$, for any $K\subset X$ compact.

\item $\mu(E)=\inf\{\mu(U)|E\subset U\ {\rm open}\}$, for any $E$ measurable.

\item $\mu(E)=\sup\{\mu(K)|K\subset E\ {\rm compact}\}$, for any $E$ open, or of finite measure.
\end{enumerate}
\end{fact}

Which looks a bit scary, but quite reasonable. Indeed, the main assertion, with $C_c(X)$ standing as usual for the continuous, compactly supported functions on $X$, is more or less the equivalence between (1) and (2) in our guiding Principle 6.1. As for the various properties (1,2,3), which are part of the whole story too, because they sort of tell us how to construct the measure $\mu$, from the knowledge of the integration $I$, and more on this in a moment, these are certainly very reasonable properties, that we can expect to hold.

\bigskip

So, very good all this, we have a concrete theorem now, waiting to be proved, and the rest will naturally come afterwards, according to Plan 6.2. Now by looking more in detail at what our theorem exactly says, we are led to the following battle plan:

\begin{strategy}
In order to prove the Riesz theorem, we must:
\begin{enumerate}
\item Have a quick discussion about null sets.

\item Have some squeezing technology for the measurable sets, $K\subset E\subset U$.

\item And then get to work, and prove the theorem.
\end{enumerate}
\end{strategy}

In short, and bad news here, not yet time to get to work and do (3), we have to deal with (1) first, and then especially with (2), which suspiciously reminds what mathematicans call ``Lemmas", and with these having the reputation of being the worst of mathematics, unless of course you are a pure mathematician yourself.

\bigskip

Nevermind. As I always say to my students, one day you will have to work, and work can be quite varied, sometimes you have to move big furniture around the office, install or debug computers, fix electricity or even plumbing problems, wait for someone at the airport, calm down a colleague who gets crazy, encourage another one who starts crying, and so on. So, doing what is to be done, and we will not be scared about some Lemmas, if they are on the menu for today, we will prove Lemmas, no problem with that.

\bigskip

Getting started now, according to our Strategy 6.4, we first need to talk about null sets. So, let us first review the material from the previous chapter, by paying particular attention to the null sets, meaning sets of zero measure. We first have here:

\begin{theorem}
Any measured space can be completed,
$$(X,M,\mu)\to(X,\bar{M},\mu)$$
by saturating everything with subsets of null sets.
\end{theorem}

\begin{proof}
This is something quite self-explanatory, the idea being as follows:

\medskip

(1) To start with, once we have a null set, $\mu(A)=0$, we can declare any subset $B\subset A$ to be measurable too, and of course of mass zero, $\mu(B)=0$.

\medskip

(2) But then, we can add the sets $B$ found above to all the measurable sets $E\in M$, and we obtain in this way a bigger algebra, $M\subset\bar{M}$.

\medskip

(3) Finally, our original measure $\mu$ extends from the original algebra $M$ to the bigger algebra $\bar{M}$ in the obvious way, according to the formula $\mu(B)=0$ from (1). 

\medskip

(4) Thus, result basically proved, modulo a number of technical details, which look quite straightforward, and whose clarification we will leave, as an instructive exercise.
\end{proof}

Along the same lines, but at the level of functions now, we have:

\begin{theorem}
At the level of functions on a measured space, we can talk about integration by avoiding null sets, based on the equivalence
$$\int_Xf(x)d\mu(x)=0\iff f=0\ {\rm a.e.}$$
for any positive function $f\geq0$, and then on $f=g-h$ with $g,h\geq0$, in general.
\end{theorem}

\begin{proof}
This is indeed something very standard, coming as a functional analytic complement to Theorem 6.5, the idea with all this being as follows:

\medskip

(1) To start with, we have indeed the equivalence in the statement, for any $f\geq0$, with this being something standard, coming from the general theory from chapter 5.

\medskip

(2) Now based on this, we can develop some theory, by neglecting the null sets, which in practice means reformulating Theorem 6.5, in functional analysis terms.

\medskip

(3) To be more precise, given $(X,M,\mu)$, we certainly know from chapter 5 how to integrate the measurable functions $f:X\to\mathbb C$. And the point is that, by using the equivalence in the statement, this tells us in fact how to integrate the measurable functions $f:X\to\mathbb C$, with respect to the saturated space $(X,\bar{M},\mu)$ from Theorem 6.5.

\medskip

(4) So, this was for the idea, and we will leave clarifying the details here, both precise statement, and few things that are needed for the proof, as an instructive exercise.
\end{proof}

As before with Fatou and Lebesgue, many other things can be said, along these lines, which are more specialized, and such knowledge can be very useful when doing applied probability, and statistics. For our purposes here, the above general theory will do.

\bigskip

We will be actually back to this in Part III of this book, when doing more systematically functional analysis, which is the good framework for such things. The idea there will be indeed that of defining various spaces of functions by neglecting the null sets.

\bigskip

Moving now towards what we want to do, namely measuring and integrating over $\mathbb R$, in full generality, the idea here will be that of squeezing the arbitrary measurable sets $E\subset X$ between compact sets $K\subset X$ and open sets $U\subset X$, as follows:
$$K\subset E\subset U$$

In order to do this, which is something quite tricky, we will need a number of technical preliminaries, as a continuation of the material from chapter 5. First, we have:

\begin{lemma}
Given $K\subset U$, compact inside open, we can find inclusions
$$K\subset V\subset\bar{V}\subset U$$
with the set $V$ being open, with compact closure $\bar{V}$.
\end{lemma}

\begin{proof}
This is something elementary, the idea being as follows:

\medskip

(1) For any $x\in K$ pick a neighborhood $W_x$ having compact closure. Since $K$ is compact, we can find finitely many such neighborhoods, covering it:
$$K\subset W_{x_1}\cup\ldots\cup W_{x_n}$$

With this done, consider now the set on the right, namely:
$$W=W_{x_1}\cup\ldots\cup W_{x_n}$$

This is then an open set containing our compact set $K$, having compact closure, that we will use in what follows, in our constructions.

\medskip

(2) In order to prove now our result, observe first that if we are in the case $U=X$ we are done, because here we can simply take $V=W$. So, assume $U\neq X$. 

\medskip

(3) Given a point $x\in U^c$, since this point is compact, and our set $K$ not containing it is compact as well, we can find an open set $V_x$ having the following properties:
$$K\subset V_x\quad,\quad x\notin\bar{V}_x$$

Now consider the following family of sets, with $W$ being the open set in (1):
$$\left\{K\cap\bar{W}\cap\bar{V}_x\Big|x\in U^c\right\}$$

This is then a family of compact sets having empty intersection, so we can find a finite subfamily having empty intersection too. That is, we can find points $x_i\in U^c$ such that:
$$K\cap\bar{W}\cap\bar{V}_{x_1}\cap\ldots\cap\bar{V}_{x_m}=\emptyset$$

(4) With this done, consider now the following open set:
$$V=W\cap V_{x_1}\cap\ldots\cap V_{x_m}$$

Then $V$ has compact closure, and we have the following inclusion:
$$\bar{V}\subset\bar{W}\cap\bar{V}_{x_1}\cap\ldots\cap\bar{V}_{x_m}$$

Thus, $V$ is the open set we are looking for, and we are done.
\end{proof}

The above result, separating $K\subset U$, compact set inside open set, via an open set $V$ with compact closure $\bar{V}$, is something quite useful, in practice. However, for our measure theory purposes here, we will need more, along the same lines. 

\bigskip

To be more precise, under the same assumptions, $K\subset U$, compact inside open, we would like to come up with a continuous, compactly supported function $f$ such that:
$$\chi_K\leq f\leq\chi_U$$

Which is something quite non-trivial, when thinking a bit, and so again, and with our apologies here, we are in need of some more technical preliminaries. 

\bigskip

To be more precise, we will need the following standard notions:

\begin{definition}
A function $f:X\to\mathbb R$, with $X$ topological space, is called:
\begin{enumerate}
\item Upper semicontinuous, if $\{x\in X|f(x)<a\}$ is open, for any $a\in\mathbb R$.

\item Lower semicontinuous, if $\{x\in X|f(x)>a\}$ is open, for any $a\in\mathbb R$.
\end{enumerate}
\end{definition}

These notions are actually quite interesting on their own, having many uses in many contexts in analysis, and as basic illustrations for them, we have:

\bigskip

(1) A characteristic function $\chi_E$ is upper semicontinuous when $E\subset X$ is closed. This follows indeed from the above definition of the upper semicontinuity.

\bigskip

(2) Also, a characteristic function $\chi_E$ is lower semicontinuous when $E\subset X$ is open. Again, this follows from definitions, or simply from (1), by using the set $E^c$.

\bigskip

(3) Observe also that a continuous function is trivially both upper and lower semicontinuous. The converse of this holds too, and more on this in a moment.

\bigskip

Many things can be said about the upper and lower semicontinuous functions, which some knowledge here being something quite useful, when doing in analysis in general. In what follows we will only need a handful of basic results on the subject, as follows:

\begin{proposition}
The upper and lower semicontinuous functions $f:X\to\mathbb R$ have the following properties:
\begin{enumerate}
\item If $F\subset X$ is closed, $\chi_F$ is upper semicontinuous.

\item If $U\subset X$ is open, $\chi_U$ is lower semicontinuous.

\item Infimum of upper semicontinuous functions is upper semicontinuous.

\item Supremum of lower semicontinuous functions is lower semicontinuous.

\item $f$ is continuous precisely when it is upper and lower semicontinuous.
\end{enumerate}
\end{proposition}

\begin{proof}
All this is elementary, the idea being as follows:

\medskip

(1) This assertion, already mentioned in the above, follows from definitions.

\medskip

(2) This assertion, also mentioned in the above, also follows from definitions.

\medskip

(3) This assertion is also trivial, also coming from definitions.

\medskip

(4) And this assertion is trivial too, also coming from definitions.

\medskip

(5) In one sense, this is clear, as already mentioned in the above. In the other sense, assuming that $f:X\to\mathbb R$ is both upper and lower semicontinuous, we can see that the preimage of any open interval $(a,b)\subset\mathbb R$ is open, due to the following formula:
$$f^{-1}(a,b)=f^{-1}(a,\infty]\cap f^{-1}[-\infty,b)$$

Now since any open set $U\subset\mathbb R$ can be written a union of open intervals $(a,b)\subset\mathbb R$, it follows that $f^{-1}(U)$ is open, and so that $f$ is continuous, as desired.
\end{proof}

Many other things can be said, about the upper and lower semicontinuous functions, notably with many examples. However, for our purposes here, the above will do.

\bigskip

Good news, we can now formulate the main technical result that we will need, for our measure theory purposes here. With the convention that the support of a function $f$ is the closure of the set $\{x|f(x)\neq0\}$, this key statement is as follows:

\begin{lemma}[Urysohn]
Given $K\subset U$, compact inside open, we can find
$$\chi_K\leq f\leq\chi_U$$
with $f$ being continuous, and compactly supported.
\end{lemma}

\begin{proof}
This is something very standard, the idea being as follows:

\medskip

(1) Given sets $K\subset U$ as in the statement, by using Lemma 6.7 we can find an open set $V_0$, having compact closure $\bar{V}_0$, such that:
$$K\subset V_0\subset\bar{V}_0\subset U$$

But then, by using again Lemma 6.7, applied this time to the inclusion $K\subset V_0$, we can find a second open set $V_1$, having compact closure $\bar{V}_1$, such that:
$$K\subset V_1\subset\bar{V}_1\subset V_0\subset\bar{V}_0\subset U$$

And so on, would be the idea. In practice now, by using the fact that the rational numbers are countable, we can construct in this way a whole family of open sets $V_r$, having compact closures $\bar{V}_r$, one for each rational number $r\in[0,1]$, such that:
$$r<s\implies \bar{V}_s\subset V_r$$

(2) Time now to construct our function $f$. Let us set, for any $r\in[0,1]$ rational:
$$f_r(x)=\begin{cases}
r&{\rm if}\ x\in V_r\\
0&{\rm otherwise}
\end{cases}$$

Then, we can construct our function $f$ in the following way:
$$f=\sup_rf_r$$

It is clear then that $f$ is lower semicontinuous, that we have $\chi_K\leq f\leq\chi_U$, and also that $f$ is compactly supported, with support included in the compact set $\bar{V}_0$. Thus, we are almost there, and it remains to prove that $f$ is upper semicontinuous as well.

\medskip

(3) For this purpose, let us set as well, for any $r\in[0,1]$ rational:
$$g_s(x)=\begin{cases}
1&{\rm if}\ x\in\bar{V}_s\\
0&{\rm otherwise}
\end{cases}$$

Then, we can construct another function $g$, in the following way:
$$g=\inf_sg_s$$

It is then clear, exactly as in (2), that $g$ is lower semicontinuous, and also that we have $\chi_K\leq g\leq\chi_U$, and that $g$ is compactly supported. Of interest for us is the lower semicontinuity of $g$, because in order to finish the proof, it is enough to show that:
$$f=g$$

(4) So, let us prove this, $f=g$. By definition of the functions $f_r$ and $g_s$, we have:
$$f_r<g_s$$

Thus $f\leq g$. Now assume that we have somewhere a strict inequality, $f(x)<g(x)$. Then, we can find two rational numbers $r,s$ in between, as follows:
$$f(x)<r<s<g(x)$$

But the first inequality tells us that $x\notin V_r$, and the last inequality tells us that $x\in\bar{V}_s$, and this contradicts the condition in (1) on the sets $V_r$, as desired.
\end{proof}

Very nice all this, but in fact, contrary to what was advertised before, we will actually need in what follows, besides the Urysohn lemma, another technical result, which is something useful too, and actually of independent interest too, as follows:

\begin{theorem}
Given a compact set inside a union of open sets
$$K\subset U_1\cup\ldots\cup U_n$$
we can find an associated partition of unity, that is, a decomposition of type
$$f_1(x)+\ldots+f_n(x)=1\quad,\quad\forall x\in K$$
with each $f_i$ being continuous, supported on $U_i$.
\end{theorem}

\begin{proof}
This follows by using the Urysohn lemma, as follows:

\medskip

(1) For any $x\in K$, let us pick a neighborhood $V_x$ such that $\bar{V}_x\subset U_i$, for some $i$. Since $K$ is compact, we can find finitely many points $x_1,\ldots,x_m\in K$ such that:
$$K\subset V_{x_1}\cup\ldots\cup V_{x_m}$$

Now for any index $i\in\{1,\ldots,n\}$ let us consider the following union:
$$K_i=\bigcup_{\bar{V}_{x_i}\subset U_i}\bar{V}_{x_i}$$

We can apply then the Urysohn lemma to the inclusion $K_i\subset U_i$, and we obtain in this way a continuous, compactly supported function $g_i$, such that:
$$\chi_{K_i}\leq g_i\leq\chi_{U_i}$$

(2) With this done, consider the following sequence of functions:
$$f_1=g_1$$
$$f_2=(1-g_1)g_2$$
$$\vdots$$
$$f_n=(1-g_1)\ldots(1-g_{n-1})g_n$$

Then each $f_i$ is continuous, supported on $U_i$, and we have:
$$f_1+\ldots+f_n=1-(1-g_1)\ldots(1-g_n)$$

(3) On the other hand, recall from the construction of $K_i$ above that we have:
$$K\subset K_1\cup\ldots\cup K_n$$

We conclude from this that we have the following formula:
$$f_1+\ldots+f_n=1$$

Thus, we have our partition of the unity, as desired.
\end{proof}

\section*{6b. Riesz theorem}

Good news, eventually, the above is all that we need, as set theoretic preliminaries. We can now formulate the key result in abstract measure theory, as follows:

\begin{theorem}[Riesz]
Any positive functional $I:C_c(X)\to\mathbb R$ comes by integrating with respect to a certain measure on $X$,
$$I(f)=\int_Xf(x)d\mu(x)$$
which has the following additional properties,
\begin{enumerate}
\item $\mu(K)<\infty$, for any $K\subset X$ compact.

\item $\mu(E)=\inf\{\mu(U)|E\subset U\ {\rm open}\}$, for any $E$ measurable.

\item $\mu(E)=\sup\{\mu(K)|K\subset E\ {\rm compact}\}$, for any $E$ open, or of finite measure.
\end{enumerate}
and which is unique with these properties, modulo the null sets.
\end{theorem}

\begin{proof}
This is something quite long and tricky, the idea being as follows:

\medskip

(1) Let us first prove the uniqueness. Assuming that our measure $\mu$ produces $I$, and has the various properties in the statement, it is clear that $\mu$ is uniquely determined by its values on the compact sets $K\subset X$. Thus, we must show that given two measures $\mu_1,\mu_2$ as in the statement, and a compact set $K\subset X$, we have:
$$\mu_1(K)=\mu_2(K)$$

For this purpose, let us pick $\varepsilon>0$. By using the properties (1,3) in the statement, for the measure $\mu_2$, we can find $K\subset U$ open such that:
$$\mu_2(U)<\mu_2(K)+\varepsilon$$

Now by using the Urysohn lemma for the inclusion $K\subset U$, we obtain a certain continuous, compactly supported function $f$, such that:
$$\chi_K\leq f\leq\chi_U$$

But, with this choice of $f$, we have the following computation:
\begin{eqnarray*}
\mu_1(K)
&\leq&\int_Xf(x)d\mu_1(x)\\
&=&\int_Xf(x)d\mu_2(x)\\
&\leq&\mu_2(U)\\
&<&\mu_2(K)+\varepsilon
\end{eqnarray*}

Thus $\mu_1(K)\leq\mu_2(K)$, and by interchanging $\mu_1,\mu_2$ we have the reverse inequality as well, so we obtain, as desired, $\mu_1(K)=\mu_2(K)$, proving the uniqueness statement.

\medskip

(2) With this done, let us get now to the main part, and discuss the construction of the algebra $M$, that we will choose to be saturated in the sense of Theorem 6.5, and of the measure $\mu$. What we have as data is a positive functional, as follows:
$$I:C_c(X)\to\mathbb R$$

We must first measure the Borel sets $E\subset X$. And here, to start with, in order to measure the open sets $U\subset X$, the formula is straightforward, namely:
$$\mu(U)=\sup\left\{I(f)\Big| f\leq\chi_U\right\}$$

Now observe that with this definition in hand, for the open sets, we have:
$$U_1\subset U_2\implies\mu(U_1)\leq\mu(U_2)$$

We deduce that the following happens, for the open sets:
$$\mu(E)=\inf\left\{\mu(U)\Big| E\subset U\ {\rm open}\right\}$$

But this can serve as the definition of $\mu$, for all the subsets $E\subset X$.

\medskip

(3) Regarding now the measurable sets, let us first consider the class $N$ of subsets $E\subset X$ which are of finite measure, $\mu(E)<\infty$, and satisfy the following condition:
$$\mu(E)=\sup\left\{\mu(K)\Big| K\subset E\ {\rm compact}\right\}$$

Then, we can define the class of measurable sets $M\subset P(X)$ as follows:
$$M=\left\{E\subset X\Big|E\cap K\in N,\forall K\subset X\ {\rm compact}\right\}$$

Summarizing, done with definitions, and it remains now to prove that $M$ is an algebra, and that $\mu$ is a measure on it, along with the other things claimed in the statement.

\medskip

(4) In order to prove that $\mu$ is indeed a measure, our first claim is that we have the following inequality, for any two open sets $U_1,U_2\subset X$:
$$\mu(U_1\cup U_2)\leq\mu(U_1)+\mu(U_2)$$ 

In order to prove this claim, choose an arbitrary function $g:X\to[0,1]$, supported on $U_1\cup U_2$. By using Theorem 6.11 we obtain a certain partition of unity, $f_1+f_2=1$ on the support of $g$, and it follows that we have the following equality:
$$g=f_1g+f_2g$$

Now by applying our integration functional $I$, we obtain from this:
\begin{eqnarray*}
I(g)
&=&I(f_1g+f_2g)\\
&=&I(f_1g)+I(f_2g)\\
&\leq&\mu(U_1)+\mu(U_2)
\end{eqnarray*}

Now since this holds for any $g:X\to[0,1]$ as above, this proves our claim.

\medskip

(5) Our second claim is that we have in fact, more generally, the following inequality, valid for any family of subsets $E_1,E_2,E_3,\ldots$ of our space $X$:
$$\mu\left(\bigcup_{i=1}^\infty E_i\right)\leq\sum_{i=1}^\infty\mu(E_i)$$

Indeed, this inequality trivially holds if $\mu(E_i)=\infty$ for some $i$, so we can assume $\mu(E_i)<\infty$ for any $i$. Now pick $\varepsilon>0$, and choose open sets $U_i\supset E_i$ such that:
$$\mu(U_i)<\mu(E_i)+\frac{\varepsilon}{2^i}$$

Consider also the union of these open sets $U_i$, chosen as above:
$$U=\bigcup_{i=1}^\infty U_i$$

Now pick an arbitrary function $f:X\to[0,1]$, supported on this open set $U$. Since $f$ has compact support, we can find a certain integer $n\in\mathbb N$ such that:
$$supp(f)\subset U_1\cup\ldots\cup U_n$$

Now by using what we found in (4), recursively, we obtain:
\begin{eqnarray*}
I(f)
&\leq&\mu(U_1\cup\ldots\cup U_n)\\
&\leq&\mu(U_1)+\ldots+\mu(U_n)\\
&\leq&\sum_{i=1}^n\mu(E_i)+\varepsilon
\end{eqnarray*}

Since this holds for any $f:X\to[0,1]$ as above, we deduce that we have:
$$\mu\left(\bigcup_{i=1}^\infty E_i\right)
\leq\mu(U)
\leq\sum_{i=1}^\infty\mu(E_i)+\varepsilon$$

Now since the number $\varepsilon>0$ was arbitrary, this proves our claim.

\medskip

(6) Our next claim, which proves in the assertion (1) in the theorem, is that any compact set $K\subset X$ is measurable, and that we have:
$$\mu(K)<\infty$$

In order to prove this claim, consider an arbitrary continuous function $f:X\to[0,1]$ satisfying $f(x)=1$ on $K$, then pick $\alpha<1$, and consider the following open set:
$$U_\alpha=\left\{x\in X\Big|f(x)>\alpha\right\}$$

We have then $K\subset U_\alpha$, and so $\alpha g\leq f$, whenever a continuous function $g:X\to[0,1]$ is supported by $U_\alpha$. By using this, we obtain the following estimate:
\begin{eqnarray*}
\mu(K)
&\leq&
\mu(U_\alpha)\\
&=&\sup\left\{I(g)\Big|g:X\to[0,1],\,supp(g)\subset U_\alpha\right\}\\
&\leq&\alpha^{-1}I(f)
\end{eqnarray*}

Now with $\alpha\to 1$, we obtain from this the following estimate:
$$\mu(K)\leq I(f)$$

Thus $\mu(K)<\infty$, as claimed, and the fact that $K$ is measurable is clear too.

\medskip

(7) Our next claim, which improves what we found in (6), is that for any compact set $K\subset X$ we have in fact the following estimate:
$$\mu(K)=\inf\left\{I(f)\Big|f:X\to[0,1],\,f(x)=1\ {\rm on}\ K\right\}$$

In order to prove this, let us go back to the proof of (6). We know from there that for any continuous function $f:X\to[0,1]$ satisfying $f(x)=1$ on $K$, as above, we have:
$$\mu(K)\leq I(f)$$

Now pick $\varepsilon>0$, and choose an open set $U\supset K$ satisfying:
$$\mu(U)<\mu(K)+\varepsilon$$

By using Lemma 6.10, we can find a compactly supported function $f$ such that:
$$\chi_K\leq f\leq\chi_U$$ 

But this gives the following estimate, using the above inequality $\mu(K)\leq I(f)$:
\begin{eqnarray*}
I(f)
&\leq&\mu(V)\\
&<&\mu(K)+\varepsilon\\
&\leq&I(f)+\varepsilon
\end{eqnarray*}

Now since the number $\varepsilon>0$ was arbitrary, this proves our claim.

\medskip

(8) Our claim now is that any open set $U\subset X$ satisfies the following condition, that we used in (3) in order to define the measurable sets:
$$\mu(U)=\sup\left\{\mu(K)\Big| K\subset U\ {\rm compact}\right\}$$

In order to prove this, pick an arbitrary number $\alpha<\mu(U)$. We can then find a continuous function $f:X\to[0,1]$ supported on $U$, such that:
$$I(f)>\alpha$$

Consider now the compact set $K=supp(f)$. Then for any open set $V$ containing $K$ we have $I(f)\leq\mu(V)$, and we deduce from this that we have:
$$I(f)\leq\mu(K)$$

In other words, given $\alpha<\mu(U)$, we have found a compact set $K$ such that:
$$\mu(K)\geq\alpha$$

Now since our number $\alpha<\mu(U)$ was arbitrary, this proves our claim.

\medskip

(9) Getting now towards the proof of the additivity of our measure $\mu$, let us first prove that for any two disjoint compact sets $K_1,K_2\subset X$, we have:
$$\mu(K_1\cup K_2)=\mu(K_1)+\mu(K_2)$$

For this purpose, pick an arbitrary $\varepsilon>0$. By using Lemma 6.10, we can find a compactly supported continuous function $f:X\to[0,1]$ satisfying:
$$f(x)=\begin{cases}
1&{\rm on}\ K_1\\
0&{\rm on}\ K_2
\end{cases}$$

On the other hand, by using (7) we can find a compactly supported continuous function $g:X\to[0,1]$ satisfying $g(x)=1$ on $K_1\cup K_2$, such that:
$$I(g)\leq\mu(K_1\cup K_2)+\varepsilon$$

By using this, and the linearity of $I$, we have the following estimate:
\begin{eqnarray*}
\mu(K_1)+\mu(K_2)
&\leq&I(fg)+I(g-fg)\\
&=&I(g)\\
&<&\mu(K_1\cup K_2)+\varepsilon
\end{eqnarray*}

Now since our number $\varepsilon>0$ was arbitrary, this proves our claim, via (5).

\medskip

(10) We are now in position of making a key verification, that of the additivity property of our measure $\mu$. To be more precise, our claim is that we have the following equality, valid for any family of pairwise disjoint subsets $E_1,E_2,E_3,\ldots$ of our space $X$:
$$\mu\left(\bigcup_{i=1}^\infty E_i\right)=\sum_{i=1}^\infty\mu(E_i)$$

In order to prove this, consider the union on the left, namely:
$$E=\bigcup_{i=1}^\infty E_i$$

In the case $\mu(E)=\infty$ our claim is proved by (5). So, assume $\mu(E)<\infty$, and pick an arbitrary number $\varepsilon>0$. We can then find compact sets $K_i\subset E_i$ such that:
$$\mu(K_i)>\mu(E_i)-\frac{\varepsilon}{2^i}$$

Now consider the following unions, which are compact sets too:
$$H_n=K_1\cup\ldots\cup K_n$$

By using the additivity formula that we found in (9), recursively, we obtain:
$$\mu(E)
\geq\mu(H_n)
=\sum_{i=1}^n\mu(K_i)
>\sum_{i=1}^n\mu(E_i)-\varepsilon$$

Now since our number $\varepsilon>0$ was arbitrary, this proves our claim.

\medskip

(11) Thus, additivity property proved, and with this in hand, the other assertions are quite standard. In order to prove these, our first claim is that for any measurable set $E$ and any $\varepsilon>0$, there is a compact set $K$, and an open set $U$, satisfying:
$$K\subset E\subset U\quad,\quad \mu(U-K)<\varepsilon$$

Indeed, in order to prove this claim, the first remark is that we can find indeed a compact set $K$ and an open set $U$, satisfying the following conditions:
$$K\subset E\subset U\quad,\quad \mu(U)-\frac{\varepsilon}{2}<\mu(E)<\mu(K)+\frac{\varepsilon}{2}$$

Since the set $U-K$ is open, by using (8) and then (10) we obtain:
$$\mu(U)=\mu(K)+\mu(U-K)$$

But with this, we can establish our inequality, as follows:
$$\mu(U-K)
=\mu(U)-\mu(K)<\varepsilon$$

(12) Our next claim is that if $A,B$ are measurable, then so are the following sets:
$$A-B\quad,\quad A\cup B\quad,\quad A\cap B$$

In order to prove this, pick an arbitrary number $\varepsilon>0$. By using (11) we can find certain compact sets $K,H$, and certain open sets $U,V$, satisfying:
$$K\subset A\subset U\quad,\quad \mu(U-K)<\varepsilon$$
$$H\subset B\subset V\quad,\quad \mu(V-H)<\varepsilon$$

Now observe that we have the following inclusion:
$$A-B
\subset U-H
\subset(U-K)\cup(K-V)\cup(V-H)$$

By using now (5), we obtain from this the following estimate:
\begin{eqnarray*}
\mu(A-B)
&\leq&\mu(U-K)+\mu(K-V)+\mu(V-H)\\
&<&\varepsilon+\mu(K-V)+\varepsilon\\
&=&\mu(K-V)+2\varepsilon
\end{eqnarray*}

Now since $K-V$ is a compact subset of $A-B$, we conclude that $A-B$ is measurable, as desired. Regarding now $A\cup B$, we can use here the following formula:
$$A\cup B=(A-B)\cup B$$

Indeed, by using (10) we obtain from this that $A\cup B$ is measurable as well. Finally, regarding  $A\cap B$, we can use here the following formula:
$$A\cap B=A-(A-B)$$

Thus $A\cap B$ is measurable as well, and this finishes the proof of our claim.

\medskip

(13) Our claim now is that the set $M$ constructed in (3) is indeed an algebra, which contains all Borel sets. In order to prove this, consider an arbitrary compact set $K\subset X$. Given $A\in M$, we can write $A^c\cap K$ as a difference of two sets in $M$, as follows:
$$A^c\cap K=K-(A-K)$$

Thus $A^c\cap K\in M$, and by using this, we conclude that we have:
$$A\in M\implies A^c\in M$$

Next, let us look at unions. Assume $A_i\in M$, and consider their union:
$$A=\bigcup_{i=1}^\infty A_i$$

With $K\subset X$ being an arbitrary compact set, as above, set $B_1=A_1\cap K$, and then define recursively the following sets:
$$B_n=(A_n\cap K)-(B_1\cup\ldots\cup B_{n-1})$$

In terms of these sets $B_n$, we have the following formula:
$$A\cap K=\bigcup_{i=1}^\infty B_n$$

On the other hand, according to our definition of the sets $B_n$, and by using (12), we have $B_n\in M$. Thus, by using (10), we obtain $A\in M$. Thus, we have proved that:
$$A_i\in M\implies \bigcup_{i=1}^\infty A_i\in M$$

Finally, if a subset $C\subset X$ is closed, then $C\cap K$ is compact, and so $C\in M$. Thus, as claimed above, $M$ is indeed an algebra, which contains all the Borel sets.

\medskip

(14) Our next claim, which together with what we found in (10) will prove that $\mu$ is indeed a measure on $M$, is that, in the context of our constructions in (3), the family $N$ constructed there consists precisely of the sets $E\in M$ satisfying:
$$\mu(E)<\infty$$

In order to prove this, let $E\in N$. By using (7) and (12) we deduce that we have $E\cap K\in N$ for any compact set $K\subset X$, and we conclude from this that we have:
$$E\in N\implies E\in M$$

Conversely, assume that $E\in M$ has finite measure, $\mu(E)<\infty$. In order to prove $E\in N$, we pick a number $\varepsilon>0$. We can then find an open set $U\subset X$ such that:
$$E\subset U\quad,\quad\mu(U)<\infty$$

Also, by (8) and (11) we can find a compact set $K\subset X$ such that:
$$K\subset V\quad,\quad\mu(U-K)<\varepsilon$$

Now since $E\cap K\in N$, we can find a compact set $H\subset X$ such that:
$$H\subset E\cap K\quad,\quad\mu(E\cap K)<\mu(H)+\varepsilon$$

Observe now that we have the following inclusion:
$$E\subset(E\cap K)\cup(U-K)$$

By using this, we obtain the following inequality:
\begin{eqnarray*}
\mu(E)
&\leq&\mu\left((E\cap K)\cup(U-K)\right)\\
&\leq&\mu(E\cap K)+\mu(U-K)\\
&<&\mu(H)+2\varepsilon
\end{eqnarray*}

Thus $E\in N$, which finishes the proof of our claim, and proves that $\mu$ is a measure.

\medskip

(15) Time now to get into the final verification, that of the following formula:
$$I(f)=\int_Xf(x)d\mu(x)$$

As a first observation, by taking real and imaginary parts, it is enough to prove this for the real functions $f$. Moreover, and assuming now that $f$ is real, by using $f\to-f$, in order to prove the above equality, it is enough to prove the following inequality:
$$I(f)\leq\int_Xf(x)d\mu(x)$$

(16) So, let us prove now this latter inequality. For this purpose, let $K\subset X$ be the support of our function $f$, choose an interval containing the range, $Im(f)\subset[a,b]$, pick an arbitrary $\varepsilon>0$, and then numbers $y_0,\ldots,y_n$ such that $y_i-y_{i-1}<\varepsilon$, and:
$$y_0<a<y_1<\ldots<y_n=b$$

With this done, consider now the following sets:
$$E_i=\left\{x\in X\Big|y_{i-1}<f(x)<y_i\right\}\cap K$$

Since $f$ is continuous, these are disjoint Borel sets, whose union is $K$. Thus, we can find open sets $U_i\subset X$, subject to the following conditions:
$$E_i\subset U_i\quad,\quad \mu(U_i)<\mu(E_i)+\frac{\varepsilon}{n}\quad,\quad f(x)<y_i+\varepsilon,\forall x\in U_i$$

By using now Theorem 6.11, we can find certain continuous functions $h_i:X\to[0,1]$, with $h_i$ supported on $U_i$, such that the following equality holds, on $K$:
$$h_1(x)+\ldots+h_n(x)=1$$

By multiplying by $f$, we conclude that the following equality holds, on $K$:
$$f(x)=h_1(x)f(x)+\ldots+h_n(x)f(x)$$

On the other hand, by using (7), we have the following estimate:
$$\mu(K)\leq I\left(\sum_{i=1}^nh_i\right)=\sum_{i=1}^nI(h_i)$$

Good news, with all these ingredients in hand, we can now finish. Since we have $h_if\leq(y_i+\varepsilon)h_i$, and since $y_i-\varepsilon<f(x)$ on $E$, we have the following estimate:
\begin{eqnarray*}
I(f)
&=&\sum_{i=1}^nI(h_if)\\
&\leq&\sum_{i=1}^n(y_i+\varepsilon)I(h_i)\\
&=&\sum_{i=1}^n(|a|+y_i+\varepsilon)I(h_i)-|a|\sum_{i=1}^nI(h_i)\\
&\leq&\sum_{i=1}^n(|a|+y_i+\varepsilon)\left(\mu(E_i)+\frac{\varepsilon}{n}\right)-|a|\mu(K)\\
&=&\sum_{i=1}^n(y_i-\varepsilon)\mu(E_i)+2\varepsilon\mu(K)+\frac{\varepsilon}{n}\sum_{i=1}^n(|a|+y_i+\varepsilon)\\
&\leq&\int_Xf(x)d\mu(x)+\varepsilon\left(2\mu(K)+|a|+b+\varepsilon\right)
\end{eqnarray*}

Now since $\varepsilon>0$ was arbitrary, this finishes the proof of our latest claim, that is, of the inequality in (15), and so finishes the proof of the present theorem.
\end{proof}

\section*{6c. Regular measures}

We proved the Riesz theorem, and that was good learning. In practice now, in order to measure the various measurable sets $E\subset X$, the tools that we have come from (2) and (3) in that theorem. So, let us have a closer look at that two properties.

\bigskip

Let us start with a definition, inspired from what we have:

\begin{definition}
Let $\mu$ be a Borel measure on a space $X$.
\begin{enumerate}
\item $E\subset X$ is called inner regular if $\mu(E)=\sup\{\mu(K)|K\subset E\ {\rm compact}\}$.

\item $E\subset X$ is called outer regular if $\mu(E)=\inf\{\mu(U)|E\subset U\ {\rm open}\}$.
\end{enumerate}
If all the Borel sets $E\subset X$ are inner and outer regular, we say that $\mu$ is regular.
\end{definition}

We know from Theorem 6.12 that, under the assumptions there, all sets $E\subset X$ are outer regular. However, in what regards inner regularity, we only have that for the sets $E\subset X$ which are open, or of finite measure. So, there is certainly room for improvement here, and in relation with this, we first have the following result:

\begin{theorem}
Assume, in the context of the Riesz theorem, that $X$ is locally compact, and is $\sigma$-compact too, in the sense that it is a countable union of compact sets.
\begin{enumerate}
\item Given a measurable set $E\subset X$ and a number $\varepsilon>0$, we can find $F\subset E\subset U$ with $F$ closed, $U$ open, and $\mu(U-F)<\varepsilon$. 

\item Also given $E\subset X$ measurable, we can find $A\subset E\subset B$, with $A$ countable union of closed sets, $B$ countable intersection of open sets, and $\mu(B-A)=0$. 

\item The measure $\mu$ is regular, in the above sense.
\end{enumerate}
\end{theorem}

\begin{proof}
This follows indeed by further building on Theorem 6.12 and its proof, under the assumptions in the statement, the idea being as follows:

\medskip

(1) Let us write $X=\cup_i K_i$, with each $K_i$ being compact. Given a measurable set $E\subset X$ and a number $\varepsilon>0$, we can find then open sets $U_i\supset(K_i\cap E)$ such that:
$$\mu\left(U_i-(K_i\cap E)\right)<\frac{\varepsilon}{2^i}$$

Now if we set $U=\cup_iU_i$, we have $U-E\subset\cup_i(U_i-(K_i\cap E))$, which gives:
$$\mu(U-E)<\frac{\varepsilon}{2}$$

We can apply this to $E^c=X-E$ too, and we get an open set $V\supset E^c$ such that:
$$\mu(V-E^c)<\frac{\varepsilon}{2}$$

But this gives the result, with $U$ being as above, and with $F=V^c$.

\medskip

(2) This follows indeed from (1), in the obvious way, applied with $\varepsilon_i=1/i$.

\medskip

(3) This follows again from (1), which proves the needed inner regularity of $E$.
\end{proof}

Along the same lines, we have as well the following result:

\begin{theorem}
Assume that $X$ is locally compact, with all its open sets $U\subset X$ being $\sigma$-compact. If $\lambda$ is a Borel measure on $X$ having the property that
$$\lambda(K)<\infty$$
for any compact subset $K\subset X$, then $\lambda$ is regular.
\end{theorem}

\begin{proof}
Consider indeed the integration with respect to the measure $\lambda$:
$$I(f)=\int_Xf(x)d\lambda(x)$$

By using our assumption, $\lambda(K)<\infty$ for any $K\subset X$ compact, we conclude that the Riesz theorem applies to this functional. Thus, we obtain a certain measure $\mu$, that we know in addition to be regular, as shown in Theorem 6.14, such that:
$$I(f)=\int_Xf(x)d\mu(x)$$

Thus, it remains to prove that we have $\lambda=\mu$, and this can be done as follows:

\medskip

(1) Let us first prove that we have $\lambda(U)=\mu(U)$, for any open set $U\subset X$. For this purpose, we can use the $\sigma$-compactness assumption in the statement, which allows us to write our set $U$ as follows, with the sets $K_i\subset X$ being compact:
$$U=\bigcup_iK_i$$

By using the Urysohn lemma, we can find continuous functions $f_i:X\to[0,1]$ which are supported on $U$, and are equal to 1 on $K_i$. With this done, let us set:
$$g_n=\max(f_1,\ldots,f_n)$$

The monotone convergence theorem applies then, and gives, as desired:
\begin{eqnarray*}
\lambda(U)
&=&\lim_{n\to\infty}\int_Xg_n(x)d\lambda(x)\\
&=&\lim_{n\to\infty}\int_Xg_n(x)d\mu(x)\\
&=&\mu(U)
\end{eqnarray*}

(2) Now let us prove that we have $\lambda(E)=\mu(E)$, for any Borel set $E\subset X$. For this purpose, pick $\varepsilon>0$. By using Theorem 6.14 (1), applied to the above measure $\mu$, coming from Riesz, we can find $F\subset E\subset U$, with $F$ closed and $U$ open, such that:
$$\mu(U-F)<\varepsilon$$

Now by using (1), applied to the open set $U-F$, we obtain from this $|\lambda(E)-\mu(E)|<\varepsilon$. But this being true for any $\varepsilon>0$, we obtain $\lambda(E)=\mu(E)$, as desired.
\end{proof}

\section*{6d. Lebesgue measure}

Good work that we did, in the above, and time now to enjoy what we learned. As a first consequence of the Riesz theorem, we can now formulate, following Lebesgue:

\begin{theorem}[Lebesgue]
We can measure all the Borel sets 
$$E\subset\mathbb R$$
and the corresponding integration theory extends the usual Riemann integral.
\end{theorem}

\begin{proof}
This follows indeed from Theorem 6.12, by using the Riemann integral, that we know well from basic calculus, as the positive functional needed there:
$$f\to\int_\mathbb Rf(x)dx$$

Indeed, it follows from basic calculus that the Riemann integral satisfies indeed all the needed properties in Theorem 6.12, and this gives the result. We will leave the various verifications here, which are all elementary, as an instructive exercise.
\end{proof}

Goind ahead with more theory, let us discuss now the case of $\mathbb R^N$. This is something quite tricky, and you would probably say here, why not simply generalizing Theorem 6.16 and its proof, by using this time the multivariable Riemann integral, that we know as well from basic calculus, as the positive functional needed there:
$$f\to\int_{\mathbb R^N}f(z)dz$$

However, let me ask you here, for instance at $N=2$, do we really master well the above integral? For instance, what is its precise definition? And here, I bet that your answer would be that we can define the multivariable integral by iterating, as follows:
$$\int_{\mathbb R^2}f(z)dz=\int_\mathbb R\int_\mathbb Rf(x,y)dxdy$$

Which looks fine, at a first glance, but there is in fact a bug, with this. Indeed, assuming so, we would have by symmetry the following formula too:
$$\int_{\mathbb R^2}f(z)dz=\int_\mathbb R\int_\mathbb Rf(x,y)dydx$$

Thus, and forgetting now about what we wanted to do, we can see that our method is based on the Fubini formula, stating that we must have:
$$\int_\mathbb R\int_\mathbb Rf(x,y)dxdy=\int_\mathbb R\int_\mathbb Rf(x,y)dydx$$

But, you might already know from calculus that Fubini does not always work, with the counterexamples being not very difficult to construct, as follows:

\begin{theorem}
The Fubini formula, namely 
$$\int_\mathbb R\int_\mathbb Rf(x,y)dxdy=\int_\mathbb R\int_\mathbb Rf(x,y)dydx$$
can fail, for certain suitably chosen functions.
\end{theorem}

\begin{proof}
Our starting point is the following differentiation formula:
$$\frac{d}{dx}\left(\frac{x}{x^2+y^2}\right)=\frac{x^2+y^2-x\cdot 2x}{(x^2+y^2)^2}=\frac{y^2-x^2}{(x^2+y^2)^2}$$

By using this formula, we have the following computation:
\begin{eqnarray*}
\int_0^1\int_0^1\frac{y^2-x^2}{(x^2+y^2)^2}\,dxdy
&=&\int_0^1\left[\frac{x}{x^2+y^2}\right]_0^1dy\\
&=&\int_0^1\frac{1}{1+y^2}\,dy\\
&=&\Big[\arctan y\Big]_0^1\\
&=&\frac{\pi}{4}
\end{eqnarray*}

On the other hand, by using this formula, and symmetry, we have as well:
\begin{eqnarray*}
\int_0^1\int_0^1\frac{y^2-x^2}{(x^2+y^2)^2}\,dydx
&=&\int_0^1\int_0^1\frac{x^2-y^2}{(x^2+y^2)^2}\,dxdy\\
&=&-\int_0^1\int_0^1\frac{y^2-x^2}{(x^2+y^2)^2}\,dxdy\\
&=&-\frac{\pi}{4}
\end{eqnarray*}

Thus Fubini fails. Please pass this knowledge to your children, and grandchildren.
\end{proof}

Summarizing, our method does not work. Which can be regarded as good news for us mathematicians, time now to get into some action, show our skills, and fix all this. Let us first formulate a general objective, inspired from the above, as follows:

\begin{objective}
We would like to measure all the Borel sets 
$$E\subset\mathbb R^N$$
with the corresponding integration theory extending and clarifying the Riemann integral, in several variables. Also, we would like to clarify the validity of the Fubini formula
$$\int_\mathbb R\int_\mathbb Rf(x,y)dxdy=\int_\mathbb R\int_\mathbb Rf(x,y)dydx$$
in this formulation, in $2$ variables, and in general $N\geq2$ variables too.
\end{objective}

Getting to work now, it is pretty much obvious that, in order to deal with this problem, rigorously, it is better to forget everything that we know from multivariable calculus, regarding the integration in several variables, and recreate everything from scratch, with our tools being Theorem 6.12, complemented by Theorem 6.14 and Theorem 6.15.

\bigskip

So, here is our first result, using our abstract measure theory techniques:

\begin{proposition}
Given a compactly supported function $f:\mathbb R^N\to\mathbb C$, if we set
$$I_k(f)=\frac{1}{2^{Nk}}\sum_{x\in\mathbb Z^N/2^k}f(x)$$
with $\mathbb Z^N/2^k\subset\mathbb R^N$ being the points having as coordinates integer multiples of $1/2^k$, then
$$I(f)=\lim_{k\to\infty}I_k(f)$$
converges, and $f\to I(f)$ satisfies the assumptions of the Riesz theorem.
\end{proposition}

\begin{proof}
In order to prove the first assertion, we can assume that our function is real, $f:\mathbb R^N\to\mathbb R$. But then, we can use the fact that $f$ is uniformly continuous, and this gives the convergence in the statement. As for the second assertion, the fact that $f\to I(f)$ satisfies indeed the assumptions of the Riesz theorem is clear from definitions.
\end{proof}

In view of the above, we can indeed apply the Riesz theorem. We obtain:

\begin{theorem}[Lebesgue]
There is a unique regular, translation-invariant measure on $\mathbb R^N$, normalized as for the volume of the unit cube to be $1$. This measure is called Lebesgue measure, and appears alternatively as the measure on the Borel sets 
$$E\subset\mathbb R$$
whose corresponding integration theory extends the usual Riemann integral, as constructed in Proposition 6.19, or equivalently, as in multivariable calculus.
\end{theorem}

\begin{proof}
Obviously, many things going on here, and as a first remark, this fulfills indeed our requirements from Objective 6.18, save for Fubini, that we will discuss later, in the next chapter. In what regards the rest, the idea is as follows:

\medskip

(1) We want to prove that the Lebesgue measure exists, and is unique, with the various properties from the statement. But here, a bit of thinking shows that this measure can only be the one coming from Proposition 6.19, via the Riesz theorem.

\medskip

(2) So, let us go back to Proposition 6.19, which allows us to apply the Riesz theorem, and apply indeed this Riesz theorem. We obtain in this way a certain regular Borel measure on $\mathbb R^N$, that we have to prove to have the properties in the statement.

\medskip

(3) But here, all the assertions follow from what we have in the Riesz theorem, complemented where needed by Theorem 6.14 and Theorem 6.15, that we can use too. We will leave the various verifications here as an instructive exercise. 
\end{proof}

Summarizing, we know now how to integrate on $\mathbb R^N$. As a continuation of this, as a first task, let us go back to the Fubini formula problematics. We have:

\begin{theorem}[Fubini, Tonelli]
Given a function $f:\mathbb R^N\to\mathbb R$ which is measurable and integrable, in the sense that the following integral is finite,
$$\int_{\mathbb R^N}|f(z)|dz<\infty$$
we have the following equalities, for any decomposition $N=N_1+N_2$:
$$\int_{\mathbb R^{N_1}}\int_{\mathbb R^{N_2}}f(x,y)dydx=\int_{\mathbb R^{N_2}}\int_{\mathbb R^{N_1}}f(x,y)dxdy=\int_{\mathbb R^N}f(z)dz$$
Moreover, the same holds when $f:\mathbb R^N\to\mathbb R$ is assumed positive, and measurable. 
\end{theorem}

\begin{proof}
In this statement, which is something quite compact, the first assertion is called Fubini theorem, and the second assertion is called Tonelli theorem. These two theorems are related, and can be also subject to various minor improvements. In what follows we will mostly focus on proving the Fubini theorem in its basic form, as stated.

\medskip

(1) In order to prove the Fubini theorem, consider the vector space $L$ consisting of the functions $f:\mathbb R^N\to\mathbb R$ which are measurable, and integrable:
$$L=\left\{f:\mathbb R^N\to\mathbb R\Big|\int_{\mathbb R^N}|f(z)|dz<\infty\right\}$$

Now fix a decomposition $N=N_1+N_2$, and consider the family $F\subset L$ of functions $f:\mathbb R^N\to\mathbb R$ which satisfy the formula in the statement, namely:
$$\int_{\mathbb R^{N_1}}\int_{\mathbb R^{N_2}}f(x,y)dydx=\int_{\mathbb R^{N_2}}\int_{\mathbb R^{N_1}}f(x,y)dxdy=\int_{\mathbb R^N}f(z)dz$$

With this conventions, we want to prove that the inclusion $F\subset L$ is an equality.

\medskip

(2) But this comes from the following sequence of observations, which all come from definitions, and from the Lebesgue dominated convergence theorem from chapter 5:

\medskip

-- $F$ is stable under taking finite linear combinations.

\medskip

-- $F$ is stable under taking pointwise limits of increasing sequences.

\medskip

-- $1_E\in F$, for any set of finite measure of type $E=E_1\times E_2\subset\mathbb R^N$.

\medskip

-- $1_E\in F$, for any open set of finite measure $E\subset\mathbb R^N$.

\medskip

-- $1_E\in F$, for any countable intersection of open sets, of finite measure $E\subset\mathbb R^N$.

\medskip

-- $1_E\in F$, for any set of null measure $E\subset\mathbb R^N$.

\medskip

-- $1_E\in F$, for any set of finite measure $E\subset\mathbb R^N$.

\medskip

Indeed, the proof of all these assertions, in the above precise order, is something quite straightforward, and once we have the last assertion, the result follows, because we can approximate any given function $f\in L$ by step functions, and we obtain $f\in F$.

\medskip

(3) Summarizing, Fubini proved, modulo some straightforward verifications left to you, and the same goes with Tonelli, whose proof is quite routine, based on Fubini.
\end{proof}

As a conclusion, we know how to rigorously integrate on $\mathbb R^N$, and as a bonus, the theory that we developed seems to have many follow-ups. More on this later.

\bigskip

Finally, let us mention that the existence and uniqueness, up to scalars, of the uniform measure on $\mathbb R^N$ has several possible generalizations, with a very useful one concerning the case of the locally compact abelian groups $G$. Moreover, this latter generalization further extends to the case of the arbitrary locally compact groups $G$, non necessarily abelian, with some care in distinguishing between the left and right invariance conditions. And then, even more generally, under suitable assumptions, we can talk about uniform integration over quotient spaces of type $X=G/H$, with $H\subset G$. Thus, many things to be learned, and we will be back to some of these questions, with details, in Part IV.

\section*{6e. Exercises}

Tough chapter that we had here, with many things explained quite quickly, and as exercises on this, all in relation with better understanding what we did, we have:

\begin{exercise}
Clarify what has been said above about saturating with null sets.
\end{exercise}

\begin{exercise}
Clarify as well what we said about integration, modulo null sets.
\end{exercise}

\begin{exercise}
Learn more about almost everywhere equality, from probabilists.
\end{exercise}

\begin{exercise}
Meditate about Fubini, its victories, and its failures.
\end{exercise}

\begin{exercise}
Fill in the details for the properties of the Lebesgue measure.
\end{exercise}

\begin{exercise}
Learn as well other methods for constructing the Lebesgue measure. 
\end{exercise}

\begin{exercise}
Learn about the Lusin theorem, and its applications.
\end{exercise}

\begin{exercise}
Learn also about the Vitali-Carath\'eodory theorem.
\end{exercise}

As bonus exercise, read some systematic measure theory, say from Rudin \cite{ru2}, with what we did here being an introduction to that, and from an alternative book too.

\chapter{Integration theory}

\section*{7a. Multiple integrals}

In this chapter and in the next one we discuss more advanced aspects of integration, on various explicit spaces $X\subset\mathbb R^N$, notably in relation with derivatives. To start with, we will be interested in the explicit computation of the integrals in $\mathbb R^N$.

\bigskip

Let us start with some basics. In what regards the area of the circle, sure you know that, but never too late to learn the truth about it, which is as follows:

\begin{theorem}
The following two definitions of $\pi$ are equivalent:
\begin{enumerate}
\item The length of the unit circle is $L=2\pi$.

\item The area of the unit disk is $A=\pi$.
\end{enumerate}
\end{theorem}

\begin{proof}
In order to prove this theorem let us cut the unit disk as a pizza, into $N$ slices, and forgetting about gastronomy, leave aside the rounded parts:
$$
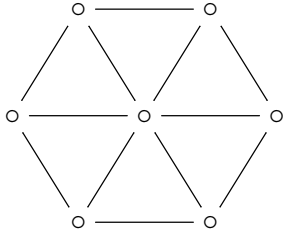
\xymatrix@R=28pt@C=12pt{
&\circ\ar@{-}[rr]\ar@{-}[dl]\ar@{-}[dr]&&\circ\ar@{-}[dl]\ar@{-}[dr]\\
\circ\ar@{-}[rr]&&\circ\ar@{-}[rr]&&\circ\\
&\circ\ar@{-}[rr]\ar@{-}[ul]\ar@{-}[ur]&&\circ\ar@{-}[ul]\ar@{-}[ur]
}$$

The area to be eaten can be then computed as follows, where $H$ is the height of the slices, $S$ is the length of their sides, and $P=NS$ is the total length of the sides:
$$A
=N\times \frac{HS}{2}
=\frac{HP}{2}
\simeq\frac{1\times L}{2}$$

Thus, with $N\to\infty$ we obtain that we have $A=L/2$, as desired.
\end{proof}

As a second computation, the area of an ellipse can be computed as follows:

\begin{theorem}
The area of an ellipse, given by the equation
$$\left(\frac{x}{a}\right)^2+\left(\frac{y}{b}\right)^2=1$$
with $a,b>0$ being half the size of a box containing the ellipse, is $A=\pi ab$.
\end{theorem}

\begin{proof}
The idea is that of cutting the ellipse into vertical slices. First observe that, according to our equation $(x/a)^2+(y/b)^2=1$, the $x$ coordinate can range as follows:
$$x\in[-a,a]$$

For any such $x$, the other coordinate $y$, satisfying $(x/a)^2+(y/b)^2=1$, is given by:
$$y=\pm b\sqrt{1-\frac{x^2}{a^2}}$$

We conclude that the area of the ellipse is given by the following formula:
\begin{eqnarray*}
A
&=&2b\int_{-a}^a\sqrt{1-\frac{x^2}{a^2}}\,dx\\
&=&4ab\int_0^1\sqrt{1-y^2}\,dy\\
&=&\pi ab
\end{eqnarray*}

Finally, as a verification, for $a=b=R$ we get $A=\pi R^2$, as we should.
\end{proof}

In what regards now the length of the ellipse, the ``pizza'' argument from the proof of Theorem 7.1 does not work, and things here get fairly complicated, as follows:

\begin{fact}
The length of an ellipse, given by $(x/a)^2+(y/b)^2=1$, is
$$L=4\int_0^{\pi/2}\sqrt{a^2\sin^2t+b^2\cos^2t}\,dt$$
and with this integral being generically not computable.
\end{fact}

To be more precise, the above formula can be deduced in the following way, and more on such things in chapter 8 below, when systematically discussing the curves:
\begin{eqnarray*}
L
&=&4\int_0^{\pi/2}\sqrt{\left(\frac{dx}{dt}\right)^2+\left(\frac{dy}{dt}\right)^2}\,dt\\
&=&4\int_0^{\pi/2}\sqrt{\left(\frac{da\cos t}{dt}\right)^2+\left(\frac{db\sin t}{dt}\right)^2}\,dt\\
&=&4\int_0^{\pi/2}\sqrt{(-a\sin t)^2+(b\cos t)^2}\,dt\\
&=&4\int_0^{\pi/2}\sqrt{a^2\sin^2t+b^2\cos^2t}\,dt
\end{eqnarray*}

As for the last assertion, when $a=b=R$ we get of course $L=2\pi R$, as we should, but in general, when $a\neq b$, there is no trick for computing the above integral.

\bigskip

Moving now to 3D, as an obvious challenge here, we can try to compute the area and volume of the sphere, and more generally of the ellipsoids. We have here:

\index{volume of sphere}

\begin{theorem}
The volume of the unit sphere in $\mathbb R^3$ is given by:
$$V=\frac{4\pi}{3}$$
More generally, the volume of an ellipsoid, given by $(x/a)^2+(y/b)^2+(z/c)^2=1$, is:
$$V=\frac{4\pi abc}{3}$$
The area of the sphere is $A=4\pi$. For ellipsoids, the area is generically not computable.
\end{theorem}

\begin{proof}
There are several things going on here, as follows:

\medskip

(1) Let us first compute the volume of the ellipsoid, which at $a=b=c=1$ will give the volume of the unit sphere. The range of the first coordinate $x$ is as follows:
$$x\in[-a,a]$$

Now when the first coordinate $x$ is fixed, the other coordinates $y,z$ vary on an ellipse, given by the equation $(y/b)^2+(z/c)^2=1-(x/a)^2$, which can be written as follows:
$$\left(\frac{y}{\beta}\right)^2+\left(\frac{z}{\gamma}\right)^2=1\quad:
\quad \beta=b\sqrt{1-\left(\frac{x}{a}\right)^2}\ ,\ \gamma=c\sqrt{1-\left(\frac{x}{a}\right)^2}$$

Thus, the vertical slice of our ellipsoid at $x$ has area as follows:
$$A(x)=\pi\beta\gamma=\pi bc\left[1-\left(\frac{x}{a}\right)^2\right]$$

We conclude that the volume of the ellipsoid is given, as claimed, by:
\begin{eqnarray*}
V
&=&\pi bc\int_{-a}^a1-\left(\frac{x}{a}\right)^2\,dx\\
&=&\pi bc\left[x-\frac{x^3}{3a^2}\right]_{-a}^a\\
&=&\pi bc\left(\frac{2a}{3}+\frac{2a}{3}\right)\\
&=&\frac{4\pi abc}{3}
\end{eqnarray*}

(2) At $a=b=c=1$ we get $V=4\pi/3$, and this gives the area of the unit sphere too, because the ``pizza'' method from the proof of Theorem 7.1 applies, and gives:
$$A=3\times V=3\times\frac{4\pi}{3}=4\pi$$

(3) Finally, the last assertion is something quite informal, inspired from Fact 7.3.
\end{proof}

In order to deal now with $N$ dimensions, we will need the Wallis formula:

\index{double factorials}
\index{trigonometric integral}
\index{Wallis formula}

\begin{theorem}[Wallis]
We have the following formulae,
$$\int_0^{\pi/2}\cos^nt\,dt=\int_0^{\pi/2}\sin^nt\,dt=\left(\frac{\pi}{2}\right)^{\varepsilon(n)}\frac{n!!}{(n+1)!!}$$
where $\varepsilon(n)=1$ if $n$ is even, and $\varepsilon(n)=0$ if $n$ is odd, and where
$$m!!=(m-1)(m-3)(m-5)\ldots$$
with the product ending at $2$ if $m$ is odd, and ending at $1$ if $m$ is even.
\end{theorem}

\begin{proof}
Let us first compute the integral on the left in the statement:
$$I_n=\int_0^{\pi/2}\cos^nt\,dt$$

We do this by partial integration. We have the following formula:
\begin{eqnarray*}
(\cos^nt\sin t)'
&=&n\cos^{n-1}t(-\sin t)\sin t+\cos^nt\cos t\\
&=&n\cos^{n+1}t-n\cos^{n-1}t+\cos^{n+1}t\\
&=&(n+1)\cos^{n+1}t-n\cos^{n-1}t
\end{eqnarray*}

By integrating between $0$ and $\pi/2$, we obtain the following formula:
$$(n+1)I_{n+1}=nI_{n-1}$$

Thus we can compute $I_n$ by recurrence, and we obtain:
\begin{eqnarray*}
I_n
&=&\frac{n-1}{n}\,I_{n-2}\\
&=&\frac{n-1}{n}\cdot\frac{n-3}{n-2}\,I_{n-4}\\
&=&\frac{n-1}{n}\cdot\frac{n-3}{n-2}\cdot\frac{n-5}{n-4}\,I_{n-6}\\
&&\vdots\\
&=&\frac{n!!}{(n+1)!!}\,I_{1-\varepsilon(n)}
\end{eqnarray*}

But $I_0=\frac{\pi}{2}$ and $I_1=1$, so we get the result. As for the second formula, this follows from the first one, with $t=\frac{\pi}{2}-s$. Thus, we have proved both formulae in the statement.
\end{proof}

We can now compute the volumes of the $N$-dimensional spheres, as follows:

\index{volume of sphere}
\index{double factorial}

\begin{theorem}
The volume of the unit sphere in $\mathbb R^N$ is given by
$$V=\left(\frac{\pi}{2}\right)^{[N/2]}\frac{2^N}{(N+1)!!}$$
with our usual convention $N!!=(N-1)(N-3)(N-5)\ldots$
\end{theorem}

\begin{proof}
If we denote by $V_N$ the volume of the unit sphere in $\mathbb R^N$, we have:
\begin{eqnarray*}
V_N
&=&\int_{-1}^1(1-x^2)^{(N-1)/2}dx\cdot V_{N-1}\\
&=&2V_{N-1}\int_0^1(1-x^2)^{(N-1)/2}dx\\
&=&2V_{N-1}\int_0^{\pi/2}(1-\sin^2 t)^{(N-1)/2}\cos tdt\\
&=&2V_{N-1}\int_0^{\pi/2}\cos^{N-1}t\cos tdt\\
&=&2V_{N-1}\int_0^{\pi/2}\cos^Ntdt
\end{eqnarray*}

Now by recurrence, and by using the formula in Theorem 7.5, we obtain:
\begin{eqnarray*}
V_N
&=&2^N\int_0^{\pi/2}\cos^Ntdt\int_0^{\pi/2}\cos^{N-1}tdt\ldots\ldots\int_0^{\pi/2}\cos tdt\\
&=&2^N\left(\frac{\pi}{2}\right)^{\varepsilon(N)+\varepsilon(N-1)+\ldots+\varepsilon(1)}\frac{N!!}{(N+1)!!}\cdot\frac{(N-1)!!}{N!!}\ldots\frac{1!!}{2!!}\\
&=&\left(\frac{\pi}{2}\right)^{\varepsilon(N)+\varepsilon(N-1)+\ldots+\varepsilon(1)}\frac{2^N}{(N+1)!!}\\
&=&\left(\frac{\pi}{2}\right)^{[N/2]}\frac{2^N}{(N+1)!!}
\end{eqnarray*}

Thus, we are led to the formula in the statement.
\end{proof}

As main particular cases of the above formula, we have:

\index{volume of sphere}

\begin{proposition}
The volumes of the low-dimensional spheres are as follows:
\begin{enumerate}
\item At $N=1$, the length of the unit interval is $V=2$.

\item At $N=2$, the area of the unit disk is $V=\pi$.

\item At $N=3$, the volume of the unit sphere is $V=\frac{4\pi}{3}$

\item At $N=4$, the volume of the corresponding unit sphere is $V=\frac{\pi^2}{2}$.
\end{enumerate}
\end{proposition}

\begin{proof}
Some of these results are well-known, but we can obtain all of them as particular cases of the general formula in Theorem 7.6, as follows:

\medskip

(1) At $N=1$ we obtain $V=1\cdot\frac{2}{1}=2$.

\medskip

(2) At $N=2$ we obtain $V=\frac{\pi}{2}\cdot\frac{4}{2}=\pi$.

\medskip

(3) At $N=3$ we obtain $V=\frac{\pi}{2}\cdot\frac{8}{3}=\frac{4\pi}{3}$.

\medskip

(4) At $N=4$ we obtain $V=\frac{\pi^2}{4}\cdot\frac{16}{8}=\frac{\pi^2}{2}$.
\end{proof}

Next, we can compute as well the volumes of general ellipsoids, as follows:

\begin{theorem}
The volume of an arbitrary ellipsoid in $\mathbb R^N$, given by
$$\left(\frac{x_1}{a_1}\right)^2+\ldots+\left(\frac{x_N}{a_N}\right)^2=1$$
is $V=a_1\ldots a_NV_0$, with $V_0$ being the volume of the unit sphere in $\mathbb R^N$.
\end{theorem}

\begin{proof}
We already know this at $N=2,3$, and the proof in general is similar, by suitably adapting the proof for the sphere. We will be back to this in a moment, with a more conceptual argument, based on Theorem 7.6, and a change of variables.
\end{proof}

Finally, let us record as well the formula of the area of the sphere, as follows:

\index{area of sphere}

\begin{theorem}
The area of the unit sphere in $\mathbb R^N$ is given by:
$$A=\left(\frac{\pi}{2}\right)^{[N/2]}\frac{2^N}{(N-1)!!}$$
In particular, at $N=2,3,4$ we obtain respectively $A=2\pi,4\pi,2\pi^2$.
\end{theorem}

\begin{proof}
As shown by the pizza argument from the proof of Theorem 7.1, which extends to arbitrary dimensions, the area and volume of the sphere in $\mathbb R^N$ are related by:
$$A=N\cdot V$$

Together with the formula in Theorem 7.6 for $V$, this gives the result.
\end{proof}

Needless to say, in what regards the area of the general ellipsoids in $\mathbb R^N$, this is in general not computable, as we already know well at $N=2,3$, from the above.

\section*{7b. Change of variables}

We know from one-variable calculus that the change of variables for integrals is related to differentiation, and we can of course expect the same to happen in several variables. So, let us first recall the basics of differentiation, in several variables. We first have:

\index{partial derivatives}

\begin{theorem}
A function $f:\mathbb R^N\to\mathbb R^M$ is continuously differentiable,
$$f(x+t)\simeq f(x)+f'(x)t$$
with $f'(x)$ linear, and $x\to f'(x)$ continuous, precisely when it has partial derivatives,
$$\frac{df_i}{dx_j}(x)=\lim_{t\to 0}\frac{f_i(x+te_j)-f_i(x)}{t}$$
which depend continuously on $x$. In this case the derivative is
$$f'(x)=\left(\frac{df_i}{dx_j}(x)\right)_{ij}\in M_{M\times N}(\mathbb R)$$ 
acting on the vectors $t\in\mathbb R^N$ by usual multiplication.
\end{theorem}

\begin{proof}
As a first observation, the formula in the statement makes sense indeed, as an equality, or rather approximation, of vectors in $\mathbb R^M$, as follows:
$$f\begin{pmatrix}x_1+t_1\\ \vdots\\ x_N+t_N\end{pmatrix}
\simeq f\begin{pmatrix}x_1\\ \vdots\\ x_N\end{pmatrix}
+\begin{pmatrix}
\frac{df_1}{dx_1}(x)&\ldots&\frac{df_1}{dx_N}(x)\\
\vdots&&\vdots\\
\frac{df_M}{dx_1}(x)&\ldots&\frac{df_M}{dx_N}(x)
\end{pmatrix}\begin{pmatrix}t_1\\ \vdots\\ t_N\end{pmatrix}$$

In order to prove now this formula, we can proceed by recurrence, as follows:

\medskip

(1) First of all, at $N=M=1$ what we have is a usual 1-variable function $f:\mathbb R\to\mathbb R$, and the formula in the statement is something that we know well, namely:
$$f(x+t)\simeq f(x)+f'(x)t$$

(2) Next, at $N=2,M=1$ the result holds as well, by using (1) twice, as follows:
\begin{eqnarray*}
f\binom{x_1+t_1}{x_2+t_2}
&\simeq&f\binom{x_1+t_1}{x_2}+\frac{df}{dx_2}(x)t_2\\
&\simeq&f\binom{x_1}{x_2}+\frac{df}{dx_1}(x)t_1+\frac{df}{dx_2}(x)t_2\\
&=&f\binom{x_1}{x_2}+\begin{pmatrix}\frac{df}{dx_1}(x)&\frac{df}{dx_2}(x)\end{pmatrix}\binom{t_1}{t_2}
\end{eqnarray*}

(3) More generally, we can deal in this way with the general case $M=1$, with the formula here, obtained via a straightforward recurrence, being as follows:
\begin{eqnarray*}
f\begin{pmatrix}x_1+t_1\\ \vdots\\ x_N+t_N\end{pmatrix}
&\simeq&f\begin{pmatrix}x_1\\ \vdots\\ x_N\end{pmatrix}+\frac{df}{dx_1}(x)t_1+\ldots+\frac{df}{dx_N}(x)t_N\\
&=&f\begin{pmatrix}x_1\\ \vdots\\ x_N\end{pmatrix}+
\begin{pmatrix}\frac{df}{dx_1}(x)&\ldots&\frac{df}{dx_N}(x)\end{pmatrix}
\begin{pmatrix}t_1\\ \vdots\\ t_N\end{pmatrix}
\end{eqnarray*}

(4) But this gives the result in the general case too. Indeed, given $f:\mathbb R^N\to\mathbb R^M$, we can apply (3) to each of the components $f_i:\mathbb R^N\to\mathbb R$, and we obtain:
$$f_i\begin{pmatrix}x_1+t_1\\ \vdots\\ x_N+t_N\end{pmatrix}
\simeq f_i\begin{pmatrix}x_1\\ \vdots\\ x_N\end{pmatrix}+
\begin{pmatrix}\frac{df_i}{dx_1}(x)&\ldots&\frac{df_i}{dx_N}(x)\end{pmatrix}
\begin{pmatrix}t_1\\ \vdots\\ t_N\end{pmatrix}$$

But these $M$ formulae tell us precisely that the equality in the statement happens.
\end{proof}

Generally speaking, Theorem 7.10 is all you need to know for upgrading from calculus to multivariable calculus. As a standard result here, we have:

\begin{theorem}
We have the chain derivative formula
$$(f\circ g)'(x)=f'(g(x))\cdot g'(x)$$
as an equality of matrices.
\end{theorem}

\begin{proof}
Consider indeed a composition of functions, as follows:
$$f:\mathbb R^N\to\mathbb R^M\quad,\quad 
g:\mathbb R^K\to\mathbb R^N\quad,\quad 
f\circ g:\mathbb R^K\to\mathbb R^M$$

According to Theorem 7.10, the derivatives of these functions are certain linear maps, corresponding to certain rectangular matrices, as follows:
$$f'(g(x))\in M_{M\times N}(\mathbb R)\quad,\quad 
g'(x)\in M_{N\times K}(\mathbb R)\quad\quad
(f\circ g)'(x)\in M_{M\times K}(\mathbb R)$$

Thus, our formula makes sense indeed. As for proof, this comes from:
\begin{eqnarray*}
(f\circ g)(x+t)
&=&f(g(x+t))\\
&\simeq&f(g(x)+g'(x)t)\\
&\simeq&f(g(x))+f'(g(x))g'(x)t
\end{eqnarray*}

Thus, we are led to the conclusion in the statement.
\end{proof}

Getting now to integration matters, let us first recall that we have:

\begin{proposition}
We have the change of variable formula
$$\int_a^bf(x)dx=\int_c^df(\varphi(t))\varphi'(t)dt$$
where $c=\varphi^{-1}(a)$ and $d=\varphi^{-1}(b)$.
\end{proposition}

\begin{proof}
This follows with $f=F'$, from the following differentiation rule:
$$(F\varphi)'(t)=F'(\varphi(t))\varphi'(t)$$

Indeed, by integrating between $c$ and $d$, we obtain the result.
\end{proof}

In several variables now, things are quite similar, the result being as follows:

\index{change of variables}
\index{Jacobian}
\index{matrix determinant}
\index{volume inflation}

\begin{theorem}
Given a transformation $\varphi=(\varphi_1,\ldots,\varphi_N)$, we have
$$\int_Ef(x)dx=\int_{\varphi^{-1}(E)}f(\varphi(t))|J_\varphi(t)|dt$$
with the $J_\varphi$ quantity, called Jacobian, being given by
$$J_\varphi(t)=\det\left[\left(\frac{d\varphi_i}{dx_j}(x)\right)_{ij}\right]$$ 
and with this generalizing the usual formula from one variable calculus.
\end{theorem}

\begin{proof}
This is something quite tricky, the idea being as follows:

\medskip

(1) Observe first that this generalizes indeed the change of variable formula in 1 dimension, from Proposition 7.12, the point here being that the absolute value on the derivative appears as to compensate for the lack of explicit bounds for the integral.

\medskip 

(2) As a second observation, we can assume if we want, by linearity, that we are dealing with the constant function $f=1$. For this function, our formula reads:
$$vol(E)=\int_{\varphi^{-1}(E)}|J_\varphi(t)|dt$$

In terms of $D={\varphi^{-1}(E)}$, this amounts in proving that we have:
$$vol(\varphi(D))=\int_D|J_\varphi(t)|dt$$

Now since this latter formula is additive with respect to $D$, it is enough to prove it for small cubes $D$. And here, as a first remark, our formula is clear for the linear maps $\varphi$, by using the definition of the determinant of real matrices, as a signed volume.

\medskip

(3) However, the extension of this to the case of non-linear maps $\varphi$ is something which looks non-trivial, so we will not follow this path, in what follows. So, while the above $f=1$ discussion is certainly something nice, our theorem is still in need of a proof.
 
\medskip

(4) In order to prove the theorem, as stated, let us rather focus on the transformations used $\varphi$, instead of the functions to be integrated $f$. Our first claim is that the validity of the theorem is stable under taking compositions of such transformations $\varphi$.

\medskip

(5) In order to prove this claim, consider a composition, as follows:
$$\varphi:E\to F\quad,\quad 
\psi:D\to E\quad,\quad 
\varphi\circ\psi:D\to F$$

Assuming that the theorem holds for $\varphi,\psi$, we have the following computation:
\begin{eqnarray*}
\int_Ff(x)dx
&=&\int_Ef(\varphi(s))|J_\varphi(s)|ds\\
&=&\int_Df(\varphi\circ\psi(t))|J_\varphi(\psi(t))|\cdot|J_\psi(t)|dt\\
&=&\int_Df(\varphi\circ\psi(t))|J_{\varphi\circ\psi}(t)|dt
\end{eqnarray*}

Thus, our theorem holds as well for $\varphi\circ\psi$, and we have proved our claim.

\medskip

(6) Next, as a key ingredient, let us examine the case where we are in $N=2$ dimensions, and our transformation $\varphi$ has one of the following special forms:
$$\varphi(x,y)=(\psi(x,y),y)\quad,\quad\varphi(x,y)=(x,\psi(x,y))$$

By symmetry, it is enough to deal with the first case. Here the Jacobian is $d\psi/dx$, and by replacing if needed $\psi\to-\psi$, we can assume that this Jacobian is positive, $d\psi/dx>0$. Now by assuming as before that $D=\varphi^{-1}(E)$ is a rectangle, $D=[a,b]\times[c,d]$, we can prove our formula by using the change of variables in 1 dimension, as follows:
\begin{eqnarray*}
\int_Ef(s)ds
&=&\int_{\varphi(D)}f(x,y)dxdy\\
&=&\int_c^d\int_{\psi(a,y)}^{\psi(b,y)}f(x,y)dxdy\\
&=&\int_c^d\int_a^bf(\psi(x,y),y)\frac{d\psi}{dx}\,dxdy\\
&=&\int_Df(\varphi(t))J_\varphi(t)dt
\end{eqnarray*}

(7) But with this, we can now prove the theorem, in $N=2$ dimensions. Indeed, given a transformation $\varphi=(\varphi_1,\varphi_2)$, consider the following two transformations:
$$\phi(x,y)=(\varphi_1(x,y),y)\quad,\quad \psi(x,y)=(x,\varphi_2\circ\phi^{-1}(x,y))$$

We have then $\varphi=\psi\circ\phi$, and by using (6) for $\psi,\phi$, which are of the special form there, and then (3) for composing, we conclude that the theorem holds for $\varphi$, as desired.

\medskip

(8) Thus, theorem proved in $N=2$ dimensions, and the extension of the above proof to arbitrary $N$ dimensions is straightforward, that we will leave this as an exercise.
\end{proof}

As a basic application of this technology, we can recover Theorem 7.8, as follows:

\begin{theorem}
The volume of an arbitrary ellipsoid in $\mathbb R^N$, given by
$$\left(\frac{x_1}{a_1}\right)^2+\ldots+\left(\frac{x_N}{a_N}\right)^2=1$$
is $V=a_1\ldots a_NV_0$, with $V_0$ being the volume of the unit sphere in $\mathbb R^N$.
\end{theorem}

\begin{proof}
This is indeed something clear, with the change of variables $x_i=a_iy_i$, whose Jacobian is constant, $J=a_1\ldots a_N$. Thus, we are led to the formula in the statement.
\end{proof}

In order to discuss now some other applications, in 2 dimensions, let us start with:

\index{polar coordinates}

\begin{theorem}
We have polar coordinates in $2$ dimensions,
$$\begin{cases}
x\!\!\!&=\ r\cos t\\
y\!\!\!&=\ r\sin t
\end{cases}$$
the corresponding Jacobian being $J=r$.
\end{theorem}

\begin{proof}
This is something elementary, with the Jacobian being as follows:
\begin{eqnarray*}
J
&=&\begin{vmatrix}
\frac{d(r\cos t)}{dr}&&\frac{d(r\cos t)}{dt}\\
\\
\frac{d(r\sin t)}{dr}&&\frac{d(r\sin t)}{dt}
\end{vmatrix}\\
&=&\begin{vmatrix}
\cos t&-r\sin t\\
\sin t&r\cos t
\end{vmatrix}\\
&=&r\cos^2t+r\sin^2t\\
&=&r
\end{eqnarray*}

Thus, we have indeed the formula in the statement.
\end{proof}

We can now compute the Gauss integral, which is the best calculus formula ever:

\index{Gauss integral}

\begin{theorem}
We have the following formula,
$$\int_\mathbb Re^{-x^2}dx=\sqrt{\pi}$$
called Gauss integral formula.
\end{theorem}

\begin{proof}
Let $I$ be the above integral. By using polar coordinates, we obtain:
\begin{eqnarray*}
I^2
&=&\int_\mathbb R\int_\mathbb Re^{-x^2-y^2}dxdy\\
&=&\int_0^{2\pi}\int_0^\infty e^{-r^2}rdrdt\\
&=&2\pi\int_0^\infty\left(-\frac{e^{-r^2}}{2}\right)'dr\\
&=&2\pi\left[0-\left(-\frac{1}{2}\right)\right]\\
&=&\pi
\end{eqnarray*}

Thus, we are led to the formula in the statement.
\end{proof}

Moving now to 3 dimensions, we have here the following result:

\index{spherical coordinates}
\index{latitude}
\index{longitude}

\begin{theorem}
We have spherical coordinates in $3$ dimensions,
$$\begin{cases}
x\!\!\!&=\ r\cos s\\
y\!\!\!&=\ r\sin s\cos t\\
z\!\!\!&=\ r\sin s\sin t
\end{cases}$$
the corresponding Jacobian being $J(r,s,t)=r^2\sin s$.
\end{theorem}

\begin{proof}
The fact that we have indeed spherical coordinates is clear. Regarding now the Jacobian, this is given by the following formula:
\begin{eqnarray*}
&&J(r,s,t)\\
&=&\begin{vmatrix}
\cos s&-r\sin s&0\\
\sin s\cos t&r\cos s\cos t&-r\sin s\sin t\\
\sin s\sin t&r\cos s\sin t&r\sin s\cos t
\end{vmatrix}\\
&=&r^2\sin s\sin t
\begin{vmatrix}\cos s&-r\sin s\\ \sin s\sin t&r\cos s\sin t\end{vmatrix}
+r\sin s\cos t\begin{vmatrix}\cos s&-r\sin s\\ \sin s\cos t&r\cos s\cos t\end{vmatrix}\\
&=&r\sin s\sin^2 t
\begin{vmatrix}\cos s&-r\sin s\\ \sin s&r\cos s\end{vmatrix}
+r\sin s\cos^2 t\begin{vmatrix}\cos s&-r\sin s\\ \sin s&r\cos s\end{vmatrix}\\
&=&r\sin s(\sin^2t+\cos^2t)\begin{vmatrix}\cos s&-r\sin s\\ \sin s&r\cos s\end{vmatrix}\\
&=&r\sin s\times 1\times r\\
&=&r^2\sin s
\end{eqnarray*}

Thus, we have indeed the formula in the statement.
\end{proof}

Let us work out now the general spherical coordinate formula, in arbitrary $N$ dimensions. The formula here, which generalizes those at $N=2,3$, is as follows:

\index{spherical coordinates}

\begin{theorem}
We have spherical coordinates in $N$ dimensions,
$$\begin{cases}
x_1\!\!\!&=\ r\cos t_1\\
x_2\!\!\!&=\ r\sin t_1\cos t_2\\
\vdots\\
x_{N-1}\!\!\!&=\ r\sin t_1\sin t_2\ldots\sin t_{N-2}\cos t_{N-1}\\
x_N\!\!\!&=\ r\sin t_1\sin t_2\ldots\sin t_{N-2}\sin t_{N-1}
\end{cases}$$
the corresponding Jacobian being given by the following formula,
$$J(r,t)=r^{N-1}\sin^{N-2}t_1\sin^{N-3}t_2\,\ldots\,\sin^2t_{N-3}\sin t_{N-2}$$
and with this generalizing the known formulae at $N=2,3$.
\end{theorem}

\begin{proof}
As before, the fact that we have spherical coordinates is clear. Regarding now the Jacobian, also as before, by developing over the last column, we have:
\begin{eqnarray*}
J_N
&=&r\sin t_1\ldots\sin t_{N-2}\sin t_{N-1}\times \sin t_{N-1}J_{N-1}\\
&+&r\sin t_1\ldots \sin t_{N-2}\cos t_{N-1}\times\cos t_{N-1}J_{N-1}\\
&=&r\sin t_1\ldots\sin t_{N-2}(\sin^2 t_{N-1}+\cos^2 t_{N-1})J_{N-1}\\
&=&r\sin t_1\ldots\sin t_{N-2}J_{N-1}
\end{eqnarray*}

Thus, we obtain the formula in the statement, by recurrence.
\end{proof}

Let us discuss now the computation of the arbitrary integrals over the sphere. We will need a technical result extending Theorem 7.5, as follows:

\index{trigonometric integral}
\index{Wallis formula}

\begin{theorem}[Wallis]
We have the following formula,
$$\int_0^{\pi/2}\cos^pt\sin^qt\,dt=\left(\frac{\pi}{2}\right)^{\varepsilon(p)\varepsilon(q)}\frac{p!!q!!}{(p+q+1)!!}$$
where $\varepsilon(p)=1$ if $p$ is even, and $\varepsilon(p)=0$ if $p$ is odd, and where
$$m!!=(m-1)(m-3)(m-5)\ldots$$
with the product ending at $2$ if $m$ is odd, and ending at $1$ if $m$ is even.
\end{theorem}

\begin{proof}
We use the same idea as in Theorem 7.5. Let $I_{pq}$ be the integral in the statement. In order to do the partial integration, observe that we have:
\begin{eqnarray*}
(\cos^pt\sin^qt)'
&=&p\cos^{p-1}t(-\sin t)\sin^qt\\
&+&\cos^pt\cdot q\sin^{q-1}t\cos t\\
&=&-p\cos^{p-1}t\sin^{q+1}t+q\cos^{p+1}t\sin^{q-1}t
\end{eqnarray*}

By integrating between $0$ and $\pi/2$, we obtain, for $p,q>0$:
$$pI_{p-1,q+1}=qI_{p+1,q-1}$$

Thus, we can compute $I_{pq}$ by recurrence. When $q$ is even we have:
\begin{eqnarray*}
I_{pq}
&=&\frac{q-1}{p+1}\,I_{p+2,q-2}\\
&=&\frac{q-1}{p+1}\cdot\frac{q-3}{p+3}\,I_{p+4,q-4}\\
&=&\vdots\\
&=&\frac{p!!q!!}{(p+q)!!}\,I_{p+q}
\end{eqnarray*}

But the last term comes from Theorem 7.5, and we obtain the result:
\begin{eqnarray*}
I_{pq}
&=&\frac{p!!q!!}{(p+q)!!}\,I_{p+q}\\
&=&\frac{p!!q!!}{(p+q)!!}\left(\frac{\pi}{2}\right)^{\varepsilon(p+q)}\frac{(p+q)!!}{(p+q+1)!!}\\
&=&\left(\frac{\pi}{2}\right)^{\varepsilon(p)\varepsilon(q)}\frac{p!!q!!}{(p+q+1)!!}
\end{eqnarray*}

Observe that this gives the result for $p$ even as well, by using $I_{pq}=I_{qp}$, coming via $t=\frac{\pi}{2}-s$. In the remaining case now, where both $p,q$ are odd, we can use once again the formula $pI_{p-1,q+1}=qI_{p+1,q-1}$ established above, and the recurrence goes as follows:
\begin{eqnarray*}
I_{pq}
&=&\frac{q-1}{p+1}\,I_{p+2,q-2}\\
&=&\frac{q-1}{p+1}\cdot\frac{q-3}{p+3}\,I_{p+4,q-4}\\
&=&\vdots\\
&=&\frac{p!!q!!}{(p+q-1)!!}\,I_{p+q-1,1}
\end{eqnarray*}

In order to compute the last term, observe that we have:
$$I_{p1}
=\int_0^{\pi/2}\cos^pt\sin t\,dt
=-\frac{1}{p+1}\int_0^{\pi/2}(\cos^{p+1}t)'\,dt
=\frac{1}{p+1}$$

Thus, we can finish our computation in the case where $p,q$ are odd, as follows:
$$I_{pq}
=\frac{p!!q!!}{(p+q-1)!!}\cdot\frac{1}{p+q}
=\frac{p!!q!!}{(p+q+1)!!}$$

Thus, we obtain the formula in the statement, the exponent of $\pi/2$ appearing there being $\varepsilon(p)\varepsilon(q)=0\cdot 0=0$ in the present case, and this finishes the proof.
\end{proof}

We can now integrate over the spheres, as follows:

\index{spherical integral}
\index{double factorials}

\begin{theorem}
The polynomial integrals over the unit sphere $S^{N-1}_\mathbb R\subset\mathbb R^N$, with respect to the normalized, mass $1$ measure, are given by the following formula,
$$\int_{S^{N-1}_\mathbb R}x_1^{k_1}\ldots x_N^{k_N}\,dx=\frac{(N-1)!!k_1!!\ldots k_N!!}{(N+\sum k_i-1)!!}$$
valid when all exponents $k_i$ are even. If an exponent is odd, the integral vanishes.
\end{theorem}

\begin{proof}
If one exponent $k_i$ is odd, with $x_i\to-x_i$ we can see that the integral in the statement vanishes. So, assume that all $k_i$ are even. As a first observation, the result holds indeed at $N=2$, due to the formula from Theorem 7.19, which reads:
$$\int_0^{\pi/2}\cos^pt\sin^qt\,dt
=\left(\frac{\pi}{2}\right)^{\varepsilon(p)\varepsilon(q)}\frac{p!!q!!}{(p+q+1)!!}
=\frac{p!!q!!}{(p+q+1)!!}$$

In the general case now, where the dimension $N\in\mathbb N$ is arbitrary, the integral in the statement can be written in spherical coordinates, as follows:
$$I=\frac{2^N}{A}\int_0^{\pi/2}\ldots\int_0^{\pi/2}x_1^{k_1}\ldots x_N^{k_N}J\,dt_1\ldots dt_{N-1}$$

Here $A$ is the area of the sphere, $J$ is the Jacobian, and the $2^N$ factor comes from the restriction to the $1/2^N$ part of the sphere where all the coordinates are positive. According to Theorem 7.9, the normalization constant in front of the integral is:
$$\frac{2^N}{A}=\left(\frac{2}{\pi}\right)^{[N/2]}(N-1)!!$$

As for the unnormalized integral, this is given by:
\begin{eqnarray*}
I'=\int_0^{\pi/2}\ldots\int_0^{\pi/2}
&&(\cos t_1)^{k_1}
(\sin t_1\cos t_2)^{k_2}\\
&&\vdots\\
&&(\sin t_1\sin t_2\ldots\sin t_{N-2}\cos t_{N-1})^{k_{N-1}}\\
&&(\sin t_1\sin t_2\ldots\sin t_{N-2}\sin t_{N-1})^{k_N}\\
&&\sin^{N-2}t_1\sin^{N-3}t_2\ldots\sin^2t_{N-3}\sin t_{N-2}\\
&&dt_1\ldots dt_{N-1}
\end{eqnarray*}

By rearranging the terms, we obtain:
\begin{eqnarray*}
I'
&=&\int_0^{\pi/2}\cos^{k_1}t_1\sin^{k_2+\ldots+k_N+N-2}t_1\,dt_1\\
&&\int_0^{\pi/2}\cos^{k_2}t_2\sin^{k_3+\ldots+k_N+N-3}t_2\,dt_2\\
&&\vdots\\
&&\int_0^{\pi/2}\cos^{k_{N-2}}t_{N-2}\sin^{k_{N-1}+k_N+1}t_{N-2}\,dt_{N-2}\\
&&\int_0^{\pi/2}\cos^{k_{N-1}}t_{N-1}\sin^{k_N}t_{N-1}\,dt_{N-1}
\end{eqnarray*}

Now by using the above-mentioned formula at $N=2$, this gives:
\begin{eqnarray*}
I'
&=&\frac{k_1!!(k_2+\ldots+k_N+N-2)!!}{(k_1+\ldots+k_N+N-1)!!}\left(\frac{\pi}{2}\right)^{\varepsilon(N-2)}\\
&&\frac{k_2!!(k_3+\ldots+k_N+N-3)!!}{(k_2+\ldots+k_N+N-2)!!}\left(\frac{\pi}{2}\right)^{\varepsilon(N-3)}\\
&&\vdots\\
&&\frac{k_{N-2}!!(k_{N-1}+k_N+1)!!}{(k_{N-2}+k_{N-1}+l_N+2)!!}\left(\frac{\pi}{2}\right)^{\varepsilon(1)}\\
&&\frac{k_{N-1}!!k_N!!}{(k_{N-1}+k_N+1)!!}\left(\frac{\pi}{2}\right)^{\varepsilon(0)}
\end{eqnarray*}

By simplifying this, and putting everything together, we obtain:
$$I=\left(\frac{2}{\pi}\right)^{[N/2]}(N-1)!!\cdot\left(\frac{\pi}{2}\right)^{[N/2]}
\frac{k_1!!\ldots k_N!!}{(N+\sum k_i-1)!!}
=\frac{(N-1)!!k_1!!\ldots k_N!!}{(N+\sum k_i-1)!!}$$

Thus, we have obtained the formula in the statement.
\end{proof}

In the case of the complex spheres, we have the following result:

\index{spherical integral}

\begin{theorem}
We have the following integration formula over the complex sphere $S^{N-1}_\mathbb C\subset\mathbb C^N$, with respect to the normalized uniform measure, 
$$\int_{S^{N-1}_\mathbb C}|z_1|^{2k_1}\ldots|z_N|^{2k_N}\,dz=\frac{(N-1)!k_1!\ldots k_n!}{(N+\sum k_i-1)!}$$
valid for any exponents $k_i\in\mathbb N$. As for the other polynomial integrals in $z_1,\ldots,z_N$ and their conjugates $\bar{z}_1,\ldots,\bar{z}_N$, these all vanish.
\end{theorem}

\begin{proof}
The last assertion being clear by using $z\to\lambda z$, assume that we are in the non-vanishing case. By using the standard identification $S^{N-1}_\mathbb C\simeq S^{2N-1}_\mathbb R$, we obtain:
\begin{eqnarray*}
I
&=&\int_{S^{N-1}_\mathbb C}|z_1|^{2k_1}\ldots|z_N|^{2k_N}\,dz\\
&=&\int_{S^{2N-1}_\mathbb R}(x_1^2+y_1^2)^{k_1}\ldots(x_N^2+y_N^2)^{k_N}\,d(x,y)\\
&=&\sum_{r_1\ldots r_N}\binom{k_1}{r_1}\ldots\binom{k_N}{r_N}\int_{S^{2N-1}_\mathbb R}x_1^{2k_1-2r_1}y_1^{2r_1}\ldots x_N^{2k_N-2r_N}y_N^{2r_N}\,d(x,y)
\end{eqnarray*}

By using now our integration formula for the real spheres, we obtain:
\begin{eqnarray*}
&&I\\
&=&\sum_{r_1\ldots r_N}\binom{k_1}{r_1}\ldots\binom{k_N}{r_N}\frac{(2N-1)!!(2r_1)!!\ldots(2r_N)!!(2k_1-2r_1)!!\ldots (2k_N-2r_N)!!}{(2N+2\sum k_i-1)!!}\\
&=&\sum_{r_1}\binom{2r_1}{r_1}\binom{2k_1-2r_1}{k_1-r_1}\ldots\sum_{r_N}\binom{2r_N}{r_N}\binom{2k_N-2r_N}{k_N-r_N}\frac{(N-1)!k_1!\ldots k_N!}{4^{\sum k_i}(N+\sum k_i-1)!}\\
&=&4^{k_1}\times\ldots\times 4^{k_N}\times\frac{(N-1)!k_1!\ldots k_N!}{4^{\sum k_i}(N+\sum k_i-1)!}\\
&=&\frac{(N-1)!k_1!\ldots k_N!}{(N+\sum k_i-1)!}
\end{eqnarray*}

Thus, we are led to the formula in the statement.
\end{proof}

\section*{7c. Probability, revised}

Good news, with our accumulated measure theory knowledge, we can now lay down some solid foundations for probability theory, both discrete and continuous. With the idea in mind of doing things a bit abstractly, our starting point will be:

\begin{definition}
Let $X$ be a probability space, that is, a space with a probability measure, and with the corresponding integration denoted $E$, and called expectation.
\begin{enumerate}
\item The random variables are the real functions $f\in L^\infty(X)$.

\item The moments of such a variable are the numbers $M_k(f)=E(f^k)$.

\item The law of such a variable is the measure given by $M_k(f)=\int_\mathbb Rx^kd\mu_f(x)$.
\end{enumerate}
\end{definition}

Here the fact that $\mu_f$ exists indeed is well-known. By linearity, we would like to have a real probability measure making hold the following formula, for any $P\in\mathbb R[X]$:
$$E(P(f))=\int_\mathbb RP(x)d\mu_f(x)$$

By using a standard continuity argument, it is enough to have this formula for the characteristic functions $\chi_I$ of the measurable sets of real numbers $I\subset\mathbb R$:
$$E(\chi_I(f))=\int_\mathbb R\chi_I(x)d\mu_f(x)$$

But this latter formula, which reads $P(f\in I)=\mu_f(I)$, can serve as a definition for $\mu_f$, and we are done. Next in line, we need to talk about independence:

\index{independence}

\begin{definition}
Two variables $f,g\in L^\infty(X)$ are called independent when
$$E(f^kg^l)=E(f^k)\,E(g^l)$$
happens, for any $k,l\in\mathbb N$.
\end{definition}

Again, this definition hides some non-trivial things. Indeed, by linearity, we would like to have a formula as follows, valid for any polynomials $P,Q\in\mathbb R[X]$:
$$E[P(f)Q(g)]=E[P(f)]\,E[Q(g)]$$

By using a continuity argument, it is enough to have this formula for characteristic functions $\chi_I,\chi_J$ of the measurable sets of real numbers $I,J\subset\mathbb R$:
$$E[\chi_I(f)\chi_J(g)]=E[\chi_I(f)]\,E[\chi_J(g)]$$

Thus, we are led to the usual definition of independence, namely:
$$P(f\in I,g\in J)=P(f\in I)\,P(g\in J)$$

All this might seem a bit abstract, but in practice, the idea is of course that $f,g$ must be independent, in an intuitive, real-life sense. As a first result now, we have:

\index{convolution}

\begin{proposition}
Assuming that $f,g\in L^\infty(X)$ are independent, we have
$$\mu_{f+g}=\mu_f*\mu_g$$
where $*$ is the convolution of real probability measures.
\end{proposition}

\begin{proof}
We have the following formula, coming from the independence of $f,g$:
$$M_k(f+g)
=\sum_r\binom{k}{r}M_r(f)M_{k-r}(g)$$

On the other hand, by using the Fubini theorem, we have as well:
\begin{eqnarray*}
\int_\mathbb Rx^kd(\mu_f*\mu_g)(x)
&=&\int_{\mathbb R\times\mathbb R}(x+y)^kd\mu_f(x)d\mu_g(y)\\
&=&\sum_r\binom{k}{r}\int_\mathbb Rx^rd\mu_f(x)\int_\mathbb Ry^{k-r}d\mu_g(y)\\
&=&\sum_r\binom{k}{r}M_r(f)M_{k-r}(g)
\end{eqnarray*}

Thus $\mu_{f+g}$ and $\mu_f*\mu_g$ have the same moments, so they coincide, as claimed.
\end{proof}

Here is now a second result on independence, which is something more advanced:

\index{independence}
\index{Fourier transform}

\begin{theorem}
Assuming that $f,g\in L^\infty(X)$ are independent, we have
$$F_{f+g}=F_fF_g$$
where $F_f(x)=E(e^{ixf})$ is the Fourier transform.
\end{theorem}

\begin{proof}
We have the following computation, using Proposition 7.24 and Fubini:
\begin{eqnarray*}
F_{f+g}(x)
&=&\int_\mathbb Re^{ixz}d\mu_{f+g}(z)\\
&=&\int_\mathbb Re^{ixz}d(\mu_f*\mu_g)(z)\\
&=&\int_{\mathbb R\times\mathbb R}e^{ix(z+t)}d\mu_f(z)d\mu_g(t)\\
&=&\int_\mathbb Re^{ixz}d\mu_f(z)\int_\mathbb Re^{ixt}d\mu_g(t)\\
&=&F_f(x)F_g(x)
\end{eqnarray*}

Thus, we are led to the conclusion in the statement.
\end{proof}

But with this in hand, and with the Gauss formula helping too, we can now have a new, final look at the normal variables, discussed in chapter 4. Let us start with:

\begin{definition}
The normal law of parameter $t>0$ is the following measure:
$$g_t=\frac{1}{\sqrt{2\pi t}}\,e^{-x^2/2t}dx$$
This is also called Gaussian distribution, with ``g'' standing for Gauss.
\end{definition}

Observe that the above laws have indeed mass 1, as they should, due to:
$$\int_\mathbb R e^{-x^2/2t}dx
=\int_\mathbb R e^{-y^2}\sqrt{2t}\,dy
=\sqrt{2\pi t}$$

Generally speaking, the normal laws appear as bit everywhere, in real life, the reasons for this coming from the Central Limit Theorem (CLT). As a first result, we have:

\index{Fourier transform}
\index{convolution semigroup}

\begin{theorem}
We have the following formula, valid for any $t>0$:
$$F_{g_t}(x)=e^{-tx^2/2}$$
In particular, the normal laws satisfy $g_s*g_t=g_{s+t}$, for any $s,t>0$.
\end{theorem}

\begin{proof}
The Fourier transform formula can be established as follows:
\begin{eqnarray*}
F_{g_t}(x)
&=&\frac{1}{\sqrt{2\pi t}}\int_\mathbb Re^{-z^2/2t+ixz}dz\\
&=&\frac{1}{\sqrt{2\pi t}}\int_\mathbb Re^{-(z/\sqrt{2t}-\sqrt{t/2}\,iz)^2-tx^2/2}dz\\
&=&\frac{1}{\sqrt{2\pi t}}\int_\mathbb Re^{-y^2-tx^2/2}\sqrt{2t}\,dy\\
&=&\frac{1}{\sqrt{\pi}}e^{-tx^2/2}\int_\mathbb Re^{-y^2}dy\\
&=&e^{-tx^2/2}
\end{eqnarray*}

As for $g_s*g_t=g_{s+t}$, this follows via Theorem 7.25, $\log F_{g_t}$ being linear in $t$.
\end{proof}

We are now ready to state and prove the CLT, as follows:

\index{CLT}
\index{Central Limit Theorem}

\begin{theorem}[CLT]
Given real variables $f_1,f_2,f_3,\ldots\in L^\infty(X)$ which are i.i.d., centered, and with common variance $t>0$, we have
$$\frac{1}{\sqrt{n}}\sum_{i=1}^nf_i\sim g_t$$
with $n\to\infty$, in moments.
\end{theorem}

\begin{proof}
The Fourier transform of the variable in the statement is:
$$F(x)
=\left[F_f\left(\frac{x}{\sqrt{n}}\right)\right]^n
=\left[1-\frac{tx^2}{2n}+O(n^{-2})\right]^n
\simeq e^{-tx^2/2}$$

But this function being the Fourier transform of $g_t$, we obtain the result.
\end{proof}

Finally, we will need the following result, regarding the moments:

\begin{theorem}
The even moments of the normal law are the numbers
$$M_k(g_t)=t^{k/2}\times k!!$$
where $k!!=(k-1)(k-3)(k-5)\ldots\,$, and the odd moments vanish. 
\end{theorem}

\begin{proof}
We have the following computation, valid for any integer $k\in\mathbb N$:
\begin{eqnarray*}
M_k
&=&\frac{1}{\sqrt{2\pi t}}\int_\mathbb Ry^ke^{-y^2/2t}dy\\
&=&\frac{1}{\sqrt{2\pi t}}\int_\mathbb R(ty^{k-1})\left(-e^{-y^2/2t}\right)'dy\\
&=&\frac{1}{\sqrt{2\pi t}}\int_\mathbb Rt(k-1)y^{k-2}e^{-y^2/2t}dy\\
&=&t(k-1)\times\frac{1}{\sqrt{2\pi t}}\int_\mathbb Ry^{k-2}e^{-y^2/2t}dy\\
&=&t(k-1)M_{k-2}
\end{eqnarray*}

Thus, by recurrence, we are led to the formula in the statement.
\end{proof}

In relation now with geometry, we first have the following result:

\index{hyperspherical law}
\index{asymptotic independence}
\index{asymptotic law}

\begin{theorem}
The moments of the hyperspherical variables are
$$\int_{S^{N-1}_\mathbb R}x_i^pdx=\frac{(N-1)!!p!!}{(N+p-1)!!}$$
and the rescaled variables $y_i=\sqrt{N}x_i$ become normal and independent with $N\to\infty$.
\end{theorem}

\begin{proof}
The moment formula in the statement follows from our general integration formula over the real spheres, from Theorem 7.20, which was as follows:
$$\int_{S^{N-1}_\mathbb R}x_1^{k_1}\ldots x_N^{k_N}\,dx=\frac{(N-1)!!k_1!!\ldots k_N!!}{(N+\sum k_i-1)!!}$$

Indeed, as a consequence, with $N\to\infty$ we have the following estimate:
$$\int_{S^{N-1}_\mathbb R}x_i^pdx
\simeq N^{-p/2}\times p!!
=N^{-p/2}M_p(g_1)$$

Thus, the variables $\sqrt{N}x_i$ become normal with $N\to\infty$, as claimed. As for the proof of the asymptotic independence, this is standard too, once again by using our general integration formula above. Indeed, the joint moments of $x_1,\ldots,x_N$ satisfy:
$$\int_{S^{N-1}_\mathbb R}x_1^{k_1}\ldots x_N^{k_N}\,dx
\simeq N^{-\sum k_i}\times k_1!!\ldots k_N!!$$

By rescaling, the joint moments of the variables $y_i=\sqrt{N}x_i$ are given by:
$$\int_{S^{N-1}_\mathbb R}y_1^{k_1}\ldots y_N^{k_N}\,dx\simeq k_1!!\ldots k_N!!$$

Thus, we have multiplicativity, and so independence with $N\to\infty$, as claimed.
\end{proof}

In the case of the complex spheres, we have a similar result, as follows:

\index{complex sphere}
\index{complex hyperspherical laws}
\index{unitary group}
\index{rotation group}

\begin{theorem}
The rescalings $\sqrt{N}z_i$ of the unit complex sphere coordinates
$$z_i:S^{N-1}_\mathbb C\to\mathbb C$$
become complex Gaussian and independent with $N\to\infty$. 
\end{theorem}

\begin{proof}
As before, we have the following estimate, for a single coordinate, which shows that  the rescaled variables $\sqrt{N}z_i$ become normal with $N\to\infty$:
$$\int_{S^{N-1}_\mathbb C}|z_i|^{2k}\,dz=\frac{(N-1)!k!}{(N+k-1)!}\simeq N^{-k}\times k!$$

As for the asymptotic independence, this is standard too, again by using our general integration formulae. Indeed, the joint moments of $z_1,\ldots,z_N$ are given by:
\begin{eqnarray*}
\int_{S^{N-1}_\mathbb R}|z_1|^{2k_1}\ldots|z_N|^{2k_N}\,dx
&=&\frac{(N-1)!k_1!\ldots k_n!}{(N+\sum k_i-1)!}\\
&\simeq&N^{-\sum k_i}\times k_1!\ldots k_N!
\end{eqnarray*}

Thus, we have multiplicativity, and so independence with $N\to\infty$, as claimed.
\end{proof}

As a comment now, the above results are just the tip of the iceberg, because the hyperspherical laws and related variables can be systematically investigated at various values of $N\in\mathbb N$, and many interesting things can be said about them, notably at the special values $N=1,2,3,4$. For more on all this, have a look at a probability book.

\section*{7d. Radon-Nikodym}

How many probability measures are there? In order to discuss this question, we need to compare abstract measures, and the main theorem here is a very useful decomposition result, due to Radon-Nikodym, that we would like to explain now. 

\bigskip

Let us start with some abstract notions, in relation with this, as follows:

\index{absolutely continuous}

\begin{definition}
Let $\lambda,\mu$ be two measures on a measurable space $X$.
\begin{enumerate}
\item We say that $\lambda$ is absolutely continuous with respect to $\mu$, and write $\lambda<\mu$, when $\mu(E)=0\implies\lambda(E)=0$, for any measurable set $E$.

\item We say that $\lambda,\mu$ are orthogonal, and write $\lambda\perp\mu$, when we can find disjoint sets $A,B$ such that $\lambda$ is concentrated on $A$, and $\mu$ is concentrated on $B$.
\end{enumerate}
\end{definition}

As illustrations for this, consider the case where $X=\mathbb R^N$, and $\mu$ is the Lebesgue measure. In this case it is easy to construct mesures $\lambda<\mu$, because any integrable function $f:\mathbb R^N\to\mathbb R$ produces such a measure, according to the following formula:
$$\lambda(E)=\int_Ef(x)d\mu(x)$$

As for the measures $\lambda\perp\mu$, there are many of them too, because any discrete measure $\lambda$, appearing by definition as follows, with $\{x_i\}\subset\mathbb R^N$ being a sequence of points, and $\{c_i\}\subset[0,\infty)$ being a sequence of positive numbers, has this orthogonality property:
$$\lambda=\sum_ic_i\delta_{x_i}$$

Getting back now to the general case, where the measurable space $X$ is arbitrary, we have the following remarkable decomposition result, due to Radon-Nikodym:

\index{Randon-Nikodym}

\begin{theorem}
Consider a measured space $(X,\mu)$, with $\mu$ being $\sigma$-finite, and let $\lambda$ be another measure on $X$. We can then find a decomposition as follows,
$$\lambda=\nu+\eta\quad,\quad\nu<\mu\quad,\quad\eta\perp\mu$$
and in what regards the absolutely continuous part $\nu$, this appears as
$$\nu(E)=\int_Ef(x)d\mu(x)$$
for a certain function $f$, integrable with respect to $\mu$, called density of $\nu$.
\end{theorem}

\begin{proof}
This is something quite standard, based on the material that we know, but instead of explaining this, which is quite long, here is a quick proof due to von Neumann, using some functional analysis material that we will learn later in this book:

\medskip

(1) Pick a function $w:X\to\mathbb(0,1)$ which is integrable with respect to $\mu$, and set:
$$\rho=\lambda+w\mu$$

The integration with respect to $\lambda$ being a bounded linear functional on the space of square-integrable functions with respect to $\rho$, we must have a formula as follows for this functional, for a certain function $g$, which is square-integrable with respect to $\rho$:
$$\int_Xf(x)d\lambda(x)=\int_Xf(x)g(x)d\rho(x)$$

We have then $g:X\to[0,1]$, and the above formula can be rewritten as follows:
$$\int_X(1-g(x))f(x)d\lambda(x)=\int_Xf(x)g(x)w(x)d\mu(x)$$

As already mentioned, we have used here some functional analysis material, that we will learn later in this book. So, don't forget to come back later to review this proof.

\medskip

(2) But this gives both the assertions. Indeed, consider first the following sets:
$$A=\left\{x\in X\Big|g(x)<1\right\}\quad,\quad B=\left\{x\in X\Big|g(x)=1\right\}$$ 

We can then construct the needed decomposition $\lambda=\nu+\eta$ by setting:
$$\nu(E)=\lambda(E\cap A)\quad,\quad\eta(E)=\lambda(E\cap B)$$

Indeed, we have $\lambda=\nu+\eta$, and the conditions $\nu<\mu$ and $\eta\perp\mu$ are both clear.

\medskip

(3) Moreover, with $f=1+g+\ldots+g^{n-1}$ in the formula from (1), we obtain:
$$\int_X(1-g^n(x))d\lambda(x)=\int_X(g(x)+g^2(x)+\ldots+g^n(x))w(x)d\mu(x)$$

But with this, we can set the density function $f$ to be the limit of the integrands on the right, and the density function property is satisfied indeed, as desired.
\end{proof}

Getting back now to $\mathbb R^N$, the Radon-Nikodym theorem has the following consequence:

\begin{theorem}
Any probability measure on $\mathbb R^N$ is of the following form,
$$d\lambda(x)=f(x)dx+d\eta(x)$$
with $f:\mathbb R^N\to\mathbb R_+$ being a certain density, and $\eta$ being supported on a measure $0$ set.
\end{theorem}

\begin{proof}
This follows indeed from Theorem 7.33, when applied to the case of the probability measures, that is, of the positive measures of mass 1.
\end{proof}

Observe that, as basic examples of probability measures on $\mathbb R^N$, we have the measures of the following type, with $f\geq0$ and $c_i\geq0$, subject to $\int_{\mathbb R^N}f(x)dx+\sum_ic_i=1$:
$$\lambda=f(x)dx+\sum_ic_i\delta_{x_i}$$

However, while certainly very frequent, these are just the basic examples, and this because exotic measures do exist, as shown by the following result, at $N=1$: 

\begin{theorem}
Exotic probability measures on $\mathbb R$ exist, with a standard example being the Cantor distribution, supported by the Cantor set $K\subset\mathbb R$.
\end{theorem}

\begin{proof}
Consider indeed the Cantor set $K\subset\mathbb R$, appearing by definition as the intersection of the following unions of intervals:
$$\xymatrix@R=10pt@C=10pt{
K_0:&0\ar@{-}[rrrrrrrrr]&&&&&&&&&1\\
K_1:&0\ar@{-}[rrr]&&&\frac{1}{3}&&&\frac{2}{3}\ar@{-}[rrr]&&&1\\
K_2:&0\ar@{-}[r]&\frac{1}{9}&\frac{2}{9}\ar@{-}[r]&\frac{1}{3}&&&\frac{2}{3}\ar@{-}[r]&\frac{7}{9}&\frac{8}{9}\ar@{-}[r]&1\\
&&&&&\vdots&\vdots}$$

As a first observation, the Lebesgue measure of this Cantor set $K$ is given by:
\begin{eqnarray*}
\lambda(K)
&=&1-\frac{1}{3}-\frac{2}{9}-\frac{4}{27}-\ldots\\
&=&1-\frac{1}{3}\left(1+\frac{2}{3}+\frac{4}{9}+\ldots\right)\\
&=&1-\frac{1}{3}\cdot\frac{1}{1-2/3}\\
&=&0
\end{eqnarray*}

The point now is that we can construct a natural probability measure on $K$, which is however exotic, meaning that it is not a linear combination of Dirac masses, by stating that $\mu$ must be uniform with respect to the 2 intervals of $K_1$, then with respect to the 4 intervals of $K_2$, and so on. And, exercise for you, to learn more about this.
\end{proof}

As a conclusion to all this, the measure theory developed in chapters 5-6 brings a new light on probability theory, by providing the good formalism which is needed, in general. However, if you are really into applied probability, and are not particularly excited about exotic measures, well, things are a bit complicated, and in the lack of a good axiomatics fitting your needs, you will still have to learn that material from chapters 5-6.

\section*{7e. Exercises}

We had a quite exciting chapter here, and as exercises on all this, we have:

\begin{exercise}
Meditate some more on the various definitions of $\pi$.
\end{exercise}

\begin{exercise}
Do some computations for the lengths of the ellipses.
\end{exercise}

\begin{exercise}
Do some computations for the areas of ellipsoids, too.
\end{exercise}

\begin{exercise}
Learn if needed the theory of partial derivatives, with full details.
\end{exercise}

\begin{exercise}
Fill in the details, in the proof of the change of variable formula.
\end{exercise}

\begin{exercise}
Learn about the stereographic projection, and cartography in general.
\end{exercise}

\begin{exercise}
Compute the laws of hyperspherical variables at small values of $N$.
\end{exercise}

\begin{exercise}
Learn, with full details, the Radon-Nikodym theorem.
\end{exercise}

As bonus exercise, find and solve 100 good exercises on multivariable calculus.

\chapter{Contour integrals}

\section*{8a. Curves, surfaces}

Our goal in this chapter is to integrate over curves, surfaces, and other manifolds. But, as a question coming before that, what are the manifolds themselves?

\bigskip

In answer, things can be quite complicated, but everything comes from conics. We have indeed the following result, which is at the origin of modern mathematics:

\index{degree 2 curve}
\index{algebraic curve}

\begin{theorem}
The conics, which are the algebraic curves of degree $2$ in the plane,
$$C=\left\{(x,y)\in\mathbb R^2\Big|P(x,y)=0\right\}$$
with $\deg P\leq 2$, appear modulo degeneration by cutting a $2$-sided cone with a plane, and can be classified into ellipses, parabolas and hyperbolas.
\end{theorem}

\begin{proof}
This is something very standard, the idea being as follows:

\medskip

(1) Let us first classify the conics up to non-degenerate linear transformations of the plane, which are by definition transformations as follows, with $\det A\neq0$:
$$\binom{x}{y}\to A\binom{x}{y}$$

Our claim is that as solutions we have the circles, parabolas, hyperbolas, along with some degenerate solutions, namely $\emptyset$, points, lines, pairs of lines, $\mathbb R^2$.

\medskip

(2) As a first remark, it looks like we forgot precisely the ellipses, but via linear transformations these become circles, so things fine. As a second remark, all our claimed solutions can appear. Indeed, the circles, parabolas, hyperbolas can appear as follows:
$$x^2+y^2=1\quad,\quad x^2=y\quad,\quad xy=1$$

As for $\emptyset$, points, lines, pairs of lines, $\mathbb R^2$, these can appear too, as follows, and with our polynomial $P$ chosen, whenever possible, to be of degree exactly 2:
$$x^2=-1\quad,\quad x^2+y^2=0\quad,\quad x^2=0\quad,\quad xy=0\quad,\quad 0=0$$

Observe here that, when dealing with these degenerate cases, assuming $\deg P=2$ instead of $\deg P\leq 2$ would only rule out $\mathbb R^2$ itself, which is not worth it. 

\medskip

(3) Getting now to the proof of our claim in (1), classification up to linear transformations, consider an arbitrary conic, written as follows, with $a,b,c,d,e,f\in\mathbb R$:
$$ax^2+by^2+cxy+dx+ey+f=0$$

Assume first $a\neq0$. By making a square out of $ax^2$, up to a linear transformation in $(x,y)$, we can get rid of the term $cxy$, and we are left with:
$$ax^2+by^2+dx+ey+f=0$$

In the case $b\neq0$ we can make two obvious squares, and again up to a linear transformation in $(x,y)$, we are left with an equation as follows:
$$x^2\pm y^2=k$$

In the case of positive sign, $x^2+y^2=k$, the solutions are the circle, when $k\geq0$, the point, when $k=0$, and $\emptyset$, when $k<0$. As for the case of negative sign, $x^2-y^2=k$, which reads $(x-y)(x+y)=k$, here once again by linearity our equation becomes $xy=l$, which is a hyperbola when $l\neq0$, and two lines when $l=0$.

\medskip

(4) In the case $b\neq0$ the study is similar, with the same solutions, so we are left with the case $a=b=0$. Here our conic is as follows, with $c,d,e,f\in\mathbb R$:
$$cxy+dx+ey+f=0$$

If $c\neq 0$, by linearity our equation becomes $xy=l$, which produces a hyperbola or two lines, as explained before. As for the remaining case, $c=0$, here our equation is:
$$dx+ey+f=0$$

But this is generically the equation of a line, unless we are in the case $d=e=0$, where our equation is $f=0$, having as solutions $\emptyset$ when $f\neq0$, and $\mathbb R^2$ when $f=0$.

\medskip

(5) Thus, done with the classification, up to linear transformations as in (1). But this classification leads to the classification in general too, by applying now linear transformations to the solutions that we found. So, done with this, and very good.

\medskip

(6) It remains to discuss the cone cutting. But here, what we have to do is to see how the cone equation $x^2+y^2=kz^2$ changes, under a linear change of coordinates, and then set $z=0$, as to get the $(x,y)$ equation of the intersection. But this leads, via some thinking or computations, to the conclusion that the cone equation $x^2+y^2=kz^2$ becomes in this way a degree 2 equation in $(x,y)$, which can be arbitrary, as desired.
\end{proof}

Ready for some physics? And here, good news, not only what we did in the above is relevant, but is actually at the origin of the whole modern physics, thanks to:

\index{conic}
\index{Kepler laws}
\index{Netwon law}
\index{gravity}
\index{classical mechanics}
\index{celestial mechanics}
\index{equation of motion}
\index{polar coordinates}
\index{Newton theorem}
\index{differential equation}

\begin{theorem}
Planets and other celestial bodies move around the Sun on conics,
$$C=\left\{(x,y)\in\mathbb R^2\Big|P(x,y)=0\right\}$$
with $P\in\mathbb R[x,y]$ being of degree $2$, which can be ellipses, parabolas or hyperbolas.
\end{theorem}

\begin{proof}
This is something quite long, due to Kepler and Newton, as follows:

\medskip

(1) The force of attraction between two bodies of masses $M,m$ is given by:
$$||F||=G\cdot\frac{Mm}{d^2}$$

Here $d$ is the distance between the two bodies, and $G\simeq 6.674\times 10^{-11}$ is a constant. Now assuming that $M$ is fixed at the origin $0\in\mathbb R^2$, the force exterted on $m$ positioned at $z=(x,y)\in\mathbb R^2$, regarded as a vector $F\in\mathbb R^2$, is given by the following formula:
$$F
=-||F||\cdot\frac{z}{||z||}
=-\frac{GMm}{||z||^2}\cdot\frac{z}{||z||}
=-\frac{GMmz}{||z||^3}$$

But $F=ma=m\ddot{z}$, with $a=\ddot{z}$ being the acceleration, so the equation of motion of $m$, assuming that $M$ is fixed at $0$, and with the notation $K=GM$, is:
$$\ddot{z}=-\frac{Kz}{||z||^3}$$

(2) Let us write now our vector $z=(x,y)$ in polar coordinates, as follows:
$$x=r\cos\theta\quad,\quad 
y=r\sin\theta$$

We have then $||z||=r$, and our equation of motion becomes:
$$\ddot{z}=-\frac{Kz}{r^3}$$

Let us differentiate now $x,y$. By using the standard calculus rules, we have:
$$\ddot{x}=\ddot{r}\cos\theta-2\dot{r}\sin\theta\cdot\dot{\theta}-r\cos\theta\cdot\dot{\theta}^2-r\sin\theta\cdot\ddot\theta$$
$$\ddot{y}=\ddot{r}\sin\theta+2\dot{r}\cos\theta\cdot\dot{\theta}-r\sin\theta\cdot\dot{\theta}^2+r\cos\theta\cdot\ddot\theta$$

Consider now the following two quantities, appearing as coefficients in the above:
$$a=\ddot{r}-r\dot{\theta}^2\quad,\quad b=2\dot{r}\dot{\theta}+r\ddot{\theta}$$

In terms of these quantities, our second derivative formulae read:
$$\ddot{x}=a\cos\theta-b\sin\theta\quad,\quad 
\ddot{y}=a\sin\theta+b\cos\theta$$

(3) We can now solve the equation of motion. Indeed, with the formulae that we found above for $\ddot{x},\ddot{y}$, our equation of motion takes the following form:
$$a\cos\theta-b\sin\theta=-\frac{K}{r^2}\cos\theta\quad,\quad
a\sin\theta+b\cos\theta=-\frac{K}{r^2}\sin\theta$$

But these two formulae can be written in the following way:
$$\left(a+\frac{K}{r^2}\right)\cos\theta=b\sin\theta\quad,\quad 
\left(a+\frac{K}{r^2}\right)\sin\theta=-b\cos\theta$$

By making now the product, and assuming that we are in a non-degenerate case, where the angle $\theta$ varies indeed, we obtain by positivity that we must have:
$$a+\frac{K}{r^2}=b=0$$

(4) Let us first examine the second equation, $b=0$. This can be solved as follows:
$$b=0
\iff 2\dot{r}\dot{\theta}+r\ddot{\theta}=0
\iff\dot{\theta}=\frac{\lambda}{r^2}$$

As for the first equation, namely $a+K/r^2=0$, with $K=\lambda^2/c$ this becomes:
$$\ddot{r}=\frac{\lambda^2}{r^2}\left(\frac{1}{r}-\frac{1}{c}\right)$$

Now set $f=1/r$. With the convention that dots mean as usual derivatives with respect to $t$, and that primes denote derivatives with respect to $\theta=\theta(t)$, we have:
$$\dot{r}=-\frac{f'\dot{\theta}}{f^2}=-\frac{f'}{f^2}\cdot\frac{\lambda}{r^2}=-\lambda f'$$

By differentiating one more time with respect to $t$, we obtain:
$$\ddot{r}=-\lambda f''\dot{\theta}=-\lambda f''\cdot\frac{\lambda}{r^2}=-\frac{\lambda^2}{r^2}f''$$

Thus, in terms of $f=1/r$ as above, our equation for $\ddot{r}$ simply reads:
$$f''+f=\frac{1}{c}$$

But this latter equation is easy to solve, and by inverting, we obtain:
$$r=\frac{1}{f}=\frac{c}{1+\varepsilon\cos\theta+\delta\sin\theta}$$

(5) But this leads to the conclusion that the trajectory is a conic. Indeed, in terms of the parameter $\theta$, the formulae of the coordinates are:
$$x=\frac{c\cos\theta}{1+\varepsilon\cos\theta+\delta\sin\theta}\quad,\quad 
y=\frac{c\sin\theta}{1+\varepsilon\cos\theta+\delta\sin\theta}$$

We conclude from this that our coordinates $x,y$ satisfy the following equation:
$$x^2+y^2=(\varepsilon x+\delta y-c)^2$$

But what we have here is an equation of a conic, and we are done.
\end{proof}

All the above was very nice, all good old things, but the continuation of the story is more complicated, because beyond conics, things ramify. A first idea, in order to generalize the conics, is to look at the algebraic manifolds, in the following sense:

\begin{definition}
An algebraic manifold $X\subset\mathbb R^N$ is a space of the form
$$X=\left\{x\in\mathbb R^N\Big|P_i(x)=0\right\}$$
with $P_i\in\mathbb R[x_1,\ldots,x_N]$ being certain polynomials.
\end{definition}

Manhy things can be said about such manifolds, both of mathematical and physical nature, and as a fundamental result here, generalizing Theorem 8.1, let us record:

\begin{theorem}
The degree $2$ hypersurfaces $S\subset\mathbb R^N$, called quadrics, are up to degeneracy and to linear transformations the hypersurfaces of the following form,
$$\pm x_1^2\pm\ldots\pm x_N^2=0,1$$
and with the sphere being the only compact one.
\end{theorem}

\begin{proof}
We have two statements here, the idea being as follows:

\medskip

(1) The equations for a quadric $S\subset\mathbb R^N$ are best written as follows, with $A\in M_N(\mathbb R)$ being a matrix, $B\in M_{1\times N}(\mathbb R)$ being a row vector, and $C\in\mathbb R$ being a constant:
$$<Ax,x>+Bx+C=0$$

(2) By doing the linear algebra, or by invoking the theorem of Sylvester on quadratic forms, we are left, modulo linear transformations, with signed sums of squares:
$$\pm x_1^2\pm\ldots\pm x_N^2=0,1$$

(3) To be more precise, with linear algebra, by evenly distributing the terms $x_ix_j$ above and below the diagonal, we can assume that our matrix $A\in M_N(\mathbb R)$ is symmetric. Thus $A$ must be diagonalizable, and by changing the basis of $\mathbb R^N$, as to have it diagonal, our equation becomes as follows, with $D\in M_N(\mathbb R)$ being now diagonal:
$$<Dx,x>+Ex+F=0$$

(4) But now, by making squares in the obvious way, which amounts in applying yet another linear transformation to our quadric, the equation takes the following form, with $G\in M_N(-1,0,1)$ being diagonal, and with $H\in\{0,1\}$ being a constant:
$$<Gx,x>=H$$

(5) Now barring the degenerate cases, we can further assume $G\in M_N(-1,1)$, and we are led in this way to the equation claimed in (2) above, namely:
$$\pm x_1^2\pm\ldots\pm x_N^2=0,1$$

(6) In particular we see that, up to some degenerate cases, namely emptyset and point, the only compact quadric, up to linear transformations, is the one given by:
$$x_1^2+\ldots+x_N^2=1$$

(7) But this is the unit sphere, so are led to the conclusions in the statement.
\end{proof}

Summarizing, we have some beginning of algebraic geometry theory going on. However, for our purposes here, which are integration, along with applications to physics, although many interesting algebraic manifolds appear at the advanced level, in what concerns the basics, here we are mostly in need of a different definition, namely:

\index{differential manifold}
\index{smooth manifold}
\index{chart}
\index{coordinates}

\begin{definition}
A smooth manifold is a space $X$ which is locally isomorphic to $\mathbb R^N$. To be more precise, this space $X$ must be covered by charts, bijectively mapping open pieces of it to open pieces of $\mathbb R^N$, with the changes of charts being $C^\infty$ functions.
\end{definition}

As a basic example, we have $\mathbb R^N$ itself, or any open subset $X\subset\mathbb R^N$. Another example is the circle, or curves like ellipses and so on, for obvious reasons. However, this might seem a bit vague, and here is a more precise statement in this sense:

\begin{theorem}
The following are smooth manifolds, in the plane:
\begin{enumerate}
\item The circles.

\item The ellipses.

\item The non-degenerate conics.

\item Smooth deformations of these.
\end{enumerate}
\end{theorem}

\begin{proof}
All this is quite intuitive, the idea being as follows:

\medskip

(1) Consider the unit circle, $x^2+y^2=1$. We can write then $x=\cos t$, $y=\sin t$, with $t\in[0,2\pi)$, and we seem to have here the solution to our problem, just using 1 chart. But this is of course wrong, because $[0,2\pi)$ is not open, and we have a problem at $0$. In practice we need to use 2 such charts, say with the first one being with $t\in(0,3\pi/2)$, and the second one being with $t\in(\pi,5\pi/2)$. As for the fact that the change of charts is indeed smooth, this comes by writing down the formulae, or just thinking a bit, and arguing that this change of chart being actually a translation, it is automatically linear.

\medskip

(2) This follows from (1), by pulling the circle in both the $Ox$ and $Oy$ directions, and the  formulae here, based on our various formulae above, are left to you reader.

\medskip

(3) We already have the ellipses, and the case of the parabolas and hyperbolas is elementary as well, and in fact simpler than the case of the ellipses. Indeed, a parablola is clearly homeomorphic to $\mathbb R$, and a hyperbola, to two copies of $\mathbb R$.

\medskip

(4) This is something which is clear too, depending of course on what exactly we mean by ``smooth deformation'', and by using a bit of multivariable calculus if needed.
\end{proof}

In higher dimensions now, as basic examples here, we have the unit sphere in $\mathbb R^N$, and smooth deformations of it, once again, somehow by obvious reasons, and with the formal proof of this fact coming for instance by using the spherical coordinate formulae from chapter 7. Alternatively, we can use the stereographic projection, which provides us with a proof that the sphere $S^{N-1}_\mathbb R$ is indeed a smooth manifold, with just 2 charts.

\section*{8b. Contour integrals}

We would like to discuss now the integration over curves, surfaces, and other smooth manifolds. Let us start with something very basic, regarding the curves, namely:

\begin{theorem}
The length of a curve $\gamma:[a,b]\to\mathbb R^N$ is given by
$$L(\gamma)=\int_a^b||\gamma'(t)||dt$$
with $||.||$ being the usual norm of $\mathbb R^N$.
\end{theorem}

\begin{proof}
This is something quite intuitive, that can even stand as a definition for the length of the curves, and that we will not prove in detail here, the idea being as follows:

\medskip

(1) To start with, what is the length of a curve? Good question, and in answer, a physicist would say that this is the quantity obtained by integrating the magnitude of the velocity vector over the curve, with respect to time. But this velocity vector is $\gamma'(t)$, having magnitude $||\gamma'(t)||$, so we are led to the formula in the statement.

\medskip

(2) Regarding now mathematicians, these would say that the length of a curve is the following quantity, with $(t_1=a,t_2,\ldots,t_{n-1},t_n=b)$ being a uniform division of $(a,b)$:
$$L(\gamma)=\lim_{n\to\infty}\sum_{i=1}^n||\gamma(t_i)-\gamma(t_{i-1})||$$

But, by using the fundamental theorem of calculus, we can write this as follows:
$$L(\gamma)=\lim_{n\to\infty}\sum_{i=1}^n\left|\left|\int_{t_{i-1}}^{t_i}\gamma'(t)dt\right|\right|$$

And the point now is that, by doing some standard analysis, that we will leave here as an instructive exercise, we are led to the formula in the statement.

\medskip

(3) So, we have our definition-theorem for the length of the curves, and for the discussion to be complete, we just need an illustration for this. But here, we have the following standard computation, for the length of the ellipses $(x/a)^2+(y/b)^2=1$:
\begin{eqnarray*}
L
&=&4\int_0^{\pi/2}\sqrt{\left(\frac{da\cos t}{dt}\right)^2+\left(\frac{db\sin t}{dt}\right)^2}\,dt\\
&=&4\int_0^{\pi/2}\sqrt{a^2\sin^2t+b^2\cos^2t}\,dt
\end{eqnarray*}

There are of course many other applications of the formula in the statement, and we will leave some thinking and computations here as an instructive exercise.
\end{proof}

Next, let us talk about the contour integrals over curves. We have here:

\begin{theorem}
Given a path $\gamma\subset\mathbb R^3$, we can talk about integrals of type
$$I=\int_\gamma f(x)dx_1+g(x)dx_2+h(x)dx_3$$ 
with $f,g,h:\mathbb R^3\to\mathbb R$, which are independent on the chosen parametrization of the path.
\end{theorem}

\begin{proof}
This is something quite straightforward, the idea being as follows:

\medskip

(1) Regarding the statement itself, assume indeed that we have a path in $\mathbb R^3$, which can be best thought of as corresponding to a function as follows:
$$\gamma:[a,b]\to\mathbb R^3$$

Observe that this function $\gamma$ is not exactly the path itself, for instance because the following functions produce the same path, parametrized differently:
$$\delta:[0,b-a]\to\mathbb R^3\quad,\quad \delta(t)=\gamma(t+a)$$
$$\varepsilon:[0,1]\to\mathbb R^3\quad,\quad \varepsilon(t)=\delta((b-a)t)$$
$$\varphi:[0,1]\to\mathbb R^3\quad,\quad \varphi(t)=\varepsilon(t^2)$$
$$\psi:[0,1]\to\mathbb R^3\quad,\quad \psi(t)=\varepsilon(1-t)$$
$$\vdots$$

Our claim, however, is that we can talk about integrals as follows, with $f,g,h:\mathbb R^3\to\mathbb R$, which are independent on the chosen parametrization of our path:
$$I=\int_\gamma f(x)dx_1+g(x)dx_2+h(x)dx_3$$ 

(2) In order to prove this, let us choose a parametrization $\gamma:[a,b]\to\mathbb R^3$ as above. This parametrization has as components three functions $\gamma_1,\gamma_2,\gamma_3$, given by:
$$\gamma=(\gamma_1,\gamma_2,\gamma_3):[a,b]\to\mathbb R^3$$

In order to construct the integral $I$, it is quite clear, by suitably cutting our path into pieces, that we can restrict the attention to the case where all three components $\gamma_1,\gamma_2,\gamma_3:[a,b]\to\mathbb R$ are increasing, or decreasing. Thus, we can assume that these three components are as follows, increasing or decreasing, and bijective on their images:
$$\gamma_i:[a,b]\to[a_i,b_i]$$

(3) Moreover, by using the obvious symmetry between the coordinates $x_1,x_2,x_3$, in order to construct $I$, we just need to construct integrals of the following type:
$$I_1=\int_\gamma f(x)dx_1$$ 

(4) So, let us construct this latter integral $I_1$, under the assumptions in (2). The simplest case is when the first path, $\gamma_1:[a,b]\to[a_1,b_1]$, is the identity:
$$\gamma_1:[a,b]\to[a,b]\quad,\quad\gamma_1(x)=x$$

In other words, the simplest case is when our path is of the following form, with $\gamma_2,\gamma_3:[a,b]\to\mathbb R$ being certain functions, that should be increasing or decreasing, as per our conventions (2) above, but in what follows we will not need this assumption:
$$\gamma(x_1)=(x_1,\gamma_2(x_1),\gamma_3(x_1))$$ 

But now, with this convention made, we can define our contour integral, or rather its first component, as explained above, as a usual one-variable integral, as follows:
$$I_1=\int_a^bf(x_1,\gamma_2(x_1),\gamma_3(x_1))dx_1$$

(5) With this understood, let us examine now the general case, where the first path, $\gamma_1:[a,b]\to[a_1,b_1]$, is arbitrary, increasing or decreasing, and bijective on its image. In this case we can reparametrize our curve, as to have it as in (4) above, as follows:
$$\tilde{\gamma}=(id,\gamma_2\gamma_1^{-1},\gamma_3\gamma_1^{-1}):[a_1,b_1]\to\mathbb R^3$$

Now since we want our integral $I_1=\int_\gamma f(x)dx_1$ to be independent of the parametrization, we are led to the following formula for it, coming from the formula in (4):
\begin{eqnarray*}
I_1
&=&\int_{\tilde{\gamma}}f(x)dx_1\\
&=&\int_{a_1}^{b_1}f(x_1,\gamma_2\gamma_1^{-1}(x_1),\gamma_3\gamma_1^{-1}(x_1))dx_1\\
&=&\int_a^bf(\gamma_1(y_1),\gamma_2(y_1),\gamma_3(y_1))\gamma_1'(y_1)dy_1
\end{eqnarray*}

Here we have used at the end the change of variable formula, with $x_1=\gamma_1(y_1)$.

\medskip

(6) Thus, job done, we have our definition for the contour integrals, with the formula being as follows, obtained by using (5) for all three coordinates $x_1,x_2,x_3$:
\begin{eqnarray*}
I
&=&\int_a^bf(\gamma_1(y_1),\gamma_2(y_1),\gamma_3(y_1))\gamma_1'(y_1)dy_1\\
&&+\int_a^bg(\gamma_1(y_2),\gamma_2(y_2),\gamma_3(y_2))\gamma_2'(y_2)dy_2\\
&&+\int_a^bh(\gamma_1(y_3),\gamma_2(y_3),\gamma_3(y_3))\gamma_3'(y_3)dy_3
\end{eqnarray*}

And with this, we are led to the conclusion in the statement.
\end{proof}

Let us record as well the following more compact form of Theorem 8.8:

\begin{theorem}
The contour integrals over a curve $\gamma:[a,b]\to\mathbb R^3$ are given by
$$\int_\gamma<F(x),dx>=\int_a^b<F(\gamma(y)),\gamma'(y)dy>$$
valid for any $F:\mathbb R^3\to\mathbb R^3$, where on the right $\gamma'(y)dy=(\gamma_i'(y_i)dy_i)_i$.
\end{theorem}

\begin{proof}
This is a fancy reformulation of what we did in Theorem 8.8 and its proof. Indeed, with the notation $F=(F_1,F_2,F_3)=(f,g,h)$, the integral computed there is:
$$\int_\gamma<F(x),dx>=\int_\gamma F_1(x)dx_1+F_2(x)dx_2+F_3(x)dx_3$$

As for the value of this integral, according to the proof of Theorem 8.8, this is:
\begin{eqnarray*}
\int_\gamma<F(x),dx>
&=&\int_a^bF_1(\gamma_1(y_1),\gamma_2(y_1),\gamma_3(y_1))\gamma_1'(y_1)dy_1\\
&&+\int_a^bF_2(\gamma_1(y_2),\gamma_2(y_2),\gamma_3(y_2))\gamma_2'(y_2)dy_2\\
&&+\int_a^bF_3(\gamma_1(y_3),\gamma_2(y_3),\gamma_3(y_3))\gamma_3'(y_3)dy_3
\end{eqnarray*}

Now observe that we can write this in a more compact way, as follows:
\begin{eqnarray*}
\int_\gamma<F(x),dx>
&=&\int_a^bF_1(\gamma(y_1))\gamma'(y_1)dy_1\\
&&+\int_a^bF_2(\gamma(y_2))\gamma'(y_2)dy_2\\
&&+\int_a^bF_3(\gamma(y_3))\gamma'(y_3)dy_3
\end{eqnarray*}

And we can do even better. Indeed, we have only one integral here, $\int_a^b$, and in order to best express the integrand, consider the formal vector in the statement, namely:
$$\gamma'(y)dy=\begin{pmatrix}
\gamma_1'(y_1)dy_1\\
\gamma_2'(y_2)dy_2\\
\gamma_3'(y_3)dy_3
\end{pmatrix}$$

Our integrand appears then as the scalar product of $F(\gamma(y))$ with this latter vector $\gamma'(y)dy$, so our formula above for the contour integral takes the following form:
$$\int_\gamma<F(x),dx>=\int_a^b<F(\gamma(y)),\gamma'(y)dy>$$

Thus, we are led to the conclusion in the statement. 
\end{proof}

More concretely now, let us temporarily forget about the paths $\gamma$, and have a look at the quantities which are to be integrated, namely:
$$\alpha=F_1(x)dx_1+F_2(x)dx_2+F_3(x)dx_3$$

Obviously, these are something rather mathematical, and many things can be said here. However, we can have some physical intuition on them. Assume indeed that we are given a function as follows, that you can think for instance as corresponding to an external force, with $F(x)\in\mathbb R^3$ being the force vector applied at a given point $x\in\mathbb R^3$:
$$F:\mathbb R^3\to\mathbb R^3$$

By writing $F=(F_1,F_2,F_3)$, we can then consider the following quantity, and when $F:\mathbb R^3\to\mathbb R^3$ varies, we obtain exactly the abstract quantities $\alpha$ considered above:
$$<F(x),dx>=F_1(x)dx_1+F_2(x)dx_2+F_3(x)dx_3$$

Thus, all in all, what we have done in the above with our construction of contour integrals, was to define quantities as follows, with $\gamma$ being a path in $\mathbb R^3$, and with $F:\mathbb R^3\to\mathbb R^3$ being a certain function, that we can think of, if we want, as being a force:
$$I=\int_\gamma<F(x),dx>$$

Which brings us into physics. Indeed, by assuming now that $F:\mathbb R^3\to\mathbb R^3$ does correspond to a force, we can formulate the following definition:

\index{work}
\index{conservative force}
\index{friction}

\begin{definition}
The work done by a force $F=F(x)$ for moving a particle from point $p\in\mathbb R^3$ to point $q\in\mathbb R^3$ via a given path $\gamma:p\to q$ is the following quantity:
$$W(\gamma)=\int_\gamma<F(x),dx>$$
We say that $F$ is conservative if this work quantity $W(\gamma)$ does not depend on the chosen path $\gamma:p\to q$, and in this case we denote this quantity by $W(p,q)$.
\end{definition}

This definition is something quite subtle, and as a first comment, assume that we have two paths $\gamma:p\to q$ and $\delta:p\to q$. We can then consider the path $\circ:p\to p$ obtained by going along $\gamma:p\to q$, and then along $\delta$ reversed, $\delta^{-1}:q\to p$, and we have:
$$W(\circ)=W(\gamma)-W(\delta)$$ 

Thus $F$ is conservative precisely when, for any loop $\circ:p\to p$, we have: 
$$W(\circ)=0$$

Intuitively, this means that $F$ is some sort of ``clean'', ideal force, with no dirty things like friction involved. As we will soon see, gravity is such a clean force, with a simple example coming from throwing a rock up in the sky. That rock will travel on a loop $p\to q\to p$, and will come back here to $p$ unchanged, save for the fact that its speed vector is reversed. Thus, and assuming now that work has something to do with energy, which is intuitive, there has been no overall work of gravity on this loop, $W(\circ)=0$.

\bigskip

As a first result now, regarding the conservative forces, we have:

\begin{theorem}
The work done by a conservative force $F$ on a mass $m$ object is
$$W(p,q)=T(q)-T(p)$$
with $T=m||v||^2/2$ standing as usual for the kinetic energy of the object.
\end{theorem}

\begin{proof}
Assuming that $F$ is conservative, and acts via the usual formula $F=ma$ on our object of mass $m$, we have the following computation, as desired:
\begin{eqnarray*}
W(p,q)
&=&\int_p^q<F(x),dx>\\
&=&m\int_p^q<a(x),dx>\\
&=&m\int_p^q\left<\frac{dv(x)}{dt},v(x)dt\right>\\
&=&\frac{m}{2}\int_p^q\frac{d<v(x),v(x)>}{dt}\,dt\\
&=&\frac{m}{2}\int_p^q\frac{d||v(x)||^2}{dt}\,dt\\
&=&\frac{m}{2}\left(||v(q)||^2-||v(p)||^2\right)\\
&=&T(q)-T(p)
\end{eqnarray*}

Here we have used in the middle the fact that the time derivative of a scalar product of functions $<v,w>$ consists of two terms, which are equal when $v=w$.
\end{proof}

In order to formulate our next result, observe that we have the following computation for the contour integrals of gradients, which is independent on the chosen path:
\begin{eqnarray*}
\int_p^q<\nabla V(x),dx>
&=&\int_p^q\frac{dV}{dx_1}\cdot dx_1+\frac{dV}{dx_2}\cdot dx_2+\frac{dV}{dx_3}\cdot dx_3\\
&=&\int_p^qdV\\
&=&V(q)-V(p)
\end{eqnarray*}

To be more precise, this computation certainly works when $V$ is a function of just one variable, $x_1$, $x_2$ or $x_3$, thanks to the fundamental theorem of calculus, and the general case follows from this, by using the chain rule for derivatives.

\bigskip

Now with this formula in hand, we can formulate the following result, which is quite conceptual, and which includes some basic gravitation physics too, at the end:

\begin{theorem}
A force $F$ is conservative precisely when it is of the form
$$F=-\nabla V$$
for a certain function $V$, and in this case the work done by it is given by:
$$W(p,q)=V(p)-V(q)$$
Also, the gravitation force is conservative, coming from $V=-C/||x||$, with $C>0$.
\end{theorem}

\begin{proof}
This is something quite tricky, the idea being as follows:

\medskip

(1) In one sense, assume that $F$ is conservative. Since the work $W(p,q)=W(\gamma)$ is independent of the chosen path $\gamma:p\to q$, we can find a function $V$ such that:
$$W(p,q)=V(p)-V(q)$$

Observe that this function $V$ is well-defined up to an additive constant. Now with this formula in hand, we further obtain, as desired:
\begin{eqnarray*}
W(p,q)=V(p)-V(q)
&\implies&\int_p^q<F(x),dx>=-\int_p^q<\nabla V(x),dx>\\
&\implies&<F(x),dx>=-<\nabla V(x),dx>\\
&\implies&F_i(x)dx_i=-\frac{dV}{dx_i}\cdot dx_i\\
&\implies&F_i(x)=-\frac{dV}{dx_i}\\
&\implies&F=-\nabla V
\end{eqnarray*}

(2) In the other sense now, assuming $F=-\nabla V$, we have the following computation, valid for any loop $\circ:p\to p$, which shows that $F$ is indeed conservative:
$$W(\circ)
=-\int_\circ\nabla V
=0$$

More generally, regarding the work done by such a force $F=-\nabla V$, along a path $\gamma:p\to q$, which is independent on this path $\gamma$, this is given by:
$$W(p,q)
=-\int_p^q\nabla V
=V(p)-V(q)$$

(3) According to the Newton principles, and with $K=GM$ as usual, the gravitational force applied to $m$ by an object $M$ positioned at 0 is given by:
$$F=-||F||\cdot\frac{x}{||x||}=-G\cdot\frac{mM}{||x||^2}\cdot\frac{x}{||x||}=-\frac{Kmx}{||x||^3}$$

Now let us compute $\nabla V$, with $V=-Km/||x||$. For this purpose, we first have:
\begin{eqnarray*}
\frac{d  V}{d  x_i}
&=&-\frac{d  Km/\sqrt{x_1^2+x_2^2+x_3^2}}{d  x_i}\\
&=&-Km\cdot\frac{d (x_1^2+x_2^2+x_3^2)^{-1/2}}{d  x_i}\\
&=&-Km\cdot\left(-\frac{1}{2}\right)\cdot\frac{2x_i}{(x_1^2+x_2^2+x_3^2)^{3/2}}\\
&=&\frac{Kmx_i}{||x||^3}
\end{eqnarray*}

Thus we have $\nabla V=Kmx/||x||^3$, which by the above equals $-F$, as desired.
\end{proof}

We can now put everything together, and we have the following result:

\index{potential}

\begin{theorem}
Given a conservative force $F$, appearing as follows, with $V$ being uniquely determined up to an additive constant,
$$F=-\nabla V$$
the movements of a particle under $F$ preserve the total energy, given by
$$E=T+V$$
with $T=m||v||^2/2$ being the kinetic energy, and with $V$ being called potential energy.
\end{theorem}

\begin{proof}
By using Theorem 8.11 and Theorem 8.12, we have:
$$W(p,q)=T(q)-T(p)\quad,\quad 
W(p,q)=V(p)-V(q)$$

Now observe that these equalities give the following formula:
$$T(p)+V(p)=T(q)+V(q)$$

Thus, the total energy $E=T+V$ is conserved, as claimed.
\end{proof}

\section*{8c. Green and Stokes} 

At the level of mathematics now, in relation with the above, a useful result is:

\begin{theorem}[Green]
Given a plane curve $C\subset\mathbb R^2$, we have the formula
$$\int_CPdx+Qdy=\int_D\left(\frac{dQ}{dx}-\frac{dP}{dy}\right)dxdy$$
where $D\subset\mathbb R^2$ is the domain enclosed by $C$.
\end{theorem}

\begin{proof}
Assume indeed that we have a plane curve $C\subset\mathbb R^2$, without self-intersections, which is piecewise $C^1$, and is assumed to be counterclockwise oriented. In order to prove the formula regarding $P$, we can parametrize the enclosed domain $D$ as follows:
$$D=\left\{(x,y)\Big|a\leq x\leq b,f(x)\leq y\leq g(x)\right\}$$

We have then the following computation, which gives the result for $P$:
\begin{eqnarray*}
\int_D-\frac{dP}{dy}\,dxdy
&=&\int_a^b\left(\int_{f(x)}^{g(x)}-\frac{dP}{dy}(x,y)dy\right)dx\\
&=&\int_a^bP(x,f(x))-P(x,g(x))dx\\
&=&\int_a^bP(x,f(x))dx+\int_b^aP(x,g(x))dx\\
&=&\int_CPdx
\end{eqnarray*}

As for the result for the $Q$ term, the computation here is similar.
\end{proof}

Moving forward to 3D, let us start with a standard definition, as follows:

\begin{definition}
The vector product of two vectors in $\mathbb R^3$ is given by
$$x\times y=||x||\cdot||y||\cdot\sin\theta\cdot n$$
where $n\in\mathbb R^3$ with $n\perp x,y$ and $||n||=1$ is constructed using the right-hand rule:
$$\begin{matrix}
\ \ \ \ \ \ \ \ \ \uparrow_{x\times y}\\
\leftarrow_x\\
\swarrow_y
\end{matrix}$$
Alternatively, in usual vertical linear algebra notation for all vectors,
$$\begin{pmatrix}x_1\\ x_2\\x_3\end{pmatrix}
\times\begin{pmatrix}y_1\\ y_2\\y_3\end{pmatrix}
=\begin{pmatrix}x_2y_3-x_3y_2\\ x_3y_1-x_1y_3\\x_1y_2-x_2y_1\end{pmatrix}$$
the rule being that of computing $2\times2$ determinants, and adding a middle sign.
\end{definition}

In practice now, in order to get familiar with the vector products, nothing better than doing some classical mechanics. We have here the following key result:

\begin{theorem}
In the gravitational $2$-body problem, the angular momentum
$$J=x\times p$$
with $p=mv$ being the usual momentum, is conserved.
\end{theorem}

\begin{proof}
Consider indeed a particle $m$ moving under the gravitational force of a particle $M$, assumed, as usual, to be fixed at 0. By using the fact, coming from definitions, that for two proportional vectors, $p\sim q$, we have $p\times q=0$, we obtain:
\begin{eqnarray*}
\dot{J}
&=&\dot{x}\times p+x\times\dot{p}\\
&=&v\times mv+x\times ma\\
&=&m(v\times v+x\times a)\\
&=&m(0+0)\\
&=&0
\end{eqnarray*}

Now since the derivative of $J$ vanishes, this quantity is constant, as stated.
\end{proof}

At the mathematical level now, we have the following key result:

\begin{theorem}
The area of a surface $S\subset\mathbb R^3$, parametrized as $S=r(D)$ with $r:D\to\mathbb R^3$ and $D\subset\mathbb R^2$, is given by
$$A(S)=\int_D||r_x\times r_y||\,dxdy$$
with $r_x,r_y:D\to\mathbb R^3$ being the partial derivatives of $r$, and $||.||$ being the norm of $\mathbb R^3$.
\end{theorem}

\begin{proof}
Thus is something quite similar to Theorem 8.7, and many things can be said here, at the theoretical level, in analogy with those for the curves. Among others, let us mention that we can talk, more generally, about surface integrals, defined as follows:
$$\int_Sf(s)ds=\int_Df(r(x,y))||r_x\times r_y||\,dxdy$$

As for the basic illustrations of the formula in the statement, consider a surface of type $z=f(x,y)$. Here we have $r(x,y)=(x,y,f(x,y))$, and we obtain:
\begin{eqnarray*}
A(S)
&=&\int_D\left|\left|\begin{pmatrix}1\\0\\f_x\end{pmatrix}\times\begin{pmatrix}0\\1\\f_y\end{pmatrix}\right|\right|dxdy\\
&=&\int_D\left|\left|\begin{pmatrix}-f_x\\-f_y\\1\end{pmatrix}\right|\right|dxdy\\
&=&\int_D\sqrt{f_x^2+f_y^2+1}\,dxdy
\end{eqnarray*}

There are of course many other applications of the formula in the statement, and we will leave some thinking and computations here as an instructive exercise.
\end{proof}

Many other things can be said, as a continuation of the above, but let us not deviate too much, from what we wanted to do here. Next, comes the following definition:

\begin{definition}
We can talk about the divergence of $\varphi:\mathbb R^3\to\mathbb R^3$, as being
$$<\nabla,\varphi>=
\left<\begin{pmatrix}
\frac{d}{dx}\\
\frac{d}{dy}\\
\frac{d}{dz}
\end{pmatrix},\begin{pmatrix}
\varphi_x\\
\varphi_y\\
\varphi_z
\end{pmatrix}\right>=
\frac{d\varphi_x}{dx}+\frac{d\varphi_y}{dy}+\frac{d\varphi_z}{dz}$$
as well as about the curl of the same function $\varphi:\mathbb R^3\to\mathbb R^3$, as being
$$\nabla\times\varphi=
\begin{vmatrix}
u_x&\frac{d}{dx}&\varphi_x\\
u_y&\frac{d}{dy}&\varphi_y\\
u_z&\frac{d}{dz}&\varphi_z
\end{vmatrix}
=\begin{pmatrix}
\frac{d\varphi_z}{dy}-\frac{d\varphi_y}{dz}\\
\frac{d\varphi_x}{dz}-\frac{d\varphi_z}{dx}\\
\frac{d\varphi_y}{dx}-\frac{d\varphi_x}{dy}
\end{pmatrix}
$$
where $u_x,u_y,u_z$ are the unit vectors along the coordinate directions $x,y,z$.
\end{definition}

Getting back now to calculus tools, in 3 dimensions, we have the following result:

\begin{theorem}[Stokes]
Given a smooth oriented surface $S\subset\mathbb R^3$, with boundary $C\subset\mathbb R^3$, and a vector field $F$, we have the following formula:
$$\int_S<(\nabla\times F)(x),n(x)>dx=\int_C<F(x),dx>$$
In other words, the line integral of a vector field $F$ over a loop $C$ equals the surface integral of the curl of the vector field $\varphi$, over the enclosed surface $S$. 
\end{theorem}

\begin{proof}
This basically follows from the Green theorem, the idea being as follows:

\medskip

(1) Let us first parametrize our surface $S$, and its boundary $C$. We can assume that we are in the situation where we have a closed oriented curve $\gamma:[a,b]\to\mathbb R^2$, with interior $D\subset\mathbb R^2$, and where the surface appears as $S=\psi(D)$, with $\psi:D\to\mathbb R^3$. In this case, the function $\delta=\psi\circ\gamma$ parametrizes the boundary of our surface, $C=\delta[a,b]$.

\medskip

(2) Let us first look at the integral on the right in the statement. We have the following formula, with $J_y(\psi)$ standing for the Jacobian of $\psi$ at the point $y=\gamma(t)$:
\begin{eqnarray*}
\int_C<F(x),dx>
&=&\int_\gamma<F(\psi(\gamma)),d\psi(\gamma)>\\
&=&\int_\gamma<F(\psi(y)),J_y(\psi)d\gamma>
\end{eqnarray*}

In order to further process this formula, let us introduce the following function:
$$P(u,v)=\left<F(\psi(u,v)),\frac{d\psi}{du}(u,v)\right>e_u+\left<F(\psi(u,v)),\frac{d\psi}{dv}(u,v)\right>e_v$$

In terms of this function, we have the following formula for our line integral:
$$\int_C<F(x),dx>=\int_\gamma<P(y),dy>$$

(3) In order to compute now the other integral in the statement, we first have:
\begin{eqnarray*}
\frac{dP_v}{du}-\frac{dP_u}{dv}
&=&\left<\frac{d(F\psi)}{du},\frac{d\psi}{dv}\right>+\left<F\psi,\frac{d^2\psi}{dudv}\right>
-\left<\frac{d(F\psi)}{dv},\frac{d\psi}{du}\right>-\left<F\psi,\frac{d^2\psi}{dvdu}\right>\\
&=&\left<\frac{d(F\psi)}{du},\frac{d\psi}{dv}\right>-\left<\frac{d(F\psi)}{dv},\frac{d\psi}{du}\right>\\
&=&\left<\frac{d\psi}{dv},\left(J_{\psi(u,v)}F-(J_{\psi(u,v)}F)^t\right)\frac{d\psi}{du}\right>\\
&=&\left<\frac{d\psi}{dv},(\nabla\times F)\times\frac{d\psi}{du}\right>\\
&=&\left<\nabla\times F,\frac{d\psi}{du}\times\frac{d\psi}{dv}\right>
\end{eqnarray*}

We conclude that the integral on the left in the statement is given by:
\begin{eqnarray*}
&&\int_S<(\nabla\times F)(x),n(x)>dx\\
&=&\int_D\left<(\nabla\times F)(\psi(u,v)),\frac{d\psi}{du}(u,v)\times\frac{d\psi}{dv}(u,v)\right>dudv\\
&=&\int_D\left(\frac{dP_v}{du}-\frac{dP_u}{dv}\right)dudv
\end{eqnarray*}

(4) But with this, we are done, because the integrals computed in (2) and (3) are indeed equal, due to the Green theorem. Thus, the Stokes formula holds indeed.
\end{proof}

As a conclusion to what we did so far, we have the following statement:

\begin{theorem}
The following results hold, in $3$ dimensions:
\begin{enumerate}
\item Fundamental theorem for gradients, namely
$$\int_a^b<\nabla f,dx>=f(b)-f(a)$$

\item Fundamental theorem for divergences, or Gauss or Green formula,
$$\int_B<\nabla,\varphi>=\int_S<\varphi(x),n(x)>dx$$

\item Fundamental theorem for curls, or Stokes formula,
$$\int_A<(\nabla\times\varphi)(x),n(x)>dx=\int_P<\varphi(x),dx>$$
\end{enumerate}
where $S$ is the boundary of the body $B$, and $P$ is the boundary of the area $A$.
\end{theorem}

\begin{proof}
This follows indeed from the various formulae established above.
\end{proof}

\section*{8d. Gauss, Maxwell}

We would like to discuss now, following Gauss and others, some applications of the above to electrostatics, and why not to electrodynamics too. Let us start with:

\begin{fact}[Coulomb law]
Any pair of charges $q_1,q_2\in\mathbb R$ is subject to a force as follows, which is attractive if $q_1q_2<0$ and repulsive if $q_1q_2>0$,
$$||F||=K\cdot\frac{|q_1q_2|}{d^2}$$
where $d>0$ is the distance between the charges, and $K>0$ is a certain constant.
\end{fact}

Observe the amazing similarity with the Newton law for gravity. However, as we will soon see, passed a few simple facts, things will be more complicated here.

\bigskip

We will be interested in the electric fields of various charge configurations, such as charged curves, surfaces and solid bodies. So, let us formulate:

\begin{definition}
The electric field of a charge configuration $L\subset\mathbb R^3$, with charge density function $\rho:L\to\mathbb R$, is the vector function
$$E(x)=K\int_L\frac{\rho(z)(x-z)}{||x-z||^3}\,dz$$
so that the force of $L$ applied to a charge $Q$ positioned at $x$ is given by $F=QE$.
\end{definition}

With this in hand, it is most convenient now to forget about the charges, and focus on the study of the corresponding electric fields $E$. These fields are by definition vector functions $E:\mathbb R^3\to\mathbb R^3$, with the convention that they take $\pm\infty$ values at the places where the charges are located, and intuitively, are best represented by their field lines:

\begin{definition}
The field lines of an electric field $E:\mathbb R^3\to\mathbb R^3$ are the oriented curves $\gamma\subset\mathbb R^3$ pointing at every point $x\in\mathbb R^3$ at the direction of the field, $E(x)\in\mathbb R^3$.
\end{definition}

As a basic example here, for one charge the field lines are the half-lines emanating from its position, oriented according to the sign of the charge:
$$\begin{matrix}
\nwarrow&\uparrow&\nearrow\\
\leftarrow&\oplus&\rightarrow\\
\swarrow&\downarrow&\searrow
\end{matrix}\qquad\qquad\qquad\qquad
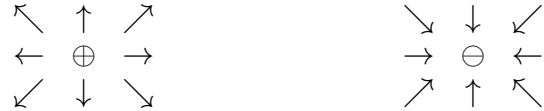
\begin{matrix}
\searrow&\downarrow&\swarrow\\
\rightarrow&\ominus&\leftarrow\\
\nearrow&\uparrow&\nwarrow
\end{matrix}$$

For two charges now, if these are of opposite signs, $+$ and $-$, you get a picture that you are very familiar with, namely that of the field lines of a bar magnet:
$$\begin{matrix}
\nearrow&\ \ \nearrow&\rightarrow&\rightarrow&\rightarrow&\rightarrow&\searrow\ \ \ &\!\!\!\searrow\\
\nwarrow&\!\!\!\uparrow&\nearrow&\rightarrow&\rightarrow&\searrow&\downarrow&\!\!\!\swarrow\\
\leftarrow&\!\!\!\oplus&\rightarrow&\rightarrow&\rightarrow&\rightarrow&\ominus&\!\!\!\leftarrow\\
\swarrow&\!\!\!\downarrow&\searrow&\rightarrow&\rightarrow&\nearrow&\uparrow&\!\!\!\nwarrow\\
\searrow&\ \ \searrow&\rightarrow&\rightarrow&\rightarrow&\rightarrow&\nearrow\ \ \ &\!\!\!\nearrow
\end{matrix}$$

If the charges are $+,+$ or $-,-$, you get something of similar type, but repulsive this time, with the field lines emanating from the charges being no longer shared:
$$\begin{matrix}
\leftarrow\ &\!\!\!\!\!\nwarrow&\nwarrow&&&\nearrow&\ \ \ \nearrow&\rightarrow\\
&\uparrow&\nearrow&&&\nwarrow&\uparrow&\\
\leftarrow&\oplus&\ &\ &\ \ \ \ \ \ \ &\ &\oplus&\rightarrow\\
&\downarrow&\searrow&&&\swarrow&\downarrow&\\
\leftarrow\ &\!\!\!\!\!\swarrow&\swarrow&&&\searrow&\ \ \ \searrow&\rightarrow
\end{matrix}$$

With this discussed, let us introduce now a key definition, as follows:

\begin{definition}
The flux of an electric field $E:\mathbb R^3\to\mathbb R^3$ through a surface $S\subset\mathbb R^3$, assumed to be oriented, is the quantity
$$\Phi_E(S)=\int_S<E(x),n(x)>dx$$
with $n(x)$ being unit vectors orthogonal to $S$, following the orientation of $S$. Intuitively, the flux measures the signed number of field lines crossing $S$.
\end{definition}

Here by orientation of $S$ we mean precisely the choice of unit vectors $n(x)$ as above, orthogonal to $S$, which must vary continuously with $x$. As a first illustration, we have:

\begin{proposition}
For a point charge $q\in\mathbb R$ at the center of a sphere $S$,
$$\Phi_E(S)=\frac{q}{\varepsilon_0}$$
where the constant is $\varepsilon_0=1/(4\pi K)$, independently of the radius of $S$.
\end{proposition}

\begin{proof}
Assuming that $S$ has radius $r$, we have the following computation:
\begin{eqnarray*}
\Phi_E(S)
&=&\int_S\left<\frac{Kqx}{r^3},\frac{x}{r}\right>dx\\
&=&\int_S\frac{Kq}{r^2}\,dx\\
&=&\frac{Kq}{r^2}\times 4\pi r^2\\
&=&4\pi Kq
\end{eqnarray*} 

Thus with $\varepsilon_0=1/(4\pi K)$ as above, we obtain the result.
\end{proof}

More generally now, we have the following result:

\index{flux}
\index{Jacobian}
\index{flux through sphere}
\index{enclosed charge}

\begin{theorem}
The flux of a field $E$ through a sphere $S$ is given by
$$\Phi_E(S)=\frac{Q_{enc}}{\varepsilon_0}$$
where $Q_{enc}$ is the total charge enclosed by $S$, and $\varepsilon_0=1/(4\pi K)$.
\end{theorem}

\begin{proof}
We can assume that we are dealing with a single charge, positioned inside the unit sphere $S$. Moreover, by rotating we can assume that our charge $q$ lies on the $Ox$ axis, that is, that we have $v=(r,0,0)$ with $r\geq0$, $r\neq1$. With $u=(x,y,z)$, we have: 
\begin{eqnarray*}
\Phi_E(S)
&=&\int_S<E(u),u>du\\
&=&\int_S\left<\frac{Kq(u-v)}{||u-v||^3},u\right>du\\
&=&Kq\int_S\frac{<u-v,u>}{||u-v||^3}\,du\\
&=&Kq\int_S\frac{1-<v,u>}{||u-v||^3}\,du\\
&=&Kq\int_S\frac{1-rx}{(1-2xr+r^2)^{3/2}}\,du
\end{eqnarray*}

By using now spherical coordinates, the above integral becomes:
\begin{eqnarray*}
\Phi_E(S)
&=&Kq\int_S\frac{1-rx}{(1-2xr+r^2)^{3/2}}\,du\\
&=&Kq\int_0^{2\pi}\int_0^\pi\frac{1-r\cos s}{(1-2r\cos s+r^2)^{3/2}}\cdot\sin s\,ds\,dt\\
&=&2\pi Kq\int_0^\pi\frac{(1-r\cos s)\sin s}{(1-2r\cos s+r^2)^{3/2}}\,ds\\
&=&\frac{q}{2\varepsilon_0}\int_0^\pi\frac{(1-r\cos s)\sin s}{(1-2r\cos s+r^2)^{3/2}}\,ds
\end{eqnarray*}

The point now is that the integral on the right can be computed with the change of variables $x=\cos s$. Indeed, we have $dx=-\sin s\,ds$, and we obtain:
\begin{eqnarray*}
\int_0^\pi\frac{(1-r\cos s)\sin s}{(1-2r\cos s+r^2)^{3/2}}\,ds
&=&\int_{-1}^1\frac{1-rx}{(1-2rx+r^2)^{3/2}}\,dx\\
&=&\left[\frac{x-r}{\sqrt{1-2rx+r^2}}\right]_{-1}^1\\
&=&\frac{1-r}{\sqrt{1-2r+r^2}}-\frac{-1-r}{\sqrt{1+2r+r^2}}\\
&=&\frac{1-r}{|1-r|}+1\\
&=&2\delta_{r<1}
\end{eqnarray*}

Thus, we are led to the conclusion in the statement.
\end{proof}

Even more generally now, we have the following result, due to Gauss:

\index{Gauss law}
\index{flux}
\index{charge enclosed}
\index{flux through surface}
\index{Feynman}

\begin{theorem}[Gauss law]
The flux of a field $E$ through a surface $S$ is given by
$$\Phi_E(S)=\frac{Q_{enc}}{\varepsilon_0}$$
where $Q_{enc}$ is the total charge enclosed by $S$, and $\varepsilon_0=1/(4\pi K)$.
\end{theorem}

\begin{proof}
This basically follows from Theorem 8.26, or even from Proposition 8.25, by adding to the results there a number of new ingredients, as follows:

\medskip

(1) Our first claim is that given a closed surface $S$, with no charges inside, the flux through it of any choice of external charges vanishes:
$$\Phi_E(S)=0$$

This claim is indeed supported by the intuitive interpretation of the flux, as corresponding to the signed number of field lines crossing $S$. Indeed, any field line entering as $+$ must exit somewhere as $-$, and vice versa, so when summing we get $0$.

\medskip

(2) In practice now, in order to prove this rigorously, there are several ways. A first argument, which is quite elementary, is the one used by Feynman in \cite{fe2}, based on the fact that, due to $F\sim 1/d^2$, local deformations of $S$ will leave invariant the flux, and so in the end we are left with a rotationally invariant surface, where the result is clear.

\medskip

(3) A second argument, which basically uses the same idea, but is perhaps a bit more robust, is by redoing the computations in the proof of Theorem 8.26, by assuming this time that the integration takes place on an arbitrary surface as follows:
$$S_\lambda=\left\{\lambda(u)u\Big|u\in S\right\}$$

And I will leave the computations here to you, as an interesting exercise.

\medskip

(4) But with this, we can finish by using a standard math trick. We can assume indeed that our system of charges is discrete, consisting of enclosed charges $q_1,\ldots,q_k\in\mathbb R$, and an exterior total charge $Q_{ext}$. Then, we can surround each of $q_1,\ldots,q_k$ by small disjoint spheres $U_1,\ldots,U_k$, chosen such that their interiors do not touch $S$, and we have:
\begin{eqnarray*}
\Phi_E(S)
&=&\Phi_E(S-\cup U_i)+\Phi_E(\cup U_i)\\
&=&\sum_i\Phi_E(U_i)\\
&=&\sum_i\frac{q_i}{\varepsilon_0}\\
&=&\frac{Q_{enc}}{\varepsilon_0}
\end{eqnarray*}

Thus, we are led to the conclusion in the statement. We will be actually back to the present theorem in a moment, with a second proof as well, even more conceptual.
\end{proof}

In relation now with our previous mathematics, we have the following result:

\begin{theorem}
Given an electric potential $E$, its divergence is given by
$$<\nabla,E>=\frac{\rho}{\varepsilon_0}$$
where $\rho$ denotes as usual the charge distribution. Also, we have
$$\nabla\times E=0$$
meaning that the curl of $E$ vanishes.
\end{theorem}

\begin{proof}
We have several assertions here, the idea being as follows:

\medskip

(1) The first formula, called Gauss law in differential form, follows from:
\begin{eqnarray*}
\int_B<\nabla,E>
&=&\int_S<E(x),n(x)>dx\\
&=&\Phi_E(S)\\
&=&\frac{Q_{enc}}{\varepsilon_0}\\
&=&\int_B\frac{\rho}{\varepsilon_0}
\end{eqnarray*}

Now since this must hold for any $B$, this gives the formula in the statement.

\medskip

(2) Regarding the curl, by discretizing and linearity we can assume that we are dealing with a single charge $q$, positioned at $0$. We have, by using spherical coordinates $r,s,t$:
\begin{eqnarray*}
\int_a^b<E(x),dx>
&=&\int_a^b\left<\frac{Kqx}{||x||^3},dx\right>\\
&=&\int_a^b\left<\frac{Kq}{r^2}\cdot\frac{x}{||x||},dx\right>\\
&=&\int_a^b\frac{Kq}{r^2}\,dr\\
&=&\left[-\frac{Kq}{r}\right]_a^b\\
&=&Kq\left(\frac{1}{r_a}-\frac{1}{r_b}\right)
\end{eqnarray*}

In particular the integral of $E$ over any closed loop vanishes, and by using now the Stokes formula, we conclude that the curl of $E$ vanishes, as stated.
\end{proof}

So long for electrostatics, which provide a good motivation and illustration for our mathematics. When upgrading to electrodynamics, things become even more interesting, because our technology can be used in order to understand the Maxwell equations:

\begin{theorem}
Electrodynamics is governed by the formulae
$$<\nabla,E>=\frac{\rho}{\varepsilon_0}\quad,\quad 
<\nabla,B>=0$$
$$\nabla\times E=-\dot{B}\quad,\quad 
\nabla\times B=\mu_0J+\mu_0\varepsilon_0\dot{E}$$
called Maxwell equations.
\end{theorem}

\begin{proof}
This is something fundamental, the idea being as follows:

\medskip

(1) To start with, electrodynamics is the science of moving electrical charges. And this is something quite complicated, because moving charges produce magnetism, and with this being visible when putting together two electric wires, which will attract or repel, depending on orientation. Thus, in contrast with classical mechanics, where static or dynamic problems are described by a unique field, the gravitational one, in electrodynamics we have two fields, namely the electric field $E$, and the magnetic field $B$.

\medskip

(2) Fortunately, there is a full set of equations relating $E$ and $B$, those above. To be more precise, the first formula is the Gauss law, $\rho$ being the charge, and $\varepsilon_0$ being a constant, and with this Gauss law more or less replacing the Coulomb law from electrostatics. The second formula is something basic, and anonymous. The third formula is the Faraday law. As for the fourth formula, this is the Amp\`ere law, as modified by Maxwell, with $J$ being the volume current density, and $\mu_0$ being a constant.
\end{proof}

However, the above is not all. Quite remarkably, the constants $\mu_0,\varepsilon_0$ are related by the formula $\mu_0\varepsilon_0=1/c^2$, with $c$ being the speed of light. And, more on this later.

\section*{8e. Exercises}

We had an exciting geometric chapter here, and as exercises on this, we have:

\begin{exercise}
Clarify all the details in the classification of the conics.
\end{exercise}

\begin{exercise}
Learn about the stereographic projection, and cartography in general.
\end{exercise}

\begin{exercise}
Compute the lengths of some curves, of your choice.
\end{exercise}

\begin{exercise}
Learn more about conservative forces, and their properties.
\end{exercise}

\begin{exercise}
Compute the areas of some surfaces, of your choice.
\end{exercise}

\begin{exercise}
Clarify all the details in the proof of the Stokes formula.
\end{exercise}

\begin{exercise}
Try to compute the field lines, for simple charge configurations.
\end{exercise}

\begin{exercise}
Learn more about the Gauss law, and the Maxwell equations.
\end{exercise}

As bonus exercise, read some systematic algebraic and differential geometry.

\part{Function spaces}

\ \vskip50mm

\begin{center}
{\em Wanted man in California

Wanted man in Buffalo

Wanted man in Kansas City

Wanted man in Ohio}
\end{center}

\chapter{Banach spaces}

\section*{9a. Normed spaces}

Welcome to function space theory, also known as functional analysis. Although, for being fully honest, the basics here, which are quite algebraic, rather deserve the name ``functional algebra''. But do not worry, we will keep things as analytic as possible.

\bigskip

The idea is that various types of functions form various types of infinite dimensional complex spaces, that we can study with the usual methods from linear algebra, complemented for dealing with $\infty$ by our favorite logic weapon, which is the Zorn lemma.

\bigskip

Let us start with something very general, as follows:

\begin{definition}
A normed space is a complex vector space $V$, which can be finite or infinite dimensional, together with a map
$$||.||:V\to\mathbb R_+$$
called norm, subject to the following conditions:
\begin{enumerate}
\item $||x||=0$ implies $x=0$.

\item $||\lambda x||=|\lambda|\cdot||x||$, for any $x\in V$, and $\lambda\in\mathbb C$.

\item $||x+y||\leq||x||+||y||$, for any $x,y\in V$.
\end{enumerate} 
\end{definition}

As a basic example here, which is finite dimensional, we have the space $V=\mathbb C^N$, with the norm on it being the usual length of the vectors, namely:
$$||x||=\sqrt{\sum_i|x_i|^2}$$

Indeed, for this space (1) is clear, (2) is clear too, and (3) is something well-known, which is equivalent to the triangle inequality in $\mathbb C^N$, and which can be deduced from the Cauchy-Schwarz inequality. More on this, with some generalizations, in a moment.

\bigskip

Getting back now to the general case, we have the following result:

\begin{proposition}
Any normed vector space $V$ is a metric space, with
$$d(x,y)=||x-y||$$
as distance. If this metric space is complete, we say that $V$ is a Banach space.
\end{proposition}

\begin{proof}
This follows from the definition of the metric spaces, as follows:

\medskip

(1) The first distance axiom, $d(x,y)\geq0$, and $d(x,y)=0$ precisely when $x=y$, follows from the fact that the norm takes values in $\mathbb R_+$, and from $||x||=0\implies x=0$.

\medskip

(2) The second distance axiom, which is the symmetry one, $d(x,y)=d(y,x)$, follows from our condition $||\lambda x||=|\lambda|\cdot||x||$, with $\lambda=-1$.

\medskip

(3) As for the third distance axiom, which is the triangle inequality $d(x,y)\leq d(x,z)+d(y,z)$, this follows from our third norm axiom, namely $||x+y||\leq||x||+||y||$.
\end{proof}

Very nice all this, and it is possible to develop some general theory here, but before everything, however, we need more examples, besides $\mathbb C^N$ with its usual norm. 

\bigskip

However, these further examples are actually quite tricky to construct, needing some inequality know-how. Let us start with a very basic result, as follows:

\index{convex function}
\index{concave function}
\index{Jensen inequality}

\begin{theorem}[Jensen]
Given a convex function $f:\mathbb R\to\mathbb R$, we have the following inequality, for any $x_1,\ldots,x_N\in\mathbb R$, and any $\lambda_1,\ldots,\lambda_N>0$ summing up to $1$,
$$f(\lambda_1x_1+\ldots+\lambda_Nx_N)\leq\lambda_1f(x_1)+\ldots+\lambda_Nx_N$$
with equality when $x_1=\ldots=x_N$. In particular, by taking the weights $\lambda_i$ to be all equal, we obtain the following inequality, valid for any $x_1,\ldots,x_N\in\mathbb R$,
$$f\left(\frac{x_1+\ldots+x_N}{N}\right)\leq\frac{f(x_1)+\ldots+f(x_N)}{N}$$
and once again with equality when $x_1=\ldots=x_N$. We have a similar statement holds for the concave functions, with all the inequalities being reversed.
\end{theorem}

\begin{proof}
This is indeed something quite routine, the idea being as follows:

\medskip

(1) First, we can talk about convex functions in a usual, intuitive way, with this meaning by definition that the following inequality must be satisfied:
$$f\left(\frac{x+y}{2}\right)\leq\frac{f(x)+f(y)}{2}$$

(2) But this means, via a simple argument, by approximating numbers $t\in[0,1]$ by sums of powers $2^{-k}$, that for any $t\in[0,1]$ we must have:
$$f(tx+(1-t)y)\leq tf(x)+(1-t)f(y)$$

Alternatively, via yet another simple argument, this time by doing some geometry with triangles, this means that we must have:
$$f\left(\frac{x_1+\ldots+x_N}{N}\right)\leq\frac{f(x_1)+\ldots+f(x_N)}{N}$$

But then, again alternatively, by combining the above two simple arguments, the following must happen, for any $\lambda_1,\ldots,\lambda_N>0$ summing up to $1$:
$$f(\lambda_1x_1+\ldots+\lambda_Nx_N)\leq\lambda_1f(x_1)+\ldots+\lambda_Nx_N$$

(3) Summarizing, all our Jensen inequalities, at $N=2$ and at $N\in\mathbb N$ arbitrary, are equivalent. The point now is that, if we look at what the first Jensen inequality, that we took as definition for the convexity, means, this is simply equivalent to:
$$f''(x)\geq0$$

(4) Thus, we are led to the conclusions in the statement, regarding the convex functions. As for the concave functions, the proof here is similar. Alternatively, we can say that $f$ is concave precisely when $-f$ is convex, and get the results from what we have.
\end{proof}

As a basic application of the Jensen inequality, we have:

\index{H\"older inequality}
\index{Cauchy-Schwarz}

\begin{proposition}
For $p\in(1,\infty)$ we have the following inequality,
$$\left|\frac{x_1+\ldots+x_N}{N}\right|^p\leq\frac{|x_1|^p+\ldots+|x_N|^p}{N}$$
and for $p\in(0,1)$ we have the following reverse inequality,
$$\left|\frac{x_1+\ldots+x_N}{N}\right|^p\geq\frac{|x_1|^p+\ldots+|x_N|^p}{N}$$
with in both cases equality precisely when $|x_1|=\ldots=|x_N|$.
\end{proposition}

\begin{proof}
This follows indeed from Theorem 9.3, because we have:
$$(x^p)''=p(p-1)x^{p-2}$$

Thus $x^p$ is convex for $p>1$ and concave for $p<1$, which gives the results.
\end{proof}

Observe that, more generally, Theorem 9.3 applied with $f(x)=x^p$ with $p\in(1,\infty)$ shows that we have the following inequality, whenever $\lambda_i>0$ sum up to 1:
$$(\lambda_1x_1+\ldots+\lambda_Nx_N)^p\leq\lambda_1x_1^p+\ldots+\lambda_Nx_N^p$$

For exponents $p\in(0,1)$ we have the same inequality, reversed, namely:
$$(\lambda_1x_1+\ldots+\lambda_Nx_N)^p\geq\lambda_1x_1^p+\ldots+\lambda_Nx_N^p$$

Observe also that at $p=2$ we obtain as particular case of the above inequality the Cauchy-Schwarz inequality, or rather something equivalent to it, namely:
$$\left(\frac{x_1+\ldots+x_N}{N}\right)^2\leq\frac{x_1^2+\ldots+x_N^2}{N}$$

We will be back to this later on in this book, when talking scalars products and Hilbert spaces, with some more conceptual proofs for such inequalities.

\bigskip

As another basic application of the Jensen inequality, we have:

\begin{theorem}[Young]
We have the following inequality,
$$ab\leq \frac{a^p}{p}+\frac{b^q}{q}$$
valid for any $a,b\geq0$, and any exponents $p,q>1$ satisfying $\frac{1}{p}+\frac{1}{q}=1$. 
\end{theorem}

\begin{proof}
We use the logarithm function, which is concave on $(0,\infty)$, due to:
$$(\log x)''=\left(-\frac{1}{x}\right)'=-\frac{1}{x^2}$$

Thus we can apply the Jensen inequality, and we obtain in this way:
\begin{eqnarray*}
\log\left(\frac{a^p}{p}+\frac{b^q}{q}\right)
&\geq&\frac{\log(a^p)}{p}+\frac{\log(b^q)}{q}\\
&=&\log(a)+\log(b)\\
&=&\log(ab)
\end{eqnarray*}

Now by exponentiating, we obtain the Young inequality.
\end{proof}

Observe that for the simplest exponents, namely $p=q=2$, the Young inequality gives something which is trivial, but is very useful and basic, namely:
$$ab\leq\frac{a^2+b^2}{2}$$

In general, the Young inequality is something non-trivial, and the idea with it is that ``when stuck with a problem, and with $ab\leq\frac{a^2+b^2}{2}$ not working, try Young''. We will be back to this general principle, later in this book, with some illustrations.

\bigskip

Moving forward now, as a consequence of the Young inequality, we have:

\index{H\"older inequality}

\begin{theorem}[H\"older]
Assuming that $p,q\geq 1$ are conjugate, in the sense that 
$$\frac{1}{p}+\frac{1}{q}=1$$
we have the following inequality, valid for any two vectors $x,y\in\mathbb C^N$,
$$\sum_i|x_iy_i|\leq\left(\sum_i|x_i|^p\right)^{1/p}\left(\sum_i|y_i|^q\right)^{1/q}$$
with the convention that an $\infty$ exponent produces a $\max|x_i|$ quantity.
\end{theorem}

\begin{proof}
This is something very standard, the idea being as follows:

\medskip

(1) Assume first that we are dealing with finite exponents, $p,q\in(1,\infty)$. By linearity we can assume that $x,y$ are normalized, in the following way:
$$\sum_i|x_i|^p=\sum_i|y_i|^q=1$$

In this case, we want to prove that the following inequality holds:
$$\sum_i|x_iy_i|\leq1$$

For this purpose, we use the Young inequality, which gives, for any $i$:
$$|x_iy_i|\leq\frac{|x_i|^p}{p}+\frac{|y_i|^q}{q}$$

By summing now over $i=1,\ldots,N$, we obtain from this, as desired:
\begin{eqnarray*}
\sum_i|x_iy_i|
&\leq&\sum_i\frac{|x_i|^p}{p}+\sum_i\frac{|y_i|^q}{q}\\
&=&\frac{1}{p}+\frac{1}{q}\\
&=&1
\end{eqnarray*}

(2) In the case $p=1$ and $q=\infty$, or vice versa, the inequality holds too, trivially, with the convention that an $\infty$ exponent produces a max quantity, according to:
$$\lim_{p\to\infty}\left(\sum_i|x_i|^p\right)^{1/p}=\max|x_i|$$

Thus, we are led to the conclusion in the statement.
\end{proof}

As a consequence now of the H\"older inequality, we have:

\index{Minkowski inequality}

\begin{theorem}[Minkowski]
Assuming $p\in[1,\infty]$, we have the inequality
$$\left(\sum_i|x_i+y_i|^p\right)^{1/p}\leq \left(\sum_i|x_i|^p\right)^{1/p}+\left(\sum_i|y_i|^p\right)^{1/p}$$
for any two vectors $x,y\in\mathbb C^N$, with our usual conventions at $p=\infty$.
\end{theorem}

\begin{proof}
We have indeed the following estimate, using the H\"older inequality, and the conjugate exponent $q\in[1,\infty]$, given by $1/p+1/q=1$:
\begin{eqnarray*}
\sum_i|x_i+y_i|^p
&=&\sum_i|x_i+y_i|\cdot|x_i+y_i|^{p-1}\\
&\leq&\sum_i|x_i|\cdot|x_i+y_i|^{p-1}+\sum_i|y_i|\cdot|x_i+y_i|^{p-1}\\
&\leq&\left(\sum_i|x_i|^p\right)^{1/p}\left(\sum_i|x_i+y_i|^{(p-1)q}\right)^{1/q}\\
&+&\left(\sum_i|y_i|^p\right)^{1/p}\left(\sum_i|x_i+y_i|^{(p-1)q}\right)^{1/q}\\
&=&\left[\left(\sum_i|x_i|^p\right)^{1/p}+\left(\sum_i|y_i|^p\right)^{1/p}\right]\left(\sum_i|x_i+y_i|^p\right)^{1-1/p}
\end{eqnarray*}

Here we have used the following fact, at the end:
$$\frac{1}{p}+\frac{1}{q}=1\implies\frac{1}{q}=\frac{p-1}{p}\implies(p-1)q=p$$

Now by dividing both sides by the last quantity at the end, we obtain:
$$\left(\sum_i|x_i+y_i|^p\right)^{1/p}\leq \left(\sum_i|x_i|^p\right)^{1/p}+\left(\sum_i|y_i|^p\right)^{1/p}$$

Thus, we are led to the conclusion in the statement.
\end{proof}

Good news, done with inequalities, and as a consequence of the above results, and more specifically of the Minkowski inequality obtained above, we can formulate:

\index{p-norm}
\index{normed space}

\begin{theorem}
Given an exponent $p\in[1,\infty]$, the formula
$$||x||_p=\left(\sum_i|x_i|^p\right)^{1/p}$$
with usual conventions at $p=\infty$, defines a norm on $\mathbb C^N$, making it a Banach space.
\end{theorem}

\begin{proof}
Here the normed space assertion follows from the Minkowski inequality, established above, and the Banach space assertion is trivial, because our space being finite dimensional, by standard linear algebra all the Cauchy sequences converge.
\end{proof}

We will see later in this chapter that Theorem 9.8 has several generalizations, with some of them dealing with functions. Before that, however, you might have another question regarding Theorem 9.8, namely, what is the underlying geometry there.

\bigskip

So, let us discuss this latter question. We will actually see that, with a bit of geometric know-how, not only we will reach to a better understanding of the $p$-norm on $\mathbb C^N$, but we will also be able to recover the H\"older inequality for it, without any computation. 

\bigskip

It is convenient to talk about $\mathbb R^N$ instead of $\mathbb C^N$. Let us start with:

\begin{proposition}
For any exponent $p>1$, the following set
$$S_p=\left\{x\in\mathbb R^N\Big|\sum_i|x_i|^p=1\right\}$$
is a smooth submanifold of $\mathbb R^N$.
\end{proposition}

\begin{proof}
We know from chapter 8 that the unit sphere in $\mathbb R^N$ is a smooth manifold. In our terms, this solves our problem at $p=2$, because this unit sphere is:
$$S_2=\left\{x\in\mathbb R^N\Big|\sum_ix_i^2=1\right\}$$

Now observe that we have a bijection $S_p\simeq S_2$, at least on the part where all the coordinates are positive, $x_i>0$, given by the following function:
$$x_i\to x_i^{2/p}$$

Thus we obtain that $S_p$ is indeed a manifold, as claimed.
\end{proof}

We already know that the manifold $S_p$ constructed above is the unit sphere, in the case $p=2$. In order to have a better geometric picture of what is going on, in general, observe that $S_p$ can be constructed as well at $p=1$, as follows:
$$S_1=\left\{x\in\mathbb R^N\Big|\sum_i|x_i|=1\right\}$$

However, this is no longer a manifold, as we can see for instance at $N=2$, where we obtain a square. Now observe that we can talk as well about $p=\infty$, as follows:
$$S_\infty=\left\{x\in\mathbb R^N\Big|\sup_i|x_i|=1\right\}$$

This letter set is no longer a manifold either, as we can see for instance at $N=2$, where we obtain a again a square, containing the previous square, the one at $p=1$.

\bigskip

With these limiting constructions in hand, we can now have a better geometric picture of what is going on, in the general context of Proposition 9.9. Indeed, let us draw, at $N=2$ for simplifying, our sets $S_p$ at the values $p=1,2,\infty$ of the exponent:
$$
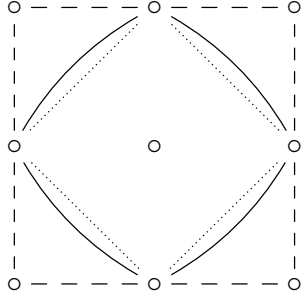
\xymatrix@R=40pt@C=40pt{
\circ\ar@{--}[r]\ar@{--}[d]&\circ\ar@{--}[r]\ar@{.}[dr]\ar@{.}[dl]\ar@{-}@/^/[dr]&\circ\ar@{--}[d]\\
\circ\ar@{--}[d]\ar@{.}[dr]\ar@{-}@/^/[ur]&\circ&\circ\ar@{--}[d]\ar@{.}[dl]\ar@{-}@/^/[dl]\\
\circ\ar@{--}[r]&\circ\ar@{--}[r]\ar@{-}@/^/[ul]&\circ
}$$

We can see that what we have is a small square, at $p=1$, becoming smooth and inflating towards the circle, in the parameter range $p\in(1,2]$, and then further inflating, in the parameter range $p\in[2,\infty)$, towards the big square appearing at $p=\infty$.

\bigskip

Now let us get to our second objective, proving H\"older, without computations. We will use Lagrange multipliers, whose basic theory can be summarized as follows:

\begin{fact}
In order for a function $f:X\to\mathbb R$ defined on a manifold $X$ to have a local extremum at $x\in X$, we must have, as usual 
$$f'(x)=0$$
but with this taking into account the fact that the equations defining the manifold count as well as ``zero'', and so must be incorporated into the formula $f'(x)=0$. 
\end{fact}

To be more precise, we must have a formula as follows, with $g_i$ being the constraint functions for $X$, and with $\lambda_i\in\mathbb R$ being certain scalars, called Lagrange multipliers:
$$f'(x)=\sum_i\lambda_ig_i'(x)$$

With these preliminaries in hand, we can formulate our result, as follows:

\begin{theorem}
The local extrema over $S_p$ of the function
$$f(x)=\sum_ix_iy_i$$ 
can be computed by using Lagrange multipliers, and this gives
$$\left|\sum_ix_iy_i\right|\leq\left(\sum_i|x_i|^p\right)^{1/p}\left(\sum_i|y_i|^q\right)^{1/q}$$
with $1/p+1/q=1$, that is, the H\"older inequality, with a purely geometric proof.
\end{theorem}

\begin{proof}
We can restrict the attention to the case where all the coordinates are positive, $x_i>0$ and  $y_i>0$. The derivative of the function in the statement is:
$$f'(x)=(y_1,\ldots,y_N)$$

On the other hand, we know that the manifold $S_p$ appears by definition as the set of zeroes of the function $\varphi(x)=\sum_ix_i^p-1$, having derivative as follows:
$$\varphi'(x)=p(x_1^{p-1},\ldots,x_N^{p-1})$$

Thus, by using Lagrange multipliers, the critical points of $f$ must satisfy:
$$(y_1,\ldots,y_N)\sim(x_1^{p-1},\ldots,x_N^{p-1})$$

In other words, the critical points must satisfy $x_i=\lambda y_i^{1/(p-1)}$, for some $\lambda>0$, and by using now $\sum_ix_i^p=1$ we can compute the precise value of $\lambda$, and we get:
$$\lambda=\left(\sum_iy_i^{p/(p-1)}\right)^{-1/p}$$ 

Now let us see what this means. Since the critical point is unique, this must be a maximum of our function, and we conclude that for any $x\in S_p$, we have:
\begin{eqnarray*}
\sum_ix_iy_i
&\leq&\sum_i\lambda y_i^{1/(p-1)}\cdot y_i\\
&=&\left(\sum_iy_i^{p/(p-1)}\right)^{1-1/p}\\
&=&\left(\sum_iy_i^q\right)^{1/q}
\end{eqnarray*}

Thus we have H\"older, and the general case follows from this, by rescaling.
\end{proof}

\section*{9b. Banach spaces}

Very nice all the above, but you might wonder at this point, what is the relation of all this with functions. In answer, Theorem 9.8 can be reformulated as follows:

\begin{theorem}
Given an exponent $p\in[1,\infty]$, the formula
$$||f||_p=\left(\int|f(x)|^p\right)^{1/p}$$
with usual conventions at $p=\infty$, defines a norm on the space of functions 
$$f:\{1,\ldots,N\}\to\mathbb C$$
making it a Banach space.
\end{theorem}

\begin{proof}
This is a just fancy reformulation of Theorem 9.8, by using the fact that the space formed by the functions $f:\{1,\ldots,N\}\to\mathbb C$ is canonically isomorphic to $\mathbb C^N$, in the obvious way, and by replacing the sums from the $\mathbb C^N$ context with integrals with respect to the counting measure on $\{1,\ldots,N\}$, in the function context.
\end{proof}

Moving now towards infinite dimensions and more standard analysis, the idea will be that of extending Theorem 9.12 to the arbitrary measured spaces. Let us start with:

\index{p-norm}
\index{normed space}

\begin{theorem}
Given an exponent $p\in[1,\infty]$, the formula
$$||x||_p=\left(\sum_i|x_i|^p\right)^{1/p}$$
with usual conventions at $p=\infty$, defines a norm on the space of sequences
$$l^p=\left\{(x_i)_{i\in\mathbb N}\Big|\sum_i|x_i|^p<\infty\right\}$$
making it a Banach space.
\end{theorem}

\begin{proof}
As before with the finite sequences, the normed space assertion follows from the Minkowski inequality, established above, which extends without problems to the case of the infinite sequences, and with the Banach space assertion being clear too.
\end{proof}

We can unify and generalize what we have, in the following way:

\begin{theorem}
Given a discrete measured space $X$, and an exponent $p\in[1,\infty]$,
$$||f||_p=\left(\int_X|f(x)|^p\right)^{1/p}$$
with usual conventions at $p=\infty$, defines a norm on the space of functions 
$$l^p(X)=\left\{f:X\to\mathbb C\Big|\int_X|f(x)|^p<\infty\right\}$$
making it a Banach space.
\end{theorem}

\begin{proof}
This is just a fancy reformulation of what we have:

\medskip

(1) The case where $X$ is finite corresponds to Theorem 9.12.

\medskip

(2) The case where $X$ is countable corresponds to Theorem 9.13.

\medskip

(3) Finally, the case where $X$ is uncountable is easy to deal with too, by using the same arguments as in the countable case.
\end{proof}

Before going further, let us mention that the $l^p$ spaces above are not the only interesting Banach spaces of sequences. Indeed, we have as well the following result:

\begin{theorem}
The following space of complex sequences,
$$c_0=\left\{(x_i)_{i\in\mathbb N}\Big|x_i\to0\right\}$$
is a Banach space, with the sup norm, and we have $c_0\subset l^\infty$, proper inclusion.
\end{theorem}

\begin{proof}
We can define indeed $c_0$ as above, as being the space of sequences $x_i\in\mathbb C$ satisfying $x_i\to0$, with the sup norm. We have then an inclusion as follows:
$$c_0\subset l^\infty$$

It is then clear that $c_0$ is closed, so it is indeed a Banach space, as claimed. 
\end{proof}

Now back to the $l^p$ spaces, in order to further extend the above results, to the case of the arbitrary measured spaces $X$, which are not necessarily discrete, let us start with:

\index{Minkowski inequality}
\index{H\"older inequality}

\begin{theorem}
Given two functions $f,g:X\to\mathbb C$ and an exponent $p\geq1$, we have
$$\left(\int_X|f+g|^p\right)^{1/p}\leq\left(\int_X|f|^p\right)^{1/p}+\left(\int_X|g|^p\right)^{1/p}$$
called Minkowski inequality. Also, assuming that $p,q\geq1$ satisfy $1/p+1/q=1$, we have
$$\int_X|fg|\leq \left(\int_X|f|^p\right)^{1/p}\left(\int_X|g|^q\right)^{1/q}$$
called H\"older inequality. These inequalities hold as well for $\infty$ values of the exponents.
\end{theorem}

\begin{proof}
This is very standard, exactly as in the case of sequences, finite or not, but since the above inequalities are really very general and final, here are the details:

\medskip

(1) Let us first prove H\"older, in the case of finite exponents, $p,q\in(1,\infty)$. By linearity we can assume that $f,g$ are normalized, in the following way:
$$\int_X|f|^p=\int_X|g|^q=1$$

In this case, we want to prove that the following inequality holds:
$$\int_X|fg|\leq1$$

For this purpose, we use the Young inequality, which gives, for any $x\in X$:
$$|f(x)g(x)|\leq\frac{|f(x)|^p}{p}+\frac{|g(x)|^q}{q}$$

By integrating now over $x\in X$, we obtain from this, as desired:
\begin{eqnarray*}
\int_X|fg|
&\leq&\int_X\frac{|f(x)|^p}{p}+\int_X\frac{|g(x)|^q}{q}\\
&=&\frac{1}{p}+\frac{1}{q}\\
&=&1
\end{eqnarray*}

(2) Let us prove now Minkowski, again in the finite exponent case, $p\in(1,\infty)$. We have the following estimate, using the H\"older inequality, and the conjugate exponent:
\begin{eqnarray*}
\int_X|f+g|^p
&=&\int_X|f+g|\cdot|f+g|^{p-1}\\
&\leq&\int_X|f|\cdot|f+g|^{p-1}+\int_X|g|\cdot|f+g|^{p-1}\\
&\leq&\left(\int_X|f|^p\right)^{1/p}\left(\int_X|f+g|^{(p-1)q}\right)^{1/q}\\
&+&\left(\int_X|g|^p\right)^{1/p}\left(\int_X|f+g|^{(p-1)q}\right)^{1/q}\\
&=&\left[\left(\int|f|^p\right)^{1/p}+\left(\int_X|g|^p\right)^{1/p}\right]\left(\int_X|f+g|^p\right)^{1-1/p}
\end{eqnarray*}

Thus, we are led to the Minkowski inequality in the statement.

\medskip

(3) Finally, in the infinite exponent cases we have similar results, with the convention that an $\infty$ exponent produces an essential supremum, according to:
$$\lim_{p\to\infty}\left(\int_X|f|^p\right)^{1/p}=\ {\rm ess\;sup}|f|$$

Thus, we are led to the conclusion in the statement.
\end{proof}

We can now extend Theorem 9.14, into something very general, as follows:

\index{equal almost everywhere}
\index{function space}
\index{p-norm}
\index{normed space}

\begin{theorem}
Given a measured space $X$, and $p\in[1,\infty]$, the following space, with the convention that functions are identified up to equality almost everywhere,
$$L^p(X)=\left\{f:X\to\mathbb C\Big|\int_I|f(x)|^pdx<\infty\right\}$$
is a vector space, and the following quantity
$$||f||_p=\left(\int_X|f(x)|^p\right)^{1/p}$$
is a norm on it, making it a Banach space.
\end{theorem}

\begin{proof}
This follows indeed from Theorem 9.16, with due attention to the null sets, and this because of the first normed space axiom, namely:
$$||x||=0\implies x=0$$

To be more precise, in order for this axiom to hold, we must identify the functions up to equality almost everywhere, as indicated in the statement.
\end{proof}

Summarizing, very nice all this. So, we have our examples of Banach spaces, which look definitely interesting, and related to analysis. In the remainder of this chapter we will develop some general Banach space theory, and apply it to the above $L^p$ spaces.

\section*{9c. Linear forms}

Getting now to work, as a first result about the abstract normed spaces, we would like to talk about the linear maps $T:V\to W$, which can be thought of as being some kind of infinite matrices, when $V,W$ are infinite dimensional. We first have here:

\begin{proposition}
For a linear map $T:V\to W$, the following conditions are equivalent, and if they hold, we say that $T$ is bounded:
\begin{enumerate}
\item $T$ is continuous.

\item $T$ is continuous at $0$.

\item $T$ maps the unit ball of $V$ into something bounded.

\item $T$ is bounded, in the sense that $||T||=\sup_{||x||=1}||Tx||$ is finite.
\end{enumerate}
\end{proposition}

\begin{proof}
This is something elementary, the idea being as follows:

\medskip

$(1)\iff(2)$ This is indeed clear from definitions, by using the linearity of $T$, and performing various rescalings.

\medskip

$(2)\iff(3)$ This is clear from definitions too, again by using the linearity of $T$, and performing various rescalings.

\medskip

$(3)\iff(4)$ This is clear from definitions too, with the number $||T||$ needed in (4) being the bound coming from (3).
\end{proof}

With the above result in hand, we can now formulate:

\begin{theorem}
Given two Banach spaces $V,W$, the bounded linear maps 
$$T:V\to W$$
form a linear space $B(V,W)$, on which the following quantity is a norm,
$$||T||=\sup_{||x||=1}||Tx||$$
making $B(V,W)$ a Banach space. When $V=W$, we obtain a Banach algebra.
\end{theorem}

\begin{proof}
All this is very standard, and in the case $V=W$, the proof goes as follows:

\medskip

(1) The fact that we have indeed an algebra, satisfying the product condition in the statement, follows from the following estimates, which are all elementary:
$$||S+T||\leq||S||+||T||$$
$$||\lambda T||=|\lambda|\cdot||T||$$
$$||ST||\leq||S||\cdot||T||$$

(2) Regarding now the last assertion, if $\{T_n\}\subset B(V)$ is Cauchy then $\{T_nx\}$ is Cauchy for any $x\in V$, so we can define the limit $T=\lim_{n\to\infty}T_n$ by setting:
$$Tx=\lim_{n\to\infty}T_nx$$

Let us first check that the application $x\to Tx$ is linear. We have:
\begin{eqnarray*}
T(x+y)
&=&\lim_{n\to\infty}T_n(x+y)\\
&=&\lim_{n\to\infty}T_n(x)+T_n(y)\\
&=&\lim_{n\to\infty}T_n(x)+\lim_{n\to\infty}T_n(y)\\
&=&T(x)+T(y)
\end{eqnarray*}

Similarly, we have as well the following computation:
\begin{eqnarray*}
T(\lambda x)
&=&\lim_{n\to\infty}T_n(\lambda x)\\
&=&\lambda\lim_{n\to\infty}T_n(x)\\
&=&\lambda T(x)
\end{eqnarray*}

Thus we have a linear map $T:A\to A$. It remains to prove that we have $T\in B(V)$, and that we have $T_n\to T$ in norm. For this purpose, observe that we have:
\begin{eqnarray*}
&&||T_n-T_m||\leq\varepsilon\ ,\ \forall n,m\geq N\\
&\implies&||T_nx-T_mx||\leq\varepsilon\ ,\ \forall||x||=1\ ,\ \forall n,m\geq N\\
&\implies&||T_nx-Tx||\leq\varepsilon\ ,\ \forall||x||=1\ ,\ \forall n\geq N\\
&\implies&||T_Nx-Tx||\leq\varepsilon\ ,\ \forall||x||=1\\
&\implies&||T_N-T||\leq\varepsilon
\end{eqnarray*}

As a first consequence, we obtain $T\in B(V)$, because we have:
\begin{eqnarray*}
||T||
&=&||T_N+(T-T_N)||\\
&\leq&||T_N||+||T-T_N||\\
&\leq&||T_N||+\varepsilon\\
&<&\infty
\end{eqnarray*}

As a second consequence, we obtain $T_N\to T$ in norm, and we are done.
\end{proof}

As a basic example for the above construction, in the case where both our spaces are finite dimensional, $V=\mathbb C^N$ and $W=\mathbb C^M$, with $N,M<\infty$, we obtain a matrix space:
$$B(\mathbb C^N,\mathbb C^M)=M_{M\times N}(\mathbb C)$$

More on this later. On the other hand, of particular interest is as well the case $W=\mathbb C$ of the above construction, which leads to the following result:

\index{dual Banach space}

\begin{theorem}
Given a Banach space $V$, its dual space, constructed as
$$V^*=\Big\{f:V\to\mathbb C,\ {\rm linear\ and\ bounded}\Big\}$$
is a Banach space too, with norm given by:
$$||f||=\sup_{||x||=1}|f(x)|$$
When $V$ is finite dimensional, we have $V\simeq V^*$.
\end{theorem}

\begin{proof}
This is clear indeed from Theorem 9.19, because we have:
$$V^*=B(V,\mathbb C)$$

Thus, we are led to the conclusions in the statement.
\end{proof}

In order to better understand the linear forms, we will need:

\begin{theorem}[Hahn-Banach]
Given a Banach space $V$, the following happen:
\begin{enumerate}
\item Given $x\in V-\{0\}$, there exists $f\in V^*$ with $f(x)\neq0$.

\item Given a subspace $W\subset V$, any $f\in W^*$ extends into a $\tilde{f}\in V^*$, of same norm.
\end{enumerate}
\end{theorem}

\begin{proof}
This is something quite tricky, the idea being as follows:

\medskip

(1) As a first observation, the assertion (1) is a particular case of the assertion (2). Indeed, given a nonzero vector $x\in V-\{0\}$, we can consider the linear space $W=\mathbb Cx$, and construct a linear form $f\in W^*$ by the following formula:
$$f(\lambda x)=\lambda$$

But with this done, any extension $\tilde{f}\in V^*$ of this linear form $f\in W^*$, as provided by (2), does the job for (1), and this because we have $\tilde{f}(x)=f(x)=1$.

\medskip

(2) As a comment now, you might wonder why keeping both (1,2) in the statement of the theorem. In answer, the Hahn-Banach theorem is (2), while (1) is the main consequence of this theorem, which is something very useful, in practice. So, with our theorem formulated as above, we have there all we need to know, about Hahn-Banach.

\medskip

(3) As a second remark now, assuming that we are in finite dimensions, given a subspace $W\subset V$ as in (2), we can always find a direct sum decomposition, as follows:
$$V=W\oplus U$$

But once we have such a decomposition, given a linear form $f\in W^*$ as in (2), we can define an extension $\tilde{f}\in V^*$  by the following formula:
$$\tilde{f}(w,u)=f(w)$$

Thus we have proved a weak form of (2), without reference to the norm of the extension, which is still stronger than (1), in finite dimensions. 

\medskip

(4) Still staying in finite dimensions, it is possible to further build on this, as to have the equality $||\tilde{f}||=||f||$ happening too, as required by (2). For instance in the case of the space $V=\mathbb C^N$, with its usual norm, we can use the following decomposition, and we will leave the proof of the equality $||\tilde{f}||=||f||$ in this case as an easy exercise:
$$V=W\oplus W^\perp$$

However, in the general finite dimensional case, that of the space $V=\mathbb C^N$, but with an arbitrary norm, this simple, geometric argument is no longer valid. We will be back to this in a moment, with explanations on how to do this, directly in a more general setting, that where the codimension of $W\subset V$ is 1, or more generally, is finite.

\medskip

(5) Getting now to the general case, where $V$ can be infinite dimensional, we are facing two problems here, which can be summarized as follows:

\medskip

-- First, there is a problem coming from infinite dimensionality. Indeed, inspired by the above, we can try to extend linear forms by adding $1,2,3,\ldots$ dimensions to our space. But if we want to add $\infty$ dimensions, we will probably need the Zorn lemma.

\medskip

-- Second, at the technical level, when trying to compute norms, and arrange as for our extensions to satisfy $||\tilde{f}||=||f||$, we have a problem with real vs complex. Indeed, and you will have to believe me here, the real problem is easier than the complex one.

\medskip

In short, we are in need of a good plan here. In what follows we will first focus on the real problem, and solve it by adding 1 dimension, and then invoking the Zorn lemma. And then, we will use a trick, in order to convert our real result into a complex result.

\medskip

(6) So, consider a real Banach space $V$, having a subspace $W\subset V$, and assume in addition that this subspace is of codimension one, in the sense that we have:
$$V=W\oplus\mathbb Rx$$

Our claim is that any real linear form $f:W\to\mathbb R$ can be extended into a real linear form $f:V\to\mathbb R$, having the same norm. In order to prove this, let us set:
$$\tilde{f}(w+\lambda x)=f(w)+\lambda\alpha$$

This linear form $\tilde{f}:W\to\mathbb R$ extends then $f:V\to\mathbb R$, and the problem is that of finding the correct scalar $\alpha\in\mathbb R$, making the equality $||\tilde{f}||=||f||$ hold. 

\medskip

(7) In order to solve this latter question, we can assume, by linearity, that we have $||f||=1$. In this case, we want to find $\alpha\in\mathbb R$ such that the following holds:
$$|f(w)+\lambda\alpha|\leq||w+\lambda x||$$

By replacing $w\to-\lambda w$, we would like the following inequality to hold:
$$|-\lambda f(w)+\lambda\alpha|\leq||-\lambda w+\lambda x||$$

Equivalently, by dividing by $\lambda$, we would like the following inequality to hold:
$$|f(w)-\alpha|\leq||w-x||$$

But this is the same as asking for the following inequality to hold:
$$-||w-x||\leq f(w)-\alpha\leq||w-x||$$

Thus, our scalar $\alpha\in\mathbb R$ must be subject to the following inequalities:
$$f(w)-||w-x||\leq \alpha\leq f(w)+||w-x||$$

(8) To summarize, we can find indeed our scalar $\alpha\in\mathbb R$, making our extension $\tilde{f}$ to have norm 1 as desired, provided that the following inequality holds:
$$\sup_{w\in W}f(w)-||w-x||\leq\inf_{w\in W}f(w)+||w-x||$$

But, in order to establish this latter inequality, observe that our assumption $||f||=1$, and then the triangle inequality give, for any two vectors $v,w\in W$:
\begin{eqnarray*}
f(v)-f(w)
&=&f(v-w)\\
&\leq&||v-w||\\
&\leq&||v-x||+||w-x||
\end{eqnarray*}

The point now is that this estimate can be written in the following way:
$$f(v)-||v-x||\leq f(w)+||w-x||$$

Thus we can indeed pick $\alpha\in\mathbb R$ to be between the quantities on the left, and those on the right, with $v,w\in W$ varying, and with this choice, we have, as needed:
$$||\tilde{f}||=||f||$$

(9) Next, with this done, we can now prove Hahn-Banach in general, in the real case, by invoking the Zorn lemma. Assume indeed that we have a real Banach space $V$, a real subspace $W\subset V$, and a real linear form $f:W\to\mathbb R$. We can then consider the family of extensions $\tilde{f}:\tilde{W}\to\mathbb R$ of our linear form $f$, having the same norm as it, $||\tilde{f}||=||f||$. This family is then partially ordered, in the obvious way, and by using the Zorn lemma, we conclude that it must have a maximal element. But, in view of what we found in (8), this maximal element $\tilde{f}:\tilde{W}\to\mathbb R$ must be such that $\tilde{W}=V$, and we are done.

\medskip

(10) Finally, getting back now to the complex case, assume that we have a complex Banach space $V$, a complex subspace $W\subset V$, and a linear form $f\in W^*$. We can then apply the real result, established above, to the following real linear form:
$$g=Re(f):W\to\mathbb R$$

We obtain in this way a certain extension $\tilde{g}:V\to\mathbb R$, and then we can set:
$$\tilde{f}(x)=\tilde{g}(x)-i\tilde{g}(ix)$$

And with this, our claim is that we are done, with this being the needed extension. 

\medskip

(11) Indeed, as a first observation, $\tilde{f}:V\to\mathbb C$ as constructed above is indeed a linear form, in the complex sense, with this coming from the following computation:
\begin{eqnarray*}
\tilde{f}((a+ib) x)
&=&\tilde{g}((a+ib)x)-i\tilde{g}(i(a+ib)x)\\
&=&\tilde{g}((a+ib)x)-i\tilde{g}((ia-b)x)\\
&=&a\tilde{g}(x)+b\tilde{g}(ix)-i(a\tilde{g}(ix)-b\tilde{g}(x))\\
&=&a(\tilde{g}(x)-i\tilde{g}(ix))+ib(\tilde{g}(x)-i\tilde{g}(ix))\\
&=&(a+ib)(\tilde{g}(x)-i\tilde{g}(ix))\\
&=&(a+ib)\tilde{f}(x)
\end{eqnarray*}

(12) Next, since we have $z=Re(z)-iRe(iz)$ for any complex number $z\in\mathbb C$, by linearity we obtain that $\tilde{f}\in V^*$ extends indeed the original linear form $f\in W^*$. As for the same norm assertion, this follows from the following equalities:
$$||\tilde{f}||=||\tilde{g}||=||g||=||f||$$

Thus, theorem proved, in its general complex form, the one in the statement.
\end{proof}

\section*{9d. Duality results} 

With the above understood, we can now formulate a key result, as follows:

\index{bidual}
\index{reflexivity}

\begin{theorem}
Given a Banach space $V$, we have an embedding as follows,
$$V\subset V^{**}$$
which is an isomorphism in finite dimensions, and for the $l^p$ and $L^p$ spaces too. 
\end{theorem}

\begin{proof}
There are several things going on here, the idea being as follows:

\medskip

(1) The fact that we have indeed a vector space embedding $V\subset V^{**}$ is clear from definitions, the formula of this embedding being as follows:
$$i(v)[f]=f(v)$$

\medskip

(2) However, the fact that this embedding $V\subset V^{**}$ is isometric is something more subtle, which requires the use of the Hahn-Banach result from Theorem 9.21.

\medskip

(3) Next, the fact that we have $V=V^{**}$ in finite dimensions is clear.

\medskip

(4) Regarding now the formula $V=V^{**}$ for the various $l^p$ and $L^p$ spaces, this is something quite tricky. Let us start with the simplest case, that of the space $V=l^2$. We know that this space is given by definition by the following formula:
$$l^2=\left\{(x_i)_{i\in\mathbb N}\Big|\sum_ix_i^2<\infty\right\}$$

Now let us look for linear forms $f:l^2\to\mathbb C$. By linearity such a linear form must appear as follows, for certain scalars $a_i\in\mathbb C$, which must be such that $f$ is well-defined:
$$f\left((x_i)_i\right)=\sum_ia_ix_i$$

But, what does the fact that $f$ is well-defined mean? In answer, this means that the values of $f$ must all converge, which in practice means that we must have:
$$\sum_ix_i^2<\infty\implies\left|\sum_ia_ix_i\right|<\infty$$

Moreover, we would like our linear form $f:l^2\to\mathbb C$ to be bounded, and by denoting by $A=||f||<\infty$ the minimal bound, this means that we must have:
$$\left|\sum_ia_ix_i\right|\leq A\sqrt{\sum_ix_i^2}$$

Now recall that the Cauchy-Schwarz inequality tells us that we have:
$$\left|\sum_ia_ix_i\right|\leq\sqrt{\sum_ia_i^2}\cdot\sqrt{\sum_ix_i^2}$$

Thus, the linear form $f:l^2\to\mathbb C$ associated to any $a=(a_i)\in l^2$ will do. Moreover, conversely, by examining the proof of Cauchy-Schwarz, we conclude that this condition $a=(a_i)\in l^2$ is in fact necessary. Thus, we have proved that we have:
$$(l^2)^*=l^2$$

But this gives the $V=V^{**}$ result in the statement for our space $V=l^2$, because by dualizing one more time we obtain, as desired:
$$(l^2)^{**}=(l^2)^*=l^2$$

(5) Getting now to more complicated spaces, let us look, more generally, at $L^2(X)$. We know that this space is given by definition by the following formula:
$$L^2(X)=\left\{f:X\to\mathbb C\Big|\int_Xf(x)^2dx<\infty\right\}$$

As before, when looking for linear forms $\varphi:L^2(X)\to\mathbb C$, by linearity, and with some mesure theory helping, our forms must appear via a formula as follows:
$$\varphi(f)=\int_Xf(x)\varphi(x)dx$$

Now in order for this integral to converge, as for our map $\varphi:L^2(X)\to\mathbb C$ to be well-defined, and with the additional requirement that $\varphi$ must be actually bounded, we must have an inequality as follows, for a certain positive constant $A<\infty$:
$$\left|\int_Xf(x)\varphi(x)dx\right|\leq A\sqrt{\int_Xf(x)^2dx}$$

Now recall that the Cauchy-Schwarz inequality tells us that we have:
$$\left|\int_Xf(x)\varphi(x)dx\right|\leq\sqrt{\int_X\varphi(x)^2dx}\cdot\sqrt{\int_Xf(x)^2dx}$$

Thus, the linear form $\varphi:L^2(X)\to\mathbb C$ associated to any $\varphi\in L^2(X)$ will do. Moreover, conversely, by examining the proof of Cauchy-Schwarz, we conclude that this condition $\varphi\in L^2(X)$ is in fact necessary. Thus, we have proved that we have:
$$(L^2)^*=L^2$$

But this gives the $V=V^{**}$ result in the statement for our space $V=L^2$, because by dualizing one more time we obtain, as desired:
$$(L^2)^{**}=(L^2)^*=L^2$$

(6) Moving ahead now, let us go back to the $l^p$ spaces, as in (4), but now with general exponents $p\in[1,\infty]$, instead of $p=2$. The space $l^p$ is by definition given by:
$$l^p=\left\{(x_i)_{i\in\mathbb N}\Big|\sum_i|x_i|^p<\infty\right\}$$

Now by arguing as in (4), a linear form $f:l^p\to\mathbb C$ must come as follows:
$$f\left((x_i)_i\right)=\sum_ia_ix_i$$

To be more precise, here $a_i\in\mathbb C$ are certain scalars, which are subject to an inequality as follows, for a certain constant $A<\infty$, making $f$ well-defined, and bounded:
$$\left|\sum_ia_ix_i\right|\leq A\left(\sum_i|x_i|^p\right)^{1/p}$$

Now recall that the H\"older inequality tells us that we have, with $\frac{1}{p}+\frac{1}{q}=1$:
$$\left|\sum_ia_ix_i\right|\leq\left(\sum_i|x_i|^p\right)^{1/p}\left(\sum_i|a_i|^p\right)^{1/q}$$

Thus, the linear form $f:l^2\to\mathbb C$ associated to any element $a=(a_i)\in l^q$ will do. Moreover, conversely, by examining the proof of H\"older, we conclude that this condition $a=(a_i)\in l^q$ is in fact necessary. Thus, we have proved that we have:
$$(l^p)^*=l^q$$

But this gives the $V=V^{**}$ result in the statement for our space $V=l^p$, because by dualizing one more time we obtain, as desired:
$$(l^p)^{**}=(l^q)^*=l^q$$

(7) All this is very nice, and time now to generalize everything that we know, by looking at the general spaces $L^p(X)$, with $p\in[1,\infty]$. These spaces are given by:
$$L^p(X)=\left\{f:X\to\mathbb C\Big|\int_X|f(x)|^pdx<\infty\right\}$$

As before in (5), when looking for linear forms $\varphi:L^p(X)\to\mathbb C$, by linearity, and with some mesure theory helping, our forms must appear via a formula as follows:
$$\varphi(f)=\int_Xf(x)\varphi(x)dx$$

Now in order for this integral to converge, as for our map $\varphi:L^p(X)\to\mathbb C$ to be well-defined, and with the additional requirement that $\varphi$ must be actually bounded, we must have an inequality as follows, for a certain positive constant $A<\infty$:
$$\left|\int_Xf(x)\varphi(x)dx\right|\leq A\left(\int_X|f(x)|^pdx\right)^{1/p}$$

Now recall that the H\"older inequality tells us that we have, with $\frac{1}{p}+\frac{1}{q}=1$:
$$\left|\int_Xf(x)\varphi(x)dx\right|\leq\left(\int_X|f(x)|^pdx\right)^{1/p}\left(\int_X|\varphi(x)|^qdx\right)^{1/q}$$

Thus, the linear form $\varphi:L^p(X)\to\mathbb C$ associated to any function $\varphi\in L^q(X)$ will do. Moreover, conversely, by examining the proof of H\"older, we conclude that this condition $\varphi\in L^q(X)$ is in fact necessary. Thus, we have proved that we have:
$$(L^p)^*=L^q$$

But this gives the $V=V^{**}$ result in the statement for our space $V=L^p$, because by dualizing one more time we obtain, as desired:
$$(L^p)^{**}=(L^q)^*=L^p$$

(8) Finally, let us mention that not all Banach spaces satisfy $V=V^{**}$, with a basic counterexample here being the space $c_0$ of sequences $x_n\in\mathbb C$ satisfying $x_n\to0$, with the sup norm. Indeed, computations show that we have the following formulae:
$$c_0^*=l^1\quad,\quad (l^1)^*=l^\infty$$

Thus, in this case $V\subset V^{**}$ is the embedding $c_0\subset l^\infty$, which is not an isomorphism. 

\medskip

(9) To be more precise here, let us define  $c_0$ as above, as being the space of sequences $x_n\in\mathbb C$ satisfying $x_n\to0$, with the sup norm. We have then an inclusion as follows:
$$c_0\subset l^\infty$$

It is then clear that $c_0$ is closed, so it is indeed a Banach space. Now regarding the duals, we can use here the standard fact that given an inclusion of Banach spaces $V\subset W$, by restricting the linear forms from $W$ to $V$ we obtain a surjection $W^*\to V^*$. By using this for the above inclusion $c_0\subset l^\infty$, we obtain a surjection as follows:
$$(l^\infty)^*=l^1\to c_0^*$$

But this surjection is clearly injective, so we have indeed $c_0^*=l^1$, as claimed above.
\end{proof}

\section*{9e. Exercises}

This was a mixed algebraic and analytic chapter, and as exercises on this, we have:

\begin{exercise}
Review if needed the theory of convex functions, and Jensen.
\end{exercise}

\begin{exercise}
Learn about the Lagrange multipliers, that we used in the above.
\end{exercise}

\begin{exercise}
Find some other Banach space norms, in finite dimensions.
\end{exercise}

\begin{exercise}
Find some other examples of Banach spaces of sequences.
\end{exercise}

\begin{exercise}
Learn about bases of Banach spaces, theorems and questions.
\end{exercise}

\begin{exercise}
Learn about the tensor products of Banach spaces too.
\end{exercise}

\begin{exercise}
Learn more about the reflexive Banach spaces.
\end{exercise}

\begin{exercise}
Learn about Banach algebras, and what can be done with them.
\end{exercise}

As bonus exercise, find a good Banach space book, and start reading it.

\chapter{Hilbert spaces}

\section*{10a. Linear algebra}

We discuss in this chapter the Hilbert spaces, which are the most important types of Banach spaces. The idea is very simple, namely that the most basic Banach spaces should be those whose norm comes from a scalar product, according to the following formula:
$$||x||=\sqrt{<x,x>}$$

Summarizing, we have an idea here, and it is tempting at this point to go ahead, axiomatize the scalar products and the Hilbert spaces, and then develop the theory of these Hilbert spaces, as a continuation of what we did in chapter 9.

\bigskip

However, this is just one point of view on the Hilbert spaces. Alternatively, we can build everything from scratch, simply by starting with what we know about $\mathbb C^N$ and its scalar product $<\,,>$, from linear algebra, and generalizing this to infinite dimensions. And with this being a quite nice approach, because everything that we learned in chapter 9 about the general Banach spaces heavily simplifies in the case of the Hilbert spaces.

\bigskip

Well, hope you get my point, we are here in front of a dillema, as follows:

\begin{dillema}
We have two ways of viewing the theory of Hilbert spaces, as a generalization of linear algebra, or as a particular case of Banach space theory.
\end{dillema}

So, what to do, in this kind of situation? Ask the cats of course, that we have not seen since chapter 4. And the cats, both more in shape than ever, declare:

\begin{cats}
We cats view the theory of Hilbert spaces as being part of quantum mechanics. But for a human introduction, better go with the linear algebra way.
\end{cats}

Okay, thanks cats, I was sort of expecting that, but the advice at the end is really useful. So, we will go this way, and forgetting now everything that we learned in chapter 9 about Banach spaces, and no worries about this, we will manage to reprove all that things, in the Hilbert space case, let us start with a discussion in finite dimensions. 

\bigskip

You certainly know from linear algebra and calculus many things about $\mathbb C^N$ and its scalar product $<\,,>$, and we will review now this. Let us start with:

\begin{definition}
We endow $\mathbb C^N$ with its usual scalar product, namely
$$<x,y>=\sum_ix_i\bar{y}_i$$
so that the norm of the vectors is given by $||x||=\sqrt{<x,x>}$.
\end{definition}

Here the convention for linearity, which is at left, and the notation $<\,,>$ for the scalar product are both taken from physics, and more specifically from cat quantum mechanics. Not to be confused with human quantum mechanics, where the conventions are usually different, with the scalar product being denoted $<|>$, and taken linear at right.

\bigskip

Getting now to what can be done with scalar products, as a basic result, we have:

\begin{theorem}
The linear maps $T:\mathbb C^N\to\mathbb C^N$ are in correspondence with the square matrices $A\in M_N(\mathbb C)$, with the linear map associated to such a matrix being
$$Tx=Ax$$
and with the matrix associated to a linear map being $A_{ij}=<Te_j,e_i>$.
\end{theorem}

\begin{proof}
The first assertion is clear, because a linear map $T:\mathbb C^N\to\mathbb C^N$ must send a vector $x\in\mathbb C^N$ to a certain vector $Tx\in\mathbb C^N$, all whose components are linear combinations of the components of $x$. Thus, we can write, for certain complex numbers $A_{ij}\in\mathbb C$:
$$T\begin{pmatrix}
x_1\\
\vdots\\
x_N
\end{pmatrix}
=\begin{pmatrix}
A_{11}x_1+\ldots+A_{1N}x_N\\
\vdots\\
A_{N1}x_1+\ldots+A_{NN}x_N
\end{pmatrix}$$

But this means $Tx=Ax$, as claimed. Regarding now the second assertion, with $Tx=Ax$, if we denote by $e_1,\ldots,e_N$ the standard basis of $\mathbb C^N$, we have:
$$Te_j
=\begin{pmatrix}
A_{1j}\\
\vdots\\
A_{Nj}
\end{pmatrix}$$

But this gives the second formula, $<Te_j,e_i>=A_{ij}$, as desired.
\end{proof}

Our claim now is that, no matter what we want to do with $T$ or $A$, of advanced type, we will run at some point into their adjoints $T^*$ and $A^*$, constructed as follows:

\index{adjoint operator}

\begin{proposition}
The adjoint operator $T^*:\mathbb C^N\to\mathbb C^N$, which is given by
$$<Tx,y>=<x,T^*y>$$
corresponds to the adjoint matrix $A^*\in M_N(\mathbb C)$, given by
$$(A^*)_{ij}=\bar{A}_{ji}$$
via the correspondence between linear maps and matrices constructed above.
\end{proposition}

\begin{proof}
Given a linear map $T:\mathbb C^N\to\mathbb C^N$, fix $y\in\mathbb C^N$, and consider the linear form $\varphi(x)=<Tx,y>$. This form must be as follows, for a certain vector $T^*y\in\mathbb C^N$:
$$\varphi(x)=<x,T^*y>$$

Thus, we have constructed a map $y\to T^*y$ as in the statement, which is obviously linear, and that we can call $T^*$. Now by taking the vectors $x,y\in\mathbb C^N$ to be elements of the standard basis of $\mathbb C^N$, our defining formula for $T^*$ reads:
$$<Te_i,e_j>=<e_i,T^*e_j>$$

By reversing the scalar product on the right, this formula can be written as:
$$<T^*e_j,e_i>=\overline{<Te_i,e_j>}$$

But this means that the matrix of $T^*$ is given by $(A^*)_{ij}=\bar{A}_{ji}$, as desired.
\end{proof}

Getting back to our claim, the adjoints $*$ are indeed ubiquitous, as shown by:

\index{scalar product}
\index{unitary}
\index{projection}

\begin{theorem}
The following happen:
\begin{enumerate}
\item $T(x)=Ux$ with $U\in M_N(\mathbb C)$ is an isometry precisely when $U^*=U^{-1}$.

\item $T(x)=Px$ with $P\in M_N(\mathbb C)$ is a projection precisely when $P^2=P^*=P$.
\end{enumerate}
\end{theorem}

\begin{proof}
We can recover the scalar products out of the norms, by using the following difficult to remember formula, called complex polarization identity:
$$4<x,y>
=||x+y||^2-||x-y||^2+i||x+iy||^2-i||x-iy||^2$$

We will also use Proposition 10.5, and more specifically the following formula coming from there, valid for any matrix $A\in M_N(\mathbb C)$ and any two vectors $x,y\in\mathbb C^N$:
$$<Ax,y>=<x,A^*y>$$

(1) Given a matrix $U\in M_N(\mathbb C)$, we have indeed the following equivalences, with the first one coming from the polarization identity, and the other ones being clear:
\begin{eqnarray*}
||Ux||=||x||
&\iff&<Ux,Uy>=<x,y>\\
&\iff&<x,U^*Uy>=<x,y>\\
&\iff&U^*Uy=y\\
&\iff&U^*=U^{-1}
\end{eqnarray*}

(2) Given a matrix $P\in M_N(\mathbb C)$, in order for $x\to Px$ to be an oblique projection, we must have $P^2=P$. Now observe that this projection is orthogonal when:
\begin{eqnarray*}
<Px-x,Py>=0
&\iff&<P^*Px-P^*x,y>=0\\
&\iff&P^*Px-P^*x=0\\
&\iff&P^*P-P^*=0\\
&\iff&P^*P=P^*
\end{eqnarray*}

By conjugating, $P^*P=P$, so we must have $P=P^*$, which gives the result. 
\end{proof}

At a more advanced level now, diagonalization, we first have the following result:

\begin{theorem}
Any matrix $A\in M_N(\mathbb C)$ which is self-adjoint, $A=A^*$, is diagonalizable, with the diagonalization being of the following type,
$$A=UDU^*$$
with $U\in U_N$, and with $D\in M_N(\mathbb R)$ diagonal. The converse holds too.
\end{theorem}

\begin{proof}
As a first remark, the converse trivially holds, because if we take a matrix of the form $A=UDU^*$, with $U$ unitary and $D$ diagonal and real, then we have:
$$A^*
=(UDU^*)^*
=UDU^*
=A$$

In the other sense now, assume that $A$ is self-adjoint, $A=A^*$.  Our first claim is that the eigenvalues are real. Indeed, assuming $Av=\lambda v$, we have:
$$\lambda<v,v>
=<Av,v>
=<v,Av>
=\bar{\lambda}<v,v>$$

Thus we obtain $\lambda\in\mathbb R$, as claimed. Our next claim now is that the eigenspaces corresponding to different eigenvalues are pairwise orthogonal. Assume indeed that we have $Av=\lambda v$, $Aw=\mu w$. We have then the following computation, using $\lambda,\mu\in\mathbb R$:
$$\lambda<v,w>
=<Av,w>
=<v,Aw>
=\mu<v,w>$$

Thus $\lambda\neq\mu$ implies $v\perp w$, as claimed. In order now to finish, it remains to prove that the eigenspaces of $A$ span the whole $\mathbb C^N$. For this purpose, let us pick an eigenvector, $Av=\lambda v$. Assuming that we have a vector $w$ orthogonal to it, $v\perp w$, we have:
$$<Aw,v>
=<w,Av>
=<w,\lambda v>
=\lambda<w,v>
=0$$

Thus, if $v$ is an eigenvector, then the vector space $v^\perp$ is invariant under $A$. Moreover, since a matrix $A$ is self-adjoint precisely when $<Av,v>\in\mathbb R$ for any vector $v\in\mathbb C^N$, as one can see by expanding the scalar product, the restriction of $A$ to the subspace $v^\perp$ is self-adjoint. Thus, we can proceed by recurrence, and we obtain the result.
\end{proof}

Let us discuss as well the case of the unitary matrices. We have here:

\index{unitary}

\begin{theorem}
Any matrix $U\in M_N(\mathbb C)$ which is unitary, $U^*=U^{-1}$, is diagonalizable, with the eigenvalues on $\mathbb T$. More precisely we have
$$U=VDV^*$$
with $V\in U_N$, and with $D\in M_N(\mathbb T)$ diagonal. The converse holds too.
\end{theorem}

\begin{proof}
As a first remark, the converse trivially holds, because given a matrix of type $U=VDV^*$, with $V\in U_N$, and with $D\in M_N(\mathbb T)$ being diagonal, we have:
$$U^*
=(VDV^*)^*
=VD^{-1}V^{-1}
=(VDV^*)^{-1}
=U^{-1}$$

Let us prove now the first assertion, stating that the eigenvalues of a unitary matrix $U\in U_N$ belong to the unit circle $\mathbb T$. Assuming $Uv=\lambda v$, we have:
\begin{eqnarray*}
<v,v>
&=&<U^*Uv,v>\\
&=&<Uv,Uv>\\
&=&<\lambda v,\lambda v>\\
&=&|\lambda|^2<v,v>
\end{eqnarray*}

Thus $\lambda\in\mathbb T$, as claimed. Our next claim now is that the eigenspaces corresponding to different eigenvalues are pairwise orthogonal. Assume indeed that $Uv=\lambda v$, $Uw=\mu w$. We have then the following computation, using $U^*=U^{-1}$ and $\lambda,\mu\in\mathbb T$:
\begin{eqnarray*}
\lambda<v,w>
&=&<Uv,w>\\
&=&<v,U^*w>\\
&=&<v,U^{-1}w>\\
&=&<v,\mu^{-1}w>\\
&=&\mu<v,w>
\end{eqnarray*}

Thus $\lambda\neq\mu$ implies $v\perp w$, as claimed. In order now to finish, it remains to prove that the eigenspaces of $U$ span the whole $\mathbb C^N$. For this purpose, let us pick an eigenvector, $Uv=\lambda v$. Assuming that we have a vector $w$ orthogonal to it, $v\perp w$, we have:
\begin{eqnarray*}
<Uw,v>
&=&<w,U^*v>\\
&=&<w,U^{-1}v>\\
&=&<w,\lambda^{-1}v>\\
&=&\lambda<w,v>\\
&=&0
\end{eqnarray*}

Thus, if $v$ is an eigenvector, then the vector space $v^\perp$ is invariant under $U$. Now since $U$ is an isometry, so is its restriction to this space $v^\perp$. Thus this restriction is a unitary, and so we can proceed by recurrence, and we obtain the result.
\end{proof}

The self-adjoint matrices and the unitary matrices are particular cases of the general notion of a ``normal matrix'', and we have here the following result:

\index{normal operator}

\begin{theorem}
Any matrix $A\in M_N(\mathbb C)$ which is normal, $AA^*=A^*A$, is diagonalizable, with the diagonalization being of the following type,
$$A=UDU^*$$
with $U\in U_N$, and with $D\in M_N(\mathbb C)$ diagonal. The converse holds too.
\end{theorem}

\begin{proof}
As a first remark, the converse trivially holds, because if we take a matrix of the form $A=UDU^*$, with $U$ unitary and $D$ diagonal, then we have:
$$AA^*
=UDD^*U^*
=UD^*DU^*
=A^*A$$

In the other sense now, this is something more technical. Our first claim is that a matrix $A$ is normal precisely when the following happens, for any vector $v$:
$$||Av||=||A^*v||$$

Indeed, the above equality can be written as follows:
$$<AA^*v,v>=<A^*Av,v>$$

But this is equivalent to $AA^*=A^*A$, by expanding the scalar products. Our next claim is that $A,A^*$ have the same eigenvectors, with conjugate eigenvalues:
$$Av=\lambda v\implies A^*v=\bar{\lambda}v$$

Indeed, this follows from the following computation, and from the trivial fact that if a matrix $A$ is normal, then so is any matrix of type $A-\lambda 1_N$:
\begin{eqnarray*}
||(A^*-\bar{\lambda}1_N)v||
&=&||(A-\lambda 1_N)^*v||\\
&=&||(A-\lambda 1_N)v||\\
&=&0
\end{eqnarray*}

Let us prove now that the eigenspaces of $A$ are pairwise orthogonal. Assume that we have two eigenvectors, $Av=\lambda v$, $Aw=\mu w$, corresponding to different eigenvalues, $\lambda\neq\mu$. We have the following computation, which shows that $\lambda\neq\mu$ implies $v\perp w$:
\begin{eqnarray*}
\lambda<v,w>
&=&<Av,w>\\
&=&<v,A^*w>\\
&=&<v,\bar{\mu}w>\\
&=&\mu<v,w>
\end{eqnarray*}

In order to finish, it remains to prove that the eigenspaces of $A$ span the whole $\mathbb C^N$. For this purpose, let us pick two eigenvectors $v,w$ of the matrix $AA^*$, corresponding to different eigenvalues, $\lambda\neq\mu$. The eigenvalue equations are then as follows:
$$AA^*v=\lambda v\quad,\quad 
AA^*w=\mu w$$

We have the following computation, using the normality condition $AA^*=A^*A$, and the fact that the eigenvalues of $AA^*$, and in particular $\mu$, are real:
\begin{eqnarray*}
\lambda<Av,w>
&=&<\lambda Av,w>\\
&=&<A\lambda v,w>\\
&=&<AAA^*v,w>\\
&=&<AA^*Av,w>\\
&=&<Av,AA^*w>\\
&=&<Av,\mu w>\\
&=&\mu<Av,w>
\end{eqnarray*}

We conclude that we have $<Av,w>=0$. But this reformulates as follows:
$$\lambda\neq\mu\implies A(E_\lambda)\perp E_\mu$$

Now since the eigenspaces of $AA^*$ are pairwise orthogonal, and span the whole $\mathbb C^N$, we deduce from this that these eigenspaces are invariant under $A$:
$$A(E_\lambda)\subset E_\lambda$$

But with this result in hand, we can finish. Indeed, we can decompose the problem, and the matrix $A$ itself, following these eigenspaces of $AA^*$, which in practice amounts in saying that we can assume that we only have 1 eigenspace. Now by rescaling, this is the same as assuming that we have $AA^*=1$. But with this, we are now into the unitary case, that we know how to solve, as explained in Theorem 10.8, and so done.
\end{proof}

\section*{10b. Hilbert spaces}

Getting now to infinite dimensions, there are many things to be done here, in order to extend the above linear algebra material. Let us start with:

\index{Hilbert space}
\index{scalar product}

\begin{definition}
A Hilbert space is a complex vector space $H$ with a scalar product $<x,y>$, which will be linear at left and antilinear at right,
$$<\lambda x,y>=\lambda<x,y>\quad,\quad <x,\lambda y>=\bar{\lambda}<x,y>$$
which is complete with respect to corresponding norm
$$||x||=\sqrt{<x,x>}$$
in the sense that any sequence which is Cauchy with respect to $||.||$ converges.
\end{definition}

In other words, in terms of the Banach space theory from chapter 9, a Hilbert space is a Banach space whose norm comes from a scalar product, via $||x||=\sqrt{<x,x>}$. However, for reasons discussed in the beginning of this chapter, we would rather prefer to develop the Hilbert space theory independently from what we know from chapter 9.

\bigskip

Getting now to Definition 10.10 as stated, this is something quite subtle, and there is some mathematics encapsulated there, needing some discussion. Indeed, we must prove that $||x||=\sqrt{<x,x>}$ is a norm. But this comes from the following result:

\index{Cauchy-Schwarz}
\index{Minkowski inequality}

\begin{theorem}
Consider a complex vector space $H$ with a scalar product $<\,,>$, and set $||x||=\sqrt{<x,x>}$. The following happen:
\begin{enumerate}
\item We have the Cauchy-Schwarz inequality $|<x,y>|\leq||x||\cdot||y||$.

\item We have the Minkowski inequality $||x+y||\leq||x||+||y||$.

\item Thus, $||.||$ is indeed a norm, and $d(x,y)=||x-y||$ is a distance.
\end{enumerate}
\end{theorem}

\begin{proof}
This is something very standard, the idea being as follows:

\medskip

(1) In order to prove Cauchy-Schwarz, consider the following quantity, depending on a real variable $t\in\mathbb R$, and on a variable on the unit circle, $w\in\mathbb T$:
$$f(t)=||twx+y||^2$$

By developing $f$, we see that this is a degree 2 polynomial in $t$:
\begin{eqnarray*}
f(t)
&=&<twx+y,twx+y>\\
&=&t^2<x,x>+tw<x,y>+t\bar{w}<y,x>+<y,y>\\
&=&t^2||x||^2+2tRe(w<x,y>)+||y||^2
\end{eqnarray*}

Since $f$ is obviously positive, its discriminant must be negative. Now the point is that we can arrange for the number $w\in\mathbb T$ to be such that the quantity $w<x,y>$ is real. Thus, we obtain the Cauchy-Schwarz inequality $|<x,y>|\leq||x||\cdot||y||$, as desired.

\medskip

(2) This follows indeed from the Cauchy-Schwarz inequality, as follows:
\begin{eqnarray*}
&&||x+y||\leq||x||+||y||\\
&\iff&||x+y||^2\leq(||x||+||y||)^2\\
&\iff&||x||^2+||y||^2+2Re<x,y>\leq||x||^2+||y||^2+2||x||\cdot||y||\\
&\iff&Re<x,y>\leq||x||\cdot||y||
\end{eqnarray*}

(3) This follows indeed from the Minkowski inequality, which corresponds to the triangle inequality, the other two axioms for a distance being trivially satisfied.
\end{proof}

Summarizing, Definition 10.10 is fully justified. Before getting into basic examples and theory, let us record as well the following result, which further clarifies the relation between Hilbert and Banach spaces, and which is something useful on its own:

\begin{theorem}
The scalar products can be recovered from distances, via the following formula, called complex polarization identity:
$$4<x,y>
=||x+y||^2-||x-y||^2
+i||x+iy||^2-i||x-iy||^2$$
Thus, a Banach space is a Hilbert space precisely when this polarization formula defines, out of the norm $||.||$, a scalar product $<\,,>$, satisfying $<x,x>=||x||^2$.
\end{theorem}

\begin{proof}
In what regards the polarization formula in the statement, this is certainly difficult to memorize, but easy to prove, as follows:
\begin{eqnarray*}
&&||x+y||^2-||x-y||^2+i||x+iy||^2-i||x-iy||^2\\
&=&||x||^2+||y||^2-||x||^2-||y||^2+i||x||^2+i||y||^2-i||x||^2-i||y||^2\\
&&+2Re(<x,y>)+2Re(<x,y>)+2iIm(<x,y>)+2iIm(<x,y>)\\
&=&4<x,y>
\end{eqnarray*}

As for the last assertion, that is something rather philosophical, obtained by carefully examining Definition 10.10, with the complex polarization identity in mind.
\end{proof}

At the level of examples now, we first have the following result:

\index{square-summable}

\begin{theorem}
Given an index set $I$, the space of square-summable sequences
$$l^2(I)=\left\{(x_i)_{i\in I}\Big|\sum_i|x_i|^2<\infty\right\}$$
is a Hilbert space, with $<x,y>=\sum_ix_i\bar{y}_i$. When $|I|=N<\infty$, we have $l^2(I)=\mathbb C^N$.
\end{theorem}

\begin{proof}
We have already met such things before, but let us quickly recall this:

\medskip

(1) We know that $l^2(I)\subset\mathbb C^I$ is the space of vectors satisfying $||x||<\infty$. We want to prove that $l^2(I)$ is a vector space, that $<x,y>$ is a scalar product on it, that $l^2(I)$ is complete with respect to $||.||$, and finally that for $|I|<\infty$ we have $l^2(I)=\mathbb C^{|I|}$.

\medskip

(2) The last assertion, $l^2(I)=\mathbb C^{|I|}$ for $|I|<\infty$, is clear, because in this case the sums are finite, so the condition $||x||<\infty$ is automatic. So, we know at least one thing.

\medskip

(3) Regarding the other assertions, everything follows from Cauchy-Schwarz. We first obtain, by raising to the square and expanding, that for any $x,y\in l^2(I)$ we have:
$$||x+y||\leq||x||+||y||$$

Thus $l^2(I)$ is indeed a vector space, the other vector space conditions being trivial. Also, $<x,y>$ is surely a scalar product on this vector space, all axioms being clear.

\medskip

(4) Finally, the fact that our space $l^2(I)$ is indeed complete with respect to its norm $||.||$ follows in the obvious way, the limit of a Cauchy sequence $\{x_n\}$ being the vector $y=(y_i)$ given by $y_i=\lim_{n\to\infty}x_{ni}$, with all the verifications here being trivial.
\end{proof}

Going now a bit abstract, we have, more generally, the following result:

\begin{theorem}
Given a measured space $X$, the space of square-summable functions
$$L^2(X)=\left\{f:X\to\mathbb C\Big|\int_X|f(x)|^2\,d\mu(x)<\infty\right\}$$
is a Hilbert space, with $<f,g>=\int_Xf(x)\overline{g(x)}\,d\mu(x)$. When $X$ is discrete, $L^2(X)=l^2(X)$.
\end{theorem}

\begin{proof}
This is something routine, remake of Theorem 10.13, as follows:

\medskip

(1) The proof of the first, and main assertion is something perfectly similar to the proof of Theorem 10.13, by replacing everywhere the sums by integrals. 

\medskip

(2) With the remark that we forgot to say in the statement that the $L^2$ functions are by definition taken up to equality almost everywhere, $f=g$ when $||f-g||=0$.

\medskip

(3) As for the last assertion, when $\mu$ is the counting measure all our integrals here become usual sums, and so we recover in this way Theorem 10.13.
\end{proof}

As a conclusion to what we did so far, the Hilbert spaces are now axiomatized, and the main examples discussed. In order to do now some geometry on our spaces, in analogy with what we know from finite dimensions, let us start with the following definition:

\begin{definition}
Let $H$ be a Hilbert space.
\begin{enumerate}
\item We call two vectors orthogonal, $x\perp y$, when $<x,y>=0$.

\item Given a subset $S\subset H$, we set $S^\perp=\left\{x\in H\big|x\perp y,\forall y\in S\right\}$.
\end{enumerate}
\end{definition}

As a comment here, in relation with (1), it follows from Cauchy-Schwarz that given any two vectors $x,y$ we have $<x,y>=||x||\cdot||y||\cdot\alpha$, with $\alpha\in[-1,1]$. In the case of the real Hilbert spaces we have $\alpha=\cos t$, with $t$ being the angle between $x,y$, but here we will only deal with complex Hilbert spaces, where this formula is not available.

\bigskip

Getting now to what can be done with orthogonality, we have here:

\begin{theorem}
Let $H$ be a Hilbert space, and $E\subset H$ be a closed subspace.
\begin{enumerate}
\item Given $x\in H$, we can find a unique $y\in E$, minimizing $||x-y||$.

\item With $x,y$ as above, we have $x=y+z$, for a certain $z\in E^\perp$.

\item Thus, we have a direct sum decomposition $H=E\oplus E^\perp$.

\item In terms of $H=E\oplus E^\perp$, the projection $x\to y$ is given by $P(x,y)=x$.
\end{enumerate}
\end{theorem}

\begin{proof}
This is something very standard, the idea being as follows:

\medskip

(1) Given $x\in H$ and two vectors $v,w\in E$, we have the following estimate:
\begin{eqnarray*}
||x-v||^2+||x-w||^2
&=&2\left(\left|\left|x-\frac{v+w}{2}\right|\right|^2+\left|\left|\frac{v-w}{2}\right|\right|^2\right)\\
&\geq&2d(x,E)^2+\frac{||v-w||^2}{2}
\end{eqnarray*}

But this shows that any sequence in $E$ realizing the inf in the definition of $d(x,E)$ is Cauchy, so it converges to a vector $y$. Since $E$ is closed we have $y\in E$, so $y$ realizes the inf. Moreover, again from the above inequality, such a $y$ realizing the inf is unique.

\medskip

(2) In order to prove $x-y\in E^\perp$, let $v\in E$ and choose $w\in\mathbb T$ such that $\bar{w}<x-y,v>$ is a real number. For any $t\in\mathbb R$ we have the following equality:
$$||x-y+twv||=||x-y||^2+2t\bar{w}<x-y,v>+t^2||v||^2$$

By construction of the vector $y$ we know that this function has a minimum at $t=0$. But this function is a degree 2 polynomial, so the middle term must vanish:
$$2\bar{w}<x-y,v>=0$$

Now since this must hold for any $v\in E$, we must have $x-y\in E^\perp$, as desired.

\medskip

(3) This is consequence of what we found in (1,2).

\medskip

(4) This is also a consequence of what we found in (1,2).
\end{proof}

Many other things can be said, as a continuation of the above, as for instance with the formula $E^{\perp\perp}=E$, when $E\subset H$ is a closed subspace, which must be adjusted to $E^{\perp\perp}=\bar{E}$, in the case of the arbitrary subspaces $E\subset H$. Also, importantly, we have:

\begin{theorem}
Given a Hilbert space $H$ and a closed subspace $E\subset H$, any linear form $f:E\to\mathbb C$ can be extended into a linear form $\tilde{f}:H\to\mathbb C$, having the same norm, and this simply by using $H=E\oplus E^\perp$, and setting $\tilde{f}=0$ on $E^\perp$.
\end{theorem}

\begin{proof}
This is indeed something self-explanatory. Observe that what we have here is the Hahn-Banach theorem, for the Hilbert spaces, coming with a trivial proof.
\end{proof}

Still talking abstract functional analysis, the few Banach space findings left from chapter 9 trivialize in the case of Hilbert spaces, as shown by the following result:

\begin{theorem}
Let $H$ be a Hilbert space.
\begin{enumerate}
\item Any linear form $f:H\to\mathbb C$ must be of type $f(y)=<y,z>$, with $z\in H$. 

\item Thus, we have a Banach space isomorphism $H^*\simeq\bar{H}$.

\item In particular, $H$ is reflexive as Banach space, $H^{**}=H$.
\end{enumerate}
\end{theorem}

\begin{proof}
This is something that we already know from chapter 9, coming from our general discussion there, but we have an elementary proof for this, as follows:

\medskip

(1) Consider a linear form $f:H\to\mathbb C$. Choose $v\in H$ such that $f(v)\neq 0$. By linearity we may assume $f(v)=1$. Then each $z\in H$ decomposes in the following way:
$$z=(z-f(z)v)+f(z)v$$

This shows that we have a direct sum decomposition of $H$, as follows:
$$H=\ker(f)\oplus \mathbb C v$$

Now pick $z\in\ker(f)^\perp$ and consider the kernel of the linear form $f_z(y)=<y,z>$:
$$Ker(f_z)=\{y\in H| <y,z>=0\} \supset Ker (f)$$

The linear forms $f_z$ and $f$ are then given by the following formulae:
$$f_z(a+\lambda v)=\lambda f_z(v)\quad,\quad f(a+\lambda v)=\lambda$$

It follows that we have $f=\mu f_z$, with $\mu=f_z(v)^{-1}$, and so that we have, as desired:
$$f=f_{\overline{\mu} z}$$

(2) This is just an abstract reformulation of what we found in (1).

\medskip

(3) This follows from (2), because we have $H^{**}=\bar{H}^*=\bar{\bar{H}}=H$.
\end{proof}

As a conclusion to all this, the various general Banach space results form chapter 9 are all clear in the Hilbert space setting. However, do not worry, the Hilbert spaces have their own amount of mystery, that we will explore in what follows.

\section*{10c. Bases, separability}

At a more advanced level now, we can talk about orthonormal bases, and the related notion of dimension of a Hilbert space. We have here the following result:

\index{orthonormal basis}
\index{Gram-Schmidt}
\index{separable space}

\begin{theorem}
Any Hilbert space $H$ has an orthonormal basis $\{e_i\}_{i\in I}$, which is by definition a set of vectors whose span is dense in $H$, and which satisfy
$$<e_i,e_j>=\delta_{ij}$$
with $\delta$ being a Kronecker symbol. The cardinality $|I|$ of the index set, which can be finite, countable, or uncountable, depends only on $H$, and is called dimension of $H$. We have
$$H\simeq l^2(I)$$
in the obvious way, mapping $\sum\lambda_ie_i\to(\lambda_i)$. The Hilbert spaces with $\dim H=|I|$ being countable, such as $l^2(\mathbb N)$, are all isomorphic, and are called separable.
\end{theorem}

\begin{proof}
We have many assertions here, the idea being as follows:

\medskip

(1) In finite dimensions an orthonormal basis $\{e_i\}_{i\in I}$ can be constructed by starting with any vector space basis $\{f_i\}_{i\in I}$, and using the Gram-Schmidt procedure. As for the other assertions, these are all clear, from basic linear algebra.

\medskip

(2) In general, the same method works, namely Gram-Schmidt, with a subtlety coming from the fact that the basis $\{e_i\}_{i\in I}$ will not span in general the whole $H$, but just a dense subspace of it, as it is in fact obvious by looking at the standard basis of $l^2(\mathbb N)$. 

\medskip

(3) And there is a second subtlety as well, coming from the fact that the recurrence procedure needed for Gram-Schmidt must be replaced by some sort of ``transfinite recurrence'', using standard tools from logic, and more specifically the Zorn lemma.

\medskip

(4) Finally, everything at the end, regarding our notion of separability for the Hilbert spaces,  is clear from definitions, and from our various results above.
\end{proof}

According to Theorem 10.19, there is only one separable Hilbert space, up to isomorphism. In order to further comment on this unique space, we will need:

\begin{theorem}[Weierstrass]
Any continuous function on a closed interval
$$f:[a,b]\to\mathbb R$$
can be uniformly approximated by polynomials.
\end{theorem}

\begin{proof}
This is something very classical, with a well-known, constructive proof being by using an approximation by suitable Bernstein polynomials, as follows:

\medskip

(1) We can assume by linearity $[a,b]=[0,1]$. Consider the following polynomials:
$$p_{kn}(x)=\binom{n}{k}x^k(1-x)^{n-k}$$

Then, given a continuous function $f:[0,1]\to\mathbb R$, consider the following polynomials:
$$f_n(x)=\sum_{k=0}^nf\left(\frac{k}{n}\right)p_{kn}(x)$$

Our claim is that we have $f_n\to_uf$, uniform convergence on $[0,1]$.

\medskip

(2) In order to prove this, observe that the above polynomials $p_{kn}$ encode the densities of the binomial laws $b_{xn}$. Thus, we have the following formulae:
$$\sum_{k=0}^np_{kn}(x)=1\quad,\quad 
\sum_{k=0}^n\frac{k}{n}\cdot p_{kn}(x)=x\quad,\quad 
\sum_{k=0}^n\left(x-\frac{k}{n}\right)^2p_{kn}(x)=\frac{x(1-x)}{n}$$

(3) In order to estimate now the error $|f_n-f|$, we can use the uniform continuity property of $f$. So, pick $\varepsilon>0$, and then $\delta>0$ such that the following happens:
$$|x-y|<\delta\implies|f(x)-f(y)|<\varepsilon$$

(4) We have then the following estimate, using this, and with $M=\sup|f|$:
\begin{eqnarray*}
|f_n(x)-f(x)|
&=&\left|\sum_{k=0}^nf\left(\frac{k}{n}\right)p_{kn}(x)-\sum_{k=0}^nf(x)p_{kn}(x)\right|\\
&\leq&\sum_{k=0}^n\left|f\left(\frac{k}{n}\right)-f(x)\right|p_{kn}(x)\\
&=&\sum_{\left|x-\frac{k}{n}\right|<\delta}\left|f\left(\frac{k}{n}\right)-f(x)\right|p_{kn}(x)
+\sum_{\left|x-\frac{k}{n}\right|\geq\delta}\left|f\left(\frac{k}{n}\right)-f(x)\right|p_{kn}(x)\\
&\leq&\varepsilon+M\sum_{\left|x-\frac{k}{n}\right|\geq\delta}p_{kn}(x)
\end{eqnarray*}

(5) In order to deal with the sum on the right, we will need some standard estimates. Let us first recall the Markov inequality, which is something trivial, as follows:
$$P\big(|\varphi|\geq b\big)\leq\frac{E(\varphi)}{b}$$

By using this Markov inequality with $\varphi=(\psi-E)^2$, with $E=E(\psi)$, we obtain the Chebycheff inequality, which is something more subtle, as follows:
$$P\big(|\psi-E|\geq a\big)
\leq\frac{E((\psi-E)^2)}{a^2}
=\frac{V}{a^2}$$

(6) The point now is that this latter inequality applies to the last sum in (4), with $\psi$ being a variable following the binomial law $b_{xn}$, rescaled to $[0,1]$, and gives:
\begin{eqnarray*}
\sum_{\left|x-\frac{k}{n}\right|\geq\delta}p_{kn}(x)
&\leq&\sum_{k=0}^n\delta^{-2}\left(x-\frac{k}{n}\right)^2p_{kn}(x)\\
&=&\delta^{-2}\,\frac{x(1-x)}{n}\\
&\leq&\frac{\delta^{-2}}{4n}
\end{eqnarray*}

(7) Now by putting everything together, we obtain the following estimate:
$$|f_n(x)-f(x)|\leq \varepsilon+\frac{\delta^{-2}M}{4n}$$

Thus we have indeed $|f_n-f|\to0$, uniform convergence, as desired.
\end{proof}

Now with the Weierstrass approximation theorem in hand, getting back to our Hilbert space questions, as formulated in Theorem 10.19, as a main theoretical result regarding the unique space that we are basically interested in, the separable one, we have:

\index{orthogonal polynomials}
\index{Weierstrass basis}

\begin{theorem}
The following happen, in relation with separability:
\begin{enumerate}
\item The Hilbert space $H=L^2[-1,1]$ is separable, with orthonormal basis coming by applying Gram-Schmidt to the basis $\{x^k\}_{k\in\mathbb N}$, coming from Weierstrass.

\item In fact, any $H=L^2(\mathbb R,\mu)$, with $d\mu(x)=f(x)dx$, is separable, and the same happens in higher dimensions, for $H=L^2(\mathbb R^N,\mu)$, with $d\mu(x)=f(x)dx$.

\item More generally, given a separable abstract measured space $X$, the associated Hilbert space of square-summable functions $H=L^2(X)$ is separable.
\end{enumerate}
\end{theorem}
 
\begin{proof}
Many things can be said here, the idea being as follows:

\medskip

(1) The fact that $H=L^2[-1,1]$ is separable is clear indeed from the Weierstrass density theorem, which provides us with the algebraic basis $g_k=x^k$, which can be orthogonalized by using the Gram-Schmidt procedure, as explained in Theorem 10.19.

\medskip

(2) Regarding now more general spaces, of type $H=L^2(\mathbb R,\mu)$, we can use here the same argument, after modifying if needed our measure $\mu$, in order for the functions $g_k=x^k$ to be indeed square-summable. As for higher dimensions, the situation here is similar, because we can use here the multivariable polynomials $g_k(x)=x_1^{k_1}\ldots x_N^{k_N}$.

\medskip

(3) Finally, the last assertion, regarding the general spaces of type $H=L^2(X)$, which generalizes all this, comes as a consequence of the general measure theory developed in Part II, and we will leave working out the details here as an instructive exercise.
\end{proof}

As a conclusion to all this, which is a bit philosophical, we have:

\begin{conclusion}
We are interested in one space, namely the unique separable Hilbert space $H$, but due to various technical reasons, it is often better to forget that we have $H=l^2(\mathbb N)$, and say instead that we have $H=L^2(X)$, with $X$ being a separable measured space, or simply say that $H$ is an abstract separable Hilbert space.
\end{conclusion}

It is also possible to make some physics comments here, with the unique separable Hilbert space $H$ from Conclusion 10.22, that we will be presumably obsessed with, in what follows, being, and no surprise here, the space that we live in.

\bigskip

Let us go back now to Theorem 10.21 and its proof, which is something quite subtle. That material leads us into orthogonal polynomials, which are defined as follows:

\index{orthogonal polynomials}

\begin{definition}
The orthogonal polynomials with respect to $d\mu(x)=f(x)dx$ are polynomials $P_k\in\mathbb R[x]$ of degree $k\in\mathbb N$, which are orthogonal inside $H=L^2(\mathbb R,\mu)$:
$$\int_\mathbb RP_k(x)P_l(x)f(x)dx=0\quad,\quad\forall k\neq l$$
Equivalently, these orthogonal polynomials $\{P_k\}_{k\in\mathbb N}$, which are each unique modulo scalars, appear from the Weierstrass basis $\{x^k\}_{k\in\mathbb N}$, by doing Gram-Schmidt.
\end{definition}

As a first observation, the orthogonal polynomials exist indeed for any real measure $d\mu(x)=f(x)dx$, because we can obtain them from the monomials $x^k$ via Gram-Schmidt, as indicated above. It is possible to be a bit more explicit here, as follows:

\begin{theorem}
The orthogonal polynomials with respect to $\mu$ are given by
$$P_k=c_k\begin{vmatrix}
M_0&M_1&\ldots&M_k\\
M_1&M_2&\ldots&M_{k+1}\\
\vdots&\vdots&&\vdots\\
M_{k-1}&M_k&\ldots&M_{2k-1}\\
1&x&\ldots&x^k
\end{vmatrix}$$
where $M_k=\int_\mathbb Rx^kd\mu(x)$ are the moments of $\mu$, and $c_k\in\mathbb R^*$ can be any numbers.
\end{theorem}

\begin{proof}
Let us first see what happens at small values of $k\in\mathbb N$. At $k=0,1$ things are clear. At $k=2$ things get more complicated, with the formula being as follows:
$$P_2=c_2\begin{vmatrix}
M_0&M_1&M_2\\
M_1&M_2&M_3\\
1&x&x^2
\end{vmatrix}$$

However, no need for big computations here, in order to check the orthogonality, because by using the fact that $x^k$ integrates up to $M_k$, we obtain:
$$<1,P_2>=\int_\mathbb RP_2(x)d\mu(x)=c_2\begin{vmatrix}
M_0&M_1&M_2\\
M_1&M_2&M_3\\
M_0&M_1&M_2
\end{vmatrix}=0$$

Similarly, again by using the fact that $x^k$ integrates up to $M_k$, we have as well:
$$<x,P_2>=\int_\mathbb RxP_2(x)d\mu(x)=c_2\begin{vmatrix}
M_0&M_1&M_2\\
M_1&M_2&M_3\\
M_1&M_2&M_3
\end{vmatrix}=0$$

Thus, result proved at $k=0,1,2$, and the proof in general is similar.
\end{proof}

In practice now, all this leads us to a lot of interesting combinatorics, and countless things can be said. For the simplest measured space $X\subset\mathbb R$, which is the interval $[-1,1]$, with its uniform measure, the orthogonal basis problem can be solved as follows:

\index{Legendre polynomials}
\index{Rodrigues formula}
\index{Legendre equation}

\begin{theorem}
The orthonormal polynomials for $L^2[-1,1]$, subject to
$$\int_{-1}^1P_k(x)P_l(x)\,dx=\delta_{kl}$$
and called Legendre polynomials, satisfy the following differential equation,
$$(1-x^2)P_k''(x)-2xP_k'(x)+k(k+1)P_k(x)=0$$
which is the Legendre equation from physics. Moreover, we have the formula
$$(k+1)P_{k+1}(x)=(2k+1)xP_k(x)-kP_{k-1}(x)$$
called Bonnet recurrence formula, as well as the formula
$$P_k(x)=\frac{1}{2^kk!}\cdot\frac{d^k}{dx^k}\left(1-x^2\right)^k$$
called Rodrigues formula for the Legendre polynomials.
\end{theorem}

\begin{proof}
Thus is something very standard, the idea being as follows:

\medskip

(1) The first assertion is clear, because the Gram-Schmidt procedure applied to the Weierstrass basis $\{x^k\}$ can only lead to a certain family of polynomials $\{P_k\}$, with each $P_k$ being of degree $k$, and also unique, if we assume that it has positive leading coefficient, with this $\pm$ choice being needed, as usual, at each step of Gram-Schmidt.

\medskip

(2) Now thinking about what Gram-Schmidt does, this is certainly something by recurrence. And examining the recurrence leads to the Legendre equation, as stated. As for the Bonnet recurrence formula, the story here is similar.

\medskip

(3) Regarding the Rodrigues formula, by uniqueness no need to try to understand where this formula comes from, and we have two choices here, either by verifying that $\{P_k\}$ is orthonormal, or by verifying the Legendre equation. And both methods work.
\end{proof}

The above result is just the tip of the iceberg, and as a continuation, we have:

\index{Jacobi polynomials}
\index{Chebycheff polynomials}

\begin{theorem}
The orthogonal polynomials for $L^2[-1,1]$, with measure
$$d\mu(x)=(1-x)^a(1+x)^bdx$$
called Jacobi polynomials, satisfy as well a degree $2$ equation, namely
$$(1-x^2)P_k''(x)+(b-a-(a+b+2)x)P_k'(x)+k(k+a+b+1)P_k(x)=0$$
as well as an order $2$ recurrence relation, and are given by the following formula:
$$P_k(x)=\frac{(-1)^k}{2^kk!}(1-x)^{-a}(1+x)^{-b}\frac{d^k}{dx^k}\left[(1-x)^a(1+x)^b(1-x^2)^k\right]$$
At $a=b=0$ we recover the Legendre polynomials, and at $a=b=\pm\frac{1}{2}$ we recover the Chebycheff polynomials of the first and second kind, from trigonometry. 
\end{theorem}

\begin{proof}
There are many things going on here, the idea being as follows:

\medskip

(1) To start with, in what regards the precise statement, the order 2 recurrence relation mentioned there is something quite complicated, as follows:
\begin{eqnarray*}
&&2k(k+a+b)(2k+a+b-2)P_k(x)\\
&=&(2k+a+b-1)\left[(2k+a+b)(2k+a+b-2)x+a^2-b^2\right]P_{k-1}(x)\\
&-&2(k+a-1)(k+b-1)(2k+a+b)P_{k-2}(x)
\end{eqnarray*}

(2) Regarding now the proof, the statement appears as a generalization of Theorem 10.25, which corresponds to the particular case $a=b=0$, and the proof is similar.

\medskip

(3) Regarding now the main particular cases of the Jacobi polynomials, these are the Gegenbauer polynomials, appearing at $a=b$. However, there is not that much of a simplification when passing from general parameters $a,b$ to equal parameters, $a=b$, so in practice, the main particular cases are those indicated in the statement, namely:

\medskip

-- The Legendre polynomials, that we know well from Theorem 10.25, appearing at the simplest values of the parameters, namely $a=b=0$.

\medskip

-- The Chebycheff polynomials of the first kind $T_k$, which are given by the formula $T_k(\cos t)=\cos(kt)$ from trigonometry, appearing at $a=b=-\frac{1}{2}$.

\medskip

-- The Chebycheff polynomials of the second kind $U_k$, which are given by the formula $U_k(\cos t)\sin t=\sin((k+1)t)$, appearing at $a=b=\frac{1}{2}$.

\medskip

(4) So, this was for the story of the Jacobi polynomials, and their main particular cases, and in practice, we will leave some further learning here as an exercise, coming as a continuation of the further learning of Theorem 10.25, and its details.
\end{proof}

Getting now to other spaces $X\subset\mathbb R$, of particular interest here is the following result, which complements well Theorem 10.25, for the needs of basic quantum mechanics:

\index{Laguerre polynomials}

\begin{theorem}
The orthogonal polynomials for $L^2[0,\infty)$, with scalar product
$$<f,g>=\int_0^\infty f(x)g(x)e^{-x}\,dx$$
are the Laguerre polynomials $\{P_k\}$, satisfying the following differential equation,
$$xP_k''(x)+(1-x)P_k'(x)+kP_k(x)=0$$
as well as the following order $2$ recurrence relation,
$$(k+1)P_{k+1}(x)=(2k+1-x)P_k(x)-kP_{k-1}(x)$$
and which are given by the following formula,
$$P_k(x)=\frac{e^x}{k!}\cdot\frac{d^k}{dx^k}\left(e^{-x}x^k\right)$$
called Rodrigues formula for the Laguerre polynomials.
\end{theorem}

\begin{proof}
The story here is very similar to that of the Legendre and Jacobi polynomials, and many further things can be said here, with exercise for you to learn a bit about all this. Finally, for the story to be complete, no discussion about the Laguerre polynomials would be complete without a word about their use, in quantum mechanics. And here, as usual, we will leave some exploration of all this as an instructive exercise.
\end{proof}

Finally, regarding the space $X=\mathbb R$ itself, we have here the following result:

\index{Hermite polynomials}

\begin{theorem}
The orthogonal polynomials for $L^2(\mathbb R)$, with scalar product
$$<f,g>=\int_0^\infty f(x)g(x)e^{-x^2}\,dx$$
are the Hermite polynomials $\{P_k\}$, satisfying the following differential equation,
$$P_k''(x)-2xP_k'(x)+P_k(x)=0$$
as well as the following order $2$ recurrence relation,
$$P_{k+1}(x)=2xP_k(x)-2kP_{k-1}(x)$$
and which are given by the following formula,
$$P_k(x)=(-1)^k e^{x^2}\cdot\frac{d^k}{dx^k}\big(e^{-x^2}\big)$$
called Rodrigues formula for the Hermite polynomials.
\end{theorem}

\begin{proof}
As before, the story here is quite similar to that of the Legendre and other orthogonal polynomials, and exercise for you to learn a bit about all this.
\end{proof}

And with this, good news, end of the story with the orthogonal polynomials, at least at the very introductory level, and this due to the following fact, which is something quite technical, and that we will not attempt to prove, or even explain in detail here:

\begin{fact}
From an abstract point of view, coming from degree $2$ equations, and Rodrigues formulae for the solutions, there are only three types of ``classical'' orthogonal polynomials, namely the Jacobi, Laguerre and Hermite ones, discussed above.
\end{fact}

As a comment now, all the above material is very useful in advanced probability, with the orthogonal polynomials being more advanced technology than the moments, that we are so used to. With exercise of course for you, to learn more about all this.

\bigskip

Finally, as already mentioned, the above results are very useful in the context of basic quantum mechanics, and more specifically, for solving the hydrogen atom, following Heisenberg and Schr\"odinger. Again, exercise for you to learn a bit about this.

\section*{10d. Fourier series} 

Getting back to Theorem 10.21 as stated, heavily relying in the Weierstrass approximation theorem, the story is not over with that, and with the subsequent theory of orthogonal polynomials developed above. Indeed, as a rival to the Weierstrass approximation theorem we have the Fourier approximation theorem, leading to its own load of interesting mathematics and consequences, which are independent from Weierstrass.

\bigskip

The claim of Fourier, which is something very useful in practice, is as follows:

\begin{claim}
The space of square-summable functions on the unit circle,
$$L^2(\mathbb T)=\left\{f:\mathbb T\to\mathbb C\,\Big|\,\int_\mathbb T|f(z)|^2dz<\infty\right\}$$
with respect to the usual mass $1$ measure, has $\{z^n\}_{n\in\mathbb Z}$ as orthonormal basis. 
\end{claim}

As a first observation, this reminds a bit what we said before regarding the Weierstrass basis $\{x^n\}_{n\in\mathbb N}$ for the various spaces of functions $f:X\to\mathbb C$, with $X\subset\mathbb R$, and the related notion of orthogonal polynomials. That is, what Claim 10.30 says is that we have some sort of Weierstrass approximation theorem on the circle, but with respect to the 2-norm, and with the standard basis which is used, $\{z^n\}_{n\in\mathbb Z}$, being orthonormal for free.

\bigskip

In practice now, Claim 10.30 remains something which, while being certainly simple, beautiful and understandable, is a bit compact and abstract. For most purposes, it is better to replace it by the following equivalent formulation, in terms of real functions:

\begin{theorem}
The space of $2\pi$-periodic square-summable functions on $\mathbb R$,
$$L^2(\mathbb R)_{per}=\left\{f:\mathbb R\to\mathbb C\,\Big|\,f(t)=f(t+2\pi),\,\int_{-\pi}^\pi|f(t)|^2dt<\infty\right\}$$
has $\{e^{int}\}_{n\in\mathbb Z}$ as orthonormal basis, with respect to the normalized mass $1$ measure.
\end{theorem}

\begin{proof}
This is something quite tricky, which came as a big surprise at the time of its discovery by Fourier, the idea with all this being as follows:

\medskip

(1) As a first observation, which is philosophical, as already mentioned above, this is a real function reformulation of Claim 10.30, and for various reasons, in relation with both the proof and the future applications, this is the version that we will prefer.

\medskip

(2) Next, still talking generalities, according to what the statement says at the end, the scalar product on $L^2(\mathbb R)_{per}$ is by definition given by the following formula:
$$<f,g>=\frac{1}{2\pi}\int_{-\pi}^\pi f(t)\overline{g(t)}\,dt$$

As for the corresponding norm on $L^2(\mathbb R)_{per}$, this is given by the following formula:
$$||f||=\frac{1}{2\pi}\int_{-\pi}^\pi|f(t)|^2dt$$

Observe that we could have used $[0,2\pi]$ for integrating, or more generally, any interval $I\subset\mathbb R$ having length $2\pi$. For certain technical reasons, we prefer to use $I=[-\pi,\pi]$.

\medskip

(3) Getting now to the proof, as a first basic computation that we can do, coming from the $2\pi$-periodicity of $e^{it}=\cos t+i\sin t$, we have the following formula:
$$<e^{int},e^{imt}>
=\frac{1}{2\pi}\int_{-\pi}^\pi e^{i(n-m)t}dt
=\delta_{nm}$$

Thus $\{e^{int}\}_{n\in\mathbb Z}$ is as orthonormal system, and our theorem says that this is a basis.

\medskip

(4) But, how to prove this? The first thought goes to the Weierstrass approximation theorem, that we know well, but after thinking a bit, that is not of much use. Indeed, Weierstrass approximates the functions $f:[-\pi,\pi]\to\mathbb C$ by polynomials $\sum_nc_nt^n$, and what Fourier says is that the same functions $f:[-\pi,\pi]\to\mathbb C$ can be approximated by trigonometric polynomials $\sum_nc_ne^{int}$, which is a different story.

\medskip

(5) In fact, the difference between Fourier and Weierstrass is even more visible with Theorem 10.21 in mind. Indeed, up to a rescaling, Theorem 10.21 tells us that what we get from Weierstrass is the basis of $L^2(\mathbb R)_{per}$ formed by Legendre polynomials $\{L_k\}_{k\in\mathbb N}$, and the Fourier basis $\{e^{int}\}_{n\in\mathbb Z}$ is obviously something of different nature.

\medskip

(6) Time now for the proof? For reasons that will become clear in a moment, consider the following trigonometric polynomials, with $c_k\in\mathbb R$ being chosen for having mass $1$:
$$E_k(t)=c_k\left(\frac{1+\cos t}{2}\right)^k$$

Observe that $c_k$ can be computed explicitly, by using $1+\cos t=2\cos^2(t/2)$ and the Wallis formulae from chapter 7, but in what follows, we will not need this. Our claim, which is the reason for introducing these functions $E_k$, is that for any $\delta>0$ we have:
$$\lim_{k\to\infty}\ \sup_{\delta<|t|<\pi}E_k(t)=0$$

In other words, our claim is that $E_k\to 0$ uniformly on any $[-\pi,-\delta]\cup[\delta,\pi]$.

\medskip

(7) So, let us prove this claim. As mentioned, $c_k$ can be computed explicitly, but in what follows we will only need the following elementary estimate:
\begin{eqnarray*}
1
&=&\frac{c_k}{\pi}\int_0^\pi\left(\frac{1+\cos t}{2}\right)^kdt\\
&>&\frac{c_k}{\pi}\int_0^\pi\left(\frac{1+\cos t}{2}\right)^k\sin tdt\\
&=&\frac{c_k}{\pi}\left[-\frac{2}{k+1}\left(\frac{1+\cos t}{2}\right)^{k+1}\right]_0^\pi\\
&=&\frac{2c_k}{\pi(k+1)}
\end{eqnarray*}

Now since $E_k$ is decreasing on $[0,\pi]$, we obtain from this, for $\delta<|t|<\pi$:
$$E_k(t)<E_k(\delta)<\frac{\pi(k+1)}{2}\left(\frac{1+\cos\delta}{2}\right)^k$$

But this proves our claim, because for $\delta>0$ we have $(1+\cos\delta)/2<1$, as needed.

\medskip

(8) Getting now to what we wanted to do, we must prove that $\{e^{int}\}_{n\in\mathbb Z}$ spans a dense subset of $L^2(\mathbb R)_{per}$. Since $C(\mathbb R)_{per}\subset L^2(\mathbb R)_{per}$ is dense, it is enough to prove that any $f\in C(\mathbb R)_{per}$ can be approximated by trigonometric polynomials $\sum_nc_ne^{int}$. Moreover, since $||.||_2\leq||.||_\infty$, it is enough to prove our approximation with respect to $||.||_\infty$.

\medskip

(9) All in all, it remains to prove that given a function $f\in C(\mathbb R)_{per}$ and a number $\varepsilon>0$, we can always come with a trigonometric polynomial $\sum_nc_ne^{int}$, such that:
$$\left|f(t)-\sum_nc_ne^{int}\right|<\varepsilon\quad,\quad\forall t\in[-\pi,\pi]$$

(10) But for this, we can use the polynomials $E_k$ from (6). Let us set indeed:
$$Q_k(t)=\frac{1}{2\pi}\int_{-\pi}^\pi f(t-s)E_k(s)\,ds$$

As a first observation, with the change of variables $s\to t-s$, we have the following alternative formula, which shows that $Q_k(t)$ are indeed trigonometric polynomials:
$$Q_k(t)=\frac{1}{2\pi}\int_{-\pi}^\pi f(s)E_k(t-s)\,ds$$

(11) Now given $\varepsilon>0$, let us prove that the estimate in (9) holds indeed, with the trigonometric polynomial there being $Q_k(t)$, for $k>>0$ large enough. For this purpose, we use the uniform continuity of $f$, which tells us that we can find $\delta>0$ such that:
$$|s-t|<\delta\implies|f(s)-f(t)|<\varepsilon$$

Indeed, by using this, we have the following estimate, for the error in (9):
\begin{eqnarray*}
|Q_k(t)-f(t)|
&=&\frac{1}{2\pi}\left|\int_{-\pi}^\pi(f(t-s)-f(t))E_k(s)\,ds\right|\\
&\leq&\frac{1}{2\pi}\int_{-\pi}^\pi|f(t-s)-f(t)|E_k(s)\,ds
\end{eqnarray*}

(12) Now let us split the last integral into three parts, according to:
$$[-\pi,\pi]=[-\pi,-\delta]\cup[-\delta,\delta]\cup[\delta,\pi]$$

On the middle part the integrand is $<\varepsilon$, so the middle integral is $<\varepsilon$. As for the other two integrals, on $[-\pi,-\delta]\cup[\delta,\pi]$, we can use here (6), telling us that $E_k(t)\to0$ uniformly, on that domain. Indeed, with $k>>0$ big enough the other two integrals are $<\varepsilon$ too, so we have obtained (9) as desired, with $\varepsilon\to3\varepsilon$, which finishes the proof. 
\end{proof}

Still with me I hope, after all these computations. In practice now, a bit as Theorem 10.31 itself was a useful version of Claim 10.30, there are several other useful versions of Theorem 10.31, which can be of great use, in practice. We notably have here:

\begin{theorem}
We have an isomorphism $L^2(\mathbb R)_{per}\simeq l^2(\mathbb Z)$, as follows:
\begin{enumerate}
\item Associated to $f\in L^2(\mathbb R)_{per}$ are its Fourier coefficients, given by:
$$\widehat{f}(n)=\frac{1}{2\pi}\int_{-\pi}^\pi f(t)e^{int}dt$$

\item Associated to $g\in l^2(\mathbb Z)$ is the series $S_g(t)=\sum_{n\in\mathbb Z}g(n)e^{-int}$.
\end{enumerate}
\end{theorem}

\begin{proof}
This is something self-explanatory, based on the general orthonormal basis theory from Theorem 10.19, with the Fourier coefficients of $f\in L^2(\mathbb R)_{per}$ being its coefficients $\widehat{f}(n)=<f,e^{-int}>$ with respect to the basis $\{e^{int}\}$ from Theorem 10.31.
\end{proof}

In practice, it is possible to talk as well about real Fourier coefficients, given by $\widehat{f}(n)=a_n+ib_n$, and with the formula of these, for $f:\mathbb R\to\mathbb R$, being as follows:
$$a_n=\frac{1}{2\pi}\int_{-\pi}^\pi f(t)\cos(nt)dt\quad,\quad 
b_n=\frac{1}{2\pi}\int_{-\pi}^\pi f(t)\sin(nt)dt$$

And exercise of course for you, to learn more about all this. Finally, as yet another reformulation, or rather consequence of what we have, we have the Parseval formula:

\begin{theorem}
The Fourier coefficients $\widehat{f}(n)=\frac{1}{2\pi}\int_{-\pi}^\pi f(t)e^{int}dt$ satisfy
$$\sum_{n\in\mathbb Z}\widehat{f}(n)\overline{\widehat{g}(n)}=\frac{1}{2\pi}\int_{-\pi}^\pi f(t)\overline{g(t)}\,dt$$
for any $f,g\in L^2(\mathbb R)_{per}$, and in particular satisfy the formula
$$\sum_{n\in\mathbb Z}|\widehat{f}(n)|^2=\frac{1}{2\pi}\int_{-\pi}^\pi|f(t)|^2dt$$
for any $f\in L^2(\mathbb R)_{per}$, with this being called Parseval formula.
\end{theorem}

\begin{proof}
This is indeed yet another reformulation of what we have, coming from the fact that the scalar products and norms are invariant under $L^2(\mathbb R)_{per}\simeq l^2(\mathbb Z)$.
\end{proof}

With this discussed, time perhaps for some applications? And here, there are countless of them, because the above technology can be used in order to decompose various signals, such as mechanical, electromagnetic, seismic or acoustic waves, or even solutions of more complicated differential equations, somewhat of wave type, into sinusoids. 

\bigskip

Thus, many things to be learned here, and for more, have a look at any advanced calculus book. In what concerns us, let us just present an application to arithmetic:

\begin{theorem}
We have the following formula of Euler,
$$1+\frac{1}{4}+\frac{1}{9}+\frac{1}{16}+\frac{1}{25}+\ldots=\frac{\pi^2}{6}$$
computing $\zeta(2)$, and solving the Basel problem.
\end{theorem}

\begin{proof}
To start with, we have the following pioneering computation of Euler, obtained by factorizing the function $\sin x$, having zeroes at $\mathbb Z/2$, a bit like a polynomial, which gives the result, by looking at the coefficient of $x^2$:
\begin{eqnarray*}
\frac{\sin x}{x}
&=&1-\frac{x^2}{3!}+\frac{x^4}{5!}-\frac{x^6}{7!}+\ldots\\
&=&\left(1-\frac{x}{\pi}\right)\left(1+\frac{x}{\pi}\right)\left(1-\frac{x}{2\pi}\right)\left(1+\frac{x}{2\pi}\right)\ldots\\
&=&\left(1-\frac{x^2}{\pi^2}\right)\left(1-\frac{x^2}{4\pi^2}\right)\left(1-\frac{x^2}{9\pi^2}\right)\ldots\\
&=&1-\frac{1}{\pi^2}\sum_{n=1}^\infty\frac{1}{n^2}\,x^2+\ldots
\end{eqnarray*}

Of course this is far from being rigorous, but following Weierstrass, it is possible to fix this, with the above formula for $\sin x/x$ being indeed true. But, this is not trivial.

\medskip

(2) As an alternative approach, much quicker, we can use our Fourier series knowledge. Indeed, the nonzero Fourier coefficients of the function $f(t)=t$ on $[-\pi,\pi]$ are:
$$\widehat{f}(n)
=\frac{1}{2\pi}\int_{-\pi}^\pi te^{int}\,dt
=\frac{1}{2\pi}\left[\frac{1-int}{n^2}\,e^{int}\right]_{-\pi}^\pi
=\frac{(-1)^{n+1}}{n}\,i$$

Thus, the Parseval formula for the function $f(t)=t$ gives:
$$\sum_{n\in\mathbb Z^*}\frac{1}{n^2}=\frac{1}{2\pi}\int_{-\pi}^\pi t^2dt=\frac{1}{2\pi}\cdot\frac{2\pi^3}{3}=\frac{\pi^2}{3}$$

And we therefore solved the Basel problem, just like that. Amazing.
\end{proof}

\section*{10e. Exercises}

We had an exciting chapter here, and as exercises on this, we have:

\begin{exercise}
What is the formula of the rank one projection on $x\in\mathbb C^N$?
\end{exercise}

\begin{exercise}
Prove that the diagonalizable matrices in $M_N(\mathbb C)$ are dense.
\end{exercise}

\begin{exercise}
Work out the equality case in Cauchy-Schwarz, and Minkowski.
\end{exercise}

\begin{exercise}
Review the full basic Banach space theory, for the Hilbert spaces.
\end{exercise}

\begin{exercise}
Learn if needed the full proof of the Zorn lemma.
\end{exercise}

\begin{exercise}
Learn more about various families of orthogonal polynomials.
\end{exercise}

\begin{exercise}
Learn more about Fourier series, and their various applications.
\end{exercise}

\begin{exercise}
Learn about the Riemann zeta function, and its special values.
\end{exercise}

As bonus exercise, as suggested by the cats, learn some quantum mechanics.

\chapter{Function spaces}

\section*{11a. Distributions}

Welcome to advanced functional analysis. We have seen in chapters 9-10 the basic theory of Banach spaces and Hilbert spaces, and normally, with the knowledge of the $L^p$ spaces, and in particular of the $L^2$ spaces, we can do many things. However, there are far more things that must be known, and this even at the beginner level, and this chapter and the next one will be an introduction to this. Our plan will be as follows:

\bigskip

(1) In this chapter, which will be basically a continuation of chapter 9, we will discuss a number of fundamental functional analysis topics, namely convolution, distributions, complex analysis,  Fourier analysis, and some basic applications to physics.

\bigskip

(2) And in the next chapter, which will be basically a continuation of chapter 10, we will discuss a fundamental question that we still have left from there, namely the diagonalization of the self-adjoint operators, and of the normal operators.

\bigskip

Getting started, let us first talk about convolution. We have already talked in Part I about the convolution of Dirac masses, according to the formula $\delta_x*\delta_y=\delta_{x+y}$, and of more general discrete measures, according to this formula, and linearity. In the continuous measure setting, at the level of densities, the relevant definition is as follows:

\index{convolution}

\begin{definition}
The convolution of two functions $f,g:\mathbb R\to\mathbb C$ is the function
$$(f*g)(x)=\int_\mathbb R f(x-y)g(y)dy$$
provided that the function $y\to f(x-y)g(y)$ is indeed integrable, for any $x$.
\end{definition}

There are many reasons for introducing this operation, that we will gradually discover, in what follows. As a basic example, let us take $g=\chi_{[0,1]}$. We have then:
$$(f*g)(x)=\int_0^1f(x-y)dy$$

Thus, with this choice of $g$, the operation $f\to f*g$ has some sort of ``regularizing effect'', that can be useful for many purposes. We will be back to this, later.

\bigskip

Goinh ahead with more theory, let us try to understand when the convolution operation is well-defined. We have here the following basic result:

\index{compact support}
\index{support}

\begin{proposition}
The convolution operation is well-defined on the space
$$C_c(\mathbb R)=\left\{f\in C(\mathbb R)\Big|supp(f)=\ {\rm compact}\right\}$$
of continuous functions $f:\mathbb R\to\mathbb C$ having compact support.
\end{proposition}

\begin{proof}
We have several things to be proved, the idea being as follows:

\medskip

(1) First we must show that given two functions $f,g\in C_c(\mathbb R)$, their convolution $f*g$ is well-defined, as a function $f*g:\mathbb R\to\mathbb C$. But this follows from the following estimate, where $l$ denotes the length of the compact subsets of $\mathbb R$:
\begin{eqnarray*}
\int_\mathbb R|f(x-y)g(y)|dy
&=&\int_{supp(g)}|f(x-y)g(y)|dy\\
&\leq&\max(g)\int_{supp(g)}|f(x-y)|dy\\
&\leq&\max(g)\cdot l(supp(g))\cdot\max(f)\\
&<&\infty
\end{eqnarray*}

(2) Next, we must show that the function $f*g:\mathbb R\to\mathbb C$ that we constructed is indeed continuous. But this follows from the following estimate, where $K_f$ is the constant of uniform continuity for the function $f\in C_c(\mathbb R)$:
\begin{eqnarray*}
|(f*g)(x+\varepsilon)-(f*g)(x)|
&=&\left|\int_\mathbb R f(x+\varepsilon-y)g(y)dy-\int_\mathbb R f(x-y)g(y)dy\right|\\
&=&\left|\int_\mathbb R\left(f(x+\varepsilon-y)-f(x-y)\right)g(y)dy\right|\\
&\leq&\int_\mathbb R|f(x+\varepsilon-y)-f(x-y)|\cdot|g(y)|dy\\
&\leq&K_f\cdot\varepsilon\cdot\int_\mathbb R|g|
\end{eqnarray*}

(3) Finally, we must show that the function $f*g\in C(\mathbb R)$ that we constructed has indeed compact support. For this purpose, our claim is that we have:
$$supp(f+g)\subset supp(f)+supp(g)$$

In order to prove this claim, observe that we have, by definition of $f*g$:
$$(f*g)(x)
=\int_\mathbb R f(x-y)g(y)dy
=\int_{supp(g)}f(x-y)g(y)dy$$

But this latter quantity being 0 for $x\notin supp(f)+supp(g)$, this gives the result.
\end{proof}

Here are now a few remarkable properties of the convolution operation:

\begin{proposition}
The following hold, for the functions in $C_c(\mathbb R)$:
\begin{enumerate}
\item $f*g=g*f$.

\item $f*(g*h)=(f*g)*h$.

\item $f*(\lambda g+\mu h)=\lambda f*g+\mu f*h$.
\end{enumerate}
\end{proposition}

\begin{proof}
These formulae are all elementary, the idea being as follows:

\medskip

(1) This follows from the following computation, with $y=x-t$:
\begin{eqnarray*}
(f*g)(x)
&=&\int_\mathbb R f(x-y)g(y)dy\\
&=&\int_\mathbb R f(t)g(x-t)dt\\
&=&\int_\mathbb R g(x-t)f(t)dt\\
&=&(g*f)(x)
\end{eqnarray*}

(2) This is clear from definitions.

\medskip

(3) Once again, this is clear from definitions.
\end{proof}

In relation with derivatives, and with the ``regularizing effect'' of the convolution operation mentioned after Definition 11.1, we have the following result:

\index{convolution}
\index{regularization}

\begin{theorem}
Given two functions $f,g\in C_c(\mathbb R)$, assuming that $g$ is differentiable, then so is $f*g$, with derivative given by the following formula:
$$(f*g)'=f*g'$$
More generally, given $f,g\in C_c(\mathbb R)$, and assuming that $g$ is $k$ times differentiable, then so is $f*g$, with $k$-th derivative given by $(f*g)^{(k)}=f*g^{(k)}$.
\end{theorem}

\begin{proof}
In what regards the first assertion, with $y=x-t$, then $t=x-y$, we get:
\begin{eqnarray*}
(f*g)'(x)
&=&\frac{d}{dx}\int_\mathbb R f(x-y)g(y)dy\\
&=&\frac{d}{dx}\int_\mathbb R f(t)g(x-t)dt\\
&=&\int_\mathbb R f(t)g'(x-t)dt\\
&=&\int_\mathbb R f(x-y)g'(y)dy\\
&=&(f*g')(x)
\end{eqnarray*}

As for the second assertion, this follows form the first one, by recurrence.
\end{proof}

Finally, getting beyond the compactly supported continuous functions, we have the following result, which is of particular theoretical importance:

\begin{theorem}
The convolution operation is well-defined on $L^1(\mathbb R)$, and we have:
$$||f*g||_1\leq||f||_1||g||_1$$
Thus, if $f\in L^1(\mathbb R)$ and $g\in C_c^k(\mathbb R)$, then $f*g$ is well-defined, and $f*g\in C_c^k(\mathbb R)$.
\end{theorem}

\begin{proof}
In what regards the first assertion, this follows from:
\begin{eqnarray*}
\int_\mathbb R|(f*g)(x)|dx
&\leq&\int_\mathbb R\int_\mathbb R|f(x-y)g(y)|dydx\\
&=&\int_\mathbb R\int_\mathbb R|f(x-y)g(y)|dxdy\\
&=&\int_\mathbb R|f(x)|dx\int_\mathbb R|g(y)|dy
\end{eqnarray*}

As for the second assertion, this follows from this, and from Theorem 11.4.
\end{proof}

Let us discuss now a related topic, the mathematical distributions. These are something quite smart, and as an advertisement for what we will be doing, we have:

\begin{advertisement}
With a suitable theory of distributions, covering both the functions and the Dirac masses, the basic step function, namely
$$H(x)=\begin{cases}
0&(x\leq0)\\
1&(x>0)
\end{cases}$$
is differentiable when viewed as distribution, with derivative $H'=\delta_0$.
\end{advertisement}

And isn't this crazy, hope you agree with me. Getting started now, there is a price to pay for doing such things, namely formulating a technical definition, as follows:

\index{distribution}

\begin{definition}
A distribution on an open interval $I\subset\mathbb R$ is a functional
$$\varphi:C_c^\infty(I)\to\mathbb C$$
such that for any $K\subset I$ compact, there exist $n\in\mathbb N$ and $c>0$ such that
$$|\varphi(f)|\leq c||f||_{C^n(K)}$$
for any $f\in C_c^\infty(I)$ having support in $K$, where $||f||_{C^n(K)}=\sup_{x\in K}\sum_{i=0}^n|f^{(i)}(x)|$.
\end{definition}

At the level of main examples of distributions, we have the integration functionals associated to the measures, and in particular to the measures having a density. In view of this, we can consider any function $f\in L^1(I)$, viewed as density, as being a distribution. Other basic examples include the Dirac masses $\delta_x$ at the points $x\in I$. 

\bigskip

Regarding the general theory of distributions, that is quite similar to the theory of functions. Algebrically, the distributions form a vector space, and are subject to a number of supplementary operations, such as dilations, translations and so on, and multiplication by functions too. Analytically, we can talk about convergence of distributions, $\varphi_n\to\varphi$, and about their support too, $supp(\varphi)\subset I$, in a quite straightforward way. Leaving all this aside, and getting now to what we wanted to do, derivatives, we have here:

\begin{theorem}
We can talk about the derivatives of distributions, given by
$$\varphi'(f)=-\varphi(f')$$
and with this notion in hand, the following happen:
\begin{enumerate}
\item When $\varphi$ is a usual differentiable function, $\varphi'$ is the usual derivative.

\item For the basic step function we have $H'=\delta_0$, as previously advertised.

\item In fact, for a function $\varphi=g$ with jumps at $\{x_i\}$, we have $\varphi'=g'+\sum_iJ_g(x_i)\delta_{x_i}$.
\end{enumerate}
\end{theorem}

\begin{proof}
The first assertion, which by the way explains the need for the $-$ sign, follows from the Leibnitz rule for derivatives. Regarding the second assertion, this comes from:
\begin{eqnarray*}
H'(f)
&=&-H(f')\\
&=&-\int_0^\infty f'(x)dx\\
&=&-f(\infty)+f(0)\\
&=&f(0)\\
&=&\delta_0(f)
\end{eqnarray*}

As for the third assertion, which generalizes (1,2), we will leave this as an exercise.
\end{proof}

Summarizing, job done. Many other things can be said about distributions, and we would like to end with a physics trick. We have the following computation, for $a>0$:
\begin{eqnarray*}
\int_\mathbb Rf(x)\delta(x^2-a^2)dx
&=&\int_{-\infty}^0f(x)\delta(x^2-a^2)dx+\int_0^\infty f(x)\delta(x^2-a^2)dx\\
&=&\int_{-\infty}^af(y-a)\delta(y^2-2ay)dy+\int_{-a}^\infty f(y+a)\delta(y^2+2ay)dy\\
&\simeq&\int_{-\infty}^af(y-a)\delta(-2ay)dy+\int_{-a}^\infty f(y+a)\delta(2ay)dy\\
&=&\int_{-\infty}^{2a^2}f\left(\frac{z}{2a}-a\right)\delta(-z)\frac{dz}{2a}+\int_{-2a^2}^\infty f\left(\frac{z}{2a}+a\right)\delta(z)\frac{dz}{2a}\\
&=&\frac{f(-a)}{2a}+\frac{f(a)}{2a}\\
&=&\int_Rf(x)\,\frac{\delta(x-a)+\delta(x+a)}{2a}\,dx
\end{eqnarray*}

Sounds like physics, you would say, and in answer, yes physics that is, but in any case, we have in this way a definition for the quadratic Dirac masses, as follows:
$$\delta(x^2-a^2)=\frac{\delta(x-a)+\delta(x+a)}{2a}$$

And we will end our discussion here. Good learning this was, and exercise for you to learn more about distributions, and their applications to mathematics and physics.

\section*{11b. Complex analysis} 

Time now for some complex analysis. You certainly know about holomorphic functions, and perhaps about harmonic functions too, and time to review all this. At the beginning, we have the Cauchy formula for the complex functions, as follows:

\begin{theorem}
The Cauchy formula, namely
$$f(x)=\frac{1}{2\pi i}\int_\gamma\frac{f(y)}{y-x}\,dy$$
holds for any holomorphic function $f:X\to\mathbb C$.
\end{theorem}

\begin{proof}
This is something standard, which can be proved as follows:

\medskip

(1) Our first claim is that given $f\in Hol(X)$, with $f'\in C(X)$, the integral of $f'$ vanishes on any path. Indeed, by using the change of variable formula, we have:
\begin{eqnarray*}
\int_\gamma f'(z)dz
&=&\int_a^bf'(\gamma(t))\gamma'(t)dt\\
&=&f(\gamma(b))-f(\gamma(a))\\
&=&0
\end{eqnarray*}

(2) Our second claim is that given $f\in Hol(X)$ and a triangle $\Delta\subset X$, we have:
$$\int_\Delta f(z)dz=0$$

Indeed, let us call $\Delta=ABC$ our triangle. Now consider the midpoints $A',B',C'$ of the edges $BC,CA,AB$, and then consider the following smaller triangles:
$$\Delta_1=AC'B'\quad,\quad\Delta_2=BA'C'\quad,\quad\Delta_3=CB'A'\quad,\quad\Delta_4=A'B'C'$$

These smaller triangles partition then $\Delta$, and due to our above conventions for the vertex ordering, which produce cancellations when integrating over them, we have:
$$\int_\Delta f(z)dz=\sum_{i=1}^4\int_{\Delta_i}f(z)dz$$

Thus we can pick, among the triangles $\Delta_i$, a triangle $\Delta^{(1)}$ such that:
$$\left|\int_\Delta f(z)dz\right|\leq 4\left|\int_{\Delta^{(1)}}f(z)dz\right|$$

In fact, by repeating the procedure, we obtain triangles $\Delta^{(n)}$ such that:
$$\left|\int_\Delta f(z)dz\right|\leq 4^n\left|\int_{\Delta^{(n)}}f(z)dz\right|$$

(3) Now let $x$ be the limiting point of these triangles $\Delta^{(n)}$, and fix $\varepsilon>0$. By using the fact that the functions $1,z$ integrate over paths up to 0, coming from (1), we obtain the following estimate, with $n\in\mathbb N$ being big enough, and $L$ being the perimeter of $\Delta$:
\begin{eqnarray*}
\left|\int_{\Delta^{(n)}}f(z)dz\right|
&=&\left|\int_{\Delta^{(n)}}f(z)-f(x)-f'(x)(z-x)dz\right|\\
&\leq&\int_{\Delta^{(n)}}\left|f(z)-f(x)-f'(x)(z-x)\right|dz\\
&\leq&\int_{\Delta^{(n)}}\varepsilon|z-x|dz\\
&\leq&\varepsilon(2^{-n}L)^2
\end{eqnarray*}

Now by combining this with the estimate in (2), this proves our claim.

\medskip

(4) The rest is quite routine. First, we can pass from triangles to boundaries of convex sets, in a straightforward way, with the same conclusion as in (2), namely:
$$\int_\gamma f(z)dz=0$$

Getting back to what we want to prove, namely the Cauchy formula for an arbitrary holomorphic function $f\in Hol(X)$, let $z\in X$, and consider the following function:
$$g(y)=\begin{cases}
\frac{f(y)-f(z)}{y-z}&(y\neq z)\\
f'(z)&(y=z)
\end{cases}$$

Now assuming that $\gamma$ encloses a convex set, we can apply what we found, namely vanishing of the integral, to this function $g$, and we obtain the Cauchy formula for $f$.

\medskip

(5) Finally, the extension to general curves is standard. See Rudin \cite{ru2}.
\end{proof}

As a main application of the Cauchy formula, we have:

\index{holomorphic function}
\index{infinitely differentiable}
\index{analytic function}
\index{Cauchy formula}

\begin{theorem}
The following conditions are equivalent, for a function $f:X\to\mathbb C$:
\begin{enumerate}
\item $f$ is holomorphic.

\item $f$ is infinitely differentiable.

\item $f$ is analytic.

\item The Cauchy formula holds for $f$.
\end{enumerate}
\end{theorem}

\begin{proof}
This is routine from what we have, the idea being as follows:

\medskip

$(1)\implies(4)$ is non-trivial, but we know this from Theorem 11.9.

\medskip

$(4)\implies(3)$ is trivial, by expanding the series in the Cauchy formula.

\medskip

$(3)\implies(2)\implies(1)$ are both elementary, and good exercises for you.
\end{proof}

As another application of the Cauchy formula, we have:

\index{maximum principle}

\begin{theorem}
Any holomorphic function $f:X\to\mathbb C$ satisfies the maximum principle, in the sense that given a domain $D$, with boundary $\gamma$, we have:
$$\exists x\in\gamma\quad,\quad |f(x)|=\max_{y\in D}|f(y)|$$
That is, the maximum of $|f|$ over a domain is attained on its boundary.
\end{theorem}

\begin{proof}
This follows indeed from the Cauchy formula. Observe that the converse is not true, for instance because functions like $\bar{x}$ satisfy too the maximum principle.
\end{proof}

Conceptually, a good explanation for the fact that the maximum principle holds could be that the values of our function inside a disk can be recovered from its values on the boundary. And fortunately, this is indeed the case, and we have:

\index{main value formula}

\begin{theorem}
Given an holomorphic function $f:X\to\mathbb C$, and a disk $D$, with boundary $\gamma$, the following happen:
\begin{enumerate}
\item $f$ satisfies the plain mean value formula $f(x)=\int_Df(y)dy$.

\item $f$ satisfies the boundary mean value formula $f(x)=\int_\gamma f(y)dy$.
\end{enumerate}
\end{theorem}

\begin{proof}
As usual, this follows from the Cauchy formula, which is something truly powerful, and with working out the details being a good exercise for you.
\end{proof}

Finally, as yet another application of the Cauchy formula, which is something nice-looking and conceptual, we have the following statement, called Liouville theorem:

\index{Liouville theorem}

\begin{theorem}
An entire, bounded holomorphic function
$$f:\mathbb C\to\mathbb C\quad,\quad |f|\leq M$$
must be constant. In particular, knowing $f\to0$ with $z\to\infty$ gives $f=0$.
\end{theorem}

\begin{proof}
This follows as usual from the Cauchy formula, in the obvious way. Alternatively, we can view this as a consequence of Theorem 11.12, because given two points $x\neq y$, we can view the values of $f$ at these points as averages over big disks centered at these points, say $D=D_x(R)$ and $E=D_y(R)$, with $R>>0$:
$$f(x)=\int_Df(z)dz\quad,\quad f(y)=\int_Ef(z)dz$$

Indeed, the point is that when the radius goes to $\infty$, these averages tend to be equal, and so we have $f(x)\simeq f(y)$, which gives $f(x)=f(y)$ in the limit.
\end{proof}

In view of a continuation of this, let us formulate the following definition:

\index{Laplace operator}
\index{harmonic function}

\begin{definition}
The Laplace operator in $2$ dimensions is:
$$\Delta f=\frac{d^2f}{dx^2}+\frac{d^2f}{dy^2}$$
A function $f:\mathbb R^2\to\mathbb C$ satisfying $\Delta f=0$ will be called harmonic.
\end{definition}

Here the formula of $\Delta$ is the one coming from physics, and more on this later. As for the notion of harmonic function, this is something quite natural. Indeed, we can think of $\Delta$ as being a linear operator on the space of functions $f:\mathbb R^2\to\mathbb C$, and previous experience with linear operators, and linear algebra in general, suggests looking first into the eigenvectors of $\Delta$. But the simplest such eigenvectors are those corresponding to the eigenvalue $\lambda=0$, and these are exactly our harmonic functions, satisfying:
$$\Delta f=0$$

So, let us try to find the functions $f:\mathbb R^2\to\mathbb C$ which are harmonic. And here, as a good surprise, we have an interesting link with the holomorphic functions:

\index{complex conjugate}

\begin{theorem}
Any holomorphic function $f:\mathbb C\to\mathbb C$, when regarded as real function
$$f:\mathbb R^2\to\mathbb C$$
is harmonic. Moreover, the conjugates $\bar{f}$ of holomorphic functions are harmonic too.
\end{theorem}

\begin{proof}
The first assertion comes from the following computation, with $z=x+iy$:
\begin{eqnarray*}
\Delta z^n
&=&\frac{d^2z^n}{dx^2}+\frac{d^2z^n}{dy^2}\\
&=&\frac{d(nz^{n-1})}{dx}+\frac{d(inz^{n-1})}{dy}\\
&=&n(n-1)z^{n-2}-n(n-1)z^{n-2}\\
&=&0
\end{eqnarray*}

As for the second assertion, this follows from $\Delta\bar{f}=\overline{\Delta f}$, which is itself clear from definitions, and which shows that if $f$ is harmonic, then so is its conjugate $\bar{f}$.
\end{proof}

In view of the above, we can expect the harmonic functions to share many interesting properties with the holomorphic functions. We will be back to this, which holds indeed, in a moment. Moving now to arbitrary $N\in\mathbb N$ dimensions, let us formulate:

\index{Laplace operator}
\index{harmonic function}

\begin{definition}
The Laplace operator in $N$ dimensions is:
$$\Delta f=\sum_{i=1}^N\frac{d^2f}{dx_i^2}$$
A function $f:\mathbb R^N\to\mathbb C$ satisfying $\Delta f=0$ will be called harmonic.
\end{definition}

Again, these are notions coming from physics, and more on this later, at the end of the present chapter, where we will discuss the wave equation, involving $\Delta$. In view of Theorem 11.15, and in analogy with the theory of holomorphic functions, we have:

\index{maximum modulus}
\index{mean value formula}
\index{Liouville theorem}

\begin{theorem}
The harmonic functions in $N$ dimensions obey to the same general principles as the holomorphic functions, namely:
\begin{enumerate}
\item The plain mean value formula. 

\item The boundary mean value formula.

\item The maximum modulus principle.

\item The Liouville theorem.
\end{enumerate}
\end{theorem}

\begin{proof}
This is something quite tricky, the idea being as follows:

\medskip

(1) Regarding the plain mean value formula, here the statement is that given an harmonic function $f:X\to\mathbb C$, and a ball $B$, the following happens:
$$f(x)=\int_Bf(y)dy$$

In order to prove this, we can assume that $B$ is centered at $0$, of radius $r>0$. If we denote by $\chi_r$ the characteristic function of this ball, nomalized as to integrate up to 1, in terms of the usual convolution operation for functions, we want to prove that we have:
$$f=f*\chi_r$$

For proving this formula, pick $0<s<r$, and a solution $w$ of the following equation, on $B$, which can be constructed explicitly, say as a radial function:
$$\Delta w=\chi_r-\chi_s$$

By using the basic properties of the convolution operation $*$, we have:
\begin{eqnarray*}
f*\chi_r-f*\chi_s
&=&f*(\chi_r-\chi_s)\\
&=&f*\Delta w\\
&=&\Delta f*w\\
&=&0
\end{eqnarray*}

Thus $f*\chi_r=f*\chi_s$, and by letting now $s\to0$, we get $f*\chi_r=f$, as desired.

\medskip

(2) Regarding the boundary mean value formula, here the statement is that given an harmonic function $f:X\to\mathbb C$, and a ball $B$, with boundary $\gamma$, the following happens:
$$f(x)=\int_\gamma f(y)dy$$

But this follows as a consequence of the plain mean value formula in (1), with our two mean value formulae, the one there and the one here, being in fact equivalent, by using annuli and radial integration for the proof of the equivalence, in the obvious way. 

\medskip

(3) Regarding the maximum modulus principle, the statement here is that any holomorphic function $f:X\to\mathbb C$ has the property that the maximum of $|f|$ over a domain is attained on its boundary. That is, given a domain $D$, with boundary $\gamma$, we have:
$$\exists x\in\gamma\quad,\quad |f(x)|=\max_{y\in D}|f(y)|$$

But this is something which follows again from the mean value formula in (1), first for the balls, and then in general, by using a standard division argument.

\medskip

(4) Finally, regarding the Liouville theorem, we can view this as a consequence of (1), because given two points $x\neq y$, we can view the values of $f$ at these points as averages over big balls centered at these points, say $B=B_x(R)$ and $C=B_y(R)$, with $R>>0$:
$$f(x)=\int_Bf(z)dz\quad,\quad f(y)=\int_Cf(z)dz$$

Indeed, the point is that when the radius goes to $\infty$, these averages tend to be equal, and so we have $f(x)\simeq f(y)$, which gives $f(x)=f(y)$ in the limit, as desired.
\end{proof}

In addition to the above, let us mention too that at $N=2$, locally, the real harmonic functions are known to be the real parts of holomorphic functions. However, this is something non-trivial, and for the proof here, we refer to Rudin \cite{ru2}. 

\bigskip

Also, there is an interesting relation between harmonic functions and soap fims. And exercise for you, to learn more about this, with all this being first class mathematics.

\section*{11c. Fourier analysis}

Let us discuss now the construction and main properties of the Fourier transform, which is the main tool in analysis, and mathematics in general. We first have:

\index{Fourier transform}

\begin{definition}
Given $f\in L^1(\mathbb R)$, we define a function $\widehat{f}:\mathbb R\to\mathbb C$ by
$$\widehat{f}(x)=\int_\mathbb R f(t)e^{ixt}dt$$
and call it Fourier transform of $f$.
\end{definition}

As a first observation, even if $f$ is a real function, $\widehat{f}$ is a complex function, which is not necessarily real. Also, $\widehat{f}$ is obviously well-defined, because $f\in L^1(\mathbb R)$ and $|e^{ixt}|=1$. Also, the condition $f\in L^1(\mathbb R)$ is basically needed for construcing $\widehat{f}$, because:
$$\widehat{f}(0)=\int_\mathbb Rf(t)dt$$

Generally speaking, the Fourier transform is there for helping with various computations, with the above formula $\widehat{f}(0)=\int f$ being something quite illustrating. Here are some basic properties of the Fourier transform, all providing some good motivations:

\begin{proposition}
The Fourier transform has the following properties:
\begin{enumerate}
\item Linearity: $\widehat{f+g}=\widehat{f}+\widehat{g}$, $\widehat{\lambda f}=\lambda\widehat{f}$.

\item Regularity: $\widehat{f}$ is continuous and bounded.

\item If $f$ is even then $\widehat{f}$ is even. 

\item If $f$ is odd then $\widehat{f}$ is odd. 
\end{enumerate}
\end{proposition}

\begin{proof}
These results are all elementary, as follows:

\medskip

(1) The linearity formulae are indeed both clear from definitions.

\medskip

(2) The continuity of $\widehat{f}$ comes indeed from the following estimate:
$$\left|\widehat{f}(x+\varepsilon)-\widehat{f}(x)\right|
\leq\int_\mathbb R|f(t)|\cdot|e^{i\varepsilon t}-1|dt$$

As for the boundedness of $\widehat{f}$, this is something which is clear as well.

\medskip

(3) This comes from the following computation, assuming that $f$ is even:
\begin{eqnarray*}
\widehat{f}(-x)
&=&\int_\mathbb R f(t)e^{-ixt}dt\\
&=&\int_\mathbb R f(-t)e^{ixt}dt\\
&=&\int_\mathbb R f(t)e^{ixt}dt\\
&=&\widehat{f}(x)
\end{eqnarray*}

(4) The proof here is similar to the proof of (3), by changing some signs.
\end{proof}

We will be back to more theory in a moment, but let us explore now the examples. Here are some basic computations of Fourier transforms:

\begin{proposition}
We have the following Fourier transform formulae,
$$f=\chi_{[-a,a]}\implies\widehat{f}(x)=\frac{2\sin(ax)}{x}$$
$$f=e^{-ax}\chi_{[0,\infty]}(x)\implies\widehat{f}(x)=\frac{1}{a-ix}$$
$$f=e^{ax}\chi_{[-\infty,0]}(x)\implies\widehat{f}(x)=\frac{1}{a+ix}$$
$$f=e^{-a|x|}\implies\widehat{f}(x)=\frac{2a}{a^2+x^2}$$
$$f=sgn(x)e^{-a|x|}\implies\widehat{f}(x)=\frac{2ix}{a^2+x^2}$$
valid for any number $a>0$.
\end{proposition}

\begin{proof}
All this comes from some basic calculus, as follows:

\medskip

(1) In what regards first formula, for the function $f=\chi_{[-a,a]}$, we have indeed:
$$\widehat{f}(x)
=\int_{-a}^ae^{ixt}dt
=\left[\frac{e^{ixt}}{xt}\right]_{-a}^a
=\frac{2\sin(ax)}{x}$$

(2) Next, with $f(x)=e^{-ax}\chi_{[0,\infty]}(x)$, the computation goes as follows:
$$\widehat{f}(x)
=\int_0^\infty e^{ixt-at}dt
=\left[\frac{e^{(ix-a)t}}{ix-a}\right]_0^\infty
=\frac{1}{a-ix}$$

(3) Regarding the third formula, this follows from the second one, by using the following fact, generalizing the parity computations from Proposition 11.19:
$$F(x)=f(-x)\implies\widehat{F}(x)=\widehat{f}(-x)$$

(4) The last 2 formulae follow from what we have, by making sums and differences, and using the linearity properties of the Fourier transform, from Proposition 11.19.
\end{proof}

Getting back to theory, we have the following result, adding to the various properties in Proposition 11.19, and providing more motivations for the Fourier transform:

\index{exchange of hat}

\begin{proposition}
Given $f,g\in L^1(\mathbb R)$ we have $\widehat{f}g,f\widehat{g}\in L^1(\mathbb R)$ and
$$\int_\mathbb R f(x)\widehat{g}(x)dx=\int_\mathbb R\widehat{f}(x)g(x)dx$$
called ``exchange of hat'' formula.
\end{proposition}

\begin{proof}
Regarding the fact that we have indeed $\widehat{f}g,f\widehat{g}\in L^1(\mathbb R)$, this is actually a bit non-trivial, but we will be back to this later. Assuming this, we have:
$$\int_\mathbb R f(x)\widehat{g}(x)dx=\int_\mathbb R\int_\mathbb R f(x)g(y)e^{ixy}dxdy$$

On the other hand, we have as well the following formula:
$$\int_\mathbb R\widehat{f}(x)g(x)dx=\int_\mathbb R\int_\mathbb R f(y)e^{iyx}g(x)dydx$$

Thus, with $x\leftrightarrow y$, we are led to the formula in the statement.
\end{proof}

As a key result now, showing the power of the Fourier transform, we have:

\begin{theorem}
Given $f:\mathbb R\to\mathbb C$ such that $f,f'\in L^1(\mathbb R)$, we have:
$$\widehat{f'}(x)=-ix\widehat{f}(x)$$
More generally, assuming $f,f',f'',\ldots,f^{(n)}\in L^1(\mathbb R)$, we have
$$\widehat{f^{(k)}}(x)=(-ix)^k\widehat{f}(x)$$
for any $k=1,2,\ldots,n$.
\end{theorem}

\begin{proof}
Assuming that $f:\mathbb R\to\mathbb C$ has compact support, we have indeed:
\begin{eqnarray*}
\widehat{f'}(x)
&=&\int_\mathbb R f'(t)e^{ixt}dt\\
&=&-\int_\mathbb R f(t)\cdot ixe^{ixt}dt\\
&=&-ix\int_\mathbb R f(t)e^{ixt}dt\\
&=&-ix\widehat{f}(x)
\end{eqnarray*}

As for the higher derivatives, the formula here follows by recurrence.
\end{proof}

Importantly, we have a converse statement as well, as follows:

\begin{theorem}
Assuming that $f\in L^1(\mathbb R)$ is such that $F(t)=tf(t)$ belongs to $L^1(\mathbb R)$ too, the function $\widehat{f}$ is differentiable, with derivative given by:
$$(\widehat{f})'(x)=i\widehat{F}(x)$$
More generally, if $F_k(t)=t^kf(t)$ belongs to $L^1(\mathbb R)$, for $k=0,1,\ldots,n$, we have
$$(\widehat{f})^{(k)}(x)=i^k\widehat{F_k}(x)$$
for any $k=1,2,\ldots,n$.
\end{theorem}

\begin{proof}
Regarding the first assertion, the computation here is as follows:
\begin{eqnarray*}
(\widehat{f})'(x)
&=&\frac{d}{dx}\int_\mathbb R f(t)e^{ixt}dt\\
&=&\int_\mathbb R f(t)\cdot ite^{ixt}dt\\
&=&i\int_\mathbb R tf(t)e^{ixt}dt\\
&=&i\widehat{F}(x)
\end{eqnarray*}

As for the second assertion, this follows from the first one, by recurrence.
\end{proof}

Here is another useful result, of the same type, this time regarding convolutions:

\begin{theorem}
Assuming $f,g\in L^1(\mathbb R)$, the following happens:
$$\widehat{f*g}=\widehat{f}\cdot\widehat{g}$$
Conversely, we have $\widehat{fg}=\frac{1}{2\pi}\,\widehat{f}*\widehat{g}$, which holds almost everywhere.
\end{theorem}

\begin{proof}
The first assertion is something elementary, coming as follows:
\begin{eqnarray*}
\widehat{f*g}(x)
&=&\int_\mathbb R (f*g)(t)e^{ixt}dt\\
&=&\int_\mathbb R\int_\mathbb R f(t-s)g(s)e^{ixt}dsdt\\
&=&\int_\mathbb R \left(\int_\mathbb R f(t-s)e^{ix(t-s)}dt\right)g(s)e^{ixs}ds\\
&=&\int_\mathbb R \left(\int_\mathbb R f(r)e^{ixr}dr\right)g(s)e^{ixs}ds\\
&=&\int_\mathbb R \widehat{f}(x)g(s)e^{ixs}ds\\
&=&\widehat{f}(x)\widehat{g}(x)
\end{eqnarray*}

As for the second assertion, this is something more tricky, which follows from the first one by using the Fourier inversion formula, that we will soon learn.
\end{proof}

Let us develop now more theory for the Fourier transform. We first have:

\index{Riemann-Lebesgue property}

\begin{theorem}
Given $f\in L^1(\mathbb R)$, its Fourier transform satisfies
$$\lim_{x\to\pm\infty}\widehat{f}(x)=0$$
called Riemann-Lebesgue property of $\widehat{f}$.
\end{theorem}

\begin{proof}
This is something quite technical, as follows:

\medskip

(1) Given a function $f:\mathbb R\to\mathbb C$ and a number $r\in\mathbb R$, let us set:
$$f_r(t)=f(t-r)$$

Our claim is then is that if $f\in L^p(\mathbb R)$, then the following function is uniformly continuous, with respect to the usual $p$-norm on the right:
$$\mathbb R\to L^p(\mathbb R)\quad,\quad r\to f_r$$

(2) In order to prove this, fix $\varepsilon>0$. Since $f\in L^p(\mathbb R)$, we can find a function of type $g:[-K,K]\to\mathbb C$ which is continuous, such that:
$$||f-g||_p<\varepsilon$$

Now since $g$ is uniformly continuous, we can find $\delta\in(0,K)$ such that:
$$|u-v|<\delta\implies|g(u)-g(v)|<(3K)^{-1/p}\varepsilon$$

But this shows that we have the following estimate:
\begin{eqnarray*}
||g_r-g_s||_p
&=&\left(\int_\mathbb R\big|g(t-r)-g(t-s)\big|^pdt\right)^{1/p}\\
&<&\left[(3K)^{-1}\varepsilon^p(2k+\delta)\right]^{1/p}\\
&<&\varepsilon
\end{eqnarray*}

By using now the formula $||f||_p=||f_r||_p$, which is clear, we obtain:
\begin{eqnarray*}
||f_r-f_s||_p
&\leq&||f_r-g_r||_p+||g_r-g_s||_p+||g_s-f_s||_p\\
&<&\varepsilon+\varepsilon+\varepsilon\\
&=&3\varepsilon
\end{eqnarray*}

But this being true for any $|r-s|<\delta$, we have proved our claim.

\medskip

(3) Let us prove now the Riemann-Lebesgue property of $\widehat{f}$, as formulated in the statement. By using $e^{\pi i}=-1$, and the change of variables $t\to t-\pi/x$, we have:
\begin{eqnarray*}
\widehat{f}(x)
&=&\int_\mathbb Rf(t)e^{ixt}dt\\
&=&-\int_\mathbb Re^{\pi i}f(t)e^{ixt}dt\\
&=&-\int_\mathbb Rf(t)e^{ix(t+\pi/x)}dt\\
&=&-\int_\mathbb Rf\left(t-\frac{\pi}{x}\right)e^{ixt}dt
\end{eqnarray*}

On the other hand, we have as well the following formula:
$$\widehat{f}(x)=\int_\mathbb Rf(t)e^{ixt}dt$$

Thus by summing, we obtain the following formula:
$$2\widehat{f}(x)=\int_\mathbb R\left(f(t)-f\left(t-\frac{\pi}{x}\right)\right)e^{ixt}dt$$

But this gives the following estimate, with notations from (1):
$$2|\widehat{f}(x)|\leq||f-f_{\pi/x}||_1$$

Since by (1) this goes to $0$ with $x\to\pm\infty$, this gives the result.
\end{proof}

Quite remarkably, and as a main result now regarding Fourier transforms, a function $f:\mathbb R\to\mathbb C$ can be recovered from its Fourier transform $\widehat{f}:\mathbb R\to\mathbb C$, as follows:

\index{Fourier inversion}

\begin{theorem}
Assuming $f,\widehat{f}\in L^1(\mathbb R)$, we have
$$f(t)=\frac{1}{2\pi}\int_\mathbb R \widehat{f}(x)e^{-itx}dx$$
almost everywhere, called Fourier inversion formula.
\end{theorem}

\begin{proof}
Consider the following function, depending on a parameter $\lambda>0$:
$$\varphi_\lambda(s)=\frac{1}{2\pi}\int_\mathbb Re^{-isx-\lambda|x|}dx$$

We have then the following convolution computation:
\begin{eqnarray*}
(f*\varphi_\lambda)(t)
&=&\int_\mathbb Rf(t-s)\varphi_\lambda(s)ds\\
&=&\frac{1}{2\pi}\int_\mathbb R\int_\mathbb Rf(t-s)e^{-isx-\lambda|x|}dxds\\
&=&\frac{1}{2\pi}\int_\mathbb R\left(\int_\mathbb Rf(t-s)e^{-isx}ds\right)e^{-\lambda|x|}dx\\
&=&\frac{1}{2\pi}\int_\mathbb R\widehat{f}(x)e^{-itx}e^{-\lambda|x|}dx
\end{eqnarray*}

By letting now $\lambda\to0$, we obtain from this the following formula:
$$\lim_{\lambda\to0}(f*\varphi_\lambda)(t)=\frac{1}{2\pi}\int_\mathbb R \widehat{f}(x)e^{-itx}dx$$

On the other hand, by using Theorem 11.25 we obtain that, almost everywhere:
$$\lim_{\lambda\to0}(f*\varphi_\lambda)(t)=f(t)$$

We are therefore led to the conclusion in the statement. For more details on all this, and for some basic consequences too, we refer as usual to Rudin \cite{ru2}.
\end{proof}

As a first application, we have now the fix for Theorem 11.24, according to:
$$\widehat{f*g}=\widehat{f}\cdot\widehat{g}\ \implies\ \widehat{fg}=\frac{1}{2\pi}\,\widehat{f}*\widehat{g}$$

There are many more things that can be said about Fourier transforms, a key result being the Plancherel formula, over $L^2(\mathbb R)$. Also, we can talk about the Fourier transform over the Schwartz space $\mathcal S$. Also, we can talk about other types of Fourier transforms. As usual, if very interested in mathematical analysis, have a look at all this.

\bigskip

Finally, the Fourier transform can be used a bit like the Fourier series, for dealing with the differential equations, and exercise for you, to learn more about all this.

\section*{11d. Into the waves}

We have now enough accumulated material for talking about waves. Hang on, tough mathematics to come next. To start with, the truth about waves is as follows:

\begin{fact}
Waves can be of many types, and basically fall into two classes:
\begin{enumerate}
\item Mechanical waves, such as the usual water waves, but also the sound waves, or the seismic waves. In all these cases, the wave propagates mechanically, via a certain elastic medium, which can be solid, liquid or gaseous.

\item Electromagnetic waves, coming via a more complicated mechanism, namely an accelerating charge in the context of electromagnetism. These are the radio waves, microwaves, IR, visible light, UV, X-rays and $\gamma$-rays.
\end{enumerate}
\end{fact}

Quite remarkably, the behavior of all the above waves is basically described by the same wave equation, which can be mathematically recovered as follows:

\index{wave equation}
\index{Laplace operator}
\index{lattice model}
\index{Hooke law}
\index{Newton law}

\begin{theorem}
The wave equation in $\mathbb R^N$ is
$$\ddot{\varphi}=v^2\Delta\varphi$$
where $v>0$ is the propagation speed.
\end{theorem}

\begin{proof}
Leaving aside the electromagnetic waves, whose study requires going into the Maxwell equations, let us focus on the mechanical waves. The equation for them in the statement is of course what comes out of experiments, but allowing us a bit of imagination, we can mathematically ``prove'' this equation, by discretizing, as follows:

\medskip

(1) Let us first consider the 1D case. In order to understand the propagation of waves, we will model $\mathbb R$ as a network of balls, with springs between them, as follows:
$$\cdots\times\!\!\!\times\!\!\!\times\bullet\times\!\!\!\times\!\!\!\times\bullet\times\!\!\!\times\!\!\!\times\bullet\times\!\!\!\times\!\!\!\times\bullet\times\!\!\!\times\!\!\!\times\bullet\times\!\!\!\times\!\!\!\times\cdots$$

Now let us send an impulse, and see how the balls will be moving. For this purpose, we zoom on one ball. The situation here is as follows, $l$ being the spring length:
$$\cdots\cdots\bullet_{\varphi(x-l)}\times\!\!\!\times\!\!\!\times\bullet_{\varphi(x)}\times\!\!\!\times\!\!\!\times\bullet_{\varphi(x+l)}\cdots\cdots$$

We have two forces acting at $x$. First is the Newton motion force, mass times acceleration, which is as follows, with $m$ being the mass of each ball:
$$F_n=m\cdot\ddot{\varphi}(x)$$

And second is the Hooke force, displacement of the spring, times spring constant. Since we have two springs at $x$, this is as follows, $k$ being the spring constant:
\begin{eqnarray*}
F_h
&=&F_h^r-F_h^l\\
&=&k(\varphi(x+l)-\varphi(x))-k(\varphi(x)-\varphi(x-l))\\
&=&k(\varphi(x+l)-2\varphi(x)+\varphi(x-l))
\end{eqnarray*}

We conclude that the equation of motion, in our model, is as follows:
$$m\cdot\ddot{\varphi}(x)=k(\varphi(x+l)-2\varphi(x)+\varphi(x-l))$$

(2) Now let us take the limit of our model, as to reach to continuum. For this purpose we will assume that our system consists of $B>>0$ balls, having a total mass $M$, and spanning a total distance $L$. Thus, our previous infinitesimal parameters are as follows, with $K$ being the spring constant of the total system, which is of course lower than $k$:
$$m=\frac{M}{B}\quad,\quad k=KB\quad,\quad l=\frac{L}{B}$$

With these changes, our equation of motion found in (1) reads:
$$\ddot{\varphi}(x)=\frac{KB^2}{M}(\varphi(x+l)-2\varphi(x)+\varphi(x-l))$$

Now observe that this equation can be written, more conveniently, as follows:
$$\ddot{\varphi}(x)=\frac{KL^2}{M}\cdot\frac{\varphi(x+l)-2\varphi(x)+\varphi(x-l)}{l^2}$$

With $N\to\infty$, and therefore $l\to0$, we obtain in this way:
$$\ddot{\varphi}(x)=\frac{KL^2}{M}\cdot\frac{d^2\varphi}{dx^2}(x)$$

We are therefore led to the wave equation in the statement, which is $\ddot{\varphi}=v^2\varphi''$ in our present $N=1$ dimensional case, the propagation speed being $v=\sqrt{K/M}\cdot L$.

\medskip

(3) In $2$ dimensions now, the same argument carries on. Indeed, we can use here a lattice model as follows, with all the edges standing for small springs:
$$
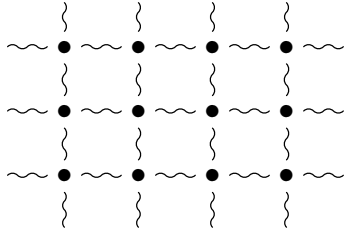
\xymatrix@R=12pt@C=15pt{
&\ar@{~}[d]&\ar@{~}[d]&\ar@{~}[d]&\ar@{~}[d]\\
\ar@{~}[r]&\bullet\ar@{~}[r]\ar@{~}[d]&\bullet\ar@{~}[r]\ar@{~}[d]&\bullet\ar@{~}[r]\ar@{~}[d]&\bullet\ar@{~}[r]\ar@{~}[d]&\\
\ar@{~}[r]&\bullet\ar@{~}[r]\ar@{~}[d]&\bullet\ar@{~}[r]\ar@{~}[d]&\bullet\ar@{~}[r]\ar@{~}[d]&\bullet\ar@{~}[r]\ar@{~}[d]&\\
\ar@{~}[r]&\bullet\ar@{~}[r]\ar@{~}[d]&\bullet\ar@{~}[r]\ar@{~}[d]&\bullet\ar@{~}[r]\ar@{~}[d]&\bullet\ar@{~}[r]\ar@{~}[d]&\\
&&&&&}$$

As before in one dimension, we send an impulse, and we zoom on one ball. The situation here is as follows, with $l$ being the spring length:
$$
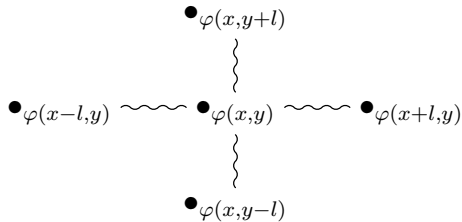
\xymatrix@R=20pt@C=20pt{
&\bullet_{\varphi(x,y+l)}\ar@{~}[d]&\\
\bullet_{\varphi(x-l,y)}\ar@{~}[r]&\bullet_{\varphi(x,y)}\ar@{~}[r]\ar@{~}[d]&\bullet_{\varphi(x+l,y)}\\
&\bullet_{\varphi(x,y-l)}}$$

We have two forces acting at $(x,y)$. First is the Newton motion force, mass times acceleration, which is as follows, with $m$ being the mass of each ball:
$$F_n=m\cdot\ddot{\varphi}(x,y)$$

And second is the Hooke force, displacement of the spring, times spring constant. Since we have four springs at $(x,y)$, this is as follows, $k$ being the spring constant:
\begin{eqnarray*}
F_h
&=&F_h^r-F_h^l+F_h^u-F_h^d\\
&=&k(\varphi(x+l,y)-\varphi(x,y))-k(\varphi(x,y)-\varphi(x-l,y))\\
&+&k(\varphi(x,y+l)-\varphi(x,y))-k(\varphi(x,y)-\varphi(x,y-l))\\
&=&k(\varphi(x+l,y)-2\varphi(x,y)+\varphi(x-l,y))\\
&+&k(\varphi(x,y+l)-2\varphi(x,y)+\varphi(x,y-l))
\end{eqnarray*}

We conclude that the equation of motion, in our model, is as follows:
\begin{eqnarray*}
m\cdot\ddot{\varphi}(x,y)
&=&k(\varphi(x+l,y)-2\varphi(x,y)+\varphi(x-l,y))\\
&+&k(\varphi(x,y+l)-2\varphi(x,y)+\varphi(x,y-l))
\end{eqnarray*}

(4) Now let us take the limit of our model, as to reach to continuum. For this purpose we will assume that our system consists of $B^2>>0$ balls, having a total mass $M$, and spanning a total area $L^2$. Thus, our previous infinitesimal parameters are as follows, with $K$ being the spring constant of the total system, taken to be equal to $k$:
$$m=\frac{M}{B^2}\quad,\quad k=K\quad,\quad l=\frac{L}{B}$$

With these changes, our equation of motion found in (3) reads:
\begin{eqnarray*}
\ddot{\varphi}(x,y)
&=&\frac{KB^2}{M}(\varphi(x+l,y)-2\varphi(x,y)+\varphi(x-l,y))\\
&+&\frac{KB^2}{M}(\varphi(x,y+l)-2\varphi(x,y)+\varphi(x,y-l))
\end{eqnarray*}

Now observe that this equation can be written, more conveniently, as follows:
\begin{eqnarray*}
\ddot{\varphi}(x,y)
&=&\frac{KL^2}{M}\times\frac{\varphi(x+l,y)-2\varphi(x,y)+\varphi(x-l,y)}{l^2}\\
&+&\frac{KL^2}{M}\times\frac{\varphi(x,y+l)-2\varphi(x,y)+\varphi(x,y-l)}{l^2}
\end{eqnarray*}

With $N\to\infty$, and therefore $l\to0$, we obtain in this way:
$$\ddot{\varphi}(x,y)=\frac{KL^2}{M}\left(\frac{d^2\varphi}{dx^2}+\frac{d^2\varphi}{dy^2}\right)(x,y)$$

Thus, we are led in this way to the following wave equation in two dimensions, with $v=\sqrt{K/M}\cdot L$ being the propagation speed of our wave:
$$\ddot{\varphi}(x,y)=v^2\left(\frac{d^2\varphi}{dx^2}+\frac{d^2\varphi}{dy^2}\right)(x,y)$$

But we recognize at right the Laplace operator, and we are done. As before in 1D, there is of course some discussion to be made here, arguing that our spring model in (3) is indeed the correct one. But do not worry, experiments confirm our findings.

\medskip

(5) In 3 dimensions now, which is the case of the main interest, corresponding to our real-life world, the same argument carries over, and the wave equation is as follows:
$$\ddot{\varphi}(x,y,z)=v^2\left(\frac{d^2\varphi}{dx^2}+\frac{d^2\varphi}{dy^2}+\frac{d^2\varphi}{dz^2}\right)(x,y,z)$$

(6) Finally, the same argument, namely a lattice model, carries on in arbitrary $N$ dimensions, and the wave equation here is as follows:
$$\ddot{\varphi}(x_1,\ldots,x_N)=v^2\sum_{i=1}^N\frac{d^2\varphi}{dx_i^2}(x_1,\ldots,x_N)$$

Thus, we are led to the conclusion in the statement.
\end{proof}

Getting now to mathematics, we will need a standard calculus result, as follows:

\begin{proposition}
The derivative of a function of type
$$\varphi(x)=\int_{g(x)}^{h(x)}f(s)ds$$
is given by the formula $\varphi'(x)=f(h(x))h'(x)-f(g(x))g'(x)$.
\end{proposition}

\begin{proof}
Consider a primitive of the function that we integrate, $F'=f$. We have:
\begin{eqnarray*}
\varphi(x)
&=&\int_{g(x)}^{h(x)}f(s)ds\\
&=&\int_{g(x)}^{h(x)}F'(s)ds\\
&=&F(h(x))-F(g(x))
\end{eqnarray*}

By using now the chain rule for derivatives, we obtain from this:
\begin{eqnarray*}
\varphi'(x)
&=&F'(h(x))h'(x)-F'(g(x))g'(x)\\
&=&f(h(x))h'(x)-f(g(x))g'(x)
\end{eqnarray*}

Thus, we are led to the formula in the statement.
\end{proof}

Now back to the waves, the general result in 1D, due to d'Alembert, along with a little more, in relation with our lattice models above, is as follows:

\begin{theorem}
The solution of the 1D wave equation with initial value conditions $\varphi(x,0)=f(x)$ and $\dot{\varphi}(x,0)=g(x)$ is given by the d'Alembert formula, namely:
$$\varphi(x,t)=\frac{f(x-vt)+f(x+vt)}{2}+\frac{1}{2v}\int_{x-vt}^{x+vt}g(s)ds$$
In the context of our lattice model discretization, what happens is more or less that the above d'Alembert integral gets computed via Riemann sums.
\end{theorem}

\begin{proof}
There are several things going on here, the idea being as follows:

\medskip

(1) Let us first check that the d'Alembert solution is indeed a solution of the wave equation $\ddot{\varphi}=v^2\varphi''$. The first time derivative is computed as follows:
$$\dot{\varphi}(x,t)=\frac{-vf'(x-vt)+vf'(x+vt)}{2}+\frac{1}{2v}(vg(x+vt)+vg(x-vt))$$

The second time derivative is computed as follows:
$$\ddot{\varphi}(x,t)=\frac{v^2f''(x-vt)+v^2f(x+vt)}{2}+\frac{vg'(x+vt)-vg'(x-vt)}{2}$$

Regarding now space derivatives, the first one is computed as follows:
$$\varphi'(x,t)=\frac{f'(x-vt)+f'(x+vt)}{2}+\frac{1}{2v}(g(x+vt)-g(x-vt))$$

As for the second space derivative, this is computed as follows:
$$\varphi''(x,t)=\frac{f''(x-vt)+f''(x+vt)}{2}+\frac{g'(x+vt)-g'(x-vt)}{2v}$$

Thus we have indeed $\ddot{\varphi}=v^2\varphi''$. As for the initial conditions, $\varphi(x,0)=f(x)$ is clear from our definition of $\varphi$, and $\dot{\varphi}(x,0)=g(x)$ is clear from our above formula of $\dot{\varphi}$.

\medskip

(2) Conversely now, we must show that our solution is unique, but instead of going here into abstract arguments, we will simply solve our equation, which among others will doublecheck the computations in (1). Let us make the following change of variables:
$$y=x-vt\quad,\quad z=x+vt$$

Observe that this is something quite tricky, mixing space and time variables. You would probably  even say that this is crazy, but after thinking a bit, with the basic waves, which are sinusoids, tending to mix space and time, this is, after all, not that crazy. 

\medskip

(3) In any case, with this change of variables done, we have, using the chain rule:
$$\frac{d\varphi}{dt}
=\frac{d\varphi}{dy}\cdot\frac{dy}{dt}+\frac{d\varphi}{dz}\cdot\frac{dz}{dt}
=-v\,\frac{d\varphi}{dy}+v\,\frac{d\varphi}{dz}$$

By using the chain rule again, the second time derivative is given by:
\begin{eqnarray*}
\frac{d^2\varphi}{dt^2}
&=&-v\left(\frac{d^2\varphi}{dy^2}\cdot\frac{dy}{dt}+\frac{d^2\varphi}{dydz}\cdot\frac{dz}{dt}\right)
+v\left(\frac{d^2\varphi}{dzdy}\cdot\frac{dy}{dt}+\frac{d^2\varphi}{dz^2}\cdot\frac{dz}{dt}\right)\\
&=&-v\left(-v\,\frac{d^2\varphi}{dy^2}+v\,\frac{d^2\varphi}{dydz}\right)
+v\left(-v\,\frac{d^2\varphi}{dzdy}+v\,\frac{d^2\varphi}{dz^2}\right)\\
&=&v^2\left(\frac{d^2\varphi}{dy^2}+\frac{d^2\varphi}{dz^2}-2\frac{d^2\varphi}{dydz}\right)
\end{eqnarray*}

(4) Regarding now the first space derivative, this can be computed as follows:
$$\frac{d\varphi}{dx}
=\frac{d\varphi}{dy}\cdot\frac{dy}{dx}+\frac{d\varphi}{dz}\cdot\frac{dz}{dx}
=\frac{d\varphi}{dy}+\frac{d\varphi}{dz}$$

By using the chain rule again, the second space derivative is given by:
\begin{eqnarray*}
\frac{d^2\varphi}{dt^2}
&=&\left(\frac{d^2\varphi}{dy^2}\cdot\frac{dy}{dx}+\frac{d^2\varphi}{dydz}\cdot\frac{dz}{dx}\right)
+\left(\frac{d^2\varphi}{dzd\xi}\cdot\frac{dy}{dx}+\frac{d^2\varphi}{dz^2}\cdot\frac{dz}{dx}\right)\\
&=&\left(\frac{d^2\varphi}{dy^2}+\frac{d^2\varphi}{dydz}\right)
+\left(\frac{d^2\varphi}{dzdy}+\frac{d^2\varphi}{dz^2}\right)\\
&=&\frac{d^2\varphi}{dy^2}+\frac{d^2\varphi}{dz^2}+2\frac{d^2\varphi}{dydz}
\end{eqnarray*}

(5) Thus, our wave equation $\ddot{\varphi}=v^2\varphi''$ reformulates in a very simple way, as follows:
$$\frac{d^2\varphi}{dydz}=0$$

But this latter equation tells us that our new $y,z$ variables get separated, and we conclude from this that the solution must be of the following special form:
$$\varphi(x,t)=H(y)+K(z)=H(x-vt)+K(x+vt)$$

(6) In order now to finish, we must take into account the intial conditions for our problem, namely $\varphi(x,0)=f(x)$ and $\dot{\varphi}(x,0)=g(x)$. The first condition reads:
$$\varphi(x,0)=f(x)\iff H(x)+K(x)=f(x)$$

As for the second condition, this can be processed as follows, with $G'=g$:
\begin{eqnarray*}
\dot{\varphi}(x,0)=g(x)
&\iff&-vH'(x)+vK'(x)=g(x)\\
&\iff&K'(x)-H'(x)=\frac{g(x)}{v}\\
&\iff&K'(x)-H'(x)=\frac{G'(x)}{v}\\
&\iff&K(x)-H(x)=\frac{G(x)}{v}+c
\end{eqnarray*}

(7) Now by putting our two equations for $H,K$ together, we obtain:
$$H(x)=\frac{f(x)}{2}-\frac{G(x)}{2v}-\frac{c}{2}\quad,\quad 
K(x)=\frac{f(x)}{2}+\frac{G(x)}{2v}+\frac{c}{2}$$

We conclude that the solution of our equation is given by:
\begin{eqnarray*}
\varphi(x,t)
&=&H(x-vt)+K(x+vt)\\
&=&\frac{f(x-vt)}{2}-\frac{G(x-vt)}{2v}+\frac{f(x+vt)}{2}+\frac{G(x+vt)}{2v}\\
&=&\frac{f(x-vt)+f(x+vt)}{2}+\frac{G(x+vt)-G(x-vt)}{2v}\\
&=&\frac{f(x-vt)+f(x+vt)}{2}+\frac{1}{2v}\int_{x-vt}^{x+vt}g(s)ds
\end{eqnarray*}

Thus, we are led to the d'Alembert formula in the statement.

\medskip

(8) Finally, regarding the discretization questions, by using a 1D lattice model with balls and springs as before, what happens to all the above is more or less that the above d'Alembert integral gets computed via Riemann sums, in our model, as stated.
\end{proof}

In $N\geq2$ dimensions things get more complicated, among others requiring the use of spherical coordinates, and we refer here to any reasonably advanced physics book.

\section*{11e. Exercises}

We had a quite broad chapter here, and as exercises on this, we have:

\begin{exercise}
Learn more about the theory of distributions, on $\mathbb R$.
\end{exercise}

\begin{exercise}
Learn as well about distributions in higher dimensions.
\end{exercise}

\begin{exercise}
Fill in the details in the proof of the Cauchy formula.
\end{exercise}

\begin{exercise}
Learn more about harmonic functions, in $2$ or more dimensions.
\end{exercise}

\begin{exercise}
Learn the Plancherel formula, over $L^2(\mathbb R)$, and its applications.
\end{exercise}

\begin{exercise}
Learn about the Fourier transform over the Schwartz space $\mathcal S$.
\end{exercise}

\begin{exercise}
Learn about the other possible types of Fourier transforms.
\end{exercise}

\begin{exercise}
Read some physics, on the wave equation in $2D$, and $3D$.
\end{exercise}

As bonus exercise, now that we know about waves, learn about the heat equation too.

\chapter{Linear operators}

\section*{12a. Linear operators}

We discuss in this chapter the diagonalization of the linear operators $T:H\to H$ over a Hilbert space $H$, usually taken to be separable. Our motivation comes from both mathematics, because what we will be doing here will be a natural infinite dimensional extension of basic linear algebra, and physics, more specifically quantum mechanics.

\bigskip

Let us begin our study with a number of general results regarding the operators that we are targeting. At the beginning of everything, we have the following fact:

\index{infinite matrix}
\index{linear operator}
\index{bounded operator}
\index{continuous operator}

\begin{theorem}
Given a Hilbert space $H$, consider the linear operators $T:H\to H$, and for each such operator define its norm by the following formula:
$$||T||=\sup_{||x||=1}||Tx||$$
The operators which are bounded, $||T||<\infty$, form then a complex algebra $B(H)$, which is complete with respect to $||.||$. When $H$ comes with a basis $\{e_i\}_{i\in I}$, we have
$$B(H)\subset M_I(\mathbb C)$$
with the correspondence $T\to M$ coming via the usual linear algebra formulae, namely:
$$T(x)=Mx\quad,\quad M_{ij}=<Te_j,e_i>$$
In infinite dimensions, the inclusion $B(H)\subset M_I(\mathbb C)$ is not an equality.
\end{theorem}

\begin{proof}
This is something straightforward, the idea being as follows:

\medskip

(1) The fact that we have indeed an algebra, satisfying the product condition in the statement, follows from the following estimates, which are all elementary:
$$||S+T||\leq||S||+||T||\quad,\quad 
||\lambda T||=|\lambda|\cdot||T||\quad,\quad 
||ST||\leq||S||\cdot||T||$$

(2) Regarding now the completness assertion, if $\{T_n\}\subset B(H)$ is Cauchy then $\{T_nx\}$ is Cauchy for any $x\in H$, so we can define the limit $T=\lim_{n\to\infty}T_n$ by setting:
$$Tx=\lim_{n\to\infty}T_nx$$

Let us first check that the application $x\to Tx$ is linear. We have:
\begin{eqnarray*}
T(x+y)
&=&\lim_{n\to\infty}T_n(x+y)\\
&=&\lim_{n\to\infty}T_n(x)+T_n(y)\\
&=&\lim_{n\to\infty}T_n(x)+\lim_{n\to\infty}T_n(y)\\
&=&T(x)+T(y)
\end{eqnarray*}

Similarly, we have $T(\lambda x)=\lambda T(x)$, and we conclude that $x\to Tx$ is linear.

\medskip

(3) With this done, it remains to prove now that we have $T\in B(H)$, and that $T_n\to T$ in norm. For this purpose, observe that we have:
\begin{eqnarray*}
||T_n-T_m||\leq\varepsilon\ ,\ \forall n,m\geq N
&\implies&||T_nx-T_mx||\leq\varepsilon\ ,\ \forall||x||=1\ ,\ \forall n,m\geq N\\
&\implies&||T_nx-Tx||\leq\varepsilon\ ,\ \forall||x||=1\ ,\ \forall n\geq N\\
&\implies&||T_Nx-Tx||\leq\varepsilon\ ,\ \forall||x||=1\\
&\implies&||T_N-T||\leq\varepsilon
\end{eqnarray*}

But this gives both $T\in B(H)$, and $T_N\to T$ in norm, and we are done.

\medskip

(4) Regarding the embedding, the correspondence $T\to M$ in the statement is indeed linear, and its kernel is $\{0\}$, so we have indeed an embedding as follows, as claimed:
$$B(H)\subset M_I(\mathbb C)$$

In finite dimensions we have an isomorphism, because any matrix $M\in M_N(\mathbb C)$ determines a linear operator $T:\mathbb C^N\to\mathbb C^N$, given by the following formula:
$$<Te_j,e_i>=M_{ij}$$

However, in infinite dimensions we have matrices not producing operators, as for instance the all-one matrix, so the embedding $B(H)\subset M_I(\mathbb C)$ is not an isomorphism.
\end{proof}

As a second and last basic result regarding the operators, we will need:

\index{adjoint operator}
\index{adjoint matrix}

\begin{theorem}
Each operator $T\in B(H)$ has an adjoint $T^*\in B(H)$, given by: 
$$<Tx,y>=<x,T^*y>$$
The operation $T\to T^*$ is antilinear, antimultiplicative, involutive, and satisfies:
$$||T||=||T^*||\quad,\quad ||TT^*||=||T||^2$$
When $H$ comes with a basis $\{e_i\}_{i\in I}$, the operation $T\to T^*$ corresponds to
$$(M^*)_{ij}=\overline{M}_{ji}$$ 
at the level of the associated matrices $M\in M_I(\mathbb C)$.
\end{theorem}

\begin{proof}
This is standard too, and can be proved in 3 steps, as follows:

\medskip

(1) The existence of the adjoint operator $T^*$, given by the formula in the statement, comes from the fact that the function $\varphi(x)=<Tx,y>$ being a linear map $H\to\mathbb C$, we must have a formula as follows, for a certain vector $T^*y\in H$:
$$\varphi(x)=<x,T^*y>$$

Moreover, since this vector is unique, $T^*$ is unique too, and we have as well:
$$(S+T)^*=S^*+T^*\quad,\quad
(\lambda T)^*=\bar{\lambda}T^*\quad,\quad 
(ST)^*=T^*S^*\quad,\quad 
(T^*)^*=T$$

Observe also that we have indeed $T^*\in B(H)$, because:
\begin{eqnarray*}
||T||
&=&\sup_{||x||=1}\sup_{||y||=1}<Tx,y>\\
&=&\sup_{||y||=1}\sup_{||x||=1}<x,T^*y>\\
&=&||T^*||
\end{eqnarray*}

(2) Regarding now $||TT^*||=||T||^2$, which is a key formula, observe that we have:
$$||TT^*||
\leq||T||\cdot||T^*||
=||T||^2$$

On the other hand, we have as well the following estimate:
\begin{eqnarray*}
||T||^2
&=&\sup_{||x||=1}|<Tx,Tx>|\\
&=&\sup_{||x||=1}|<x,T^*Tx>|\\
&\leq&||T^*T||
\end{eqnarray*}

By replacing $T\to T^*$ we obtain from this $||T||^2\leq||TT^*||$, as desired.

\medskip

(3) Finally, when $H$ comes with a basis, the formula $<Tx,y>=<x,T^*y>$ applied with $x=e_i$, $y=e_j$ translates into the formula $(M^*)_{ij}=\overline{M}_{ji}$, as desired.
\end{proof}

As in finite dimensions, the operators $T,T^*$ can be thought of as being ``twin brothers'', and there is a lot more mathematics connecting them, as for instance:

\index{adjoint operator}

\begin{proposition}
Given a bounded operator $T\in B(H)$, the following happen:
\begin{enumerate}
\item $\ker T^*=(Im T)^\perp$.

\item $\overline{Im T^*}=(\ker T)^\perp$.
\end{enumerate}
\end{proposition}

\begin{proof}
Both these assertions are elementary, as follows:

\medskip

(1) Let us first prove ``$\subset$''. Assuming $T^*x=0$, we have indeed $x\perp ImT$, because:
$$<x,Ty>
=<T^*x,y>
=0$$

As for ``$\supset$'', assuming $<x,Ty>=0$ for any $y$, we have $T^*x=0$, because:
$$<T^*x,y>
=<x,Ty>
=0$$

(2) This can be deduced from (1), applied to the operator $T^*$, as follows:
$$(\ker T)^\perp
=(Im T^*)^{\perp\perp}
=\overline{Im T^*}$$

Here we have used the formula $K^{\perp\perp}=\bar{K}$, valid for any linear subspace $K\subset H$ of a Hilbert space, which for $K$ closed reads $K^{\perp\perp}=K$, and comes from $H=K\oplus K^\perp$, and which in general follows from $K^{\perp\perp}\subset\bar{K}^{\perp\perp}=\bar{K}$, the reverse inclusion being clear.
\end{proof}

Regarding the self-adjoint operators, we have about them the following result:

\index{self-adjoint operator}

\begin{proposition}
For an operator $T\in B(H)$, the following are equivalent:
\begin{enumerate}
\item $T$ is self-adjoint, $T=T^*$.

\item $<Tx,x>\in\mathbb R$, for any $x\in H$.
\end{enumerate}
\end{proposition}

\begin{proof}
This is something very standard, the idea being as follows:

\medskip

$(1)\implies(2)$ This implication is clear, because we have:
$$\overline{<Tx,x>}
=<x,Tx>
=<T^*x,x>
=<Tx,x>$$

$(2)\implies(1)$ In order to prove this, observe that the beginning of the above computation shows that, when assuming $<Tx,x>\in\mathbb R$, the following happens:
$$<Tx,x>=<T^*x,x>$$

Thus, in terms of the operator $S=T-T^*$, we have the following formula:
$$<Sx,x>=0$$

In order to finish, we use a polarization trick. We have the following formula:
$$<S(x+y),x+y>=<Sx,x>+<Sy,y>+<Sx,y>+<Sy,x>$$

Since the first 3 terms vanish, the sum of the 2 last terms vanishes too. But, by using $S^*=-S$, coming from $S=T-T^*$, we can process this latter vanishing as follows:
$$<Sx,y>
=-<Sy,x>
=<y,Sx>
=\overline{<Sx,y>}$$

Thus we must have $<Sx,y>\in\mathbb R$, and with $y\to iy$ we obtain $<Sx,y>\in i\mathbb R$ too, and so $<Sx,y>=0$. Thus $S=0$, which gives $T=T^*$, as desired.
\end{proof}

Regarding the normal operators, which are more general, we have:

\index{normal operator}

\begin{proposition}
For an operator $T\in B(H)$, the following are equivalent:
\begin{enumerate}
\item $T$ is normal, $TT^*=T^*T$.

\item $||Tx||=||T^*x||$, for any $x\in H$.
\end{enumerate}
\end{proposition}

\begin{proof}
Again, this is something very standard, the idea being as follows:

\medskip

$(1)\implies(2)$ This is clear, due to the following computation:
\begin{eqnarray*}
||Tx||^2
&=&<Tx,Tx>\\
&=&<T^*Tx,x>\\
&=&<TT^*x,x>\\
&=&<T^*x,T^*x>\\
&=&||T^*x||^2
\end{eqnarray*}

$(2)\implies(1)$ This is clear as well, because the above computation shows that, when assuming $||Tx||=||T^*x||$, the following happens:
$$<TT^*x,x>=<T^*Tx,x>$$

Thus, in terms of the operator $S=TT^*-T^*T$, we have:
$$<Sx,x>=0$$

In order to finish, we use a polarization trick. We have the following formula:
$$<S(x+y),x+y>=<Sx,x>+<Sy,y>+<Sx,y>+<Sy,x>$$

Since the first 3 terms vanish, the sum of the 2 last terms vanishes too. But, by using $S=S^*$, coming from $S=TT^*-T^*T$, we can process this latter vanishing as follows:
\begin{eqnarray*}
<Sx,y>
&=&-<Sy,x>\\
&=&-<y,Sx>\\
&=&-\overline{<Sx,y>}
\end{eqnarray*}

Thus we must have $<Sx,y>\in i\mathbb R$, and with $y\to iy$ we obtain $<Sx,y>\in \mathbb R$ too, and so $<Sx,y>=0$. Thus $S=0$, which gives $TT^*=T^*T$, as desired.
\end{proof}

Getting now to diagonalization work for the linear operators, as a first observation, we can talk about eigenvalues and eigenvectors of such operators, as follows:

\index{eigenvector}
\index{eigenvalue}

\begin{definition}
Given an operator $T\in B(H)$, assuming that we have
$$Tx=\lambda x$$
we say that $x\in H$ is an eigenvector of $T$, with eigenvalue $\lambda\in\mathbb C$.
\end{definition}

We know many things about eigenvalues and eigenvectors, in the finite dimensional case. However, most of these will not extend to the infinite dimensional case, or at least not extend in a straightforward way, due to a number of reasons:

\medskip

\begin{enumerate}
\item Most of basic linear algebra is based on the fact that $Tx=\lambda x$ is equivalent to $(T-\lambda)x=0$, so that $\lambda$ is an eigenvalue when $T-\lambda$ is not invertible. In the infinite dimensional setting $T-\lambda$ might be injective and not surjective, or vice versa, or invertible with $(T-\lambda)^{-1}$ not bounded, and so on.

\medskip

\item Also, in linear algebra $T-\lambda$ is not invertible when $\det(T-\lambda)=0$, and with this leading to most of the advanced results about eigenvalues and eigenvectors. In infinite dimensions, however, it is impossible to construct a determinant function $\det:B(H)\to\mathbb C$, and this even for the diagonal operators on $l^2(\mathbb N)$.
\end{enumerate}

\medskip

Summarizing, we are in trouble. Forgetting about (2), which obviously leads nowhere, let us focus on the difficulties in (1). In order to cut short the discussion there, regarding the various properties of $T-\lambda$, we can just say that $T-\lambda$ is either invertible with bounded inverse, the ``good case'', or not. We are led in this way to the following definition:

\index{spectrum}
\index{invertible operator}

\begin{definition}
The spectrum of an operator $T\in B(H)$ is the set
$$\sigma(T)=\left\{\lambda\in\mathbb C\Big|T-\lambda\not\in B(H)^{-1}\right\}$$
where $B(H)^{-1}\subset B(H)$ is the set of invertible operators.
\end{definition}

As a basic example, in the finite dimensional case, $H=\mathbb C^N$, the spectrum of a usual matrix $A\in M_N(\mathbb C)$ is the collection of its eigenvalues, taken without multiplicities. We will see many other examples. In general, the spectrum has the following properties:

\index{eigenvalue}

\begin{proposition}
The spectrum of $T\in B(H)$ contains the eigenvalue set
$$\varepsilon(T)=\left\{\lambda\in\mathbb C\Big|\ker(T-\lambda)\neq\{0\}\right\}$$
and $\varepsilon(T)\subset\sigma(T)$ is an equality in finite dimensions, but not in infinite dimensions.
\end{proposition}

\begin{proof}
We have several assertions here, the idea being as follows:

\medskip

(1) First of all, the eigenvalue set is indeed the one in the statement, because $Tx=\lambda x$ tells us precisely that $T-\lambda$ must be not injective. The fact that we have $\varepsilon(T)\subset\sigma(T)$ is clear as well, because if $T-\lambda$ is not injective, it is not bijective.

\medskip

(2) In finite dimensions we have $\varepsilon(T)=\sigma(T)$, because $T-\lambda$ is injective if and only if it is bijective, with the boundedness of the inverse being automatic. 

\medskip

(3) In infinite dimensions we can assume $H=l^2(\mathbb N)$, and the shift operator $S(e_i)=e_{i+1}$ is injective but not surjective. Thus $0\in\sigma(T)-\varepsilon(T)$.
\end{proof}

Philosophically, the best way of thinking at this is as follows: the numbers $\lambda\notin\sigma(T)$ are good, because we can invert $T-\lambda$, the numbers $\lambda\in\sigma(T)-\varepsilon(T)$ are bad, because so they are, and the eigenvalues $\lambda\in\varepsilon(T)$ are evil. Welcome to operator theory.

\bigskip

Let us develop now some general theory. As a first result, we would like to prove that the spectra are non-empty. This is something tricky, and we will need:

\index{invertible operator}

\begin{proposition}
The following happen:
\begin{enumerate}
\item $||T||<1\implies(1-T)^{-1}=1+T+T^2+\ldots$

\item The set $B(H)^{-1}$ is open.

\item The map $T\to T^{-1}$ is differentiable.
\end{enumerate}
\end{proposition}

\begin{proof}
All these assertions are elementary, as follows:

\medskip

(1) This comes as in the scalar case, the computation being as follows:
$$(1-T)(1+T+T^2+\ldots)
=1-T+T-T^2+T^2-T^3+\ldots
=1$$

(2) Assuming $T\in B(H)^{-1}$, let us pick $S\in B(H)$ such that:
$$||T-S||<\frac{1}{||T^{-1}||}$$

With this choice, we have then the following estimate:
$$||1-T^{-1}S||
=||T^{-1}(T-S)||
\leq||T^{-1}||\cdot||T-S||
<1$$

Thus we have $T^{-1}S\in B(H)^{-1}$, and so $S\in B(H)^{-1}$, as desired.

\medskip

(3) In the scalar case, the derivative of $f(t)=t^{-1}$ is $f'(t)=-t^{-2}$. In the present normed space setting the derivative is no longer a number, but rather a linear transformation, which can be found by developing $f(T)=T^{-1}$ at order 1, as follows:
\begin{eqnarray*}
(T+S)^{-1}
&=&((1+ST^{-1})T)^{-1}\\
&=&T^{-1}(1+ST^{-1})^{-1}\\
&=&T^{-1}(1-ST^{-1}+(ST^{-1})^2-\ldots)\\
&\simeq&T^{-1}(1-ST^{-1})\\
&=&T^{-1}-T^{-1}ST^{-1}
\end{eqnarray*}

Thus $f(T)=T^{-1}$ is indeed differentiable, with derivative $f'(T)S=-T^{-1}ST^{-1}$.
\end{proof}

We can now formulate our first theorem about spectra, as follows:

\index{spectrum}

\begin{theorem}
The spectrum of a bounded operator $T\in B(H)$ is:
\begin{enumerate}
\item Compact.

\item Contained in the disc $D_0(||T||)$.

\item Non-empty.
\end{enumerate}
\end{theorem}

\begin{proof}
This can be proved by using Proposition 12.9, as follows:

\medskip

(1) In view of (2) below, it is enough to prove that $\sigma(T)$ is closed. But this comes from the following computation, with $|\varepsilon|$ being small:
\begin{eqnarray*}
\lambda\notin\sigma(T)
&\implies&T-\lambda\in B(H)^{-1}\\
&\implies&T-\lambda-\varepsilon\in B(H)^{-1}\\
&\implies&\lambda+\varepsilon\notin\sigma(T)
\end{eqnarray*}

(2) This comes indeed from the following computation:
\begin{eqnarray*}
\lambda>||T||
&\implies&\Big|\Big|\frac{T}{\lambda}\Big|\Big|<1\\
&\implies&1-\frac{T}{\lambda}\in B(H)^{-1}\\
&\implies&\lambda-T\in B(H)^{-1}\\
&\implies&\lambda\notin\sigma(T)
\end{eqnarray*}

(3) Assume by contradiction $\sigma(T)=\emptyset$. Given a linear form $f\in B(H)^*$, consider the following map, which is well-defined, due to our assumption $\sigma(T)=\emptyset$:
$$\varphi:\mathbb C\to\mathbb C\quad,\quad 
\lambda\to f((T-\lambda)^{-1})$$

By using the fact that $T\to T^{-1}$ is differentiable, that we know from Proposition 12.9, we conclude that this map is differentiable, and so holomorphic. Also, we have:
\begin{eqnarray*}
\lambda\to\infty
&\implies&T-\lambda\to\infty\\
&\implies&(T-\lambda)^{-1}\to0\\
&\implies&f((T-\lambda))^{-1}\to0
\end{eqnarray*}

Thus by the Liouville theorem we obtain $\varphi=0$. But, in view of the definition of $\varphi$, this gives $(T-\lambda)^{-1}=0$, which is a contradiction, as desired.
\end{proof}

Here is now a second basic result regarding the spectra, inspired from what happens in finite dimensions, for the usual complex matrices, and which shows that things do not necessarily extend without troubles to the infinite dimensional setting:

\index{spectrum of products}

\begin{theorem}
We have the following formula, valid for any operators $S,T$:
$$\sigma(ST)\cup\{0\}=\sigma(TS)\cup\{0\}$$
In finite dimensions we have $\sigma(ST)=\sigma(TS)$, but this fails in infinite dimensions.
\end{theorem}

\begin{proof}
There are several assertions here, the idea being as follows:

\medskip

(1) This is something that we know in finite dimensions, coming from the fact that the characteristic polynomials of the associated matrices $A,B$ coincide:
$$P_{AB}=P_{BA}$$

Thus we obtain $\sigma(ST)=\sigma(TS)$ in this case, as claimed. Observe that this improves twice the general formula in the statement, first because we have no issues at 0, and second because what we obtain is actually an equality of sets with mutiplicities.

\medskip

(2) In general now, let us first prove the main assertion, stating that $\sigma(ST),\sigma(TS)$ coincide outside 0. We first prove that we have the following implication:
$$1\notin\sigma(ST)\implies1\notin\sigma(TS)$$

Assume indeed that $1-ST$ is invertible, with inverse $R=(1-ST)^{-1}$. We have then the following formulae, relating our variables $R,S,T$:
$$RST=STR=R-1$$

By using $RST=R-1$, we have the following computation:
\begin{eqnarray*}
(1+TRS)(1-TS)
&=&1+TRS-TS-TRSTS\\
&=&1+TRS-TS-TRS+TS\\
&=&1
\end{eqnarray*}

A similar computation, using $STR=R-1$, shows that we have:
$$(1-TS)(1+TRS)=1$$

Thus $1-TS$ is invertible, with inverse $1+TRS$, which proves our claim. Now by multiplying by scalars, we deduce that for any $\lambda\in\mathbb C-\{0\}$ we have, as desired:
$$\lambda\notin\sigma(ST)\implies\lambda\notin\sigma(TS)$$ 

(3) Regarding now the counterexample to the formula $\sigma(ST)=\sigma(TS)$, in general, let us take $S$ to be the shift on $H=L^2(\mathbb N)$, given by the following formula:
$$S(e_i)=e_{i+1}$$

As for $T$, we can take it to be the adjoint of $S$, which is the following operator:
$$S^*(e_i)=\begin{cases}
e_{i-1}&{\rm if}\ i>0\\
0&{\rm if}\ i=0
\end{cases}$$

Let us compose now these two operators. In one sense, we have:
$$S^*S=1\implies 0\notin\sigma(SS^*)$$

In the other sense, however, the situation is different, as follows:
$$SS^*=Proj(e_0^\perp)\implies 0\in\sigma(SS^*)$$

Thus, the spectra do not match on $0$, and we have our counterexample, as desired.
\end{proof}

\section*{12b. Spectral radius}

Let us develop now some systematic theory for the computation of the spectra, based on what we know about the eigenvalues of the usual complex matrices. As a first result, which is well-known for the usual matrices, and extends well, we have:

\index{polynomial calculus}

\begin{theorem}
We have the ``polynomial functional calculus'' formula
$$\sigma(P(T))=P(\sigma(T))$$
valid for any polynomial $P\in\mathbb C[X]$, and any operator $T\in B(H)$.
\end{theorem}

\begin{proof}
We pick a scalar $\lambda\in\mathbb C$, and we decompose the polynomial $P-\lambda$:
$$P(X)-\lambda=c(X-r_1)\ldots(X-r_n)$$

We have then the following equivalences, all coming from definitions:
\begin{eqnarray*}
\lambda\notin\sigma(P(T))
&\iff&P(T)-\lambda\in B(H)^{-1}\\
&\iff&c(T-r_1)\ldots(T-r_n)\in B(H)^{-1}\\
&\iff&T-r_1,\ldots,T-r_n\in B(H)^{-1}\\
&\iff&r_1,\ldots,r_n\notin\sigma(T)\\
&\iff&\lambda\notin P(\sigma(T))
\end{eqnarray*}

Thus, we are led to the formula in the statement.
\end{proof}

Next, we have the following useful extension of Theorem 12.12:

\index{rational calculus}

\begin{theorem}
We have the ``rational functional calculus'' formula
$$\sigma(f(T))=f(\sigma(T))$$
valid for any rational function $f\in\mathbb C(X)$ having poles outside $\sigma(T)$.
\end{theorem}

\begin{proof}
We pick a scalar $\lambda\in\mathbb C$, we write $f=P/Q$, and we set:
$$F=P-\lambda Q$$

By using now Theorem 12.12, for this polynomial, we obtain:
\begin{eqnarray*}
\lambda\in\sigma(f(T))
&\iff&F(T)\notin B(H)^{-1}\\
&\iff&0\in\sigma(F(T))\\
&\iff&0\in F(\sigma(T))\\
&\iff&\exists\mu\in\sigma(T),F(\mu)=0\\
&\iff&\lambda\in f(\sigma(T))
\end{eqnarray*}

Thus, we are led to the formula in the statement.
\end{proof}

As an application of the above methods, we can investigate certain special classes of operators, such as the self-adjoint ones, and the unitary ones. Let us start with:

\index{adjoint operator}

\begin{proposition}
The following happen:
\begin{enumerate}
\item We have $\sigma(T^*)=\overline{\sigma(T)}$, for any $T\in B(H)$.

\item If $T=T^*$ then $X=\sigma(T)$ satisfies $X=\overline{X}$.

\item If $U^*=U^{-1}$ then $X=\sigma(U)$ satisfies $X^{-1}=\overline{X}$.
\end{enumerate}
\end{proposition}

\begin{proof}
We have several assertions here, the idea being as follows:

\medskip

(1) The spectrum of the adjoint operator $T^*$ can be computed as follows:
\begin{eqnarray*}
\sigma(T^*)
&=&\left\{\lambda\in\mathbb C\Big|T^*-\lambda\notin B(H)^{-1}\right\}\\
&=&\left\{\lambda\in\mathbb C\Big|T-\bar{\lambda}\notin B(H)^{-1}\right\}\\
&=&\overline{\sigma(T)}
\end{eqnarray*}

(2) This is clear indeed from (1).

\medskip

(3) For a unitary operator, $U^*=U^{-1}$, Theorem 12.13 and (1) give:
$$\sigma(U)^{-1}=\sigma(U^{-1})=\sigma(U^*)=\overline{\sigma(U)}$$

Thus, we are led to the conclusion in the statement.
\end{proof}

In analogy with what happens for the usual matrices, we would like to improve now (2,3) above, with results stating that the spectrum $X=\sigma(T)$ satisfies $X\subset\mathbb R$ for self-adjoints, and $X\subset\mathbb T$ for unitaries. This will be tricky. Let us start with:

\index{unitary}

\begin{theorem}
The spectrum of a unitary operator 
$$U^*=U^{-1}$$
is on the unit circle, $\sigma(U)\subset\mathbb T$. 
\end{theorem}

\begin{proof}
Assuming $U^*=U^{-1}$, we have the following norm computation:
$$||U||
=\sqrt{||UU^*||}
=\sqrt{1}
=1$$

Now if we denote by $D$ the unit disk, we obtain from this:
$$\sigma(U)\subset D$$

On the other hand, once again by using $U^*=U^{-1}$, we have as well:
$$||U^{-1}||
=||U^*||
=||U||
=1$$

Thus, as before with $D$ being the unit disk in the complex plane, we have:
$$\sigma(U^{-1})\subset D$$

Now by using Theorem 12.13, we obtain $\sigma(U)
\subset D\cap D^{-1}
=\mathbb T$, as desired.
\end{proof}

We have as well a similar result for self-adjoints, as follows:

\index{self-adjoint operator}

\begin{theorem}
The spectrum of a self-adjoint operator
$$T=T^*$$
consists of real numbers, $\sigma(T)\subset\mathbb R$.
\end{theorem}

\begin{proof}
The idea is that we can deduce the result from Theorem 12.15, by using the following remarkable rational function, depending on a parameter $r\in\mathbb R$:
$$f(z)=\frac{z+ir}{z-ir}$$

Indeed, for $r>>0$ the operator $f(T)$ is well-defined, and we have:
$$\left(\frac{T+ir}{T-ir}\right)^*
=\frac{T-ir}{T+ir}
=\left(\frac{T+ir}{T-ir}\right)^{-1}$$

Thus $f(T)$ is unitary, and by using Theorem 12.15 we obtain:
$$\sigma(T)
\subset f^{-1}(f(\sigma(T)))
=f^{-1}(\sigma(f(T)))
\subset f^{-1}(\mathbb T)
=\mathbb R$$

Thus, we are led to the conclusion in the statement.
\end{proof}

Moving on, one key thing that we know about matrices, which is clear for the diagonalizable matrices, and then in general follows by density, is the following formula:
$$\sigma(e^A)=e^{\sigma(A)}$$

We would like to have such formulae for the general operators $T\in B(H)$, but this is something quite technical, needing the Cauchy formula. Let us start with:

\index{Cauchy formula}
\index{loop}
\index{winding number}

\begin{definition}
If $\gamma$ is a loop in $\mathbb C$, the number of times $\gamma$ goes around a point $z\in\mathbb C-\gamma$ is computed by the following integral, called winding number:
$$Ind(\gamma,z)=\frac{1}{2\pi i}\int_\gamma\frac{dy}{y -z}$$
We say that $\gamma$ turns around $z$ if $Ind(\gamma,z)=1$, and that it does not turn if $Ind(\gamma,z)=0$. Otherwise, we say that $\gamma$ turns around $z$ many times, or in the bad sense, or both.
\end{definition}

Next, let $f:X\to\mathbb C$ be an holomorphic function defined on $X\subset\mathbb C$ open, and $\gamma$ be a loop in $X$. If $Ind(\gamma,z)\neq 0$ for $z\in\mathbb C-X$ then $f(z)$ is given by the Cauchy formula:
$$Ind(\gamma,z)f(z)=\frac{1}{2\pi i}\int_\gamma\frac{f(y)}{y-z}\,dy$$

Also, if $Ind(\gamma,z)=0$ for $z\in\mathbb C-X$ then the integral of $f$ on $\gamma$ is zero:
$$\int_\gamma f(y)\,dy=0$$

In practice, we will need in fact something which is even more general than this. Let us introduce the formal combinations of loops, called cycles:
$$\Sigma=n_1\gamma_1+\ldots +n_r\gamma_r$$

We can define then the winding number for $\Sigma$ as being the corresponding linear combination of winding numbers of its loop components, and the Cauchy formula holds then in this setting, for the arbitrary cycles. Let us summarize this discussion as follows:

\begin{theorem}
We have the generalized Cauchy formula
$$Ind(\Sigma,z)f(z)=\frac{1}{2\pi i}\int_\Sigma\frac{f(y)}{y-z}\,dy$$
valid for any holomorphic function $f:X\to\mathbb C$. 
\end{theorem}

\begin{proof}
This follows indeed from the usual Cauchy formula, from chapter 11, according to the discussion above, first for the arbitrary loops, and then for the cycles.
\end{proof}

Now by getting back to operators, we can formulate the following key definition:

\index{holomorphic calculus}
\index{loop}
\index{cycle}

\begin{definition}
Let $T\in B(H)$ and let $f:X\to\mathbb C$ be an holomorphic function defined on an open set containing $\sigma(T)$. Define an element $f(T)$ by the formula
$$f(T)=\frac{1}{2\pi i}\int_\Sigma\frac{f(y)}{y-T}\,dy$$
where $\Sigma$ is a cycle in $X-\sigma(T)$ which turns around $\sigma(T)$ and doesn't turn around $\mathbb C-X$.
\end{definition}

The formula makes sense because $\Sigma$ is in $X-\sigma(T)$. Also, $f(T)$ is independent of the choice of $\Sigma$. Indeed, let $\Sigma_1$ and $\Sigma_2$ be two cycles. Their difference $\Sigma_1-\Sigma_2$ is a cycle which doesn't turn around $\sigma(T)$, neither around $\mathbb C-X$. The function $z\to f(z)/(z-T)$ being holomorphic $X-\sigma(T)\to B(H)$, its integral on $\Sigma_1-\Sigma_2$ must be zero:
$$\int_{\Sigma_1-\Sigma_2}\frac{f(y)}{y-T}\,dy=0$$

Thus $f(T)$ is the same with respect to $\Sigma_1$ and to $\Sigma_2$, and so Definition 12.19 is fully justified. Now with this definition in hand, we first have the following result:

\begin{proposition}
We have the formula
$$f(T)g(T)=(fg)(T)$$
whenever the equality makes sense.
\end{proposition}

\begin{proof}
Let $\Sigma_1$ be a cycle in $X-\sigma(T)$ around $\sigma(T)$ and consider the following set:
$$Int(\Sigma_1)=\left\{z\in\mathbb C-\Sigma_1\Big|Ind(\Sigma_1,z)\neq 0\right\}\bigcup\Sigma_1$$

This is a compact set, included in $X$ and containing the spectrum of $T$:
$$\sigma(T)\subset Int(\Sigma_1)\subset X$$

Let $\Sigma_2$ be a cycle in $X-Int(\Sigma_1)$ turning around $Int(\Sigma_1)$. Consider two holomorphic functions $f,g$ defined around $\sigma(T)$, so that the statement make sense. We have:
\begin{eqnarray*}
f(T)g(T)
&=&\left(\frac{1}{2\pi i}\right)^2\left(\int_{\Sigma_1}\frac{f(y)}{y -T}\,dy\right) 
\left(\int_{\Sigma_2}\frac{g(z)}{z-T}\,dz\right)\\
&=&\left(\frac{1}{2\pi i}\right)^2\int_{\Sigma_1}\int_{\Sigma_2}\frac{f(y)g(z)}{(y-T)(z-T)}\,dydz
\end{eqnarray*}

In order to integrate, we can use the following identity:
$$\frac{1}{(y-T)(z-T)}=\frac{1}{(z-y)(y-T)}+\frac{1}{(y-z)(z-T)}$$

Thus our integral splits into two terms. The first term can be computed by integrating first over $\Sigma_2$, and we obtain $(fg)(T)$. As for the second term, here we can integrate first over $\Sigma_1$, and we get $0$. It follows that $f(T)g(T)$ is equal to $(fg)(T)$, as claimed.
\end{proof}

We can now formulate our extension of rational calculus, as follows:

\index{holomorphic calculus}
\index{Cauchy formula}

\begin{theorem}
Given $T\in B(H)$, we have a morphism of algebras as follows, where $Hol(\sigma(T))$ is the algebra of functions which are holomorphic around $\sigma(T)$,
$$Hol(\sigma(T))\to B(H)\quad,\quad f\to f(T)$$
which extends the previous rational functional calculus $f\to f(T)$. We have:
$$\sigma(f(T))=f(\sigma(T))$$
Moreover, if $\sigma(T)$ is contained in an open set $X$ and $f_n,f:X\to\mathbb C$ are holomorphic functions such that $f_n\to f$ uniformly on compact subsets of $X$ then $f_n(T)\to f(T)$.
\end{theorem}

\begin{proof}
There are several things to be proved here, as follows:

\medskip

(1) Consider indeed the algebra $Hol(\sigma(T))$, with the convention that two functions are identified if they coincide on an open set containing $\sigma(T)$. We have then a construction $f\to f(T)$ as in the statement, provided by Definition 12.19 and Proposition 12.20.

\medskip

(2) Let us prove now that our construction extends the one for rational functions. Since $1,z$ generate $\mathbb C(X)$, it is enough to show that $f(z)=1$ implies $f(T)=1$, and that $f(z)=z$ implies $f(T)=T$. For this purpose, we prove that $f(z)=z^n$ implies $f(T)=T^n$ for any $n$. But this follows by integrating over a circle $\gamma$ of big radius, as follows:
\begin{eqnarray*}
f(T)
&=&\frac{1}{2\pi i}\int_\gamma\frac{y^n}{y-T}\,dy\\
&=&\frac{1}{2\pi i}\int_\gamma y^{n-1}\left(1-\frac{T}{y}\right)^{-1}dy\\
&=&\frac{1}{2\pi i}\int_\gamma y^{n-1}\left(\sum_{k=0}^\infty y^{-k}T^k\right)dy\\
&=&\sum_{k=0}^\infty\left(\frac{1}{2\pi i}\int_\gamma y^{n-k-1}dy\right)T^k\\
&=&T^n
\end{eqnarray*}

(3) Regarding now $\sigma(f(T))=f(\sigma(T))$, by translation it is enough to prove that this equality holds on the point $0$, and we can do this by double inclusion, as follows:

\medskip

``$\supset$''. Assume that $f(\sigma(T))$ contains $0$, and let $z_0\in\sigma(T)$ be such that $f(z_0)=0$. Consider the function $g(z)=f(z)/(z-z_0)$.  We have $g(T)(T-z_0)=f(T)$ by using the morphism property. Since $T-z_0$ is not invertible, $f(T)$ is not invertible either.

\medskip

``$\subset$''. Assume now that $f(\sigma(T))$ does not contain $0$. With the holomorphic function $g(z)=1/f(z)$ we get $g(T)=f(T)^{-1}$, so $f(T)$ is invertible, and we are done.

\medskip

(4) Finally, the last assertion is clear from definitions. And with the remark that this can be applied to holomorphic functions written as series, $f(z)=\sum_{n=0}^\infty a_n(z-z_0)^n$. Indeed, if this is the expansion of $f$ around $z_0$, with convergence radius $r$, and if $\sigma(T)$ is contained in the disc centered at $z_0$ of radius $r$, then $f(T)$ is given by:
$$f(T)=\sum_{n=0}^\infty a_n(T-z_0)^n$$

Summarizing, we have proved the result, and fully extended the rational calculus.
\end{proof}

In order to formulate now our next result, we will need the following notion:

\index{spectral radius}

\begin{definition}
Given an operator $T\in B(H)$, its spectral radius 
$$\rho(T)\in\big[0,||T||\big]$$
is the radius of the smallest disk centered at $0$ containing $\sigma(T)$. 
\end{definition}

Now with this notion in hand, we have the following key result, improving our key theoretical result so far about spectra, namely $\sigma(T)\neq\emptyset$, from Theorem 12.10:

\begin{theorem}
The spectral radius of an operator $T\in B(H)$ is given by
$$\rho(T)=\lim_{n\to\infty}||T^n||^{1/n}$$
and in this formula, we can replace the limit by an inf.
\end{theorem}

\begin{proof}
We have several things to be proved, the idea being as follows:

\medskip

(1) Our first claim is that the numbers $u_n=||T^n||^{1/n}$ satisfy:
$$(n+m)u_{n+m}\leq nu_n+mu_m$$

Indeed, we have the following estimate, using the Young inequality $ab\leq a^p/p+b^q/q$, with exponents $p=(n+m)/n$ and $q=(n+m)/m$:
\begin{eqnarray*}
u_{n+m}
&=&||T^{n+m}||^{1/(n+m)}\\
&\leq&||T^n||^{1/(n+m)}||T^m||^{1/(n+m)}\\
&\leq&||T^n||^{1/n}\cdot\frac{n}{n+m}+||T^m||^{1/m}\cdot\frac{m}{n+m}\\
&=&\frac{nu_n+mu_m}{n+m}
\end{eqnarray*}

(2) Our second claim is that the second assertion holds, namely:
$$\lim_{n\to\infty}||T^n||^{1/n}=\inf_n||T^n||^{1/n}$$

For this purpose, we just need the inequality found in (1). Indeed, fix $m\geq1$, let $n\geq1$, and write $n=lm+r$ with $0\leq r\leq m-1$. By using twice $u_{ab}\leq u_b$, we get:
\begin{eqnarray*}
u_n
&\leq&\frac{1}{n}( lmu_{lm}+ru_r)\\
&\leq&\frac{1}{n}( lmu_{m}+ru_1)\\
&\leq&u_{m}+\frac{r}{n}\,u_1
\end{eqnarray*}

It follows that we have $\lim\sup_nu_n\leq u_m$, which proves our claim.

\medskip

(3) Summarizing, we are left with proving the main formula, which is as follows, and with the remark that we already know that the sequence on the right converges:
$$\rho(T)=\lim_{n\to\infty}||T^n||^{1/n}$$

In one sense, we can use the polynomial calculus formula $\sigma(T^n)=\sigma(T)^n$. Indeed, this gives the following estimate, valid for any $n$, as desired:
\begin{eqnarray*}
\rho(T)
&=&\sup_{\lambda\in\sigma(T)}|\lambda|\\
&=&\sup_{\rho\in\sigma(T)^n}|\rho|^{1/n}\\
&=&\sup_{\rho\in\sigma(T^n)}|\rho|^{1/n}\\
&=&\rho(T^n)^{1/n}\\
&\leq&||T^n||^{1/n}
\end{eqnarray*}

(4) For the reverse inequality, we fix a number $\rho>\rho(T)$, and we want to prove that we have $\rho\geq\lim_{n\to\infty}||T^n||^{1/n}$. By using the Cauchy formula, we have:
\begin{eqnarray*}
\frac{1}{2\pi i}\int_{|z|=\rho}\frac{z^n}{z-T}\,dz
&=&\frac{1}{2\pi i}\int_{|z|=\rho}\sum_{k=0}^\infty z^{n-k-1}T^k\,dz\\
&=&\sum_{k=0}^\infty\frac{1}{2\pi i}\left(\int_{|z|=\rho}z^{n-k-1}dz\right)T^k\\
&=&\sum_{k=0}^\infty\delta_{n,k+1}T^k\\
&=&T^{n-1}
\end{eqnarray*}

By applying the norm we obtain from this formula:
$$||T^{n-1}||
\leq\frac{1}{2\pi}\int_{|z|=\rho}\left|\left|\frac{z^n}{z-T}\right|\right|\,dz
\leq\rho^n\cdot\sup_{|z|=\rho}\left|\left|\frac{1}{z-T}\right|\right|$$

Since the sup does not depend on $n$, by taking $n$-th roots, we obtain in the limit:
$$\rho\geq\lim_{n\to\infty}||T^n||^{1/n}$$

Now recall that $\rho$ was by definition an arbitrary number satisfying $\rho>\rho(T)$. Thus, we have obtained the following estimate, valid for any $T\in B(H)$:
$$\rho(T)\geq\lim_{n\to\infty}||T^n||^{1/n}$$

Thus, we are led to the conclusion in the statement.
\end{proof}

In the case of the normal elements, we have the following finer result:

\index{normal operator}

\begin{theorem}
The spectral radius of a normal element,
$$TT^*=T^*T$$
is equal to its norm.
\end{theorem}

\begin{proof}
We can proceed in two steps, as follows:

\medskip

\underline{Step 1}. In the case $T=T^*$ we have $||T^n||=||T||^n$ for any exponent of the form $n=2^k$, by using the formula $||TT^*||=||T||^2$, and by taking $n$-th roots we get:
$$\rho(T)\geq||T||$$

Thus, we are done with the self-adjoint case, with the result $\rho(T)=||T||$.

\medskip

\underline{Step 2}. In the general normal case $TT^*=T^*T$ we have $T^n(T^n)^*=(TT^*)^n$, and by using this, along with the result from Step 1, applied to $TT^*$, we obtain:
\begin{eqnarray*}
\rho(T)
&=&\lim_{n\to\infty}||T^n||^{1/n}\\
&=&\sqrt{\lim_{n\to\infty}||T^n(T^n)^*||^{1/n}}\\
&=&\sqrt{\lim_{n\to\infty}||(TT^*)^n||^{1/n}}\\
&=&\sqrt{\rho(TT^*)}\\
&=&\sqrt{||T||^2}\\
&=&||T||
\end{eqnarray*}

Thus, we are led to the conclusion in the statement.
\end{proof}

\section*{12c. Normal operators} 

By using Theorem 12.24 we can say a number of non-trivial things about the self-adjoint and normal operators, commonly known as ``spectral theorems''. As a first result here, we can improve the polynomial functional calculus formula, as follows:

\index{normal operator}
\index{polynomial calculus}

\begin{theorem}
Given $T\in B(H)$ normal, we have a morphism of algebras
$$\mathbb C[X]\to B(H)\quad,\quad 
P\to P(T)$$
having the properties $||P(T)||=||P_{|\sigma(T)}||$, and $\sigma(P(T))=P(\sigma(T))$.
\end{theorem}

\begin{proof}
This is an improvement of Theorem 12.12 in the normal case, with the extra assertion being the norm estimate. But the element $P(T)$ being normal, we can apply to it the spectral radius formula for normal elements, and we obtain:
\begin{eqnarray*}
||P(T)||
&=&\rho(P(T))\\
&=&\sup_{\lambda\in\sigma(P(T))}|\lambda|\\
&=&\sup_{\lambda\in P(\sigma(T))}|\lambda|\\
&=&||P_{|\sigma(T)}||
\end{eqnarray*}

Thus, we are led to the conclusions in the statement.
\end{proof}

We can improve as well the rational calculus formula, and the holomorphic calculus formula, in the same way. Importantly now, at a more advanced level, we have:

\index{normal operator}
\index{continuous calculus}

\begin{theorem}
Given $T\in B(H)$ normal, we have a morphism of algebras
$$C(\sigma(T))\to B(H)\quad,\quad 
f\to f(T)$$
which is isometric, $||f(T)||=||f||$, and has the property $\sigma(f(T))=f(\sigma(T))$.
\end{theorem}

\begin{proof}
The idea here is to ``complete'' the morphism in Theorem 12.25, namely:
$$\mathbb C[X]\to B(H)\quad,\quad 
P\to P(T)$$

Indeed, we know from Theorem 12.25 that this morphism is continuous, and is in fact isometric, when regarding the polynomials $P\in\mathbb C[X]$ as functions on $\sigma(T)$:
$$||P(T)||=||P_{|\sigma(T)}||$$

Thus, by Stone-Weierstrass, we have a unique isometric extension, as follows:
$$C(\sigma(T))\to B(H)\quad,\quad  
f\to f(T)$$

It remains to prove $\sigma(f(T))=f(\sigma(T))$, and we can do this by double inclusion:

\medskip

``$\subset$'' Given a continuous function $f\in C(\sigma(T))$, we must prove that we have:
$$\lambda\notin f(\sigma(T))\implies\lambda\notin\sigma(f(T))$$

For this purpose, consider the following function, which is well-defined:
$$\frac{1}{f-\lambda}\in C(\sigma(T))$$

We can therefore apply this function to $T$, and we obtain:
$$\left(\frac{1}{f-\lambda}\right)T=\frac{1}{f(T)-\lambda}$$

In particular $f(T)-\lambda$ is invertible, so  $\lambda\notin\sigma(f(T))$, as desired.

\medskip

``$\supset$'' Given a continuous function $f\in C(\sigma(T))$, we must prove that we have: 
$$\lambda\in f(\sigma(T))\implies\lambda\in\sigma(f(T))$$

But this is the same as proving that we have the following implication:
$$\mu\in\sigma(T)\implies f(\mu)\in\sigma(f(T))$$

For this purpose, we approximate our function by polynomials, $P_n\to f$, and we examine the following convergence, which follows from $P_n\to f$:
$$P_n(T)-P_n(\mu)\to f(T)-f(\mu)$$

We know from polynomial functional calculus that we have:
$$P_n(\mu)
\in P_n(\sigma(T))
=\sigma(P_n(T))$$

Thus, the operators $P_n(T)-P_n(\mu)$ are not invertible. On the other hand, we know that the set formed by the invertible operators is open, so its complement is closed. Thus the limit $f(T)-f(\mu)$ is not invertible either, and so $f(\mu)\in\sigma(f(T))$, as desired.
\end{proof}

As an important comment, Theorem 12.26 is not exactly in final form, because it misses an important point, namely that our correspondence maps:
$$\bar{z}\to T^*$$

However, this is something non-trivial, and we will be back to this later. Observe however that Theorem 12.26 is fully powerful for the self-adjoint operators, $T=T^*$, where the spectrum is real, so where $z=\bar{z}$ on the spectrum. We will be back to this.

\bigskip

As a second result now, along the same lines, we can further extend Theorem 12.26 into a measurable functional calculus theorem, as follows:

\index{normal operator}
\index{measurable calculus}

\begin{theorem}
Given $T\in B(H)$ normal, we have a morphism of algebras as follows, with $L^\infty$ standing for abstract measurable functions, or Borel functions,
$$L^\infty(\sigma(T))\to B(H)\quad,\quad 
f\to f(T)$$
which is isometric, $||f(T)||=||f||$, and has the property $\sigma(f(T))=f(\sigma(T))$.
\end{theorem}

\begin{proof}
As before, the idea will be that of ``completing'' what we have. To be more precise, we can use the Riesz theorem and a polarization trick, as follows:

\medskip

(1) Given a vector $x\in H$, consider the following functional:
$$C(\sigma(T))\to\mathbb C\quad,\quad 
g\to<g(T)x,x>$$

By the Riesz theorem, this functional must be the integration with respect to a certain measure $\mu$ on the space $\sigma(T)$. Thus, we have a formula as follows:
$$<g(T)x,x>=\int_{\sigma(T)}g(z)d\mu(z)$$

Now given an arbitrary Borel function $f\in L^\infty(\sigma(T))$, as in the statement, we can define a number $<f(T)x,x>\in\mathbb C$, by using exactly the same formula, namely:
$$<f(T)x,x>=\int_{\sigma(T)}f(z)d\mu(z)$$

Thus, we have managed to define numbers $<f(T)x,x>\in\mathbb C$, for all vectors $x\in H$, and in addition we can recover these numbers as follows, with $g_n\in C(\sigma(T))$:
$$<f(T)x,x>=\lim_{g_n\to f}<g_n(T)x,x>$$ 

(2) In order to define now numbers $<f(T)x,y>\in\mathbb C$, for all vectors $x,y\in H$, we can use a polarization trick. Indeed, for any operator $S\in B(H)$ we have:
$$<S(x+y),x+y>=<Sx,x>+<Sy,y>+<Sx,y>+<Sy,x>$$

By replacing $y\to iy$, we have as well the following formula:
$$<S(x+iy),x+iy>=<Sx,x>+<Sy,y>-i<Sx,y>+i<Sy,x>$$

By multiplying this formula by $i$, and summing with the first one, we obtain:
\begin{eqnarray*}
<S(x+y),x+y>+i<S(x+iy),x+iy>
&=&(1+i)[<Sx,x>+<Sy,y>]\\
&+&2<Sx,y>
\end{eqnarray*}

(3) But with this, we can now finish. Indeed, by combining (1,2), given a Borel function $f\in L^\infty(\sigma(T))$, we can define numbers $<f(T)x,y>\in\mathbb C$ for any $x,y\in H$, and it is routine to check, by using approximation by continuous functions $g_n\to f$ as in (1), that we obtain in this way an operator $f(T)\in B(H)$, having all the desired properties.
\end{proof}

As a comment here, the above result and its proof provide us with more than a Borel functional calculus, because what we got is a certain measure on the spectrum $\sigma(T)$, along with a functional calculus for the $L^\infty$ functions with respect to this measure. We will be back to this later, and for the moment we will only need Theorem 12.27 as formulated, with $L^\infty(\sigma(T))$ standing, a bit abusively, for the Borel functions on $\sigma(T)$.

\section*{12d. Diagonalization}

We can now diagonalize the normal operators. We will do this in 3 steps, first for the self-adjoint operators, then for the families of commuting self-adjoint operators, and finally for the general normal operators, by using a trick of the following type:
$$T=Re(T)+iIm(T)$$

The diagonalization in infinite dimensions is more tricky than in finite dimensions, and instead of writing a formula of type $T=UDU^*$, with $U,D\in B(H)$ being respectively unitary and diagonal, we will express our operator as $T=U^*MU$, with $U:H\to K$ being a certain unitary, and $M\in B(K)$ being a certain diagonal operator. This is how the spectral theorem is best formulated, in view of applications. 

\bigskip

In practice, the explicit construction of $U,M$, which will be actually part of the proof, is also needed. For the self-adjoint operators, the statement and proof are as follows:

\index{self-adjoint operator}
\index{diagonalization}

\begin{theorem}
Any self-adjoint operator $T\in B(H)$ can be diagonalized,
$$T=U^*M_fU$$
with $U:H\to L^2(X)$ being a unitary operator from $H$ to a certain $L^2$ space associated to $T$, with $f:X\to\mathbb R$ being a certain function, once again associated to $T$, and with
$$M_f(g)=fg$$
being the usual multiplication operator by $f$, on the Hilbert space $L^2(X)$.
\end{theorem}

\begin{proof}
The construction of $U,f$ can be done in several steps, as follows:

\medskip

(1) We first prove the result in the special case where our operator $T$ has a cyclic vector $x\in H$, with this meaning that the following holds:
$$\overline{span\left(T^kx\Big|n\in\mathbb N\right)}=H$$

For this purpose, let us go back to the proof of Theorem 12.27. We will use the following formula from there, with $\mu$ being the measure on $X=\sigma(T)$ associated to $x$:
$$<g(T)x,x>=\int_{\sigma(T)}g(z)d\mu(z)$$

Our claim is that we can define a unitary $U:H\to L^2(X)$, first on the dense part spanned by the vectors $T^kx$, by the following formula, and then by continuity:
$$U[g(T)x]=g$$

Indeed, the following computation shows that $U$ is well-defined, and isometric:
\begin{eqnarray*}
||g(T)x||^2
&=&<g(T)x,g(T)x>\\
&=&<g(T)^*g(T)x,x>\\
&=&<|g|^2(T)x,x>\\
&=&\int_{\sigma(T)}|g(z)|^2d\mu(z)\\
&=&||g||_2^2
\end{eqnarray*}

We can then extend $U$ by continuity into a unitary $U:H\to L^2(X)$, as claimed. Now observe that we have the following formula:
\begin{eqnarray*}
UTU^*g
&=&U[Tg(T)x]\\
&=&U[(zg)(T)x]\\
&=&zg
\end{eqnarray*} 

Thus our result is proved in the present case, with $U$ as above, and with $f(z)=z$.

\medskip

(2) We discuss now the general case. Our first claim is that $H$ has a decomposition as follows, with each $H_i$ being invariant under $T$, and admitting a cyclic vector $x_i$:
$$H=\bigoplus_iH_i$$

Indeed, this is something elementary, the construction being by recurrence in finite dimensions, in the obvious way, and by using the Zorn lemma in general. Now with this decomposition in hand, we can make a direct sum of the diagonalizations obtained in (1), for each of the restrictions $T_{|H_i}$, and we obtain the formula in the statement.
\end{proof}

We have the following technical generalization of the above result:

\index{commuting self-adjoint operators}
\index{diagonalization}

\begin{theorem}
Any family of commuting self-adjoint operators $T_i\in B(H)$ can be jointly diagonalized,
$$T_i=U^*M_{f_i}U$$
with $U:H\to L^2(X)$ being a unitary operator from $H$ to a certain $L^2$ space associated to $\{T_i\}$, with $f_i:X\to\mathbb R$ being certain functions, once again associated to $T_i$, and with
$$M_{f_i}(g)=f_ig$$
being the usual multiplication operator by $f_i$, on the Hilbert space $L^2(X)$.
\end{theorem}

\begin{proof}
This is similar to the proof of Theorem 12.28, by suitably modifying the measurable calculus formula, and the measure $\mu$ itself, as to have this formula working for all the operators $T_i$. With this modification done, everything extends.
\end{proof}

In order to discuss now the case of the arbitrary normal operators, we will need:

\begin{proposition}
Any operator $T\in B(H)$ can be written as
$$T=Re(T)+iIm(T)$$
with $Re(T),Im(T)\in B(H)$ being self-adjoint, and this decomposition is unique.
\end{proposition}

\begin{proof}
This is something elementary, the idea being as follows:

\medskip

(1) We can use indeed the same formulae for the real and imaginary part as in the complex number case, the decomposition formula being as follows:
$$T=\frac{T+T^*}{2}+i\cdot\frac{T-T^*}{2i}$$

To be more precise, both the operators on the right are self-adjoint, and the summing formula holds indeed, and so we have our decomposition result, as desired.

\medskip

(2) Regarding now the uniqueness, by linearity it is enough to show that $R+iS=0$ with $R,S$ both self-adjoint implies $R=S=0$. But this follows by applying the adjoint to $R+iS=0$, which gives $R-iS=0$, and so $R=S=0$, as desired.
\end{proof}

We can now discuss the case of arbitrary normal operators, as follows:

\index{normal operator}
\index{diagonalization}

\begin{theorem}
Any normal operator $T\in B(H)$ can be diagonalized,
$$T=U^*M_fU$$
with $U:H\to L^2(X)$ being a unitary operator from $H$ to a certain $L^2$ space associated to $T$, with $f:X\to\mathbb C$ being a certain function, once again associated to $T$, and with
$$M_f(g)=fg$$
being the usual multiplication operator by $f$, on the Hilbert space $L^2(X)$.
\end{theorem}

\begin{proof}
This is our main diagonalization theorem, the idea being as follows:

\medskip

(1) Consider the decomposition of $T$ into its real and imaginary parts, as constructed in the proof of Proposition 12.30, namely:
$$T=\frac{T+T^*}{2}+i\cdot\frac{T-T^*}{2i}$$

We know that the real and imaginary parts are self-adjoint operators. Now since $T$ was assumed to be normal, $TT^*=T^*T$, these real and imaginary parts commute:
$$\left[\frac{T+T^*}{2}\,,\,\frac{T-T^*}{2i}\right]=0$$

Thus Theorem 12.29 applies to these real and imaginary parts, and gives the result.
\end{proof}

This was for our series of diagonalization theorems. There is of course one more result here, regarding the families of commuting normal operators, as follows:

\index{commuting normal operators}
\index{diagonalization}

\begin{theorem}
Any family of commuting normal operators $T_i\in B(H)$ can be jointly diagonalized,
$$T_i=U^*M_{f_i}U$$
with $U:H\to L^2(X)$ being a unitary operator from $H$ to a certain $L^2$ space associated to $\{T_i\}$, with $f_i:X\to\mathbb C$ being certain functions, once again associated to $T_i$, and with
$$M_{f_i}(g)=f_ig$$
being the usual multiplication operator by $f_i$, on the Hilbert space $L^2(X)$.
\end{theorem}

\begin{proof}
This is indeed similar to the proof of Theorem 12.29 and Theorem 12.31.
\end{proof}

With the above diagonalization results in hand, we can now ``fix'' the continuous and measurable functional calculus theorems, with a key complement, as follows:

\index{polynomial calculus}
\index{rational calculus}
\index{holomorphic calculus}
\index{continuous calculus}
\index{measurable calculus}
\index{adjoint operator}

\begin{theorem}
Given a normal operator $T\in B(H)$, the following hold, for both the functional calculus and the measurable calculus morphisms:
\begin{enumerate}
\item These morphisms are $*$-morphisms.

\item The function $\bar{z}$ gets mapped to $T^*$.

\item The functions $Re(z),Im(z)$ get mapped to $Re(T),Im(T)$.

\item The function $|z|^2$ gets mapped to $TT^*=T^*T$.

\item If $f$ is real, then $f(T)$ is self-adjoint. 
\end{enumerate}
\end{theorem}

\begin{proof}
These assertions are more or less equivalent, with (1) being the main one, which obviously implies everything else. But this assertion (1) follows from the diagonalization result for normal operators, from Theorem 12.31.
\end{proof}

\section*{12e. Exercises}

This was a quite fundamental chapter, and as exercises on this, we have:

\begin{exercise}
Develop the theory of positive operators, satisfying $<Tx,x>\geq0$.
\end{exercise}

\begin{exercise}
Discuss as well the strict positivity, $<Tx,x>>0$ for $x\neq0$.
\end{exercise}

\begin{exercise}
Check the various details of the holomorphic calculus theorem.
\end{exercise}

\begin{exercise}
Check as well the details of the measurable calculus theorem.
\end{exercise}

\begin{exercise}
Learn the various alternative proofs for the spectral theorem.
\end{exercise}

\begin{exercise}
Work out a polar decomposition result, for the linear operators.
\end{exercise}

\begin{exercise}
Learn about the compact operators, and their various properties.
\end{exercise}

\begin{exercise}
Learn also about unbounded operators, and quantum mechanics.
\end{exercise}

As bonus exercise, and no surprise here, learn some systematic operator theory.

\part{Haar integration}

\ \vskip50mm

\begin{center}
{\em Down the road I look and there runs Mary

Hair of gold and lips like cherries

It's good to touch

The green, green grass of home}
\end{center}

\chapter{Random walks}

\section*{13a. Random walks}

We have learned the basics of measure theory and probability, and time now to see if this knowledge can be of any help, in relation with concrete questions. The first question that we would like to discuss, which is something very basic, is as follows:

\begin{question}
Given a graph $X$, with a distinguished vertex $*$:
\begin{enumerate}
\item What is the number $L_k$ of length $k$ loops on $X$, based at $*$? 

\item Equivalently, what is the measure $\mu$ having $L_k$ as moments?
\end{enumerate}
\end{question}

To be more precise, we are mainly interested in the first question, counting loops on graphs, with this being notoriously related to many applied mathematics questions, of discrete type. As for the second question, this is a technical, useful probabilistic reformulation of the first question, that we will usually prefer, in what follows. 

\bigskip

In order to discuss this, let us start with some graph theory basics. We first have:

\index{adjacency matrix}
\index{symmetric matrix}

\begin{proposition}
A graph $X$, with vertices labeled $1,\ldots,N$, is uniquely determined by its adjacency matrix, which is the matrix $d\in M_N(0,1)$ given by:
$$d_{ij}=\begin{cases}
1&{\rm if}\ i-j\\
0&{\rm if}\ {i\not\!\!-\,j}
\end{cases}$$
Moreover, the matrices $d\in M_N(0,1)$ which can appear in this way, from graphs, are precisely those which are symmetric, and have $0$ on the diagonal.
\end{proposition}

\begin{proof}
We have two things to be proved, the idea being as follows:

\medskip

(1) Given a graph $X$, we can construct a matrix $d\in M_N(0,1)$ as in the statement, and this matrix is obviously symmetric, and has $0$ on the diagonal.

\medskip

(2) Conversely, given a matrix $d\in M_N(0,1)$ which is symmetric, and has $0$ on the diagonal, we can associate to it the graph $X$ having as vertices the numbers $1,\ldots,N$, and with the edges between these vertices being defined as follows:
$$i-j\iff d_{ij}=1$$

It is then clear that the adjacency matrix of this graph $X$ is the matrix $d$ itself. Thus, we have established a correspondence $X\leftrightarrow d$, as in the statement.
\end{proof}

Getting now to the walks on graphs, this brings us into the mathematics of the corresponding adjacency matrix $d$, thanks to the following key observation:

\index{paths on graph}
\index{power of matrix}

\begin{proposition}
For a graph $X$, with adjacency matrix $d\in M_N(0,1)$, we have:
$$(d^k)_{ij}=\#\Big\{i=i_0-i_1-\ldots-i_{k-1}-i_k=j\Big\}$$
That is, the $k$-th power of $d$ describes the length $k$ paths on $X$.
\end{proposition}

\begin{proof}
According to the usual rule of matrix multiplication, the formula for the powers of the adjacency matrix $d\in M_N(0,1)$ is as follows:
\begin{eqnarray*}
(d^k)_{i_0i_k}
&=&\sum_{i_1,\ldots,i_{k-1}}d_{i_0i_1}d_{i_1i_2}\ldots d_{i_{k-1}i_k}\\
&=&\sum_{i_1,\ldots,i_{k-1}}\delta_{i_0-i_1}\delta_{i_1-i_2}\ldots\delta_{i_{k-1}-i_k}\\
&=&\sum_{i_1,\ldots,i_{k-1}}\delta_{i_0-i_1-\ldots-i_{k-1}-i_k}\\
&=&\#\Big\{i_0-i_1-\ldots-i_{k-1}-i_k\Big\}
\end{eqnarray*}

Thus, we are led to the conclusion in the statement.
\end{proof}

Of particular interest are the paths which begin and end at the same point. These are called loops, and in the case of loops, Proposition 13.3 particularizes as follows:

\index{loops on graph}
\index{diagonalization}

\begin{proposition}
For a graph $X$, with adjacency matrix $d\in M_N(0,1)$, we have:
$$(d^k)_{ii}=\#\Big\{k-{\rm loops\ based\ at\ }i\in I\Big\}$$
Also, the total number of $k$-loops on $X$, at various vertices, is the number
$$Tr(d^k)=\sum_i(d^k)_{ii}$$
which can be computed by diagonalizing $d$.
\end{proposition}

\begin{proof}
The first assertion follows from Proposition 13.3, with $i=j$, and the second assertion follows from this, by summing over all the vertices $i\in X$. Finally, the third assertion comes from linear algebra, the idea being that we have:
$$A=PDP^{-1}
\implies A^k=PD^kP^{-1}
\implies Tr(A^k)=Tr(D^k)$$

Indeed, in our graph case, if we denote by $\lambda_1,\ldots,\lambda_N\in\mathbb R$ the eigenvalues of $d$, then the formula of the trace of $d^k$, that we are interested in, is as follows:
$$Tr(d^k)=\lambda_1^k+\ldots+\lambda_N^k$$

Here we have used the fact that $d$, which is a real symmetric matrix, is diagonalizable, and with real eigenvalues, which is something that we know from chapter 10.
\end{proof}

Getting now to Question 13.1 as stated, we still have to talk about the associated probability measure $\mu$. So, here is the result that we will need, in what follows:

\index{adjacency matrix}
\index{random walk}

\begin{theorem}
Given a graph $X$, with adjacency matrix $d\in M_N(0,1)$, and with a distinguished vertex $*$, the number $L_k$ of loops of length $k$ based at $*$ is given by:
$$L_k=(d^k)_{**}$$
When writing $d=UDU^t$ with $U\in O_N$ and $D=diag(\lambda_1,\ldots,\lambda_N)$ with $\lambda_i\in\mathbb R$, we have
$$L_k=\sum_iU_{*i}^2\lambda_i^k$$
and the real probability measure $\mu$ having these numbers as moments is given by
$$\mu=\sum_iU_{*i}^2\delta_{\lambda_i}$$
with the delta symbols standing as usual for Dirac masses.
\end{theorem}

\begin{proof}
There are several things going on here, the idea being as follows:

\medskip

(1) According to Proposition 13.4, we have the first formula in the statement:
$$(d^k)_{**}=\#\Big\{\!*-\,i_1-\ldots-i_{k-1}-*\Big\}=L_k$$

(2) Now since $d\in M_N(0,1)$ is symmetric, this matrix is diagonalizable, with the diagonalization being as follows, with $U\in O_N$, and $D=diag(\lambda_1,\ldots,\lambda_N)$ with $\lambda_i\in\mathbb R$:
$$d=UDU^t$$

By using this formula, we obtain the second formula in the statement:
\begin{eqnarray*}
L_k
&=&(d^k)_{**}\\
&=&(UD^kU^t)_{**}\\
&=&\sum_iU_{*i}\lambda_i^k(U^t)_{i*}\\
&=&\sum_iU_{*i}^2\lambda_i^k
\end{eqnarray*}

(3) Finally, the last assertion is clear from this, because the moments of the measure in the statement, $\mu=\sum_iU_{*i}^2\delta_{\lambda_i}$, are the following numbers:
$$M_k
=\int_\mathbb Rx^kd\mu(x)
=\sum_iU_{*i}^2\lambda_i^k
=L_k$$

Observe also that $\mu$ is indeed of mass 1, because all rows of $U\in O_N$ must be of norm 1, and so $\sum_iU_{*i}^2=1$. Thus, we are led to the conclusions in the statement.
\end{proof}

At the level of examples, the finite graphs can be quite complicated, and we will discuss them later. So, let us look instead at the infinite graphs, and try for instance to count the length $k$ paths on $\mathbb Z$, based at $0$. At $k=1$ we have $2$ such paths, ending at $-1$ and $1$, and the count results can be pictured as follows, in a self-explanatory way:
$$
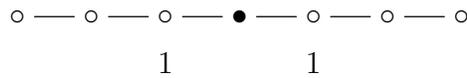
\xymatrix@R=5pt@C=15pt{
\circ\ar@{-}[r]&\circ\ar@{-}[r]&\circ\ar@{-}[r]&\bullet\ar@{-}[r]&\circ\ar@{-}[r]&\circ\ar@{-}[r]&\circ\\
&&1&&1
}$$

At $k=2$ now, we have 4 paths, one of which ends at $-2$, two of which end at 0, and one of which ends at 2. The results can be pictured as follows:
$$
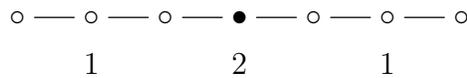
\xymatrix@R=5pt@C=15pt{
\circ\ar@{-}[r]&\circ\ar@{-}[r]&\circ\ar@{-}[r]&\bullet\ar@{-}[r]&\circ\ar@{-}[r]&\circ\ar@{-}[r]&\circ\\
&1&&2&&1
}$$

At $k=3$ now, we have 8 paths, the distribution of the endpoints being as follows:
$$
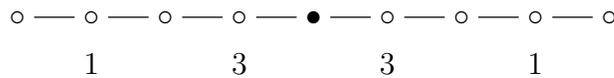
\xymatrix@R=5pt@C=15pt{
\circ\ar@{-}[r]&\circ\ar@{-}[r]&\circ\ar@{-}[r]&\circ\ar@{-}[r]&\bullet\ar@{-}[r]&\circ\ar@{-}[r]&\circ\ar@{-}[r]&\circ\ar@{-}[r]&\circ\\
&1&&3&&3&&1
}$$

As for $k=4$, here we have 16 paths, the distribution of the endpoints being as follows:
$$
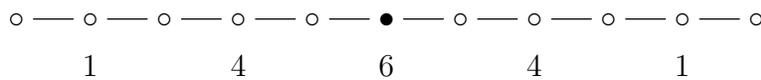
\xymatrix@R=5pt@C=15pt{
\circ\ar@{-}[r]&\circ\ar@{-}[r]&\circ\ar@{-}[r]&\circ\ar@{-}[r]&\circ\ar@{-}[r]&\bullet\ar@{-}[r]&\circ\ar@{-}[r]&\circ\ar@{-}[r]&\circ\ar@{-}[r]&\circ\ar@{-}[r]&\circ\\
&1&&4&&6&&4&&1
}$$

And good news, we can see in the above the Pascal triangle. Thus, getting back now to Question 13.1, we can answer it for the graph $\mathbb Z$, the result being as follows:

\index{central binomial coefficients}

\begin{theorem}
The paths on $\mathbb Z$ are counted by the binomial coefficients. In particular, the $2k$-paths based at $0$ are counted by the central binomial coefficients,
$$L_{2k}=\binom{2k}{k}$$
and $\mu$ is the centered measure having these numbers as even moments.
\end{theorem}

\begin{proof}
This basically follows from the above discussion, as follows:

\medskip

(1) In what regards the count, we certainly have the Pascal triangle, as discovered above, and the rest is just a matter of finishing. There are many possible ways here, a straightforward one being that of arguing that the number $C_k^l$ of length $k$ loops $0\to l$  is subject, due to the binary choice at the end, to the following recurrence relation:
$$C_k^l=C_{k-1}^{l-1}+C_{k-1}^{l+1}$$

But this is exactly the recurrence for the Pascal triangle, so done with the count. 

\medskip

(2) As for the second assertion, the first part, regarding $L_{2k}$, is clear from this, and the second part is more of an empty statement, with $\mu$ still remaining to be computed.
\end{proof}

\section*{13b. Catalan numbers}

As a second illustration, let us try to count the loops of $\mathbb N$, based at 0. This is something less obvious, and at the experimental level, the result is as follows:

\index{Catalan numbers}

\begin{proposition}
The Catalan numbers $C_k$, counting the loops on $\mathbb N$ based at $0$,
$$C_k=\#\Big\{0-i_1-\ldots-i_{2k-1}-0\Big\}$$
are numerically $1,2,5,14,42,132,429,1430,4862,16796,58786,\ldots$
\end{proposition}

\begin{proof}
To start with, we have indeed $C_1=1$, the only loop here being $0-1-0$. Then we have $C_2=2$, due to two possible loops, namely:
$$0-1-0-1-0$$
$$0-1-2-1-0$$

Then we have $C_3=5$, the possible loops here being as follows:
$$0-1-0-1-0-1-0$$
$$0-1-0-1-2-1-0$$
$$0-1-2-1-0-1-0$$
$$0-1-2-1-2-1-0$$
$$0-1-2-3-2-1-0$$

In general, the same method works, with $C_4=14$ being left to you, as an exercise, and with $C_5$ and higher to me, and I will be back with the solution, in due time.
\end{proof}

Obviously, computing the numbers $C_k$ is no easy task, and finding the formula of $C_k$, out of the data that we have, does not look as an easy task either. So, we will do what combinatorists do, let me teach you. The first step is to relax, then to look around, not with the aim of computing your numbers $C_k$, but rather with the aim of finding other objects counted by the same numbers $C_k$. With a bit of luck, among these objects some will be easier to count than the others, and this will eventually compute $C_k$.

\bigskip

This was for the strategy. In practice now, we first have the following result:

\index{Dyck paths}

\begin{theorem}
The Catalan numbers $C_k$ count:
\begin{enumerate}
\item The length $2k$ loops on $\mathbb N$, based at $0$.

\item The noncrossing pairings of $1,\ldots,2k$.

\item The noncrossing partitions of $1,\ldots,k$.

\item The length $2k$ Dyck paths in the plane.
\end{enumerate}
\end{theorem}

\begin{proof}
All this is standard combinatorics, the idea being as follows:

\medskip

(1) To start with, in what regards the various objects involved, the length $2k$ loops on $\mathbb N$ are the length $2k$ loops on $\mathbb N$ that we know, and the same goes for the noncrossing pairings of $1,\ldots,2k$, and for the noncrossing partitions of $1,\ldots,k$, the idea here being that you must be able to draw the pairing or partition in a noncrossing way. 

\medskip

(2) Regarding now the length $2k$ Dyck paths in the plane, these are by definition the paths from $(0,0)$ to $(k,k)$, marching North-East over the integer lattice $\mathbb Z^2\subset\mathbb R^2$, by staying inside the square $[0,k]\times[0,k]$, and staying as well under the diagonal of this square. As an example, here are the 5 possible Dyck paths at $n=3$:
$$
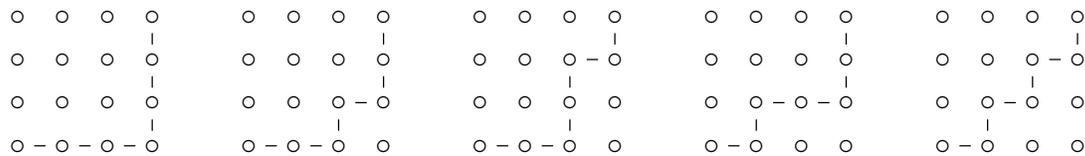
\xymatrix@R=4pt@C=4pt
{\circ&\circ&\circ&\circ\\
\circ&\circ&\circ&\circ\ar@{-}[u]\\
\circ&\circ&\circ&\circ\ar@{-}[u]\\
\circ\ar@{-}[r]&\circ\ar@{-}[r]&\circ\ar@{-}[r]&\circ\ar@{-}[u]}
\qquad
\xymatrix@R=4pt@C=4pt
{\circ&\circ&\circ&\circ\\
\circ&\circ&\circ&\circ\ar@{-}[u]\\
\circ&\circ&\circ\ar@{-}[r]&\circ\ar@{-}[u]\\
\circ\ar@{-}[r]&\circ\ar@{-}[r]&\circ\ar@{-}[u]&\circ}
\qquad
\xymatrix@R=4pt@C=4pt
{\circ&\circ&\circ&\circ\\
\circ&\circ&\circ\ar@{-}[r]&\circ\ar@{-}[u]\\
\circ&\circ&\circ\ar@{-}[u]&\circ\\
\circ\ar@{-}[r]&\circ\ar@{-}[r]&\circ\ar@{-}[u]&\circ}
\qquad
\xymatrix@R=4pt@C=4pt
{\circ&\circ&\circ&\circ\\
\circ&\circ&\circ&\circ\ar@{-}[u]\\
\circ&\circ\ar@{-}[r]&\circ\ar@{-}[r]&\circ\ar@{-}[u]\\
\circ\ar@{-}[r]&\circ\ar@{-}[u]&\circ&\circ}
\qquad
\xymatrix@R=4pt@C=4pt
{\circ&\circ&\circ&\circ\\
\circ&\circ&\circ\ar@{-}[r]&\circ\ar@{-}[u]\\
\circ&\circ\ar@{-}[r]&\circ\ar@{-}[u]&\circ\\
\circ\ar@{-}[r]&\circ\ar@{-}[u]&\circ&\circ}
$$

(3) Thus, we have definitions for all the objects involved, and in each case, if you start counting them, as we did in Proposition 13.7 with the loops on $\mathbb N$, you always end up with the same sequence of numbers, namely those found in Proposition 13.7:
$$1,2,5,14,42,132,429,1430,4862,16796,58786,\ldots$$

(4) In order to prove now that (1-4) produce indeed the same numbers, many things can be said. The idea is that, leaving aside mathematical brevity, and more specifically abstract reasonings of type $a=b,b=c\implies a=c$, what we have to do, in order to fully understand what is going on, is to etablish $\binom{4}{2}=6$ equalities, via bijective proofs.

\medskip

(5) But this can be done, indeed. As an example here, the noncrossing pairings of $1,\ldots,2k$ from (2) are in bijection with the noncrossing partitions of $1,\ldots,k$ from (3), via  fattening the pairings and shrinking the partitions. We will leave the details here as an instructive exercise, and exercise as well, to add (1) and (4) to the picture.

\medskip

(6) However, matter of having our theorem formally proved, I mean by me professor and not by you student, here is a less elegant argument, which is however very quick, and does the job. The point is that, in each of the cases (1-4) under consideration, the numbers $C_k$ that we get are easily seen to be subject to the following recurrence:
$$C_{k+1}=\sum_{a+b=k}C_aC_b$$ 

The initial data being the same, namely $C_1=1$ and $C_2=2$, in each of the cases (1-4) under consideration, we get indeed the same numbers.
\end{proof}

Now we can pass to the second step, namely selecting in the above list the objects that we find the most convenient to count, and count them. This leads to:

\begin{theorem}
The Catalan numbers are given by the formula
$$C_k=\frac{1}{k+1}\binom{2k}{k}$$
with this being best seen by counting the length $2k$ Dyck paths in the plane.
\end{theorem}

\begin{proof}
This is something quite tricky, the idea being as follows:

\medskip

(1) Let us count indeed the Dyck paths in the plane. For this purpose, we use a trick. Indeed, if we ignore the assumption that our path must stay under the diagonal of the square, we have $\binom{2k}{k}$ such paths. And among these, we have the ``good'' ones, those that we want to count, and then the ``bad'' ones, those that we want to ignore.

\medskip

(2) So, let us count the bad paths, those crossing the diagonal of the square, and reaching the higher diagonal next to it, the one joining $(0,1)$ and $(k,k+1)$. In order to count these, the trick is to ``flip'' their bad part over that higher diagonal, as follows:
$$
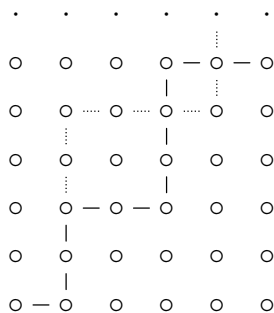
\xymatrix@R=6pt@C=6pt
{\cdot&\cdot&\cdot&\cdot&\cdot&\cdot\\
\circ&\circ&\circ&\circ\ar@{-}[r]&\circ\ar@{-}[r]\ar@{.}[u]&\circ\\
\circ&\circ\ar@{.}[r]&\circ\ar@{.}[r]&\circ\ar@{.}[r]\ar@{-}[u]&\circ\ar@{.}[u]&\circ\\
\circ&\circ\ar@{.}[u]&\circ&\circ\ar@{-}[u]&\circ&\circ\\
\circ&\circ\ar@{-}[r]\ar@{.}[u]&\circ\ar@{-}[r]&\circ\ar@{-}[u]&\circ&\circ\\
\circ&\circ\ar@{-}[u]&\circ&\circ&\circ&\circ\\
\circ\ar@{-}[r]&\circ\ar@{-}[u]&\circ&\circ&\circ&\circ}$$

(3) Now observe that, as it is obvious on the above picture, due to the flipping, the flipped bad path will no longer end in $(k,k)$, but rather in $(k-1,k+1)$. Moreover, more is true, in the sense that, by thinking a bit, we see that the flipped bad paths are precisely those ending in $(k-1,k+1)$. Thus, we can count these flipped bad paths, and so the bad paths, and so the good paths too, and so good news, we are done.

\medskip

(4) To finish now, by putting everything together, we have:
\begin{eqnarray*}
C_k
&=&\binom{2k}{k}-\binom{2k}{k-1}\\
&=&\binom{2k}{k}-\frac{k}{k+1}\binom{2k}{k}\\
&=&\frac{1}{k+1}\binom{2k}{k}
\end{eqnarray*}

Thus, we are led to the formula in the statement.
\end{proof}

We have as well another approach to all this, computation of the Catalan numbers, this time based on rock-solid standard calculus, as follows:

\index{Catalan numbers}

\begin{theorem}
The Catalan numbers have the following properties:
\begin{enumerate}
\item They satisfy $C_{k+1}=\sum_{a+b=k}C_aC_b$.

\item The series $f(z)=\sum_{k\geq0}C_kz^k$ satisfies $zf^2-f+1=0$.

\item This series is given by $f(z)=\frac{1-\sqrt{1-4z}}{2z}$.

\item We have the formula $C_k=\frac{1}{k+1}\binom{2k}{k}$.
\end{enumerate}
\end{theorem}

\begin{proof}
This is best viewed by using noncrossing pairings, as follows: 

\medskip

(1) Let us count the noncrossing pairings of $\{1,\ldots,2k+2\}$. Such a pairing appears by pairing 1 to an odd number, $2a+1$, and then inserting a noncrossing pairing of $\{2,\ldots,2a\}$, and a noncrossing pairing of $\{2a+2,\ldots,2k+2\}$. Thus we have, as claimed:
$$C_{k+1}=\sum_{a+b=k}C_aC_b$$ 

(2) Consider now the generating series of the Catalan numbers, $f(z)=\sum_{k\geq0}C_kz^k$. In terms of this generating series, the above recurrence gives, as desired:
\begin{eqnarray*}
zf^2
&=&\sum_{a,b\geq0}C_aC_bz^{a+b+1}\\
&=&\sum_{k\geq1}\sum_{a+b=k-1}C_aC_bz^k\\
&=&\sum_{k\geq1}C_kz^k\\
&=&f-1
\end{eqnarray*}

(3) By solving the equation $zf^2-f+1=0$ found above, and choosing the solution which is bounded at $z=0$, we obtain the following formula, as claimed:
$$f(z)=\frac{1-\sqrt{1-4z}}{2z}$$ 

(4) In order to compute this function, we use the generalized binomial formula, which is as follows, with $p\in\mathbb R$ being an arbitrary exponent, and with $|t|<1$:
$$(1+t)^p=\sum_{k=0}^\infty\binom{p}{k}t^k$$

To be more precise, this formula, which generalizes the usual binomial formula, holds indeed by some calculus, with the binomial coefficients being given by:
$$\binom{p}{k}=\frac{p(p-1)\ldots(p-k+1)}{k!}$$

(5) For the exponent $p=1/2$, the generalized binomial coefficients are:
\begin{eqnarray*}
\binom{1/2}{k}
&=&\frac{1/2(-1/2)(-3/2)\ldots(3/2-k)}{k!}\\
&=&(-1)^{k-1}\frac{1\cdot 3\cdot 5\ldots(2k-3)}{2^kk!}\\
&=&(-1)^{k-1}\frac{(2k-2)!}{2^{k-1}(k-1)!2^kk!}\\
&=&\frac{(-1)^{k-1}}{2^{2k-1}}\cdot\frac{1}{k}\binom{2k-2}{k-1}\\
&=&-2\left(\frac{-1}{4}\right)^k\cdot\frac{1}{k}\binom{2k-2}{k-1}
\end{eqnarray*}

(6) Thus the generalized binomial formula at exponent $p=1/2$ reads:
$$\sqrt{1+t}=1-2\sum_{k=1}^\infty\frac{1}{k}\binom{2k-2}{k-1}\left(\frac{-t}{4}\right)^k$$

With $t=-4z$ we obtain from this the following formula:
$$\sqrt{1-4z}=1-2\sum_{k=1}^\infty\frac{1}{k}\binom{2k-2}{k-1}z^k$$

(7) Now back to our series $f$, we obtain the following formula for it:
\begin{eqnarray*}
f(z)
&=&\frac{1-\sqrt{1-4z}}{2z}\\
&=&\sum_{k=1}^\infty\frac{1}{k}\binom{2k-2}{k-1}z^{k-1}\\
&=&\sum_{k=0}^\infty\frac{1}{k+1}\binom{2k}{k}z^k
\end{eqnarray*}

(8) Thus the Catalan numbers are given by the formula the statement, namely:
$$C_k=\frac{1}{k+1}\binom{2k}{k}$$

So done, and note in passing that I kept my promise, from the proof of Proposition 13.7. Indeed, with the above final formula, the numerics are easily worked out.
\end{proof}

Many other things can be said about the Catalan numbers, as a continuation of the above, and about the central binomial coefficients too. We will be back to this.

\section*{13c. Stieltjes inversion}

According to the above, the problem is now, how to recover a probability measure out of its moments. And the answer here, which is something non-trivial, is as follows:

\index{Stieltjes inversion}
\index{Cauchy transform}

\begin{theorem}
The density of a real probability measure $\mu$ can be recaptured from the sequence of moments $\{M_k\}_{k\geq0}$ via the Stieltjes inversion formula
$$d\mu (x)=\lim_{t\searrow 0}-\frac{1}{\pi}\,Im\left(G(x+it)\right)\cdot dx$$
where the function on the right, given in terms of moments by
$$G(\xi)=\xi^{-1}+M_1\xi^{-2}+M_2\xi^{-3}+\ldots$$
is the Cauchy transform of the measure $\mu$.
\end{theorem}

\begin{proof}
The Cauchy transform of our measure $\mu$ is given by:
\begin{eqnarray*}
G(\xi)
&=&\xi^{-1}\sum_{k=0}^\infty M_k\xi^{-k}\\\
&=&\int_\mathbb R\frac{\xi^{-1}}{1-\xi^{-1}y}\,d\mu(y)\\
&=&\int_\mathbb R\frac{1}{\xi-y}\,d\mu(y)
\end{eqnarray*}

Now with $\xi=x+it$, we obtain the following formula:
\begin{eqnarray*}
Im(G(x+it))
&=&\int_\mathbb RIm\left(\frac{1}{x-y+it}\right)d\mu(y)\\
&=&\int_\mathbb R\frac{1}{2i}\left(\frac{1}{x-y+it}-\frac{1}{x-y-it}\right)d\mu(y)\\
&=&-\int_\mathbb R\frac{t}{(x-y)^2+t^2}\,d\mu(y)
\end{eqnarray*}

By integrating over $[a,b]$ we obtain, with the change of variables $x=y+tz$:
\begin{eqnarray*}
\int_a^bIm(G(x+it))dx
&=&-\int_\mathbb R\int_a^b\frac{t}{(x-y)^2+t^2}\,dx\,d\mu(y)\\
&=&-\int_\mathbb R\int_{(a-y)/t}^{(b-y)/t}\frac{t}{(tz)^2+t^2}\,t\,dz\,d\mu(y)\\
&=&-\int_\mathbb R\int_{(a-y)/t}^{(b-y)/t}\frac{1}{1+z^2}\,dz\,d\mu(y)\\
&=&-\int_\mathbb R\left(\arctan\frac{b-y}{t}-\arctan\frac{a-y}{t}\right)d\mu(y)
\end{eqnarray*}

Now observe that with $t\searrow0$ we have:
$$\lim_{t\searrow0}\left(\arctan\frac{b-y}{t}-\arctan\frac{a-y}{t}\right)
=\begin{cases}
\frac{\pi}{2}-\frac{\pi}{2}=0& (y<a)\\
\frac{\pi}{2}-0=\frac{\pi}{2}& (y=a)\\
\frac{\pi}{2}-(-\frac{\pi}{2})=\pi& (a<y<b)\\
0-(-\frac{\pi}{2})=\frac{\pi}{2}& (y=b)\\
-\frac{\pi}{2}-(-\frac{\pi}{2})=0& (y>b)
\end{cases}$$

We therefore obtain the following formula:
$$\lim_{t\searrow0}\int_a^bIm(G(x+it))dx=-\pi\left(\mu(a,b)+\frac{\mu(a)+\mu(b)}{2}\right)$$

Thus, we are led to the conclusion in the statement.
\end{proof}

As a comment on this, Theorem 13.11 assumes of course that the measure $\mu$ to be found has indeed a density, the point being that, in the opposite case, the density found there will not integrate up to 1. However, we can deal with atoms too, as follows:
$$\mu=\nu+\sum_ic_i\delta_{a_i}\implies G_\mu(\xi)=G_{\nu}(\xi)+\sum_i\frac{c_i}{\xi-a_i}$$

Before getting further, let us mention that the above result does not fully solve the moment problem, because we still have the question of understanding when a sequence of numbers $M_1,M_2,M_3,\ldots$ can be the moments of a measure $\mu$.  We have here:

\index{Hankel determinant}

\begin{theorem}
A sequence of numbers $M_0,M_1,M_2,M_3,\ldots\in\mathbb R$, with $M_0=1$, is the series of moments of a real probability measure $\mu$ precisely when:
$$\begin{vmatrix}M_0\end{vmatrix}\geq0\quad,\quad 
\begin{vmatrix}
M_0&M_1\\
M_1&M_2
\end{vmatrix}\geq0\quad,\quad 
\begin{vmatrix}
M_0&M_1&M_2\\
M_1&M_2&M_3\\
M_2&M_3&M_4\\
\end{vmatrix}\geq0\quad,\quad 
\ldots$$
That is, the associated Hankel determinants must be all positive.
\end{theorem}

\begin{proof}
To start with, the positivity conditions in the statement reformulate as:
$$\sum_{i,j=1}^nc_i\bar{c}_jM_{i+j}\geq0$$

But this is something very classical, in one sense the result being elementary, coming from the following computation, which shows that we have positivity indeed:
$$\int_\mathbb R\left|\sum_{i=1}^nc_ix^i\right|^2d\mu(x)
=\int_\mathbb R\sum_{i,j=1}^nc_i\bar{c}_jx^{i+j}d\mu(x)
=\sum_{i,j=1}^nc_i\bar{c}_jM_{i+j}$$

As for the other sense, here the result comes once again from the above formula, this time via some functional analysis. With exercise for you, to learn more about all this.
\end{proof}

As a basic application of the Stieltjes formula, let us solve the moment problem for the Catalan numbers $C_k$, and for the central binomial coefficients $D_k$. We first have:

\index{semicircle law}
\index{Wigner law}

\begin{theorem}
The real measure having as even moments the Catalan numbers, $C_k=\frac{1}{k+1}\binom{2k}{k}$, and having all odd moments $0$ is the measure
$$\gamma_1=\frac{1}{2\pi}\sqrt{4-x^2}\,dx$$
called Wigner semicircle law on $[-2,2]$.
\end{theorem}

\begin{proof}
In order to apply the inversion formula, our starting point will be the formula from Theorem 13.10 for the generating series of the Catalan numbers, namely:
$$\sum_{k=0}^\infty C_kz^k=\frac{1-\sqrt{1-4z}}{2z}$$

By using this formula with $z=\xi^{-2}$, we obtain the following formula:
\begin{eqnarray*}
G(\xi)
&=&\xi^{-1}\sum_{k=0}^\infty C_k\xi^{-2k}\\
&=&\xi^{-1}\cdot\frac{1-\sqrt{1-4\xi^{-2}}}{2\xi^{-2}}\\
&=&\frac{\xi}{2}\left(1-\sqrt{1-4\xi^{-2}}\right)\\
&=&\frac{\xi}{2}-\frac{1}{2}\sqrt{\xi^2-4}
\end{eqnarray*}

Now let us apply Theorem 13.11. The study here goes as follows:

\medskip

(1) According to the general philosophy of the Stieltjes formula, the first term, namely $\xi/2$, which is ``trivial'', will not contribute to the density. 

\medskip

(2) As for the second term, which is something non-trivial, this will contribute to the density, the rule here being that the square root $\sqrt{\xi^2-4}$ will be replaced by the ``dual'' square root $\sqrt{4-x^2}\,dx$, and that we have to multiply everything by $-1/\pi$. 

\medskip

(3) As a conclusion, by Stieltjes inversion we obtain the following density:
$$d\mu(x)
=-\frac{1}{\pi}\cdot-\frac{1}{2}\sqrt{4-x^2}\,dx
=\frac{1}{2\pi}\sqrt{4-x^2}\,dx$$

Thus, we have obtained the mesure in the statement, and we are done.
\end{proof}

Next, we have the following interesting version of the above result:

\index{Marchenko-Pastur law}

\begin{theorem}
The real measure having as sequence of moments the Catalan numbers, $C_k=\frac{1}{k+1}\binom{2k}{k}$, is the measure
$$\pi_1=\frac{1}{2\pi}\sqrt{4x^{-1}-1}\,dx$$
called Marchenko-Pastur law on $[0,4]$.
\end{theorem}

\begin{proof}
As before, we can use the standard formula for the generating series of the Catalan numbers, that we know from Theorem 13.10, namely:
$$\sum_{k=0}^\infty C_kz^k=\frac{1-\sqrt{1-4z}}{2z}$$

Indeed, we obtain from this the following formula, for the Cauchy transform:
\begin{eqnarray*}
G(\xi)
&=&\xi^{-1}\sum_{k=0}^\infty C_k\xi^{-k}\\
&=&\xi^{-1}\cdot\frac{1-\sqrt{1-4\xi^{-1}}}{2\xi^{-1}}\\
&=&\frac{1}{2}\left(1-\sqrt{1-4\xi^{-1}}\right)\\
&=&\frac{1}{2}-\frac{1}{2}\sqrt{1-4\xi^{-1}}
\end{eqnarray*}

With this in hand, let us apply now the Stieltjes inversion formula, from Theorem 13.11. We obtain, a bit as before in Theorem 13.13, the following density:
$$d\mu(x)
=-\frac{1}{\pi}\cdot-\frac{1}{2}\sqrt{4x^{-1}-1}\,dx
=\frac{1}{2\pi}\sqrt{4x^{-1}-1}\,dx$$

Thus, we are led to the conclusion in the statement.
\end{proof}

Getting now to the central binomial coefficients, we have two potential computations to do for them, in the spirit of those done in Theorems 13.13 and 13.14, for the Catalan numbers. It is interesting to do both these computations, and the result is as follows:

\index{arcsine law}

\begin{theorem}
The real probability measures having as even moments/moments the central binomial coefficients, $D_k=\binom{2k}{k}$, are the measures
$$\alpha_1=\frac{1}{\pi\sqrt{4-x^2}}\,dx\qquad,\qquad\alpha_1'=\frac{1}{\pi\sqrt{x(4-x)}}\,dx$$
called arcsine laws on $[-2,2]$, and on $[0,4]$.
\end{theorem}

\begin{proof}
We use the following formula for the generating series of the central binomial coefficients, coming from the generalized binomial formula, with exponent $-1/2$:
$$\sum_{k=0}^\infty D_kz^k=\frac{1}{\sqrt{1-4z}}$$

For the first measure, the Cauchy transform is given by the following formula:
$$G(\xi)
=\xi^{-1}\sum_{k=0}^\infty D_k\xi^{-2k}
=\frac{1}{\xi}\cdot\frac{1}{\sqrt{1-4/\xi^2}}
=\frac{1}{\sqrt{\xi^2-4}}$$

As for the second measure, here the Cauchy transform is given by:
$$G(\xi)
=\xi^{-1}\sum_{k=0}^\infty D_k\xi^{-k}
=\frac{1}{\xi}\cdot\frac{1}{\sqrt{1-4/\xi}}
=\frac{1}{\sqrt{\xi(\xi-4)}}$$

But this gives the densities in the statement, via Theorem 13.11. 
\end{proof}

\section*{13d. Finite graphs}

With the above done, we can come back now to walks on finite graphs, that we know from the above to be related to the eigenvalues of the adjacency matrix $d\in M_N(0,1)$. But here, we are led to the following philosophical question, to start with:

\begin{question}
What are the most important finite graphs, that we should do our computations for?
\end{question}

Not an easy question, you have to agree with me, with the answer to this obviously depending on your previous experience with mathematics, or physics, or chemistry, or computer science, or other branch of science that you are interested in, and also, on the specific problems that you are the most in love with, in that part of science.

\bigskip

So, we have to be subjective here. And with me writing this book, and doing some sort of complicated quantum physics, as daytime job, I will choose the ADE graphs. It is beyond our scope here to explain where these ADE graphs exactly come from, and what they are good for, but as a piece of advertisement for them, we have:

\begin{fact}
The ADE graphs classify the following:
\begin{enumerate}
\item Basic Lie groups and algebras.

\item Subgroups of $SU_2$ and of $SO_3$.

\item Singularities of algebraic manifolds.

\item Basic invariants of knots and links.

\item Subfactors and planar algebras of small index.

\item Subgroups of the quantum permutation group $S_4^+$.

\item Basic quantum field theories, and other physics beasts.
\end{enumerate}
\end{fact}

Which sounds exciting, doesn't it. So, have a look at this, and with the comment that some heavy learning work is needed, in order to understand how all this works. And with the extra comment that, in view of (7), tough physics, no one really understands how all this works. For an introduction to all this, read Jones \cite{jo1}, \cite{jo2}, \cite{jo3}, \cite{jo4}.

\bigskip

Getting to work now, we first need to know what the ADE graphs are. The A graphs, which are the simplest, are as follows, with the distinguished vertex being denoted $\bullet$, and with $A_n$ having $n\geq2$ vertices, and $\tilde{A}_{2n}$ having $2n\geq2$ vertices:
$$A_n=\bullet-\circ-\circ\cdots\circ-\circ-\circ\hskip18mm 
A_{\infty}=\bullet-\circ-\circ-\circ\cdots\hskip7mm$$
\vskip-3mm
$$\ \ \ \ \ \ \ \tilde{A}_{2n}=
\begin{matrix}
\circ&\!\!\!\!-\circ-\circ\cdots\circ-\circ-&\!\!\!\!\circ\\
|&&\!\!\!\!|\\
\bullet&\!\!\!\!-\circ-\circ-\circ-\circ-&\!\!\!\!\circ\\
\\
\\
\end{matrix}\hskip20mm 
\tilde{A}_\infty=
\begin{matrix}
\circ&\!\!\!\!-\circ-\circ-\circ\cdots\\
|&\\
\bullet&\!\!\!\!-\circ-\circ-\circ\cdots\\
\\
\\
\end{matrix}
\hskip15mm$$
\vskip-7mm

You might probably say, why not stopping here, and doing some computations for the segment and the circle, with whatever new ideas that we might have. Good point, but in answer, these ideas will apply as well, with minimal changes, to the D graphs, which are as follows, with $D_n$ having $n\geq3$ vertices, and $\tilde{D}_n$ having $n+1\geq5$ vertices:
$$D_n=\bullet-\circ-\circ\dots\circ-
\begin{matrix}\ \circ\\
\ |\\
\ \circ \\
\ \\
\  \end{matrix}-\circ\hskip71mm$$
\vskip-7mm
$$\hskip7mm\tilde{D}_n=\bullet-
\begin{matrix}\circ\\
|\\
\circ\\
\ \\
\ \end{matrix}-\circ\dots\circ-
\begin{matrix}\ \circ\\
\ |\\
\ \circ \\
\ \\
\  \end{matrix}-\circ\hskip18mm$$
\vskip-7mm
$$\hskip50mm D_\infty=\bullet-
\begin{matrix}\circ\\
|\\
\circ\\
\ \\
\ \end{matrix}-\circ-\circ\cdots$$
\vskip-7mm

As a comment here, the labeling conventions for the AD graphs, while very standard, can be a bit confusing. The first graph in each series is by definition as follows:
$$A_2=\bullet-\circ\hskip13mm 
\tilde{A}_2=\begin{matrix}
\circ\\
||\\
\bullet\\
&\\
&\\
\end{matrix}\hskip13mm 
D_3=\begin{matrix}\ \circ\\
\ |\\
\ \bullet \\
\ \\
\  \end{matrix}-\circ \hskip13mm
\tilde{D}_4=\bullet-\!\!\!\!\!\begin{matrix}
\circ\hskip5mm \circ\\
\backslash\ \,\slash\\
\circ\\
&\\
&\\
\end{matrix}\!\!\!\!\!\!\!\!\!\!-\circ$$
\vskip-7mm

Finally, there are also a number of exceptional ADE graphs. First we have:
$$E_6=\bullet-\circ-
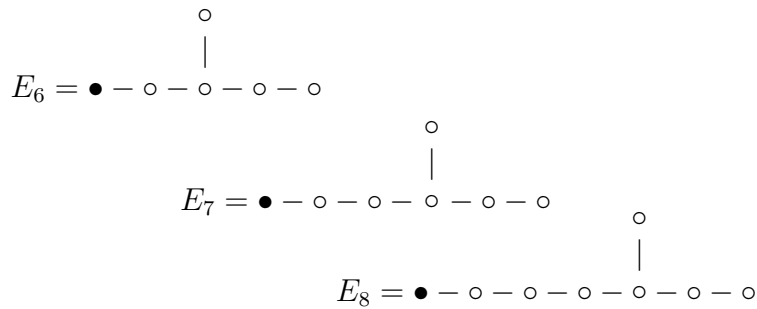
\begin{matrix}\circ\\
|\\
\circ\\
\ \\
\ \end{matrix}-
\circ-\circ\hskip71mm$$
\vskip-13mm
$$E_7=\bullet-\circ-\circ-
\begin{matrix}\circ\\
|\\
\circ\\
\ \
\\
\ \end{matrix}-
\circ-\circ\hskip18mm$$
\vskip-15mm
$$\hskip30mm E_8=\bullet-\circ-\circ-\circ-
\begin{matrix}\circ\\
|\\
\circ\\
\ \\
\ \end{matrix}-
\circ-\circ$$
\vskip-5mm

Then, we have extended versions of the above exceptional graphs, as follows:
$$\tilde{E}_6=\bullet-\circ-\begin{matrix}
\circ\\
|
\\
\circ\\
|&\\
\circ&\!\!\!\!-\ \circ\\
\ \\
\   \\
\ \\
\ \end{matrix}-\circ\hskip71mm$$
\vskip-22mm
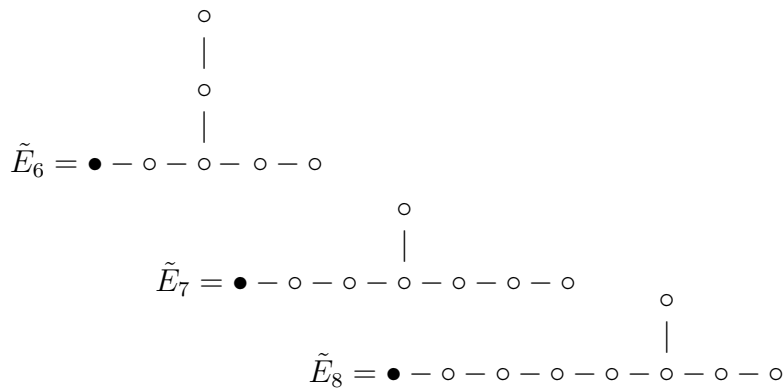
$$\tilde{E}_7=\bullet-\circ-\circ-
\begin{matrix}\circ\\
|\\
\circ\\
\ \\
\ \end{matrix}-
\circ-\circ-\circ\hskip18mm$$
\vskip-15mm
$$\hskip30mm \tilde{E}_8=\bullet-\circ-\circ-\circ-\circ-
\begin{matrix}\circ\\
|\\
\circ\\
\ \\
\ \end{matrix}-
\circ-\circ$$
\vskip-5mm

And good news, according to the general theory, that is all. Getting now to work, we have plenty of graphs here, and the problem that we would like to solve is:

\begin{problem}
How to count loops on the ADE graphs?
\end{problem}

In answer, we know that $A_\infty$ and $\tilde{A}_\infty$ are respectively the graphs that we previously called $\mathbb N$ and $\mathbb Z$. So, based on our previous work for these graphs, where the combinatorics naturally led us into generating series, let us formulate the following definition:

\begin{definition}
The Poincar\'e series of a rooted bipartite graph $X$ is
$$f(z)=\sum_{k=0}^\infty L_{2k}z^k$$
where $L_{2k}$ is the number of $2k$-loops based at the root.
\end{definition}

To be more precise, observe that all the above ADE graphs are indeed bipartite. Now the point is that, for a bipartite graph, the loops based at any point must have even length. Thus, in order to study the loops on the ADE graphs, based at the root, we just have to count the above numbers $L_{2k}$. And then, considering the generating series $f(z)$ of these numbers, and calling this Poincar\'e series, is something very standard.

\bigskip

Before getting into computations, let us introduce as well:

\begin{definition}
The positive spectral measure $\mu$ of a rooted bipartite graph $X$ is the real probability measure having the numbers $L_{2k}$ as moments:
$$\int_\mathbb Rx^kd\mu(x)=L_{2k}$$
Equivalently, we must have the following formula,
$$f(z)=\int_\mathbb R\frac{1}{1-xz}\,d\mu(x)$$
where $f$ is the Poincar\'e series of $X$.
\end{definition}

Here the existence of $\mu$, and the fact that this is indeed a positive measure, meaning a measure supported on $[0,\infty)$, comes from the following simple fact:

\begin{theorem}
The positive spectral measure of a rooted bipartite graph $X$ is given by the following formula, with $d$ being the adjacency matrix of the graph,
$$\mu=law(d^2)$$
and with the probabilistic computation being with respect to the expectation 
$$A\to<A>$$
with $<A>$ being the $(*,*)$-entry of a matrix $A$, where $*$ is the root.
\end{theorem}

\begin{proof}
With the above conventions, we have the following computation:
\begin{eqnarray*}
f(z)
&=&\sum_{k=0}^\infty L_{2k}z^k\\
&=&\sum_{k=0}^\infty\left<d^{2k}\right>z^k\\
&=&\left<\frac{1}{1-d^2z}\right>
\end{eqnarray*}

But this shows that we have $\mu=law(d^2)$, as desired.
\end{proof}

The above result shows that computing $\mu$ might be actually a simpler problem than computing $f$, and in practice, this is indeed the case. So, in what follows we will rather forget about loops and Definition 13.19, and use Definition 13.20 instead, with our computations to follow being based on the concrete interpretation from Theorem 13.21.

\bigskip

However, even with this probabilistic trick in our bag, things are not exactly trivial. Let us introduce as well the following notion:

\index{circular measure}

\begin{definition}
The circular measure $\varepsilon$ of a rooted bipartite graph $X$ is given by
$$d\varepsilon(q)=d\mu((q+q^{-1})^2)$$
where $\mu$ is the associated positive spectral measure.
\end{definition}

To be more precise, we know from Theorem 13.21 that the positive measure $\mu$ is the spectral measure of a certain positive matrix, $d^2\geq0$, and it follows from this, and from basic spectral theory, that this measure is supported by the positive reals:
$$supp(\mu)\subset\mathbb R_+$$

But then, with this observation in hand, we can define indeed the circular measure $\varepsilon$ as above, as being the pullback of $\mu$ via the following map:
$$\mathbb R\cup\mathbb T\to\mathbb R_+\quad,\quad 
q\to (q+q^{-1})^2$$

As a basic example for this, to start with, assume that $\mu$ is a discrete measure, supported by $n$ positive numbers $x_1<\ldots<x_n$, with corresponding densities $p_1,\ldots,p_n$:
$$\mu=\sum_{i=1}^n p_i\delta_{x_i}$$

For each $i\in\{1,\ldots,n\}$ the equation $(q+q^{-1})^2=x_i$ has then four solutions, that we can denote $q_i,q_i^{-1},-q_i,-q_i^{-1}$. And with this notation, we have:
$$\varepsilon=\frac{1}{4}\sum_{i=1}^np_i\left(\delta_{q_i}+\delta_{q_i^{-1}}+\delta_{-q_i}+\delta_{-q_i^{-1}}\right)$$

In general, the basic properties of $\varepsilon$ can be summarized as follows:

\begin{theorem}
The circular measure has the following properties:
\begin{enumerate}
\item $\varepsilon$ has equal density at $q,q^{-1},-q,-q^{-1}$.

\item The odd moments of $\varepsilon$ are $0$.

\item The even moments of $\varepsilon$ are half-integers.

\item When $X$ has norm $\leq 2$, $\varepsilon$ is supported by the unit circle.

\item When $X$ is finite, $\varepsilon$ is discrete.

\item If $K$ is a solution of $d=K+K^{-1}$, then $\varepsilon=law(K)$. 
\end{enumerate}
\end{theorem}

\begin{proof}
These results can be deduced from definitions, the idea being that (1-5) are trivial, and that (6) follows from the formula of $\mu$ from Theorem 13.21.
\end{proof}

Getting now to computations, we first have the following result:

\begin{theorem}
The circular measure of the basic index $4$ graph, namely 
$$\begin{matrix}
&\circ&\!\!\!\!-\circ-\circ\cdots\circ-\circ-&\!\!\!\!\circ\cr
\tilde{A}_{2n}=&|&&\!\!\!\!|\cr
&\bullet&\!\!\!\!-\circ-\circ-\circ-\circ-&\!\!\!\!\circ\cr\cr\cr\end{matrix}$$
\vskip-7mm

\noindent is the uniform measure on the $2n$-roots of unity.
\end{theorem}

\begin{proof}
Let us identify the vertices of $X=\tilde{A}_{2n}$ with the group $\{w^k\}$ formed by the $2n$-th roots of unity in the complex plane, where $w=e^{\pi i/n}$. The adjacency matrix of $X$ acts then on the functions $f\in C(X)$ in the following way:
$$df(w^s)=f(w^{s-1})+f(w^{s+1})$$

But this shows that we have $d=K+K^{-1}$, where $K$ is given by:
$$Kf(w^s)=f(w^{s+1})$$

Thus we can use Theorem 13.21 and Theorem 13.23 (6), and we get:
$$\varepsilon=law(K)$$

But this is the uniform measure on the $2n$-roots of unity, as claimed.
\end{proof}

All this is nice, but before going ahead with more computations, let us have a look into subfactor theory, and explain what is behind this. Following Jones \cite{jo4}, we have:

\begin{definition}
The theta series of a rooted bipartite graph $X$ is
$$\Theta(q)=q+\frac{1-q}{1+q}\,f\left(\frac{q}{(1+q)^2}\right)$$
where $f$ is the Poincar\'e series.
\end{definition}

The theta series can be written as $\Theta(q)=\sum a_rq^r$, and it follows from the above formula, via some simple manipulations, that its coefficients are integers:
$$a_r\in\mathbb Z$$

When $X$ is the principal graph of a subfactor of index $N>4$, it is known that the numbers $a_r$ are certain multiplicities associated to the planar algebra inclusion $TL_N\subset P$. In particular, the coefficients of the theta series are in this case positive integers:
$$a_r\in\mathbb N$$

In relation now with the circular measure, the result here, which is quite similar to the functional transform formula from Definition 13.20, is as follows:

\begin{theorem}
We have the functional transform formula
$$2\int\frac{1}{1-qu^2}\,d\varepsilon(u)=1+T(q)(1-q)$$
where the $T$ series of a rooted bipartite graph $X$ is by definition given by
$$T(q)=\frac{\Theta(q)-q}{1-q}$$
with $\Theta$ being the associated theta series.
\end{theorem}

\begin{proof}
This follows indeed by applying the change of variables $q\to (q+q^{-1})^2$ to the fact that $f$ is the moment generating function of $\mu$.
\end{proof}

In order to discuss all this more systematically, let us introduce as well:

\begin{definition}
The series of the form
$$\xi(n_1,\ldots,n_s:m_1,\ldots,m_t)=\frac{(1-q^{n_1})\ldots(1-q^{n_s})}{(1-q^{m_1})\ldots(1-q^{m_t})}$$
with $n_i,m_i\in\mathbb N$ are called cyclotomic.
\end{definition}

It is technically convenient to allow as well $1+q^n$ factors, to be designated by $n^+$ symbols in the above writing. For instance we have, by definition:
$$\xi(2^+:3)=\xi(4:2,3)$$

Also, it is convenient in what follows to use the following notations:
$$\xi'=\frac{\xi}{1-q}\quad,\quad \xi''=\frac{\xi}{1-q^2}$$

The Poincar\'e series of the ADE graphs are given by quite complicated formulae. However, the corresponding $T$ series are all cyclotomic, as follows:

\begin{theorem}
The $T$ series of the ADE graphs are as follows:
\begin{enumerate}
\item For $A_{n-1}$ we have $T=\xi(n-1:n)$.

\item For $D_{n+1}$ we have $T=\xi(n-1^+:n^+)$.

\item For $\tilde{A}_{2n}$ we have $T=\xi'(n^+:n)$.

\item For $\tilde{D}_{n+2}$ we have $T=\xi''(n+1^+:n)$.

\item For $E_6$ we have $T=\xi(8:3,6^+)$.

\item For $E_7$ we have $T=\xi(12:4,9^+)$.

\item For $E_8$ we have $T=\xi(5^+,9^+:15^+)$.

\item For $\tilde{E}_6$ we have $T=\xi(6^+:3,4)$.

\item For $\tilde{E}_7$ we have $T=\xi(9^+:4,6)$.

\item For $\tilde{E}_8$ we have $T=\xi(15^+:6,10)$.
\end{enumerate}
\end{theorem}

\begin{proof}
These formulae can be obtained by counting loops, and then by making the following change of variables, and factorizing the resulting series:
$$z^{-1/2}=q^{1/2}+q^{-1/2}$$

An alternative proof can be obtained by using planar algebra methods.
\end{proof}

Our purpose now will be that of converting the above technical results, regarding the $T$ series, into some final results, regarding the corresponding circular measures $\varepsilon$. In order to formulate our results, we will need some more theory. First, we have:

\begin{definition}
A cyclotomic measure is a probability measure $\varepsilon$ on the unit circle, having the following properties:
\begin{enumerate}
\item  $\varepsilon$ is supported by the $2n$-roots of unity, for some $n\in\mathbb N$.

\item $\varepsilon$ has equal density at $q,q^{-1},-q,-q^{-1}$.
\end{enumerate}
\end{definition}

As a first observation, it follows from Theorem 13.23 and from Theorem 13.28 that the circular measures of the finite ADE graphs are supported by certain roots of unity, hence are cyclotomic. We will be back to this in a moment, with details, and computations.

\bigskip

At the general level now, let us introduce as well the following notion:

\begin{definition}
The $T$ series of a cyclotomic measure $\varepsilon$ is given by
$$1+T(q)(1-q)=2\int\frac{1}{1-qu^2}\,d\varepsilon(u)$$
with $\varepsilon$ being as usual the circular spectral measure.
\end{definition}

Observe that this formula is nothing but the one in Theorem 13.26, written now in the other sense. In other words, if the cyclotomic measure $\varepsilon$ happens to be the circular measure of a rooted bipartite graph, then the $T$ series as defined above coincides with the $T$ series as defined before. This is useful for explicit computations.

\bigskip

Good news, with this technology in hand, and with a computation already done, in Theorem 13.24, we are now ready to discuss the circular measures of all ADE graphs. 

\bigskip

The idea will be that these measures are all cyclotomic, of level $\leq 3$, and can be expressed in terms of the basic polynomial densities of degree $\leq 6$, namely:
$$\alpha=Re(1-q^2)$$
$$\beta=Re(1-q^4)$$
$$\gamma=Re(1-q^6)$$

To be more precise, we have the following final result on the subject, with $\alpha,\beta,\gamma$ being as above, with $d_n$ being the uniform measure on the $2n$-th roots of unity, and with $d_n'=2d_{2n}-d_n$ being the uniform measure on the odd $4n$-roots of unity:

\index{ADE graph}
\index{circular measure}

\begin{theorem}
The circular measures of the ADE graphs are given by:
\begin{enumerate}
\item $A_{n-1}\to\alpha_n$.

\item $\tilde{A}_{2n}\to d_n$.

\item $D_{n+1}\to\alpha_n'$.

\item $\tilde{D}_{n+2}\to (d_n+d_1')/2$.

\item $E_6\to\alpha_{12}+(d_{12}-d_6-d_4+d_3)/2$.

\item $E_7\to\beta_9'+(d_1'-d_3')/2$.

\item $E_8\to\alpha_{15}'+\gamma_{15}'-(d_5'+d_3')/2$.

\item $\tilde{E}_{n+3}\to (d_n+d_3+d_2-d_1)/2$.
\end{enumerate}
\end{theorem}

\begin{proof}
This is something which can be proved in three steps, as follows:

\medskip

(1) For the simplest graph, namely the circle $\tilde{A}_{2n}$, we already have the result, from Theorem 13.24, with the proof there being something elementary.

\medskip

(2) For the other non-exceptional graphs, that is, of type A and D, the same method works, namely direct loop counting, with some matrix tricks.

\medskip

(3) In general, this follows from the $T$ series formulae in Theorem 13.28, via some manipulations based on the general conversion formulae given above. See \cite{bbi}.
\end{proof}

Finally, let us point out that all this leads to a more conceptual understanding of what we did before, for $\mathbb N$ and $\mathbb Z$. Indeed, even for these very basic graphs, using the unit circle and circular measures as above leads to a better understanding of the combinatorics.

\section*{13e. Exercises}

This was a rather elementary chapter, and as exercises on this, we have:

\begin{exercise}
Work out loop count results for various products of graphs.
\end{exercise}

\begin{exercise}
Learn more about Catalan numbers, the more, the better.
\end{exercise}

\begin{exercise}
Work out some other applications of the Stieltjes formula.
\end{exercise}

\begin{exercise}
Learn more about Hankel determinants, and the moment problem.
\end{exercise}

\begin{exercise}
Learn about the Wigner law, and its various remarkable properties.
\end{exercise}

\begin{exercise}
Learn about the Marchenko-Pastur law, and its various properties.
\end{exercise}

\begin{exercise}
Learn as well about the arcsine law, and its various properties.
\end{exercise}

\begin{exercise}
Learn about various ADE classifications, in math and physics.
\end{exercise}

As bonus exercise, learn about subfactors, from Jones \cite{jo1}, \cite{jo2}, \cite{jo3}, \cite{jo4}.

\chapter{Haar integration}

\section*{14a. Compact groups}

We have seen so far the foundations and basic results of measure theory and probability. Before stepping into more complicated things, such as random matrices and free probability, we would like to clarify one important question which appeared several times, namely the computation of integrals over the compact groups of unitary matrices $G\subset U_N$, and its probabilistic consequences. The precise question that we have in mind is:

\begin{question}
Given a compact group $G\subset U_N$, how to compute the integrals
$$I_{ij}^e=\int_Gg_{i_1j_1}^{e_1}\ldots g_{i_kj_k}^{e_k}\,dg$$
and then, how to use their formula in order to compute the laws of variables of type
$$f_P=P\Big(\{g_{ij}\}_{i,j=1,\ldots,N}\Big)$$
depending on a polynomial $P$? What about the $N\to\infty$ asymptotics of such laws?
\end{question}

All this is quite subtle, and as a basic illustration for this, we have a fundamental result from chapter 3, stating that for $G=S_N$ the law of the variable $\chi=\sum_ig_{ii}$ can be explicitly computed, and becomes Poisson (1) with $N\to\infty$. This is something truly remarkable, and it is this kind of result that we would like to systematically have.

\bigskip

We will discuss this in this whole chapter, and later on too. This might seem of course quite long, but believe me, it is worth the effort, because it is quite hard to do any type of advanced probability theory without knowing the answer to Question 14.1. But probably enough advertisement, let us get to work. Following Weyl, we first have:

\index{compact group}
\index{representation}
\index{character}

\begin{definition}
A unitary representation of a compact group $G$ is a continuous group morphism into a unitary group
$$u:G\to U_N\quad,\quad g\to u_g$$
which can be faithful or not. The character of such a representation is the function
$$\chi:G\to\mathbb C\quad,\quad g\to Tr(u_g)$$
where $Tr$ is the usual, unnormalized trace of the $N\times N$ matrices.
\end{definition}

At the level of examples, most of the compact groups that we met so far, finite or continuous, naturally appear as closed subgroups $G\subset U_N$. In this case, the embedding $G\subset U_N$ is of course a representation, called fundamental representation. In general now, let us first discuss the various operations on the representations. We have here:

\index{sum of representations}
\index{product of representations}
\index{conjugate representation}

\begin{proposition}
The representations of a compact group $G$ are subject to:
\begin{enumerate}
\item Making sums. Given representations $u,v$, of dimensions $N,M$, 
their sum is the $N+M$-dimensional representation $u+v=diag(u,v)$.

\item Making products. Given representations $u,v$, of dimensions $N,M$, their product is the $NM$-dimensional representation $(u\otimes v)_{ia,jb}=u_{ij}v_{ab}$.

\item Taking conjugates. Given a $N$-dimensional representation $u$, its conjugate is the $N$-dimensional representation $(\bar{u})_{ij}=\bar{u}_{ij}$.

\item Spinning by unitaries. Given a $N$-dimensional representation $u$, and a unitary $W\in U_N$, we can spin $u$ by this unitary, $u\to WuW^*$.
\end{enumerate}
\end{proposition}

\begin{proof}
The fact that the operations in the statement are indeed well-defined, among morphisms from $G$ to unitary groups, is indeed clear from definitions.
\end{proof}

In relation now with characters, we have the following result:

\begin{proposition}
We have the following formulae, regarding characters
$$\chi_{u+v}=\chi_u+\chi_v\quad,\quad 
\chi_{u\otimes v}=\chi_u\chi_v\quad,\quad 
\chi_{\bar{u}}=\bar{\chi}_u\quad,\quad
\chi_{WuW^*}=\chi_u$$
in relation with the basic operations for the representations.
\end{proposition}

\begin{proof}
All these assertions are elementary, by using the following well-known trace formulae, valid for any square matrices $U,V$, and any unitary $W$:
$$Tr(diag(U,V))=Tr(U)+Tr(V)\quad,\quad 
Tr(U\otimes V)=Tr(U)Tr(V)$$
$$Tr(\bar{U})=\overline{Tr(U)}\quad,\quad 
Tr(WUW^*)=Tr(U)$$

Thus, we are led to the formulae in the statement.
\end{proof}

Assume now that we are given a closed subgroup $G\subset U_N$. By using the above operations, we can construct a whole family of representations of $G$, as follows:

\index{Peter-Weyl representations}

\begin{definition}
Given a closed subgroup $G\subset U_N$, its Peter-Weyl representations are the various tensor products between the fundamental representation and its conjugate:
$$u:G\subset U_N\quad,\quad 
\bar{u}:G\subset U_N$$ 
We denote these tensor products $u^{\otimes k}$, with $k=\circ\bullet\bullet\circ\ldots$ being a colored integer, with the colored tensor powers being defined according to the rules 
$$u^{\otimes\circ}=u\quad,\quad
u^{\otimes\bullet}=\bar{u}\quad,\quad
u^{\otimes kl}=u^{\otimes k}\otimes u^{\otimes l}$$
and with the convention that $u^{\otimes\emptyset}$ is the trivial representation $1:G\to U_1$.
\end{definition}

Here are a few examples of such representations, namely those coming from the colored integers of length 2, which will often appear in what follows:
$$u^{\otimes\circ\circ}=u\otimes u\quad,\quad 
u^{\otimes\circ\bullet}=u\otimes\bar{u}$$
$$u^{\otimes\bullet\circ}=\bar{u}\otimes u\quad,\quad
u^{\otimes\bullet\bullet}=\bar{u}\otimes\bar{u}$$

In relation now with characters, we have the following result:

\index{colored powers}

\begin{proposition}
The characters of the Peter-Weyl representations are given by
$$\chi_{u^{\otimes k}}=(\chi_u)^k$$
with the colored powers being given by $\chi^\circ=\chi$, $\chi^\bullet=\bar{\chi}$ and multiplicativity.
\end{proposition}

\begin{proof}
This follows indeed from the additivity, multiplicativity and conjugation formulae from Proposition 14.4, via the conventions in Definition 14.5.
\end{proof}

Getting back now to our motivations, we can see the interest in the above constructions. Indeed, the joint moments of the main character $\chi=\chi_u$ and its adjoint $\bar{\chi}=\chi_{\bar{u}}$ are the expectations of the characters of various Peter-Weyl representations: 
$$\int_G\chi^k=\int_G \chi_{u^{\otimes k}}$$

In order to advance, we must develop some general theory. Let us start with:

\index{Hom space}
\index{End space}
\index{Fix space}
\index{intertwiners}

\begin{definition}
Given a compact group $G$, and two of its representations,
$$v:G\to U_N\quad,\quad 
w:G\to U_M$$
we define the space of intertwiners between these representations as being 
$$Hom(v,w)=\left\{T\in M_{M\times N}(\mathbb C)\Big|Tv_g=w_gT,\forall g\in G\right\}$$
and we use the following conventions:
\begin{enumerate}
\item We use the notations $Fix(v)=Hom(1,v)$, and $End(v)=Hom(v,v)$.

\item We write $v\sim w$ when $Hom(v,w)$ contains an invertible element.

\item We say that $v$ is irreducible, and write $v\in Irr(G)$, when $End(v)=\mathbb C1$.
\end{enumerate}
\end{definition}

The terminology here is standard, with Fix, Hom, End standing for fixed points, homomorphisms and endomorphisms. We will see later that irreducible means indecomposable, in a suitable sense. Here are now a few basic results, regarding these spaces:

\index{tensor category}

\begin{proposition}
The spaces of intertwiners have the following properties:
\begin{enumerate}
\item $T\in Hom(v,w),S\in Hom(w,z)\implies ST\in Hom(v,z)$.

\item $S\in Hom(v,w),T\in Hom(z,t)\implies S\otimes T\in Hom(v\otimes z,w\otimes t)$.

\item $T\in Hom(v,w)\implies T^*\in Hom(w,v)$.
\end{enumerate}
In abstract terms, we say that the Hom spaces form a tensor $*$-category.
\end{proposition}

\begin{proof}
All the formulae in the statement are indeed clear from definitions, via elementary computations. As for the last assertion, this is something coming from (1,2,3). We will be back to tensor categories later on, with more details on this latter fact.
\end{proof}

As a main consequence of the above result, we have:

\begin{proposition}
Given a representation $v:G\to U_N$, the linear space
$$End(v)\subset M_N(\mathbb C)$$
is a $*$-algebra, with respect to the usual involution of the matrices.
\end{proposition}

\begin{proof}
By definition, $End(v)$ is a linear subspace of $M_N(\mathbb C)$. We know from Proposition 14.8 (1) that this subspace $End(v)$ is a subalgebra of $M_N(\mathbb C)$, and then we know as well from Proposition 14.8 (3) that this subalgebra is stable under the involution $*$. Thus, what we have here is a $*$-subalgebra of $M_N(\mathbb C)$, as claimed.
\end{proof}

In order to exploit the above fact, we will need a basic result from linear algebra, stating that any $*$-algebra $A\subset M_N(\mathbb C)$ decomposes as a direct sum, as follows:
$$A\simeq M_{N_1}(\mathbb C)\oplus\ldots\oplus M_{N_k}(\mathbb C)$$

Indeed, let us write the unit $1\in A$ as $1=p_1+\ldots+p_k$, with $p_i\in A$ being central minimal projections. Then each of the spaces $A_i=p_iAp_i$ is a subalgebra of $A$, and we have a decomposition $A=A_1\oplus\ldots\oplus A_k$. But since each central projection $p_i\in A$ was chosen minimal, we have $A_i\simeq M_{N_i}(\mathbb C)$, with $N_i=rank(p_i)$, as desired.

\bigskip

We can now formulate our first Peter-Weyl type theorem, as follows:

\index{Peter-Weyl}

\begin{theorem}[Peter-Weyl 1]
Let $v:G\to U_N$ be a representation, consider the algebra $A=End(v)$, and write its unit $1=p_1+\ldots+p_k$ as above. We have then 
$$v=v_1+\ldots+v_k$$
with each $v_i$ being an irreducible representation, obtained by restricting $v$ to $Im(p_i)$.
\end{theorem}

\begin{proof}
This basically follows from Proposition 14.9, as follows:

\medskip

(1) We first associate to our representation $v:G\to U_N$ the corresponding action map on $\mathbb C^N$. If a linear subspace $W\subset\mathbb C^N$ is invariant, the restriction of the action map to $W$ is an action map too, which must come from a subrepresentation $w\subset v$.

\medskip

(2) Consider now a projection $p\in End(v)$. From $pv=vp$ we obtain that the linear space $W=Im(p)$ is invariant under $v$, and so this space must come from a subrepresentation $w\subset v$. It is routine to check that the operation $p\to w$ maps subprojections to subrepresentations, and minimal projections to irreducible representations.

\medskip

(3) With these preliminaries in hand, let us decompose the algebra $End(v)$ as above, by using the decomposition $1=p_1+\ldots+p_k$ into central minimal projections. If we denote by $v_i\subset v$ the subrepresentation coming from the vector space $V_i=Im(p_i)$, then we obtain in this way a decomposition $v=v_1+\ldots+v_k$, as in the statement.
\end{proof}

Here is now our second Peter-Weyl theorem, complementing Theorem 14.10:

\index{Peter-Weyl}
\index{coefficients of representations}
\index{smooth representation}

\begin{theorem}[Peter-Weyl 2]
Given a closed subgroup $G\subset_uU_N$, any of its irreducible smooth representations 
$$v:G\to U_M$$
appears inside a tensor product of the fundamental representation $u$ and its adjoint $\bar{u}$.
\end{theorem}

\begin{proof}
Given a representation $v:G\to U_M$, we define the space of coefficients $C_v\subset C(G)$ of this representation as being the following linear space:
$$C_v=span\Big[g\to v(g)_{ij}\Big]$$

With this notion in hand, the result can be deduced as follows:

\medskip

(1) The construction $v\to C_v$ is functorial, in the sense that it maps subrepresentations into linear subspaces. This is indeed something which is routine to check.

\medskip

(2) A closed subgroup $G\subset_uU_N$ is a Lie group, and a representation $v:G\to U_M$ is smooth when we have an inclusion $C_v\subset<C_u>$. This is indeed well-known.

\medskip

(3) By definition of the Peter-Weyl representations, as arbitrary tensor products between the fundamental representation $u$ and its conjugate $\bar{u}$, we have:
$$<C_u>=\sum_kC_{u^{\otimes k}}$$

(4) Now by putting together the above observations (2,3) we conclude that we must have an inclusion as follows, for certain exponents $k_1,\ldots,k_p$:
$$C_v\subset C_{u^{\otimes k_1}\oplus\ldots\oplus u^{\otimes k_p}}$$

(5) By using now (1), we deduce that we have an inclusion $v\subset u^{\otimes k_1}\oplus\ldots\oplus u^{\otimes k_p}$, and by applying Theorem 14.10, this leads to the conclusion in the statement.
\end{proof}

\section*{14b. Haar integration}

In order to further advance with Peter-Weyl theory, we need to talk about integration over $G$. In the finite group case the situation is trivial, as follows:

\begin{proposition}
Any finite group $G$ has a unique probability measure which is invariant under left and right translations,
$$\mu(E)=\mu(gE)=\mu(Eg)$$
and this is the normalized counting measure on $G$, given by $\mu(E)=|E|/|G|$.
\end{proposition}

\begin{proof}
This is indeed something trivial, which follows from definitions.
\end{proof}

In the general, continuous case, let us begin with the following key result:

\begin{proposition}
Given a unital positive linear form $\psi:C(G)\to\mathbb C$, the limit
$$\int_\varphi f=\lim_{n\to\infty}\frac{1}{n}\sum_{k=1}^n\psi^{*k}(f)$$
exists, and for a coefficient of a representation $f=(\tau\otimes id)v$ we have
$$\int_\varphi f=\tau(P)$$
where $P$ is the orthogonal projection onto the $1$-eigenspace of $(id\otimes\psi)v$.
\end{proposition}

\begin{proof}
By linearity it is enough to prove the first assertion for functions of the following type, where $v$ is a Peter-Weyl representation, and $\tau$ is a linear form:
$$f=(\tau\otimes id)v$$

Thus we are led into the second assertion, and more precisely we can have the whole result proved if we can establish the following formula, with $f=(\tau\otimes id)v$:
$$\lim_{n\to\infty}\frac{1}{n}\sum_{k=1}^n\psi^{*k}(f)=\tau(P)$$

In order to prove this latter formula, observe that we have:
$$\psi^{*k}(f)
=(\tau\otimes\psi^{*k})v
=\tau((id\otimes\psi^{*k})v)$$

Let us set $M=(id\otimes\psi)v$. In terms of this matrix, we have:
$$((id\otimes\psi^{*k})v)_{i_0i_{k+1}}
=\sum_{i_1\ldots i_k}M_{i_0i_1}\ldots M_{i_ki_{k+1}}
=(M^k)_{i_0i_{k+1}}$$

Thus we have the following formula, valid for any $k\in\mathbb N$:
$$(id\otimes\psi^{*k})v=M^k$$

It follows that our Ces\`aro limit is given by the following formula:
$$\lim_{n\to\infty}\frac{1}{n}\sum_{k=1}^n\psi^{*k}(f)
=\lim_{n\to\infty}\frac{1}{n}\sum_{k=1}^n\tau(M^k)
=\tau\left(\lim_{n\to\infty}\frac{1}{n}\sum_{k=1}^nM^k\right)$$

Now since $v$ is unitary we have $||v||=1$, and so $||M||\leq1$. Thus the last Ces\`aro limit converges, and equals the orthogonal projection onto the $1$-eigenspace of $M$:
$$\lim_{n\to\infty}\frac{1}{n}\sum_{k=1}^nM^k=P$$

Thus our initial Ces\`aro limit converges as well, to $\tau(P)$, as desired.
\end{proof}

When the linear form $\psi\in C(G)^*$ is faithful, we have the following finer result:

\begin{proposition}
Given a faithful unital linear form $\psi\in C(G)^*$, the limit
$$\int_\psi f=\lim_{n\to\infty}\frac{1}{n}\sum_{k=1}^n\psi^{*k}(f)$$
exists, and is independent of $\psi$, given on coefficients of representations by
$$\left(id\otimes\int_\psi\right)v=P$$
where $P$ is the orthogonal projection onto the space $Fix(v)=\left\{\xi\in\mathbb C^n\big|v\xi=\xi\right\}$.
\end{proposition}

\begin{proof}
In view of Proposition 14.13, it remains to prove that when $\psi$ is faithful, the $1$-eigenspace of the matrix $M=(id\otimes\psi)v$ equals the space $Fix(v)$.

\medskip

``$\supset$'' This is clear, and for any $\psi$, because we have the following implication:
$$v\xi=\xi\implies M\xi=\xi$$

``$\subset$'' Here we must prove that, when $\psi$ is faithful, we have:
$$M\xi=\xi\implies v\xi=\xi$$

For this purpose, assume that we have $M\xi=\xi$, and consider the following function:
$$f=\sum_i\left(\sum_jv_{ij}\xi_j-\xi_i\right)\overline{\left(\sum_kv_{ik}\xi_k-\xi_i\right)}$$

We must prove that we have $f=0$. Since $v$ is unitary, we have:
\begin{eqnarray*}
f
&=&\sum_{ijk}v_{ij}\bar{v}_{ik}\xi_j\bar{\xi}_k-\frac{1}{N}v_{ij}\xi_j\bar{\xi}_i-\frac{1}{N}\bar{v}_{ik}\xi_i\bar{\xi}_k+\frac{1}{N^2}\xi_i\bar{\xi}_i\\
&=&\sum_j|\xi_j|^2-\sum_{ij}v_{ij}\xi_j\bar{\xi}_i-\sum_{ik}\bar{v}_{ik}\xi_i\bar{\xi}_k+\sum_i|\xi_i|^2\\
&=&||\xi||^2-<v\xi,\xi>-\overline{<v\xi,\xi>}+||\xi||^2\\
&=&2(||\xi||^2-Re(<v\xi,\xi>))
\end{eqnarray*}

By using now our assumption $M\xi=\xi$, we obtain from this:
\begin{eqnarray*}
\psi(f)
&=&2\psi(||\xi||^2-Re(<v\xi,\xi>))\\
&=&2(||\xi||^2-Re(<M\xi,\xi>))\\
&=&2(||\xi||^2-||\xi||^2)\\
&=&0
\end{eqnarray*}

Now since $\psi$ is faithful, this gives $f=0$, and so $v\xi=\xi$, as claimed.
\end{proof}

We can now formulate a main result about integration, as follows:

\index{Haar measure}
\index{Ces\`aro limit}
\index{Haar integration}

\begin{theorem}
Any compact group $G$ has a unique Haar integration, which can be constructed by starting with any faithful positive unital form $\psi\in C(G)^*$, and setting:
$$\int_G=\lim_{n\to\infty}\frac{1}{n}\sum_{k=1}^n\psi^{*k}$$
Moreover, for any representation $v$ we have the formula
$$\left(id\otimes\int_G\right)v=P$$
where $P$ is the orthogonal projection onto $Fix(v)=\left\{\xi\in\mathbb C^n\big|v\xi=\xi\right\}$.
\end{theorem}

\begin{proof}
Let us first go back to the general context of Proposition 14.13. Since convolving one more time with $\psi$ will not change the Ces\`aro limit appearing there, the functional $\int_\psi\in C(G)^*$ constructed there has the following invariance property:
$$\int_\psi*\,\psi=\psi*\int_\psi=\int_\psi$$

In the case where $\psi$ is assumed to be faithful, as in Proposition 14.14, our claim is that we have the following formula, valid this time for any $\varphi\in C(G)^*$:
$$\int_\psi*\,\varphi=\varphi*\int_\psi=\varphi(1)\int_\psi$$

Indeed, it is enough to prove this formula on a coefficient of a corepresentation:
$$f=(\tau\otimes id)v$$

In order to do so, consider the following two matrices:
$$P=\left(id\otimes\int_\psi\right)v\quad,\quad 
Q=(id\otimes\varphi)v$$

We have then the following formulae, which all follow from definitions:
$$\left(\int_\psi*\,\varphi\right)f=\tau(PQ)\quad,\quad 
\left(\varphi*\int_\psi\right)f=\tau(QP)\quad,\quad 
\varphi(1)\int_\psi f=\varphi(1)\tau(P)$$

Thus, in order to prove our claim, it is enough to establish the following formula:
$$PQ=QP=\psi(1)P$$

But this follows from the fact, from Proposition 14.14, that $P=(id\otimes\int_\psi)v$ is the orthogonal projection onto $Fix(v)$. Thus, we proved our claim. Now observe that, with $\Delta f(g\otimes h)=f(gh)$, this formula that we proved can be written as follows:
$$\varphi\left(\int_\psi\otimes\,id\right)\Delta
=\varphi\left(id\otimes\int_\psi\right)\Delta
=\varphi\int_\psi(.)1$$

This formula being true for any $\varphi\in C(G)^*$, we can simply delete $\varphi$, and we conclude that $\int_G=\int_\psi$ has the required left and right invariance property, namely:
$$\left(\int_G\otimes\,id\right)\Delta
=\left(id\otimes\int_G\right)\Delta
=\int_G(.)1$$

Finally, the uniqueness is clear as well, because if we have two invariant integrals $\int_G,\int_G'$, then their convolution equals on one hand $\int_G$, and on the other hand, $\int_G'$.
\end{proof}

Summarizing, we know how to integrate over $G$. Before getting into probabilistic applications, let us develop however more Peter-Weyl theory. We will need:

\index{Frobenius isomorphism}

\begin{proposition}
We have a Frobenius type isomorphism
$$Hom(v,w)\simeq Fix(v\otimes\bar{w})$$
valid for any two representations $v,w$.
\end{proposition}

\begin{proof}
According to definitions, we have the following equivalences:
\begin{eqnarray*}
T\in Hom(v,w)
&\iff&Tv=wT\\
&\iff&\sum_iT_{ai}v_{ij}=\sum_bw_{ab}T_{bj},\forall a,j
\end{eqnarray*}

On the other hand, we have as well the following equivalences:
\begin{eqnarray*}
T\in Fix(v\otimes\bar{w})
&\iff&(v\otimes\bar{w})T=\xi\\
&\iff&\sum_{bi}v_{ji}\bar{w}_{ab}T_{bi}=T_{aj}\forall a,j
\end{eqnarray*}

With these formulae in hand, both inclusions follow from the unitarity of $v,w$.
\end{proof}

We can now formulate a third Peter-Weyl theorem, as follows:

\index{Peter-Weyl}

\begin{theorem}[Peter-Weyl 3]
The dense subalgebra $\mathcal C(G)\subset C(G)$ generated by the coefficients of the fundamental representation decomposes as a direct sum 
$$\mathcal C(G)=\bigoplus_{v\in Irr(G)}M_{\dim(v)}(\mathbb C)$$
with the summands being pairwise orthogonal with respect to the scalar product
$$<f,g>=\int_Gf\bar{g}$$
where $\int_G$ is the Haar integration over $G$.
\end{theorem}

\begin{proof}
By combining the previous two Peter-Weyl results, Theorems 14.10 and 14.11, we deduce that we have a linear space decomposition as follows:
$$\mathcal C(G)
=\sum_{v\in Irr(G)}C_v
=\sum_{v\in Irr(G)}M_{\dim(v)}(\mathbb C)$$

Thus, in order to conclude, it is enough to prove that for any two irreducible representations $v,w\in Irr(G)$, the corresponding spaces of coefficients are orthogonal:
$$v\not\sim w\implies C_v\perp C_w$$ 

But this follows from Theorem 14.15, via Proposition 14.16. Let us set indeed:
$$P_{ia,jb}=\int_Gv_{ij}\bar{w}_{ab}$$

Then $P$ is the orthogonal projection onto the following vector space:
$$Fix(v\otimes\bar{w})
\simeq Hom(v,w)
=\{0\}$$

Thus we have $P=0$, and this gives the result.
\end{proof}

Finally, we have the following result, completing the Peter-Weyl theory:

\index{Peter-Weyl}

\begin{theorem}[Peter-Weyl 4]
The characters of irreducible representations belong to the algebra
$$\mathcal C(G)_{central}=\left\{f\in\mathcal C(G)\Big|f(gh)=f(hg),\forall g,h\in G\right\}$$
called algebra of central functions on $G$, and form an orthonormal basis of it.
\end{theorem}

\begin{proof}
Observe first that $\mathcal C(G)_{central}$ is indeed an algebra, which contains all the characters. Conversely, consider a function $f\in\mathcal C(G)$, written as follows:
$$f=\sum_{v\in Irr(G)}f_v$$

The condition $f\in\mathcal C(G)_{central}$ states then that for any $v\in Irr(G)$, we must have:
$$f_v\in\mathcal C(G)_{central}$$

But this means that $f_v$ must be a scalar multiple of $\chi_v$, so the characters form a basis of $\mathcal C(G)_{central}$, as stated. Also, the fact that we have an orthogonal basis follows from Theorem 14.17. As for the fact that the characters have norm 1, this follows from:
$$\int_G\chi_v\bar{\chi}_v
=\sum_{ij}\int_Gv_{ii}\bar{v}_{jj}
=\sum_i\frac{1}{\dim(v)}
=1$$

Here we have used the fact, coming from Theorem 14.15 and Proposition 14.16, that the integrals $\int_Gv_{ij}\bar{v}_{kl}$ form the orthogonal projection onto the following vector space:
$$Fix(v\otimes\bar{v})\simeq End(v)=\mathbb C1$$

Thus, the proof of our theorem is now complete.
\end{proof}

\section*{14c. Diagrams, easiness}

In view of the above results, no matter on what we want to do with our group, we must compute the spaces $Fix(u^{\otimes k})$. It is technically convenient to slightly enlarge the class of spaces to be computed, by talking about Tannakian categories, as follows:

\index{tensor category}
\index{Tannakian category}
\index{Peter-Weyl representations}

\begin{definition}
The Tannakian category associated to a closed subgroup $G\subset_uU_N$ is the collection $C_G=(C_G(k,l))$ of vector spaces
$$C_G(k,l)=Hom(u^{\otimes k},u^{\otimes l})$$
where the representations $u^{\otimes k}$ with $k=\circ\bullet\bullet\circ\ldots$ colored integer, defined by
$$u^{\otimes\emptyset}=1\quad,\quad
u^{\otimes\circ}=u\quad,\quad
u^{\otimes\bullet}=\bar{u}$$
and multiplicativity, $u^{\otimes kl}=u^{\otimes k}\otimes u^{\otimes l}$, are the Peter-Weyl representations.
\end{definition}

Let us make a summary of what we have so far, regarding these spaces $C_G(k,l)$. In order to formulate our result, let us start with the following definition:

\index{tensor category}

\begin{definition}
Let $H$ be a finite dimensional Hilbert space. A tensor category over $H$ is a collection $C=(C(k,l))$ of linear spaces 
$$C(k,l)\subset\mathcal L(H^{\otimes k},H^{\otimes l})$$
satisfying the following conditions:
\begin{enumerate}
\item $S,T\in C$ implies $S\otimes T\in C$.

\item If $S,T\in C$ are composable, then $ST\in C$.

\item $T\in C$ implies $T^*\in C$.

\item $C(k,k)$ contains the identity operator.

\item $C(\emptyset,k)$ with $k=\circ\bullet,\bullet\circ$ contain the operator $R:1\to\sum_ie_i\otimes e_i$.

\item $C(kl,lk)$ with $k,l=\circ,\bullet$ contain the flip operator $\Sigma:a\otimes b\to b\otimes a$.
\end{enumerate}
\end{definition}

Here the tensor power Hilbert spaces $H^{\otimes k}$, with $k=\circ\bullet\bullet\circ\ldots$ being a colored integer, are defined by the following formulae, and multiplicativity:
$$H^{\otimes\emptyset}=\mathbb C\quad,\quad
H^{\otimes\circ}=H\quad,\quad
H^{\otimes\bullet}=\bar{H}\simeq H$$

With these conventions, we have the following result, summarizing our knowledge on the subject, coming from the results established in the above:

\begin{theorem}
For a closed subgroup $G\subset_uU_N$, the associated Tannakian category
$$C_G(k,l)=Hom(u^{\otimes k},u^{\otimes l})$$
is a tensor category over the Hilbert space $H=\mathbb C^N$.
\end{theorem}

\begin{proof}
We know that the fundamental representation $u$ acts on the Hilbert space $H=\mathbb C^N$, and that its conjugate $\bar{u}$ acts on the Hilbert space $\bar{H}=\mathbb C^N$. Now by multiplicativity we conclude that any Peter-Weyl representation $u^{\otimes k}$ acts on the Hilbert space $H^{\otimes k}$, and so that we have embeddings as in Definition 14.20, as follows:
$$C_G(k,l)\subset\mathcal L(H^{\otimes k},H^{\otimes l})$$

Regarding now the fact that the axioms (1-6) in Definition 14.20 are indeed satisfied, this is something that we basically already know. To be more precise, (1-4) are clear, and (5) follows from the fact that each element $g\in G$ is a unitary, which gives:
$$R\in Hom(1,g\otimes\bar{g})\quad,\quad 
R\in Hom(1,\bar{g}\otimes g)$$

As for (6), this is something trivial, coming from the fact that the matrix coefficients $g\to g_{ij}$ and their complex conjugates $g\to\bar{g}_{ij}$ commute with each other.
\end{proof}

Our purpose now will be that of showing that any closed subgroup $G\subset U_N$ is uniquely determined by its Tannakian category $C_G=(C_G(k,l))$. This result, known as Tannakian duality, is something quite deep, and extremely useful. Indeed, the idea is that what we would have here is a ``linearization'' of $G$, allowing us to do combinatorics, and to ultimately reach to concrete and powerful results, regarding $G$ itself. We first have:

\begin{theorem}
Given a tensor category $C=(C(k,l))$ over a finite dimensional Hilbert space $H\simeq\mathbb C^N$, the following construction,
$$G_C=\left\{g\in U_N\Big|Tg^{\otimes k}=g^{\otimes l}T\ ,\ \forall k,l,\forall T\in C(k,l)\right\}$$
produces a closed subgroup $G_C\subset U_N$.
\end{theorem}

\begin{proof}
This is something elementary, with the fact that the closed subset $G_C\subset U_N$ constructed in the statement is indeed stable under the multiplication, unit and inversion operation for the unitary matrices $g\in U_N$ being clear from definitions.
\end{proof}

We can now formulate the Tannakian duality result, as follows:

\index{Tannakian duality}

\begin{theorem}
The above Tannakian constructions 
$$G\to C_G\quad,\quad 
C\to G_C$$
are bijective, and inverse to each other.
\end{theorem}

\begin{proof}
This is something quite technical, obtained by doing some abstract algebra, and for details here, we refer to the Tannakian duality literature. The whole subject is actually, in modern times, for the most part of quantum algebra, and you can consult here various quantum group papers and books, such as \cite{ba2}, for details on the above.
\end{proof}

In order to reach now to more concrete things, following Brauer, we have:

\index{category of partitions}

\begin{definition}
Let $P(k,l)$ be the set of partitions between an upper colored integer $k$, and a lower colored integer $l$. A collection of subsets 
$$D=\bigsqcup_{k,l}D(k,l)$$
with $D(k,l)\subset P(k,l)$ is called a category of partitions when it has the following properties:
\begin{enumerate}
\item Stability under the horizontal concatenation, $(\pi,\sigma)\to[\pi\sigma]$.

\item Stability under vertical concatenation $(\pi,\sigma)\to[^\sigma_\pi]$, with matching middle symbols.

\item Stability under the upside-down turning $*$, with switching of colors, $\circ\leftrightarrow\bullet$.

\item Each set $P(k,k)$ contains the identity partition $||\ldots||$.

\item The sets $P(\emptyset,\circ\bullet)$ and $P(\emptyset,\bullet\circ)$ both contain the semicircle $\cap$.

\item The sets $P(k,\bar{k})$ with $|k|=2$ contain the crossing partition $\slash\hskip-2.0mm\backslash$.
\end{enumerate}
\end{definition} 

There are many examples of such categories, as for instance the category of all pairings $P_2$, or of all matching pairings $\mathcal P_2$. We will be back to examples in a moment.

\bigskip

Let us formulate as well the following definition:

\index{Kronecker symbols}
\index{maps associated to partitions}

\begin{definition}
Given a partition $\pi\in P(k,l)$ and an integer $N\in\mathbb N$, we can construct a linear map between tensor powers of $\mathbb C^N$,
$$T_\pi:(\mathbb C^N)^{\otimes k}\to(\mathbb C^N)^{\otimes l}$$
by the following formula, with $e_1,\ldots,e_N$ being the standard basis of $\mathbb C^N$,
$$T_\pi(e_{i_1}\otimes\ldots\otimes e_{i_k})=\sum_{j_1\ldots j_l}\delta_\pi\begin{pmatrix}i_1&\ldots&i_k\\ j_1&\ldots&j_l\end{pmatrix}e_{j_1}\otimes\ldots\otimes e_{j_l}$$
and with the coefficients on the right being Kronecker type symbols,
$$\delta_\pi\begin{pmatrix}i_1&\ldots&i_k\\ j_1&\ldots&j_l\end{pmatrix}\in\{0,1\}$$
whose values depend on whether the indices fit or not.
\end{definition}

To be more precise, we put the indices of $i,j$ on the legs of $\pi$, in the obvious way. In case all the blocks of $\pi$ contain equal indices of $i,j$, we set $\delta_\pi(^i_j)=1$. Otherwise, we set $\delta_\pi(^i_j)=0$. The relation with the Tannakian categories comes from:

\begin{proposition}
The assignement $\pi\to T_\pi$ is categorical, in the sense that
$$T_\pi\otimes T_\nu=T_{[\pi\nu]}\quad,\quad 
T_\pi T_\nu=N^{c(\pi,\nu)}T_{[^\nu_\pi]}\quad,\quad 
T_\pi^*=T_{\pi^*}$$
where $c(\pi,\nu)$ are certain integers, coming from the erased components in the middle.
\end{proposition}

\begin{proof}
This is something elementary, the computations being as follows:

\medskip

(1) The concatenation axiom can be checked as follows:
\begin{eqnarray*}
&&(T_\pi\otimes T_\nu)(e_{i_1}\otimes\ldots\otimes e_{i_p}\otimes e_{k_1}\otimes\ldots\otimes e_{k_r})\\
&=&\sum_{j_1\ldots j_q}\sum_{l_1\ldots l_s}\delta_\pi\begin{pmatrix}i_1&\ldots&i_p\\j_1&\ldots&j_q\end{pmatrix}\delta_\nu\begin{pmatrix}k_1&\ldots&k_r\\l_1&\ldots&l_s\end{pmatrix}e_{j_1}\otimes\ldots\otimes e_{j_q}\otimes e_{l_1}\otimes\ldots\otimes e_{l_s}\\
&=&\sum_{j_1\ldots j_q}\sum_{l_1\ldots l_s}\delta_{[\pi\nu]}\begin{pmatrix}i_1&\ldots&i_p&k_1&\ldots&k_r\\j_1&\ldots&j_q&l_1&\ldots&l_s\end{pmatrix}e_{j_1}\otimes\ldots\otimes e_{j_q}\otimes e_{l_1}\otimes\ldots\otimes e_{l_s}\\
&=&T_{[\pi\nu]}(e_{i_1}\otimes\ldots\otimes e_{i_p}\otimes e_{k_1}\otimes\ldots\otimes e_{k_r})
\end{eqnarray*}

(2) The composition axiom can be checked as follows:
\begin{eqnarray*}
&&T_\pi T_\nu(e_{i_1}\otimes\ldots\otimes e_{i_p})\\
&=&\sum_{j_1\ldots j_q}\delta_\nu\begin{pmatrix}i_1&\ldots&i_p\\j_1&\ldots&j_q\end{pmatrix}
\sum_{k_1\ldots k_r}\delta_\pi\begin{pmatrix}j_1&\ldots&j_q\\k_1&\ldots&k_r\end{pmatrix}e_{k_1}\otimes\ldots\otimes e_{k_r}\\
&=&\sum_{k_1\ldots k_r}N^{c(\pi,\nu)}\delta_{[^\nu_\pi]}\begin{pmatrix}i_1&\ldots&i_p\\k_1&\ldots&k_r\end{pmatrix}e_{k_1}\otimes\ldots\otimes e_{k_r}\\
&=&N^{c(\pi,\nu)}T_{[^\nu_\pi]}(e_{i_1}\otimes\ldots\otimes e_{i_p})
\end{eqnarray*}

(3) Finally, the involution axiom can be checked as follows:
\begin{eqnarray*}
&&T_\pi^*(e_{j_1}\otimes\ldots\otimes e_{j_q})\\
&=&\sum_{i_1\ldots i_p}<T_\pi^*(e_{j_1}\otimes\ldots\otimes e_{j_q}),e_{i_1}\otimes\ldots\otimes e_{i_p}>e_{i_1}\otimes\ldots\otimes e_{i_p}\\
&=&\sum_{i_1\ldots i_p}\delta_\pi\begin{pmatrix}i_1&\ldots&i_p\\ j_1&\ldots& j_q\end{pmatrix}e_{i_1}\otimes\ldots\otimes e_{i_p}\\
&=&T_{\pi^*}(e_{j_1}\otimes\ldots\otimes e_{j_q})
\end{eqnarray*}

Summarizing, our correspondence is indeed categorical.
\end{proof}

In relation now with the groups, following \cite{bsp}, we have the following result:

\index{Tannakian duality}

\begin{theorem}
Each category of partitions $D=(D(k,l))$ produces a family of compact groups $G=(G_N)$, with $G_N\subset_uU_N$, via the formula
$$Hom(u^{\otimes k},u^{\otimes l})=span\left(T_\pi\Big|\pi\in D(k,l)\right)$$
and the Tannakian duality correspondence.
\end{theorem}

\begin{proof}
Given an integer $N\in\mathbb N$, consider the correspondence $\pi\to T_\pi$ constructed in Definition 14.25, and then the collection of linear spaces in the statement, namely:
$$C(k,l)=span\left(T_\pi\Big|\pi\in D(k,l)\right)$$

According to Proposition 14.26, and to our axioms for the categories of partitions, from Definition 14.24, this collection of spaces $C=(C(k,l))$ satisfies the axioms for the Tannakian categories, from Definition 14.20. Thus the Tannakian duality result, Theorem 14.23, applies, and provides us with a closed subgroup $G_N\subset_uU_N$ such that:
$$C(k,l)=Hom(u^{\otimes k},u^{\otimes l})$$

Thus, we are led to the conclusion in the statement.
\end{proof}

Still following \cite{bsp}, we can now formulate a key definition, as follows:

\index{easy group}

\begin{definition}
A closed subgroup $G\subset_uU_N$ is called easy when we have
$$Hom(u^{\otimes k},u^{\otimes l})=span\left(T_\pi\Big|\pi\in D(k,l)\right)$$
for any colored integers $k,l$, for a certain category of partitions $D\subset P$.
\end{definition}

The notion of easiness goes back to the results of Brauer regarding the orthogonal group $O_N$, and the unitary group $U_N$, which can be formulated as follows:

\index{easy group}
\index{orthogonal group}
\index{unitary group}
\index{Brauer theorem}
\index{matching pairings}

\begin{theorem}
We have the following results:
\begin{enumerate}
\item $U_N$ is easy, coming from the category of matching pairings $\mathcal P_2$.

\item $O_N$ is easy too, coming from the category of all pairings $P_2$.
\end{enumerate}
\end{theorem}

\begin{proof}
This is something very standard, the idea being as follows:

\medskip

(1) The group $U_N$ being defined via the relations $u^*=u^{-1}$, $u^t=\bar{u}^{-1}$, the associated Tannakian category is $C=span(T_\pi|\pi\in D)$, with:
$$D
=<{\ }^{\,\cap}_{\circ\bullet}\,\,,{\ }^{\,\cap}_{\bullet\circ}>
=\mathcal P_2$$

(2) The group $O_N\subset U_N$ being defined by imposing the relations $u_{ij}=\bar{u}_{ij}$, the associated Tannakian category is $C=span(T_\pi|\pi\in D)$, with:
$$D
=<\mathcal P_2,|^{\hskip-1.32mm\circ}_{\hskip-1.32mm\bullet},|_{\hskip-1.32mm\circ}^{\hskip-1.32mm\bullet}>
=P_2$$
  
Thus, we are led to the conclusion in the statement.
\end{proof}

Beyond this, a first natural question is that of computing the easy group associated to the category $P$ itself, and we have here the following Brauer type theorem:

\index{symmetric group}
\index{Brauer theorem}

\begin{theorem}
The symmetric group $S_N$, regarded as group of unitary matrices,
$$S_N\subset O_N\subset U_N$$
via the permutation matrices, is easy, coming from the category of all partitions $P$.
\end{theorem}

\begin{proof}
Consider the easy group $G\subset O_N$ coming from the category of all partitions $P$. Since $P$ is generated by the one-block partition $Y\in P(2,1)$, we have:
$$C(G)=C(O_N)\Big/\Big<T_Y\in Hom(u^{\otimes 2},u)\Big>$$

The linear map associated to $Y$ is given by the following formula:
$$T_Y(e_i\otimes e_j)=\delta_{ij}e_i$$

Thus, the relation defining the above group $G\subset O_N$ reformulates as follows:
$$T_Y\in Hom(u^{\otimes 2},u)\iff u_{ij}u_{ik}=\delta_{jk}u_{ij},\forall i,j,k$$

In other words, the elements $u_{ij}$ must be projections, and these projections must be pairwise orthogonal on the rows of $u=(u_{ij})$. We conclude that $G\subset O_N$ is the subgroup of matrices $g\in O_N$ having the property $g_{ij}\in\{0,1\}$. Thus we have $G=S_N$, as claimed. 
\end{proof}

In fact, we have the following general easiness result, regarding the series of complex reflection groups $H_N^s=\mathbb Z_s\wr S_N\subset U_N$, that we talked about in chapter 3:

\index{reflection group}
\index{hyperoctahedral group}
\index{complex reflection group}

\begin{theorem}
The group $H_N^s=\mathbb Z_s\wr S_N$ is easy, the corresponding category $P^s$ consisting of the partitions satisfying $\#\circ=\#\bullet(s)$ in each block. In particular:
\begin{enumerate}
\item $S_N=H_N^1$ is easy, coming from the category $P$.

\item $H_N=H_N^2$ is easy, coming from the category $P_{even}$.

\item $K_N=H_N^\infty$ is easy, coming from the category $\mathcal P_{even}$.
\end{enumerate}
\end{theorem}

\begin{proof}
This is something that we already know at $s=1$, from Theorem 14.30. In general, the proof is similar, based on Tannakian duality. To be more precise, in what regards the main assertion, the idea here is that the one-block partition $\pi\in P(s)$, which generates the category $P^s$, implements the relations producing the subgroup $H_N^s\subset U_N$. As for the last assertions, these follow from the following observations:

\medskip

(1) At $s=1$ we know that we have $H_N^1=S_N$. Regarding now the corresponding category, here the condition $\#\circ=\#\bullet(1)$ is automatic, and so $P^1=P$.

\medskip

(2) At $s=2$ we know that we have $H_N^2=H_N$. Regarding now the corresponding category, here the condition $\#\circ=\#\bullet(2)$ reformulates as follows:
$$\#\circ+\,\#\bullet=0(2)$$

Thus each block must have even size, and we obtain, as claimed, $P^2=P_{even}$.

\medskip

(3) At $s=\infty$ we know that we have $H_N^\infty=K_N$. Regarding now the corresponding category, here the condition $\#\circ=\#\bullet(\infty)$ reads:
$$\#\circ=\#\bullet$$

But this is the condition defining $\mathcal P_{even}$, and so $P^\infty=\mathcal P_{even}$, as claimed.
\end{proof}

Let us go back now to probability questions, with the aim of applying the above abstract theory, to questions regarding characters. The situation here is as follows:

\bigskip

(1) Given a closed subgroup $G\subset_uU_N$, we know from Peter-Weyl that the moments of the main character count the fixed points of the representations $u^{\otimes k}$. 

\bigskip

(2) On the other hand, assuming that our group $G\subset_uU_N$ is easy, coming from a category of partitions $D=(D(k,l))$, the space formed by these fixed points is spanned by the following vectors, indexed by partitions $\pi$ belonging to the set $D(k)=D(0,k)$:
$$\xi_\pi=\sum_{i_1\ldots i_k}\delta_\pi\begin{pmatrix}i_1&\ldots&i_k\end{pmatrix}e_{i_1}\otimes\ldots\otimes e_{i_k}$$

(3) Thus, we are left with investigating linear independence questions for the vectors $\xi_\pi$, and once these questions solved, to compute the moments of $\chi$.

\bigskip

In order to investigate linear independence questions for the vectors $\xi_\pi$, we will use the Gram matrix of these vectors. Let us begin with some standard definitions:

\index{order on partitions}

\begin{definition}
Let $P(k)$ be the set of partitions of $\{1,\ldots,k\}$, and let $\pi,\nu\in P(k)$.
\begin{enumerate}
\item We write $\pi\leq\nu$ if each block of $\pi$ is contained in a block of $\nu$.

\item We let $\pi\vee\nu\in P(k)$ be the partition obtained by superposing $\pi,\nu$.
\end{enumerate}
\end{definition}

As an illustration here, at $k=2$ we have $P(2)=\{||,\sqcap\}$, and the order is:
$$||\leq\sqcap$$

At $k=3$ we have $P(3)=\{|||,\sqcap|,\sqcap\hskip-3.2mm{\ }_|\,,|\sqcap,\sqcap\hskip-0.7mm\sqcap\}$, and the order relation is as follows:
$$|||\leq\sqcap|,\sqcap\hskip-3.2mm{\ }_|\,,|\sqcap\leq\sqcap\hskip-0.7mm\sqcap$$

Observe also that we have $\pi,\nu\leq\pi\vee\nu$. In fact, $\pi\vee\nu$ is the smallest partition with this property, called supremum of $\pi,\nu$. Now back to the easy groups, we have:

\index{Gram matrix}

\begin{proposition}
The Gram matrix $G_{kN}(\pi,\nu)=<\xi_\pi,\xi_\nu>$ is given by
$$G_{kN}(\pi,\nu)=N^{|\pi\vee\nu|}$$
where $|.|$ is the number of blocks.
\end{proposition}

\begin{proof}
According to our formula of the vectors $\xi_\pi$, we have:
\begin{eqnarray*}
<\xi_\pi,\xi_\nu>
&=&\sum_{i_1\ldots i_k}\delta_\pi(i_1,\ldots,i_k)\delta_\nu(i_1,\ldots,i_k)\\
&=&\sum_{i_1\ldots i_k}\delta_{\pi\vee\nu}(i_1,\ldots,i_k)\\
&=&N^{|\pi\vee\nu|}
\end{eqnarray*}

Thus, we have obtained the formula in the statement.
\end{proof}

In order to study the Gram matrix, and more specifically to compute its determinant, we will need several standard facts about the partitions. We first have:

\index{M\"obius function}

\begin{definition}
The M\"obius function of any lattice, and so of $P$, is given by
$$\mu(\pi,\nu)=\begin{cases}
1&{\rm if}\ \pi=\nu\\
-\sum_{\pi\leq\tau<\nu}\mu(\pi,\tau)&{\rm if}\ \pi<\nu\\
0&{\rm if}\ \pi\not\leq\nu
\end{cases}$$
with the construction being performed by recurrence.
\end{definition}

As an illustration here, let us go back to the set of 2-point partitions, $P(2)=\{||,\sqcap\}$. Here we have by definition:
$$\mu(||,||)=\mu(\sqcap,\sqcap)=1$$

Also, we know that we have $||<\sqcap$, with no intermediate partition in between, and so the above recurrence procedure gives the following formular:
$$\mu(||,\sqcap)=-\mu(||,||)=-1$$

Finally, we have $\sqcap\not\leq||$, which gives $\mu(\sqcap,||)=0$. Thus, as a conclusion, the M\"obius matrix $M_{\pi\nu}=\mu(\pi,\nu)$ of the lattice $P(2)=\{||,\sqcap\}$ is as follows:
$$M=\begin{pmatrix}1&-1\\ 0&1\end{pmatrix}$$

The interest in the M\"obius function comes from the M\"obius inversion formula:
$$f(\nu)=\sum_{\pi\leq\nu}g(\pi)\implies g(\nu)=\sum_{\pi\leq\nu}\mu(\pi,\nu)f(\pi)$$

In linear algebra terms, the statement and proof of this formula are as follows:

\index{M\"obius inversion}

\begin{theorem}
The inverse of the adjacency matrix of $P$, given by
$$A_{\pi\nu}=\begin{cases}
1&{\rm if}\ \pi\leq\nu\\
0&{\rm if}\ \pi\not\leq\nu
\end{cases}$$
is the M\"obius matrix of $P$, given by $M_{\pi\nu}=\mu(\pi,\nu)$.
\end{theorem}

\begin{proof}
This is well-known, coming for instance from the fact that $A$ is upper triangular. Thus, when inverting, we are led into the recurrence from Definition 14.34.
\end{proof}

As an illustration here, for $P(2)$ the formula $M=A^{-1}$ appears as follows:
$$\begin{pmatrix}1&-1\\ 0&1\end{pmatrix}=
\begin{pmatrix}1&1\\ 0&1\end{pmatrix}^{-1}$$

Now back to our Gram matrix considerations, we have the following result:

\index{Gram matrix}

\begin{proposition}
The Gram matrix is given by $G_{kN}=AL$, where
$$L(\pi,\nu)=
\begin{cases}
N(N-1)\ldots(N-|\pi|+1)&{\rm if}\ \nu\leq\pi\\
0&{\rm otherwise}
\end{cases}$$
and where $A=M^{-1}$ is the adjacency matrix of $P(k)$.
\end{proposition}

\begin{proof}
We have the following computation:
\begin{eqnarray*}
N^{|\pi\vee\nu|}
&=&\#\left\{i_1,\ldots,i_k\in\{1,\ldots,N\}\Big|\ker i\geq\pi\vee\nu\right\}\\
&=&\sum_{\tau\geq\pi\vee\nu}\#\left\{i_1,\ldots,i_k\in\{1,\ldots,N\}\Big|\ker i=\tau\right\}\\
&=&\sum_{\tau\geq\pi\vee\nu}N(N-1)\ldots(N-|\tau|+1)
\end{eqnarray*}

According to Proposition 14.33 and to the definition of $A,L$, this formula reads:
$$(G_{kN})_{\pi\nu}
=\sum_{\tau\geq\pi}L_{\tau\nu}
=\sum_\tau A_{\pi\tau}L_{\tau\nu}
=(AL)_{\pi\nu}$$

Thus, we obtain the formula in the statement.
\end{proof}

With the above result in hand, we can now investigate the linear independence properties of the vectors $\xi_\pi$. To be more precise, we have the following result:

\index{Gram determinant}

\begin{theorem}[Lindst\"om]
The determinant of the Gram matrix $G_{kN}$ is
$$\det(G_{kN})=\prod_{\pi\in P(k)}\frac{N!}{(N-|\pi|)!}$$
and in particular, for $N\geq k$, the vectors $\{\xi_\pi|\pi\in P(k)\}$ are linearly independent.
\end{theorem}

\begin{proof}
According to the formula in Proposition 14.36, we have:
$$\det(G_{kN})=\det(A)\det(L)$$

Now if we order the set $P(k)$ as usual, with respect to the number of blocks, and then lexicographically, we can see that $A$ is upper triangular, and that $L$ is lower triangular. Thus $\det(A)$ can be computed simply by making the product on the diagonal, and we obtain $1$. As for $\det(L)$, this can computed as well by making the product on the diagonal, and we obtain the number in the statement, with the technical remark that in the case $N<k$ the convention is that we obtain a vanishing determinant. 
\end{proof}

Now back to the laws of characters, we can formulate:

\index{moments of characters}
\index{asymptotic characters}

\begin{proposition}
For an easy group $G=(G_N)$, coming from a category of partitions $D=(D(k,l))$, the asymptotic moments of the main character are given by
$$\lim_{N\to\infty}\int_{G_N}\chi^k=\# D(k)$$
where $D(k)=D(\emptyset,k)$, with the limiting sequence on the left consisting of certain integers, and being stationary at least starting from the $k$-th term.
\end{proposition}

\begin{proof}
This follows indeed from the Peter-Weyl theory, by using the linear independence result for the vectors $\xi_\pi$ coming from Theorem 14.37.
\end{proof}

With these preliminaries in hand, we can now state and prove:

\index{normal law}
\index{Bessel law}
\index{real Bessel law}
\index{complex Bessel law}

\begin{theorem}
In the $N\to\infty$ limit, the laws of the main character for the main easy groups, real and complex, discrete and continuous, are as follows, with $P_1=p^\infty_1$,
$$\xymatrix@R=51.5pt@C=50pt{
K_N\ar[r]&U_N\\
H_N\ar[u]\ar[r]&O_N\ar[u]}\qquad
\xymatrix@R=25pt@C=50pt{\\:}
\qquad
\xymatrix@R=50pt@C=51pt{
P_1\ar[r]&G_1\\
p^2_1\ar[u]\ar[r]&g_1\ar[u]}$$
with these laws, namely the real and complex Gaussian and Bessel laws, being the main limiting laws in real and complex, and discrete and continuous probability.
\end{theorem}

\begin{proof}
This follows from the above results. To be more precise, we know that the above groups are all easy, the corresponding categories of partitions being as follows:
$$\xymatrix@R=16mm@C=16mm{
\mathcal P_{even}\ar[d]&\mathcal P_2\ar[l]\ar[d]\\
P_{even}&P_2\ar[l]}$$

Thus, we can use Proposition 14.38, are we are led into counting partitions, and then recovering the measures via their moments, and this leads to the result.
\end{proof}

\section*{14d. Weingarten formula}

Our aim now is to go beyond what we have, with results regarding the truncated characters. Let us start with a general formula coming from Peter-Weyl, namely:

\index{Weingarten formula}
\index{Gram matrix}
\index{Weingarten matrix}

\begin{theorem}
The Haar integration over a closed subgroup $G\subset_uU_N$ is given on the dense subalgebra of smooth functions by the Weingarten type formula
$$\int_Gg_{i_1j_1}^{e_1}\ldots g_{i_kj_k}^{e_k}\,dg=\sum_{\pi,\nu\in D(k)}\delta_\pi(i)\delta_\sigma(j)W_k(\pi,\nu)$$
valid for any colored integer $k=e_1\ldots e_k$ and any multi-indices $i,j$, where $D(k)$ is a linear basis of $Fix(u^{\otimes k})$, the associated generalized Kronecker symbols are given by
$$\delta_\pi(i)=<\pi,e_{i_1}\otimes\ldots\otimes e_{i_k}>$$
and $W_k=G_k^{-1}$ is the inverse of the Gram matrix, $G_k(\pi,\nu)=<\pi,\nu>$.
\end{theorem}

\begin{proof}
This is something very standard, coming from the fact that the above integrals form altogether the orthogonal projection $P^k$ onto the following space:
$$Fix(u^{\otimes k})=span(D(k))$$

Consider now the following linear map, with $D(k)=\{\xi_k\}$ being as in the statement:
$$E(x)=\sum_{\pi\in D(k)}<x,\xi_\pi>\xi_\pi$$

By a standard linear algebra computation, it follows that we have $P=WE$, where $W$ is the inverse of the restriction of $E$ to the following space:
$$K=span\left(T_\pi\Big|\pi\in D(k)\right)$$

But this restriction is the linear map given by the matrix $G_k$, and so $W$ is the linear map given by the inverse matrix $W_k=G_k^{-1}$, and this gives the result.
\end{proof}

The point now is that in the easy group case, following \cite{bsp}, \cite{csn}, \cite{wei} and related work, we have the following more concrete result, which is something fully usable:

\index{Weingarten formula}
\index{Gram matrix}
\index{easy group}

\begin{theorem}
For an easy group $G\subset U_N$, coming from a category of partitions $D=(D(k,l))$, we have the Weingarten formula
$$\int_Gg_{i_1j_1}^{e_1}\ldots g_{i_kj_k}^{e_k}\,dg=\sum_{\pi,\nu\in D(k)}\delta_\pi(i)\delta_\nu(j)W_{kN}(\pi,\nu)$$
for any $k=e_1\ldots e_k$ and any $i,j$, where $D(k)=D(\emptyset,k)$, $\delta$ are usual Kronecker type symbols, checking whether the indices match, and $W_{kN}=G_{kN}^{-1}$, with 
$$G_{kN}(\pi,\nu)=N^{|\pi\vee\nu|}$$
where $|.|$ is the number of blocks. 
\end{theorem}

\begin{proof}
We can use the abstract Weingarten formula, from Theorem 14.40. Indeed, the Kronecker type symbols there are then the usual ones, as shown by:
\begin{eqnarray*}
\delta_{\xi_\pi}(i)
&=&<\xi_\pi,e_{i_1}\otimes\ldots\otimes e_{i_k}>\\
&=&\left<\sum_j\delta_\pi(j_1,\ldots,j_k)e_{j_1}\otimes\ldots\otimes e_{j_k},e_{i_1}\otimes\ldots\otimes e_{i_k}\right>\\
&=&\delta_\pi(i_1,\ldots,i_k)
\end{eqnarray*}

The Gram matrix being as well the correct one, we obtain the result.
\end{proof}

Let us go back now to the general easy groups $G\subset U_N$, with the idea in mind of computing the laws of truncated characters. First, we have the following formula:

\index{truncated characters}

\begin{proposition}
The moments of truncated characters are given by the formula
$$\int_G(g_{11}+\ldots +g_{ss})^kdg=Tr(W_{kN}G_{ks})$$
where $G_{kN}$ and $W_{kN}=G_{kN}^{-1}$ are the associated Gram and Weingarten matrices.
\end{proposition}

\begin{proof}
We have indeed the following computation:
\begin{eqnarray*}
\int_G(g_{11}+\ldots +g_{ss})^kdg
&=&\sum_{i_1=1}^{s}\ldots\sum_{i_k=1}^s\int_Gg_{i_1i_1}\ldots g_{i_ki_k}\,dg\\
&=&\sum_{\pi,\nu\in D(k)}W_{kN}(\pi,\nu)\sum_{i_1=1}^{s}\ldots\sum_{i_k=1}^s\delta_\pi(i)\delta_\nu(i)\\
&=&\sum_{\pi,\nu\in D(k)}W_{kN}(\pi,\nu)G_{ks}(\nu,\pi)\\
&=&Tr(W_{kN}G_{ks})
\end{eqnarray*}

Thus, we have reached to the formula in the statement.
\end{proof}

In order to process now the above formula, and reach to concrete results, we must impose on our group a uniformity condition. Let us start with:

\index{uniform group}

\begin{proposition}
For an easy group $G=(G_N)$, coming from a category of partitions $D\subset P$, the following conditions are equivalent:
\begin{enumerate}
\item $G_{N-1}=G_N\cap U_{N-1}$, via the embedding $U_{N-1}\subset U_N$ given by $g\to diag(g,1)$.

\item $G_{N-1}=G_N\cap U_{N-1}$, via the $N$ possible diagonal embeddings $U_{N-1}\subset U_N$.

\item $D$ is stable under the operation which consists in removing blocks.
\end{enumerate}
If these conditions are satisfied, we say that $G=(G_N)$ is uniform.
\end{proposition}

\begin{proof}
The equivalence $(1)\iff(2)$ comes from the inclusion $S_N\subset G_N$, which makes everything $S_N$-invariant. Regarding $(1)\iff(3)$, given a subgroup $K\subset_uU_{N-1}$, consider the matrix $v=diag(u,1)$. Our claim is that for any $\pi\in P(k)$ we have:
$$\xi_\pi\in Fix(v^{\otimes k})\iff\xi_{\pi'}\in Fix(v^{\otimes k'}),\,\forall\pi'\in P(k'),\pi'\subset\pi$$

In order to prove this claim, we must study the condition on the left. We have:
\begin{eqnarray*}
\xi_\pi\in Fix(v^{\otimes k})
&\iff&(v^{\otimes k}\xi_\pi)_{i_1\ldots i_k}=(\xi_\pi)_{i_1\ldots i_k},\forall i\\
&\iff&\sum_j(v^{\otimes k})_{i_1\ldots i_k,j_1\ldots j_k}(\xi_\pi)_{j_1\ldots j_k}=(\xi_\pi)_{i_1\ldots i_k},\forall i\\
&\iff&\sum_j\delta_\pi(j_1,\ldots,j_k)v_{i_1j_1}\ldots v_{i_kj_k}=\delta_\pi(i_1,\ldots,i_k),\forall i
\end{eqnarray*}

Now let us recall that our representation has the special form $v=diag(u,1)$. We conclude from this that for any index $a\in\{1,\ldots,k\}$, we have:
$$i_a=N\implies j_a=N$$

With this observation in hand, if we denote by $i',j'$ the multi-indices obtained from $i,j$ obtained by erasing all the above $i_a=j_a=N$ values, and by $k'\leq k$ the common length of these new multi-indices, our condition becomes:
$$\sum_{j'}\delta_\pi(j_1,\ldots,j_k)(v^{\otimes k'})_{i'j'}=\delta_\pi(i_1,\ldots,i_k),\forall i$$

Here the index $j$ is by definition obtained from the index $j'$ by filling with $N$ values. In order to finish now, we have two cases, depending on $i$, as follows:

\medskip

\underline{Case 1}. Assume that the index set $\{a|i_a=N\}$ corresponds to a certain subpartition $\pi'\subset\pi$. In this case, the $N$ values will not matter, and our formula becomes:
$$\sum_{j'}\delta_\pi(j'_1,\ldots,j'_{k'})(v^{\otimes k'})_{i'j'}=\delta_\pi(i'_1,\ldots,i'_{k'})$$

\underline{Case 2}. Assume now the opposite, namely that the set $\{a|i_a=N\}$ does not correspond to a subpartition $\pi'\subset\pi$. In this case the indices mix, and our formula reads $0=0$. Thus we have $\xi_{\pi'}\in Fix(v^{\otimes k'})$ in both cases, for any subpartition $\pi'\subset\pi$, as desired.
\end{proof}

Now back to the laws of truncated characters, we have the following result:

\index{easy group}
\index{uniform group}
\index{truncated character}

\begin{theorem}
For a uniform easy group $G=(G_N)$, we have the formula
$$\lim_{N\to\infty}\int_{G_N}\chi_t^k=\sum_{\pi\in D(k)}t^{|\pi|}$$
with $D\subset P$ being the associated category of partitions.
\end{theorem}

\begin{proof}
We use Proposition 14.42. Since in the uniform case the Gram matrix, and so the Weingarten matrix too, is asymptotically diagonal, we obtain:
\begin{eqnarray*}
\int_{G_N}\chi_t^k
&=&Tr(W_{kN}G_{k[tN]})\\
&\simeq&\sum_{\pi\in D(k)}N^{-|\pi|}(tN)^{|\pi|}\\
&=&\sum_{\pi\in D(k)}t^{|\pi|}
\end{eqnarray*}

Thus, we are led to the formula in the statement.
\end{proof}

We can now enlarge our collection of character results, and we have:

\begin{theorem}
With $N\to\infty$, the laws of truncated characters are as follows:
\begin{enumerate}
\item For $O_N$ we obtain the Gaussian law $g_t$.

\item For $U_N$ we obtain the complex Gaussian law $G_t$.

\item For $S_N$ we obtain the Poisson law $p_t$.

\item For $H_N$ we obtain the Bessel law $p^2_t$.

\item For $H_N^s$ we obtain the generalized Bessel law $p_t^s$.

\item For $K_N$ we obtain the purely complex Bessel law $P_t$.
\end{enumerate}
\end{theorem}

\begin{proof}
We already know these results at $t=1$. In the general case, $t>0$, these follow via some standard combinatorics, from the formula in Theorem 14.44.
\end{proof}

\section*{14e. Exercises}

This was a technical and exciting chapter, and as exercises on this, we have:

\begin{exercise}
Learn the decomposition result for the $*$-algebras $A\subset M_N(\mathbb C)$.
\end{exercise}

\begin{exercise}
Learn more about central functions, on finite and compact groups.
\end{exercise}

\begin{exercise}
Find and read one of the available proofs of Tannakian duality.
\end{exercise}

\begin{exercise}
Learn about the theorems of Deligne, and Doplicher-Roberts.
\end{exercise}

\begin{exercise}
Work out all the details for the easiness property of $H_N^s$.
\end{exercise}

\begin{exercise}
Compute the Gram determinants for other easy groups.
\end{exercise}

\begin{exercise}
Work out all the details for our main character computations.
\end{exercise}

\begin{exercise}
Work out too the details for our truncated character computations.
\end{exercise}

As bonus exercise, learn as well about Haar integration over locally compact groups.

\chapter{Random matrices}

\section*{15a. Spectral measures}

Welcome to the random matrices, which are first class mathematics and physics. In order to talk about such matrices and their spectral measures, we need to do some more linear algebra in infinite dimensions, as a continuation of our operator theory discussion from chapter 12. It is convenient to upgrade our formalism, as follows:

\index{operator algebra}
\index{normed algebra}
\index{Banach algebra}

\begin{definition}
An abstract operator algebra, or $C^*$-algebra, is a complex algebra $A$ having a norm $||.||$ and an involution $*$, subject to the following conditions:
\begin{enumerate}
\item $A$ is closed with respect to the norm.

\item We have $||aa^*||=||a||^2$, for any $a\in A$.
\end{enumerate}
\end{definition}

In other words, what we did here is to axiomatize the abstract properties of the operator algebras $A\subset B(H)$, coming from the various general results about linear operators from chapter 12, without any reference to the ambient Hilbert space $H$. 

\bigskip

As basic examples here, we have the usual matrix algebras $M_N(\mathbb C)$, with the norm and the involution being the usual matrix norm and involution, given by:
$$||A||=\sup_{||x||=1}||Ax||\quad,\quad 
(A^*)_{ij}=\overline{A}_{ji}$$

Some other basic examples are the algebras $L^\infty(X)$ of essentially bounded functions $f:X\to\mathbb C$ on a measured space $X$, with the usual norm and involution, namely:
$$||f||=\sup_{x\in X}|f(x)|\quad,\quad 
f^*(x)=\overline{f(x)}$$

We can put these two basic classes of examples together, as follows:

\index{random matrix algebra}

\begin{proposition}
The random matrix algebras $A=M_N(L^\infty(X))$ are $C^*$-algebras, with their usual norm and involution, given by:
$$||Z||=\sup_{x\in X}||Z_x||\quad,\quad 
(Z^*)_{ij}=\overline{Z}_{ij}$$
These algebras generalize both the algebras $M_N(\mathbb C)$, and the algebras $L^\infty(X)$.
\end{proposition}

\begin{proof}
The fact that the $C^*$-algebra axioms are satisfied is clear from definitions. As for the last assertion, this follows by taking $X=\{.\}$ and $N=1$, respectively.
\end{proof}

We can in fact say more about the above algebras, as follows:

\begin{theorem}
Any algebra of type $L^\infty(X)$ is an operator algebra, as follows:
$$L^\infty(X)\subset B(L^2(X))\quad,\quad 
f\to(g\to fg)$$
More generally, any random matrix algebra is an operator algebra, as follows,
$$M_N(L^\infty(X))\subset B\left(\mathbb C^N\otimes L^2(X)\right)$$
with the embedding being the above one, tensored with the identity.
\end{theorem}

\begin{proof}
We have two assertions to be proved, the idea being as follows:

\medskip

(1) Given $f\in L^\infty(X)$, consider the following operator, acting on $H=L^2(X)$:
$$T_f(g)=fg$$

Observe that $T_f$ is indeed well-defined, and bounded as well, because:
$$||fg||_2
=\sqrt{\int_X|f(x)|^2|g(x)|^2d\mu(x)}
\leq||f||_\infty||g||_2$$

The application $f\to T_f$ being linear, involutive, continuous, and injective as well, we obtain in this way a $C^*$-algebra embedding $L^\infty(X)\subset B(H)$, as desired.

\medskip

(2) Regarding the second assertion, this is best viewed in the following way:
\begin{eqnarray*}
M_N(L^\infty(X))
&=&M_N(\mathbb C)\otimes L^\infty(X)\\
&\subset&M_N(\mathbb C)\otimes B(L^2(X))\\
&=&B\left(\mathbb C^N\otimes L^2(X)\right)
\end{eqnarray*}

Here we have used (1), and some standard tensor product identifications.
\end{proof}

Our purpose in what follows is to develop the spectral theory of the $C^*$-algebras, and in particular that of the random matrix algebras $A=M_N(L^\infty(X))$ that we are interested in, one of our objectives being that of talking about spectral measures, in the normal case, in analogy with what we know about the usual matrices. Let us start with:

\index{spectrum}
\index{spectral radius}
\index{unitary operator}
\index{self-adjoint operator}
\index{normal operator}
\index{closed subalgebra}
\index{functional calculus}

\begin{theorem}
Given an element $a\in A$ of a $C^*$-algebra, define its spectrum as:
$$\sigma(a)=\left\{\lambda\in\mathbb C\Big|a-\lambda\notin A^{-1}\right\}$$
The following spectral theory results hold, exactly as in the $A=B(H)$ case:
\begin{enumerate}
\item We have $\sigma(ab)\cup\{0\}=\sigma(ba)\cup\{0\}$.

\item We have $\sigma(f(a))=f(\sigma(a))$, for any $f\in\mathbb C(X)$ having poles outside $\sigma(a)$.

\item The spectrum $\sigma(a)$ is compact, non-empty, and contained in $D_0(||a||)$.

\item The spectra of unitaries $(u^*=u^{-1})$ and self-adjoints $(a=a^*)$ are on $\mathbb T,\mathbb R$.

\item The spectral radius of normal elements $(aa^*=a^*a)$ is given by $\rho(a)=||a||$.
\end{enumerate}
In addition, assuming $a\in A\subset B$, the spectra of $a$ with respect to $A$ and to $B$ coincide.
\end{theorem}

\begin{proof}
Here the assertions (1-5), which are of course formulated a bit informally, are well-known for the full operator algebra $A=B(H)$, and the proof in general is similar:

\medskip

(1) Assuming that $1-ab$ is invertible, with inverse $c$, we have $abc=cab=c-1$, and it follows that $1-ba$ is invertible too, with inverse $1+bca$. Thus $\sigma(ab),\sigma(ba)$ agree on $1\in\mathbb C$, and by linearity, it follows that $\sigma(ab),\sigma(ba)$ agree on any point $\lambda\in\mathbb C^*$.

\medskip

(2) The formula $\sigma(f(a))=f(\sigma(a))$ is clear for polynomials, $f\in\mathbb C[X]$, by factorizing $f-\lambda$, with $\lambda\in\mathbb C$. Then, the extension to the rational functions is straightforward, because $P(a)/Q(a)-\lambda$ is invertible precisely when $P(a)-\lambda Q(a)$ is.

\medskip

(3) By using $1/(1-b)=1+b+b^2+\ldots$ for $||b||<1$ we obtain that $a-\lambda$ is invertible for $|\lambda|>||a||$, and so $\sigma(a)\subset D_0(||a||)$. It is also clear that $\sigma(a)$ is closed, so what we have is a compact set. Finally, assuming $\sigma(a)=\emptyset$ the function $f(\lambda)=\varphi((a-\lambda)^{-1})$ is well-defined, for any $\varphi\in A^*$, and by Liouville we get $f=0$, contradiction.

\medskip

(4) Assuming $u^*=u^{-1}$ we have $||u||=1$, and so $\sigma(u)\subset D_0(1)$. But with $f(z)=z^{-1}$ we obtain via (2) that we have as well $\sigma(u)\subset f(D_0(1))$, and this gives $\sigma(u)\subset\mathbb T$. As for the result regarding the self-adjoints, this can be obtained from the result for the unitaries, by using (2) with functions of type $f(z)=(z+it)/(z-it)$, with $t\in\mathbb R$.

\medskip

(5) It is routine to check, by integrating quantities of type $z^n/(z-a)$ over circles centered at the origin, and estimating, that the spectral radius is given by $\rho(a)=\lim||a^n||^{1/n}$. But in the self-adjoint case, $a=a^*$, this gives $\rho(a)=||a||$, by using exponents of type $n=2^k$, and then the extension to the general normal case is straightforward.

\medskip 

(6) Regarding now the last assertion, the inclusion $\sigma_B(a)\subset\sigma_A(a)$ is clear. For the converse, assume $a-\lambda\in B^{-1}$, and set $b=(a-\lambda )^*(a-\lambda )$. We have then:
$$\sigma_A(b)-\sigma_B(b)=\left\{\mu\in\mathbb C-\sigma_B(b)\Big|(b-\mu)^{-1}\in B-A\right\}$$

Thus this difference in an open subset of $\mathbb C$. On the other hand $b$ being self-adjoint, its two spectra are both real, and so is their difference. Thus the two spectra of $b$ are equal, and in particular $b$ is invertible in $A$, and so $a-\lambda\in A^{-1}$, as desired.
\end{proof}

We can now prove a key result, as follows:

\index{Gelfand theorem}
\index{commutative algebra}
\index{spectrum of algebra}
\index{algebra character}
\index{compact space}

\begin{theorem}[Gelfand]
If $X$ is a compact space,  the algebra $C(X)$ of continuous functions on it $f:X\to\mathbb C$ is a $C^*$-algebra, with usual norm and involution, namely:
$$||f||=\sup_{x\in X}|f(x)|\quad,\quad 
f^*(x)=\overline{f(x)}$$
Conversely, any commutative $C^*$-algebra is of this form, $A=C(X)$, with 
$$X=\Big\{\chi:A\to\mathbb C\ ,\ {\rm normed\ algebra\ character}\Big\}$$
with topology making continuous the evaluation maps $ev_a:\chi\to\chi(a)$.
\end{theorem}

\begin{proof}
There are several things going on here, the idea being as follows:

\medskip

(1) The first assertion is clear from definitions. Observe that we have indeed:
$$||ff^*||
=\sup_{x\in X}|f(x)|^2
=||f||^2$$

Observe also that the algebra $C(X)$ is commutative, because $fg=gf$.

\medskip

(2) Conversely, given a commutative $C^*$-algebra $A$, let us define $X$ as in the statement. Then $X$ is compact, and $a\to ev_a$ is a morphism of algebras, as follows:
$$ev:A\to C(X)$$

(3) We first prove that $ev$ is involutive. We use the following formula, which is similar to the $z=Re(z)+iIm(z)$ decomposition formula for usual complex numbers:
$$a=\frac{a+a^*}{2}+i\cdot\frac{a-a^*}{2i}$$

Thus it is enough to prove $ev_{a^*}=ev_a^*$ for the self-adjoint elements $a$. But this is the same as proving that $a=a^*$ implies that $ev_a$ is a real function, which is in turn true, by Theorem 15.4, because $ev_a(\chi)=\chi(a)$ is an element of $\sigma(a)$, contained in $\mathbb R$.

\medskip

(4) Since $A$ is commutative, each element is normal, so $ev$ is isometric:
$$||ev_a||
=\rho(a)
=||a||$$

It remains to prove that $ev$ is surjective. But this follows from the Stone-Weierstrass theorem, because $ev(A)$ is a closed subalgebra of $C(X)$, which separates the points.
\end{proof}

As a main consequence of the Gelfand theorem, we have:

\index{continuous calculus}

\begin{theorem}
For any normal element $a\in A$ we have an identification as follows:
$$<a>=C(\sigma(a))$$
In addition, given a function $f\in C(\sigma(a))$, we can apply it to $a$, and we have
$$\sigma(f(a))=f(\sigma(a))$$
which generalizes the previous rational calculus formula, in the normal case.
\end{theorem}

\begin{proof}
Since $a$ is normal, the $C^*$-algebra $<a>$ that is generates is commutative, so if we denote by $X$ the space of the characters $\chi:<a>\to\mathbb C$, we have:
$$<a>=C(X)$$

Now since the map $X\to\sigma(a)$ given by evaluation at $a$ is bijective, we obtain:
$$<a>=C(\sigma(a))$$

Thus, we are dealing here with usual functions, and this gives all the assertions.
\end{proof}

In order to get now towards noncommutative probability, we first have to develop the theory of positive elements, and linear forms. First, we have the following result:

\index{positive element}

\begin{proposition}
For an element $a\in A$, the following are equivalent:
\begin{enumerate}
\item $a$ is positive, in the sense that $\sigma(a)\subset[0,\infty)$.

\item $a=b^2$, for some $b\in A$ satisfying $b=b^*$.

\item $a=cc^*$, for some $c\in A$.
\end{enumerate}
\end{proposition}

\begin{proof}
This is something very standard, as follows:

\medskip

$(1)\implies(2)$ Observe first that $\sigma(a)\subset\mathbb R$ implies $a=a^*$. Thus the algebra $<a>$ is commutative, and by using Theorem 15.6, we can set $b=\sqrt{a}$.

\medskip

$(2)\implies(3)$ This is trivial, because we can simply set $c=b$. 

\medskip

$(3)\implies(1)$ We proceed by contradiction. By multiplying $c$ by a suitable element of the algebra $<cc^*>$, we are led to the existence of an element $d\neq0$ satisfying $-dd^*\geq0$. By writing now $d=x+iy$ with $x=x^*,y=y^*$ we have:
$$dd^*+d^*d
=2(x^2+y^2)
\geq0$$

Thus $d^*d\geq0$, which is easily seen to contradict the condition $-dd^*\geq0$.
\end{proof}

Next, we can talk as well about positive linear forms, as follows:

\begin{definition}
Consider a linear map $\varphi:A\to\mathbb C$.
\begin{enumerate}
\item $\varphi$ is called positive when $a\geq0\implies\varphi(a)\geq0$.

\item $\varphi$ is called faithful and positive when $a\geq0,a\neq0\implies\varphi(a)>0$.
\end{enumerate}
\end{definition}

In the commutative case, $A=C(X)$, the positive linear forms appear as follows, with $\mu$ being positive, and strictly positive if we want $\varphi$ to be faithful and positive:
$$\varphi(f)=\int_Xf(x)d\mu(x)$$

In general, the positive linear forms can be thought of as being integration functionals with respect to some underlying ``positive measures''. Next, let us formulate:

\index{random variable}
\index{moments}
\index{colored moments}
\index{distribution}
\index{law}

\begin{definition}
Let $A$ be a $C^*$-algebra, given with a positive trace $tr:A\to\mathbb C$.
\begin{enumerate}
\item The elements $a\in A$ are called random variables.

\item The moments of such a variable are the numbers $M_k(a)=tr(a^k)$.

\item The law of such a variable is the functional $\mu_a:P\to tr(P(a))$.
\end{enumerate}
\end{definition}

Here the exponent $k=\circ\bullet\bullet\circ\ldots$ is by definition a colored integer, and the powers $a^k$ are defined by the following formulae, and multiplicativity: 
$$a^\emptyset=1\quad,\quad
a^\circ=a\quad,\quad
a^\bullet=a^*$$ 

As for the polynomial $P$, this is a noncommuting $*$-polynomial in one variable: 
$$P\in\mathbb C<X,X^*>$$

Observe that the law is uniquely determined by the moments, because we have:
$$P(X)=\sum_k\lambda_kX^k
\implies\mu_a(P)=\sum_k\lambda_kM_k(a)$$

At the level of the general theory, we have the following key result, extending the various results that we have, regarding the self-adjoint and normal matrices:

\index{normal element}
\index{spectral measure}

\begin{theorem}
Let $A$ be a $C^*$-algebra, with a trace $tr$, and consider an element $a\in A$ which is normal, in the sense that $aa^*=a^*a$.
\begin{enumerate}
\item $\mu_a$ is a complex probability measure, satisfying $supp(\mu_a)\subset\sigma(a)$.

\item In the self-adjoint case, $a=a^*$, this measure $\mu_a$ is real.

\item Assuming that $tr$ is faithful, we have $supp(\mu_a)=\sigma(a)$.
\end{enumerate}
\end{theorem}

\begin{proof}
In the normal case, $aa^*=a^*a$, the Gelfand theorem, or rather the subsequent continuous functional calculus theorem, tells us that we have: 
$$<a>=C(\sigma(a))$$

Thus the functional $f(a)\to tr(f(a))$ can be regarded as an integration functional on the algebra $C(\sigma(a))$, and by the Riesz theorem, this latter functional must come from a probability measure $\mu$ on the spectrum $\sigma(a)$, in the sense that we must have:
$$tr(f(a))=\int_{\sigma(a)}f(z)d\mu(z)$$

We are therefore led to the conclusions in the statement, with the uniqueness assertion coming from the fact that the elements $a^k$, taken as usual with respect to colored integer exponents, $k=\circ\bullet\bullet\circ\ldots$\,, generate the whole $C^*$-algebra $C(\sigma(a))$.
\end{proof}

Getting back to the random matrices, we have, as an application of Theorem 15.10:

\begin{theorem}
Given a random matrix $Z\in M_N(L^\infty(X))$, assumed to be normal, $ZZ^*=Z^*Z$, its law, which is by definition the following abstract functional,
$$\mu:\mathbb C<X,X^*>\to\mathbb C\quad,\quad 
P\to\frac{1}{N}\int_Xtr(P(Z))$$
when restricted to the usual polynomials in two variables,
$$\mu:\mathbb C[X,X^*]\to\mathbb C\quad,\quad 
P\to\frac{1}{N}\int_Xtr(P(Z))$$
must come from a probability measure on the spectrum $\sigma(Z)\subset\mathbb C$, as follows:
$$\mu(P)=\int_{\sigma(T)}P(z)d\mu(z)$$
We agree to use the symbol $\mu$ for all these notions.
\end{theorem}

\begin{proof}
This follows indeed from what we know from Theorem 15.10, applied to the normal element $a=Z$, belonging to the $C^*$-algebra $A=M_N(L^\infty(X))$. 
\end{proof}

\section*{15b. Gaussian matrices}

We have now all the needed ingredients for launching some explicit random matrix computations. Our goal will be that of computing the asymptotic moments, and then the asymptotic laws, with $N\to\infty$, for the main classes of large random matrices. 

\bigskip

Let us begin by specifying the precise classes of matrices that we are interested in. First we have the complex Gaussian matrices, which are constructed as follows:

\index{Gaussian matrix}

\begin{definition}
A complex Gaussian matrix is a random matrix of type
$$Z\in M_N(L^\infty(X))$$
which has i.i.d. centered complex normal entries.
\end{definition}

To be more precise, the assumption in this definition is that all the matrix entries $Z_{ij}$ are independent, and follow the same complex normal law $G_t$, for a fixed value of $t>0$. We will see that the above matrices have an interesting, and ``central'' combinatorics, among all kinds of random matrices, with the study of the other random matrices being usually obtained as a modification of the study of the Gaussian matrices.

\bigskip

As a somewhat surprising remark, using real normal variables in Definition 15.12, instead of the complex ones appearing there, leads nowhere. The correct real versions of the Gaussian matrices are the Wigner random matrices, constructed as follows: 

\index{Wigner matrix}

\begin{definition}
A Wigner matrix is a random matrix of type
$$Z\in M_N(L^\infty(X))$$
which has i.i.d. centered complex normal entries, up to the constraint $Z=Z^*$.
\end{definition}

This definition is something a bit compacted, and to be more precise, a Wigner matrix is by definition a random matrix as follows, with the diagonal entries being real normal variables, $a_i\sim g_t$, for some $t>0$, the upper diagonal entries being complex normal variables, $b_{ij}\sim G_t$, the lower diagonal entries being the conjugates of the upper diagonal entries, as indicated, and with all the variables $a_i,b_{ij}$ being independent: 
$$Z=\begin{pmatrix}
a_1&b_{12}&\ldots&\ldots&b_{1N}\\
\bar{b}_{12}&a_2&\ddots&&\vdots\\
\vdots&\ddots&\ddots&\ddots&\vdots\\
\vdots&&\ddots&a_{N-1}&b_{N-1,N}\\
\bar{b}_{1N}&\ldots&\ldots&\bar{b}_{N-1,N}&a_N
\end{pmatrix}$$

As a comment here, for many concrete applications the Wigner matrices are in fact the central objects in random matrix theory, and in particular, they are often more important than the Gaussian matrices. In fact, these are the random matrices which were first considered and investigated, a long time ago, by Wigner himself \cite{wig}.

\bigskip

However, as we will soon discover, the Gaussian matrices are somehow more fundamental than the Wigner matrices, at least from an abstract point of view, and this will be the point of view that we will follow here, with the Gaussian matrices coming first.

\bigskip

Finally, we will be interested as well in the complex Wishart matrices, which are the positive versions of the above random matrices, constructed as follows: 

\index{Wishart matrix}

\begin{definition}
A complex Wishart matrix is a random matrix of type
$$Z=YY^*\in M_N(L^\infty(X))$$
with $Y$ being a complex Gaussian matrix.
\end{definition}

As before with the Gaussian and Wigner matrices, there are many possible comments that can be made here, of technical or historical nature. As a first key fact, using real Gaussian variables instead of complex ones leads to a less interesting combinatorics. Also, these matrices were introduced and studied by Marchenko and Pastur in \cite{mpa}, not long after Wigner, and so historically came second. Finally, in what regards their combinatorics and applications, these matrices quite often come first, before both the Gaussian and the Wigner ones, with all this being of course a matter of knowledge and taste.

\bigskip

Summarizing, we have three main types of random matrices, which can be thought of as being ``complex'', ``real'' and ``positive'', that we will study in what follows. Let us also mention that there are many other interesting classes of random matrices, which are more specialized, usually appearing as modifications of the above. More on these later.

\bigskip

Getting to work now, let us first study the Gaussian matrices. These matrices are not normal, so we will be interested in computing the moments, and depending on what we find, speculate on the corresponding asymptotic law $\mu:\mathbb C<X,X^*>\to\mathbb C$. Now thinking a bit, for the moment computation we will certainly need the Wick formula, that we learned in chapter 4. So, let us first recall that formula:

\begin{theorem}[Wick formula]
Given independent variables $f_i$, each following the complex normal law $G_t$, with $t>0$ being a fixed parameter, we have the formula
$$E\left(f_{i_1}^{k_1}\ldots f_{i_s}^{k_s}\right)=t^{s/2}\#\left\{\pi\in\mathcal P_2(k)\Big|\pi\leq\ker i\right\}$$
with $\mathcal P_2(k)$ standing for the pairings of $k$ which are matching, pairing $\circ-\bullet$ symbols, and where $\pi\leq\ker i$ means that the indices of $i$ must fit into the blocks of $\pi$.
\end{theorem}

\begin{proof}
This is something from chapter 4, the idea being that for single variables the Wick formula is the moment formula for $G_t$, which was as follows:
$$M_k(G_t)=\sum_{\pi\in\mathcal P_2(k)}t^{|\pi|}$$

As for the general case, that of several variables, this follows from this. 
\end{proof}

Next, it is also pretty much clear that for doing our computations, we will need a good knowledge of the partitions. In practice, here is what we need to know: 

\index{cycling inside blocks}

\begin{proposition}
For a permutation $\sigma\in S_k$, let $|\sigma|\in\{1,\ldots,k\}$ be its number of cycles, and $l(\sigma)=k-|\sigma|\in\{0,\ldots,k-1\}$ be its length, that is, the minimal number of transpositions needed for decomposing $\sigma$ as a product of transpositions.
\begin{enumerate}
\item $d(\sigma,\pi)=l(\sigma^{-1}\pi)$ is a distance function on $S_k$.

\item If $\gamma\in S_k$ is a full cycle, $l(\sigma)+l(\sigma^{-1}\gamma)\geq k-1$.

\item Equivalently, we have $|\sigma|+|\sigma^{-1}\gamma|\leq k+1$.

\item Equality happens when $\sigma\in NC(k)\subset S_k$, by cycling inside each block.
\end{enumerate} 
\end{proposition}

\begin{proof}
To start with, given a permutation $\sigma\in S_k$, we can certainly talk about the number of cycles $|\sigma|\in\{1,\ldots,k\}$, and about the length $l(\sigma)\in\{0,\ldots,k-1\}$ too, with the convention $l(id)=0$. In order to prove now $l(\sigma)=k-|\sigma|$, let us decompose $\sigma$ as a product of distinct cycles, having lengths $k_1,\ldots,k_r\geq1$, as follows:
$$\sigma=\gamma_{k_1}\ldots\gamma_{k_r}$$

We can then compute the length of $\sigma$, in the obvious way, and we get, as desired:
\begin{eqnarray*}
l(\sigma)
&=&(k_1-1)+\ldots+(k_r-1)\\
&=&(k_1+\ldots+k_r)-r\\
&=&k-|\sigma|
\end{eqnarray*}

Thus we have the basics, and regarding the proof of the rest, this goes as follows:

\medskip

(1) First, we have $d(\sigma,\sigma)=l(id)=0$. Next, the symmetry axiom comes from:
$$d(\sigma,\pi)=l(\sigma^{-1}\pi)=l\left((\sigma^{-1}\pi)^{-1}\right)=l(\pi^{-1}\sigma)=d(\pi,\sigma)$$

Finally, the triangle inequality $d(\sigma,\pi)+d(\pi,\nu)\geq d(\sigma,\nu)$ amounts in saying that we have $l(\sigma^{-1}\pi)+l(\pi^{-1}\nu)\geq l(\sigma^{-1}\nu)$. But this is clear, coming from $\sigma^{-1}\pi\cdot \pi^{-1}\nu=\sigma^{-1}\nu$.

\medskip

(2) This is the triangle inequality, applied to the triangle $id-\sigma-\gamma$.

\medskip

(3) This comes indeed from (2), via the formula $l(\sigma)=k-|\sigma|$.

\medskip

(4) Consider the embedding $P(k)\subset S_k$, obtained by cycling inside each block, in the obvious way. As a first observation, for $\sigma\in P(k)$ our previous notation $|\sigma|$ for the number of blocks matches with our new notation $|\sigma|$ for the number of cycles. Thus, our formalism is fine, and what happens now is that equality happens in (2,3), meaning that $id-\sigma-\gamma$ must be aligned, precisely when $\sigma\in NC(k)\subset P(k)\subset S_k$. We will leave the details as an exercice, and for this, and more, you can have a look at Biane \cite{bia}.
\end{proof}

With this discussed, here is the moment computation for the Gaussian matrices:

\index{Gaussian matrix}

\begin{theorem}
Given a sequence of Gaussian random matrices
$$Z_N\in M_N(L^\infty(X))$$
having independent $G_t$ variables as entries, for some fixed $t>0$, we have
$$M_k\left(\frac{Z_N}{\sqrt{N}}\right)\simeq t^{|k|/2}|\mathcal{NC}_2(k)|$$
for any colored integer $k=\circ\bullet\bullet\circ\ldots\,$, in the $N\to\infty$ limit.
\end{theorem}

\begin{proof}
This is something standard, which can be done as follows:

\medskip

(1) We fix $N\in\mathbb N$, and we let $Z=Z_N$. Let us first compute the trace of $Z^k$. With $k=k_1\ldots k_s$, and with the convention $(ij)^\circ=ij,(ij)^\bullet=ji$, we have:
\begin{eqnarray*}
Tr(Z^k)
&=&Tr(Z^{k_1}\ldots Z^{k_s})\\
&=&\sum_{i_1=1}^N\ldots\sum_{i_s=1}^N(Z^{k_1})_{i_1i_2}(Z^{k_2})_{i_2i_3}\ldots(Z^{k_s})_{i_si_1}\\
&=&\sum_{i_1=1}^N\ldots\sum_{i_s=1}^N(Z_{(i_1i_2)^{k_1}})^{k_1}(Z_{(i_2i_3)^{k_2}})^{k_2}\ldots(Z_{(i_si_1)^{k_s}})^{k_s}
\end{eqnarray*}

(2) Next, we rescale our variable $Z$ by a $\sqrt{N}$ factor, as in the statement, and we also replace the usual trace by its normalized version, $tr=Tr/N$. Our formula becomes:
$$tr\left(\left(\frac{Z}{\sqrt{N}}\right)^k\right)=\frac{1}{N^{s/2+1}}\sum_{i_1=1}^N\ldots\sum_{i_s=1}^N(Z_{(i_1i_2)^{k_1}})^{k_1}(Z_{(i_2i_3)^{k_2}})^{k_2}\ldots(Z_{(i_si_1)^{k_s}})^{k_s}$$

Thus, the moment that we are interested in is given by:
$$M_k\left(\frac{Z}{\sqrt{N}}\right)=\frac{1}{N^{s/2+1}}\sum_{i_1=1}^N\ldots\sum_{i_s=1}^N\int_X(Z_{(i_1i_2)^{k_1}})^{k_1}(Z_{(i_2i_3)^{k_2}})^{k_2}\ldots(Z_{(i_si_1)^{k_s}})^{k_s}$$

(3) Let us apply now the Wick formula, from Theorem 15.15. We conclude that the moment that we are interested in is given by the following formula:
\begin{eqnarray*}
&&M_k\left(\frac{Z}{\sqrt{N}}\right)\\
&=&\frac{t^{s/2}}{N^{s/2+1}}\sum_{i_1=1}^N\ldots\sum_{i_s=1}^N\#\left\{\pi\in\mathcal P_2(k)\Big|\pi\leq\ker\left((i_1i_2)^{k_1},(i_2i_3)^{k_2},\ldots,(i_si_1)^{k_s}\right)\right\}\\
&=&t^{s/2}\sum_{\pi\in\mathcal P_2(k)}\frac{1}{N^{s/2+1}}\#\left\{i\in\{1,\ldots,N\}^s\Big|\pi\leq\ker\left((i_1i_2)^{k_1},(i_2i_3)^{k_2},\ldots,(i_si_1)^{k_s}\right)\right\}
\end{eqnarray*}

(4) Our claim now is that in the $N\to\infty$ limit the combinatorics of the above sum simplifies, with only the noncrossing partitions contributing to the sum, and with each of them contributing precisely with a 1 factor, so that we will have, as desired:
\begin{eqnarray*}
M_k\left(\frac{Z}{\sqrt{N}}\right)
&=&t^{s/2}\sum_{\pi\in\mathcal P_2(k)}\Big(\delta_{\pi\in NC_2(k)}+O(N^{-1})\Big)\\
&\simeq&t^{s/2}\sum_{\pi\in\mathcal P_2(k)}\delta_{\pi\in NC_2(k)}\\
&=&t^{s/2}|\mathcal{NC}_2(k)|
\end{eqnarray*}

(5) In order to prove this, the first observation is that when $k$ is not uniform, in the sense that it contains a different number of $\circ$, $\bullet$ symbols, we have $\mathcal P_2(k)=\emptyset$, and so:
$$M_k\left(\frac{Z}{\sqrt{N}}\right)=t^{s/2}|\mathcal{NC}_2(k)|=0$$

(6) Thus, we are left with the case where $k$ is uniform. Let us examine first the case where $k$ consists of an alternating sequence of $\circ$ and $\bullet$ symbols, as follows:
$$k=\underbrace{\circ\bullet\circ\bullet\ldots\ldots\circ\bullet}_{2p}$$

In this case it is convenient to relabel our multi-index $i=(i_1,\ldots,i_s)$, with $s=2p$, in the form $(j_1,l_1,j_2,l_2,\ldots,j_p,l_p)$. With this done, our moment formula becomes:
$$M_k\left(\frac{Z}{\sqrt{N}}\right)
=t^p\sum_{\pi\in\mathcal P_2(k)}\frac{1}{N^{p+1}}\#\left\{j,l\in\{1,\ldots,N\}^p\Big|\pi\leq\ker\left(j_1l_1,j_2l_1,j_2l_2,\ldots,j_1l_p\right)\right\}$$

Now observe that, with $k$ being as above, we have an identification $\mathcal P_2(k)\simeq S_p$, obtained in the obvious way. With this done too, our moment formula becomes:
$$M_k\left(\frac{Z}{\sqrt{N}}\right)
=t^p\sum_{\pi\in S_p}\frac{1}{N^{p+1}}\#\left\{j,l\in\{1,\ldots,N\}^p\Big|j_r=j_{\pi(r)+1},l_r=l_{\pi(r)},\forall r\right\}$$

(7) We are now ready to do our asymptotic study, and prove the claim in (4). Let indeed $\gamma\in S_p$ be the full cycle, which is by definition the following permutation:
$$\gamma=(1 \, 2 \, \ldots \, p)$$

In terms of $\gamma$, the conditions $j_r=j_{\pi(r)+1}$ and $l_r=l_{\pi(r)}$ found above read:
$$\gamma\pi\leq\ker j\quad,\quad 
\pi\leq\ker l$$

Counting the number of free parameters in our moment formula, we obtain:
$$M_k\left(\frac{Z}{\sqrt{N}}\right)
=\frac{t^p}{N^{p+1}}\sum_{\pi\in S_p}N^{|\pi|+|\gamma\pi|}
=t^p\sum_{\pi\in S_p}N^{|\pi|+|\gamma\pi|-p-1}$$

(8) Now by Proposition 15.16 the last exponent is $\leq 0$, with equality precisely when $\pi\in S_p$ comes from a noncrossing partition. Thus, with $N\to\infty$, we obtain:
$$M_k\left(\frac{Z}{\sqrt{N}}\right)\simeq t^p|\mathcal{NC}_2(k)|$$

But this finishes the proof for the exponents $k$ which are alternating, and the case where $k$ is an arbitrary uniform exponent is similar, by permuting everything.
\end{proof}

We can reformulate what we found in Theorem 15.17, in the following way:

\begin{theorem}
Given a sequence of complex Gaussian matrices $Z_N\in M_N(L^\infty(X))$ having independent $G_t$ variables as entries, with $t>0$, we have
$$\frac{Z_N}{\sqrt{N}}\sim\Gamma_t$$
in the $N\to\infty$ limit, with the limiting law being defined via its colored moments by
$$M_k(\Gamma_t)=\sum_{\pi\in\mathcal{NC}_2(k)}t^{|\pi|}$$
and with this being a law which is not a usual measure.
\end{theorem}

\begin{proof}
According to what we found in Theorem 15.17, the limiting law $\Gamma_t$, that of the sequence of matrices $(Z_N/\sqrt{N})_{N\in\mathbb N}$, has moments as follows:
$$M_k(\Gamma_t)=t^{|k|/2}|\mathcal{NC}_2(k)|$$

Now since the number of blocks of a pairing $\pi\in\mathcal P_2(k)$ is always $|k|/2$, independently of the pairing, we are led to the moment formula for $\Gamma_t$ in the statement.
\end{proof}

Next, we can make the following speculation on the law $\Gamma_t$ that we found:

\begin{speculation}
In view of the moment formulae for $G_t$ and $\Gamma_t$, namely
$$M_k(G_t)=\sum_{\pi\in\mathcal P_2(k)}t^{|\pi|}\quad,\quad 
M_k(\Gamma_t)=\sum_{\pi\in\mathcal{NC}_2(k)}t^{|\pi|}$$
and assuming that the crossings correspond somehow to commutation
$$\slash\hskip-2.1mm\backslash\quad\longleftrightarrow\quad ab=ba$$
we can say that $\Gamma_t$ is the free analogue of the complex normal law $G_t$.
\end{speculation}

And wild speculation this is, but if you have something better, at this stage of things, let me know. As some sort of rationale for this, we can regard $\slash\hskip-2.1mm\backslash$ as a Feynman diagram, which will make the incoming particles $a,b$ switch position at the arrival:
$$\xymatrix@R=35pt@C=25pt{
a\ar[dr]&b\ar[dl]\\
b&a}$$

Now since obviously nothing happened, I mean there was no collision or scattering or something, we conclude that having $\slash\hskip-2.1mm\backslash$ in our theory means to assume $ab=ba$. Which sounds sweet, in the hope that I convinced you guys of the validity of my theory. 

\bigskip

More seriously, Voiculescu's free probability theory \cite{vo1}, \cite{vo2}, \cite{vo3}, \cite{vdn} provides a conceptual approach to all this, with the free analogue of $G_t$ being indeed the above law $\Gamma_t$, called Voiculescu circular law. As for the above $\slash\hskip-2.1mm\backslash\longleftrightarrow ab=ba$ speculation, this can be made into something rigorous too, and we refer here to \cite{bco}, \cite{bsp}, \cite{bpa}, \cite{spe}.

\bigskip

As a second comment now on Theorem 15.18, which is advanced too, we have:

\begin{comment}
Given a family of sequences of complex Gaussian matrices, 
$$Z^i_N\in M_N(L^\infty(X))\quad,\quad i\in I$$
having independent $G_t$ variables as entries, with $t>0$, the rescaled sequences
$$\frac{Z^i_N}{\sqrt{N}}\in M_N(L^\infty(X))\quad,\quad i\in I$$
become with $N\to\infty$ circular, each following the Voiculescu law $\Gamma_t$, and free.
\end{comment}

To be more precise, what is new here with respect to Theorem 15.18, reformulated as in Speculation 15.19, taking into account the fact that we are now dealing with a family of Gaussian matrices, is the word ``free'' at the end, and with this coming from a famous theorem of Voiculescu \cite{vo3}. So, this is how probabilistic freeness basically appears, in the random matrix context, and for more on this, have a look at \cite{agz}, \cite{msp}, \cite{vdn}.

\section*{15c. Wigner matrices}
 
Regarding now the Wigner matrices, we have here the following result, coming as a consequence of Theorem 15.17, via some simple algebraic manipulations:

\begin{theorem}
Given a sequence of Wigner random matrices
$$Z_N\in M_N(L^\infty(X))$$
having independent $G_t$ variables as entries, with $t>0$, up to $Z_N=Z_N^*$, we have
$$M_k\left(\frac{Z_N}{\sqrt{N}}\right)\simeq t^{k/2}|NC_2(k)|$$
for any integer $k\in\mathbb N$, in the $N\to\infty$ limit.
\end{theorem}

\begin{proof}
We have two possible proofs here, as follows:

\medskip

(1) The formula in the statement can be certainly established via a direct computation based on the Wick formula, similar to that from the proof of Theorem 15.17. We will leave the calculations here, featuring no particular difficulties, as an exercise.

\medskip

(2) However, the best is to deduce this result from Theorem 15.17 itself. Indeed, we know from there that for Gaussian matrices $Y_N\in M_N(L^\infty(X))$ we have the following formula, valid for any colored integer $K=\circ\bullet\bullet\circ\ldots\,$, in the $N\to\infty$ limit:
$$M_K\left(\frac{Y_N}{\sqrt{N}}\right)\simeq t^{|K|/2}|\mathcal{NC}_2(K)|$$

We deduce from this that we have the following formula for the moments of the matrices $Re(Y_N)$, with respect to usual exponents, $k\in\mathbb N$:
\begin{eqnarray*}
M_k\left(\frac{Re(Y_N)}{\sqrt{N}}\right)
&=&2^{-k}\cdot M_k\left(\frac{Y_N}{\sqrt{N}}+\frac{Y_N^*}{\sqrt{N}}\right)\\
&=&2^{-k}\sum_{|K|=k}M_K\left(\frac{Y_N}{\sqrt{N}}\right)\\
&\simeq&2^{-k}\sum_{|K|=k}t^{k/2}|\mathcal{NC}_2(K)|\\
&=&2^{-k}\cdot t^{k/2}\cdot 2^{k/2}|\mathcal{NC}_2(k)|\\
&=&2^{-k/2}\cdot t^{k/2}|NC_2(k)|
\end{eqnarray*}

In terms of the matrices $Z_N=\sqrt{2}Re(Y_N)$, which are of Wigner type, this reads:
$$M_k\left(\frac{Z_N}{\sqrt{2N}}\right)\simeq 2^{-k/2}\cdot t^{k/2}|NC_2(k)|$$

Thus, we are led to the moment formula in the statement.
\end{proof}

We can reformulate what we found in Theorem 15.21, in the following way:

\begin{theorem}
Given a sequence of Wigner random matrices $Z_N\in M_N(L^\infty(X))$ having independent $G_t$ variables as entries, with $t>0$, up to $Z_N=Z_N^*$, we have
$$\frac{Z_N}{\sqrt{N}}\sim\gamma_t$$
in the $N\to\infty$ limit, with the limiting law being defined via its moments by
$$M_k(\gamma_t)=\sum_{\pi\in NC_2(k)}t^{|\pi|}$$
and with this law being this time a usual real probability measure.
\end{theorem}

\begin{proof}
According to what we found in Theorem 15.21, the limiting law $\gamma_t$, that of the sequence of matrices $(Z_N/\sqrt{N})_{N\in\mathbb N}$, has moments as follows:
$$M_k(\gamma_t)=t^{k/2}|NC_2(k)|$$

Now since the number of blocks of a pairing $\pi\in P_2(k)$ is always $k/2$, independently of the pairing, we are led to the moment formula for $\gamma_t$ in the statement.
\end{proof}

Next, in analogy with our previous Speculation 15.19, we can formulate:

\begin{speculation}
In view of the moment formulae for $g_t$ and $\gamma_t$, namely
$$M_k(g_t)=\sum_{\pi\in P_2(k)}t^{|\pi|}\quad,\quad 
M_k(\gamma_t)=\sum_{\pi\in NC_2(k)}t^{|\pi|}$$
we can say that $\gamma_t$ is the free analogue of the normal law $g_t$.
\end{speculation}

To be more precise, this is something quite self-explanatory, which can be made into a theorem, by free probability \cite{vo1}, \cite{vo2}, \cite{vo3}, \cite{vdn}. And with the further comment that we have as well asymptotic freeness in this setting, along the lines of Comment 15.20. 

\bigskip

With this discussed, what is next? Solving I guess the moment problem for the Catalan numbers, with a parameter $t>0$ added, as above, and this can be done as follows:

\begin{theorem}
The centered measure having $M_{2k}=t^kC_k$ as even moments is
$$\gamma_t=\frac{1}{2\pi t}\sqrt{4t-x^2}\,dx$$
the Wigner semicircle law on $[-2\sqrt{t},2\sqrt{t}]$.
\end{theorem}

\begin{proof}
We have two possible proofs for this, as follows:

\medskip

(1) Stieltjes inversion. With $M_{2k}=t^kC_k$, the Cauchy transform is: 
$$G(\xi)
=\xi^{-1}\sum_{k=0}^\infty\frac{1}{k+1}\binom{2k}{k}(t\xi^{-2})^k
=\frac{\xi-\sqrt{\xi^2-4t}}{2t}$$

Now let us apply the Stieltjes inversion formula. The study here goes as follows:

\medskip

-- According to the general philosophy of the Stieltjes formula, the first term, namely $\xi/2t$, which is ``trivial'', will not contribute to the density. 

\medskip

-- As for the second term, which is something non-trivial, this will contribute to the density, the rule here being that the square root $\sqrt{\xi^2-4t}$ will be replaced by the ``dual'' square root $\sqrt{4t-x^2}\,dx$, and that we have to multiply everything by $-1/\pi$. 

\medskip

As a conclusion, by Stieltjes inversion we obtain the following density, as claimed:
$$d\mu(x)
=-\frac{1}{\pi}\cdot-\frac{\sqrt{4t-x^2}}{2t}\,dx
=\frac{1}{2\pi t}\sqrt{4t-x^2}dx$$

(2) Cheating. Assuming that a cat or a good physicist suggested you to look at the semicircle law $\gamma_t$, you can prove that this law $\gamma_t$ does indeed the job, with the moment computation being as follows, using the change of variables $x=2\sqrt{t}\cos s$:
\begin{eqnarray*}
M_{2k}
&=&\frac{1}{\pi t}\int_0^{2\sqrt{t}}x^{2k}\sqrt{4t-x^2}\,dx\\
&=&\frac{1}{\pi t}\int_0^{\pi/2}(2\sqrt{t}\cos s)^{2k}\sqrt{4t-4t\cos^2s}\cdot2\sqrt{t}\sin s\,ds\\
&=&\frac{4^{k+1}t^k}{\pi}\int_0^{\pi/2}\cos^{2s}t\sin^2s\,ds\\
&=&\frac{4^{k+1}t^k}{\pi}\cdot\frac{\pi}{2}\cdot\frac{(2k)!!2!!}{(2k+3)!!}\\
&=&2\cdot 4^kt^k\cdot\frac{(2k)!/2^kk!}{2^{k+1}(k+1)!}\\
&=&t^kC_k
\end{eqnarray*}

To be more precise, we have used here at the end the Wallis formula from chapter 7. As for the odd moments, these all vanish, because $\gamma_t$ is an even function. 
\end{proof}

Good news, we can now formulate the Wigner theorem, as follows:

\begin{theorem}
Given a sequence of Wigner random matrices 
$$Z_N\in M_N(L^\infty(X))$$
having independent $G_t$ variables as entries, with $t>0$, up to $Z_N=Z_N^*$, we have
$$\frac{Z_N}{\sqrt{N}}\sim\gamma_t$$
in the $N\to\infty$ limit, where $\gamma_t$ is the Wigner semicircle law.
\end{theorem}

\begin{proof}
This comes indeed from Theorems 15.22 and 15.24.
\end{proof}

So long for the Wigner theorem. Of course, this was just a beginning, and far more things can be said, regarding the precise nature of the convergence, and the fluctuations, with the problematics here being highly non-trivial, and still subject to research. For more on all this, you can consult the various random matrix books and articles referenced at the end of this book. And by the way, do not forget to learn, either from Wigner himself \cite{wig}, or from a book like Mehta \cite{meh}, how Wigner came upon these matrices, in relation with the physics of the heavy atoms. First class science, all this.

\section*{15d. Marchenko-Pastur}

Let us discuss now the Wishart matrices, which are the positive analogues of the Wigner matrices. Quite surprisingly, the computation here leads to the Catalan numbers, but not in the same way as for the Wigner matrices, the result being as follows:

\begin{theorem}
Given a sequence of complex Wishart matrices
$$Z_N=Y_NY_N^*\in M_N(L^\infty(X))$$
with $Y_N$ being $N\times N$ complex Gaussian of parameter $t>0$, we have
$$M_k\left(\frac{Z_N}{N}\right)\simeq t^kC_k$$
for any exponent $k\in\mathbb N$, in the $N\to\infty$ limit.
\end{theorem}

\begin{proof}
There are several possible proofs for this result, as follows:

\medskip

(1) A first method is by using the formula that we have in Theorem 15.17, for the Gaussian matrices $Y_N$. Indeed, we know from there that we have the following formula, valid for any colored integer $K=\circ\bullet\bullet\circ\ldots\,$, in the $N\to\infty$ limit:
$$M_K\left(\frac{Y_N}{\sqrt{N}}\right)\simeq t^{|K|/2}|\mathcal{NC}_2(K)|$$

With $K=\circ\bullet\circ\bullet\ldots\,$, alternating word of length $2k$, with $k\in\mathbb N$, this gives:
$$M_k\left(\frac{Y_NY_N^*}{N}\right)\simeq t^k|\mathcal{NC}_2(K)|$$

Thus, in terms of the Wishart matrix $Z_N=Y_NY_N^*$ we have, for any $k\in\mathbb N$:
$$M_k\left(\frac{Z_N}{N}\right)\simeq t^k|\mathcal{NC}_2(K)|$$

The point now is that, by doing some combinatorics, we have:
$$|\mathcal{NC}_2(K)|=|NC_2(2k)|=C_k$$

Thus, we are led to the formula in the statement.

\medskip

(2) A second method, that we will explain now as well, is by proving the result directly, starting from definitions. The matrix entries of our matrix $Z=Z_N$ are given by:
$$Z_{ij}=\sum_{r=1}^NY_{ir}\bar{Y}_{jr}$$

Thus, the normalized traces of powers of $Z$ are given by the following formula:
\begin{eqnarray*}
tr(Z^k)
&=&\frac{1}{N}\sum_{i_1=1}^N\ldots\sum_{i_k=1}^NZ_{i_1i_2}Z_{i_2i_3}\ldots Z_{i_ki_1}\\
&=&\frac{1}{N}\sum_{i_1=1}^N\ldots\sum_{i_k=1}^N\sum_{r_1=1}^N\ldots\sum_{r_k=1}^NY_{i_1r_1}\bar{Y}_{i_2r_1}Y_{i_2r_2}\bar{Y}_{i_3r_2}\ldots Y_{i_kr_k}\bar{Y}_{i_1r_k}
\end{eqnarray*}

By rescaling now $Z$ by a $1/N$ factor, as in the statement, we obtain:
$$tr\left(\left(\frac{Z}{N}\right)^k\right)=\frac{1}{N^{k+1}}\sum_{i_1=1}^N\ldots\sum_{i_k=1}^N\sum_{r_1=1}^N\ldots\sum_{r_k=1}^NY_{i_1r_1}\bar{Y}_{i_2r_1}Y_{i_2r_2}\bar{Y}_{i_3r_2}\ldots Y_{i_kr_k}\bar{Y}_{i_1r_k}$$

By using now the Wick rule, we obtain the following formula for the moments, with $K=\circ\bullet\circ\bullet\ldots\,$, alternating word of lenght $2k$, and with $I=(i_1r_1,i_2r_1,\ldots,i_kr_k,i_1r_k)$:
\begin{eqnarray*}
M_k\left(\frac{Z}{N}\right)
&=&\frac{t^k}{N^{k+1}}\sum_{i_1=1}^N\ldots\sum_{i_k=1}^N\sum_{r_1=1}^N\ldots\sum_{r_k=1}^N\#\left\{\pi\in\mathcal P_2(K)\Big|\pi\leq\ker(I)\right\}\\
&=&\frac{t^k}{N^{k+1}}\sum_{\pi\in\mathcal P_2(K)}\#\left\{i,r\in\{1,\ldots,N\}^k\Big|\pi\leq\ker(I)\right\}
\end{eqnarray*}

In order to compute this quantity, we use the standard bijection $\mathcal P_2(K)\simeq S_k$. By identifying the pairings $\pi\in\mathcal P_2(K)$ with their counterparts $\pi\in S_k$, we obtain:
\begin{eqnarray*}
M_k\left(\frac{Z}{N}\right)
&=&\frac{t^k}{N^{k+1}}\sum_{\pi\in S_k}\#\left\{i,r\in\{1,\ldots,N\}^k\Big|i_s=i_{\pi(s)+1},r_s=r_{\pi(s)},\forall s\right\}
\end{eqnarray*}

As before in our study of the Gaussian matrices, let $\gamma=(1 \, 2 \, \ldots \, k)\in S_k$ be the full cycle. By counting the number of free parameters in our moment formula, we get:
$$M_k\left(\frac{Z}{N}\right)
=t^k\sum_{\pi\in S_k}N^{|\pi|+|\gamma\pi|-k-1}$$

The point now is that the last exponent is $\leq 0$, with equality precisely when $\pi$ comes from a noncrossing partition. Thus we obtain, in the $N\to\infty$ limit:
$$M_k\left(\frac{Z}{N}\right)\simeq t^kC_k$$

We are therefore led to the conclusion in the statement.
\end{proof}

We can reformulate what we found in Theorem 15.26, in the following way:

\begin{theorem}
Given a sequence of complex Wishart matrices
$$Z_N=Y_NY_N^*\in M_N(L^\infty(X))$$
with $Y_N$ being $N\times N$ complex Gaussian of parameter $t>0$, we have
$$\frac{Z_N}{tN}\sim\pi_1$$
in the $N\to\infty$ limit, with the limiting law being defined via its moments by
$$M_k(\pi_1)=|NC(k)|$$
and with this law being a usual real positive probability measure.
\end{theorem}

\begin{proof}
This follows indeed from our formula from Theorem 15.26.
\end{proof}

Next, in analogy with our previous study of the Gaussian and Wigner matrices, we have the following speculation, corresponding to a theorem in free probability:

\begin{speculation}
In view of the moment formulae for $p_1$ and $\pi_1$, namely
$$M_k(p_1)=|P(k)|\quad,\quad 
M_k(\pi_1)=|NC(k)|$$
we can say that $\pi_1$ is the free analogue of the Poisson law $p_1$.
\end{speculation}

Finally, by solving the moment problem for the Catalan numbers, we are led to the following result, regarding the asymptotic laws of the complex Wishart matrices: 

\begin{theorem}
Given a sequence of complex Wishart matrices
$$Z_N=Y_NY_N^*\in M_N(L^\infty(X))$$
with $Y_N$ being $N\times N$ complex Gaussian of parameter $t>0$, we have
$$\frac{Z_N}{tN}\sim\frac{1}{2\pi}\sqrt{4x^{-1}-1}\,dx$$
with $N\to\infty$, with the limiting measure being the Marchenko-Pastur law $\pi_1$.
\end{theorem}

\begin{proof}
We have three possible methods here, all working:

\medskip

(1) Stieltjes inversion. This is something that we know from chapter 13, based on:
$$M(z)=\frac{1-\sqrt{1-4z}}{2z}\quad ,\quad G(\xi)=\frac{1-\sqrt{1-4\xi^{-1}}}{2}$$

Indeed, by Stieltjes inversion we obtain the following density, as stated:
$$d\mu(x)
=-\frac{1}{\pi}\cdot-\frac{\sqrt{4x^{-1}-1}}{2}\,dx
=\frac{1}{2\pi}\sqrt{4x^{-1}-1}\,dx$$

(2) Squaring. Alternatively, we can capture the density from that of $\gamma_1$, by using:
$$f\sim\gamma_1\ \implies\ f^2\sim\pi_1$$

To be more precise, the computation here, which is straightforward, is as follows:
\begin{eqnarray*}
E(\varphi(f^2))
&=&\frac{1}{2\pi}\int_{-2}^2\varphi(x^2)\sqrt{4-x^2}\,dx\\
&=&\frac{1}{\pi}\int_0^2\varphi(x^2)\sqrt{4-x^2}\,dx\\
&=&\frac{1}{\pi}\int_0^4\varphi(y)\sqrt{4-y}\cdot\frac{dy}{2\sqrt{y}}\\
&=&\frac{1}{2\pi}\int_0^4\varphi(y)\sqrt{4y^{-1}-1}\,dy
\end{eqnarray*}

(3) Cheating. Again this is standard, because the moments of the law $\pi_1$ in the statement can be computed with the change of variable $x=4\cos^2t$, as follows:
\begin{eqnarray*}
M_k
&=&\frac{1}{2\pi}\int_0^4x^k\sqrt{4x^{-1}-1}dx\\
&=&\frac{1}{2\pi}\int_0^{\pi/2}(4\cos^2t)^k\cdot\frac{\sin t}{\cos t}\cdot2\cos t\sin t\,dt\\
&=&\frac{4^{k+1}}{\pi}\int_0^{\pi/2}\cos^{2k}t\sin^2t\,dt\\
&=&\frac{4^{k+1}}{\pi}\cdot\frac{\pi}{2}\cdot\frac{(2k)!!2!!}{(2k+3)!!}\\
&=&2\cdot 4^k\cdot\frac{(2k)!/2^kk!}{2^{k+1}(k+1)!}\\
&=&C_k
\end{eqnarray*}

Thus, one way or another, we are led to the density in the statement. 
\end{proof}

As a comment now, what we have in Theorem 15.29 looks good at $t=1$, but what we have at general $t>0$ is not very satisfactory, because the limiting law $\pi_1$ stays the same, modulo a trivial rescaling. The correct result at $t>0$ is in fact as follows:

\index{Marchenko-Pastur}

\begin{theorem}
Given a sequence of general complex Wishart matrices
$$Z_N=Y_NY_N^*\in M_N(L^\infty(X))$$
with $Y_N$ being $N\times M$ complex Gaussian of parameter $1$, we have
$$M_k\left(\frac{Z}{N}\right)
\simeq\sum_{\pi\in NC(k)}t^{|\pi|}$$
for any exponent $k\in\mathbb N$, in the $M=tN\to\infty$ limit.
\end{theorem}

\begin{proof}
This is again something which is very standard, as follows:

\medskip

(1) In order to prove the formula in the statement, we can proceed as usual, by using the Wick formula. The matrix entries of our Wishart matrix $Z=Z_N$ are given by:
$$Z_{ij}=\sum_{r=1}^MY_{ir}\bar{Y}_{jr}$$

Thus, the normalized traces of powers of $Z$ are given by the following formula:
\begin{eqnarray*}
tr(Z^k)
&=&\frac{1}{N}\sum_{i_1=1}^N\ldots\sum_{i_k=1}^NZ_{i_1i_2}Z_{i_2i_3}\ldots Z_{i_ki_1}\\
&=&\frac{1}{N}\sum_{i_1=1}^N\ldots\sum_{i_k=1}^N\sum_{r_1=1}^M\ldots\sum_{r_k=1}^MY_{i_1r_1}\bar{Y}_{i_2r_1}Y_{i_2r_2}\bar{Y}_{i_3r_2}\ldots Y_{i_kr_k}\bar{Y}_{i_1r_k}
\end{eqnarray*}

By rescaling now $Z$ by a $1/N$ factor, as in the statement, we obtain:
$$tr\left(\left(\frac{Z}{N}\right)^k\right)=\frac{1}{N^{k+1}}\sum_{i_1=1}^N\ldots\sum_{i_k=1}^N\sum_{r_1=1}^M\ldots\sum_{r_k=1}^MY_{i_1r_1}\bar{Y}_{i_2r_1}Y_{i_2r_2}\bar{Y}_{i_3r_2}\ldots Y_{i_kr_k}\bar{Y}_{i_1r_k}$$

(2) By using now the Wick rule, we obtain the following formula for the moments, with $K=\circ\bullet\circ\bullet\ldots\,$, alternating word of lenght $2k$, and $I=(i_1r_1,i_2r_1,\ldots,i_kr_k,i_1r_k)$:
\begin{eqnarray*}
M_k\left(\frac{Z}{N}\right)
&=&\frac{1}{N^{k+1}}\sum_{i_1=1}^N\ldots\sum_{i_k=1}^N\sum_{r_1=1}^M\ldots\sum_{r_k=1}^M\#\left\{\pi\in\mathcal P_2(K)\Big|\pi\leq\ker I\right\}\\
&=&\frac{1}{N^{k+1}}\sum_{\pi\in\mathcal P_2(K)}\#\left\{i\in\{1,\ldots,N\}^k,r\in\{1,\ldots,M\}^k\Big|\pi\leq\ker I\right\}
\end{eqnarray*}

(3) In order to compute this quantity, we use the standard bijection $\mathcal P_2(K)\simeq S_k$. By identifying the pairings $\pi\in\mathcal P_2(K)$ with their counterparts $\pi\in S_k$, we obtain:
\begin{eqnarray*}
M_k\left(\frac{Z}{N}\right)
&=&\frac{1}{N^{k+1}}\sum_{\pi\in S_k}\#\left\{i\in\{1,\ldots,N\}^k,r\in\{1,\ldots,M\}^k\Big|i_s=i_{\pi(s)+1},r_s=r_{\pi(s)}\right\}
\end{eqnarray*}

With $\gamma=(1 \, 2 \, \ldots \, k)\in S_k$ being the full cycle, by counting the number of free parameters in our expectation formula, as before for other random matrices, we obtain:
$$M_k\left(\frac{Z}{N}\right)
=\frac{1}{N^{k+1}}\sum_{\pi\in S_k}N^{|\gamma\pi|}M^{|\pi|}
=\sum_{\pi\in S_k}N^{|\gamma\pi|-k-1}M^{|\pi|}$$

(4) Now by using the same arguments as in the case $M=N$, from the proof of Theorem 15.26, we conclude that in the $M=tN\to\infty$ limit the permutations $\pi\in S_k$ which matter are those coming from noncrossing partitions, and so that we have:
$$M_k\left(\frac{Z}{N}\right)
\simeq\sum_{\pi\in NC(k)}N^{-|\pi|}M^{|\pi|}
=\sum_{\pi\in NC(k)}t^{|\pi|}$$

Thus, we are led to the conclusion in the statement.
\end{proof}

We can reformulate what we found in Theorem 15.30, in the following way:

\begin{theorem}
Given a sequence of general complex Wishart matrices
$$Z_N=Y_NY_N^*\in M_N(L^\infty(X))$$
with $Y_N$ being $N\times M$ complex Gaussian of parameter $1$, we have
$$\frac{Z_N}{N}\sim\pi_t$$
in the $M=tN\to\infty$ limit, with the limiting law being given by
$$M_k(\pi_t)=\sum_{\pi\in NC(k)}t^{|\pi|}$$
and with this law being a usual real positive probability measure.
\end{theorem}

\begin{proof}
This follows indeed from our formula from Theorem 15.30.
\end{proof}

Next, in analogy with our previous speculations, we can formulate:

\begin{speculation}
In view of the moment formulae for $p_t$ and $\pi_t$, namely
$$M_k(p_t)=\sum_{\pi\in P(k)}t^{|\pi|}\quad,\quad 
M_k(\pi_t)=\sum_{\pi\in NC(k)}t^{|\pi|}$$
we can say that $\pi_t$ is the free analogue of the Poisson law $p_t$.
\end{speculation}

To be more precise, this is something quite self-explanatory, which can be made into a theorem, by free probability \cite{vo1}, \cite{vo2}, \cite{vo3}, \cite{vdn}. And with the further comment that we have as well asymptotic freeness in this setting, at $t\geq1$. See \cite{agz}, \cite{msp}, \cite{vdn}.

\bigskip

Finally, by solving the moment problem, we are led to the following result: 

\begin{theorem}
Given a sequence of general complex Wishart matrices
$$Z_N=Y_NY_N^*\in M_N(L^\infty(X))$$
with $Y_N$ being $N\times M$ complex Gaussian of parameter $1$, we have
$$\frac{Z_N}{N}\sim\max(1-t,0)\delta_0+\frac{\sqrt{4t-(x-1-t)^2}}{2\pi x}\,dx$$
with $M=tN\to\infty$, with the limiting measure being the Marchenko-Pastur law $\pi_t$.
\end{theorem}

\begin{proof}
This is something quite tricky, the idea being as follows:

\medskip

(1) Let us first review the $t=1$ case. Here the moment generating function is:
$$M(z)=\sum_{k\geq0}C_kz^k$$

For $\pi\in NC(k+1)$ we can consider the last leg of the first block, and make a cut at right of it. This gives a decomposition of $\pi$ into $2$ partitions in $NC$, and we get:
$$C_{k+1}=\sum_{k=k_0+k_1}C_{k_0}C_{k_1}$$

By multiplying by $z^{k+1}$, then summing over $k$, we obtain the following equation:
$$M-1=zM^2$$

But this gives $M$, then $G$, then $\pi_1$, as explained in the proof of Theorem 15.29.

\medskip

(2) At $t>0$ now, let us denote by $F_{kb}$ the number of partitions in $NC(k)$ having $b$ blocks, and we set $F_{kb}=0$ for other integer values of $k,b$. Also, let us make the convention that, in what follows, all sums will be over integer indices $\geq 0$. With these conventions, the moment generating function for our measure is given by the following formula:
$$M(z)=\sum_{kb}F_{kb}z^kt^b$$

Now observe that the recurrence formula for the numbers $C_k$ in (1) becomes:
$$\sum_bF_{k+1,b}=\sum_{k=k_0+k_1}\sum_{b_0,b_1}F_{k_0b_0}F_{k_1b_1}$$

In this formula, each term contributes to $F_{k+1,b}$ with a contribution $b=\sum b_i$, except for those of the form $F_{00}F_{k_1b_1}$, which contribute to $F_{k+1,b+1}$. Thus, we have:
$$F_{k+1,b}
=\sum_{k=k_0+k_1}\sum_{b=b_0+b_1}F_{k_0b_0}F_{k_1b_1}
+\sum_{k=k_0+k_1}F_{k_1,b-1}
-\sum_{k=k_0+k_1}F_{k_1,b-1}$$

But this gives the following formula for the polynomials $P_k=\sum_bF_{kb}t^b$:
$$P_{k+1}=\sum_{k=k_0+k_1}P_{k_0}P_{k_1}+(t-1)\sum_{k=k_0+k_1}P_{k_1}$$

In terms of $M(z)=\sum_kP_kz^k$, we obtain the following equation:
$$M-1=zM^2+(t-1)zM$$

(3) But with this, we can go ahead like we did before at $t=1$. Our equation is:
$$zM^2+((t-1)z-1)M+1=0$$

Thus, the moment generating series is given by the following formula:
$$M(z)=\frac{(1-t)z+1-\sqrt{((t-1)z-1)^2-4z}}{2z}$$

Equivalently, the Cauchy transform is given by the following formula:
\begin{eqnarray*}
G(\xi)
&=&\frac{1-t+\xi-\sqrt{(t-1-\xi)^2-4\xi}}{2\xi}\\
&=&\frac{1-t+\xi-\sqrt{(\xi-1-t)^2-4t}}{2\xi}
\end{eqnarray*}

Now by Stieltjes inversion we obtain the following density:
$$d\mu(x)
=\max(1-t,0)\delta_0+\frac{\sqrt{4t-(x-1-t)^2}}{2\pi x}\,dx$$

Summarizing, we have obtained the density in the statement of $\pi_t$.
\end{proof}

So long for the basic theory of the random matrices. For more on all this, random matrices and free probability, check the various books and articles referenced below.

\section*{15e. Exercises}

This was a quite exciting chapter, and as exercises on this, we have:

\begin{exercise}
Compute the spectral measures of various matrices, of your choice.
\end{exercise}

\begin{exercise}
Learn more about $C^*$-algebras, including the GNS theorem.
\end{exercise}

\begin{exercise}
Importantly, learn as well about the von Neumann algebras.
\end{exercise}

\begin{exercise}
Learn as well some other approaches to the spectral measures.
\end{exercise}

\begin{exercise}
Why is the asymptotic law of Gaussian matrices a ``circular law''?
\end{exercise}

\begin{exercise}
Why is the Wigner semicircle law a ``free Gaussian law''?
\end{exercise}

\begin{exercise}
Why is the Marchenko-Pastur law a ``free Poisson law''?
\end{exercise}

\begin{exercise}
Learn also, from Aubrun and others, about block modifications.
\end{exercise}

As bonus exercise, learn of course more random matrices, as much as you can.

\chapter{Free integration}

\section*{16a. Free tori} 

Welcome to freeness. We will be interested here in developing free geometry and analysis, with the hope that all this might be related to physics, at very small scales, quarks and below. The idea being very simple, based on old findings of Heisenberg and others, if it is true indeed that the more you zoom down, the more commutativity dissapears, then, logically, if you zoom hard enough, things will become free.

\bigskip

So, what is free? The simplest free object in mathematics is the free group $F_N$:

\index{free group}

\begin{definition}
The free group $F_N$ is the infinite group
$$F_N=\left<g_1,\ldots,g_N\Big|\,\emptyset\right>$$
generated by $N$ variables $g_1,\ldots,g_N$, with no relations between them.
\end{definition}

This might look a bit abstract, but no worries, $F_N$ has some interesting mathematics, coming right away, if you have some knowledge in discrete groups, and know how to look for interesting questions. For instance if you want to draw the Cayley graph of $F_N$, whose vertices are the elements of $F_N$, with edges $h-k$ drawn when $h=g_i^{\pm1}k$ for some $i$, you will end up with an interesting picture, which at $N=2$ looks like this:
$$
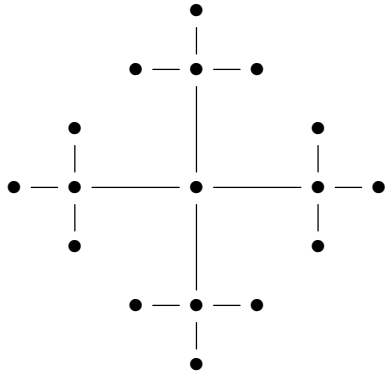
\xymatrix@R=10pt@C=10pt{
&&&\bullet\ar@{-}[d]\\
&&\bullet\ar@{-}[r]&\bullet\ar@{-}[dd]\ar@{-}[r]&\bullet\\
&\bullet\ar@{-}[d]&&&&\bullet\ar@{-}[d]\\
\bullet\ar@{-}[r]&\bullet\ar@{-}[rr]&&\bullet\ar@{-}[rr]\ar@{-}[dd]&&\bullet\ar@{-}[r]&\bullet\\
&\bullet\ar@{-}[u]&&&&\bullet\ar@{-}[u]\\
&&\bullet\ar@{-}[r]&\bullet\ar@{-}[d]\ar@{-}[r]&\bullet\\
&&&\bullet
}$$

And this type of graph certainly has interesting mathematics. One good question for instance is that of computing the number of length $2k$ loops based at the root. Another question, which is in fact equivalent, via moments, is that of computing the Kesten measure of $F_N$, which is that of the following variable in the group algebra of $F_N$:
$$\chi=g_1+\ldots+g_N$$

All this looks very good, we most likely have here our first object of free geometry. In order now to formally understand this, let us recall the following formula, with $\mathbb T_N=\mathbb T^N$ being the usual torus, and with $\mathbb Z^N$ being the free abelian group:
$$\mathbb T_N=\widehat{\mathbb Z^N}$$

Thus, getting back now to our free group $F_N$, which is the free analogue of $\mathbb Z^N$, it is in fact its dual $\widehat{F_N}$ which is a free manifold, and more specifically the free analogue of $\mathbb T^N$. Which is a nice finding, so let us formulate our conclusions as follows:

\index{free torus}

\begin{definition}
The free torus $\mathbb T_N^+$ is the dual of the free group $F_N$,
$$\mathbb T_N^+=\widehat{F_N}$$
in analogy with the fact that the usual torus $\mathbb T_N=\mathbb T^N$ appears as
$$\mathbb T_N=\widehat{\mathbb Z^N}$$
with on the right the group $\mathbb Z^N$ being the free abelian group.
\end{definition}

It is of course possible to formulate things more precisely, and we will be back to this in a moment, but before that, isn't this a bit too abstract? But the point here is that no, at the level of questions to be solved, these remain the same, as for instance the computation of the Kesten measure, which is now a ``function'' on the free torus:
$$\chi\in C(\mathbb T_N^+)$$

In fact, this function is the main character of $\mathbb T_N^+$, regarded as a compact quantum group, and so our Kesten problem suddenly becomes something very conceptual, namely the computation of the law of the main character of $\mathbb T_N^+$. Which is very nice.

\bigskip

Before getting into details regarding all this, recall that $\mathbb R^N$ is as interesting as $\mathbb C^N$. So, let us formulate as well the real version of Definition 16.2, as follows:

\index{free cube}

\begin{definition}
The free real torus, or free cube, $T_N^+$ is the dual 
$$T_N^+=\widehat{L_N}$$
of the group $L_N=F_N/<g_i^2=1>$, in analogy with the fact that the usual cube is
$$T_N=\widehat{\mathbb Z_2^N}$$
with on the right the group $\mathbb Z_2^N$ being the free real abelian group.
\end{definition}

Here the ``real'' at the end stands for the fact that the generators must satisfy the real reflection condition $g^2=1$. As for the fact that ``real torus = cube'', as stated, this needs some thinking, and in the hope that, after such thinking, you will agree with me that there is indeed a standard torus inside $\mathbb R^N$, and that is the unit cube.

\bigskip

Summarizing, all this sounds good, we have a beginning of free geometry, both real and complex, worth developing, by knowing at least what the torus of each theory is. In practice now, at the level of details, in order to talk about $\mathbb T_N^+=\widehat{F_N}$ and $T_N^+=\widehat{L_N}$ we need an extension of the usual Pontrjagin duality theory for the abelian groups, and this is best done via operator algebras, and the related notion of compact quantum group. 

\bigskip

In order to understand all this, let us start with operator algebras. As explained in chapter 15, in relation with our random matrix needs there, the starting point is:

\begin{definition}
A $C^*$-algebra is a complex algebra $A$, having:
\begin{enumerate}
\item A norm $a\to||a||$, making it a Banach algebra.

\item An involution $a\to a^*$, satisfying $||aa^*||=||a||^2$.
\end{enumerate}
\end{definition}

As basic examples, we have $B(H)$ itself, as well as any norm closed $*$-subalgebra $A\subset B(H)$. It is possible to prove that any $C^*$-algebra appears in this way, but we will not need in what follows this deep result, called GNS theorem after Gelfand, Naimark, Segal. So, let us simply agree that, by definition, the $C^*$-algebras $A$ are some sort of ``generalized operator algebras'', and their elements $a\in A$ can be thought of as being some kind of ``generalized operators'', on some Hilbert space which is not present.

\bigskip

In practice, this vague idea is all that we need. Indeed, by taking some inspiration from linear algebra, we can emulate spectral theory in our setting, as follows:

\begin{proposition}
Given $a\in A$, define its spectrum as being the set
$$\sigma(a)=\left\{\lambda\in\mathbb C\Big|a-\lambda\not\in A^{-1}\right\}$$
and its spectral radius $\rho(a)$ as the radius of the smallest centered disk containing $\sigma(a)$.
\begin{enumerate}
\item The spectrum of a norm one element is in the unit disk.

\item The spectrum of a unitary element $(a^*=a^{-1}$) is on the unit circle. 

\item The spectrum of a self-adjoint element ($a=a^*$) consists of real numbers. 

\item The spectral radius of a normal element ($aa^*=a^*a$) is equal to its norm.
\end{enumerate}
\end{proposition}

\begin{proof}
This is something that we know well from chapter 12 in the case $A\subset B(H)$, and from chapter 15 in general, and we refer to the material there for details.
\end{proof}

Generally speaking, Proposition 16.5 is all you need for doing some further operator algebras. As a main application, that we already met in chapter 15, we have:

\begin{theorem}[Gelfand]
If $X$ is a compact space, the algebra $C(X)$ of continuous functions $f:X\to\mathbb C$ is a commutative $C^*$-algebra, with structure as follows:
\begin{enumerate}
\item The norm is the usual sup norm, $||f||=\sup_{x\in X}|f(x)|$.

\item The involution is the usual involution, $f^*(x)=\overline{f(x)}$.
\end{enumerate}
Conversely, any commutative $C^*$-algebra is of the form $C(X)$, with its ``spectrum'' $X=Spec(A)$ appearing as the space of characters $\chi :A\to\mathbb C$.
\end{theorem}

\begin{proof}
Given a commutative $C^*$-algebra $A$, we can define indeed $X$ to be the set of characters $\chi :A\to\mathbb C$, with the topology making continuous all the evaluation maps $ev_a:\chi\to\chi(a)$. Then $X$ is a compact space, and $a\to ev_a$ is a morphism of algebras:
$$ev:A\to C(X)$$

We first prove that $ev$ is involutive. We use the following formula:
$$a=\frac{a+a^*}{2}-i\cdot\frac{i(a-a^*)}{2}$$

Thus it is enough to prove the equality $ev_{a^*}=ev_a^*$ for self-adjoint elements $a$. But this is the same as proving that $a=a^*$ implies that $ev_a$ is a real function, which is in turn true, because $ev_a(\chi)=\chi(a)$ is an element of $\sigma(a)$, contained in $\mathbb R$. So, claim proved. Also, since $A$ is commutative, each element is normal, so $ev$ is isometric:
$$||ev_a||=\rho(a)=||a||$$

It remains to prove that $ev$ is surjective. But this follows from the Stone-Weierstrass theorem, because $ev(A)$ is a closed subalgebra of $C(X)$, which separates the points.
\end{proof}

The Gelfand theorem suggests formulating the following definition:

\index{quantum space}
\index{compact quantum space}

\begin{definition}
Given a $C^*$-algebra $A$, not necessarily commutative, we write
$$A=C(X)$$
and call the abstract object $X$ a ``compact quantum space''.
\end{definition}

This might look quite revolutionary, but in practice, this definition changes nothing to what we have been doing so far, namely studying the $C^*$-algebras. So, we will keep studying the $C^*$-algebras, but by using the above fancy quantum space terminology. For instance whenever we have a morphism $\Phi:A\to B$, we will write $A=C(X),B=C(Y)$, and rather speak of the corresponding morphism $\phi:Y\to X$. And so on.

\bigskip

Let us also mention that, at a closer look, there are in fact some problems with the above definition, coming from amenability and others. More on this later.

\bigskip

Now that we have our notion of quantum spaces, good time to get back towards Definitions 16.2 and 16.3. In order to understand what that free tori are, we will need:

\index{group algebra}
\index{Pontrjagin dual}

\begin{theorem}
Let $\Gamma$ be a discrete group, and consider the complex group algebra $\mathbb C[\Gamma]$, with involution given by the fact that all group elements are unitaries, $g^*=g^{-1}$.
\begin{enumerate}
\item The maximal $C^*$-seminorm on $\mathbb C[\Gamma]$ is a $C^*$-norm, and the closure of $\mathbb C[\Gamma]$ with respect to this norm is a $C^*$-algebra, denoted $C^*(\Gamma)$.

\item When $\Gamma$ is abelian, we have an isomorphism $C^*(\Gamma)\simeq C(G)$, where $G=\widehat{\Gamma}$ is its Pontrjagin dual, formed by the characters $\chi:\Gamma\to\mathbb T$.
\end{enumerate}
\end{theorem}

\begin{proof}
All this is very standard, the idea being as follows:

\medskip

(1) In order to prove the result, we must find a $*$-algebra embedding $\mathbb C[\Gamma]\subset B(H)$, with $H$ being a Hilbert space. For this purpose, consider the space $H=l^2(\Gamma)$, having $\{h\}_{h\in\Gamma}$ as orthonormal basis. Our claim is that we have an embedding, as follows:
$$\pi:\mathbb C[\Gamma]\subset B(H)\quad,\quad 
\pi(g)(h)=gh$$

Indeed, since $\pi(g)$ maps the basis $\{h\}_{h\in\Gamma}$ into itself, this operator is well-defined, bounded, and is an isometry. It is also clear from the formula $\pi(g)(h)=gh$ that $g\to\pi(g)$ is a morphism of algebras, and since this morphism maps the unitaries $g\in\Gamma$ into isometries, this is a morphism of $*$-algebras. Finally, the faithfulness of $\pi$ is clear.

\medskip

(2) Since $\Gamma$ is abelian, the corresponding group algebra $A=C^*(\Gamma)$ is commutative. Thus, we can apply the Gelfand theorem, and we obtain $A=C(X)$, with:
$$X=Spec(A)$$

But the spectrum $X=Spec(A)$, consisting of the characters $\chi:C^*(\Gamma)\to\mathbb C$, can be identified with the Pontrjagin dual $G=\widehat{\Gamma}$, and this gives the result.
\end{proof}

The above result suggests the following definition:

\begin{definition}
Given a discrete group $\Gamma$, the compact quantum space $G$ given by
$$C(G)=C^*(\Gamma)$$
is called abstract dual of $\Gamma$, and is denoted $G=\widehat{\Gamma}$.
\end{definition}

Good news, this definition is exactly what we need, in order to understand the meaning of Definitions 16.2 and 16.3. To be more precise, we have the following result:

\begin{proposition}
The basic tori are all group duals, as follows,
$$\xymatrix@R=16.2mm@C=16.2mm{
T_N^+\ar[r]&\mathbb T_N^+\\
T_N\ar[r]\ar[u]&\mathbb T_N\ar[u]
}
\qquad
\xymatrix@R=8mm@C=15mm{\\ =}
\qquad
\xymatrix@R=15mm@C=15mm{
\widehat{L_N}\ar[r]&\widehat{F_N}\\
\mathbb Z_2^N\ar[r]\ar[u]&\mathbb T^N\ar[u]
}$$
where $F_N=\mathbb Z^{*N}$ is the free group on $N$ generators, and $L_N=\mathbb Z_2^{*N}$ is its real version.
\end{proposition}

\begin{proof}
The basic tori appear indeed as group duals, and together with the Fourier transform identifications from Theorem 16.8 (2), this gives the result.
\end{proof}

Moving ahead, now that we have our formalism, we can start developing free geometry. As a first objective, we would like to better understand the relation between the classical and free tori. In order to discuss this, let us introduce the following notion:

\index{classical version}
\index{commutator ideal}

\begin{definition}
Given a compact quantum space $X$, its classical version is the usual compact space $X_{class}\subset X$ obtained by dividing $C(X)$ by its commutator ideal:
$$C(X_{class})=C(X)/I\quad,\quad 
I=<[a,b]>$$
In this situation, we also say that $X$ appears as a ``liberation'' of $X$.
\end{definition}

In other words, the space $X_{class}$ appears as the Gelfand spectrum of the commutative $C^*$-algebra $C(X)/I$. Observe in particular that $X_{class}$ is indeed a classical space.

\bigskip

In relation now with our tori, we have the following result:

\begin{theorem}
We have inclusions between the various tori, as follows,
$$\xymatrix@R=16.2mm@C=16.2mm{
T_N^+\ar[r]&\mathbb T_N^+\\
T_N\ar[r]\ar[u]&\mathbb T_N\ar[u]
}$$
and the free tori on top appear as liberations of the tori on the bottom.
\end{theorem}

\begin{proof}
This is indeed clear from definitions, because commutativity of a group algebra means precisely that the group in question is abelian.
\end{proof}

\section*{16b. Free spheres}

In order to extend now the free geometries that we have, real and complex, let us begin with the spheres. Following \cite{bgo}, let us formulate:

\index{free sphere}

\begin{definition}
We have free real and complex spheres, defined via
$$C(S^{N-1}_{\mathbb R,+})=C^*\left(x_1,\ldots,x_N\Big|x_i=x_i^*,\sum_ix_i^2=1\right)$$
$$C(S^{N-1}_{\mathbb C,+})=C^*\left(x_1,\ldots,x_N\Big|\sum_ix_ix_i^*=\sum_ix_i^*x_i=1\right)$$
where the symbol $C^*$ stands for universal enveloping $C^*$-algebra.
\end{definition}

Here the fact that these algebras are indeed well-defined comes from the following estimate, which shows that the biggest $C^*$-norms on these $*$-algebras are bounded:
$$||x_i||^2
=||x_ix_i^*||
\leq\left|\left|\sum_ix_ix_i^*\right|\right|
=1$$

As a first result now, regarding the above free spheres, we have:

\begin{theorem}
We have embeddings of compact quantum spaces, as follows,
$$\xymatrix@R=15mm@C=15mm{
S^{N-1}_{\mathbb R,+}\ar[r]&S^{N-1}_{\mathbb C,+}\\
S^{N-1}_\mathbb R\ar[r]\ar[u]&S^{N-1}_\mathbb C\ar[u]
}$$
and the spaces on top appear as liberations of the spaces on the bottom.
\end{theorem}

\begin{proof}
The first assertion, regarding the inclusions, comes from the fact that at the level of the associated $C^*$-algebras, we have surjective maps, as follows:
$$\xymatrix@R=15mm@C=15mm{
C(S^{N-1}_{\mathbb R,+})\ar[d]&C(S^{N-1}_{\mathbb C,+})\ar[d]\ar[l]\\
C(S^{N-1}_\mathbb R)&C(S^{N-1}_\mathbb C)\ar[l]
}$$

For the second assertion, we must establish the following isomorphisms, where the symbol $C^*_{comm}$ stands for ``universal commutative $C^*$-algebra generated by'':
$$C(S^{N-1}_\mathbb R)=C^*_{comm}\left(x_1,\ldots,x_N\Big|x_i=x_i^*,\sum_ix_i^2=1\right)$$
$$C(S^{N-1}_\mathbb C)=C^*_{comm}\left(x_1,\ldots,x_N\Big|\sum_ix_ix_i^*=\sum_ix_i^*x_i=1\right)$$

It is enough to establish the second isomorphism. So, consider the second universal commutative $C^*$-algebra $A$ constructed above. Since the standard coordinates on $S^{N-1}_\mathbb C$ satisfy the defining relations for $A$, we have a quotient map of as follows:
$$A\to C(S^{N-1}_\mathbb C)$$

Conversely, let us write $A=C(S)$, by using the Gelfand theorem. Then $x_1,\ldots,x_N$ become in this way true coordinates, providing us with an embedding as follows:
$$S\subset\mathbb C^N$$

Also, the quadratic relations become $\sum_i|x_i|^2=1$, so we have $S\subset S^{N-1}_\mathbb C$. Thus, we have a quotient map $C(S^{N-1}_\mathbb C)\to A$, as desired, and this gives all the results.
\end{proof}

By using the free spheres constructed above, we can now formulate:

\begin{definition}
A real algebraic manifold $X\subset S^{N-1}_{\mathbb C,+}$ is a closed quantum subspace defined, at the level of the corresponding $C^*$-algebra, by a formula of type
$$C(X)=C(S^{N-1}_{\mathbb C,+})\Big/\Big<f_i(x_1,\ldots,x_N)=0\Big>$$
for certain family of noncommutative polynomials, as follows:
$$f_i\in\mathbb C<x_1,\ldots,x_N>$$
We denote by $\mathcal C(X)$ the $*$-subalgebra of $C(X)$ generated by the coordinates $x_1,\ldots,x_N$. 
\end{definition}

As a basic example here, we have the free real sphere $S^{N-1}_{\mathbb R,+}$. The classical spheres $S^{N-1}_\mathbb C,S^{N-1}_\mathbb R$, and their real submanifolds, are covered as well by this formalism. At the level of the general theory, we have the following version of the Gelfand theorem:

\begin{theorem}
If $X\subset S^{N-1}_{\mathbb C,+}$ is an algebraic manifold, as above, we have
$$X_{class}=\left\{x\in S^{N-1}_\mathbb C\Big|f_i(x_1,\ldots,x_N)=0\right\}$$
and $X$ appears as a liberation of $X_{class}$.
\end{theorem}

\begin{proof}
This is something that we already met, in the context of the free spheres. In general, the proof is similar, by using the Gelfand theorem. Indeed, if we denote by $X_{class}'$ the manifold constructed in the statement, then we have a quotient map of $C^*$-algebras as follows, mapping standard coordinates to standard coordinates:
$$C(X_{class})\to C(X_{class}')$$

Conversely now, from $X\subset S^{N-1}_{\mathbb C,+}$ we obtain $X_{class}\subset S^{N-1}_\mathbb C$. Now since the relations defining $X_{class}'$ are satisfied by $X_{class}$, we obtain an inclusion $X_{class}\subset X_{class}'$. Thus, at the level of algebras of continuous functions, we have a quotient map of $C^*$-algebras as follows, mapping standard coordinates to standard coordinates:
$$C(X_{class}')\to C(X_{class})$$

Thus, we have constructed a pair of inverse morphisms, and we are done.
\end{proof}

Finally, once again at the level of the general theory, we have:

\begin{definition}
We agree to identify two real algebraic submanifolds $X,Y\subset S^{N-1}_{\mathbb C,+}$ when we have a $*$-algebra isomorphism between $*$-algebras of coordinates
$$f:\mathcal C(Y)\to\mathcal C(X)$$
mapping standard coordinates to standard coordinates.
\end{definition}

We will see later the reasons for making this convention, coming from amenability. Now back to the tori, as constructed before, we can see that these are examples of algebraic manifolds, in the sense of Definition 16.15. In fact, we have the following result:

\begin{theorem}
The four main quantum spheres produce the main quantum tori
$$\xymatrix@R=15mm@C=15mm{
S^{N-1}_{\mathbb R,+}\ar[r]&S^{N-1}_{\mathbb C,+}\\
S^{N-1}_\mathbb R\ar[r]\ar[u]&S^{N-1}_\mathbb C\ar[u]
}\qquad
\xymatrix@R=8mm@C=15mm{\\ \to}
\qquad
\xymatrix@R=16mm@C=18mm{
T_N^+\ar[r]&\mathbb T_N^+\\
T_N\ar[r]\ar[u]&\mathbb T_N\ar[u]
}$$
via the formula $T=S\cap\mathbb T_N^+$, with the intersection being taken inside $S^{N-1}_{\mathbb C,+}$.
\end{theorem}

\begin{proof}
This comes from the above results, the situation being as follows:

\medskip

(1) Free complex case. Here the formula in the statement reads:
$$\mathbb T_N^+=S^{N-1}_{\mathbb C,+}\cap\mathbb T_N^+$$

But this is something trivial, because we have $\mathbb T_N^+\subset S^{N-1}_{\mathbb C,+}$.

\medskip

(2) Free real case. Here the formula in the statement reads:
$$T_N^+=S^{N-1}_{\mathbb R,+}\cap\mathbb T_N^+$$

But this is clear as well, the real version of $\mathbb T_N^+$ being $T_N^+$.

\medskip

(3) Classical complex case. Here the formula in the statement reads:
$$\mathbb T_N=S^{N-1}_\mathbb C\cap\mathbb T_N^+$$

But this is clear as well, the classical version of $\mathbb T_N^+$ being $\mathbb T_N$.

\medskip

(4) Classical real case. Here the formula in the statement reads:
$$T_N=S^{N-1}_\mathbb R\cap\mathbb T_N^+$$

But this follows by intersecting the formulae from the proof of (2) and (3).
\end{proof}

\section*{16c. Free rotations} 

In order to better understand the structure of the free spheres $S^{N-1}_{\mathbb R,+},S^{N-1}_{\mathbb C,+}$, we need to talk about free rotations. Following Woronowicz \cite{wor}, let us start with:

\begin{definition}
A Woronowicz algebra is a $C^*$-algebra $A$, given with a unitary matrix $u\in M_N(A)$ whose coefficients generate $A$, such that the formulae
$$\Delta(u_{ij})=\sum_ku_{ik}\otimes u_{kj}\quad,\quad
\varepsilon(u_{ij})=\delta_{ij}\quad,\quad
S(u_{ij})=u_{ji}^*$$
define morphisms of $C^*$-algebras $\Delta:A\to A\otimes A$, $\varepsilon:A\to\mathbb C$, $S:A\to A^{opp}$.
\end{definition}

We say that $A$ is cocommutative when $\Sigma\Delta=\Delta$, where $\Sigma(a\otimes b)=b\otimes a$ is the flip. We have the following result, which justifies the terminology and axioms:

\begin{theorem}
The following are Woronowicz algebras:
\begin{enumerate}
\item $C(G)$, with $G\subset U_N$ compact Lie group. Here the structural maps are:
$$\Delta(\varphi)=[(g,h)\to \varphi(gh)]\quad,\quad 
\varepsilon(\varphi)=\varphi(1)\quad,\quad 
S(\varphi)=[g\to\varphi(g^{-1})]$$

\item $C^*(\Gamma)$, with $F_N\to\Gamma$ finitely generated group. Here the structural maps are:
$$\Delta(g)=g\otimes g\quad,\quad 
\varepsilon(g)=1\quad,\quad
S(g)=g^{-1}$$
\end{enumerate}
Moreover, we obtain in this way all the commutative/cocommutative algebras.
\end{theorem}

\begin{proof}
This is something very standard, the idea being as follows:

\medskip

(1) Given $G\subset U_N$, we can set $A=C(G)$, which is a Woronowicz algebra, together with the matrix $u=(u_{ij})$ formed by coordinates of $G$, given by:
$$g=\begin{pmatrix}
u_{11}(g)&\ldots&u_{1N}(g)\\
\vdots&&\vdots\\
u_{N1}(g)&\ldots&u_{NN}(g)
\end{pmatrix}$$

Conversely, if $(A,u)$ is a commutative Woronowicz algebra, by using the Gelfand theorem we can write $A=C(X)$, with $X$ being a certain compact space. The coordinates $u_{ij}$ give then an embedding $X\subset M_N(\mathbb C)$, and since the matrix $u=(u_{ij})$ is unitary we actually obtain an embedding $X\subset U_N$, and finally by using the maps $\Delta,\varepsilon,S$ we conclude that our compact subspace $X\subset U_N$ is in fact a compact Lie group, as desired.

\medskip

(2) Consider a finitely generated group $F_N\to\Gamma$. We can set $A=C^*(\Gamma)$, which is by definition the completion of the complex group algebra $\mathbb C[\Gamma]$, with involution given by $g^*=g^{-1}$, for any $g\in\Gamma$, with respect to the biggest $C^*$-norm, and we obtain a Woronowicz algebra, together with the diagonal matrix formed by the generators of $\Gamma$:
$$u=\begin{pmatrix}
g_1&&0\\
&\ddots&\\
0&&g_N
\end{pmatrix}$$

Conversely, if $(A,u)$ is a cocommutative Woronowicz algebra, the Peter-Weyl theory of Woronowicz, to be explained below, shows that the irreducible corepresentations of $A$ are all 1-dimensional, and form a group $\Gamma$, and so we have $A=C^*(\Gamma)$, as desired.
\end{proof}

The above result makes it quite clear that what we have in Definition 16.19 is some sort of joint definition for the compact and discrete quantum groups. In order to further comment on this, let us go back to the formulae in Definition 16.19, namely:
$$\Delta(u_{ij})=\sum_ku_{ik}\otimes u_{kj}\quad,\quad
\varepsilon(u_{ij})=\delta_{ij}\quad,\quad
S(u_{ij})=u_{ji}^*$$

The morphisms $\Delta,\varepsilon,S$ are called comultiplication, counit and antipode, and they have the following properties, which are something very familiar in abstract algebra:

\begin{theorem}
Let $(A,u)$ be a Woronowicz algebra.
\begin{enumerate} 
\item $\Delta,\varepsilon$ satisfy the usual axioms for a comultiplication and a counit, namely:
\begin{eqnarray*}
(\Delta\otimes id)\Delta&=&(id\otimes \Delta)\Delta\\
(\varepsilon\otimes id)\Delta&=&(id\otimes\varepsilon)\Delta=id
\end{eqnarray*}

\item $S$ satisfies the antipode axiom, on the $*$-subalgebra generated by entries of $u$: 
$$m(S\otimes id)\Delta=m(id\otimes S)\Delta=\varepsilon(.)1$$

\item In addition, the square of the antipode is the identity, $S^2=id$.
\end{enumerate}
\end{theorem}

\begin{proof}
Observe first that the result holds indeed in the case where the algebra $A$ is commutative. Indeed, by using Theorem 16.20 we can write:
$$\Delta=m^t\quad,\quad
\varepsilon=u^t\quad,\quad
S=i^t$$

The above 3 conditions come then by transposition from the basic 3 group theory conditions satisfied by $m,u,i$, which are as follows, with $\delta(g)=(g,g)$:
$$m(m\times id)=m(id\times m)$$
$$m(id\times u)=m(u\times id)=id$$
$$m(id\times i)\delta=m(i\times id)\delta=1$$

Observe that $S^2=id$ is satisfied as well, coming from $i^2=id$, which is a consequence of the group axioms. Finally, observe that the result holds also, trivially, when $A$ is cocommutative, again by Theorem 16.20. In general now, the proof goes as follows:

\medskip

(1) We have indeed the following computation:
$$(\Delta\otimes id)\Delta(u_{ij})
=\sum_l\Delta(u_{il})\otimes u_{lj}
=\sum_{kl}u_{ik}\otimes u_{kl}\otimes u_{lj}$$

On the other hand, we have as well the following computation:
$$(id\otimes\Delta)\Delta(u_{ij})
=\sum_ku_{ik}\otimes\Delta(u_{kj})
=\sum_{kl}u_{ik}\otimes u_{kl}\otimes u_{lj}$$

The proof for the counit axiom is quite similar. We first have:
$$(id\otimes\varepsilon)\Delta(u_{ij})
=\sum_ku_{ik}\otimes\varepsilon(u_{kj})
=u_{ij}$$

On the other hand, we have as well the following computation:
$$(\varepsilon\otimes id)\Delta(u_{ij})
=\sum_k\varepsilon(u_{ik})\otimes u_{kj}
=u_{ij}$$

(2) By using the fact that the matrix $u=(u_{ij})$ is unitary, we obtain:
\begin{eqnarray*}
m(id\otimes S)\Delta(u_{ij})
&=&\sum_ku_{ik}S(u_{kj})\\
&=&\sum_ku_{ik}u_{jk}^*\\
&=&(uu^*)_{ij}\\
&=&\delta_{ij}
\end{eqnarray*}

Similarly, we have the following computation, which gives the antipode axiom:
\begin{eqnarray*}
m(S\otimes id)\Delta(u_{ij})
&=&\sum_kS(u_{ik})u_{kj}\\
&=&\sum_ku_{ki}^*u_{kj}\\
&=&(u^*u)_{ij}\\
&=&\delta_{ij}
\end{eqnarray*}

(3) Finally, the formula $S^2=id$ holds as well on the generators, and we are done.
\end{proof}

Summarizing, the Woronowicz algebras appear to have nice properties. In view of Theorem 16.20 and Theorem 16.21, we can formulate the following definition:

\begin{definition}
Given a Woronowicz algebra $A$, we formally write
$$A=C(G)=C^*(\Gamma)$$
and call $G$ compact quantum group, and $\Gamma$ discrete quantum group.
\end{definition}

In relation with this, there are actually some analytic subtleties, coming from amenability, so our objects must be divided by a certain equivalence relation, for everything to work fine. To be more precise, we agree to write $(A,u)=(B,v)$ when there is a $*$-algebra isomorphism as follows, mapping standard coordinates to standard coordinates:
$$<u_{ij}>\simeq<v_{ij}>\quad,\quad u_{ij}\to v_{ij}$$

Moving ahead now, let us call now corepresentation of $A$ any unitary matrix $v\in M_n(A)$ satisfying the same conditions as those satisfied by $u$, namely:
$$\Delta(v_{ij})=\sum_kv_{ik}\otimes v_{kj}\quad,\quad
\varepsilon(v_{ij})=\delta_{ij}\quad,\quad
S(v_{ij})=v_{ji}^*$$

These corepresentations can be thought of as corresponding representations of the underlying compact quantum group $G$. Following Woronowicz \cite{wor}, we have:

\begin{theorem}
Any Woronowicz algebra has a unique Haar integration functional, 
$$\left(\int_G\otimes id\right)\Delta=\left(id\otimes\int_G\right)\Delta=\int_G(.)1$$
which can be constructed by starting with any faithful positive form $\varphi\in A^*$, and setting
$$\int_G=\lim_{n\to\infty}\frac{1}{n}\sum_{k=1}^n\varphi^{*k}$$
where $\phi*\psi=(\phi\otimes\psi)\Delta$. Moreover, for any corepresentation $v\in M_n(\mathbb C)\otimes A$ we have
$$\left(id\otimes\int_G\right)v=P$$
where $P$ is the orthogonal projection onto $Fix(v)=\{\xi\in\mathbb C^n|v\xi=\xi\}$.
\end{theorem}

\begin{proof}
This can be done in 3 steps, as follows:

\medskip

(1) Given $\varphi\in A^*$, our claim is that the following limit converges, for any $a\in A$:
$$\int_\varphi a=\lim_{n\to\infty}\frac{1}{n}\sum_{k=1}^n\varphi^{*k}(a)$$

Indeed, by linearity we can assume that $a$ is the coefficient of corepresentation, $a=(\tau\otimes id)v$. But in this case, an elementary computation shows that we have the following formula, where $P_\varphi$ is the orthogonal projection onto the $1$-eigenspace of $(id\otimes\varphi)v$:
$$\left(id\otimes\int_\varphi\right)v=P_\varphi$$

(2) Since $v\xi=\xi$ implies $[(id\otimes\varphi)v]\xi=\xi$, we have $P_\varphi\geq P$, where $P$ is the orthogonal projection onto the space $Fix(v)=\{\xi\in\mathbb C^n|v\xi=\xi\}$. The point now is that when $\varphi\in A^*$ is faithful, by using a positivity trick, one can prove that we have $P_\varphi=P$. Thus our linear form $\int_\varphi$ is independent of $\varphi$, and is given on coefficients $a=(\tau\otimes id)v$ by:
$$\left(id\otimes\int_\varphi\right)v=P$$

(3) With the above formula in hand, the left and right invariance of $\int_G=\int_\varphi$ is clear on coefficients, and so in general, and this gives all the assertions.
\end{proof}

Consider now the dense $*$-subalgebra $\mathcal A\subset A$ generated by the coefficients of $u$, that we met after Definition 16.22, and endow it with the following scalar product:
$$<a,b>=\int_Gab^*$$

Still following Woronowicz \cite{wor}, we have the following key result:

\begin{theorem}
We have the following Peter-Weyl type results:
\begin{enumerate}
\item Any corepresentation decomposes as a sum of irreducible corepresentations.

\item Each irreducible corepresentation appears inside a certain $u^{\otimes k}$.

\item $\mathcal A=\bigoplus_{v\in Irr(A)}M_{\dim(v)}(\mathbb C)$, the summands being pairwise orthogonal.

\item The characters of irreducible corepresentations form an orthonormal system.
\end{enumerate}
\end{theorem}

\begin{proof}
All these results are very standard, the idea being as follows:

\medskip

(1) Given $v\in M_n(A)$, its intertwiner algebra $End(v)=\{T\in M_n(\mathbb C)|Tv=vT\}$ is a finite dimensional $C^*$-algebra, and so decomposes as $End(v)=M_{n_1}(\mathbb C)\oplus\ldots\oplus M_{n_r}(\mathbb C)$. But this gives a decomposition of type $v=v_1+\ldots+v_r$, as desired.

\medskip

(2) Consider indeed the Peter-Weyl corepresentations, $u^{\otimes k}$ with $k$ colored integer, defined by $u^{\otimes\emptyset}=1$, $u^{\otimes\circ}=u$, $u^{\otimes\bullet}=\bar{u}$ and multiplicativity. The coefficients of these corepresentations span the dense algebra $\mathcal A$, and by using (1), this gives the result.

\medskip

(3) Here the direct sum decomposition, which is technically a $*$-coalgebra isomorphism, follows from (2). As for the second assertion, this follows from the fact that $(id\otimes\int_G)v$ is the orthogonal projection $P_v$ onto the space $Fix(v)$, for any corepresentation $v$.

\medskip

(4) Let us define indeed the character of $v\in M_n(A)$ to be the matrix trace, $\chi_v=Tr(v)$. Since this character is a coefficient of $v$, the orthogonality assertion follows from (3). As for the norm 1 claim, this follows once again from $(id\otimes\int_G)v=P_v$. 
\end{proof}

So long for the basic theory of the compact and discrete quantum groups. For more on all this, we refer to Woronowicz's paper \cite{wor}, which remains a classic.

\bigskip

Good news, getting back now to what we wanted to do in this chapter, namely free geometry, we can talk about free rotations. Indeed, following Wang \cite{wan}, we have:

\begin{theorem}
The following universal algebras are Woronowicz algebras,
$$C(O_N^+)=C^*\left((u_{ij})_{i,j=1,\ldots,N}\Big|u=\bar{u},u^t=u^{-1}\right)$$
$$C(U_N^+)=C^*\left((u_{ij})_{i,j=1,\ldots,N}\Big|u^*=u^{-1},u^t=\bar{u}^{-1}\right)$$
so the underlying spaces $O_N^+,U_N^+$ are compact quantum groups.
\end{theorem}

\begin{proof}
This follows from the elementary fact that if a matrix $u=(u_{ij})$ is orthogonal or biunitary, then so must be the following matrices:
$$u^\Delta_{ij}=\sum_ku_{ik}\otimes u_{kj}\quad,\quad 
u^\varepsilon_{ij}=\delta_{ij}\quad,\quad
u^S_{ij}=u_{ji}^*$$

Thus, we can indeed define morphisms $\Delta,\varepsilon,S$ as in Definition 16.19, by using the universal properties of $C(O_N^+)$, $C(U_N^+)$, and this gives the result.
\end{proof}

The basic properties of these beasts $O_N^+,U_N^+$ can be summarized as follows:

\begin{theorem}
The quantum groups $O_N^+,U_N^+$ have the following properties:
\begin{enumerate}
\item The closed subgroups $G\subset U_N^+$ are exactly the $N\times N$ compact quantum groups. As for the closed subgroups $G\subset O_N^+$, these are those satisfying $u=\bar{u}$.

\item We have liberation embeddings $O_N\subset O_N^+$ and $U_N\subset U_N^+$, obtained by dividing the algebras $C(O_N^+),C(U_N^+)$ by their respective commutator ideals.

\item We have as well embeddings $\widehat{L}_N\subset O_N^+$ and $\widehat{F}_N\subset U_N^+$, where $L_N$ is the free product of $N$ copies of $\mathbb Z_2$, and where $F_N$ is the free group on $N$ generators.
\end{enumerate}
\end{theorem}

\begin{proof}
All these assertions are elementary, as follows:

\medskip

(1) This is clear from definitions, with the remark that, in the context of Definition 16.19, the formula $S(u_{ij})=u_{ji}^*$ shows that the matrix $\bar{u}$ must be unitary too. 

\medskip

(2) This follows from the Gelfand theorem, which shows that we have presentation results for the algebras $C(O_N),C(U_N)$, similar to those in Theorem 16.25, but with the commutativity between the standard coordinates and their adjoints added:
\begin{eqnarray*}
C(O_N)&=&C^*_{comm}\left((u_{ij})_{i,j=1,\ldots,N}\Big|u=\bar{u},u^t=u^{-1}\right)\\
C(U_N)&=&C^*_{comm}\left((u_{ij})_{i,j=1,\ldots,N}\Big|u^*=u^{-1},u^t=\bar{u}^{-1}\right)
\end{eqnarray*}

(3) This follows indeed from (1) and from Theorem 16.25, with the remark that with $u=diag(g_1,\ldots,g_N)$, the condition $u=\bar{u}$ is equivalent to $g_i^2=1$, for any $i$.
\end{proof}

In order to discuss now the relation with the spheres, which can only come via some sort of ``isometric actions'', let us start with the following standard fact:

\begin{proposition}
Given a closed subset $X\subset S^{N-1}_\mathbb C$, the formula
$$G(X)=\left\{U\in U_N\Big|U(X)=X\right\}$$
defines a compact group of unitary matrices, or isometries, called affine isometry group of $X$. For the spheres $S^{N-1}_\mathbb R,S^{N-1}_\mathbb C$ we obtain in this way the groups $O_N,U_N$.
\end{proposition}

\begin{proof}
The fact that $G(X)$ as defined above is indeed a group is clear, its compactness is clear as well, and finally the last assertion is clear as well. In fact, all this works for any closed subset $X\subset\mathbb C^N$, but we are not interested here in such general spaces.
\end{proof}

Observe that in the case of the real and complex spheres, the affine isometry group $G(X)$ leaves invariant the Riemannian metric, because this metric is equivalent to the one inherited from $\mathbb C^N$, which is preserved by our isometries $U\in U_N$. 

\bigskip

Thus, we could have constructed as well $G(X)$ as being the group of metric isometries of $X$, with of course some extra care in relation with the complex structure, as for the complex sphere $X=S^{N-1}_\mathbb C$ to produce $G(X)=U_N$ instead of $G(X)=O_{2N}$. But, such things won't really work for the free spheres, and so are to be avoided. 

\bigskip

The point now is that we have the following quantum analogue of Proposition 16.27, which is a perfect analogue, save for the fact that $X$ is now assumed to be algebraic, for some technical reasons, which allows us to talk about quantum isometry groups:

\begin{theorem}
Given an algebraic manifold $X\subset S^{N-1}_{\mathbb C,+}$, the category of the closed subgroups $G\subset U_N^+$ acting affinely on $X$, in the sense that the formula
$$\Phi(x_i)=\sum_jx_j\otimes u_{ji}$$ 
defines a morphism of $C^*$-algebras $\Phi:C(X)\to C(X)\otimes C(G)$, has a universal object, denoted $G^+(X)$, and called affine quantum isometry group of $X$.
\end{theorem}

\begin{proof}
Assume indeed that our manifold $X\subset S^{N-1}_{\mathbb C,+}$ comes as follows:
$$C(X)=C(S^{N-1}_{\mathbb C,+})\Big/\Big<f_\alpha(x_1,\ldots,x_N)=0\Big>$$

In order to prove the result, consider the following variables:
$$X_i=\sum_jx_j\otimes u_{ji}\in C(X)\otimes C(U_N^+)$$

Our claim is that the quantum group in the statement $G=G^+(X)$ appears as:
$$C(G)=C(U_N^+)\Big/\Big<f_\alpha(X_1,\ldots,X_N)=0\Big>$$

In order to prove this, pick one of the defining polynomials, and write it as follows:
$$f_\alpha(x_1,\ldots,x_N)=\sum_r\sum_{i_1^r\ldots i_{s_r}^r}\lambda_r\cdot x_{i_1^r}\ldots x_{i_{s_r}^r}$$

With $X_i=\sum_jx_j\otimes u_{ji}$ as above, we have the following formula:
$$f_\alpha(X_1,\ldots,X_N)=\sum_r\sum_{i_1^r\ldots i_{s_r}^r}\lambda_r\sum_{j_1^r\ldots j_{s_r}^r}x_{j_1^r}\ldots x_{j_{s_r}^r}\otimes u_{j_1^ri_1^r}\ldots u_{j_{s_r}^ri_{s_r}^r}$$

Since the variables on the right span a certain finite dimensional space, the relations $f_\alpha(X_1,\ldots,X_N)=0$ correspond to certain relations between the variables $u_{ij}$. Thus, we have indeed a closed subspace $G\subset U_N^+$, with a universal map, as follows:
$$\Phi:C(X)\to C(X)\otimes C(G)$$

In order to show now that $G$ is a quantum group, consider the following elements:
$$u_{ij}^\Delta=\sum_ku_{ik}\otimes u_{kj}\quad,\quad u_{ij}^\varepsilon=\delta_{ij}\quad,\quad u_{ij}^S=u_{ji}^*$$

Consider as well the following elements, with $\gamma\in\{\Delta,\varepsilon,S\}$:
$$X_i^\gamma=\sum_jx_j\otimes u_{ji}^\gamma$$

From the relations $f_\alpha(X_1,\ldots,X_N)=0$ we deduce that we have:
$$f_\alpha(X_1^\gamma,\ldots,X_N^\gamma)
=(id\otimes\gamma)f_\alpha(X_1,\ldots,X_N)
=0$$

Thus we can map $u_{ij}\to u_{ij}^\gamma$ for any $\gamma\in\{\Delta,\varepsilon,S\}$, and we are done.
\end{proof}

We can now formulate a result about spheres and rotations, as follows:

\index{quantum isometry group}

\begin{theorem}
The quantum isometry groups of the basic spheres are
$$\xymatrix@R=15mm@C=14mm{
S^{N-1}_{\mathbb R,+}\ar[r]&S^{N-1}_{\mathbb C,+}\\
S^{N-1}_\mathbb R\ar[r]\ar[u]&S^{N-1}_\mathbb C\ar[u]
}
\qquad
\xymatrix@R=8mm@C=15mm{\\ \to}
\qquad
\xymatrix@R=16mm@C=18mm{
O_N^+\ar[r]&U_N^+\\
O_N\ar[r]\ar[u]&U_N\ar[u]}$$
modulo identifying, as usual, the various $C^*$-algebraic completions.
\end{theorem}

\begin{proof}
We have 4 results to be proved, the idea being as follows:

\medskip

\underline{$S^{N-1}_{\mathbb C,+}$}. Let us first construct an action $U_N^+\curvearrowright S^{N-1}_{\mathbb C,+}$. We must prove here that the variables $X_i=\sum_jx_j\otimes u_{ji}$ satisfy the defining relations for $S^{N-1}_{\mathbb C,+}$, namely:
$$\sum_ix_ix_i^*=\sum_ix_i^*x_i=1$$

By using the biunitarity of $u$, we have the following computation:
$$\sum_iX_iX_i^*
=\sum_{ijk}x_jx_k^*\otimes u_{ji}u_{ki}^*
=\sum_jx_jx_j^*\otimes1
=1\otimes1$$

Once again by using the biunitarity of $u$, we have as well:
$$\sum_iX_i^*X_i
=\sum_{ijk}x_j^*x_k\otimes u_{ji}^*u_{ki}
=\sum_jx_j^*x_j\otimes1
=1\otimes1$$

Thus we have an action $U_N^+\curvearrowright S^{N-1}_{\mathbb C,+}$, which gives $G^+(S^{N-1}_{\mathbb C,+})=U_N^+$, as desired. 

\medskip

\underline{$S^{N-1}_{\mathbb R,+}$}. Let us first construct an action $O_N^+\curvearrowright S^{N-1}_{\mathbb R,+}$. We already know that the variables $X_i=\sum_jx_j\otimes u_{ji}$ satisfy the defining relations for $S^{N-1}_{\mathbb C,+}$, so we just have to check that these variables are self-adjoint. But this is clear from $u=\bar{u}$, as follows:
$$X_i^*
=\sum_jx_j^*\otimes u_{ji}^*
=\sum_jx_j\otimes u_{ji}
=X_i$$

Conversely, assume that we have an action $G\curvearrowright S^{N-1}_{\mathbb R,+}$, with $G\subset U_N^+$. The variables $X_i=\sum_jx_j\otimes u_{ji}$ must be then self-adjoint, and the above computation shows that we must have $u=\bar{u}$. Thus our quantum group must satisfy $G\subset O_N^+$, as desired.

\medskip

\underline{$S^{N-1}_\mathbb C$}. The fact that we have an action $U_N\curvearrowright S^{N-1}_\mathbb C$ is clear. Conversely, assume that we have an action $G\curvearrowright S^{N-1}_\mathbb C$, with $G\subset U_N^+$. In order to prove $G\subset U_N$, we will use:
$$\Phi(x_i)=\sum_jx_j\otimes u_{ji}$$

Indeed, by multiplying this formula with itself, we obtain:
$$\Phi(x_ix_k)=\sum_{jl}x_jx_l\otimes u_{ji}u_{lk}$$
$$\Phi(x_kx_i)=\sum_{jl}x_lx_j\otimes u_{lk}u_{ji}$$

Since the variables $x_i$ commute, these formulae can be written as:
$$\Phi(x_ix_k)=\sum_{j<l}x_jx_l\otimes(u_{ji}u_{lk}+u_{li}u_{jk})+\sum_jx_j^2\otimes u_{ji}u_{jk}$$
$$\Phi(x_ix_k)=\sum_{j<l}x_jx_l\otimes(u_{lk}u_{ji}+u_{jk}u_{li})+\sum_jx_j^2\otimes u_{jk}u_{ji}$$

Since the tensors at left are linearly independent, we must have:
$$u_{ji}u_{lk}+u_{li}u_{jk}=u_{lk}u_{ji}+u_{jk}u_{li}$$

By applying the antipode to this formula, then applying the involution, and then relabelling the indices, we succesively obtain:
$$u_{kl}^*u_{ij}^*+u_{kj}^*u_{il}^*=u_{ij}^*u_{kl}^*+u_{il}^*u_{kj}^*$$
$$u_{ij}u_{kl}+u_{il}u_{kj}=u_{kl}u_{ij}+u_{kj}u_{il}$$
$$u_{ji}u_{lk}+u_{jk}u_{li}=u_{lk}u_{ji}+u_{li}u_{jk}$$

Now by comparing with the original formula, we obtain from this:
$$u_{li}u_{jk}=u_{jk}u_{li}$$

In order to finish, it remains to prove that the coordinates $u_{ij}$ commute as well with their adjoints. For this purpose, we use a similar method. We have:
$$\Phi(x_ix_k^*)=\sum_{jl}x_jx_l^*\otimes u_{ji}u_{lk}^*$$
$$\Phi(x_k^*x_i)=\sum_{jl}x_l^*x_j\otimes u_{lk}^*u_{ji}$$

Since the variables on the left are equal, we deduce from this that we have:
$$\sum_{jl}x_jx_l^*\otimes u_{ji}u_{lk}^*=\sum_{jl}x_jx_l^*\otimes u_{lk}^*u_{ji}$$

Thus we have $u_{ji}u_{lk}^*=u_{lk}^*u_{ji}$, and so $G\subset U_N$, as claimed.

\medskip

\underline{$S^{N-1}_\mathbb R$}. The fact that we have an action $O_N\curvearrowright S^{N-1}_\mathbb R$ is clear. In what regards the converse, this follows by combining the results that we already have, as follows:
\begin{eqnarray*}
G\curvearrowright S^{N-1}_\mathbb R
&\implies&G\curvearrowright S^{N-1}_{\mathbb R,+},S^{N-1}_\mathbb C\\
&\implies&G\subset O_N^+,U_N\\
&\implies&G\subset O_N^+\cap U_N=O_N 
\end{eqnarray*}

Thus, we conclude that we have $G^+(S^{N-1}_\mathbb R)=O_N$, as desired.
\end{proof}

There are many other things, more specialized, that can be said in relation with Theorem 16.29, and for more on all this, we refer to \cite{bgo} and related papers.

\section*{16d. Fine structure}

Let us discuss now the correspondence $U\to S$. In the classical case the situation is very simple, because the sphere $S=S^{N-1}$ appears by rotating the point $x=(1,0,\ldots,0)$ by the isometries in $U=U_N$. Moreover, the stabilizer of this action is the subgroup $U_{N-1}\subset U_N$ acting on the last $N-1$ coordinates, and so the sphere $S=S^{N-1}$ appears from the corresponding rotation group $U=U_N$ as an homogeneous space, as follows:
$$S^{N-1}=U_N/U_{N-1}$$

In functional analytic terms, all this becomes even simpler, the correspondence $U\to S$ being obtained, at the level of algebras of functions, as follows:
$$C(S^{N-1})\subset C(U_N)\quad,\quad 
x_i\to u_{1i}$$

In general now, the straightforward homogeneous space interpretation of $S$ as above fails. However, we can have some theory going by using the functional analytic viewpoint, with an embedding $x_i\to u_{1i}$ as above. Let us start with the following result:

\begin{proposition}
For the basic spheres, we have a diagram as follows,
$$\xymatrix@R=50pt@C=50pt{
C(S)\ar[r]^\Phi\ar[d]^\alpha&C(S)\otimes C(U)\ar[d]^{\alpha\otimes id}\\
C(U)\ar[r]^\Delta&C(U)\otimes C(U)
}$$
where on top $\Phi(x_i)=\sum_jx_j\otimes u_{ji}$, and on the left $\alpha(x_i)=u_{1i}$.
\end{proposition}

\begin{proof}
The diagram in the statement commutes indeed on the standard coordinates, the corresponding arrows being as follows, on these coordinates:
$$\xymatrix@R=50pt@C=50pt{
x_i\ar[r]\ar[d]&\sum_jx_j\otimes u_{ji}\ar[d]\\
u_{1i}\ar[r]&\sum_ju_{1j}\otimes u_{ji}
}$$

Thus by linearity and multiplicativity, the whole the diagram commutes.
\end{proof}

As a consequence of the above result, we can now formulate:

\begin{proposition}
We have a quotient map and an inclusion as follows,
$$U\to S_U\subset S$$
with $S_U$ being the first row space of $U$, given by 
$$C(S_U)=<u_{1i}>\subset C(U)$$
at the level of the corresponding algebras of functions.
\end{proposition}

\begin{proof}
At the algebra level, we have an inclusion and a quotient map as follows:
$$C(S)\to C(S_U)\subset C(U)$$

Thus, we obtain the result, by transposing.
\end{proof}

The above result is all that we need, for getting started with our study, and we will prove in what follows that the inclusion $S_U\subset S$ constructed above is an isomorphism. This will produce the correspondence $U\to S$ that we are currently looking for.

\bigskip

In order to do so, we will use the uniform integration over $S$, which can be introduced, in analogy with what happens in the classical case, in the following way:

\begin{definition}
We endow each of the algebras $C(S)$ with its integration functional
$$\int_S:C(S)\to C(U)\to\mathbb C$$
obtained by composing the morphism $x_i\to u_{1i}$ with the Haar integration of $C(U)$.
\end{definition}

With this done, in order now to efficiently integrate over our various spheres $S$, and in the lack of some trick like spherical coordinates, we need to know how to efficiently integrate over the corresponding quantum isometry groups $U$. 

\bigskip

Now regarding this latter question, we already have some good experience with it, in the classical case, from chapter 14. And the point is that the material there extends without much difficulties in the free case, and we have the following result:

\begin{theorem}
Assuming that a compact quantum group $G\subset U_N^+$ is easy, coming from a category of partitions $D\subset P$, we have the Weingarten formula
$$\int_Gu_{i_1j_1}^{e_1}\ldots u_{i_kj_k}^{e_k}=\sum_{\pi,\sigma\in D(k)}\delta_\pi(i)\delta_\sigma(j)W_{kN}(\pi,\sigma)$$
for any indices $i_r,j_r\in\{1,\ldots,N\}$ and any exponents $e_r\in\{\emptyset,*\}$, where $\delta$ are the usual Kronecker type symbols, and where 
$$W_{kN}=G_{kN}^{-1}$$
is the inverse of the matrix $G_{kN}(\pi,\sigma)=N^{|\pi\vee\sigma|}$.
\end{theorem}

\begin{proof}
Let us arrange indeed all the integrals to be computed, at a fixed value of the exponent $k=(e_1\ldots e_k)$, into a single matrix, of size $N^k\times N^k$, as follows:
$$P_{i_1\ldots i_k,j_1\ldots j_k}=\int_Gu_{i_1j_1}^{e_1}\ldots u_{i_kj_k}^{e_k}$$

According to the construction of the Haar measure of Woronowicz, explained in the above, this matrix $P$ is the orthogonal projection onto the following space:
$$Fix(u^{\otimes k})=span\left(\xi_\pi\Big|\pi\in D(k)\right)$$ 

In order to compute this projection, consider the following linear map:
$$E(x)=\sum_{\pi\in D(k)}<x,\xi_\pi>\xi_\pi$$

Consider as well the inverse $W$ of the restriction of $E$ to the following space:
$$span\left(T_\pi\Big|\pi\in D(k)\right)$$

By a standard linear algebra computation, it follows that we have:
$$P=WE$$

But the restriction of $E$ is the linear map corresponding to $G_{kN}$, so $W$ is the linear map corresponding to $W_{kN}$, and this gives the result. See \cite{bco}, \cite{bsp}.
\end{proof}

With this in hand, we can now integrate over the spheres $S$, as follows:
 
\begin{theorem}
The integration over the basic spheres is given by
$$\int_Sx_{i_1}^{e_1}\ldots x_{i_k}^{e_k}=\sum_\pi\sum_{\sigma\leq\ker i}W_{kN}(\pi,\sigma)$$
with $\pi,\sigma\in D(k)$, where $W_{kN}=G_{kN}^{-1}$ is the inverse of $G_{kN}(\pi,\sigma)=N^{|\pi\vee\sigma|}$. 
\end{theorem}

\begin{proof}
According to our conventions, the integration over $S$ is a particular case of the integration over $U$, via $x_i=u_{1i}$. By using now Theorem 16.33, we obtain:
\begin{eqnarray*}
\int_Sx_{i_1}^{e_1}\ldots x_{i_k}^{e_k}
&=&\int_Uu_{1i_1}^{e_1}\ldots u_{1i_k}^{e_k}\\
&=&\sum_{\pi,\sigma\in D(k)}\delta_\pi(1)\delta_\sigma(i)W_{kN}(\pi,\sigma)\\
&=&\sum_{\pi,\sigma\in D(k)}\delta_\sigma(i)W_{kN}(\pi,\sigma)
\end{eqnarray*}

Thus, we are led to the formula in the statement.
\end{proof}

Again with some inspiration from the classical case, we have the following key result:

\begin{theorem}
The integration functional of $S$ has the ergodicity property
$$\left(id\otimes\int_U\right)\Phi(x)=\int_Sx$$
where $\Phi:C(S)\to C(S)\otimes C(U)$ is the universal affine coaction map.
\end{theorem}

\begin{proof}
In the real case, $x_i=x_i^*$, it is enough to check the equality in the statement on an arbitrary product of coordinates, $x_{i_1}\ldots x_{i_k}$. The left term is as follows:
\begin{eqnarray*}
\left(id\otimes\int_U\right)\Phi(x_{i_1}\ldots x_{i_k})
&=&\sum_{j_1\ldots j_k}x_{j_1}\ldots x_{j_k}\int_Uu_{j_1i_1}\ldots u_{j_ki_k}\\
&=&\sum_{j_1\ldots j_k}\ \sum_{\pi,\sigma\in D(k)}\delta_\pi(j)\delta_\sigma(i)W_{kN}(\pi,\sigma)x_{j_1}\ldots x_{j_k}\\
&=&\sum_{\pi,\sigma\in D(k)}\delta_\sigma(i)W_{kN}(\pi,\sigma)\sum_{j_1\ldots j_k}\delta_\pi(j)x_{j_1}\ldots x_{j_k}
\end{eqnarray*}

Let us look now at the last sum on the right. The situation is as follows:

\medskip

-- In the free case we have to sum quantities of type $x_{j_1}\ldots x_{j_k}$, over all choices of multi-indices $j=(j_1,\ldots,j_k)$ which fit into our given noncrossing pairing $\pi$, and just by using the condition $\sum_ix_i^2=1$, we conclude that the sum is 1. 

\medskip

-- The same happens in the classical case. Indeed, our pairing $\pi$ can now be crossing, but we can use the commutation relations $x_ix_j=x_jx_i$, and the sum is again 1.

\medskip

Thus the sum on the right is 1, in all cases, and we obtain:
$$\left(id\otimes\int_U\right)\Phi(x_{i_1}\ldots x_{i_k})
=\sum_{\pi,\sigma\in D(k)}\delta_\sigma(i)W_{kN}(\pi,\sigma)$$

On the other hand, another application of the Weingarten formula gives:
\begin{eqnarray*}
\int_Sx_{i_1}\ldots x_{i_k}
&=&\int_Uu_{1i_1}\ldots u_{1i_k}\\
&=&\sum_{\pi,\sigma\in D(k)}\delta_\pi(1)\delta_\sigma(i)W_{kN}(\pi,\sigma)\\
&=&\sum_{\pi,\sigma\in D(k)}\delta_\sigma(i)W_{kN}(\pi,\sigma)
\end{eqnarray*}

Thus, we are done with the proof of the result, in the real case. In the complex case the proof is similar, by adding exponents everywhere.
\end{proof} 

We can now deduce a useful characterization of the integration, as follows:

\begin{theorem}
There is a unique positive unital trace $tr:C(S)\to\mathbb C$ satisfying
$$(tr\otimes id)\Phi(x)=tr(x)1$$
where $\Phi$ is the coaction map of the corresponding quantum isometry group,
$$\Phi:C(S)\to C(S)\otimes C(U)$$
and this is the canonical integration, as constructed in Definition 16.32.
\end{theorem}

\begin{proof}
First of all, it follows from the Haar integral invariance condition for $U$ that the canonical integration has indeed the invariance property in the statement, namely:
$$(tr\otimes id)\Phi(x)=tr(x)1$$

In order to prove now the uniqueness, let $tr$ be as in the statement. We have:
\begin{eqnarray*}
tr\left(id\otimes\int_U\right)\Phi(x)
&=&\int_U(tr\otimes id)\Phi(x)\\
&=&\int_U(tr(x)1)\\
&=&tr(x)
\end{eqnarray*}

On the other hand, according to Theorem 16.35, we have as well:
$$
tr\left(id\otimes\int_U\right)\Phi(x)
=tr\left(\int_Sx\right)
=\int_Sx$$

We therefore conclude that $tr$ equals the standard integration, as claimed.
\end{proof}

Getting now to probabilistic aspects, we know from chapter 7 that for the classical spheres the laws of the standard coordinates, which are called hyperspherical laws, become normal and independent, with $N\to\infty$. Here is the free analogue of this result:

\begin{theorem}
For the free spheres $S^{N-1}_{\mathbb R,+}$, $S^{N-1}_{\mathbb C,+}$, the rescaled coordinates 
$$y_i=\sqrt{N}x_i$$
become semicircular/circular and free, in the $N\to\infty$ limit.
\end{theorem}

\begin{proof}
In the real case, the Weingarten formula from Theorem 16.34, coupled with the standard fact that the Gram and Weingarten matrices are asymptotically diagonal, $G_{kN}\simeq diag(N^{k/2})$ and $W_{kN}\simeq diag(N^{-k/2})$, gives the following estimate:
$$\int_{S^{N-1}_{\mathbb R,+}}x_{i_1}\ldots x_{i_k}
=\sum_{\pi,\nu\in NC_2(k)}\delta_\nu(i)W_{kN}(\pi,\nu)
\simeq N^{-k/2}\sum_{\nu\in NC_2(k)}\delta_\nu(i)$$

With this formula in hand, we can compute the asymptotic moments of each coordinate $x_i$. Indeed, by setting $i_1=\ldots=i_k=i$, all Kronecker symbols are 1, and we obtain:
$$\int_{S^{N-1}_{\mathbb R,+}}x_i^k\,dx\simeq N^{-k/2}|NC_2(k)|=N^{-k/2}M_k(\gamma_1)$$

Thus the rescaled coordinates $y_i=\sqrt{N}x_i$ become semicircular in the $N\to\infty$ limit, and the asymptotic freeness result follows as well, from the above general joint moment estimate. As for the proof in the complex case, this is similar. See \cite{bgo}.
\end{proof}

Very nice all this, all the above is quite encouraging, we definitely have a beginning of free geometry theory here. It is possible to go much further, along these lines, and certain mathematicians have done indeed so. However, before explaining all this, we should perhaps doublecheck with the cats that all this is indeed relevant to quantum physics.

\bigskip

Unfortunately the cats are gone, so I will leave it to you, reader, clarifying such questions. And in case you believe it is so, do not hesitate to have a look at the free geometry, analysis, probability and so on literature, that might help you, in your quest.

\section*{16e. Exercises}

Congratulations for having read this book, and no exercises for this final chapter. But, of course, with the hope to hear from you sometimes soon, with some heavy integration theorems of your own. The one who knows how to integrate wins.

\baselineskip=14pt

\printindex

\end{document}